\documentclass[12pt,twoside,openright,a4paper]{book}

\usepackage{amsmath, amssymb, amsthm}
\usepackage{graphicx,color}
\usepackage{multirow}
\usepackage[left=1in, right=0.75in, top=1in, bottom=1in]{geometry}
\usepackage{epsfig}
\usepackage{color}
\usepackage{setspace}
\usepackage{latexsym}
\usepackage{mathrsfs}
\usepackage{graphicx}
\usepackage{tikz}
\usepackage[titletoc]{appendix}
\usepackage{alltt}
\graphicspath{ {images/} }
\usepackage{longtable}
\usepackage{tikz}
\usepackage{acronym}
\usepackage{pgfplots}
\pgfplotsset{compat=newest}
\usepackage{csquotes}
\usepackage[hidelinks]{hyperref}
\usepackage{titletoc,tocloft}
\usepackage{longtable}
\usepackage{subcaption}
\usepackage{gensymb}
\usepackage{epigraph} 
\usepackage{tabularx} 
\usepackage{anyfontsize}
\usepackage{pdfpages}

\usepackage{times}

\usepackage{sectsty}
\usepackage{titlesec}

\chapterfont{\centering }
\titleformat{\chapter}[display]
{\normalfont%
	\huge
	\bfseries}{\chaptertitlename\ \thechapter}{14pt}{%
	\Huge 
}

\usepackage{textcase}

\sectionfont{\normalfont}
\titleformat{\section}
{\normalfont\fontsize{14pt}{16pt}\selectfont}{\thesection}{1em}{}
\titleformat{\subsection}
{\normalfont\fontsize{12pt}{14pt}\selectfont}{\thesubsection}{1em}{}
\titleformat{\subsubsection}
{\normalfont\fontsize{12pt}{14pt}\selectfont}{\thesubsubsection}{1em}{}
\usepackage{caption}
\newtheorem{definition}{Definition}
\newcommand{\brdefinition}{\begin{definition}}
	\newcommand{\erdefinition}{\end{definition}}

\newtheorem{corollary}{Corollary}
\newcommand{\bcorollary}{\begin{corollary}}
	\newcommand{\ecorollary}{\end{corollary}}
\newtheorem{example}{Example}
\newcommand{\bexample}{\begin{example}}
	\newcommand{\eexample}{\end{example}}
\newtheorem{remark}{Remark}
\newcommand{\bremark}{\begin{remark}}
	\newcommand{\eremark}{\end{remark}}

\newtheorem{theorem}{Theorem}
\newcommand{\btheorem}{\begin{theorem}}
	\newcommand{\etheorem}{\end{theorem}}
\newtheorem{lemma}{Lemma}
\newcommand{\blemma}{\begin{lemma}}
	\newcommand{\elemma}{\end{lemma}}

\newcommand{\brdef}{\begin{defi}}
	\newcommand{\erdef}{\end{defi}}
\newcommand{\bcor}{\begin{cor}}
	\newcommand{\ecor}{\end{cor}}
\newcommand{\bth}{\begin{thm}}
	\newcommand{\ble}{\begin{lem}}
		\newcommand{\ele}{\end{lem}}
	\newcommand{\bcha}{\end{cha}}

\renewcommand{\thechapter}{\arabic{chapter}}
\renewcommand{\thesection}{\thechapter.\arabic{section}}

\renewcommand{\&}{and}

\theoremstyle{definition}

\usepackage{sectsty}
\usepackage{titlesec}
\chapterfont{\centering }
\titleformat{\chapter}[display]
{\bf\centering}
{\chaptertitlename\ \thechapter}{16pt}{\large}
\sectionfont{\normalfont}
\cftsetindents{chapter}{0em}{3em}      
\cftsetindents{section}{0em}{3em}

\usepackage{natbib}

\usepackage[acronym,nomain,nonumberlist, toc=false,section=section, nogroupskip]{glossaries}
\makeglossaries

\definecolor{butter1}{rgb}{0.988,0.914,0.310}
\definecolor{chocolate1}{rgb}{0.914,0.725,0.431}
\definecolor{chameleon1}{rgb}{0.541,0.886,0.204}
\definecolor{skyblue1}{rgb}{0.447,0.624,0.812}
\definecolor{plum1}{rgb}{0.678,0.498,0.659}
\definecolor{scarletred1}{rgb}{0.937,0.161,0.161}

\usepackage{array}
\newcolumntype{L}[1]{>{\raggedright\let\newline\\\arraybackslash\hspace{0pt}}m{#1}}
\newcolumntype{C}[1]{>{\centering\let\newline\\\arraybackslash\hspace{0pt}}m{#1}}
\newcolumntype{R}[1]{>{\raggedleft\let\newline\\\arraybackslash\hspace{0pt}}m{#1}}

\usepackage{etoolbox}

\makeatletter
\patchcmd{\ttlh@hang}{\parindent\z@}{\parindent\z@\leavevmode}{}{}
\patchcmd{\ttlh@hang}{\noindent}{}{}{}
\makeatother

\newenvironment{abstract}{\rightskip1in\itshape}{}

\newcommand{\dedicatoriaUno}[1]{
\def\dedicatoriaUnoVal{#1}
}

\newcommand{\putDedicatoria}[1]{
\ifx\marcadorDedicatorias\undefined
\ifpdf
   \pdfbookmark{Dedicatoria}{dedicatoria}
\fi
\def\marcadorDedicatorias{1}
\fi
\thispagestyle{empty}\mbox{}
\vspace*{4cm}
\begin{flushright}
#1
\end{flushright}
\newpage
\thispagestyle{empty}\mbox{}
\newpage
} 

\newcommand{\makeDedicatorias}{
\ifx\dedicatoriaUnoVal\undefined
\else
\putDedicatoria{\dedicatoriaUnoVal}
\fi
}

\newcommand{\jplus}{J-PLUS}

\newcommand{\jp}{J-PAS}

\newcommand{\js}{J-spectra}
\newcommand{\mjp}{miniJPAS}

\def\photozbest{\texttt{PHOTOZ}}

\def\magauto{\texttt{MAG\_AUTO}}

\def\magpetro{\texttt{MAG\_PETRO}}
\def\magpsfcor{\texttt{MAG\_PSFCOR}}

\def\kron{\texttt{KRON\_RADIUS}}
\def\petror{\texttt{PETRO\_RADIUS}}
\def\infot{\texttt{info\_table}}

\def\filterst{\texttt{filters\_table}}

\def\baysea{\texttt{BaySeAGal}} 
\def\muff{\texttt{MUFFIT}} 
\def\alstar{\texttt{AlStar}} 
\def\tgas{\texttt{TGASPEX}}

\def\lephare{{\sc LePhare}}
\newcommand{\pycasso}{{\sc p}y{\sc casso}}          	
\newcommand{\starlight}{{\sc starlight}}          	

\def\Py{\texttt{Python}}
\def\PyDJ{\texttt{Py2DJPAS}}
\def\Jcube{\texttt{J2DCube}}

\def\sext{\texttt{SExtractor}}

\def\mangle{\texttt{mangle}}

\def\rb{$r_\mathrm{SDSS}$}
\def\gb{$g_\mathrm{SDSS}$}
\def\ib{$i_\mathrm{SDSS}$}
\def\uja{$u_\mathrm{JAVA}$}
\def\ujp{$u_\mathrm{JPAS}$}

\newcommand\ewha{$EW(\mathrm{H}\alpha)$}
\newcommand\logMt{\log M_\star}
\newcommand\logM{$\logMt$}

\def\infot{\texttt{info\_table}}

\newacronym{ADQL}{ADQL}{Astronomical Data Query Language}
\newacronym{ADU}{ADU}{Astronomical Digital Unit}
\newacronym{AGN}{AGN}{Active Galactic Nuclei}
\newacronym{ALHAMBRA}{ALHAMBRA}{Advance Large Homogeneous Area Medium Band Redshift Astronomical}
\newacronym{AMICO}{AMICO}{Adaptive Matched Identifier of Clustered Objects}
\newacronym{ANN}{ANN}{Artificial Neural Network}
\newacronym{BatMAN}{\texttt{BatMAN}}{Bayesian Technique for Multi-image Analysis}
\newacronym{CALIFA}{CALIFA}{Calar Alto Legacy Integral Field Area}
\newacronym{CCD}{CCD}{charge-coupled device}
\newacronym{CEFCA}{CEFCA}{Centro de Estudios de Física del Cosmos de Aragón}
\newacronym{COMBO-17}{COMBO-17}{Classifying Objects by Medium-Band Observations}
\newacronym{COSMOS}{COSMOS}{Cosmic Evolution Survey}
\newacronym{DES}{DES}{Dark Energy Survey}
\newacronym{ELG}{ELG}{Emission Line Galaxy}
\newacronym{EW}{EW}{Equivalent Width}
\newacronym{FoV}{FoV}{Field of View}
\newacronym{FWHM}{FWHM}{Full Width Half Maximum}
\newacronym{GEMS}{GEMS}{Galaxy Evolution from Morphology and SEDS}
\newacronym{HMR}{HMR}{Half Mass Radius}
\newacronym{HLR}{HLR}{Half Light Radius}
\newacronym{HSC-SSP}{HSC-SSP}{Hyper Suprime-Cam Subaru Strategic Program}
\newacronym{IFS}{IFS}{Integral Field Spectroscopy}
\newacronym{IFU}{IFU}{Integral Field Unit}
\newacronym{ICM}{ICM}{Intra Cluster Medium}
\newacronym{IGM}{IGM}{Inter Galactic Medium}
\newacronym{IMF}{IMF}{Initial Mass Function}
\newacronym{ISM}{ISM}{Inter Stellar Medium}
\newacronym{JPAS}{J-PAS}{Javalambre-Physics of the Accelerated Universe Astrophysical Survey}
\newacronym{JPAS-PF}{JPAS-PF}{J-PAS Pathfinder}
\newacronym{J-PLUS}{J-PLUS}{Javalambre Photometric Local Universe Survey}
\newacronym{MaNGA}{MaNGA}{Mapping Nearby Galaxies at Apache Point Observatory}
\newacronym{MCMC}{MCMC}{Markov chain Monte Carlo}
\newacronym{OAJ}{OAJ}{Observatorio Astofísico de Javalambre}
\newacronym{Pan-STARRS1}{Pan-STARRS1}{Panoramic Survey Telescope and Rapid Response System 1}
\newacronym{PMAS}{PMAS}{Potsdam Multi Aperture Spectrograph}
\newacronym{PDF}{PDF}{Probability Density Function}
\newacronym{PSF}{PSF}{Point Spread Function}
\newacronym{QFE}{QFE}{Quenched Fraction Excess}
\newacronym{QSO}{QSO}{Quasi Stellar Object}
\newacronym{ReLU}{ReLU}{Rectified Linear Unit}
\newacronym{SAURON}{SAURON}{Spectroscopic Areal Unit for Research on Optical Nebulae}
\newacronym{SDSS}{SDSS}{Sloan Digital Sky Survey}
\newacronym{SED}{SED}{Spectral Energy Distribution}
\newacronym{SFMS}{SFMS}{Star Formation Main Sequence}
\newacronym{SFH}{SFH}{Star Formation History}
\newacronym{SFR}{SFR}{Star Formation Rate}
\newacronym{S-PLUS}{S-PLUS}{Southern Photometric Local Universe Survey}
\newacronym{sSFR}{sSFR}{specific Star Formation Rate}
\newacronym{SSP}{SSP}{Single Stellar Population}
\newacronym{TAP}{TAP}{Table Access Protocol}
\newacronym{UPAD}{UPAD}{Unidad de Procesado y Archivo de Datos}
\newacronym{VIPERS}{VIPERS}{VIMOS Public Extragalactic Redshift Survey}
\newacronym{VO}{VO}{Virtual Observatory}
\newacronym{VUDS}{VUDS}{VIMOS Ultra Deep Survey}
\newacronym{VVDS}{VVDS}{VIMOS VLT Deep Survey}
\newacronym{WCS}{WCS}{World Coordinate System}
\newacronym{ZP}{ZP}{Zero Point}
\newacronym{zPDF}{$z\mathrm{PDF}$}{Probability density function of the redshift}

\newcommand*\aap{A\&A}

\newcommand*\ao{Appl.~Opt.}

\newcommand*\apjl{ApJ}

\newcommand*\apjs{ApJS}

\begin{document}
\emergencystretch 3em

\pagenumbering{gobble}

\onehalfspacing
\begin{center}
\thispagestyle{empty}

\vskip .6cm
~\\~\\~\\~
{\huge \textbf{The environment as a driver of galaxy evolution with the miniJPAS survey}}
\vskip 3.75 cm
{\it \Large Thesis submitted for the degree of Doctor in Philosophy by }
\vskip 0.5 cm

\textbf{\Large Julio Esteban Rodríguez Martín}
\vskip 0.75 cm

{\it  Supervised by }
\vskip 0.5 cm

\textbf{\Large Rosa M. González Delgado and Luis A. Díaz García}
\vskip 1.25 cm

\centerline{\includegraphics[height=39mm,width=129mm]{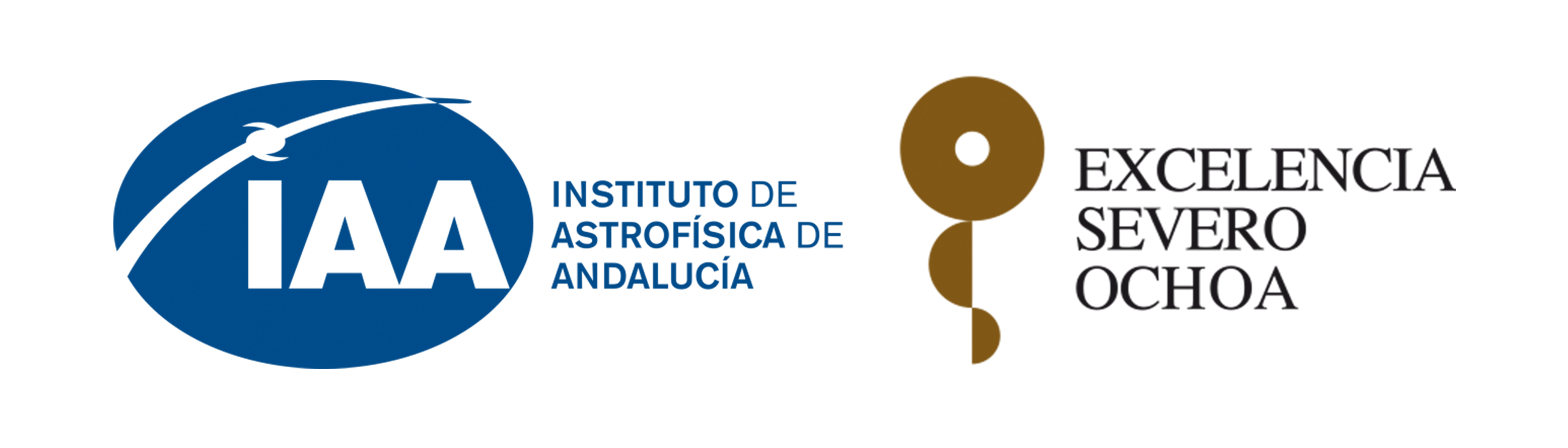}}
{\Large Instituto de Astrofísca de Andalucía, Consejo Superior de Investigaciones Científicas (IAA-CSIC)}

\centerline{\includegraphics[height=59mm,width=59mm]{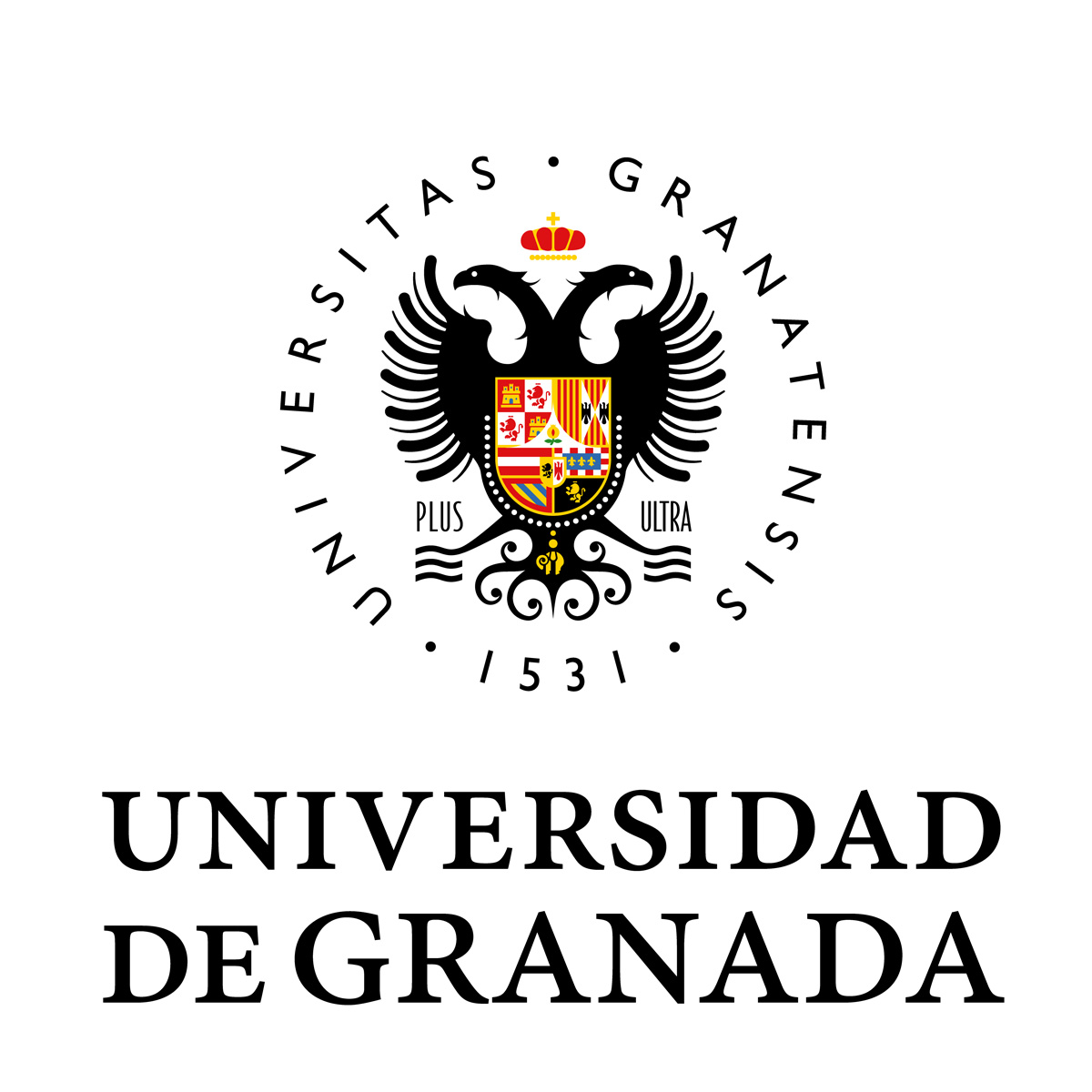}}
{\Large Programa de doctorado en Física y Matemáticas (FisyMat)}


\end{center}

\newpage
\thispagestyle{empty}\mbox{}
\newpage

\newpage
\thispagestyle{empty}\mbox{}
\newpage

\dedicatoriaUno{%
\emph{
A mi familia y amistades,\\
quienes nunca dejaron de creer en mi, \\
incluso cuando yo no lo hacía%
}%
}

\makeDedicatorias 

\newpage

\vspace*{5.5cm}
\epigraph{\textit{The beautiful thing about learning is nobody can take it away from you.}}{B.B. King}
\vspace{0.5cm}
\epigraph{\textit{Todo te lo pueden quitar en esta vida: familia, trabajo, dinero... menos la cultura}}{Antonio Esteban \\ Mi bisabuelo}

\newpage
\thispagestyle{empty}\mbox{}
\newpage

\newpage
\begin{spacing}{1.5}

\newpage
\newpage
\pagenumbering{roman}
\addcontentsline{toc}{section}{\bf ABSTRACT}
{\centerline { {\textbf{ABSTRACT}}}}
~\\

This thesis aims at discerning the effect of environment on galaxy evolution and on the properties of galaxies, which still remains under debate after decades. We use the data from the miniJPAS survey, a 1~deg$^2$ survey that uses the same photometric filter system as the incoming J-PAS survey, which is already in its scientific verification phase. This system is composed of 56 narrow-band filters that provide an spectral resolution comparable to low-resolution spectroscopy.

We study the effect of environment using two approaches. The first approach consists on the study of the properties of the spatially resolved galaxies from miniJPAS by taking advantage of its large field of view and photometric system, which brings the opportunity to perform unbiased IFU-like studies in different environments. Secondly, we studied the galaxy population in the most massive galaxy cluster detected in the miniJPAS footprint,  that is, the cluster mJPC2470--1771.

In order to study the properties of the spatially resolved galaxies, we have developed a tool that automatises all the required processes for the analysis of the data. These mainly comprise the download of scientific tables and images, the masking of nearby sources, PSF homogenisation of the miniJPAS images, the definition of galaxy regions, and the determination of the integrated fluxes and magnitudes of each region. With these magnitudes, the stellar population properties are constrained using BaySeAGal (an external and bayesian  fitting code for spectral energy distribution). The properties related to emission lines are estimated using an external algorithm based on artificial neural networks, which was specifically trained to work with data from the miniJPAS and J-PAS surveys.

After probing the accuracy and reliability of the photmetry obtained with our tool, we apply it to a miniJPAS sample of spatially resolved galaxies, divided into four sub-samples according to their spectral-type (red/quiescent and blue/star forming), and environment (field or galaxy group). With this classification we studied the stellar population properties and the emission lines of the regions of these spatially resolved galaxies, as well as the effect of environment. For this purpose, we compare the properties of the four categories of galaxies, using stellar surface mass density--colour diagrams, radial profiles of the properties, and stellar population gradients. In addition, we explore the relation between some of these properties and comparing the star formation history of inner and outer parts of the galaxies. We find that the properties of the regions of blue and red galaxies are well differentiated, but there we do not find any significant effect of the environment on them.  We find that redder, denser regions are usually older, more metal rich, and show lower values of the intensity of the star formation rate and the specific star formation rate than bluer and less dense regions. The higher  extinction values are found in blue, dense regions, as well as some of the most metal rich regions. 

Regarding the mJPC2470--1771 cluster, we  obtain their stellar population properties using BaySeAGal, and we select a sample of emission line galaxies using a contrast method along with the previous artificial neural networks. Our results point that more massive, redder and older galaxies typically populate the inner regions of the cluster. Most of the emission line galaxies are blue, star forming galaxies, which are more common in the cluster outskirts. Furthermore, active galactic nuclei host are also detected, with a greater presence in the central regions of the cluster. As a conclusion, our results suggest that  galaxies in clusters are formed at a similar epoch, but have experienced different star formation histories.

We conclude that the environment plays a role on galaxy evolution, but it is mainly reflected through the galaxy populations found in high density environments, such as galaxy clusters and groups, where the fraction of red, quiescent galaxies is larger in comparison to the field. On the other hand, the distribution of the properties of blue galaxies is shifted towards more massive, redder, and older values. However, at a fixed value of the stellar mass and colour, these properties are similar to their counterparts in the field. Similarly, the properties of the regions of the spatially resolved galaxies are well determined by their colour and stellar mass density, but the group environment is not dense enough as to show any significant difference on the properties of the regions of galaxies in groups when compared to galaxies in the field. In the spatially resolved case, this may be a consequence of the typical mass of the groups in our spatially resolved sample, since it might not be high enough to produce a significant effect on the properties of the galaxies, as opposed to the case of massive galaxy clusters. The  importance of the mass of the group or the cluster is observed for example in the quenched fraction excess, which is significantly larger in the cluster than in the low mass groups found in miniJPAS.

\newpage
\addcontentsline{toc}{section}{\bf RESUMEN}
{\centerline { {\textbf{RESUMEN}}}}
~\\

El objetivo de esta tesis doctoral es discernir el efecto del entorno en la evolución y las propiedades de las galaxias, que sigue siendo objeto de debate tras décadas de investigación. Para ello usamos los datos de miniJPAS, un sondeo de 1 grado cuadrado que utiliza el mismo sistema de filtros fotométricos que sondeo J-PAS, el cual se encuentra en su fase de verificación científica. Este sistema se compone de 56 filtros de banda estrecha que proporcionan una resolución espectral comparable a la espectroscopia de baja resolución.

Estudiamos el efecto del entorno mediante dos enfoques distintos. El primero consiste en el estudio de las propiedades espacialmente resueltas de las galaxias de miniJPAS, aprovechando su gran campo de visión y su sistema de filtros, que permiten realizar estudios de tipo IFU libres de sesgos en distintos entornos. Posteriormente, estudiamos las poblaciones de galaxias pertenecientes al cúmulo de galaxias más masivo detectado en el campo de miniJPAS, esto es, el cúmulo mJPC2470--1771.

Para poder estudiar las propiedades de las galaxias espacialmente resueltas hemos desarrollado un herramienta que automatiza todos los procesos requeridos para el análisis de los datos. Estos procesos se componen principalmente de la descarga de las tabla e imágenes científicas, el enmascaramiento de las fuentes cercanas, la homogeneización de PSF de las imágenes de miniJPAS, la segmentación de la galaxia en distintas regiones y el cálculo de los flujos y magnitudes integrados en dichas regiones. Con estas magnitudes, constreñimos las propiedades de las poblaciones estelares utilizando BaySeAGal (un código externo y bayesiano para el ajuste de la distribución de la energía espectral). Las propiedades relacionadas con las líneas de emisión son estimadas con un algoritmo externo basado en redes neuronales artificiales, entrenado específicamente para trabajar con los datos de miniJPAS y J-PAS. 

Tras demostrar la precisión y fiabilidad de la fotometría obtenida con nuestra herramienta, procedemos a aplicarla a la muestra de galaxias espacialmente resueltas en miniJPAS, divididas en cuatro subgrupos de acuerdo a su tipo espectral (rojas/pasivas y azules/con formación estelar), y entorno (campo o grupo de galaxias). Con esta clasificación, estudiamos las propiedades de las poblaciones estelares y de las líneas de emisión de las regiones de estas galaxias espacialmente resueltas, así como el efecto del entorno en ellas. Para tal propósito, comparamos las propiedades de las cuatro categorías de galaxias, usando diagramas de densidad de masa estelar--color, perfiles radiales de las propiedades y los gradientes de las propiedades de las poblaciones estelares. Adicionalmente, exploramos la relación entre algunas de estas propiedades y comparamos la historia de formación estelar de las partes internas y externas de las galaxias. Nuestros resultados muestran que las propiedades de las regiones de las galaxias azules y rojas están bien diferenciadas, pero no encontramos ningún efecto significativo del entorno en ellas. Encontramos que las regiones más rojas y más densas son en general más viejas, más metálicas y muestran valores más bajos de la intensidad de la formación estelar y  de la formación estelar específica en comparación a las regiones más azules y menos densas. Los valores más altos de la extinción se encuentran en regiones azules y densas, así como algunas de las regiones más metálicas.

En relación al cúmulo mJPC2470--1771, obtenemos las propiedades de las poblaciones estelares utilizando BaySeAGal, y seleccionamos una muestra de galaxias con líneas de emisión usando un método de contraste junto a las redes neuronales artificiales previamente mencionadas. Nuestros resultados indican que las galaxias, más masivas, rojas y viejas se encuentran típicamente en las regiones más internas del cúmulo. La mayoría de las galaxias con líneas de emisión son galaxias azules y con formación estelar, siendo más comunes en las zonas más externas del cúmulo. Además, detectamos las galaxias con núcleos activos, que se encuentran principalmente en la zona central del cúmulo. Como conclusión, nuestros resultados sugieren que las galaxias pertenecientes a los cúmulos se formaron en épocas cósmicas similares, pero han experimentado historias de formación estelar distintas. 

Concluimos que el entorno juega un papel en la evolución de galaxias, pero éste queda reflejado principalmente en las poblaciones de galaxias encontradas en entornos de alta densidad, tales como los grupos y cúmulos de galaxias, donde la fracción de galaxias rojas pasivas es mayor que la encontrada en el campo. Por otro lado, la distribución de las propiedades de las galaxias azules está desplazada hacia valores más masivos, rojos y viejos. Sin embargo, para un mismo valor de la masa y el color, las propiedades de estas galaxias son similares a sus homólogas en el campo. En el caso de las galaxias espacialmente resueltas, esto puede ser una consecuencia de la masa de los grupos en nuestra de galaxias espacialmente resueltas, ya que podría no ser lo suficientemente alta como para mostrar un efecto significativo en las propiedades de las galaxias, en contraposición a los cúmulos masivos de galaxias. La importancia de la masa del grupo o el cúmulo se observa por ejemplo en el exceso de la fracción de galaxias pasivas, que es significativamente más alto en el cúmulo que en los grupos de baja masa encontrado en miniJPAS.

\newpage
\addcontentsline{toc}{section}{\bf AGRADECIMIENTOS}
{\centerline { {\textbf{AGRADECIMIENTOS}}}}

\vspace*{1cm}

\indent\indent Una vez leí un sobrecillo de azúcar "Si quieres ir rápido ve solo, si quieres llegar lejos ve acompañado - Proverbio africano." Aunque fracasé en mi intento de encontrar un origen más concreto para este proverbio y probablemente la sabiduría contenida en los sobres de azúcar de para escribir otra tesis doctoral (no seré yo quien la haga), la frase me hizo pensar. Lo cierto es que no sé si habré llegado lejos, pero desde luego sé que si he llegado hasta donde estoy es porque he estado acompañado. De hecho, no sólo acompañado, si no muy bien acompañado. Y por eso quería parar un momento para dar las gracias. 

\indent\indent Empezando las cosas por el principio, al contrario de lo que a veces tiendo a hacer, muchas gracias Rosa, por guiarme, apoyarme y orientarme en este viaje, y muchas gracias por tu paciencia, pues sé que no ha sido una tarea fácil. Gracias Luis por ganarte a pulso el título de co-director, aguantándome hasta horas intempestivas con mis dudas, problemas y meteduras de pata. Espero que tu primera experiencia dirigiendo una tesis no te haya dejado sin ganas de más.

\indent \indent Siguiendo con nuestro grupo de investigación, gracias Rubén por estar siempre a ayudar y por esa chistera de mago llena de recursos que es tu disco duro. Gracias Ginés por nuestras conversaciones, que siempre aportaban un punto de vista distinto e interesante, y porque sabes que sin tu trabajo el mío no habría sido posible. Y gracias Ana, por ser una compañera con una inigualable conciencia de clase. Y por reírte con tanta facilidad de mis chistes malos, que siempre le levanta a uno la moral. Gracias también Roberto, por tu ayuda apoyo y experiencia, a veces en la distancia y a veces en la localidad. Guardo muy dentro de mi tus palabras: "No te preocupes que eso de escuchar pajaritos (y perros, sillas, paredes,...) nos pasa a todos en las últimas semanas de la tesis."

\indent\indent En estos agradecimientos no podrían faltar mi hermana, mi madre y mi padre. Gracias por ser mi red de apoyo constante, incondicional e inmejorable. Y gracias por ser la voz de la razón y de la esperanza cuando las voces de mi cabeza no me dedicaban ninguna palabra amable. Gracias también a mi otra red de apoyo incondicional: Adri (Arenas), Adri (Brines), Ana, Claudia, Diego, Elisa, Gadea, Jose, Loló, Luque, Manu, Natalie, Rocío, Víctor y Zaira. Os pongo por orden alfabético porque me parece la decisión más salomónica, espero que no os importe. Gracias por estar ahí siempre, por darme apoyo, ánimos y escucharme cuando más lo necesitaba, y cuando no tanto, también. Sé que soy una persona increíblemente afortunada por el mero hecho de teneros en mi vida. 

\indent\indent En su momento dije que desde que Laura Hermosa pasó a ser representante de pre-docs en el IAA, o como a mí me gusta llamarla, la Doctoranda Suprema, todas las tesis leídas aquí deberían incluirla en los agradecimientos, porque su labor desde luego fue excepcional (¡alabado sea el Manual de Supervivencia del doctorando del IAA!). Y como lo mantengo y me gusta predicar con el ejemplo, muchas gracias Laura. Por eso, y por ser una tan buena compañera.

\indent \indent Siguiendo el legado de Laura, tampoco has decepcionado como Doctoranda Suprema, Teresa. Y gracias por ser una compañera de despacho tan buena, que sabe cómo meterle la quinta marcha al día desde la primera hora de la mañana. Ciertamente, no podría haber pedido mejores compañeros de despacho a lo largo de estos años. Gracias Álvaro, Miguel y Beth por acompañarme y aguantarme todos estos años. Miguel, Teresa, nuestros coros a capella los llevaré siempre muy dentro.

\indent \indent Finalmente, quiero dar las gracias a todo el personal del IAA. Puede sonar genérico, pero es un agradecimiento genuino y sincero. No sé por qué derroteros me llevará el futuro, pero me ha quedado claro que el mayor recurso del IAA no son sus grandes mentes, que son muchas y son desde luego importantes, pero la gran fuerza del IAA es la calidad humana de su personal, siempre dispuesto a ayudar, a seguir avanzando y a hacer de este instituto un entorno laboral inmejorable. Y de forma más pragmática, gracias por todo vuestro trabajo, pues es el que llevó al merecido reconocimiento Severo Ochoa, del que salió la financiación para mi contrato. Así que gracias, porque sin todo lo que he comentado, este trabajo no habría sido posible. Que por cierto, hablando del Severo Ochoa, no podía faltar: Author Julio Esteban Rodríguez Martín acknowledges financial support from the Severo Ochoa grant CEX2021-001131-S funded by MCIN/AEI/ 10.13039/501100011033.

\end{spacing}
\newpage
\begin{spacing}{1.5}
	\tableofcontents	
\end{spacing}
\newpage
\let\origaddvspace\addvspace
\renewcommand{\addvspace}[1]{}
\begin{spacing}{1.5}
	\addcontentsline{toc}{section}{\bf LIST OF FIGURES}
	\setcounter{lofdepth}{2} \listoffigures
\end{spacing}
\newpage
\begin{spacing}{1.5}
	\addcontentsline{toc}{section}{\bf LIST OF TABLES}
	\setcounter{lotdepth}{2} \listoftables 	
\end{spacing}

\clearpage
\vspace*{0.5cm}
\begin{spacing}{1.5}
\addcontentsline{toc}{section}{\bf LIST OF TERMS AND ABBREVIATIONS}
\newacronym{URL}{URL}{Uniform Resource Locator}

\printglossary[type=\acronymtype, title={\bf ~~~~~~~~~~~~~~~~~~LIST OF TERMS AND ABBREVIATIONS}]
\end{spacing}
\renewcommand{\addvspace}[1]{\origaddvspace{#1}}

\mainmatter

\pagenumbering{arabic}

\chapter{Introduction} \label{chapter:intro}

\section{Galaxies}
Humanity has always wondered about the nature of the light sources that appeared in the night sky. As observations and technology progressed, the nature of these nebulous sources remained a topic of discussion for a long time. Today, we refer to part of these sources as galaxies, and these
are one of the most important structures in the Universe. Concerning their formation, the most accepted cosmological scenario is the Lambda-Cold Dark Matter $\Lambda \mathrm{CDM}$. This model proposes a "bottom-up" scenario in which primordial density fluctuations grow through gravitational instabilities caused by cold collisionless dark matter. As these instabilities keep growing, so do the structures of the universe, through the accretion of other dark matter haloes. The growth of these haloes produces overdensities of baryonic matter, that keep growing until they form baryonic structures, like galaxies  \citep[see e.g.][]{Peebles1982, Davis1985, White1987, Kauffmann1993, Sommerville1999,Springel2005,deLucia2006}. Galaxies are a complex mix mainly composed of stars, gas, dust, and dark matter, that show a great variability in their properties. However, they are not stationary systems, that is they change and evolve with cosmic time due to several factors. One of them, it is the environment.

\section{Bimodal distribution of properties of galaxies}
\begin{figure}
    \centering
    \includegraphics[width=0.85\textwidth]{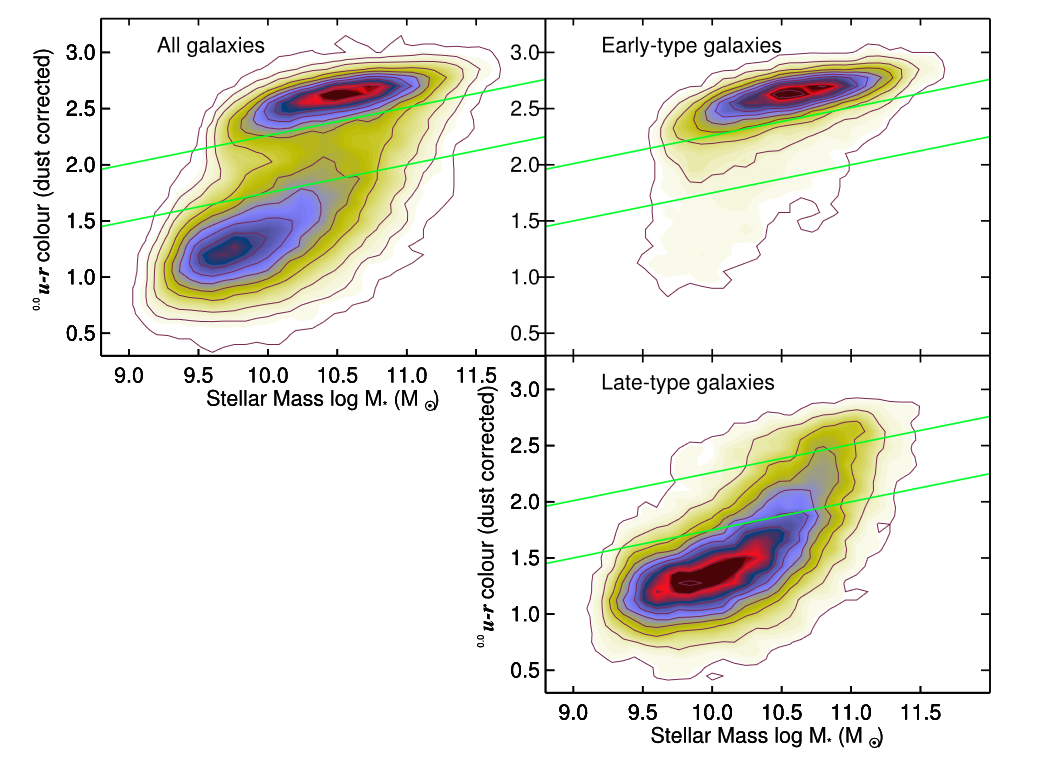}
    \caption{Example of a dust-corrected colour--mass diagram, showing the bimodality of the properties of the galaxies. Picture taken from \cite{Schawinski2014}}
    \label{fig:Schawinski2014}
\end{figure}

Galaxies are usually divided into two large groups: the red sequence and the blue cloud \citep{Strateva2001,Bell2004,Baldry2004,Williams2009,Moresco2013,Povic2013,Fritz2014,Luis2019,Rosa2021}. In fact, this bimodality has been found even up to redshift $z \sim 6$ \citep{Weaver2024}. This classification is typically reflected in colour--magnitude diagrams \citep[CMD,][]{Bell2004, Baldry2004}. Since the red sequence from CMDs suffer  of dust-reddening of galaxies in the blue cloud \citep{Moresco2013, Schawinski2014, Luis2019}, the inclusion of a second colour was proposed, in order to account for the extinction suffered in the blue part of the spectra. This way, colour-colour diagrams, such as the \textit{NUVrK} \citep{Whitaker2011,Arnouts2013} diagram and the \textit{UVJ} diagram \citep{Williams2009, Fang2018} have been used to separate these galaxies. Additionally, colour-mass have also been used for such purpose \citep{Peng2010}, producing similar results to the colour-magnitude diagrams, given the relation between the absolute magnitude and the stellar mass. For this reason, colour-extinction-corrected-mass diagrams offer a solid classification \citep{Luis2019,Luis2019b,Luis2019c}. We show and example of this type of diagrams in Fig.~\ref{fig:Schawinski2014}, taken from \cite{Schawinski2014}. This figure clearly shows two overdensities of galaxies in the colour--mass plane, corresponding to the quiescent and star-forming galaxies. However, when divided by morphological type, we find a correlation with these overdensities, further illustrating the bimodal nature of the properties of galaxies.

These two types of galaxies have properties clearly differentiated. Galaxies in the red sequence are usually old, metal-rich, with redder colours, lower \gls{SFR} and are generally more massive, while galaxies in the blue cloud are generally young, with a larger dispersion of metallicites, bluer colours, as shown by the well-known stellar mass--metallicity and stellar mass--age relations \citep{Tremonti2004,Gallazzi2005,Mendel2009,Thomas2010,Foster2012, Peng2015,Xiangcheng2016,Gao2018,DuartePuertas2022}. As mentioned, blue galaxies may sometimes appear as red galaxies due to dust reddening of their \gls{SED}, partly populating the intermediate region in the aforementioned diagrams. However, many studies point that in this region, usually referred to as the green valley, there is also  a transiting type of galaxies, and has been used to study the evolution of blue galaxies into red galaxies \citep[see e.g][]{Fang2012, Schawinski2014,McNab2021,Noirot2022}.

\begin{figure}
    \centering
    \includegraphics[width=\textwidth]{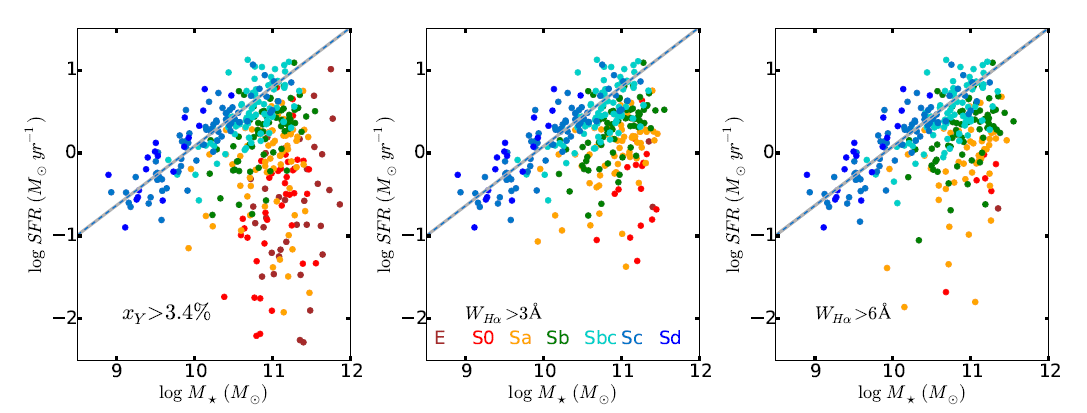}
    \caption{Star formation main sequence colour coded by galaxy morphology. Picture taken from \cite{Rosa2016}}
    \label{fig:Rosa2016}
\end{figure}

This bimodal nature of galaxies is also found in their \gls{SFR}: star-forming galaxies lie in a tight relation that reflects the correlation between the total stellar mass and the total \gls{SFR} of the galaxy \citep{Brinchmann2004,Noeske2007}. This relation is known as the \gls{SFMS} and is described by a power law \citep[see e.g.][]{Elbaz2007,Speagle2014,Sparre2015,cano2016spatially, vilella2021j}. This relation has been found at least up to $z\sim 6.5$ \citep{Rinaldi2022}. Meanwhile, quiescent galaxies remain below this relation. Results from \cite{Rosa2016} show that this bimodality is also linked to other properties of the galaxy, such as their morphology (see Fig.~\ref{fig:Rosa2016})

\section{Galaxy quenching}
Given this bimodality in the properties of galaxies, it is widely accepted that galaxies from the blue cloud move into red sequence as they evolve \citep{Faber2007}. Therefore, as time passes, blue stars  die while the red stars will remain contributing to the \gls{SED} of galaxies. If there is no available gas to form new  massive stars the \gls{SED} will shift towards redder colours. This process is commonly referred to as passive evolution. However, since $z \sim 1$, a large fraction of blue galaxies, most of them with masses lower than $10^{10}$~$M_\odot$, has seen its star formation truncated, evolving into the red sequence \citep{Luis2019b}. The timescale of this transition must be very short, since the fraction of red galaxies has almost doubled since $z \sim 1$, and the density of galaxies in the aforementioned green valley is not enough to explain the evolution of this fraction \citep[see e.g.][]{Bell2004,Faber2007,Muzzin2013}. Therefore, additional mechanisms are required, which leads to the proposal of \textit{quenching}.

\begin{figure}
    \centering
    \includegraphics[width=0.85\textwidth]{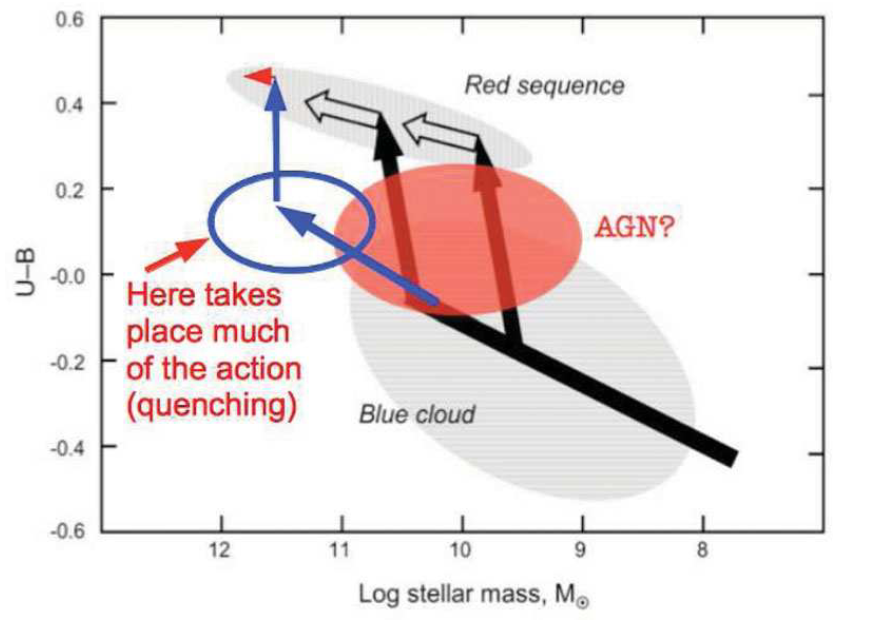}
    \caption{Cartoon representing the evolution of galaxies induced by quenching. Picture taken from \cite{Renzini2013}}
    \label{fig:renzinni2013}
\end{figure}

The term quenching is generally used to reflect the sudden cessation of ƒthe star formation \citep[see e.g.][]{Faber2007,Peng2010, Peng2012}. We show an illustration of the effect of quenching in galaxy evolution in Fig.~\ref{fig:renzinni2013}, taken from \cite{Renzini2013}. This cartoon shows how galaxies in the more massive end of the blue cloud move into the red sequence due to quenching effects. These proposals also arise because of the so-called ``cooling problem'' or ``cooling catastrophe'', this is, the predicted cooling time for the gas is far lower than the cooling time observed, thus, some mechanism must be heating the gas \citep[see e.g.][]{Ruszkowski2002, Springel2005,Croton2006, Mcnamara2007,Bower2006,Bower2008}.

Depending on how the quenching process takes place, we distinguish two scenarios \textit{mass quenching} and \textit{environmental quenching} \citep{Peng2010,Ilbert2013}. The term mass quenching is usually used to describe the increasing number of quiescent galaxies at the high mass end of the \gls{SFMS}. One candidate of these processes is the stellar feedback, since the formation of new stars and the feedback from supernovae can produce winds that remove the gas or heat it, preventing further star formation \citep[see e.g]{Larson1974,Dekel1986,Efstathiou2000,Cantalupo2010}. Similarly, the \gls{AGN} feedback may quench or slow-down the star formation of  galaxies, since the large amount of energy liberated by an active supermassive black hole in the galactic nucleus can lead to gas removal, heating and destabilisation \citep[see e.g.][and references therein]{Bluck2011,Combes2017,Piotrowska2022,Bluck2023}.

\begin{figure}
    \centering
    \includegraphics[width=0.35\textwidth]{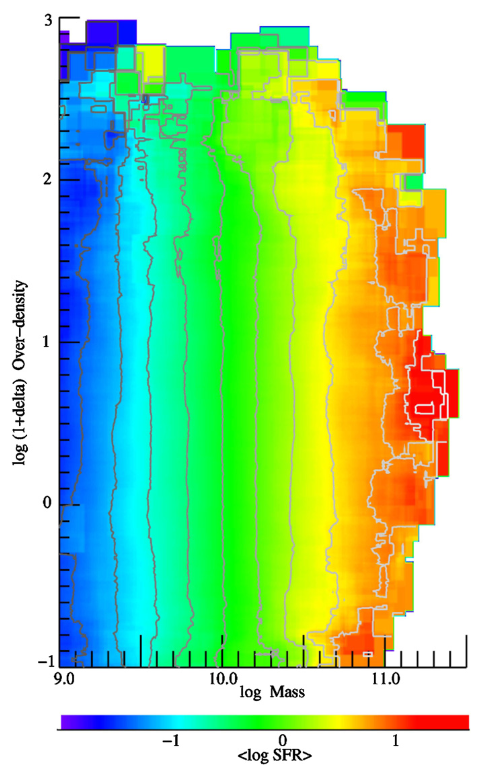}
    \caption{Mean \gls{SFR} of galaxies as a function of the stellar mass and the environment. Picture taken from \cite{Peng2010}}
    \label{fig:Peng2010}
\end{figure}

On the other hand, environmental quenching is used to label those processes related to the environment that can prevent the formation of new stars. The work by \cite{Peng2010} showed that the effects of these two types of quenching mechanisms are separable up to $z \sim 1$ (see Fig.~\ref{fig:Peng2010}). This figure illustrates that, for a fixed mass, the \gls{SFR} of galaxies also decreases as the density of nearby galaxies increases.

\section{The role of the environment on galaxy evolution}
Galaxies in high density environments, such as galaxy clusters and groups, are more likely to interact with other galaxies and with the gas among galaxies. These interactions can have an impact on the properties of the galaxies and their evolution. Some of the first works include the so-called Butcher-Oemler effect \citep{ButcherOemler1978,ButcherOemler1984}. These works shows that fraction of blue galaxies in the cores of galaxy cluster increased with redshift, indicating that as time passed, galaxies in these cores were more likely to become red. Another pioneering work worth of mention is the one by \cite{Dressler1980}, who showed that the fraction of elliptical galaxies increases with the number density of galaxies, which means that it also decreases with the distance to the cluster centre.

\begin{figure}
    \centering
    \includegraphics[width=0.65\textwidth]{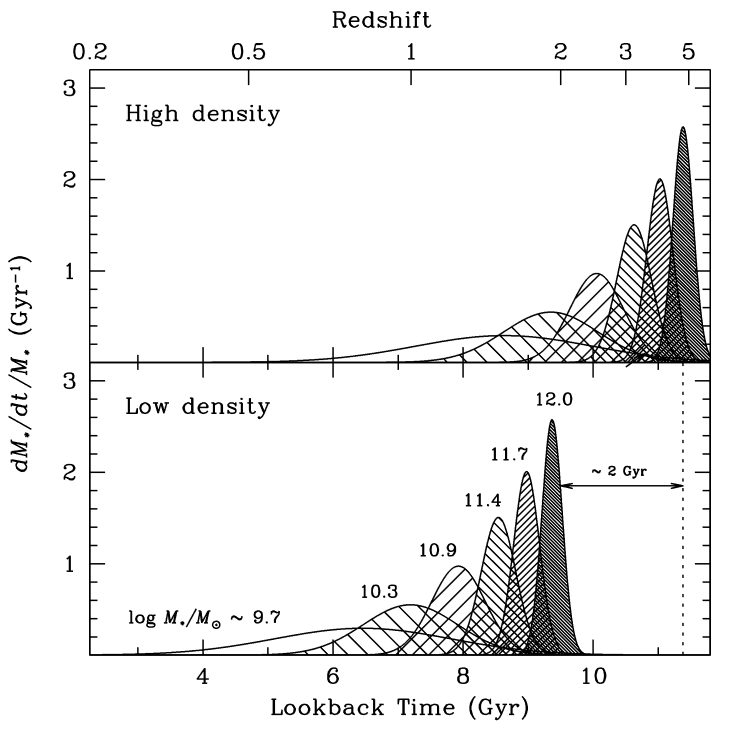}
    \caption{Cartoon representing the \gls{SFH} of early types galaxies in different environments. Picture taken from \cite{Thomas2005}}
    \label{fig:Thomas2005}
\end{figure}

The effect of the environment on galaxies is not limited to their colour or their morphology.  Properties such as the stellar mass distribution, the star formation, and the nuclear activity depend on the density too, even the galaxy populations themselves \citep{Balogh2004,Kauffmann2004,Blanton2005, Blanton2009,Pasquali2010,Lopes2014,Cappellari2016}. In general, galaxies in denser environments are older on average, which is likely a consequence of  having their star formation truncated at earlier epochs than galaxies in less dense environments.  \citep[see e.g. ][]{Bower1990, Trager2000, Thomas2005,Clemens2006, Bernardi2006, Smith2008, Cooper2010MNRAS}. We show the cartoon from \cite{Thomas2005} showing how the density of the environment affects the \gls{SFH} of early-type galaxies in Fig.~\ref{fig:Thomas2005}. This figure shows that, for a given mass, galaxies in high density environments started to form stars at earlier epochs than their counterparts in low density environments.

In order to explain these observations, several mechanisms, generally based on gravitational or hydrodynamic effect, have been proposed and studied. These mechanisms mainly prevent the formation of new stars through the removal or the heating of the gas reservoirs of galaxies, although they are not limited to these aspects. One of them is the ram-pressure stripping \citep{Gunn1972}. This mechanism is based  the interaction between the hot, dense \gls{IGM} and the \gls{ISM}  can remove the second one from the galaxy, if the ram pressure overcomes the gravitational pressure. This has been reported  in several works, both observational and simulations \citep[see e.g.][]{Abadi1999,Fujita1999,Boselli2008,Boselli2021,Book2010,Steinhauser2016,Singh2019}. Some works point to ram pressure stripping as the main quenching mechanism \citep[see e.g.][]{Muzzin2014,Boselli2016,Liu2021}. Ram pressure stripping is also responsible of the formation of the so-called jellyfish galaxies \citep{Bekki2009,Ebeling2014,Poggianti2017,Jaffe2018,Rohr2023} and can even trigger the nuclear activity \citep{Peluso2022}. 

Another mechanism related to  environment effects is harassment. It was originally proposed by \cite{Moore1996,Moore1998,Moore1999}, arguing that the high-speed close encounters among galaxies can heat the gas of galaxies, as well as perturbing the orbits of stars and dark matter, resulting in a morphological transformation. More recent studies have shown that this kind of interaction is plausible, but its effect is mostly limited to galaxies in the inner parts of clusters \citep{Bialas2015}.

Similarly to harassment, tidal forces resulting from galaxy-galaxy encounters, can disturb the distribution of star, gas, dust, and dark matter, which may remove effectively mass from the galaxy \citep[see e.g][ and references therein]{Read2006,Read2006b,Kampakoglou2007,Henriques2010,Dooley2016,Andersson2019,Keim2022,SalvadorSole2022,Pacucci2023}. This mechanism is usually referred to as tidal stripping and may be related to the formation of dwarf galaxies \citep[see e.g.][]{Klimentowski2009}

The starvation or strangulation is a mechanism used to describe the inability of the galaxy to replenish its gas reservoir, generally because of the removal of the outer galaxy halo \citep{Larson1980,balogh2000,Bosch2008,VandeVoort2017, Trussler2020, Kumari2021, Wright2022}. These works point to starvation as one of the main mechanisms driving galaxy evolution. In this regard, the work by \cite{Gutcke2017} proposes a model for galaxy formation were only starvation is required to reproduce the fraction of quiescent galaxies at $z=0$ without the need of fast-quenching, although its relevance depends on the initial conditions of the formation of the galaxy \citep{Girichidis2012}.
 
Thermal evaporation is another mechanism mainly found in clusters, where the intergalactic medium temperature is high and rapidly evaporates the gas reservoirs of a galaxy, which can not be retained by the gravitational field \citep[see e.g.][]{Cowie1977,Nipoti2007}

Alongside these mechanisms, several works propose a pre-processing scenario, where satellites galaxies and galaxies in smaller groups are quenched during their in-fall into larger clusters, which might take several orbits to occur \citep{Fujita2004,BravoAlfaro2011,deLucia2012,Mahajan2013,Haines2015,Donnari2021,Ando2022,Pallero2022,Werner2022,Lokas2023,Lopes2024,PirainoCerda2024}. This scenario is also supported by observations and simulations that show differences in the properties of central and satellite galaxies \citep[see, e.g.][]{Bosch2008,Kovac2010,Peng2010,Peng2012,Woo2017,Zinger2018,Pasquali2019,Bluck2020,Gallazzi2021}. These studies show that, in general, satellite galaxies are more prone to suffer the effects of  environment, showing redder colours and lower \gls{SFR} than their central counterparts.

\section{Galaxy surveys for galaxy evolution}
Much of our current knowledge about galaxy evolution is due to galaxy surveys. Some of them have provided data for a large number of galaxies at different redshifts. This allows us to study the properties of galaxies at different epochs with a great statistical significance, which is key to understand the evolution of galaxies. Other surveys may have studied fewer galaxies, but using an spatially resolved approach, which has shed light into the structure of galaxies and the processes that drive their evolution at smaller scales.

Certainly, the last decades have witnessed a substantial increase in the number of galactic surveys, whose contributions to the science are invaluable. Such is the case of the 2dF Galaxy Redshift Survey \citep{Folkes1999}, SDSS\footnote{\gls{SDSS}} \citep[][]{York2000}, GEMS\footnote{\gls{GEMS}} \citep[][]{Rix2004}, VVDS\footnote{\gls{VVDS}} \citep[][]{VVDS2005},  DES\footnote{\gls{DES}}\citep[][]{Wester2005},  COSMOS\footnote{\gls{COSMOS}} \citep[][]{Scoville2007},  CALIFA\footnote{\gls{CALIFA}} \citep[][]{CALIFA2012}, MaNGA\footnote{\gls{MaNGA}} \citep[][]{MANGA2015}, Pan-STARRS1\footnote{\gls{Pan-STARRS1}} \citep[][]{Chambers2016}, the ALHAMBRA\footnote{\gls{ALHAMBRA}} survey \citep[][]{Moles2008,Molino2014}, VIPERS\footnote{\gls{VIPERS}} \citep[][]{Haines2017},  VUDS\footnote{\gls{VUDS}} \citep[][]{Tasca2017}, SAMI \citep{Green2018SAMI,Scott2018,Croom2021} or HSC-SSP\footnote{\gls{HSC-SSP}} \citep[][]{Aihara2018Presentation},  to name a few.

\subsection{Photometric surveys vs specectroscopic surveys}

Galaxy surveys can be generally divided in two large classes: spectroscopic and photometric surveys. They are different by construction. Simplifying, in spectroscopic surveys, the light goes through a slit or an fiber, then is led into a diffracting device that splits the light by wavelength and leads the split light into a photo-detector, usually a \gls{CCD}. On the other hand, photometric surveys direct the light directly into the photodetector, that is usually behind a filter. The filter only allows the light within a wavelength range to go into the detector, reflecting the rest light outside of this wavelength range.   

The main advantage of spectroscopic surveys is that they provide data with much higher spectral resolution that can be used for spectral fitting with a greater accuracy. This greater accuracy comes from the ability to fit spectral features that are sensitive to the properties of the source. In the case of galaxies, to properties of the stellar or gas content. This technique has been widely used \citep[see e.g.][]{Trager1998,Jorgensen1999,Kuntschner2001,Thomas2005,Bernardi2006,SanchezBlazquez2006b}

On the other hand, the spectral resolution of photometric surveys is usually much lower. However, they have numerous advantages that are worth noting:

\begin{itemize}
    \item There is no selection bias other than the image depth. Spectroscpic surveys usually need to point the fibre or slit into one single galaxy, while photometric surveys capture everything that falls within the area of the CCD, as long as they are brighter than the limiting magnitude.
    \item Photometry is not affected by aperture bias, because the direct imaging  allows for defining different apertures which are not limited by the constant size of a fibre, and these are only limited by the \gls{FoV}.
    \item The photometric calibration of each band is independent of the other ones, which allows for a great calibration, free of systematic colour terms.
    \item For the same telescope, instrument and integration time, photometry is usually deeper, due to the efficiency of the direct imaging. This allows to reach further redshifts.
    \item Photometry allows for fast spatially resolved studies, since from a same observation it is possible to obtain the \gls{SED} of each pixel. 
\end{itemize}

Taking into account this advantages and disadvantages, two type of surveys have arisen: the \gls{IFS} and the multiband photometric surveys. These surveys are performed using \gls{IFS}, which work like traditional spectograph, but a set of fibres is used in order to cover a larger area. The goal of this methodology is to obtain spectra from different regions of the target, allowing for spatially resolved studies. This is the case of surveys like SAURON\footnote{\gls{SAURON}} \citep[][]{SAURON2001, Sauron2002}, CALIFA \citep{CALIFA2012, CALIFA2015, CALIFA2016, CALIFA2023}, MaNGA \citep{MANGA2015, MANGA2018, MANGA2019, MANGA2022}, the MUSE-Wide survey \citep{MuseWide2019}, the PHANGS-MUSE survey \citep{PhangsMuse2022} or the WEAVE-Apertif survey \citep{Hess2020}.

Multi-band photometric surveys are like regular photometric surveys, but with a higher number of filters than usual. Some examples are the COMBO-14\footnote{\gls{COMBO-17}} \citep[][]{COMBO2001, COMBO2003}, the ALHAMBRA survey \citep{Moles2008,Molino2014}, the COSMOS survey  \citep[COSMOS][]{Scoville2007},  J-PLUS\footnote{\gls{J-PLUS}} \citep[][]{Cenarro2019}, or the S-PLUS\footnote{\gls{S-PLUS}} \citep[][]{SPLUS2019}. These surveys provide a much better spectral resolution than the traditional broad band photometric surveys, since each filter provides another point to sample the observed wavelength range, increasing the quality of the spectral fitting while keeping all the advantages from the traditional photometric surveys. They have also proven to be successful at retrieving the properties of galaxies through spectral fitting techniques \citep[see e.g.][and references therein]{Luis2015,Luis2019,Luis2019b, Luis2019c, Rosa2021}. This way, multi-band photometric surveys can be equivalent to a low-resolution \gls{IFU} survey that provides spatially resolved information.

\subsection{Properties of the spatially resolved galaxies}
IFU-like surveys have greatly improved our knowledge of the properties and structures of galaxies. A general result is that colour gradients in galaxies, which have been observed since a long time ago, as long as gradients in the stellar population properties of galaxies, such as the stellar mass density, the stellar age, or the stellar metallicity, both in disk-dominated and bulge-dominated galaxies \citep{Peletier1990,deJong1996,Peletier1996,Silva1998,Peletier1998,Bell2000,LaBarbera2004,McArthur2004,Menanteau2004,Wu2005,Moorthy2006,MunozMateos2007,Bakos2008,Roche2010,Tortora2010,LaBarbera2012,Rosa2014, SanchezBlazquez2014,Ruben2017,Rosa2017,Zheng2017,Bluck2020}. The results of these studies are of great value for understanding the formation and structure of galaxies. In general, larger mass densities, older, redder, and more metal rich regions are found in the central parts of the galaxy compared to the outermost regions, supporting an inside-out formation scenario. However, some works find different results. For example, \cite{Goddard2017b} found positive age gradients in their analysis of MaNGA data for early-type galaxies, \cite{Costa-Souza2024} also finds a positive gradient in the ages of the young and intermediate populations of their Seyfert 2 sample, and  some authors  even suggest the existence of rejuvenating  galaxies \citep[see e.g.][]{Trayford2016,Cleland2021, Zhang2023,Tanaka2024}. 

Similarly, gradients in the intensity of the SFR and sSFR have been found within galaxies, suggesting an inside-out quenching scenario  \citep[see e.g.][]{Rosa2016,Medling2018,Sanchez2020}. However, \cite{Bluck2020} suggest that AGN-driven quenching is inside-out, while environmental quenching is outside-in. In this line, \cite{Lin2019b} find that the fraction of inside-out quenched galaxies increases with halo mass, and that it is larger for high-stellar-mass than for low-stellar-mass ones in all environments, but their results also suggest that the inside-out quenching is the dominant quenching mode in all environments. These results are compatible with the findings of \cite{Ge2024}, whose work shows that massive galaxies are more likely to experience an inside-out quenching mode, while low-mass galaxies are more likely to show an outside-in quenching mode.

Some other results on the spatially resolved properties of galaxies  include that the star formation density or the local star formation, along with the local mass density follows a trend very similar to the \gls{SFMS}, usually known as the local main sequence of the star formation or the spatially resolved main sequence \citep[see e.g.][]{Sanchez2013,wuyts2013,Hsieh2017,Ellison2018,Lin2019a, cano2019sdss}. 

\subsection{The role of environment on the properties of spatially resolved galaxies}
The effect of the environment can also be reflected on the properties of galaxies at a smaller scales. It has been shown that the  \gls{SFH} of a galaxy can be retrieved at such scales using techniques like fossil record, based on the full spectral fit of the optical stellar continuum of the spatially resolved data provided by \gls{IFU}-like surveys  \citep[see e.g.][]{Perez2013,Rosa2016,Rosa2017,Ruben2017,CortijoFerrero2017a}, which allows to gather clues on the processes leading up to the formation of the present-day galaxy population.

Therefore, the effect of the environment can be imprinted in the \gls{SFH} of galaxies. For example, the works by \cite{CortijoFerrero2017a,CortijoFerrero2017b,CortijoFerrero2017c} reveal how galaxy mergers can produce a rapid increase of the \gls{sSFR} towards the centre of the galaxy in time scales shorter than 1~Gyr, with a younger stellar population.  

\gls{IFU}-like surveys have also contributed to build a global picture of the effect of environment on galaxy evolution. Further evidence of the pre-processing of galaxies in groups can be derived from spatially resolved analysis \citep{Epinat2024}, and it has been shown that there is an enrichment of satellite galaxies through the exchange of gas in dense environments \citep{Schaefer2019}. The gradients of properties like stellar mass surface, intensity of the SFR, or age can also change from central to satellite galaxies \citep{Bluck2020}. Additionally, the star formation of galaxies is suppressed with increased density in an outside-in manner \citep{Schaefer2017}. However, the timescale required for quenching is still under debate \cite[see e.g.][]{Brough2013,Schaefer2017}.

IFU-like surveys have also shown that opposite phenomena are found in low density environments, like galaxy voids. Particularly, \cite{Ana2024} show that the \gls{HLR} of void galaxies is slightly higher than galaxies in filaments and walls, while their stellar mass surface density is lower and their stellar populations are younger than those of galaxies found in other environments, regardless of their morphological type. These results by \cite{Ana2024} show that void galaxies, undergo a more gradual evolution, especially in their outer
regions, with a more pronounced effect for low-mass galaxies. Certaily, spatially resolved studies can still be very useful to understand the role of environment on galaxy evolution.

\subsection{J-PAS: a survey for galaxy evolution}
The \gls{JPAS} is an ongoing photometric multi-narrow-band survey originally presented by \cite{Benitez2014}. It is expected to scan over $8000$~$\mathrm{deg}^2$ of the Northern sky. It will be carried out at the \gls{OAJ}. The details about the OAJ can be found in \cite{Cenarro2014}, and the results of the testing of the observing conditions of the Sierra de Javalambre, where the OAJ is located at a height of 1957~m, can be found in \cite{Moles2010}. The sky of the Sierra de Javalambre is exceptional in this matter, with a dark sky (the night-sky surface brightness are $B = 22.8$~mag~$\mathrm{arcsec}^{-2}$, $V = 22.1$~mag~$\mathrm{arcsec}^{-2}$, $R = 21.5$~mag~$\mathrm{arcsec}^{-2}$, $I = 20.4$~mag~$\mathrm{arcsec}^{-2}$) and contributions to the sky brightness of the typical pollution lines of $\sim 0.06$~mag in the $B$ band, $\sim 0.09$~mag in the $V$ band, and $\sim 0.06$~mag in the $R$ band on a moonless night. The median value of the seeing in the $V$ band is $0.71''$, with a mode of $0.58''$. at the same location, the J-PLUS survey is currently carried out.

\begin{figure}
    \centering
    \includegraphics[width=0.55\textwidth]{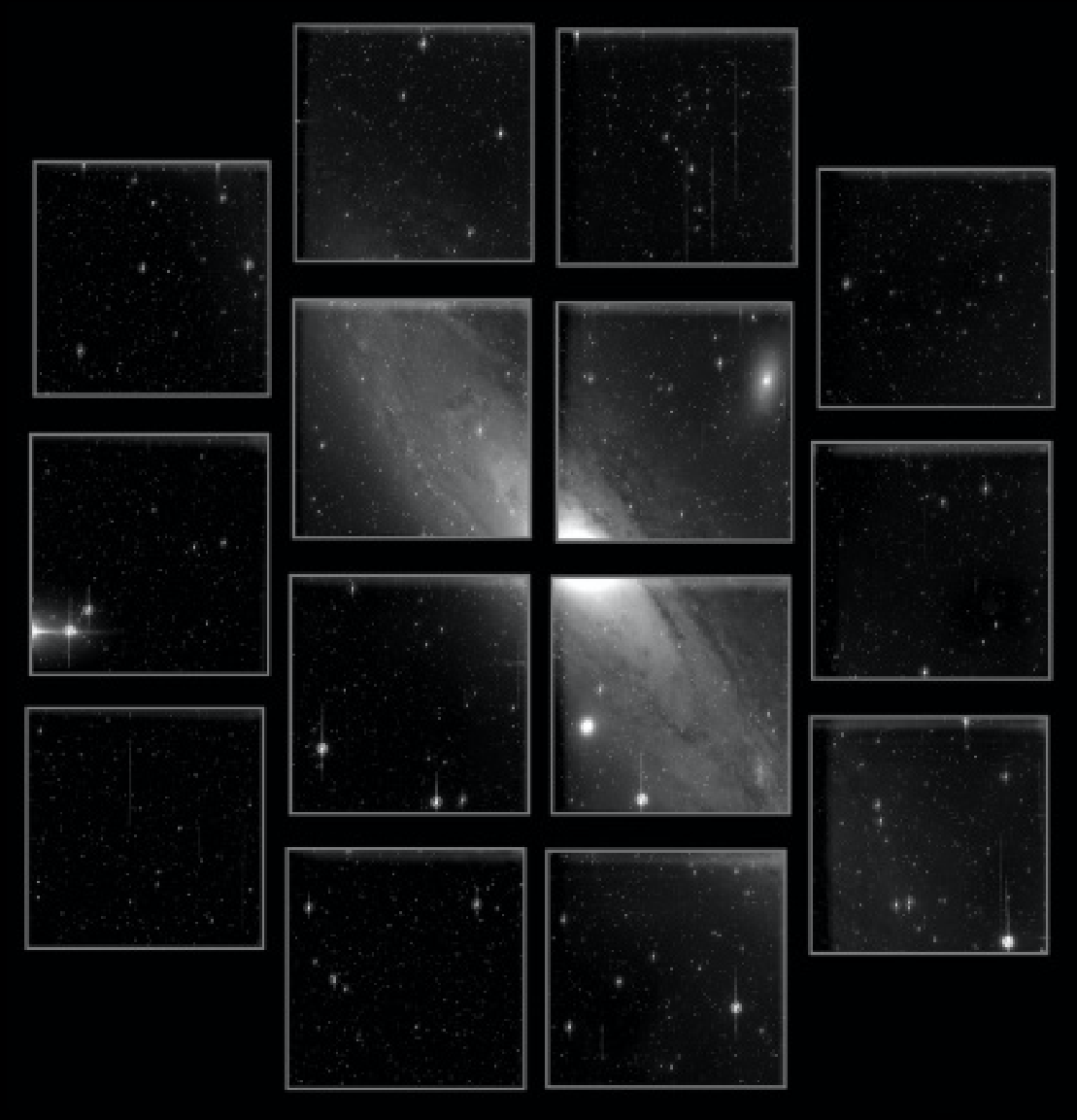}
    \caption{Adromeda galaxy (M31). Technical first light image JPCam JST/T250. Credit: Centro de estudios de física del cosmos de Aragón (CEFCA)}
    \label{fig:JPAS1stlight}
\end{figure}

The J-PAS survey will be observed using the Javalambre Survey Telescope \citep[JST/T250][]{Cenarro2018}. This telescope has an aperture of $2.55$~m, a FoV with a diameter of 3~deg, and a collecting area of $3.75$~$\mathrm{deg}^2$. The optical design was optimised to provide a good image quality in the optical 3300--11000~\AA \ wavelength range all over the focal plane. The camera that will be used is the Javalambre Panoramic Camera \citep[JPCam][]{Taylor2014,MarinFranch2017}, a camera with $1.2$~Gpixel, an effective FoV of $4.2$~$\mathrm{deg}^2$ and a pixel scale of $0.23$~arcsec~$\mathrm{pix}^{-1}$. 

The greatest strengths of the survey will be its large FoV, covered by 14 CCDs amounting to a total of $4.2$~$\mathrm{deg}^2$, capable of fitting the whole Andromeda galaxy in it (see Fig.~\ref{fig:JPAS1stlight}), and its filter system, composed of 54 narrow-band filters, plus two medium band filters (see next chapter for further details). This filter system was originally conceived to provide accurate photo-$z$ measurements up to $z \sim 1.3$ with a precision of up to $\delta z = (0.003)(1+z)$. 

With these capabilities, J-PAS offers a great strength to study the evolution of galaxies and the role of environment on them. First of all, its photometric system and the accuracy of the photo-$z$ allow to infer the stellar population properties of galaxies with great precision, as shown by \cite{Rosa2021}, as well as their \gls{SFR} and \gls{SFH}. Furthermore, it can also be used to study  the emission lines of these galaxies \citep{Gines2021,Gines2022}. In addition, the large \gls{FoV} allows us to obtain unbiased, flux-limited catalogues, which not only provide solid statistics to support the conclusions of the works performed, but are also crucial to detect and select the unbiased populations of galaxy groups and clusters, in order to study the effect of environment on galaxy evolution. 

Furthermore, the combination of its large \gls{FoV}, the size of the CCD, and the photometric system also allow for the unbiased study of spatially resolved galaxies, since larger galaxies can be fully observed. In addition, the photometric system allows to study them as an IFU-like survey, performing the aforementioned studies at smaller scales. This means that the \jp \ survey will be an excellent way to study properties of both the integrated and spatially resolved galaxies.

\section{Goals of this work}
The aim of this thesis is to study the effect of groups and cluster environment on galaxy evolution, using the available data from \mjp. In order to achieve this goal, we  develop tools and methodologies that prove the power of \jp \ and will pave the way for the analysis of the upcoming \jp \ data. Our motivation is to shed light on the relevance of environment on galaxy evolution, and contribute to  disentangle its role, which remains under debate. We take advantage of the nature of the \mjp \ and \jp \ data: the combination of the photometric filter system allows for precise estimation of  galaxy properties, and its large FoV  allows for  galaxy clusters and groups detection, with an unbiased galaxy classification. Furthermore, the large FoV also allows for unbiased spatially resolved studies, using the data as IFU-like cubes. Therefore, we shall study the effect of the environment in the properties of galaxies from an integrated perspective as well as a spatially resolved one. This general aim can be detailed in more specific goals:

\begin{itemize}
    \item The study of the spatially resolved galaxies in \mjp. In particular the comparison of the galaxies in groups and in the field with the same spectral type  allow us to study the relevance of environment on these properties. For this purpose a tool has been developed and tested, which automatises the process of analysis and that will applicable to the future \jp \ data.
    \item The study of the radial profiles of the stellar population properties and the emission lines of the spatially resolved galaxies in \mjp, classified by spectral type and environment. Not all environmental processes affect the different parts of the galaxy in the same way, and the radial profiles may play a key role disentangling the effects of the environment.
    \item The study of the star formation histories of inner and outer regions, of galaxies. This study can provide insight on the different formation and evolution scenarios that might experiment galaxies in different environments.
    \item The study of the properties of the integrated galaxies in the most massive cluster in \mjp, mJPC2470-1771, and their variation with the distance to the centre of the cluster. This cluster is an excellent laboratory to study the role of environment in galaxy evolution, due to the high density of galaxies and its radial variation. 
    \item The detection and classification of the emission line galaxies in the cluster mJPC2470--1771, in order to study the effect of high density environments in the emission lines of galaxies. 
    \item The study of the variation of the galaxy populations with cluster-centric distance, as well as the comparison of the \gls{SFH} of galaxies in inner and outer regions, in order to study possible formation and evolution scenarios.
\end{itemize}

\chapter{The \mjp \ survey} \label{chapter:minijp}

The \mjp \ survey \citep{Bonoli2020} is a $1$~$\mathrm{deg}^2$ survey that was carried out at the \gls{OAJ}, using the $2.5$~m  JST/T250 telescope. The main goals of this survey include testing and showing the potential of the \jp \ photometric filter system, exploring the capabilities of the \jp \ survey, and the first scientific exploitation of the data using this photometric system. Since all the data used in this thesis (except for a few particular cases, which are only used for testing) belongs to this survey, we dedicate this chapter to explaining the most important aspects of the survey for our work. 

\section{Technical aspects of \mjp} \label{sec:mjp:tech}

In the following subsections, we summarise the most relevant technical details of \mjp, which are mainly found in  the work by \cite{Bonoli2020}, as well as in the references therein, unless otherwise stated.

\subsection{The J-PAS Pathfinder camera}\label{sec:mjp:camera}

The camera used for the acquisition of the data was the  \gls{JPAS-PF} camera. It is equipped with a single charge-coupled device (CCD) direct imager, and with one large format, $9.2k \times 9.2k$ pxiel, low noise detector. This device reads 16 ports simultaneously. The size of image area is $92.16$~mm $\times 92.32$~mm.  It has a broad band anti-reflective coating for optimised performance from 3800~\AA \ to 8500~\AA. The miniJPAS survey has been observed with a read mode that achieves total system level noise performance of 3.4~$\mathrm{e}^-$ (rms), allowing for readout times of 12~s (full frame) and $4.3$~s ($2 \times 2$ binning). The integration times were as short as $0.1$~s with an illumination uniformity better than 1~\% over the entire FoV of the JPAS-PF. Since the filters are slightly smaller than the CCD, there is vignetting in the periphery. The resulting FoV is 0.27~$\mathrm{deg}^2$ with a pixel scale of $0.23$~arcesc~$\mathrm{pixel}^{-1}$.

\subsection{The J-PAS photometric filter system}  \label{sec:mjp:filters}

\begin{figure}
    \centering
    \includegraphics[width=\textwidth]{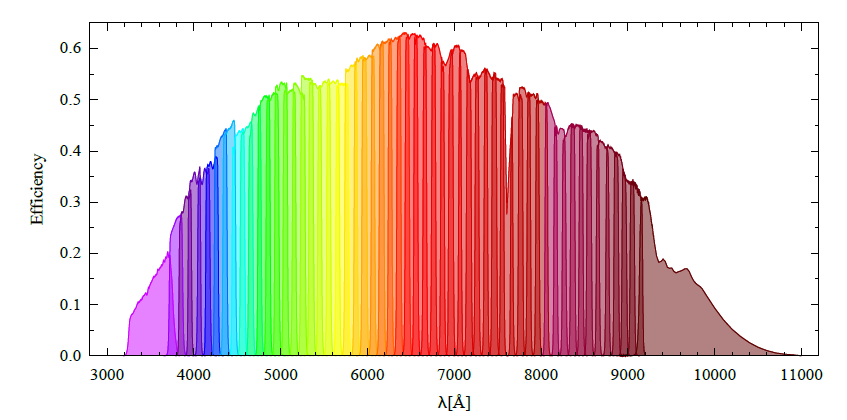}
    \caption[Efficiency of the J-PAS filters. Picture taken from  \cite{Bonoli2020}.]{Efficiency of the J-PAS filters. Effects from the CCD, telescope and sky are included.  Picture taken from  \cite{Bonoli2020}.}
    \label{fig:JPASfilters}
\end{figure}

One of the strengths of \jp \ will be its photometric filter system. One of the goals of \mjp \ is to prove the potential studies that can be achieved with this system. The system is composed of 54 narrow band filters covering the whole optical wavelength range, from 3780~\AA \ to 9100~\AA, with a FWHM of $\sim 145$~\AA, equally spaced every 100~\AA. In addition, there are two intermediate band filters, one covering the UV edge, ($u_\mathrm{JAVA}$, with a central wavelength of 3497~\AA \ and a FWHM of 495~\AA), and another one covering the redder edge, (J1007, with a central wavelength of 9316~\AA \ and a FWHM of 620~\AA). The efficiency of this system, including effects from the mirror, the atmosphere and the CCD, can be seen in Fig.~\ref{fig:JPASfilters}. This filter system provides a resolution of $R\sim 60$, which is equivalent to very low resolution spectroscopy.

In addition to these filters, \mjp \ was observed using four SDSS-like  broad bands: $u_\mathrm{JPAS}$, \gb, \rb, and \ib. In particular, \rb \ was used as the detection band during the survey imaging. Since this was the purpose of this band, it was also observed in the most favourable conditions, and we also use it as the reference filter for many aspects of our analysis.

\subsection{Data acquisition}
The \mjp \ consists of four pointings across the AEGIS field that partly overlap, resulting in four composed images: miniJPAS-AEGIS1 $(\alpha, \delta) = (214 \degree.2825, 52\degree.5143)$, miniJPAS-AEGIS2 $(\alpha, \delta) = (214\degree.8285, 52\degree.8487)$, miniJPAS-AEGIS3 $(\alpha, \delta) = (215\degree.3879, 53\degree.1832)$, and miniJPAS-AEGIS4 $(\alpha, \delta) = (213\degree.7417, 52\degree.1770)$. Each tile was observed with a minimum of four exposures per filter, with eight exposures per filter for the reddest ones. The exposure times were of 120~s for the narrow band filters and the  $u_\mathrm{JPAS}$ filter, while it was of 30~s for the broad band ones, in order to avoid saturation. These times were defined in order to reach the optimal photometric-redshift depth for the wide range of galaxies at different redshifts, according to the baseline strategy \citep{Benitez2014}. 

The readout mode chosen for the observations is a  $2 \times 2$ binning for the J-PAS filters, in order to reduce the readout noise in the pixel by a factor of 4, which in practice increases the effective photometric depth. However, this triggers an undersampling of the PSF in this images. The $u_\mathrm{JPAS}$, \gb, \rb, and \ib were read in full frame mode.

\subsection{FWHM and depth of the images} \label{sec:mjp:FWHM}

\begin{figure}
    \centering
    \includegraphics[width=\textwidth]{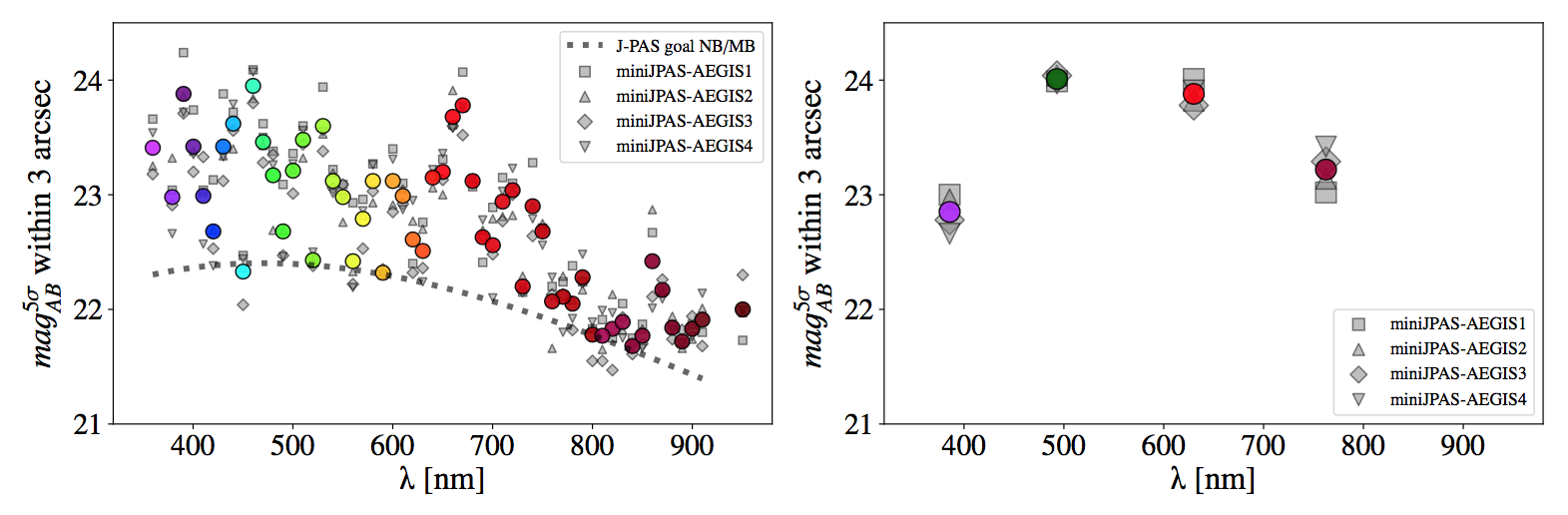}
    \caption[Image depth of the \mjp \ images. Picture taken from \cite{Bonoli2020}]{Image depths of the \mjp \ images for the narrow bands (lef panel) and broad bands (right panel). Colour symbols represent average values in each filter. Gray points represent the value measured for each filter, for each tile. Picture taken from \cite{Bonoli2020}}
    \label{fig:JPAS_depth}
\end{figure}

\begin{figure}
    \centering
    \includegraphics[width=0.6\textwidth]{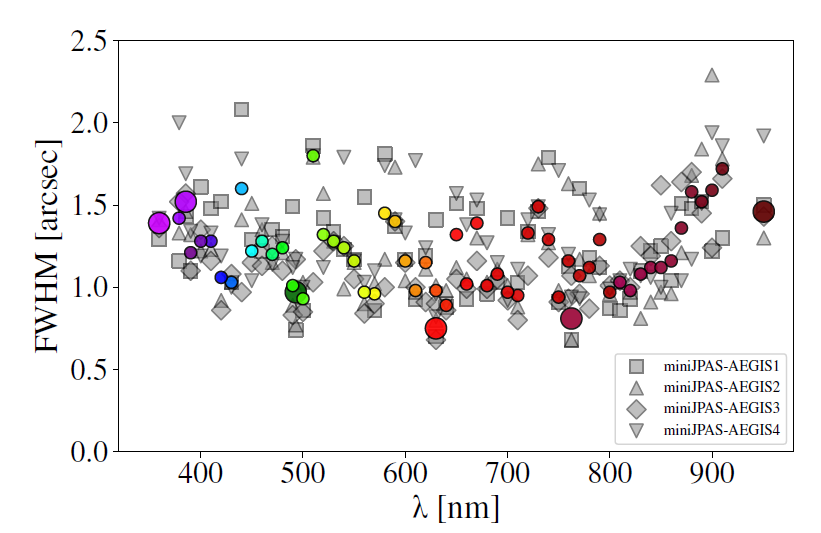}
    \caption[FWHM of the PSF of the images of \mjp. Picture taken from \cite{Bonoli2020}.]{FWHM of the PSF of the images of \mjp. Colour symbols represent average values in each filter. Gray points represent the value measured for each filter, for each tile. Picture taken from \cite{Bonoli2020}.}
    \label{fig:JPAS_PSF}
\end{figure}

Because of the large number of filters and images being observed in different nights under different sky conditions (observations were performed in groups of six filters, one filter at a time), there are variations in the depths and \gls{PSF} of each tile and band. Figures~\ref{fig:JPAS_depth} and \ref{fig:JPAS_PSF} show the distribution for the different tiles and bands. 

The difference in the depths is mostly due to the night conditions during the observations and to the final number of coadded exposures. Most of the images are above the target depth for each band, but the difference among bands is something that we will need to take into account when exploring the properties of spatially resolved galaxies, since outermost regions of galaxies may not reach the same S/N in all filters. Therefore, the number of filters with high S/N in those regions may not be enough to perform a reliable SED-fitting (see Chapter~\ref{chapter:spatiallyresolved}).

Concerning the \gls{PSF}, most of them are below 2 arcsecs. We shall take this value as an upper limit for the selection of the sample of spatially resolved galaxies (see Chapters~\ref{chapter:code} and \ref{chapter:spatiallyresolved}) and we will use the models provided for each galaxy and band. The larger values in the redder bands are due to the survey schedule, since these bands were planned to be observed last. During those campaigns, the EGS/AEGIS field reached the lowest elevations, producing larger \gls{PSF}.

\subsection{Image treatment} \label{sec:mjp:images}
The data reduction process of \mjp \ is  as the one used for J-PLUS \citep{Cenarro2019}. For the single frames, this includes the bias, prescan, and overscan subtraction, the trimming, and the flat field, illumination, and fringing corrections. However, there were three main issues that required special attention in the \mjp \ data reduction:

\begin{itemize}
    \item Vignetting: due to the larger size of the CCD in comparison to the filters, which translates into a strong gradient of efficiency. To solve this, regions with low efficiency from the images were trimmed, from $9216 \times 9232$~pixels to $7777 \times 8473$~pixels. The final FoV is $0.27$~$\mathrm{deg}^2$
    \item Background patterns: images from \mjp \ presented background patterns with strong gradients and variations on time scales of a few minutes. There are two types of patterns. The first type are straight patterns, which are due to the optics of the camera and were corrected via illumination correction. The second type shows circular patterns, likely due to small variations in the central wavelengths of the filters. These ones require a much careful subtraction that is detailed in appendix B in \cite{Bonoli2020}. The working hypothesis is that these patterns are the same in images taken close in time, as well as independent of the sky position, so a median combination of the images with objects in different positions should keep the background structure while removing the sources.
    \item Fringing:  in filters redder than J0470 ($\lambda > 4700$~\AA). To remove these effects, master fringing images were constructed using all available images for each band, since this pattern is very stable across nights. However, some residual pattern can be found in the final images of a few filters, due to the low number of available images.
\end{itemize}

After all this processing, the images were combined using the Astromatic software \texttt{Swarp} \citep{Bertin2010Swarp}. All images were resampled to the pixel size of the camera ($0.23$~arcsec~$\mathrm{pixe}^{-1}$). This images were homogenised using \texttt{PSFEx} \citep{Bertin2011}, and the models of the PSF produced for each image are available at a given position of the image.

\subsection{Zero Points}\label{sec:mjp:ZP}
The photometric  calibration of the images of \mjp \ was mainly done adopting the procedure presented in \cite{Lopez-sanJuan2019}, but using an adaptation of the last step, which implies the use of BOSS stellar spectra to calibrate the absolute colour, as summarised in \cite{Bonoli2020}. This was required because there are no high quality white dwarfs in the \mjp \ area, so an alternative is required for that final step of the calibration. 

The method presented in \cite{Lopez-sanJuan2019} is based on the stellar locus regression \citep{Covey2007,High2009,Kelly2014,Kujiken2019}. The working hypothesis of this method is that stars with different stellar parameters populate colour-colour diagrams in a particular way, which defines a well constrained region that depends on the colours used. The main advantage of this method is that it provides a consistent flux calibration for the bands, although it requires a band anchored with external calibration to be used as a reference.

The first step of the calibration process of \mjp \ is the selection of a sample of high quality stars for calibration. This first set of stars consists of all the sources in \mjp \ with $\mathrm{S/N}>10$ and a parallax measured with Gaia with  $\mathrm{S/N}>3$. These sources were corrected for galactic extinction using the 3D dust maps provided by \texttt{Bayestar17} \citep{Green2018}, using the parallaxes to derive the distance to the calibration stars. After retrieving the colour excess $E(B-V)$ with \texttt{Bayestar17}, the $G$-band absolute magnitude is estimated and the absolute $G$ band vs $G_\mathrm{BP} - G_\mathrm{RP}$ diagram is built, where the bands correspond to the $G$, $G_\mathrm{BP}$ (330--680~nm), and $G_\mathrm{RP}$ (680--10501~nm) broad bands from the \textit{Gaia} DR2 \citep{Gaia2018}. From the position in this diagram, stars are classified into giant Branch, main sequence, and white dwarfs stars. Then, the main sequence stars were selected for the next steps of the calibration.

The next step is the calibration of the broad band filters \rb, \gb, and \ib. This is done by crossmatching the calibration stars with the Pan-STARRS survey. For this purpose, the PSF corrected magnitudes from Pan-STARRS were compared to those from \mjp \ using a circular aperture of $6''$ of diameter. Then, a correction to account for the difference in apertures is applied, obtaining the zero points for these broad bands. A comparison was also made using J-PLUS broad bands, finding differences below $0.01$ mag.

The narrow bands are subsequently calibrated. This is done in two steps. The first one consists in the homogenisation of the narrow bands through the stellar locus. For each band, its instrumental magnitude $\chi_{\mathrm{inst}}$, was used to construct the de-reddened $(\chi_{\mathrm{inst}} - r)_0$ vs $(g-i)_0$ colour diagram, computing the offsets required to provide a consistent stellar locus among the four bands, an homogeneous instrumental photometry can be obtained for all \mjp. Secondly, these instrumental magnitudes are then transformed into magnitudes at the top of the atmosphere using stellar spectra from BOSS. For such purpose, each synthetic $(\xi -r)$ colour  from BOSS is compared with the instrumental magnitudes from \mjp \ to obtain the offset for each band, except for the \rb{} which is anchored to the Pan-STARRS photometry. Since the wavelength range from $u_{\mathrm{JAVA}}$ and $u_{\mathrm{JPAS}}$ is not covered by Pan-STARRS, J-PLUS was used instead for these two bands.

The zero points were also estimated using the common bands from J-PLUS, showing that they are consistent at the 4\% level. Therefore, an absolute error of $0.04$~mag was provided for all the bands as an upper limit.

\section{Parameters required for the analysis} \label{chap:mjp:catalogues}
In this section, we summarise the main catalogues from the \mjp \ survey used for this work. These mainly include the photo-$z$ catalogue from \cite{HC2021}, and the integrated magnitude catalogues of the galaxies in \mjp. The photo-$z$ will be used for the SED-fitting of the regions of the spatially resolved galaxies in \mjp \ in Chapter~\ref{chapter:spatiallyresolved}, as well as the SED-fitting of the integrated galaxies in the mJPC2470--1771 cluster in Chapter~\ref{chapter:cluster}. The integrated magnitudes will be used to test our tool for the analysis of the spatially resolved galaxies in Chapter~\ref{chapter:code}, and to obtain the properties of galaxies in Chapters~\ref{chapter:spatiallyresolved} and \ref{chapter:cluster}.

\subsection{Photometric redshifts} \label{sec:mjp:PHOTOZ}

\begin{figure}
    \centering
    \includegraphics[width=\textwidth]{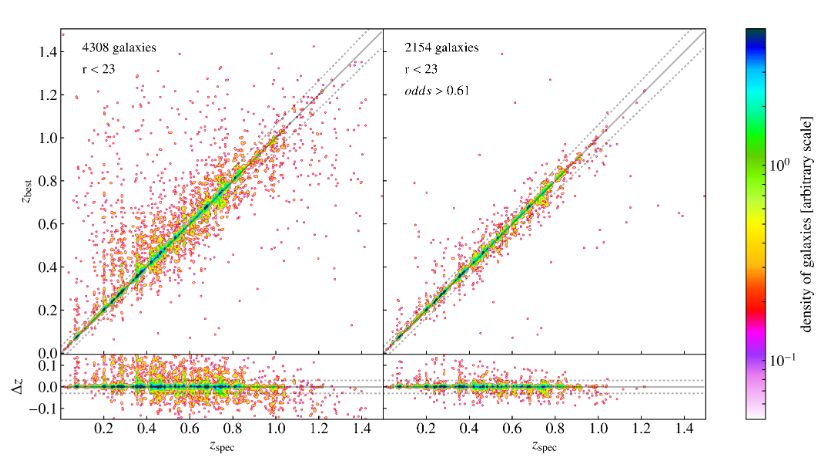}
    \caption[Comparison of the photo-$z$ and the spectroscopic redsifts of the galaxies in \mjp. Picture taken from \cite{HC2021}]{Comparison of the photo-$z$ and the spectroscopic redshifts of the galaxies in \mjp. Picture taken from \cite{HC2021}}
    \label{fig:Photoztest}
\end{figure}

The \jp \ filter system was designed in order to obtain extremely accurate photo-$z$ \citep{Bonoli2020}. The whole process for retrieving these redshifts is described and verified in \cite{HC2021}. In brief, the photo-$z$ are computed using the \texttt{JPHOTOZ} package, which is part of \texttt{JYPE}, the  data reduction pipeline \jp. This package also serves as an interface between the database and the codes for the computation of the photo-$z$, handling the required pre-processing and post-processing of the data. The code used for such computation is a modified version of \lephare \  \citep{Lephare2011}. This modification was mainly made  in order to increase the number of photometric filters used by the code, as well as the resolution of the redshift search range, from $z=0$ to $z=1.5$ with steps of $\delta z = 0.002$.

The code \lephare \ is based on a template fitting method, using $\chi^2$ as an estimator of the goodness of the fit of the observed data using synthetic photometry generated form a set of templates. The code computes the \gls{zPDF} weighting the log-likelyhood distribution,  with a prior obtained from the magnitude distribution of non-biased galaxy samples from VVDS \citep[][]{VVDS2005} at different redshifts. 

A dedicated library of templates, \texttt{CEFCA\_MINIJPAS}, was built. This library consists of  50 templates \citep[generated with \texttt{CIGALE}][]{Cigale2019} whose combinations provided the best estimations of the photo-$z$ in an iterative process \citep[see][for further details]{HC2021} and  provide solid estimations of the true redshift of the galaxies (see Fig.~\ref{fig:Photoztest} for a comparison of the estimated photo-$z$ and the spectroscopic measurements of the galaxie in \mjp).

The package provides as output not only the \gls{zPDF} of the galaxy, but also a set of scalar parameters that summarise the distribution and may be more convenient to work with, depending on the scientific case. These parameters can be found in the \texttt{PhotoZLephare\_updated} table from the \mjp \ data. The list is: 

\begin{itemize}
    \item \texttt{Z\_ML}: the median value of the \gls{zPDF}.
    \item \texttt{PHOTOZ}: the redshift at which the \gls{zPDF} reaches its absolute maximum.
    \item \texttt{CHI\_BEST}: the $\chi^2$ value of the best fitting model.
    \item \texttt{Z\_BEST68\_HIGH} and \texttt{Z\_BEST68\_LOW}: the limits of the 68\% confidence interval.
    \item \texttt{PHOTOZ\_ERR}: the 1$\sigma$ uncertainty in \texttt{PHOTOZ}, computed as 
    \begin{equation*}
        0.5 \times \left ( \texttt{Z\_BEST68\_HIGH} - \texttt{Z\_BEST68\_LOW} \right ).
    \end{equation*}
    \item \texttt{ODDS}: this parameter aims to represent the probability of the relative error in \texttt{PHOTOZ} (with respect to the spectroscopic redshift) to be smaller than 3~\%, this is: $ \frac{\left | \texttt{PHOTOZ} - z_{\mathrm{spec}} \right | }{1 + z_{\mathrm{spec}} }< 0.03 $. In practice, this quantity is calculated as the area of the redshift probability distribution
    conditional to the source being a galaxy, $P(z|\mathrm{G})$, within an interval equivalent to the 3~\% of the error, this is:
    \begin{equation*}
        \texttt{ODDS} = \int_{\texttt{PHOTOZ} - d} ^{\texttt{PHOTOZ} +d} P(z|\mathrm{G})dz
    \end{equation*}
    where $d = 0.03 \times (1+\texttt{PHOTOZ})$. 
\end{itemize}

\subsection{Flux and magnitudes catalogues} \label{sec:mjp:fluxmag}
The measurements of the fluxes and magnitudes obtained by running \sext \ on the \mjp \ can be retrieved from the CEFCA portal. There are several apertures available, for different aims. In this work, we use the following:

\begin{itemize}
    \item \magauto: this photometry aims at providing an estimation of the total flux of the galaxy. The values of the flux and the magnitudes are calculated within an adaptive, elliptical aperture that uses the Kron radius \citep{Kron1980} as a reference for its size. 
    \item \magpetro: this photometry also aims at providing an estimation of the total flux of the galaxy, using an adaptive, elliptical aperture, but the Petrosian radius \citep{Petrosian1976,Blanton2001,Yasuda2001} as a reference for the size.
    \item \magpsfcor: this photometry aims at providing reliable colour measurements, at the cost of a measurement of the flux that is lower than the total one. This photometry is calculated following the process described in \cite{Molino2019}. Very briefly, the magnitudes are calculated within an elliptical aperture with a semi-major axis equal to 1~Kron radius. Then, for each band $j$ with a broader \gls{PSF} than the reference band, a correction term is applied, this is:
    \begin{equation*}
        PSFCOR_j = REST_j + REST_r - REST(j)_r
    \end{equation*}
     where $PSFCOR_j$ is the \magpsfcor \ magnitude of the $j$ band, $REST_j$ and $REST_r$ are the magnitudes within the elliptical aperture with a semi-major axis equal to 1~Kron radius in the $j$ and reference bands, respectively, and $REST(j)_r$ is the magnitude in the reference band within the same aperture, but after degrading the image to the same \gls{PSF} as in the $j$ band. See also \cite{Molino2014,HC2021} for more details.
    \item \texttt{MAG\_APER}: These photometries correspond to fixed, circular apertures. The number in the name indicates the diameter of the aperture in arcseconds.
\end{itemize}

We will mainly use the \magauto, \magpetro, and \texttt{MAG\_APER} photometries to test our tool for the analysis of spatially resolved galaxies in Chapter~\ref{chapter:code}, we will derive the integrated properties of these galaxies using the \magauto \ photometry, and we will use the \magpsfcor \ photometry to obtain the integrated properties of the galaxies in the mJPC2470--1771 cluster in Chapter~\ref{chapter:cluster}.

\chapter{Tools, catalogues, and scientific results from the \mjp \ survey} \label{chapter:minijpresults}

The \mjp \ survey has already proven part of its capabilities as a survey for galaxy evolution and for studies related to environment through several published works. In this chapter, we summarise some of these papers, since they provide a more specific context for the science developed in this thesis. Additionally, although they are not part of the scientific results of this thesis, the author of the thesis is coauthor of these publications, because they are part of the tasks developed by the author during his research. 

\section{Determination of stellar population properties} \label{chap:mjp:tools}

\subsection{\baysea} \label{sec:mjp:BaySR}
\begin{figure}
    \centering
    \includegraphics[width=\textwidth]{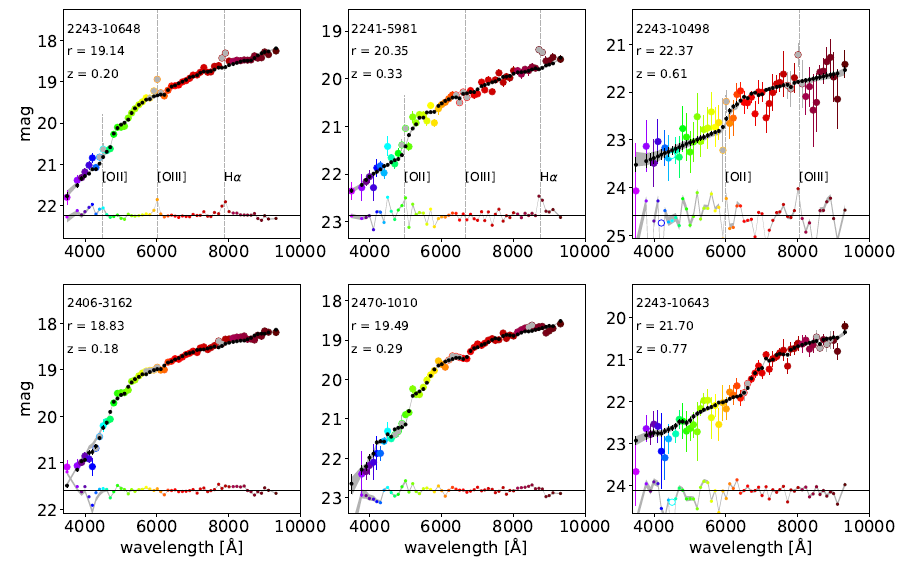}
    \caption[Example of the SED--fitting results from \baysea. Picture taken from \cite{Rosa2021}]{Example of the SED--fitting results from \baysea. Top panels show three blue galaxies. Bottom panels show three red galaxies. Coloured circles represent the observational data. Each colour represents a filter. Gray circles represent filters that are masked during fitting due to the possible presence of emission lines. The wavelength of these emission lines is shown with grey, vertical, dashed lines.  Black dots represent the magnitudes fitted by \baysea. Small coloured dots represent the difference between the observed and fitted magnitudes. The grey shade shows the $\pm \sigma$ of the difference. Black, horizontal, solid line shows the null value of the difference. Picture taken from \cite{Rosa2021}}
    \label{fig:Bayseagaltest}
\end{figure}

In order to obtain the stellar population properties of integrated galaxies and of the regions of spatially resolved galaxies, we need a SED-fitting code. Our choice here is \baysea, a parametric, Bayesian code that uses a \gls{MCMC} approach to explore the parameter space and find the set of parameters that  best fits the observed magnitudes. This code was originally developed as part of the PhD thesis by \cite{RafaTesis} and used in the work by \cite{Rafa2016} in order to fit the spectra of CALIFA galaxies along with data from GALEX. \baysea \ is the result of the adaptation of the former code in order to work with data from \jp \ and \jplus, and it will be published in a future work (de Amorim et al., in preparation). In the meantime, \baysea \ has been used to obtain the stellar population properties of the integrated galaxies in \mjp \ \citep{Rosa2021} obtaining similar results to other non-parametrical codes, such as \texttt{MUFFIT} \citep{Luis2015}, \texttt{AlStar} (Cid Fernandes, in prep.) or \texttt{TGASPEX}. It has also been used in other works in order to obtain the stellar population properties of galaxies according to different environment \citep{Rosa2022,Julio2022}. In this section we shall describe the basics of the code \citep[mainly explained by][]{RafaTesis} and the options used for these work.

Throughout this thesis, we assume a $\tau$-delayed model, this is, we assume that the \gls{SFH} of each galaxy is described as:
\begin{equation}
    \psi(t)=\frac {M_{ini}} {\tau^2 \left [ 1- e^{-\frac{t_0}{\tau}} \left ( \frac{t_0}{\tau}+ 1 \right )   \right ]}(t_0-t)e^{-\frac{t_0-t}{\tau}}
\end{equation}
where $t$ is the look-back time, $M_{ini}$ is the total mass of stars formed during the life of the galaxy (without taking into account the stellar mass loss due to stars reaching the end of their lifetime), $t_0$ is the time where the galaxy started to forming stars and $\tau$ is the e-folding time, which, intuitively, represents how fast (small $\tau$ values) or slowly (large $\tau$ values) the star formation decays. Therefore, the parameters used to fit are $t_0, \tau, A_V$ and $Z$, where $A_V$ is the dust extinction and $Z$ the stellar metallicity. The code obtains the SFH for the given set of parameters and normalises it to 1~$M_\odot$. The next step is to calculate the synthetic spectrum using \gls{SSP}:
\begin{equation}
    F(\lambda, t_0, \tau, A_V, Z) = 10^{-0.4 q_\lambda A_V} \int SFH (t, t_0, \tau, A_V, Z) \times SSP_{t,Z}(\lambda) dt,
\end{equation}
where $SSP_{t,Z}(\lambda)$ is the spectrum of a SSP of age $t$ and metallicity $Z$. Then, the magnitudes of the synthetic spectra are calculated using the response curve of the filters:
\begin{equation}
    m_{AB} = -2.5 \log \left ( \frac{\int \lambda R_X (\lambda) F(\lambda, t_0, \tau, A_V, Z)  d\lambda}{\int  R_X (\lambda) d\lambda / \lambda} \right ) -2.41 ,
\end{equation}
where $R_X (\lambda)$ is the response function of the filter. In this step, \baysea \ also uses a pre-computed matrix of magnitudes as function of the extinction and redshift, since through a Taylor expansion it is possible to find an expression for this matrix, reducing the computation of further fittings.

The sampling of the parameter space is done using bayesian statistics. Applied to our case, Baye's theorem states that the probability for a set of parameters $t_0, \tau, A_V, Z$ given an observed magnitude $M$ is 
\begin{equation}
    p(t_0, \tau, A_V, Z | M) = \frac{p(t_0, \tau, A_V, Z)p(m | t_0, \tau, A_V, Z)}{ p(M)}
\end{equation}
where $p(t_0, \tau, A_V, Z)$ is the prior and $p(m | t_0, \tau, A_V, Z)$ the likelihood function. The Bayesian approach is useful since it allows to marginalise over the parameters 
\begin{equation}
    p(X|M) = \int p(Y,X|M) dY
\end{equation}
where $X$ is one of the properties $t_0, \tau, A_V, Z$ and $Y$ represents the other parameters. Also, using the samples of the MCMC it is possible to obtain the expected value of a function that depends on our stellar population properties $f(t_0, \tau, A_V, Z)$ as 
\begin{equation}
    <f(t_0, \tau, A_V, Z)> = \int p(t_0, \tau, A_V, Z)|M)f(t_0, \tau, A_V, Z) dt_0 d\tau dA_V dZ.
\end{equation}

The prior used is an uniform prior, with a range of values to look for the solution $t_0 = [0,0.99]$ (in units of the age of the universe, taken from z), $\tau = [0.1, 10]$~Gyr, $A_V = [0,2]$~mag. The likelihood function used is 
\begin{equation}
    \chi^2 =\sum^{N_{mag}}_{i = 1} \left( \frac{M_i -M_{synt, i}}{\Delta m_i} \right )^2, 
\end{equation}
where $M_i$ is the observed magnitude, $M_{synt, i}$ is the synthetic magnitude obtained through the models and $\Delta m_i$ is the error of the magnitude. The code then aims to find the model with the best $\chi^2$. The MCMC algorithm used to sample the parameter space is the python implementation by \cite{Emcee}. The set of \gls{SSP} used for the analysis throughout this whole thesis is an updated version of the models by \cite{C&B2003}. We show an example of the fitting of the \js \ of  six galaxies, three blue and three red, in Fig.~\ref{fig:Bayseagaltest}. We note here that throughout this thesis, we use the AB magnitude system \cite{Oke1983}, unless otherwise stated. Additionally, we note that the models used in this thesis are computed using the \gls{IMF} by \cite{Chabrier2003} and the latest versions of the \cite{C&B2003} stellar population synthesis models \citep{Plat2019}.

\subsection{The stellar population properties in \mjp}

The code \baysea \ was used to obtain the stellar population properties of the galaxies in \mjp \ in the work by \cite{Rosa2021}. In total, $\sim 8000$ galaxies with $r_\mathrm{SDSS} \leqslant 22.25$ in the \magauto \ catalogue and $z\leqslant 1$ were studied using \baysea, as well as three additional non-parametric SED-fiting codes: \muff \ \citep{Luis2015,Luis2023}, \alstar, and \tgas. The results and conclusions obtained with the different codes are compatible among them. 

The median value of the mean S/N in the narrowband filters for the sample studied was $\sim 8$, with $25$~\% of the sample with $\mathrm{S/N} \leqslant 3$, obtaining a standard deviation, representing the precision of the measurement, of $0.1 \pm 0.05$~dex, $0.15 \pm 0.06$~dex,  and $0.34 \pm 0.1$ for the rest frame $(u-r)_\mathrm{res}$ colour, the stellar mass, and extinction $A_V$,  respectively and  $0.14 \pm 0.05$~dex, $0.25 \pm 0.06$~dex, for the the mass-weighted ages of red and blue galaxies. In fact, if only galaxies with mean $\mathrm{S/N} \geqslant 10$ are selected, these standard deviations improve up to  $0.04 \pm 0.02$~dex, $0.07 \pm 0.03$~dex,  and $0.20 \pm 0.09$, for the rest frame $(u-r)_\mathrm{res}$ colour, the stellar mass, and extinction $A_V$,  respectively, and to  $0.16 \pm 0.07$~dex, for the the mass--weighted ages. The standard deviation of the stellar metallicity for this S/N cut is  $0.42 \pm 0.25$~dex. The distribution of the properties is similar for all the codes, although larger discrepancies are found in the stellar ages, most likely due to the different \gls{SFH}. 

\begin{figure}
    \centering
    \includegraphics[width=0.65\textwidth]{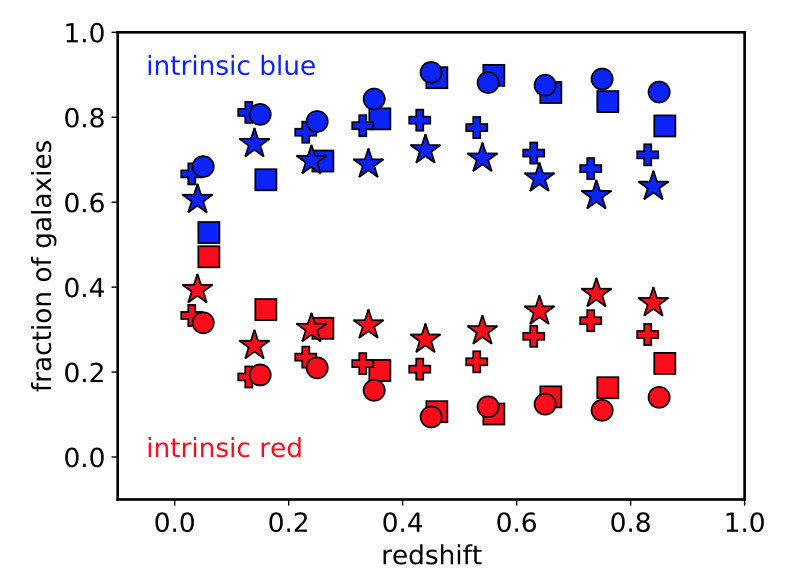}
    \caption[Evolution of the fraction of red and blue galaxies with redshift. Picture taken from \cite{Rosa2021}]{Evolution of the fraction of red and blue galaxies with redshift. Red symbols represent the fraction of red galaxies. Blue symbols represent the fraction of blue galaxies. Circles represent results obtained with \baysea, squares results from \muff, stars results from \alstar, and crosses results from \tgas. Picture taken from \cite{Rosa2021}}
    \label{fig:frac_vs_z}
\end{figure}

Using the stellar mass and the dust-corrected, rest frame $(u-r)$ to build a stellar mass--colour diagram, the red and blue galaxy populations were easily identified. The evolution of the fraction of each  population can be seen in Fig.~\ref{fig:frac_vs_z}. Results vary with the chosen code, but in general, the fraction of red galaxies decreases with the redshift, as blue galaxies become more dominant at earlier epochs. This result shows the consequence of the quenching processes in galaxy evolution. The properties of the blue and red galaxies were also found to be well characterised by their position in the mass--colour diagram.

\begin{figure}
    \centering
    \includegraphics[width=0.95\textwidth]{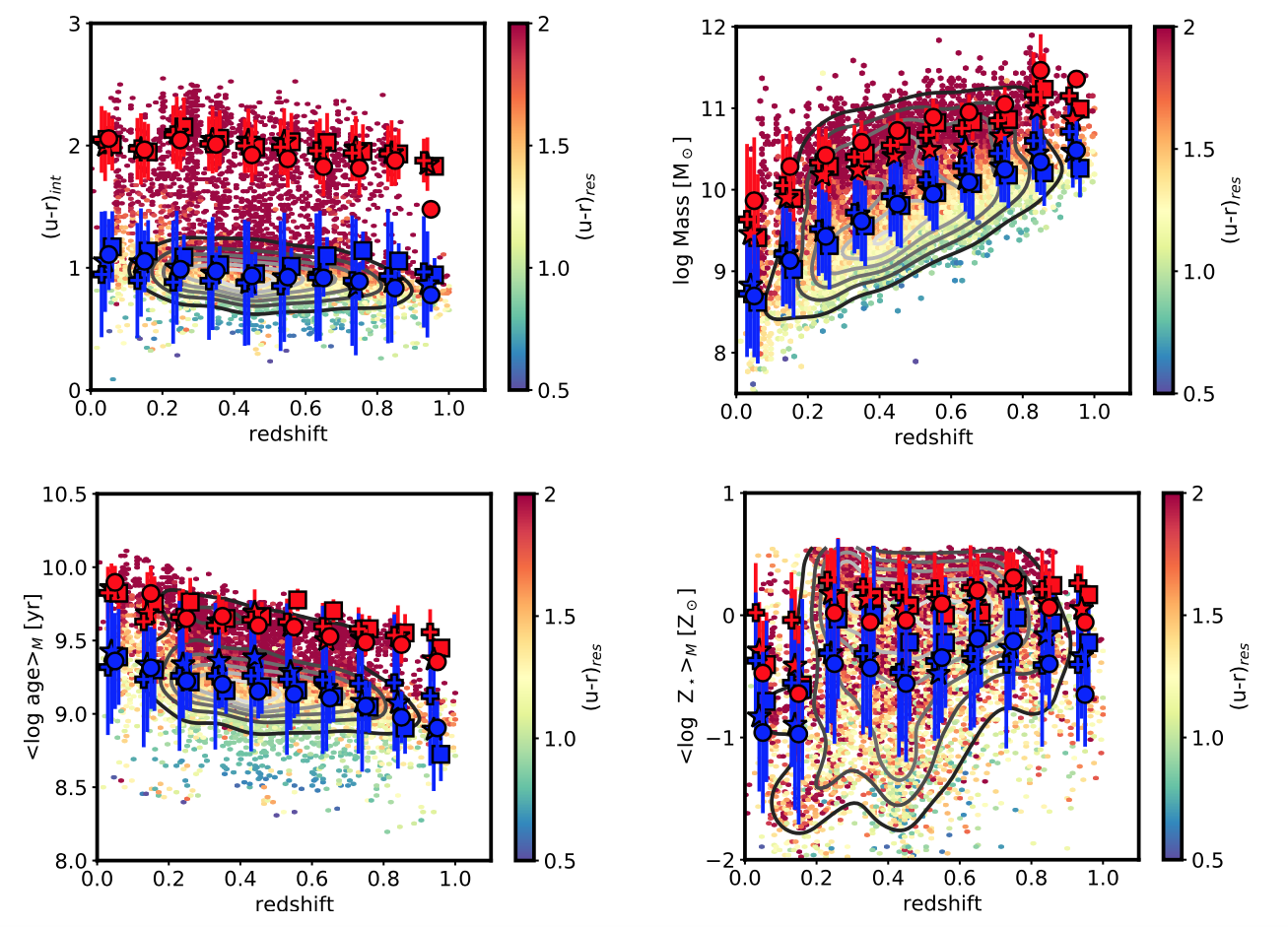}
    \caption[Evolution of the stellar population properties with the redshift. Picture taken from \cite{Rosa2021}]{Evolution of the stellar population properties with the redshift. Red symbols represent the fraction of red galaxies. Blue symbols represent the fraction of blue galaxies. Circles represent results obtained with \baysea, squares results from \muff, stars results from \alstar, and crosses results from \tgas. The colour scale indicates the rest frame $(u-r)$ colour. From left to right, top to bottom:  dust corrected rest frame $(u-r)$ colour, stellar mass, mass--weighted age,  and stellar metallicity. Picture taken from \cite{Rosa2021}}
    \label{fig:SPP_evol}
\end{figure}

Lastly, the data available from the survey allowed for studying the evolution of the stellar population properties up to $z \sim 1$ (see Fig.\ref{fig:SPP_evol}). All the properties of blue and red galaxies remain well differentiated at all epochs. The stellar mass increases with redshift, due to the selection bias in any flux-limited sample, since only the brightest and most massive galaxies are observed at higher redshifts \citep[for more information regading the completness of \mjp, see][]{Luis2023}. Probably due to the same reason, no significant metal enrichment was found since $z \sim 1$. On the other hand, the colour $(u-r)_\mathrm{int}$ is bluer for both blue and red galaxies at higher redshifts. The stellar ages  clearly decrease with redshift, indicating either ongoing star-formation, or a biased sample of low-mass blue galaxies at higher redshifts, due to the aforementioned flux-limited nature of the survey. Nonetheless, all these results served to prove the power of the \jp \ filter system for retrieving the stellar population properties and study their evolution. 

\section{Emission lines} \label{sec:mjp:ANN}

In this thesis, we also try to offer an estimation of the \gls{EW} of the most important emission lines in our wavelength range. For such purpose, we use the \gls{ANN} developed by \cite{Gines2021}. In that work, two different types of ANN were constructed: one for predicting the values of the EW of $\mathrm{H}\alpha$, $\mathrm{H}\beta$, $\mathrm{[NII]}$, and $\mathrm{[OIII]}$, shortened as $\mathrm{ANN}_R$, and the other one for classifying galaxies into galaxies with and without line emission, shortened as $\mathrm{ANN}_C$. In this section, we summarise the most important details of the first type of ANN, which is the one used for this thesis.

\subsection{Artificial Neural Networks}
The ANN is built using the \texttt{Tensorflow} \citep{Abadi2015} and \texttt{Keras} \citep{chollet2015keras} libraries. The structure consists of three type of layers: the input layer, the hidden layer and the output layer. The input layer takes as input the colours of all the filters with respect to the filter where $\mathrm{H}\alpha$ is observed, this is the colour $c_i = m_{AB}(\mathrm{H}\alpha) - m_{AB}(J_i)$, where $m_{AB}(\mathrm{H}\alpha)$ is the magnitude in the filter where $\mathrm{H}\alpha$ is detected and $ m_{AB}(J_i)$ is the magnitude in the i-th filter, different from the first filter. To take the redshift into account, several ANN are trained, one for each redshift, going from $z=0$ to $z=0.35$ with a step of $0.001$. The hidden layers are composed  of 2 layers, with 20 neurons each. Each neuron is connected to all the previous neurons, and their goal is to minimise the loss function. The function used for these ANN is a \gls{ReLU} activation function \citep{ReLU}. The output layer provides the estimations of  $\mathrm{H}\alpha$, $\mathrm{H}\beta$, $\mathrm{[NII]}$, and $\mathrm{[OIII]}$.

In order to train the ANN, a set of spectra from MANGA \citep{MANGA2015} and CALIFA \cite{CALIFA2012} were used, after applying the filters of \mjp \ system to these spectra, in order to generate synthetic \js. This provide a wide variety of spectra from different spaxels (star-forming galaxies, AGN hosts, high  and low emission spaxels, etc.) which ensures that the ANN is trained covering a wide range of possibilities, and betters the prediction of the ANN no matter what type of galaxy or region we are studying.  The same method is applied to the SDSS sample used to test the ANN. We refer the reader to the original paper \citep{Gines2021} for the details of how the errors are taken into account as well as the missing points.

\subsection{Summary of results obtained with the ANN. }
\begin{figure}
    \centering
    \includegraphics[width=\textwidth]{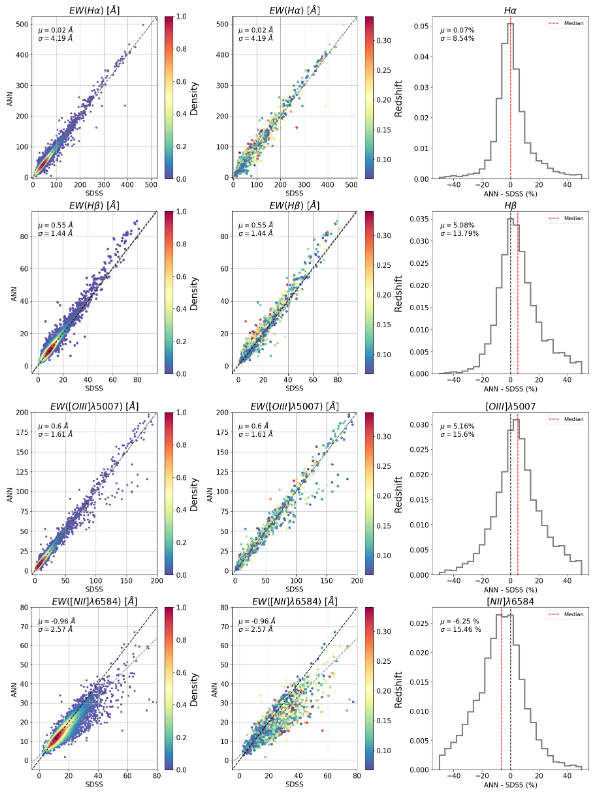}
    \caption[Comparison of the predicted EW of $\mathrm{H}\alpha$, $\mathrm{H}\beta$, $\mathrm{[NII]}$, and $\mathrm{[OIII]}$ with the actual measurement from SDSS spectra. Picture taken from \cite{Gines2021}]{Comparison of the predicted EW of $\mathrm{H}\alpha$, $\mathrm{H}\beta$, $\mathrm{[NII]}$, and $\mathrm{[OIII]}$ with the actual measurement from SDSS. Left panels shows the 1:1 relation with points colour coded with the point density. Right panels show the same relation with points colour coded with the redshift of the object. Black lines are the 1:1 relation. Grey lines represent the best fit. Right panel shows the relative difference of the predicted and measured value. Red lines represent the median value. Picture taken from \cite{Gines2021}}
    \label{fig:ANNtest}
\end{figure}

Results from \cite{Gines2021} mainly show the accuracy of the trained ANNs. The $\mathrm{ANN}_C$ is capable of classifying galaxies into galaxies with or without emission lines beyond the contrast that can be measured in \jp \ using the method by \cite{Pascual2007}, which is $\sim 16$~\AA. On the other hand, the $\mathrm{ANN}_C$ can predict the EW of $\mathrm{H}\alpha$, $\mathrm{H}\beta$, $\mathrm{[NII]}$, and $\mathrm{[OIII]}$ with a relative standard deviation of $8.4$~\%, $13.7$~\%, $14.8$~\%, and $15.7$~\%, respectively, a relative bias of $0.03$~\%, $5.0$~\%, $4.8$~\%, and $-6.4$~\%, respectively, and a minimum measurable EW of 18~\AA, 6~\AA, 40~\AA, and 13~\AA, respectively. Additionally, this ANN is capable of constraining the ratios of $\mathrm{[NII]/H}\alpha$ and $\mathrm{[OIII]/H}\beta$ within $0.092$~dex and $-0.02$~dex, respectively, proving the usefulness of the tool. A comparison among the predicted EW and their real values can be seen in Fig.~\ref{fig:ANNtest}. Additionally, the ANN trained in \cite{Gines2021} were used also used by \cite{Gines2022} to characterise the emission lines galaxies down to $z < 0.35$ in \mjp, which allowed to retrieve the \gls{SFMS}, as well as the cosmic evolution of the \gls{SFR} density.

\section{The AMICO catalogue} \label{sec:mjp:AMICO}

Since our aim is to study the effects of  environment, we need a code that allows us to discern whether a galaxy is in a dense environment (groups/clusters) or in the field. For such purpose we adopted the results in the AMICO\footnote{\gls{AMICO}} catalogues \citep{Maturi2023}. The \gls{AMICO} code \citep[][]{Maturi2005, AMICO} was adapted and applied to \mjp \ data by \cite{Maturi2023}. Here, we offer a summary of the most relevant aspects of AMICO for our work, as well as summary of the results by \cite{Maturi2023} and their implications for our work.

\subsection{The AMICO code}

This code is based in the Optimal Filtering technique. As explained by \cite{Maturi2005}, the working hypothesis of this method is that the data can be described by a model $M$, multiplied by a normalisation function $A$ plus a noise component $N$, this is, $D(x) = A(x) \times M(x) + N(x)$, where $x$ is a set of parameters. This way, the product $A \times M$ would represent the actual signal from the data. The goal is to construct a linear filter $\Psi (x)$ so that when it is convolved with the data, it produces an estimate of the amplitude 
\begin{equation}
    A_{\mathrm{est}} (x) = \int D (x) \Psi(x - x')d^{n}x',
\end{equation}
where n is the dimension of $x$. This filter must satisfy that the average error $\left < A_{\mathrm{est}} - A \right >$ vanishes and that the measurement noise $\sigma$ must be minimal, where $\sigma^2 = \left < (A_{\mathrm{est}} - A)^2 \right >$. More details about Optimal Filtering can be found in \cite{Maturi2005} and  \cite{Bellagamba2011}. In the rest of the section, we focus in our scientific case, which is described mainly in \cite{Maturi2023}, where the modification and application of AMICO to the \mjp \ data is described. Some details about the expressions of the parameters are specified in the works by \cite{AMICO} and \cite{Bellagamba2019}.

\begin{figure}
    \centering
    \includegraphics[width=\textwidth]{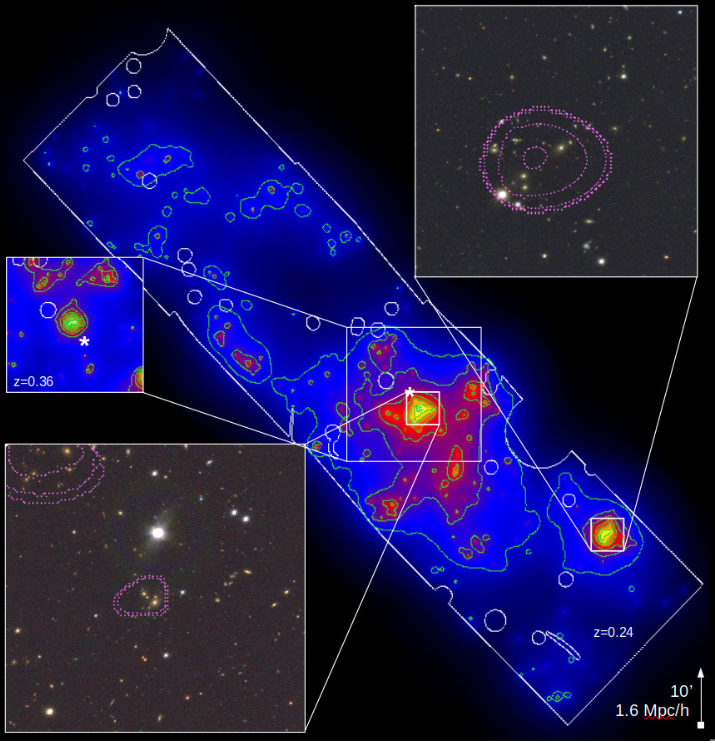}
    \caption[Amplitude response from AMICO in a slice of the \mjp \ survey at $z=0.24$. Picture taken from \cite{Maturi2023}]{Amplitude response from AMICO in a slice of the \mjp \ survey at $z=0.24$. The white lines represent the survey limits and masked areas. The boxes show the RGB images from two detections at $z=0.36$ and $z=0.24$, with the X-ray countours from \textit{Chandra} data in violet. Picture taken from \cite{Maturi2023}}
    \label{fig:AMICOtest}
\end{figure}

The data $D$ that AMICO aims to describe is the density of galaxies. The set of parameters used for the description of the model is the putative angular position of the cluster detection, $\theta_c$, its putative redshift, $z_c$, and for each galaxy, its angular position $\theta_i$, its redshift distribution $p_i(z)$ and its \rb{} magnitude, $m_i$. The amplitude is then computed as 
\begin{equation}
    A(\theta_c, z_c) = \alpha^{-1} (z_c) \sum_{i=1}^{N_\mathrm{gal}} \frac{C(z_c;\theta_i - \theta_c, m_i) p_i(z_c)}{N(m_i,z_c)} - B(z_c),
\end{equation}
where $\alpha$ is the normalisation factor, $C$ is the cluster model, and $B$ accounts for the average contribution of the field galaxies to the total signal amplitude. The used filter function is, in consequence, $\Psi_c (\theta_c - \theta, m, z) = M(\theta_c - \theta, m, z) / (m, z)$, which is the Optimal Filter under the assumption that the noise is uniform and is produced by random Poissonian counts of galaxies \citep{AMICO}. As explained by \cite{AMICO}, regions with $A > 0$ represent regions with an overdensity of galaxies, which are potential groups and cluster candidates. The cluster model for \mjp \ is $C(z_c;\theta_i - \theta_c, m_i) = R(z_i, |\theta - \theta_i|)L(z_j;m)$, where $L$ is a Schechter luminosity function in the \rb{} band, obtained combining a passive and a star-forming population, and $R$ is the radial profile that describes the projected density distribution of cluster galaxies. An example of the amplitude response by AMICO applied to \mjp \ can be seen in Fig.~\ref{fig:AMICOtest}.

The most important parameter for our study provided by AMICO is the probabilistic association, given by the expression
\begin{equation}
    P_i(j) = \Tilde{P}_{f,i} \frac{A_j C(z_j;\theta_i - \theta_j, m_i) p_i(z_j)}{A_j C(z_j;\theta_i - \theta_j, m_i) p_i(z_j) + N(m_i;z_j)}
\end{equation}
where the $i$ index account for each galaxy, the index $j$ accounts for each detection, and the term $\Tilde{P}_{f,i} = \sum^{j-1}_k P_i(k)$ accounts for the probabilistic association of previous detections or cluster candidates. This term is introduced because clusters overlap in data space and more than one cluster association can be assigned to a galaxy through an iterative approach. This way, the sum of all the probabilistic associations to a detection, $\sum_{i=1}^{N_gal} P_i(j)$, provides an estimation of the number of visible galaxies belonging to said detection.

This probabilistic association has been used in previous works in order to classify galaxies and study the effects of environment on galaxy evolution \citep{Rosa2022, Julio2022}, and we will use it for the environmental classification of the spatially resolved galaxies in \mjp \ in Chapter~\ref{chapter:spatiallyresolved}, and for the selection of the galaxies in the mJPC2470--1771 cluster in Chapter~\ref{chapter:cluster}.

\subsection{Summary of results from AMICO}
The work by \cite{Maturi2023} applied AMICO on the \mjp \ data, finding 80 structures with a S/N ratio higher than $2.5$. Among them, 30 structures had S/N ratios larger than 3 and 11 of them had a S/N above $3.$5. The structures detected had masses down to $\sim 10^{13}$~$\mathrm{M_\odot}h^{-1}$, proving the capacity of AMICO and \jp \ to detect low mass groups. With these values, \jp \ can be expected to detect $\sim 2 \times 10^5$ structures over its final footprint of 8000~$\mathrm{deg}^2$.

Using data from \textit{Chandra} and XMM-Newton to estimate the mass of the clusters and groups, several mass-proxy scaling relations were derived. The results show that the best mass-proxy is provided by the amplitude $A$, and is the most robust one with respect to the redshift, due to the filtering formalism. The other mass-proxies defined rely on the magnitudes and stellar masses of the galaxies, and are therefore limited by the  absolute magnitude cutoff. Lastly, the probabilistic membership obtained by AMICO and the membership approach applied to the spectroscopic data from DEEP3 \citep{Cooper2012} show a good agreement. Results from this work show that the precision and sensitivy achieved with \jp \ filter system place the survey between spectroscopic and photometric surveys, which might allow for studying the clustering of galaxy clusters.

We note that AMICO was not used as tool itself by the author of this thesis, but rather the catalogues produced by it. However, the author has contributed to the publication of this work mainly by collaborating in the testing of the code through the analysis of the distribution of the properties of the galaxies in the different catalogues obtained.

\section{Role of galaxy groups in quenching processes}

The data from \mjp \ has also been used to study the role that the environment of the galaxy groups plays in the quenching process of galaxies. The results of this study, co-led by the author of this thesis, were published in the paper by \cite{Rosa2022}. Since this work is not part of the results of this thesis, but it is deeply related to the topic and provides a bigger picture of the scientific context, we shall provide a summary of the most important details and results from this work in this section.

In this work we used the AMICO catalogues from \cite{Maturi2023} to classify the galaxies in \mjp \ into those that are in the field and those that are in groups. We used the probabilistic association provided by the code, classifying galaxies as galaxies in groups if their probabilistic association $P_\mathrm{assoc} \geqslant 0.7$, and galaxies in the field if $P_\mathrm{assoc} \leqslant 0.1$. We also studied the effect that using different thresholds of $P_\mathrm{assoc}$ when selecting the groups population, and of the square of the inverse of distance to fifth closest galaxy, $\Sigma_5$, had on the results obtain for some properties. In total, we studied a sample of galaxies in 80 groups, with masses in the range $M^\star _\mathrm{group}\in [10^{10.5}$,$10^{12.5}]$~$M_\odot$. We also divided galaxies into red and blue galaxies. With this classification, we studied the relation among the colour of the galaxy, its environment, and its properties.

\subsection{Properties of galaxies in groups and field}
\begin{figure}
    \centering
    \includegraphics[width=\textwidth]{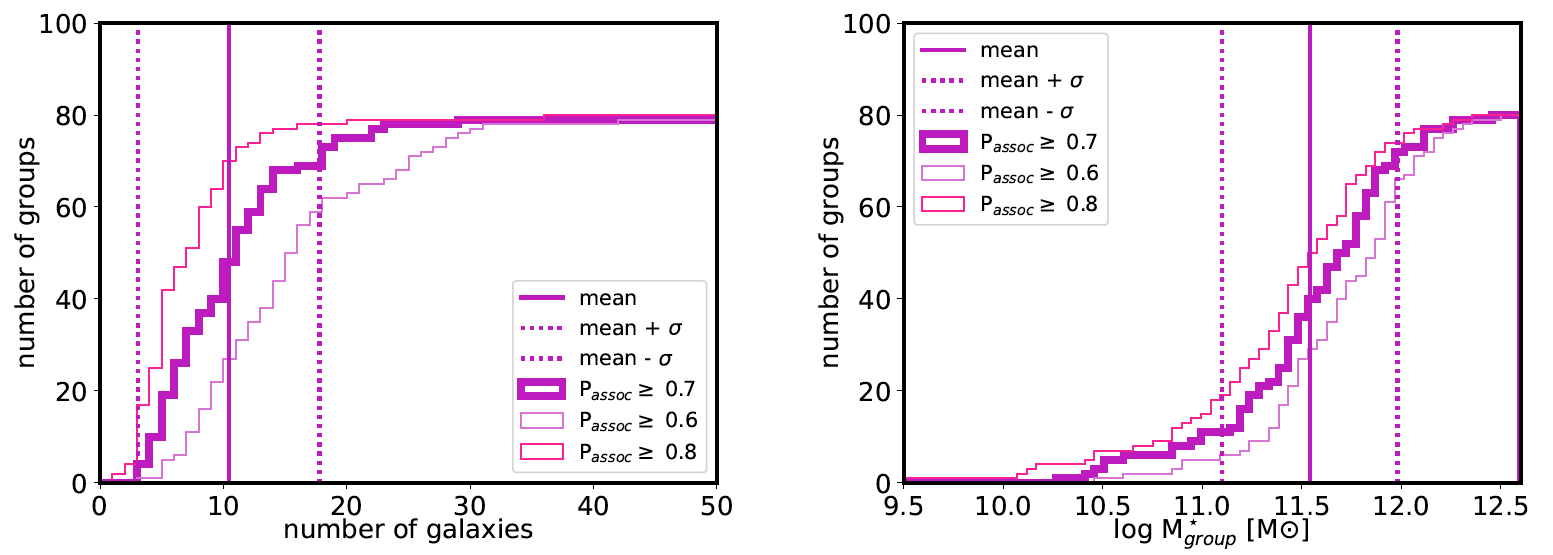}
    \caption[Cumulative distributions of the number of galaxies in the AMICO groups  and the group stellar mass. Picture taken from \cite{Rosa2022} ]{Cumulative distributions of the number of galaxies in the AMICO groups (left panel) and the group stellar mass (right panel). Picture taken from \cite{Rosa2022}}
    \label{fig:cumulativegroups}
\end{figure}

Our first finding was that the number of galaxies in groups greatly varies with the threshold used in $P_\mathrm{assoc}$ (see Fig.~\ref{fig:cumulativegroups}). Most of the groups have less than 20 galaxy members, with a mean value of $10.4$ galaxies per group ($\sim 5$ if the chosen threshold is $P_\mathrm{assoc} \geqslant 0.8$, $\sim 15$ if the chosen threshold is $P_\mathrm{assoc} \geqslant 0.6$). These values are more similar to those found for galaxy groups than for galaxy clusters, with one notable exception, the cluster mJPC2470--1771, which we studied separately in \cite{Julio2022} and that is part of the results of this thesis (see Chapter~\ref{chapter:cluster}).

Concerning the distribution of the stellar mass content of the groups, estimated by adding the stellar mass of all the galaxies in the groups, we found that there is still a dependency with the threshold of $P_\mathrm{assoc}$, although it is much weaker than in the previous case. We also found that half of the groups have masses lower that $10^{11.5}$~$M_\odot$, and that the most massive structure is the cluster mJPC2470--1771. This is the only structure more massive than $10^{12}$~$M_\odot$. These results show that the structures detected in AMICO are galaxy groups, rather than galaxy clusters. 

\begin{figure}
    \centering
    \includegraphics[width=\textwidth]{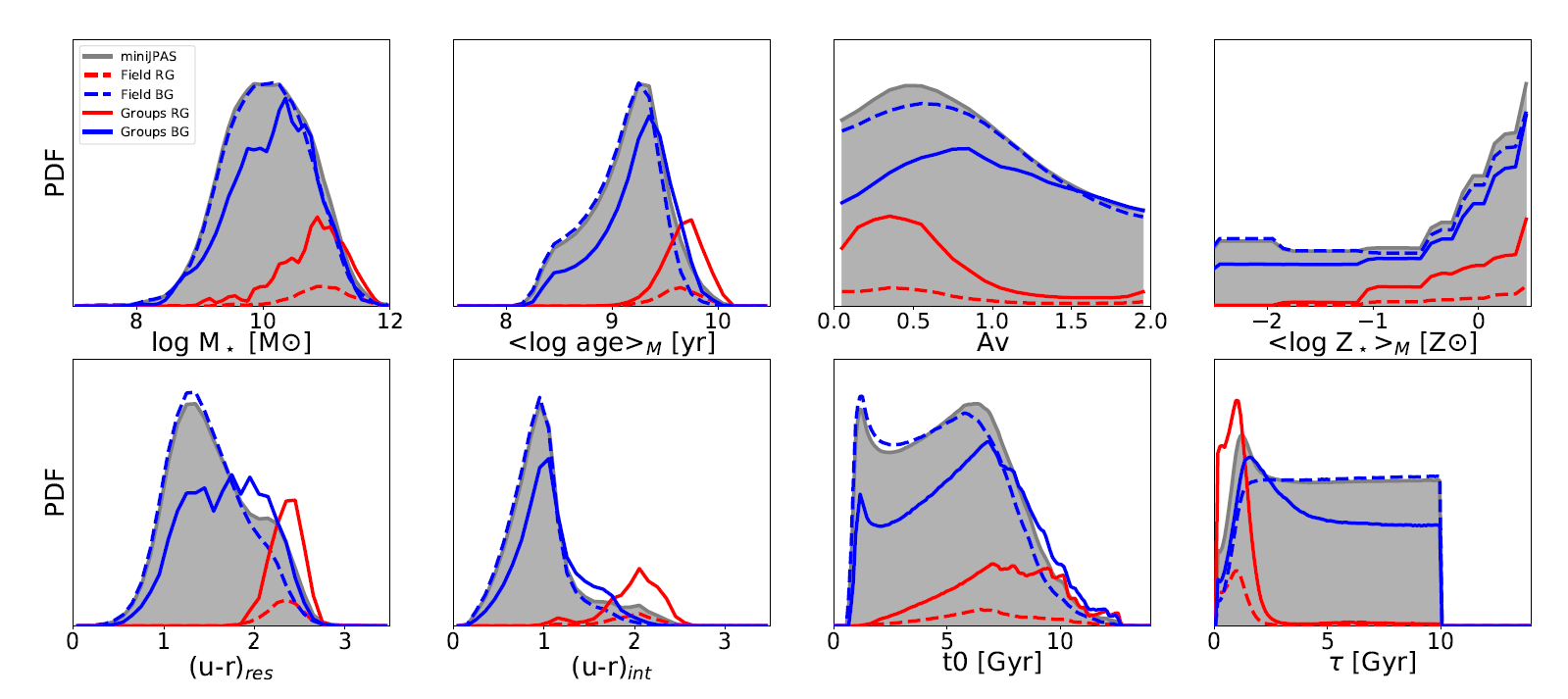}
    \caption[PDF of the stellar population properties of the galaxies in \mjp, by galaxy colour and environment. Picture taken from \cite{Rosa2022} ]{PDF of the stellar population properties of the galaxies in \mjp, by galaxy colour and environment. The grey area represents the distribution of the complete sample. Red solid lines represent the distributions of the red galaxies in the field. Red dashed lines represent the distribution the red galaxies in groups.  Blue solid lines represent the distributions of the blue galaxies in the field. Blue dashed lines represent the distribution the blue galaxies in groups. Picture taken from \cite{Rosa2022}}
    \label{fig:PDFRosa2022}
\end{figure}

We also find that the \gls{PDF} of the stellar population properties of the red galaxies shows no significant difference for red galaxies in the field and in groups, but the \gls{PDF} of blue galaxies in groups is slightly shifted towards older, redder, more metal-rich values of their properties, as well as to  lower values of the \gls{sSFR} and smaller values of $\tau / t_0$ than  blue galaxies in the field (see Fig.~\ref{fig:PDFRosa2022}).  These differences in the distributions translate into similar shifts of the distributions of the same properties of the general galaxy population in groups compared to the galaxy population in the field. 

\subsection{Fraction of red and blue galaxies}
\begin{figure}
    \centering
    \includegraphics[width=0.65\textwidth]{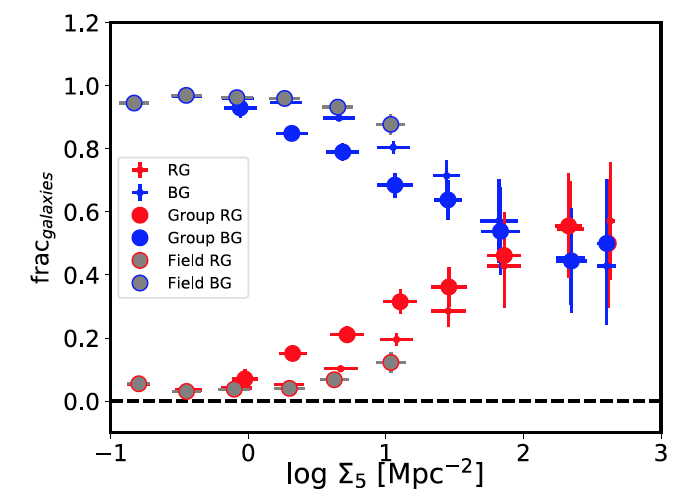}
    \caption[Fraction of blue and red galaxies as a function of the density of the environment. Picture taken from \cite{Rosa2022}]{Fraction of blue and red galaxies as a function of the density of the environment. Red symbols represent the fraction of red galaxies and blue symbols represent the fraction of blue galaxies. Coloured circles represent the fractions in groups, grey circles represent the fractions in the field, and dots represent the fractions in the complete sample.  Picture taken from \cite{Rosa2022}}
    \label{fig:frac_vs_sigma5}
\end{figure}

Based on previous works that had found relations between the density of the environment and other properties, like the pioneering work by \cite{Dressler1980} did with morphology, we studied the fraction of red and blue galaxies as a function of the density, here parameterised by the square of the inverse of the distance to the fifth closest galaxy, $\Sigma_5$, which is equivalent to the galaxy number density (see Fig.~\ref{fig:frac_vs_sigma5}). We chose to use the colour instead of the morphology since it correlates with many properties of the galaxies \citep{Kauffmann2003b, Kauffmann2003a}, and its also correlated with the environment of galaxies \citep[see e.g.][]{Lewis2002,Kauffmann2004,Rojas2005,Weinmann2006,Liu2015,Moorman2016}. We found that the fraction of red galaxies clearly increases as $\log \Sigma_5$ increases. Consequently, the fraction of blue galaxies decreases with $\log \Sigma_5$. Another interesting result is that, for a same value of $\log \Sigma_5$, the fraction of red galaxies is always higher in groups than in the field.

\begin{figure}
    \centering
    \includegraphics[width=0.65\textwidth]{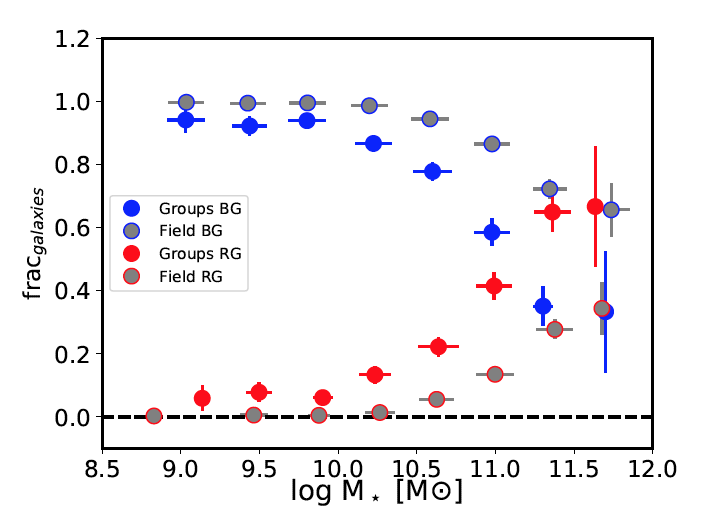}
    \caption[Fraction of blue and red galaxies as a function of the stellar mass. Picture taken from \cite{Rosa2022}]{Fraction of blue and red galaxies as a function of the stellar mass. Red symbols represent the fraction of red galaxies and blue symbols represent the fraction of blue galaxies. Coloured circles represent the fractions in groups, grey circles represent the fractions in the field, and dots represent the fractions in the complete sample.  Picture taken from \cite{Rosa2022}}
    \label{fig:frac_vs_mass}
\end{figure}

We also studied the relation of the fraction of red and blue galaxies with the stellar mass of the galaxy (see Fig.~\ref{fig:frac_vs_mass}), finding that the fraction of red galaxies in groups increases with the stellar mass, and is always higher in groups than in the field for masses $M_\star \geqslant 10^{10}$~$M_\odot$. This fraction also increases as the redshift decreases, from $z \sim 0.8$ to $0.1$, similarly to the Butcher-Oemler effect \citep{ButcherOemler1978,ButcherOemler1984}.

\subsection{Quenching fraction excess}
\begin{figure}
    \centering
    \includegraphics[width=0.65\textwidth]{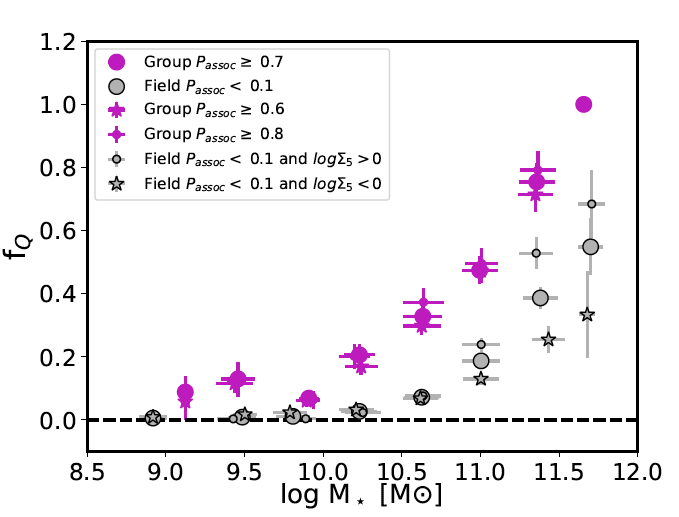}
    \caption[Fraction of quenched galaxies as a function of the stellar mass. Picture taken from \cite{Rosa2022}]{Fraction of quenched galaxies as a function of the stellar mass. The different symbols and colours represent results obtained using different criteria to select galaxies in the groups and in the field. The brown, continuous line represents results by \cite{McNab2021}, and dashed brown lines represent their 68~\% confidence interval.   Picture taken from \cite{Rosa2022}}
    \label{fig:fq_vs_mass}
\end{figure}

We studied the fraction of quenched galaxies, that we selected as those with $\mathrm{sSFR} \leqslant 0.1$~$\mathrm{Gyr}^{-1}$, following the same approach as \cite{Peng2010}. Results can be seen in Fig.~\ref{fig:fq_vs_mass}. This fraction is always higher in groups ($\sim 28$~\%) than in the field  ($\sim 5$~\%). This fraction remains approximately constant for groups regardless of the threshold value of $P_\mathrm{assoc}$, but it does vary with the population selected for field for  galaxies more massive than $10^{11}$~$M_\odot$. On the other hand, the threshold imposed in the selection of galaxies in groups does not affect greatly this fraction.

\begin{figure}
    \centering
    \includegraphics[width=0.65\textwidth]{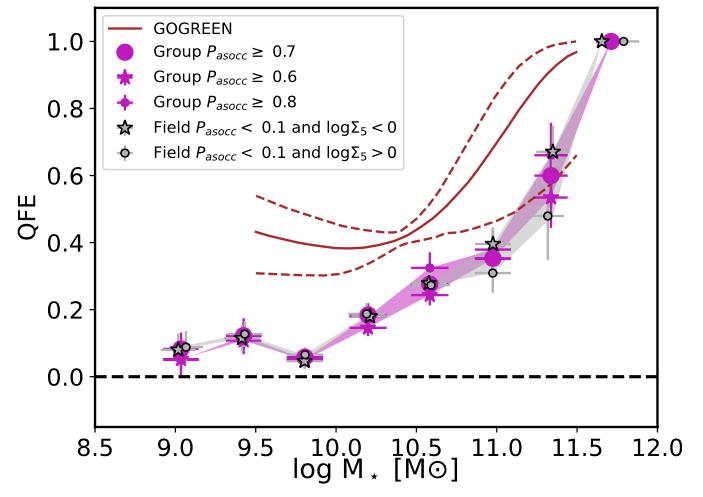}
    \caption[QFE as a function of the stellar mass. Picture taken from \cite{Rosa2022}]{QFE as a function of the stellar mass. The different symbols and colours represent results obtained using different criteria to select galaxies in the groups and in the field. The brown, continuous line represents results by \cite{McNab2021}, and dashed brown lines represent their 68~\% confidence interval.   Picture taken from \cite{Rosa2022}}
    \label{fig:QFE_vs_mass}
\end{figure}

Furthermore, in this work we also studied the \gls{QFE}, using the definition by \cite{McNab2021}:
\begin{equation}
    \mathrm{QFE} = (f^{\mathrm{F}}_{SF} - f^{\mathrm{G}}_{SF})/f^{\mathrm{F}}_{SF},
\end{equation}
where $f^{\mathrm{F}}_{SF}$ is the fraction of star--forming galaxies in the field and $f^{\mathrm{G}}_{SF}$ is the fraction of star--forming galaxies in groups. Our results show a strong dependence of the \gls{QFE} on the galaxy stellar mass for galaxies more massive than $M_\star = 10^{10}$~$M_\odot$, with the \gls{QFE} going from $0.4$ at low masses up to $0.6$ for masses larger than $10^{11.5}$~$M_\odot$  (see Fig~\ref{fig:QFE_vs_mass}) . For galaxies below that mass, the \gls{QFE} remains approximately constant, and it is negligible for masses lower than $10^9$~$M_\odot$. We also find a relation with the density of the environment (groups vs clusters) that we will comment at the end of Chapter~\ref{chapter:cluster}.

\begin{figure}
    \centering
    \includegraphics[width=0.65\textwidth]{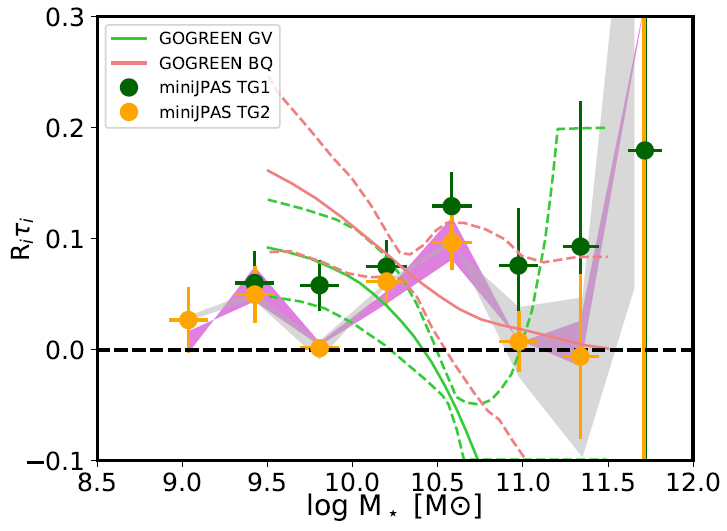}
    \caption[Relative abundance excess of transition galaxies as a function of galaxy stellar mass. Picture taken from \cite{Rosa2022}]{Relative abundance excess of transition galaxies as a function of galaxy stellar mass. Green and orange points represent values obtained for the sets of trasition galaxies TG1 and TG2, respectively. Magenta and grey shades correspond to different limits in $P_\mathrm{assoc}$ when defining the galaxy population in groups and fields, respectively. Light green and coral lines are the results from \cite{McNab2021} for the green and blue quiescent galaxy populations in GOGREEN. Dashed lines (same colours) represent the 68~\% confidence limits of their fit. Picture taken from \cite{Rosa2022}}
    \label{fig:Ritaui}
\end{figure}

We also selected transition galaxies (those that used to be star--forming galaxies and are becoming quiescent galaxies) using two different criteria: 

\begin{itemize}
    \item Galaxies with a value of the \gls{SFR} between $-0.5$~dex and $1$~dex lower than the expected value of the \gls{SFR} provided by the SFMS for the galaxy's mass. We call this set TG1 for short. This criteria is equivalent to the one used by \cite{Bluck2020}.
    \item Blue galaxies with a $\mathrm{sSFR} < 0.1$~$\mathrm{Gyr}^{-1}$, which is the limit used by \cite{Peng2010} to distinguish quiescent and star--forming galaxies. We call this set TG2 for short.
\end{itemize}

With these criteria, we found that the fraction of transition galaxies is higher for galaxies in groups than galaxies in the field. These galaxies allow us to study two additional variables related to the environment and the quenching process. The first of them is the excess in the abundance of transition galaxies. Following \cite{McNab2021} approach: 
\begin{equation}
    R_i \tau_i = (f_i ^G - f_i ^F)/ f_{SF} ^F,
\end{equation}
where $R_i$ is the fraction of field star--forming galaxies that are quenched per unit of time, $\tau_i$ is the transition time scale, this is, the time spent in the transition phase, $f_i ^G$ is the fraction of transition galaxies in groups, $f_i ^F$ is the fraction of transition galaxies in the field, and $f_{SF} ^F$ is the fraction of star-forming galaxies in the field. This equation assumes that the mass accretion rate is constant in time, that the abundance excess of transition galaxies is produced exclusively by quenching and that the number of transition galaxies produced by other factors aside from the environment is proportional to the total galaxy population. Results obtained for this parameter can be seen in Fig.~\ref{fig:Ritaui}. Values obtained vary depending on the threshold used to select group and field galaxies, as well as on the definition of transition galaxies, but this variation is reduced for mass bins lower than $10^{11}$~$M_\odot$. There is a slight dependence with the stellar mass:  $R_i \tau_i$ remains approximately constant at $0.05$ for masses lower than $10^{10}$~$M_\odot$, peaks at $0.1$ at the mass bin $10^{10.5}$~$M_\odot$  and then decreases with mass. The results are compatible with those by \cite{McNab2021} within their confidence intervals, despite the absence of the negative gradient for lower masses in our results.

\begin{figure}
    \centering
    \includegraphics[width=0.65\textwidth]{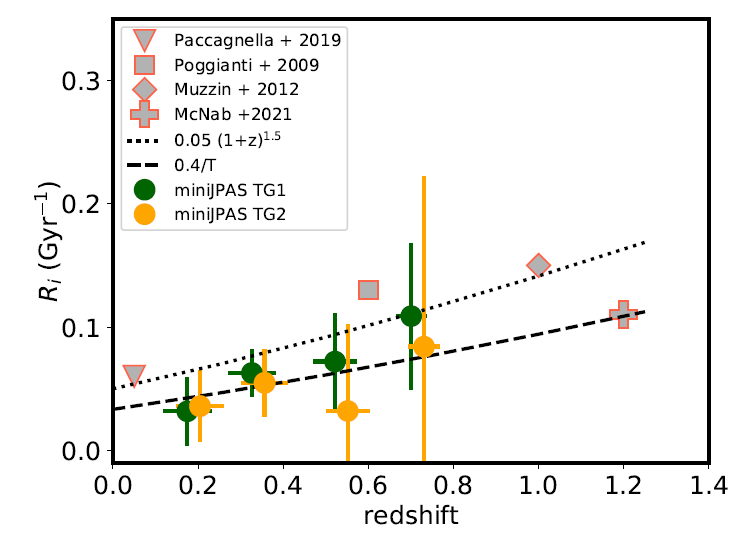}
    \caption[Rate of group galaxy quenching. Picture taken from \cite{Rosa2022}]{Rate of group galaxy quenching. Green and orange points represent the values obtained using the transition galaxies sets TG1 and TG2, respectively. Results compiled by \cite{McNab2021} are shown with grey points with red borders. Different shapes represent results obtained by \cite{McNab2021} using data from different works: triangle for \cite{Paccagnella2019}, square for \cite{Poggianti2009}, diamond for \cite{muzzin2012}, and cross for \cite{McNab2021}. The dashed line is $\mathrm{QFE/T}$ for $\mathrm{QFE} = 0.4$, and the dotted line represents the evolution of the inverse of the dynamical time, $(1 + z)^{3/2}$ scaled to $R = 0.05$~$\mathrm{Gyr}^{-1}$ at $z = 0$.  Picture taken from \cite{Rosa2022}}
    \label{fig:Ri}
\end{figure}

From $R_i \tau_i$, we inferred the rate of environmental quenching, $R_i$, by calculating $\tau_i$ as the fading timescale, $t_\mathrm{fade}$, which is the time that galaxies in groups spend in the transition phase. Using the same approach that \cite{Balogh2016}: 
\begin{equation}
    t_\mathrm{fade} / t_\mathrm{SF+trans} = \tau_i / t_q = \tau_i / T =   f_i ^F/ (f_{SF} ^G + f_i ^G),
\end{equation}
where $t_\mathrm{SF+trans}$ is the time during which all the current star
forming and transition satellite galaxies will fall into the cluster, $t_q$ is the total quenching time scale, $T$ is the lifetime of the cluster at given epoch, and $f_{SF} ^G$ is the fraction of star--forming galaxies in groups. We also used the relation found by \cite{Balogh2016} using the Millenium simulations \citep{Springel2005}:

\begin{equation}
    t_q = A \times (1+z)^{-3/2},
\end{equation}

\noindent where $A$ is the lookback time when the halo started to assemble satellites, which depends on the halo stellar mass, which we derive from Fig.~5 of \cite{Balogh2016}. This allows us to calculate the transition timescale, $\tau_i$, also defined as the fading timescale $t_\mathrm{fade}$, obtaining a value of $\tau_i = 1.5$~Gyr using the TG1 set and $\tau_i = 0.8$~Gyr  when using the TG2. We then followed a similar procedure to calculate the values of $\tau_i$ for four different redshift bins, using both sets of transition galaxies, which allowed as to obtain the values of $R_i$ in those redshift bins (see Fig.~\ref{fig:Ri}). The evolution of $R_i$ that we obtain with our results is compatible with an evolution of constant $\mathrm{QFE} = 0.4$.

The results of this work, based of the capabilities to obtain reliable stellar population properties shown in \cite{Rosa2021}, the accuracy of the photo-$z$ by \cite{HC2021}, and the capability to detect galaxy groups and the galaxies within them with AMICO \citep{Maturi2023}, show the power of the photometric filter system of \jp \ to study the effect of the environment on galaxy evolution. In this thesis, we depart from this basis and provide further insight in this topic.

\chapter{Py2DJPAS} \label{chapter:code}

\begin{abstract}
    In this chapter we present and describe in detail \PyDJ, a tool developed in \Py \ to automatise the analysis of the properties of the spatially resolved galaxies in \mjp, \jp \ and \jplus. Our goal is to provide a single code that downloads the scientific images and tables required for the analysis, performs the image treatment, the specified segmentation and, if desired, calls an external fitting code for spectral energy distribution, and a code to estimate the equivalent widths of the of $\mathrm{H}\alpha$, $\mathrm{H}\beta$, $\mathrm{[NII]}$, and $\mathrm{[OIII]}$ emission lines. We illustrate the different process of the analysis (images and table download, masking, and PSF homogenisation) and, as a sanity check, we show that we are able to retrieve the same values of the magnitudes as those provided in the \mjp \ catalogue using \sext.
\end{abstract}

\section{Introduction}
Studies using \gls{IFU} devices are one of the pillars of modern extra-galactic astrophysics. This type of device was originally proposed by \cite{Courtes1982}, with the aim of overcoming the aperture effects that traditional spectroscopy suffered since, for many galaxies, it was capable of covering only a small region of the object. By using several fibres at the same time, these instruments can observe extended objects and provide 3D-data, where two dimensions are spatial, and the other dimension is the wavelength of the light. 

In the last decade, \gls{IFS} surveys have played a key role in understanding galaxies' structure and formation. Such is the case of the  CALIFA \citep[][]{CALIFA2012} and MaNGA \citep{MANGA2015} surveys. These surveys used a bulk of fibres in the \gls{FoV} of the telescope to capture the light from the different regions of the target. This light was then diffracted using gratings (which determine the spectral resolution) and measured using a \gls{CCD}. Since the bulk of fibres did not fully cover the \gls{FoV} (there are gaps among fibres), a dithering approach was used.

The \gls{CALIFA} survey was carried out at the Observatorio de Calar Alto, using the  \gls{PMAS} instrument \citep{Roth1997,Roth2005} in its PPAK mode. This survey was carried out using two different gratings, the V500, with an spectral resolution of $R\sim 850$ at $\sim 5000$~\AA, and V1200 with a resolution of $R\sim 1250$ at $\sim 4500$~\AA, using a bulk of 382 fibres of $2.7''$ of diameter, 331 of them used to target the object, amounting for a \gls{FoV} of $74''\times 64''$, and covering a wavelength range of 3700--7000~\AA \ \citep[see][]{CALIFA2012}. Some of these fibres were used in order to perform the sky subtraction and flux calibrations processes described in \cite{CALIFA2012}. On the other hand, \gls{MaNGA} used two identical spectographs, each with a red and blue branch. The gratings provided resolutions of $R\sim 1400$ at $\sim 4000$~\AA \ and $R\sim 1800$ at $\sim 6000$~\AA \ for 1420 fibres of $2''$, distributed in 17 \gls{IFU} covering a wavelength range of 3600–-10300~\AA.  

Photometric data has long been used to obtain the stellar population properties of galaxies \citep[see e.g.][and references therein]{Wolf2003,Mathis2006,Walcher2011,Luis2015,SanRoman2018,Rosa2021} and even infer the line emission of galaxies \citep[see e.g.][and references therein]{Ly2007,Pascual2007,Takahashi2007,Villar2008,Koyama2014,VilellaRojo2015,Gines2021,Gines2022}. In order to accurately estimate these properties, reliable photometry measurements are essential. Several codes have been developed with this aim. Probably the most notable among them is \sext \ \citep{Bertin1996}, which remains extensively used up to this day. In particular, it is the code that was used to detect the sources in \mjp \ and to measure their fluxes. However we are interested in developing our own code, which allows us to extract the magnitudes and fluxes of the desired regions of the galaxies that we want to analyse. Similar codes to our purpose include \pycasso \ \citep{Pycasso2017} or  pyPipe3D \citep{Lacerda2022}.

The \jp \ survey will provide 3D-data that is ideal for IFU-like studies. Despite having a lower spectral resolution compared to spectroscopic surveys, it has been proven to retrieve the stellar population properties of galaxies with great accuracy \citep{Rosa2021}. It also has the advantage of a large \gls{FoV} ($4.2$~$\mathrm{deg}^2$ for \jp \ and  $0.27$~$\mathrm{deg}^2$ for \mjp) which has two main advantages for our purposes: it allows to detect galaxies in groups and clusters with no selection bias, and it also allows us to study larger galaxies without aperture bias. The pixel scale of $0.23$~arcesc~$\mathrm{pixel}^{-1}$ also allows for a great spatial resolution over a wavelength range of 3780--9100~\AA. The capabilities of multiband surveys to perform IFU-like studies has also been shown, in particular  using data from the ALHAMBRA \citep{Moles2008} and J-PLUS \citep{Cenarro2019} surveys, which are very related to \jp. These spatially resolved studies include the study of the stellar population properties \citep{SanRoman2018,SanRoman2019}, and the star formation rate of the regions of galaxies \citep{Logrono2019}.

\section{Data} \label{sec:code:data}
The \mjp \ survey \citep{Bonoli2020} is a $1$~$\mathrm{deg}^2$ survey that was carried out at the \gls{OAJ} \citep{OAJ}, using the $2.5$~m  JST/T250 telescope. The main goal of this survey was to show the potential of the \jp \ photometric filter system. All our data come from its public data release, and our aim is to provide a solid method to obtain the photometry of the regions of the spatially resolved galaxies in this data release.

\subsection{Sample selection} \label{sec:code:sample}
In order to test the photometry obtained with our methodology, and to study the properties of the spatially resolved galaxies in \mjp \ (see Chapters~\ref{chapter:MANGA} and \ref{chapter:spatiallyresolved}), we need to impose some criteria in our sample selection. These criteria must guarantee that the selected galaxies are suitable for such purpose. The first requirement is that their apparent size must be large enough to divide them into regions. Our reference parameter for this selection criteria is the \gls{PSF}. We need galaxies to be larger than the PSF of the images. Otherwise, our analysis could be affected by  colour terms introduced by the PSF, particularly in the inner regions \citep[see e.g.][]{Tamura2003, GonzalezPerez2011}. Another matter to take into account is whether the galaxy is face-on or edge-on. There are many studies that show that the spatial variation of the properties of the galaxy is mostly radial \citep[see e.g.][]{Mehlert2003,SanchezBlazquez2006c,SanchezBlazquez2006b,Rosa2015,Rosa2016, SanRoman2018,Bluck2020,Parikh2021}. Thus, the light extracted from a galaxy completely edge-on would be a mixture of the light emitted in inner regions, like the nucleus, and outer regions, such as the arms. The results obtained with such extractions would be really hard to interpret, so we decide to avoid very edge--on galaxies, using the ellipticity of the aperture as a proxy of their inclination. Lastly, we also desire to avoid artefacts in the images and galaxies whose photometry might be unavoidably biased by other bright nearby sources. In consequence, we also need to take the flags of the scientific images into account. Our selection criteria can be summarised in the next requirements:

\begin{itemize}
    \item The effective radius (R\_EFF) of the galaxy must be larger than $2''$. Given the typical \gls{FWHM} of the \gls{PSF} of the images in \mjp \ \citep[see Fig.~5 in ][]{Bonoli2020}, we assume a limit case of $2''$ for the \gls{FWHM} of the \gls{PSF}, to make sure that the PSF is actually better in all the filters. We aim at performing at least two extractions within $1$~$\mathrm{R\_EFF}$, and the  \gls{FWHM} of the \gls{PSF} acts as a diameter, hence the condition: $ \mathrm{R\_EFF} > 2 \times ( \mathrm{FWHM}/2 ) \Rightarrow  \mathrm{R\_EFF} > \mathrm{FWHM}  \Rightarrow \mathrm{R\_EFF} > 2'' $.
    \item The radius obtained as the square root of the isophotal area divided by $\pi$ must be at least two times larger than the \gls{FWHM} of the \gls{PSF}. This condition tries to find galaxies where we can perform at least four extractions. 
    \item The ellipticity must be smaller than $0.6$, to avoid galaxies that are not sufficiently face-on.
    \item The \texttt{MASK\_FLAGS} parameter provided by \sext \ must be 0 for all the filters. This way, we avoid galaxies that are outside the window frame, that are near a bright star, or that are masked due to nearby artefact.
    \item  The $\texttt{FLAGS}$ parameter must not contain the flag 1. This flag indicates that the object has neighbors, bright and close enough to significantly bias the photometry, or bad pixels (more than 10\% of the integrated area affected).
    \item The $\texttt{CLASS\_STAR}$ parameter must be lower than $0.1$, to filter the maximum number of stellar objects.
    
\end{itemize}

In this chapter we will focus on reproducing the magnitudes obtained with \sext \ in the \mjp \ public data release for these galaxies. Our aim is to test the validity of our methodology and proof that the \js \ of the galaxy that we will obtain and fit (see Chapters.~\ref{chapter:MANGA} and \ref{chapter:spatiallyresolved}) are accurate. Although integrated photometry can be obtained for a larger set of galaxies, this set allows us to fulfil our goals for this chapter, while also getting a deeper insight on the sample, as well as requiring a lower computing time.

\begin{figure}
    \centering
    \includegraphics[width=\textwidth]{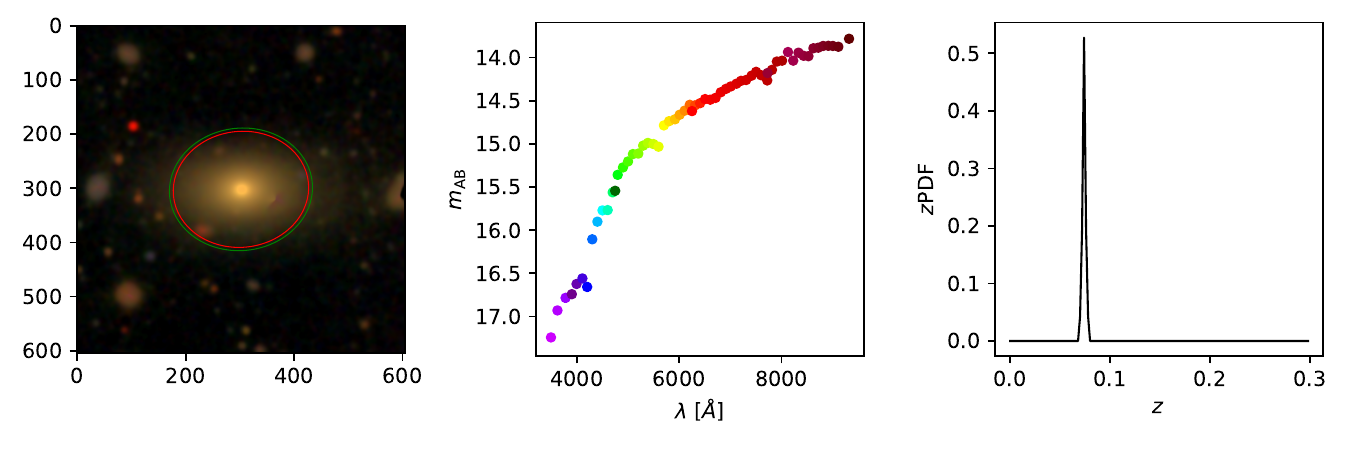}
    \caption[Observational information of the galaxy $2470-10239$. We show the RGB image, the \js, and its $z\mathrm{PDF}$]{Observational information of the galaxy $2470-10239$. First panel shows a composite RGB image of the stamp used for the analysis. Green and red ellipses show the Kron and Petrosian apertures used by \sext, respectively. Second panel shows the \magauto \ \js \ obtained through \sext. Last panel shows the $z\mathrm{PDF}$ of the galaxy.  }
    \label{fig:MANGAobs}
\end{figure}

Additionally, we will use the galaxy 2470--10239 as test example to illustrate the different parts of the methodology. This is a red, massive ($M_\star = 10^{11.51} \mathrm{M_\odot}$), elliptical galaxy with $r_{\mathrm{SDSS}} = 14.62$. The RGB image of the galaxy can be seen in Fig.~\ref{fig:MANGAobs}, as well as its J-spectrum \ and $z\mathrm{PDF}$. Its shows a typical J-spectrum  of a red galaxy, with no visible emission lines. Its $z\mathrm{PDF}$ is a narrow distribution with a very narrow peak (and no additional peaks) at $z=0.074$. The spectroscopic redshift obtained from SDSS is $0.07451\pm0.00002$. We choose this galaxy because its the largest galaxy in apparent size in \mjp, and because it has also been observed with \gls{MaNGA}, which will allow us to perform additional tests, like comparing our \js \ with the spectra obtained from \gls{MaNGA} (see Chapter~\ref{chapter:MANGA}).  

\section{Methodology} \label{sec:code:method}

\begin{figure}
    \centering
    \includegraphics[width=0.85\textwidth]{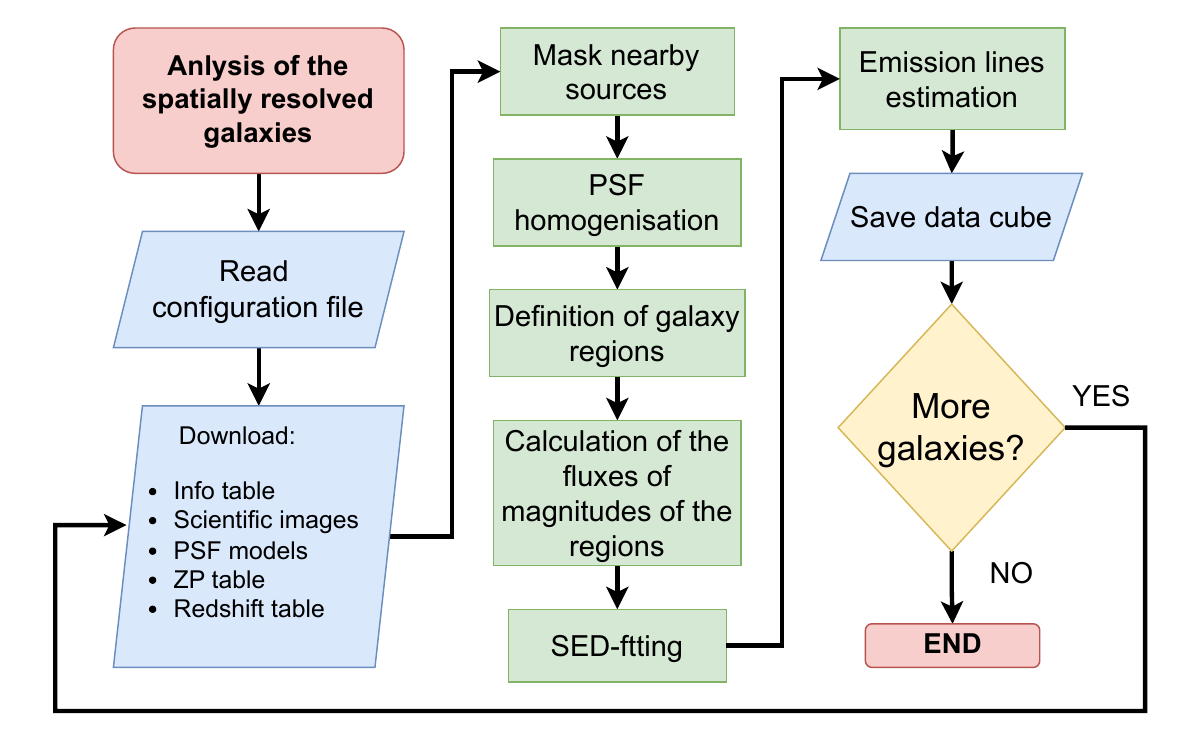}
    \caption[Flow diagram of \PyDJ.]{Flow diagram of \PyDJ. Red ovals Indicate the beginning and end of the different process, blue parallelograms indicate reading and writing input and output files, green rectangles represent computation processes and the yellow diamonds represent decision points. Further details of the processes can be found in Sects.~\ref{sec:code:download}, \ref{sec:code:flux}, \ref{sec:code:masks}, \ref{sec:code:PSF} and \ref{sec:code:photometry}, except for SED-fitting and emission-line estimate, that are described in Chapter~\ref{chapter:MANGA} and \ref{chapter:spatiallyresolved}.}
    \label{fig:flowchart}
\end{figure}

In this section we describe the working processes of \PyDJ, our tool based on \Py \ in order to automatise the steps required for the analysis of the properties of the spatially resolved galaxies. The diagram of tasks is shown in Fig.~\ref{fig:flowchart}. The main tasks can be divided in the following blocks: download of the images and tables (Sect.~\ref{sec:code:download}), flux conversion (Sect.~\ref{sec:code:flux}), masking of the image (Sect.~\ref{sec:code:masks}), PSF homogenisation (Sect.~\ref{sec:code:PSF}), segmentation and binning of the galaxy (see Sect.~\ref{sec:code:photometry} for the details of the apertures used in this chapter, and see next chapters for the details of segmentations used there) and the SED fitting and line emission analysis (see Sects.~\ref{sec:mjp:BaySR} and \ref{sec:mjp:ANN} in Chapter~\ref{chapter:MANGA} and~\ref{chapter:spatiallyresolved}). In the following subsections we explain the details of each step, specifying the most relevant parameters that control them, and provide the values of the parameters used for our work.

\subsection{Code initialisation and download of images and tables} \label{sec:code:download}

At the beginning of the process, the code creates a \Jcube. This is the \Py \ class that we have created in order to save all the relevant information for our analysis. The first table that is saved in the \Jcube \ is a table containing relevant information about the photometric filters, mainly the name of the filter and its pivot wavelength. In order to calculate the pivot wavelength we use Eq.~6 from \cite{Luis2015}, this is: 
\begin{equation}
    \lambda_{\mathrm{pivot}} = \sqrt{\frac{\int \lambda R_X (\lambda) d\lambda}{\int  R_X (\lambda) d\lambda / \lambda}}
\end{equation}
where  $R_X (\lambda)$ is the response function of the filter. This functions can be downloaded from the Filter Profile Service of the Spanish Virtual Observatory\footnote{\url{http://svo2.cab.inta-csic.es/theory/fps/}}. We shall refer to this table as \filterst. We save this information into the cube because first, the wavelengths are required for some functions and calculations, such as plotting the \js \ or transforming the \gls{ADU} into flux units; second, it also helps to keep track of the filter and survey configuration in case any modifications are made to fit the requirements of the collaboration.

The main script then begins the download of the relevant images and tables required for the analysis. All the download process are handled through \gls{ADQL} queries \citep{ADQL2008} and \gls{VO} protocols\footnote{\url{https://www.ivoa.net/}}, most of them \gls{TAP} services. The infrastructure is provided by the \gls{UPAD} at \gls{CEFCA}. For more information about how to handle the downloads, see the \gls{CEFCA} user's manual\footnote{\url{https://archive.cefca.es/doc/manuals/catalogues_portal_users_manual.pdf}}. 

The first downloaded item is what we call the \infot. This is the  \texttt{MagABDualObj} table, obtained by running \sext \ on its Dual Mode using the \rb{} filter as the reference band. This table contains useful parameters of each galaxy, such as the different integrated AB magnitudes and their errors, the sky coordinates, the ellipse parameters or the \kron \ and \petror \ radius. We also download the \texttt{FLambdaSingleObj} from the catalogues. This table contains the same information as the \texttt{MagABDualObj} table, but the photometry is in  units of $\mathrm{erg} \times \mathrm{s^{-1}} \times \mathrm{cm^{-3}} \times \mathrm{\AA^{-1}}$. These tables are stored in the \Jcube, from where we can either access the table as a whole or each parameter individually.

The next step is the download of the scientific images. We download the images for all the filters, because in this work we intend to use the maximum number of filters in order to test our performance in every band, and also to provide the largest possible number of points to the SED-fitting code (see Chapters~\ref{chapter:MANGA} and \ref{chapter:spatiallyresolved}). We can define the size according to a property  of the galaxy, such as its effective radius or its semi-major axis. Here, the code uses the \filterst \ to determine the images that must be downloaded. We found that for this work, square stamps of 30 times the size of \texttt{A\_WORLD}, the major semi-axis provided by \sext, provides an area large enough to contain the whole galaxy and to estimate the background.

From the header of the images we shall use two main type of parameters: those that provide the information about the \gls{WCS} that we will use to perform pixel-to-sky-coordinate transformations \citep[see][for a detailed description of this transformations, as well as the \texttt{astropy} documentation for its \Py \ implementation \footnote{\url{https://docs.astropy.org/en/stable/wcs/}}]{FitsWCS2002I,FitsWCS2002II,FitsWCS2006}, and the parameters from the image used to calculate the errors of the photometry. These parameters are \texttt{SNOIFIT}, \texttt{ANOIFIT}, \texttt{BNOIFIT}, and  \texttt{GAIN} and we will use them in Sect.~\ref{sec:code:flux}.

The code brings the option to download the \gls{PSF} models provided by \gls{CEFCA}. This step is not compulsory, but it is recommended, since the use of these models in order to homogenise the \gls{PSF} of all images improves the photometry of inner regions and provides more reliable data for the analysis (see Section~\ref{sec:code:PSF}. for further details). In this chapter, we will download these models, but we shall use them only when specified. The models will be used in Chapters~\ref{chapter:MANGA} and \ref{chapter:spatiallyresolved}.

The \gls{ZP} of the filters and their errors are then downloaded. These will be used in Sects.~\ref{sec:code:flux} and \ref{sec:code:photometry} to convert the images from \gls{ADU} to physical units, as well as to perform the photometry in the pertinent regions. The photometric calibration is described in \cite{Bonoli2020} and is largely based in the work by \cite{Lopez-sanJuan2019}. A  summary of this work can be found in Sect.~\ref{sec:mjp:ZP} and a revised version of this method is presented by \cite{LopezSanJuan2023}.

The next two items to be downloaded are related to redshift. The first one is the table with  information provided by the $\mathrm{JPHOTOZ}$ package \citep{HC2021}, this is, the \texttt{PhotoZLephare\_updated} table, and the table containing the complete $z\mathrm{PDF}$. The complete list of the parameters that are given in this table can be found Sect.~\ref{sec:mjp:PHOTOZ}.

The last item that we download is the cross correlation table with SDSS, from which we select the spectroscopic redshift (if provided), its error, and its quality flags. We shall use this spectroscopic redshift when available, because even though the photo-$z$ provided are in very good agreement with the spectroscopic ones \citep{HC2021}, some aspects of our analysis can benefit from the better precision of the spectroscopic redshift. For example, the estimated values of emission lines that are in the same filter, such as $\mathrm{H}\alpha$ and $\mathrm{[NII]}$ see their uncertainty reduced with a more precise value of the redshift \citep[see][for further details, and Chapter~\ref{chapter:spatiallyresolved} for our analysis of the emission lines.]{Gines2021}

\subsection{Flux conversion} \label{sec:code:flux}

The information from the scientific images can be divided into two types. The first one includes several entries from the observations, filter parameters that are needed in order to calculate the errors, values to perform pixel to sky coordinates transformations, or data from \sext. The code copies the information that is common to all the filters from the header of the first image and stores it in the header of the \Jcube, so it can be easily accessed when needed. The parameters that depend on the band and are used for the calculations of the errors (\texttt{SNOIFIT}, \texttt{ANOIFIT}, \texttt{BNOIFIT}, and  \texttt{GAIN}) are stored in a separate table. The second type is the data from the images itself.  We use the latter type to create a data cube with shape $(x,y, n_{\mathrm{filters}})$, where  $(x,y)$ are pixels units and $n_{\mathrm{filters}}$ is the number of filters. In the case of \mjp, $n_{\mathrm{filters}} = 60$.

This data is in \gls{ADU} and provides the counts for each pixel. In order to convert the \gls{ADU} into physical units, we make use of the equation:
\begin{equation}
    m_{\mathrm{AB}} = -2.5 \log_{10} (ADU) + ZP.
    \label{eq:mag_ab}
\end{equation}

If we take the equations used to convert the STMAG system \citep{Stone1996}, compiled in Eqs. 4 and 5 by \cite{Luis2015}, from \cite{Bessel2005} and \cite{Pickles2010}:
\begin{equation}
    m_{\mathrm{ST}}=-2.5\log_{10}F_\lambda - 21.1,
\end{equation}
\begin{equation}
    m_{\mathrm{AB}}=m_{\mathrm{ST}}-5\log_{10}\lambda_{\mathrm{pivot}}+18.692,
\end{equation}
we obtain that the flux is: 
\begin{equation}
    F_\lambda = 0.1088 \times ADU \frac { 10^{-0.4\times ZP}}{\lambda^2_{\mathrm{pivot}}},
    \label{eq:flux}
\end{equation}
where $ADU$ are the image counts, $ZP$ is the zero point of the image, and $\lambda_{\mathrm{pivot}}$ is the pivot wavelength of the filter.

In order to calculate the error of the flux we assume Gaussian errors:
\begin{equation}
    \sigma_{F_\lambda}=\sqrt{\left ( \frac{\partial F_\lambda}{\partial C} \right ) ^2 (\sigma_C)^2 + \left ( \frac{\partial F_\lambda}{\partial C_B} \right ) ^2 (\sigma_{C_B} )^2 + \left ( \frac{\partial F_\lambda}{\partial ZP} \right ) ^2 (\sigma_{ZP} )^2},
\end{equation}
where $\sigma_C$ and $\sigma_{C_B}$ are the error of the counts and the error in the counts in the background, respectively, which are defined in Eqs.~6 and 7 from \cite{Logrono2019} as:
\begin{equation}
    \sigma_C = \sqrt {\frac{C-C_B}{G}},
    \label{eq:counts_err}
\end{equation}
and
\begin{equation}
    \sigma_{C_B} = S_{\mathrm{fit}} \sqrt{N_{\mathrm{pix}}} \left (a_{\mathrm{fit}} +b_{\mathrm{fit}} \sqrt{N_{\mathrm{pix}}} \right ).
    \
\end{equation}

As a result we obtain that:
\begin{equation}
    \sigma_{F_\lambda}= F_\lambda \times 
      \sqrt{\frac {1} {|ADU|  G}    + \frac {S^2_{\mathrm{fit}} N_{\mathrm{pix}} \left (a_{\mathrm{fit}} +b_{\mathrm{fit}} \sqrt{N_{\mathrm{pix}}} \right )^2} {ADU^2}   +  \left (   \frac{\ln10}{2.5}  \right ) ^2 (\sigma_{ZP} )^2 },
\label{eq:flux_err}
\end{equation}
where $G$ is the gain of the detector, $N_{\mathrm{pix}}$ is the number of pixels of the integrated region, $a_{\mathrm{fit}}$,  $b_{\mathrm{fit}}$ and $s_{\mathrm{fit}}$ are parameters provided in the image headers used to calculate the background noise (\texttt{ANOIFIT}, \texttt{BNOIFIT} and \texttt{SNOIFIT}, respectively) and where we have taken into account that the images are already background subtracted, so we have used $ADU=C-C_B$ (where $C$ stands for the counts of the detector and $C_B$ for the counts of the background). The signal to noise ratio is defined as $F_\lambda / \sigma_{F_\lambda}$ and can be derived from Eq.~\ref{eq:flux_err}.

With these equations we are able to convert the \gls{ADU} data  into flux physical units. We note that, as shown in equation~\ref{eq:flux}, flux is proportional to \gls{ADU} counts, which means that in order to obtain the flux or magnitudes of one region for each filter, we can sum the counts of each filter image of the pixels in the region, this is:
\begin{equation}
    F_{\lambda, \mathrm{region}} = \sum^{N_{\mathrm{pix}}} _{i=1}  F_{\lambda,i}
\end{equation}
where $ F_{\lambda,i}$ is the flux in each pixel in the region and $F_{\lambda, \mathrm{region}}$ is the integrated flux of the region.

We also create another data cube with the same dimension as the previous one, containing the errors of the flux per pixel (by using equation~\ref{eq:flux_err} with $N_{\mathrm{pix}}=1$). This cube can be used to filter pixels with $S/N$ ratio lower than desired, or as an input for codes that require a signal image and a noise image, such as the Voronoi binning method by \cite{Voronoi}, and the \gls{BatMAN} binning by \cite{BATMAN}. However, we note that the error in the regions is computed according to Eq.~\ref{eq:flux_err} and not to the sum of the square errors, this is, using  $\sigma_{F_\lambda}^2 = \sum^{N_{\mathrm{pix}}} _{i=1} \left( \frac{\partial F_{\lambda, \mathrm{region}}}{\partial F_{\lambda,i}}\right)^2 \left( \sigma_{F_{\lambda,i}}\right)^2 =  \sum^{N_{\mathrm{pix}}} _{i=1}\left( \sigma_{F_{\lambda,i}}\right)^2 $. This last approach leads to an underestimation of the errors.

\subsection{Masks} \label{sec:code:masks}

Photometric images with a large FoV have several advantages, like unbiased object detection, depth usually larger than spectroscopic data (for a same instrument, telescope, and integration time) and the capability to study large objects without aperture bias, among others. However, this large \gls{FoV} also implies that our targets might be affected by nearby objects, such as stars or other galaxies. Also, images might suffer from artifacts, showing structures that are not real or are out of our interest of study, like cosmic rays, or artificial satellites. Therefore we require a mask, this is, a binary flag for each pixel of the stamp that tells us whether that pixel should be used when calculating the flux or magnitudes of the region, or if should not be taken into account,  (this is, the pixel should be \textit{masked}). 

With this in mind, we distinguish two types of masks: nearby sources masks and  masks computed using the \mangle\footnote{\url{https://space.mit.edu/~molly/mangle/}} software \citep{Mangle1993,Mangle2004,Mangle2008}.  This program was designed to work on spherical surfaces, such as the celestial sphere, and provide angular masks, providing polygons whose edges are part of a circle of the sphere. These masks are provided by \gls{CEFCA} and they account for image artefacts and bright stars that bias fluxes and saturate pixels around them, meaning that we cannot use them for our analysis. These masks are computed at complete image level. The use of this masks is advisable, since they account for effects that may not be easy to take into account with other methods. For example, the light of the bright stars biases the flux in more pixels than what one could appreciate at simple eyesight, and artifacts may appear only in certain filters. 

\begin{figure}
    \centering
    \includegraphics[width=\textwidth]{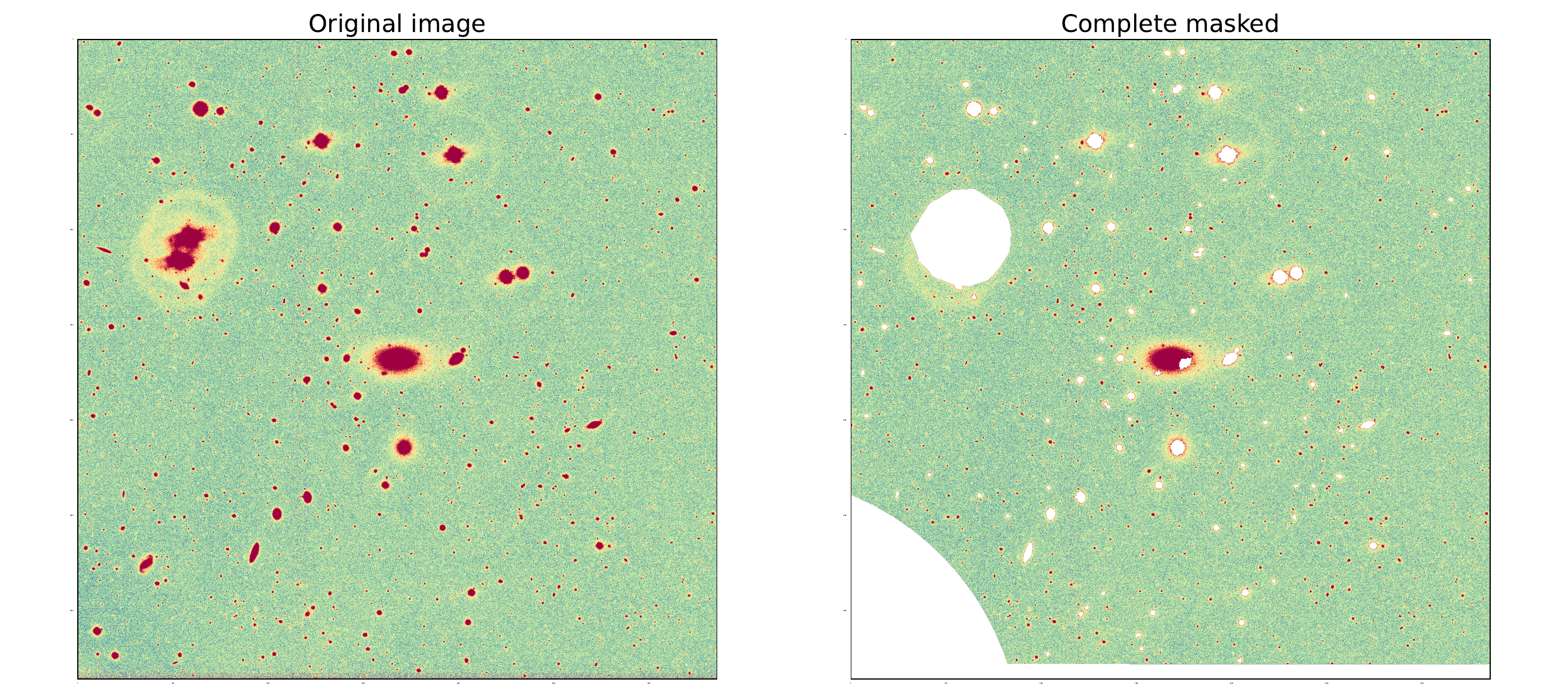}
    \caption[Example of the total mask obtained with \PyDJ.]{Example of the total mask obtained with \PyDJ. Left panel shows a stamp of the AEGIS field. Right panel shows the image with the mask obtained by detecting and deblending nearby sources and combining it with the \mangle \ output. Colour represents the flux in ADU.}
    \label{fig:complete_mask}
\end{figure}

When studying each galaxy, we want to obtain the flux of its regions and remove pixels that belong to other nearby sources. These objects are not masked with \mangle,  since they are also part of the \mjp \ catalogue and they are not biasing any flux measurement. With this aim,  we develop our own method to mask these sources. In particular, we use the \texttt{segmentation} module of the \texttt{phoutils} library\footnote{\url{https://photutils.readthedocs.io/en/stable/}}. The module allows us to distinguish between two sub-type of sources: those sources that are in the frame, but well separated from the galaxy, and those sources that might be blended with the target galaxy. We use both options in our work. This module was used in other works, such as \cite{galmask2022} but, unlike this work, that aims at providing an  object mask that delimits the target galaxy (in order to be used for morphological studies), we aim at removing the objects that are not the target galaxy. We choose this approach since our goal is to obtain solid and accurate flux and magnitude measurements that provide accurate galaxy properties through SED-fitting and ANN estimation (see Chapter~\ref{chapter:spatiallyresolved}). Since establishing the limits of a galaxy is a hard task, we prefer to be more flexible in this regard, under the assumption that if outermost pixels are still part of the galaxy, they should be included in the region and, if they are not part of it, a correct background subtraction should remove their statistical contribution to the flux. This way, we will also be able to test the maximum distance at which we are capable of retrieving the properties of the galaxy.

We show an example of the total mask obtained in Fig.~\ref{fig:complete_mask}. This stamp shows the \rb{} image of a wide field centred in galaxy the 2470--10239  $(\mathrm{R.A.}  = 213\degree.8349, \ \mathrm{Dec.} = 52\degree.3459)$. Around the galaxy on the image centre, we can find several sources, including two bright stars very close between them, at the upper left of the stamp, easy to identify because of the light halo around them. We can see that the mangle mask successfully  covers these two bright stars, and in the bottom left corner of the right image we find part of a bright star, that is still affecting the photometry of that part of the image. It also illustrates that our method masks both blended and non-blended sources near the galaxy, and serves well for our purposes.

\subsection{PSF homogenisation} \label{sec:code:PSF}
The response of an optical system has an effect on the light, spreading it, which means that the obtained image of a point source is not a point. The \gls{PSF}  describes this response of the system. It affects not only point like sources (stars) but also extended ones, such as galaxies. Taking the PSF into account has been proven to be crucial when working with photometric surveys \citep[see e.g.][]{Heasley1999, Infante-sainz2020, Massari2020}, particularly when working with multi-wavelength surveys \citep{Coe2006, Molino2014, SanRoman2018}, where the response in each band is different, due to the variation of the \gls{PSF} with wavelength, as well as in the atmospheric conditions of the observations. Therefore, we need to account for these effects in order to obtain reliable measurements of the photometry of the regions. Otherwise, we could be introducing biases in the colour gradients \citep[see e.g.][]{Tamura2003, GonzalezPerez2011} or structures that could bias our results. 

In order to avoid these problems, we decided to homogenise the images to the worst PSF among all the bands for every galaxy. This method has been used in other works such as those by \cite{Loh1986, Labbe2003, Capak2007, Enia2020} or \cite{Kiiveri2021}, since it produces homogeneous apertures in all the bands, at the cost of worsening the spatial resolution. The procedure that we follow is:

\begin{itemize}
    \item For each galaxy, for each band, we download the model of the PSF. We use the model provided for a given position of the image, using the coordinates of the galaxy in the catalogue. We assume that the spatial variation of the PSF model is negligible for the extent of the galaxy.
    \item We parameterise each model (for each galaxy and each band), using the approximation that they are two-dimensional Gaussian functions, with no correlation, since it greatly simplifies the equations and the calculations. This is, we assume that the model of the \gls{PSF} can be described as:
    \begin{equation}
        G(x,y) = \frac{1}{2 \pi \sigma_x \sigma_y} \exp{ \left  [ -  \left ( \frac{x-x_0}{\sigma_x}\right )^2 - \left ( \frac{y-y_0}{\sigma_y}\right )^2  \right ]}.
    \end{equation}
    \item From this parameterisation, we are interested in the variance in each axis of the model. These variances are calculated as the second order moments of the distribution
    \begin{equation}
        \sigma^2_x = \frac{\sum_i (x_i - \overline{x})^2 * G(x_i, \overline{y})}{\sum_i G(x_i, \overline{y})} ,
    \end{equation}
    \begin{equation}
        \sigma^2_y = \frac{\sum_i (y_i - \overline{y})^2 * G(\overline{x}, y_i)}{\sum_i G(\overline{x}, y_i)} ,
    \end{equation}
    where $x_i$ and $y_i$ \footnote{Note that, since $x_i$ and $y_i$ are pixel positions, they are integer numbers than run from 0 up to the pixel size of the model.} are the pixel positions of the model, and $\overline{x}$ and $\overline{y}$ are the first moments of the distribution, calculated as:
    \begin{equation}
        \overline{x} = \frac{\sum_{i,j}x_i* G(x_i, y_j)}{\sum_{i,j} G(x_i, y_j)}
    \end{equation}
    \begin{equation}
       \overline{y} = \frac{\sum_{i,j}y_j* G(x_i, y_j)}{\sum_{i,j} G(x_i, y_j)}
    \end{equation}

    \item Because of the Gaussian approximation, we know that the FWHM of the PSF is $\mathrm{FWHM} = 2\sqrt{2 \ln(2)} \times \sigma$. This way, we know that the filter with the worst PSF is the filter with the largest variance.

    \item We calculate the variances of the required Gaussian kernel as
    \begin{equation}
        \sigma^2_{\mathrm{kernel}_i, x} = \sigma^2_{\mathrm{worst}_i, x} - \sigma^2_{\mathrm{filter}_i, x},
    \end{equation}
    \begin{equation}
        \sigma^2_{\mathrm{kernel}_i, y} = \sigma^2_{\mathrm{worst}, y} - \sigma^2_{\mathrm{filter}_i, y},
    \end{equation}    
     where $\sigma^2_{\mathrm{worst}_i, x}$ and $\sigma^2_{\mathrm{worst}, y}$ are the variances of the filter with the worst PSF along the X and Y axis, respectively.

     \item For each image, we generate a Gaussian kernel with variances $\sigma^2_{\mathrm{kernel}, x}$ and $\sigma^2_{\mathrm{kernel}, y}$, and we convolve the image with its corresponding kernel. We find that $\sigma^2_x \simeq \sigma^2_y $, so in practice $\sigma^2_{\mathrm{kernel}, x} \simeq \sigma^2_{\mathrm{kernel}, y}$. This step of the process is skipped for the image of the filter with the worst PSF, and the same degradation is applied in both axis for the rest of the images. \footnote{If it was not the case, the assumption of no correlation allows for this simplification in the computation, since this particular case of the two-dimensional Gaussian verifies that $\sigma_{\mathrm{conv}, x}^2 = \sigma_{1,x}^2 +\sigma_{2,x}^2$ and $\sigma_{\mathrm{conv}, y}^2 = \sigma_{1,y}^2 +\sigma_{2,y}^2$}
    
\end{itemize}

With this process we can obtain homogeneous apertures for the images that provide solid magnitude measurements for our galaxies, as we will show in Sect.~\ref{sec:code:results}. We note that we are assuming that the error in the flux and magnitudes introduced with this procedure is negligible. Introducing a term that accounts for this correction in Eq.~\ref{eq:flux_err} is not trivial. Some works, such as \cite{TesisLogrono} recompute the values of $S_\mathrm{fit}$, $a_\mathrm{fit}$ and $b_\mathrm{fit}$. We argue that the regions most affected by this correction are the innermost part, and the homogenisation produces the same values of the magnitudes in the outer parts of the galaxy, where this procedure does not have a great impact (See Sect.~\ref{sec:code:results}). Inner regions are also the brightest parts of the galaxy, which imply a high $ADU$ count. In the limit $ADU \rightarrow \inf$, the first two terms of Eq.~\ref{eq:flux_err} become negligible and the error of the measurement is dominated by the error in the ZP. We therefore decide to use this assumption to simplify the calculations and reduce the computation time.

\subsection{Regions definition and photometry calculation} \label{sec:code:photometry}
The next step required to obtain the properties of the regions of galaxies, is the definition of the regions themselves. The different approaches implemented in the code can be summarised in three main types:

\begin{itemize}
    \item Standard geometrical apertures and rings. Within this type we distinguish circular apertures/rings and elliptical apertures/rings. The size of these apertures can be defined with respect to different parameters, such as the \texttt{R\_EFF}, \texttt{A\_WORLD}, or the \gls{FWHM} of the worst \gls{PSF}. The number of rings is also flexible. If the elliptical geometry is chosen, in order to calculate the parameters of the ellipse, we first create an elliptical object mask, with semi-major and semi-minor axes $a = 2 \times \texttt{A\_WORLD} \times \texttt{PETRO\_RADIUS}$ and $b= 2 \times \texttt{B\_WORLD}  \times \texttt{PETRO\_RADIUS}$, respectively, in order to limit the influence of external sources. Then, we apply the routine from \pycasso \ \citep{Pycasso2017} in order to obtain better constrains of the ellipses axes and orientation. We force the centre of the ellipse to be located in the central pixel. 
    \item The  \gls{BatMAN} binning by \cite{BATMAN}. This approach uses Bayesian statistics based on the premise that pixels belonging to a same physical region should have a similar S/N, as well as being adjacent. In very few words, the algorithm takes a pixel, joins it to an adjacent one, and checks whether the Bayesian probability of belonging to a same region is larger or lower than the probability of belonging to different regions. If the posterior probability is larger, then it joins the pixels and keeps repeating the process. This code requires a signal image and a noise image, which can actually be data cubes, this is, the segmentation can be performed taking into account several filters at the same time.
    \item The Voronoi binning method by \cite{Voronoi}. This algorithm's approach is to provide a set of regions where the S/N ratio is equal or higher that a desired target value. Simplifying, it starts with a pixel (usually the central one) and checks its S/N. If its larger than the target S/N, it moves to another pixel. If that is not the case, it joins an adjacent pixel, checks the S/N and restarts the process. This code also takes a signal image along a noise image, but not a cube.
\end{itemize}

In this chapter, we try to reproduce the available photometries in the \mjp \ catalogues for the selected galaxies as a sanity check. We shall focus on two types of apertures according to their geometry: elliptical and circular. Concerning the elliptical photometries, we shall use \magauto \ and \magpetro \ for our tests. These apertures try to estimate the complete ($\sim90\%$) flux of the galaxy, and provide the flux contained inside a Kron-like and Petrosian-like aperture\footnote{For more information, see the \sext \  user guide \url{https://sextractor.readthedocs.io/en/latest/Photom.html}}.

To calculate the circular photometries, we  use the sky coordinates of the \mjp \ catalogue to define the centre. We use the \texttt{SkyCircularAperture}\footnote{\url{https://photutils.readthedocs.io/en/stable/api/photutils.aperture.SkyEllipticalAperture.html}} function from the \texttt{photutils} \ library. We then define apertures of $0.8'', 1.0'', 1.2'', 1.5'', 2.0'', 3.0'', 4.0'',$ and $ 6.0''$ of diameter.

In order to reproduce the \magauto \ and \magpetro \ magnitudes, we first need to determine the same aperture that was used by \sext \ to compute them. We use the non PSF-homogenised images since, according to \cite{Bonoli2020}, the images were not homogenised for these photometries. From the \sext \ manual, we know that the Kron radius is given as $k \times r_{Kron}$. The maximum of the $k$ parameter is set by the \texttt{PHOT\_AUTOPARAMS} input of \sext. In Table~C.1 from \cite{Bonoli2020}, we find that this maximum was set to $2.5$. Additionally, the description of \kron \ from the CEFCA catalogues tells us that this parameter is given as a factor of \texttt{A\_WORLD} and \texttt{B\_WORLD} units. Therefore, in order to reproduce this photometry, we apply Eqs.~\ref{eq:mag_ab} and \ref{eq:mag_err} into an elliptical aperture of semi-major and semi-minor axes $a_{Kron}$ and $b_{Kron}$, calculated as 
\begin{equation}
    a_{Kron} = n \times \kron \times \texttt{A\_WORLD},
    \label{eq:akron}
\end{equation}
\begin{equation}
    b_{Kron} = n \times \kron \times \texttt{B\_WORLD},
    \label{eq:bkron}
\end{equation}
where $n \in [1, 2.5]$. We proceed similarly to retrieve the \magpetro \ photometry. The ellipse is also given by $k \times r_{Petro}$ where, according to Table C.1 from \cite{Bonoli2020}, the maximum value of $k$ is 2. The semi-major and semi-minor axes, $a_{Petro}$ and $b_{Petro}$, of the aperture calculated are as
\begin{equation}
    a_{Petro} = m \times \petror \times \texttt{A\_WORLD},
    \label{eq:apetro}
\end{equation}
\begin{equation}
    b_{Petro} = m \times \petror \times \texttt{B\_WORLD},
    \label{eq:bpetro}
\end{equation}
where $m \in [1, 2]$. In order to apply these apertures on the scientific images, we use the sky coordinates of the galaxy and the \texttt{SkyEllipticalAperture}\footnote{\url{https://photutils.readthedocs.io/en/stable/api/photutils.aperture.SkyEllipticalAperture.html}} function from the \texttt{photutils} \ library. We note here that,  to define the aperture, we also use the \texttt{THETA\_J2000} parameter from \sext, which is the angular position of the ellipse in sky coordinates\footnote{Since  \texttt{SkyEllipticalAperture} takes as origin of coordinates the vertical axis, and \sext \ the horizontal one (both counterclokwise) we actually use  \texttt{THETA\_J2000} - $90^{\circ}$}. 

In order to obtain the $n$ and $m$ factors, we follow an iterative approach, defining apertures of different sizes by changing the $n$ and $m$ value of Eqs.~\ref{eq:akron}--\ref{eq:bpetro}, in the intervals $[1,2.5]$ and  $[1,2]$, respectively, with $0.1$ steps. For each aperture, we measure the median difference in the errors provided in the catalogue and those obtained with our code, without using the error in the ZP, since it is not included in the error budget of the catalogues. We choose as the optimal aperture that providing the minimum absolute value of the median error difference. This way, we are using the errors of the magnitudes as a parameter to determine the size of the aperture, given the relation of the error with the number of pixels in the aperture (see Eq.~\ref{eq:mag_err}).

For the purpose of obtaining better results for \magauto \ and \magpetro, we also perform a local background estimation. We define an elliptical annulus, in a similar way to the apertures, but using the \texttt{SkyEllipticalAnnulus} function instead. We select as inner radii $a_{\mathrm{in}}=  4 \times \kron \times \texttt{A\_WORLD}$, and we choose $a_{\mathrm{out}}= 4.5 \times \kron \times \texttt{A\_WORLD}$  as outer radii. We obtain the average contribution of the background by summing the ADUs contained within these annuli, and we divide this counts by the number of unmasked pixels in the annulus, this is:
\begin{equation}
    \overline{ADU_\mathrm{background}} =  \frac{\sum_{i \in B} ADU_i}{NPIX_\mathrm{background} - NPIX_\mathrm{{mask, background}}},
\end{equation}
where $B$ accounts for the unmasked pixels in the annulus defined to estimate the background, $NPIX_\mathrm{background}$ is the total number of pixels within the annulus, and  $NPIX_\mathrm{mask, background}$ is the number of masked pixels within the annulus. 

We also need to account for the flux of the masked pixels within the aperture. For such purpose, we follow a similar approach as in the case of the background, by calculating the average (unmasked) counts in the region after extracting the background:
\begin{equation}
    \overline{ADU}  = \frac{\sum_{i \in S} ADU_i  }  {NPIX -NPIX_ \mathrm{mask, aperture}} - \overline{ADU_\mathrm{background}} , 
\end{equation}
where $S$ accounts for the unmasked pixels in the region, $NPIX$ is the total number of pixels within the aperture, and  $NPIX_\mathrm{mask, aperture}$ is the number of masked pixels within the aperture. Therefore, the final counts used for the calculation of the AB magintudes are 
\begin{equation}
    ADU =  \sum_{i \in S} ADU_i + \overline{ADU}  \times NPIX_\mathrm{mask, aperture}  
\end{equation}

This value of $ADU$ is that introduced in Eq.~\ref{eq:mag_ab} to obtain the AB magnitudes. We calculate the error assuming Gaussian propagation of errors and the same terms used to obtain Eq.~\ref{eq:flux_err} 
\begin{equation}
    \sigma_{m_\mathrm{AB}}= \frac{2.5}{\ln 10} \sqrt{\frac {1} {|ADU|  G}    + \frac {S^2_{\mathrm{fit}} N_{\mathrm{pix}} \left (a_{\mathrm{fit}} +b_{\mathrm{fit}} \sqrt{N_{\mathrm{pix}}} \right )^2} {ADU^2}   +  \left (  \frac{\ln10}{2.5}  \right ) ^2 (\sigma_{ZP} )^2 }
    \label{eq:mag_err}
\end{equation}

We note that we do not use this corrections from background and mask when using the circular apertures in the general case, since this apertures are very small and in the central regions. Therefore, they should not be affected by nearby objects, and the innermost region of the galaxy should be bright enough so that the contribution of the background contribution is negligible. 

\section{Results}\label{sec:code:results}

In this section we illustrate the processes of the methodology using the galaxy 2470-10239 as a test case, and compare our measurements of the magnitudes with those obtained with \sext \  provided in the catalogue. Our aim is to prove that we obtain reliable measurements of the magnitudes with our methodology, which are required to obtain results concerning the properties of the galaxies (see Chapters~\ref{chapter:MANGA} and \ref{chapter:spatiallyresolved} for more details).

\subsection{2470--10239 example}
In this section, we show how \PyDJ \ works by analysing in greater detail the galaxy 2470--10239. We choose this galaxy since it is the largest one in apparent size in the \mjp \ field.  We explain the process applied this galaxy in three blocks: the masking process, the effects of the PSF homogenisation, and the comparison of the magnitudes with the data from the \mjp \ catalogue.

\subsubsection{Masking}

\begin{figure}
    \centering
    \includegraphics[width=\textwidth]{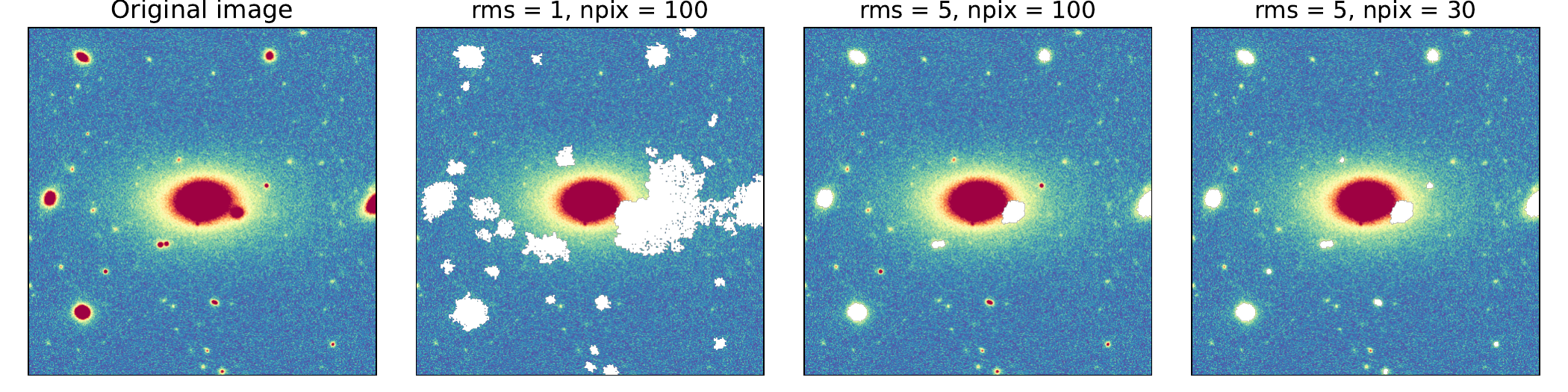}
    \caption[Example of different nearby sources and deblended sources masks obtained for $2470-10239$]{Example of different nearby sources and deblended sources masks obtained for the galaxy $2470-10239$, using different values of the threshold and pixel parameters. From left to right: original image, $\mathrm{rms} = 1, \mathrm{npix}=100$, $\mathrm{rms} = 5, \mathrm{npix}=100$, $\mathrm{rms} = 5, \mathrm{npix}=30$}
    \label{fig:sourcesparam1}
\end{figure}

We start by showing the masking process of the images. In our method, the sources masks depend on two parameters. The first one, $\mathrm{npix}$, determines the minimum number of pixels for a detection to be considered as a source. The second one, $\mathrm{rms}$, establishes the threshold level to be considered a detection. We set this threshold parameter as a function of the error background noise for one pixel, this is $\mathrm{threshold} = \mathrm{rms} \times  S_{\mathrm{fit}}  \left (a_{\mathrm{fit}} +b_{\mathrm{fit}} \right )$. In Fig.~\ref{fig:sourcesparam1} we show different masks obtained varying these two parameters. 

As we can see from the figure, setting a value too low for threshold results in the false detection from pixels of the background as sources. Setting a higher value of the minimum number of pixels does not really solve this issue, since they end up associated to other nearby sources. Setting this threshold too high would mean that dimmer sources would not be classified as such. Therefore, they would not be masked. In our test case, we find that $\mathrm{rms} = 5$ provides a nice detection of the nearby sources. However, using a value to high of the minimum number of pixels results in some smaller unmasked sources. Finding a good compromise in both values in order to obtain a proper mask is not an easy task, particularly if analysing several galaxies at the same time, and the optimal set of values might depend on the requirements or preferences of the scientific case. However, our aim is to automatise the whole process of analysis. After several close inspections and tests with different values, we find that the set of values $\mathrm{rms} = 5, \ \mathrm{npix} = 30$ using the \rb{} image as reference provides a solid masking for this particular case, as well as for the other galaxies.  These are the values that we will use throughout this work, but we shall revise them in the future \jp \ data release.

 \subsubsection{PSF homogenisation}
\begin{figure}
    \centering
    \includegraphics[width=\textwidth]{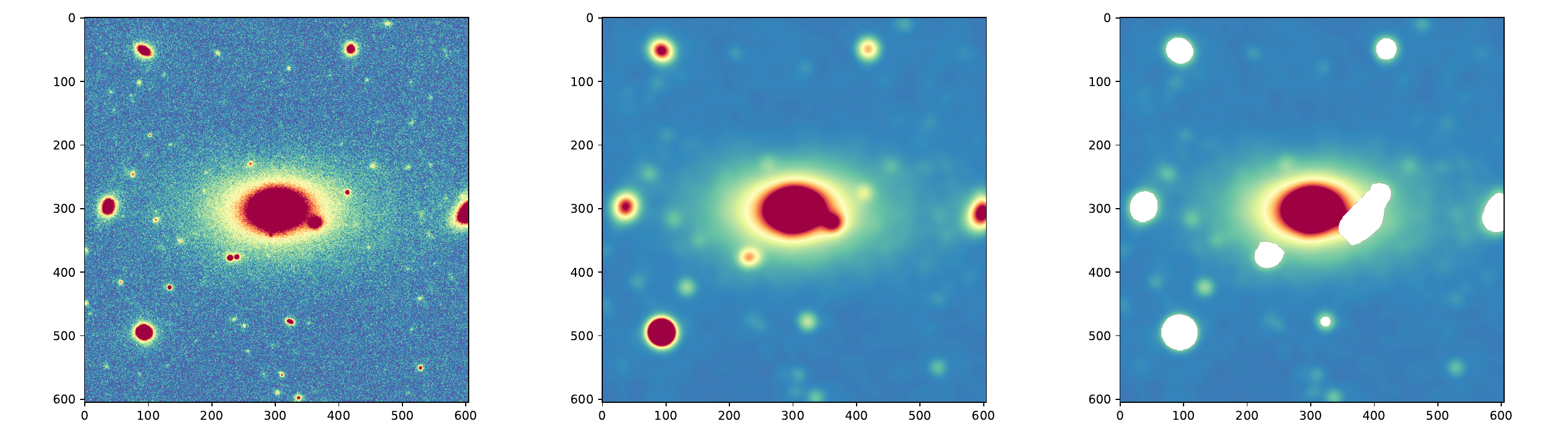}
    \caption[Summary of the image treatment process for the galaxy $2470-10239$: original image, PSF-homogenised image and  masked PSF-homogenised image.]{Summary of the image treatment process. Left panel shows the \rb{} image of the galaxy $2470-10239$ with no treatment. Middle panel shows the \rb{} image after PSF homogenisation. Last panel shows the final image, after PSF homogenisation and applying the sources mask. The colour scale indicates the flux in ADUs.}
    \label{fig:MANGA_psf_mask}
\end{figure}

We now show the effect of the PSF homogenisation in our data. Fig.~\ref{fig:MANGA_psf_mask} shows the whole process of the image treatment for the \rb{} image. The effect of the homogenisation in the image is quite clear. The image becomes blurrier, the background and shapes of the sources become smoother and the smallest sources become more extended. The effect of the PSF in the morphology has in fact been studied \citep[see e.g.][]{PaulinHenriksson2008,Lewis2009,Voigt2010}. Nonetheless, we are interested in the magnitudes measurements, and the possible effect in the morphology will be taken into account when recalculating the ellipse parameters. We also note that the nearby sources are still correctly masked after homogenisation, serving as a second proof of concept of our method.

\begin{figure}
    \centering
    \includegraphics[width=0.85\textwidth]{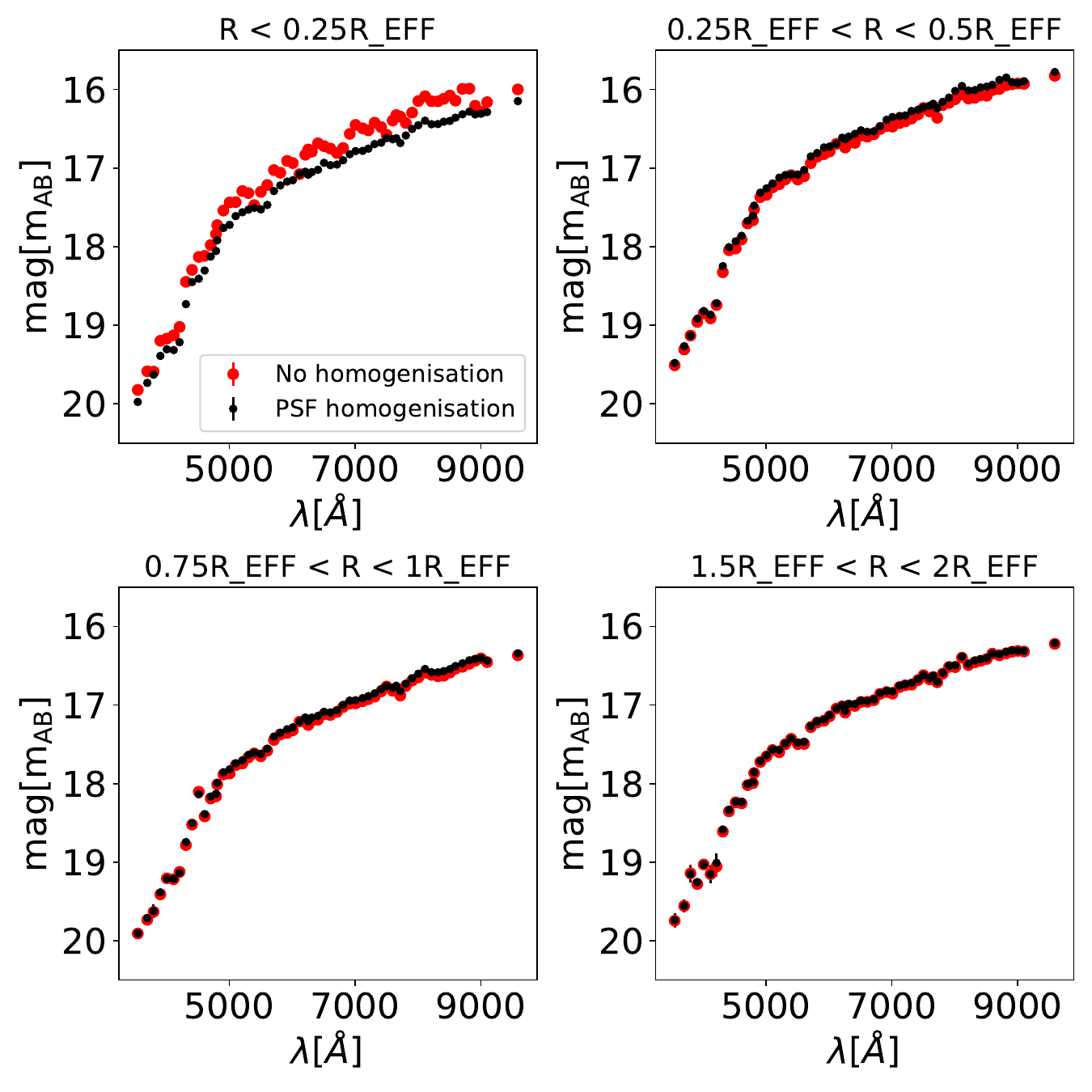}
    \caption{Flux comparison of the \js \ of different regions of the galaxy $2470-10239$ before and after applying the PSF homogenisation for different apertures.}
    \label{fig:degradationjs}
\end{figure}

In order to study the effect of the homogenisation on the photometry, we compare the \js \ retrieved at different apertures and annuli (using the same tools as in Sect.~\ref{sec:code:photometry}) for the galaxy both before and after PSF homogenisation in Fig.~\ref{fig:degradationjs}. The first panel shows the J-spectrum of the innermost region. The non-homogenised J-spectrum shows variations in the bands that are not associated with absorption or molecular bands, and are not physical. These variations either disappear or become much smoother after the PSF homogenisation, at the cost of loosing some flux in several bands. Since the mass-to-light ratio of the models is fixed by construction \citep[see e.g.][]{Conroy2013}, this flux loss could lead to a lower value of the stellar mass of the region. However, the homogenisation also affects the colour of the galaxy: red bands are, in general, more affected by the homogenisation than blue bands. Many works have pointed out that working with images where the differences in the PSF have not been taken into account leads to the introduction of undesired colour biases \citep[see e.g.][]{Cyprian2010,GonzalezPerez2011,Er2018,Liao2023}. Our SED-fitting is heavily dependent on the colour the galaxy, and the non-physical fluctuations of the \js \  could lead to incorrect results or poorer stellar population fitting. Thus, it is better for our analysis to use the homogenised \js, despite the possible mass loss. In fact, the relation of the mass-to-light ratio and the galaxy colour is well known and it has also been used to derive stellar masses \citep[see e.g.][]{Tinsley1981,Bell2001, Bell2003, Gallazzi2009, Ruben2017,LopezSanJuan2019ML}. Therefore, an incorrect galaxy colour would also lead to and incorrect mass-to-light ratio, leading to a incorrect estimation of the mass anyway.

Another effect that can be appreciated in Fig.~\ref{fig:degradationjs} is that the differences among the homogenised and not-homogenised \js \ become lesser as the distance to the galactic centre increases. This can be expected, since the homogenisation spreads the light of the filter. Brightest regions will loose more flux in comparison to dimmer regions, since these dimmer regions ``recover'' flux from other close regions with a similar brightness. Due to the usual Sérsic profile of the surface brightness \citep{Sersic1963,Sersic1968}, the innermost parts of the galaxy are noticeably brighter and the ``recovered'' flux from other regions is not enough to account for the lost flux. We can actually see this effect in the second panel. This close region to centre shows an slightly brighter \js \ after homogenisation, due to the spread light from the centre. In consequence, we can expect flatter  radial profiles of the properties of the galaxies after homogenisation.

Lastly, this Figure also illustrates our initial assumption regarding the errors introduced by the PSF homogenisation. The homogenisation is only important in the innermost regions, where the error is dominated by the error of the zero point (see Eqs.~\ref{eq:flux_err} and \ref{eq:mag_err}). On the other hand, outer (dimmer) regions are barely affected by the homogenisation, and we can assume that the error introduced is negligible.

\subsubsection{Magnitudes comparison}

We end our test case of the galaxy 2470--10239 by reproducing the magnitudes obtained using  \sext. For these comparisons, we will not use the PSF homogenisation process since, as specified in \cite{Bonoli2020}, these values of the catalogue are obtained with no homogenisation. Different tests are performed, which can be divided as follows:

\begin{table}
    \centering
    \begin{tabular}{|c|c|c|c|}
        \hline
        Photometry & median $\Delta m$    & median $\Delta m$   & median $\Delta m$   \\
        & (no correction) &  (mask correction) &  (mask and background correction) \\
        \hline
        \hline
         \magauto & $-0.019$ & $-0.03$ & $0.02$  \\
         \magpetro & $0.01$ & $0.01$ & $0.01$ \\
         \texttt{MAG\_APER\_0\_8} & $0.016$ & $0.016$ & $0.017$ \\
         \texttt{MAG\_APER\_1\_0} & $0.005$ & $0.005$ & $0.006$\\
         \texttt{MAG\_APER\_1\_2} & $-0.009$ & $-0.009$ & $-0.008$ \\
         \texttt{MAG\_APER\_1\_5} & $0.002$ & $0.002$ & $0.003$\\
         \texttt{MAG\_APER\_2\_0} & $-0.001$ & $-0.001$ & $0.0001$ \\
         \texttt{MAG\_APER\_3\_0} & $-0.001$ & $-0.001$ & $-0.0002$\\
         \texttt{MAG\_APER\_4\_0} & $-8.8\times10^{-6}$ & $-8.8\times10^{-6}$ &  $0.001$\\
         \texttt{MAG\_APER\_6\_0} & $-0.0009$ & $-0.0009$ & $0.0013$\\
         \hline

    \end{tabular}
    \caption[Median difference of the magnitudes obtained with \PyDJ \ and \sext \ for the elliptical and circular photometries available in the \mjp \ catalogues.]{Median difference of the magnitudes obtained with \PyDJ \ and \sext for the elliptical and circular photometries available in the \mjp \ catalogues. The values are expressed as $\Delta m = m_{\PyDJ} - m_{\sext}$. }
    \label{tab:MANGAphottest}
\end{table}

\begin{itemize}
    
\begin{figure}
    \centering
    \includegraphics[width=\textwidth]{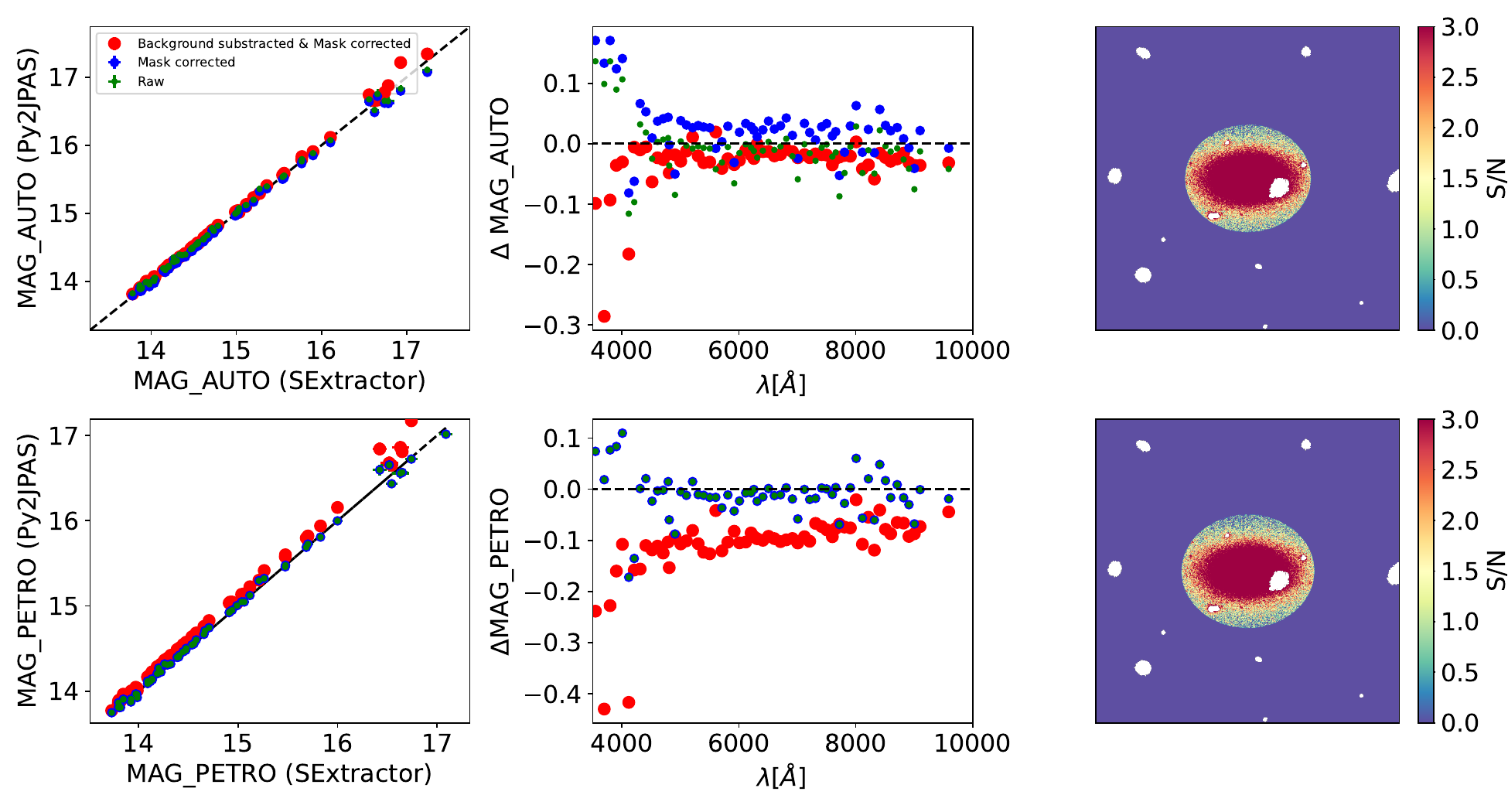}
    \caption[Comparison of the elliptical photometries available in the catalogue and the values retrieved with \PyDJ \ for the galaxy 2470--10239, using the raw image without accounting for masked pixels, the raw image accounting for masked pixels, and the raw image correcting for masked pixels and subtracting the background.  ]{Comparison of the elliptical photometries available in the catalogue and the values retrieved with \PyDJ \ for the galaxy 2470--10239. Green points show the values obtained with no image treatment and without correcting for masked pixels. Blue points represent the values obtained when we account for the masked pixels, but no background subtraction is used. The red points represent the values of the photometry obtained when accounting for the background and the masked pixels. The black solid line shows the identity relation. Left panels show the 1:1 relation. Middle panel show the difference between our calculations and the catalogue values. Right panels show the S/N image with the aperture used for each calculation.  Top row represents the comparison for \magauto. Bottom row represents the comparison for \magpetro.}
    \label{fig:MANGAellPhot}
\end{figure}

    \item \textbf{Elliptical aperture photometry}. We start by comparing our results with those obtained by \sext \ for \magauto \ and \magpetro \ (see Fig.~\ref{fig:MANGAellPhot}.). We find that, for this galaxy in particular, the background estimation is not really necessary, since we are able to reproduce the data very solidly both for \magauto \ and \magpetro. The correction of the masked pixel is negligible in both photometries. This is most likely due to the low percentage of masked pixels, and to the brightness of the galaxy, since bright, central pixels already dominate the flux contribution to the aperture. Concerning the effect of the background correction, we find that, except for two of the bluest bands, the difference of the calculated magnitudes and the magnitudes of the catalogue becomes lower for \magauto, but it increases for \magpetro. In fact, for \magpetro \ the reconstruction is almost perfect without background subtraction, and the difference becomes close to $\sim 0.1$~mag after subtraction. We note that our estimation of the background is a rough approximation, and that it can be sensitive to many factors. We also note that our estimation is not necessarily the same as that performed by \sext.  Nonetheless, we are able to reproduce the values of the catalogues with a difference well below $\sim 0.1$~mag for most magnitudes both for \magauto  \ (after background subtraction) and for \magpetro \ (without background subtraction). This differences mean a relative error below $10$~\%. We find that bluer, dimmer bands are those showing the largest offsets.

\begin{figure}
    \centering
    \includegraphics[width=\textwidth]{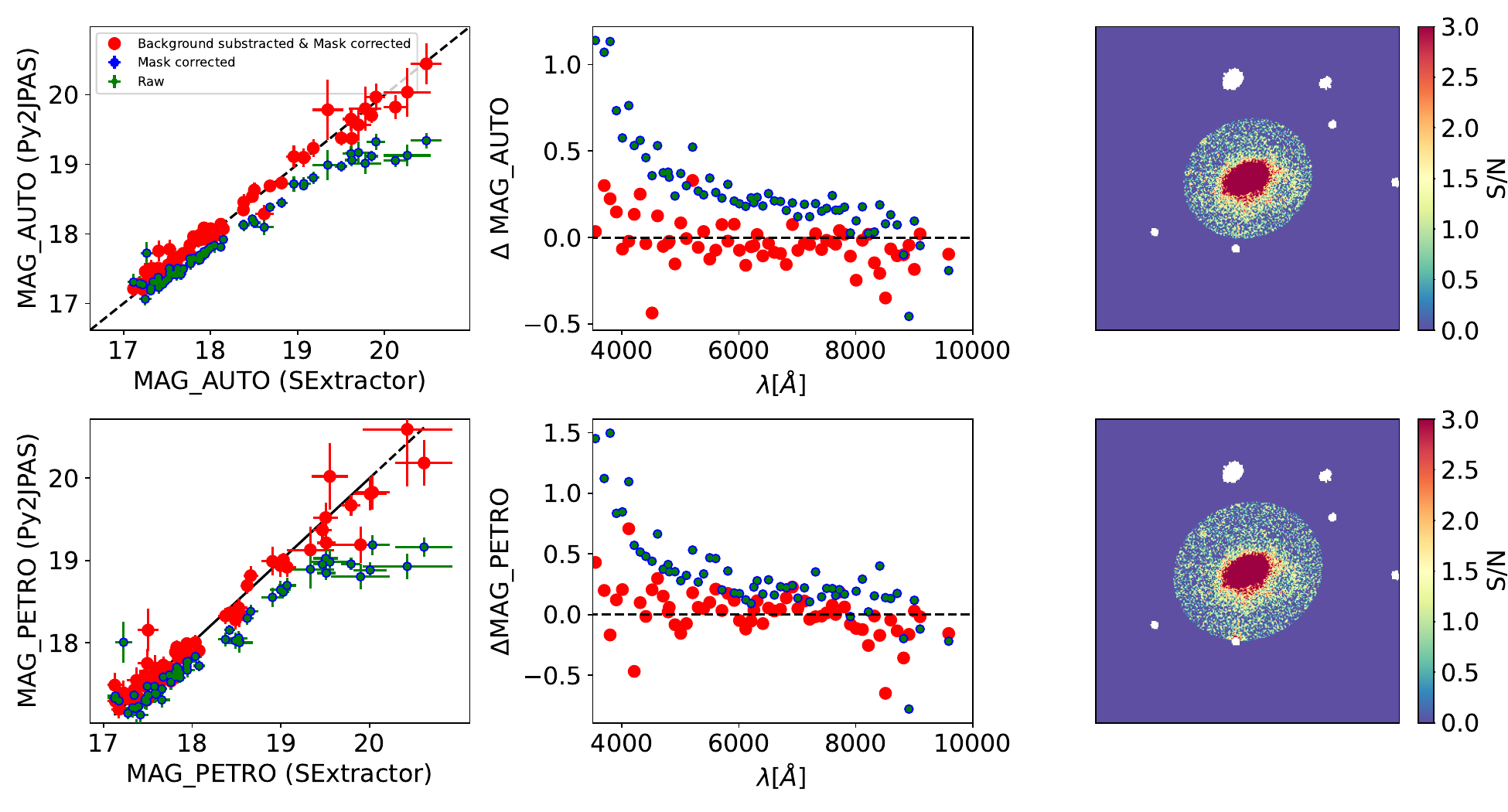}
    \caption[Comparison of the elliptical photometries available in the catalogue and the values retrieved with \PyDJ \ for the galaxy 2241--7608, using the raw image without accounting for masked pixels, the raw image accounting for masked pixels, and the raw image correcting for masked pixels and subtracting the backgroud. ]{Same as Fig.~\ref{fig:MANGAellPhot}, but for the galaxy 2241--7608.}
    \label{fig:badbackellPhot}
\end{figure}

    \item \textbf{Impact of the background subtraction}. Seeing these results, one might question the need to correct for background contribution. In Fig.~\ref{fig:badbackellPhot} we show the same plot as in the previous figure, but for the galaxy 2241--7608. In this case, we find that the differences among the calculated values and the values from the catalogue are quite significant when the background is not taken into account, reaching offsets larger than 1~mag. However, the background subtraction greatly improves the comparison for both \magauto \ and \magpetro. The main difference among both galaxies is the S/N reached close to the limit of the aperture: while in the case of 2470-10239, many of the pixels in the \rb{} managed to reach an individual S/N close or larger than 3, this is not the case for 2241--7608, where the aperture contains many pixels with very low S/N, which might actually be part of the background. After inspecting the galaxies one by one, we find a similar trend: galaxies whose apertures contain mostly pixels with high S/N ratio do not require a background correction to obtain the values of the catalogue, while galaxies with a significant fraction of pixels with low S/N require to be corrected from background. This can be expected, since the contribution of the background to pixels with high S/N (usually bright pixels) is most likely negligible, while pixels with lower S/N (usually dimmer pixels) are more affected by the background correction.

\begin{figure}
    \centering
    \includegraphics[width=0.74\textwidth]{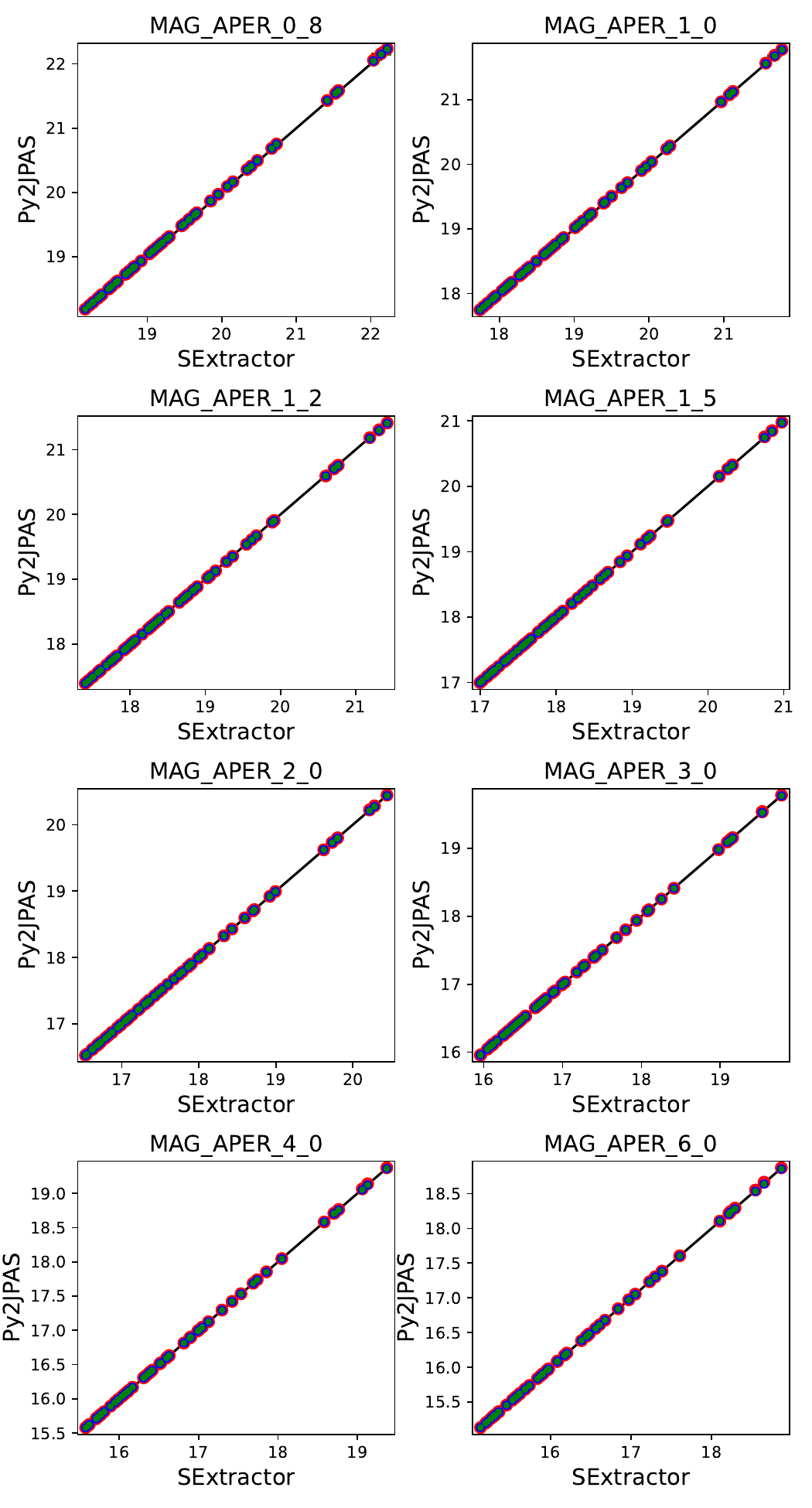}
    \caption[Comparison of the circular photometries available in the catalogue and the values retrieved with \PyDJ for the galaxy 2470--10239.]{Comparison of the circular photometries available in the catalogue and the values retrieved with \PyDJ for the galaxy 2470--10239. The red points represent the values of the photometry. The black solid line shows the identity relation.}
    \label{fig:MANGAcircPhot}
\end{figure}

    \item \textbf{Circular aperture photometry}. For the circular photometries, we use the \texttt{MAG\_APER} values provided in the catalogues. This photometries are obtained using circular apertures on the untreated \gls{ADU} images. The diameters of this apertures are $0.8'', 1.0'', 1.2'', 1.5'', 2.0'', 3.0'', 4.0'', 6.0''$.  The comparison can be seen in Figure~\ref{fig:MANGAcircPhot}. For these photometries, the agreement is even better, with a median difference generally lower than $0.01$ magnitudes (see Table~\ref{tab:MANGAphottest}). We also find, as suggested before, that there is no significant difference when using background estimation or accounting for the mask. This is most likely due to the small size of these apertures, centered in the brightest part of the galaxy. We therefore expect no nearby sources that need to be masked, and a negligible contribution from the background to these pixels. From this inspection, we can conclude that the background correction is important and it should be applied to galaxies (see next subsection to see the results for the complete selected sample). However, a good estimation of the background is highly non trivial and it might be advisable to be performed galaxy per galaxy. Nonetheless, we will discuss the automatising from a statistical point of view in the next subsection.

\end{itemize}

\subsection{Application to the \mjp \ sample}

After performing these tests and successfully retrieving the photometry of the catalogues, we proceed to apply our methodology to the selected spatially resolved galaxies from \mjp \ (see Sect~\ref{sec:code:sample} for the details of the selection).

\begin{figure}
    \centering
    \includegraphics[width=\textwidth]{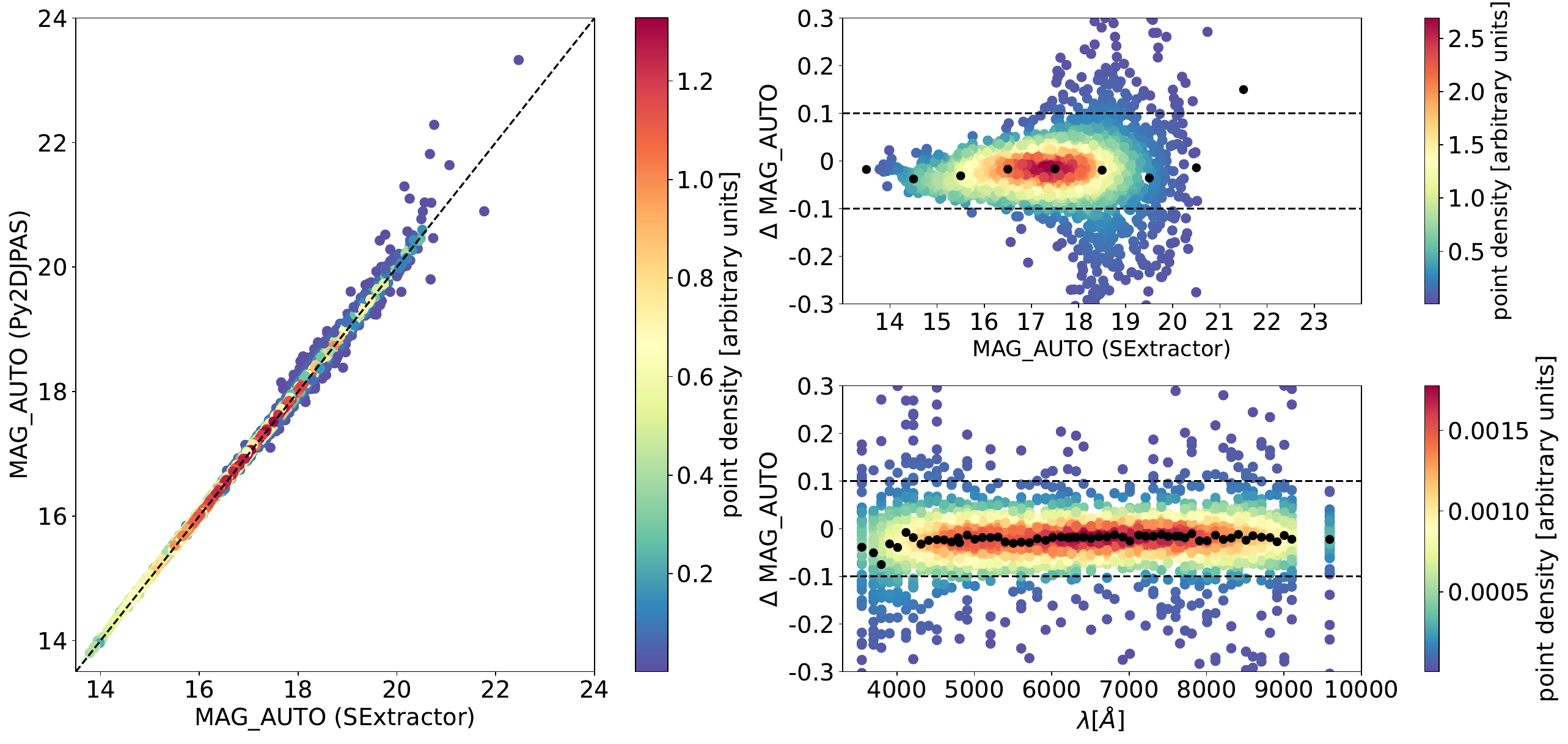}
    \caption[Comparison of the \magauto \ photometry of the catalogue and the one obtained with our methodology]{Comparison of the \magauto \ photometry of the catalogue and the one obtained with our methodology. Left panel shows the 1:1 relation. Upper right panel shows the difference of the magnitudes (\sext - \PyDJ) for each filter for each galaxy as a function of the magnitude of the band. Bottom right panel shows the difference for each filter for each galaxy as a function of the pivot wavelength of the filter. Colour scale represents the density of points. Black points represent the median value in each brightness bin and wavelength bin.}
    \label{fig:MAGAUTOdensity_corr}
\end{figure}

\begin{figure}
    \centering
    \includegraphics[width=\textwidth]{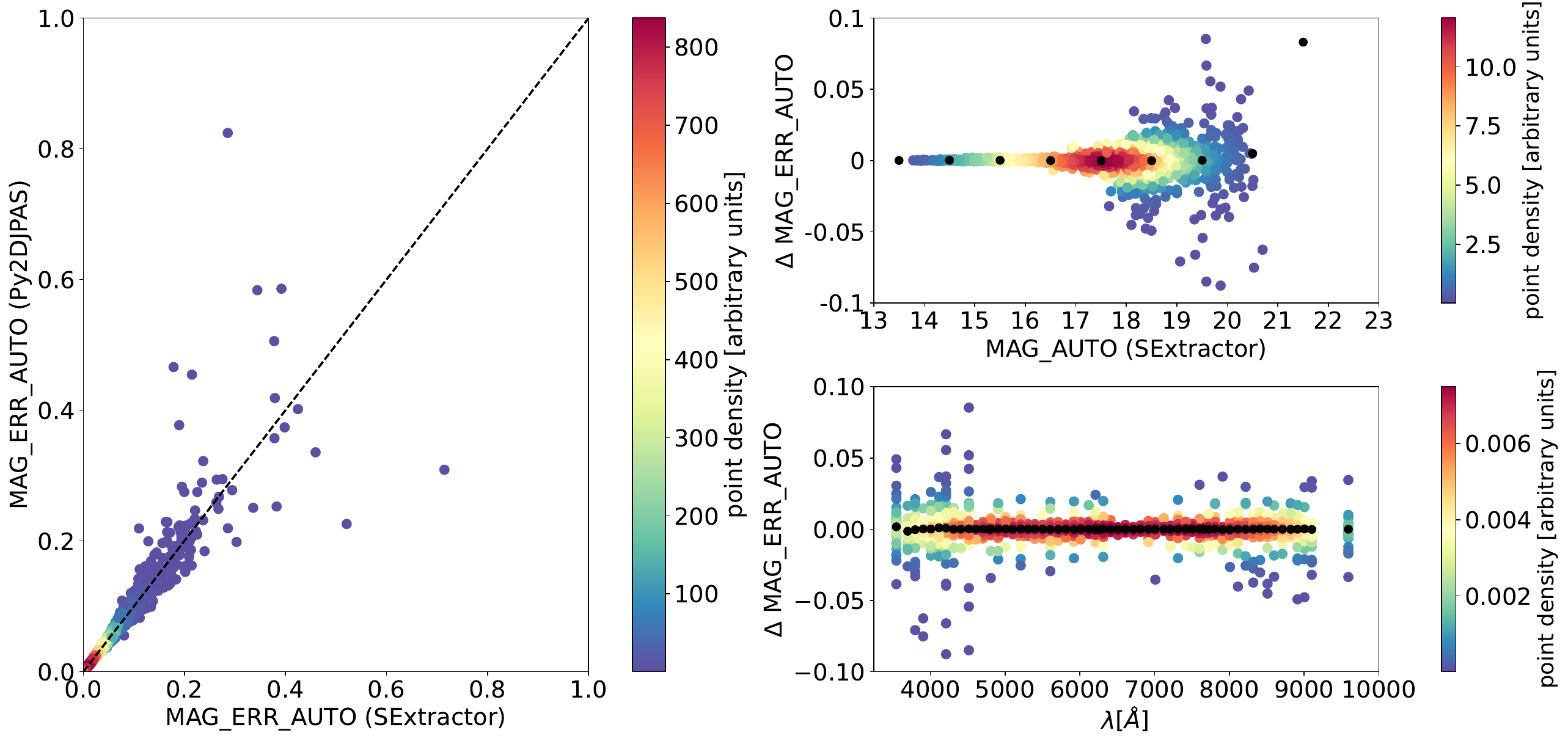}
    \caption[Comparison of the error of the \magauto \ photometry of the catalogue and the one obtained with our methodology]{Comparison of the error of the \magauto \ photometry of the catalogue and the one obtained with our methodology. Left panel shows the 1:1 relation. Upper right panel shows the difference of the magnitudes (\sext - \PyDJ) for each filter for each galaxy as a function of the magnitude of the band. Bottom right panel shows the difference for each filter for each galaxy as a function of the pivot wavelength of the filter. Colour scale represents the density of points. Black points represent the median value in each brightness bin and wavelength bin.}
    \label{fig:MAGAUTOerrdensity_corr}
\end{figure}

We fist compare the results obtained for \magauto \ (see Fig.~\ref{fig:MAGAUTOdensity_corr}). For this figure, we have used the background and masked pixels correction. We find a very tight co-relation around the 1:1 line, with most points lying perfectly in this relation. There are a few scattered points that show a larger offset. After individual inspection, we have found that these points correspond to very few, dimmer galaxies where a brighter source has not been perfectly masked or the background subtraction needs to be improved. Regarding the offsets by brightness, we find that the largest ones are found at magnitudes dimmer than 18~mag. However, the density of points is very low, and those points correspond only to very few galaxies. Additionally, the median difference is close to 0 and shows no significant bias. When studied by wavelength, we find that none of the filters shows a significant amount of larger offsets than the others and that the median difference remains close to 0 for all the filters. In both cases, the vast majority of points is contained within the $[-0.1, 0.1]$~mag range, which is a relative error below 10\%. The median values are all very close to 0. This graph shows that, even though some values show some larger offsets (of up to $\sim 0.3$~mag), from a statistical point of view our reconstruction is solid. 

We also show the comparison of the errors in Fig.~\ref{fig:MAGAUTOerrdensity_corr}. Even though the size of the aperture was fitted through the errors, we still include this graph as a sanity check, to verify that the errors obtained are compatible with the ones in the catalogues, and that there is no catastrophic failure during the iterative process, which means that the equation used to calculate the errors is actually capable of reproducing them. We find that most of the points are close to the identity relation, and that there is no significant bias in the median of the difference of the errors, with brightness or wavelength. Most of the values of the differences are below $0.1$~mag. The largest values of the difference in the errors are found in magnitudes dimmer than 19, and in some of the blue filters. These differences are likely caused by objects that are not perfectly masked, the different approach when accounting for the masked pixels, or because we are not using the weight images here. Nonetheless, the differences are small and the agreement in the magnitudes is very good.

\begin{figure}
    \centering
    \includegraphics[width=\textwidth]{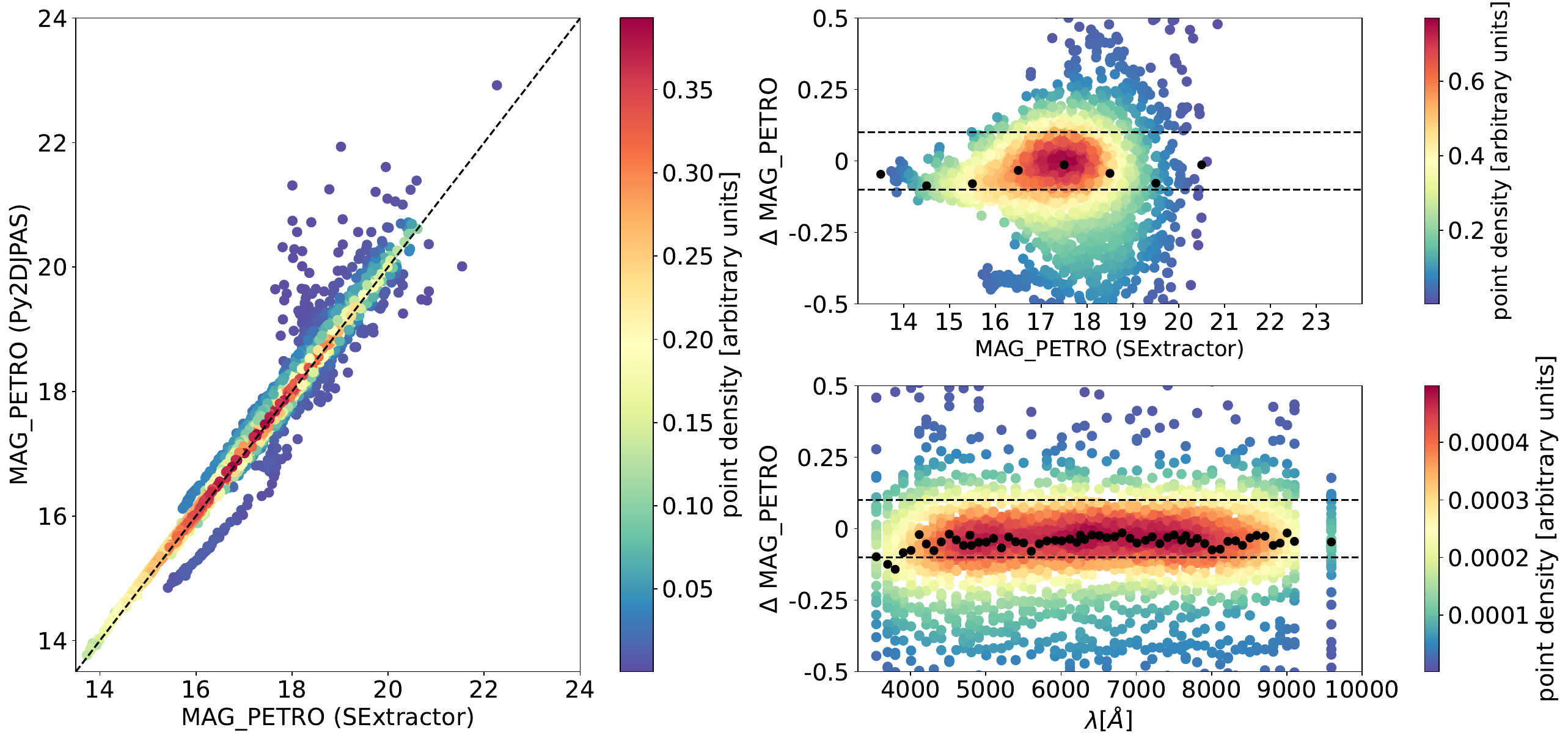}
    \caption[Comparison of the \magpetro \ photometry of the catalogue and the one obtained with our methodology]{Comparison of the \magpetro \ photometry of the catalogue and the one obtained with our methodology. Left panel shows the 1:1 relation. Upper right panel shows the difference of the magnitudes (\sext - \PyDJ) for each filter for each galaxy as a function of the magnitude of the band. Bottom right panel shows the difference for each filter for each galaxy as a function of the pivot wavelength of the filter. Colour scale represents the density of points. Black points represent the median value in each brightness bin and wavelength bin.}
    \label{fig:MAGPETROdensity_corr}
\end{figure}

\begin{figure}
    \centering
    \includegraphics[width=\textwidth]{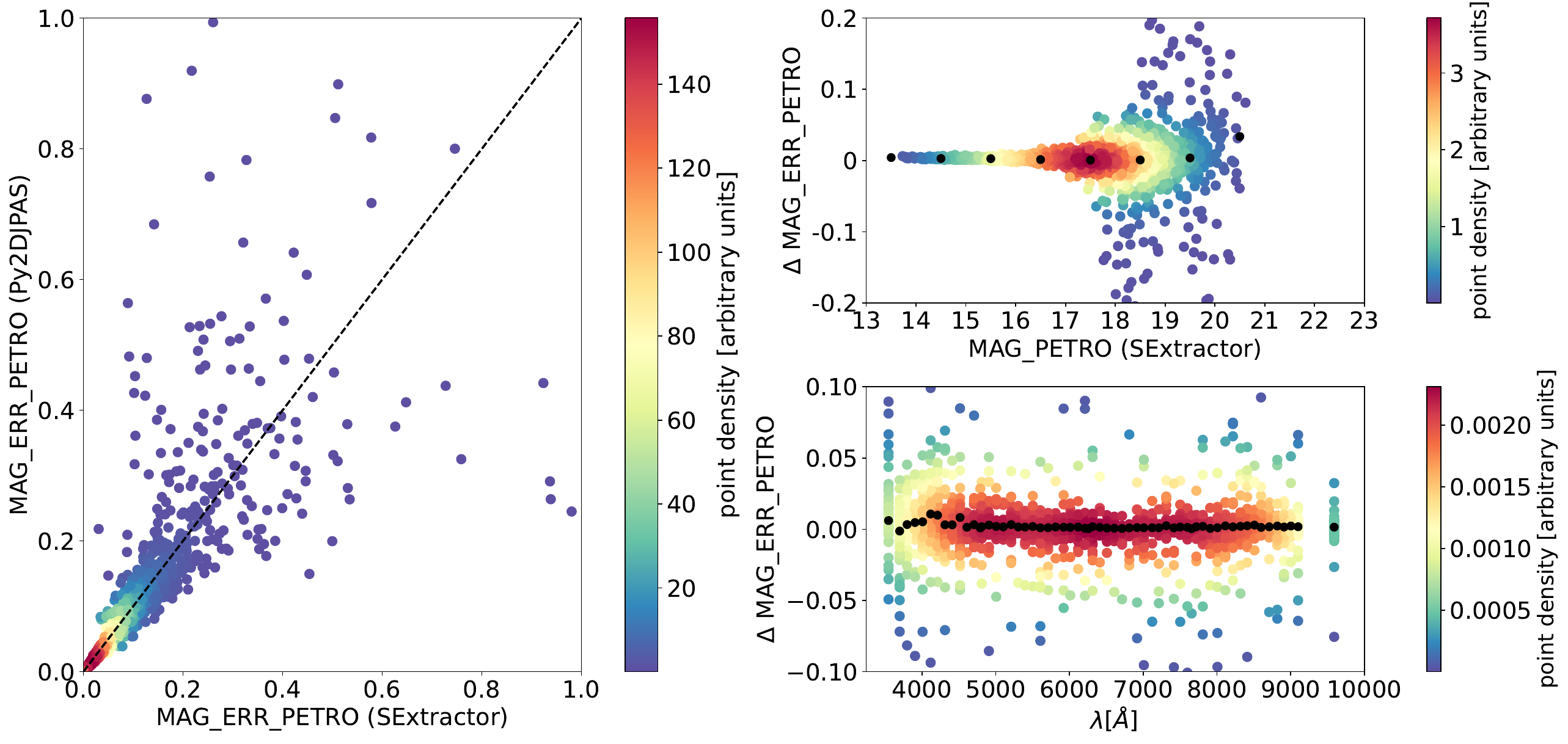}
    \caption[Comparison of the error of the \magpetro \ photometry of the catalogue and the one obtained with our methodology]{Comparison of the error of the \magpetro \ photometry of the catalogue and the one obtained with our methodology. Left panel shows the 1:1 relation. Upper right panel shows the difference of the errors of the magnitudes (\sext - \PyDJ) for each filter for each galaxy as a function of the magnitude of the band. Bottom right panel shows the difference for each filter for each galaxy as a function of the pivot wavelength of the filter. Colour scale represents the density of points. Black points represent the median value in each brightness bin and wavelength bin.}
    \label{fig:MAGPETROerrdensity_corr}
\end{figure}

The comparison of the \magpetro \ photometry yields similar results (see Fig.~\ref{fig:MAGPETROdensity_corr}). We find a larger dispersion than in the case of \magauto \ and a worse agreement in the results. Most of the points are still within the $[-0.1,0.1]$~mag range (relative error below 10\%), but we find a higher density of points than before in ranges with larger differences. There are more noticeable outliers which can be expected, since the  apertures are larger. Therefore it is  more likely to include improperly masked sources, there are a greater number of dimmer pixels, which are more sensible to the background correction. In fact, the galaxy with the largest systematic deviation from the identity relation (line of points with \magpetro \ between $\sim 16$ and $\sim 18$) is a single galaxy with a very large aperture, which contains some unmasked sources and many background pixels. Regarding the dependence on the brightness and wavelength, we also find that the median values are close to 0, without a significant bias.

We also include the comparison of the errors for this magnitudes as an additional sanity check (see Fig.~\ref{fig:MAGPETROerrdensity_corr}). The result from this check is analogous to the comparison yield in the magnitudes: most values are close to the identity relation, although the dispersion is larger than in the case of \magauto. The median value of the difference of the errors remains close to 0 both in brightness bins and in each filter, with no significant bias. The largest values of the differences are found at magnitudes dimmer than 18. Overall, the agreement of the \magauto \ magnitudes is better, but we find that our methodology is also capable of measuring the \magpetro \ magnitudes consistently with \sext, considering the sources of errors introduced by a larger aperture. 

\begin{figure}
    \centering
    \includegraphics[width=0.85\textwidth]{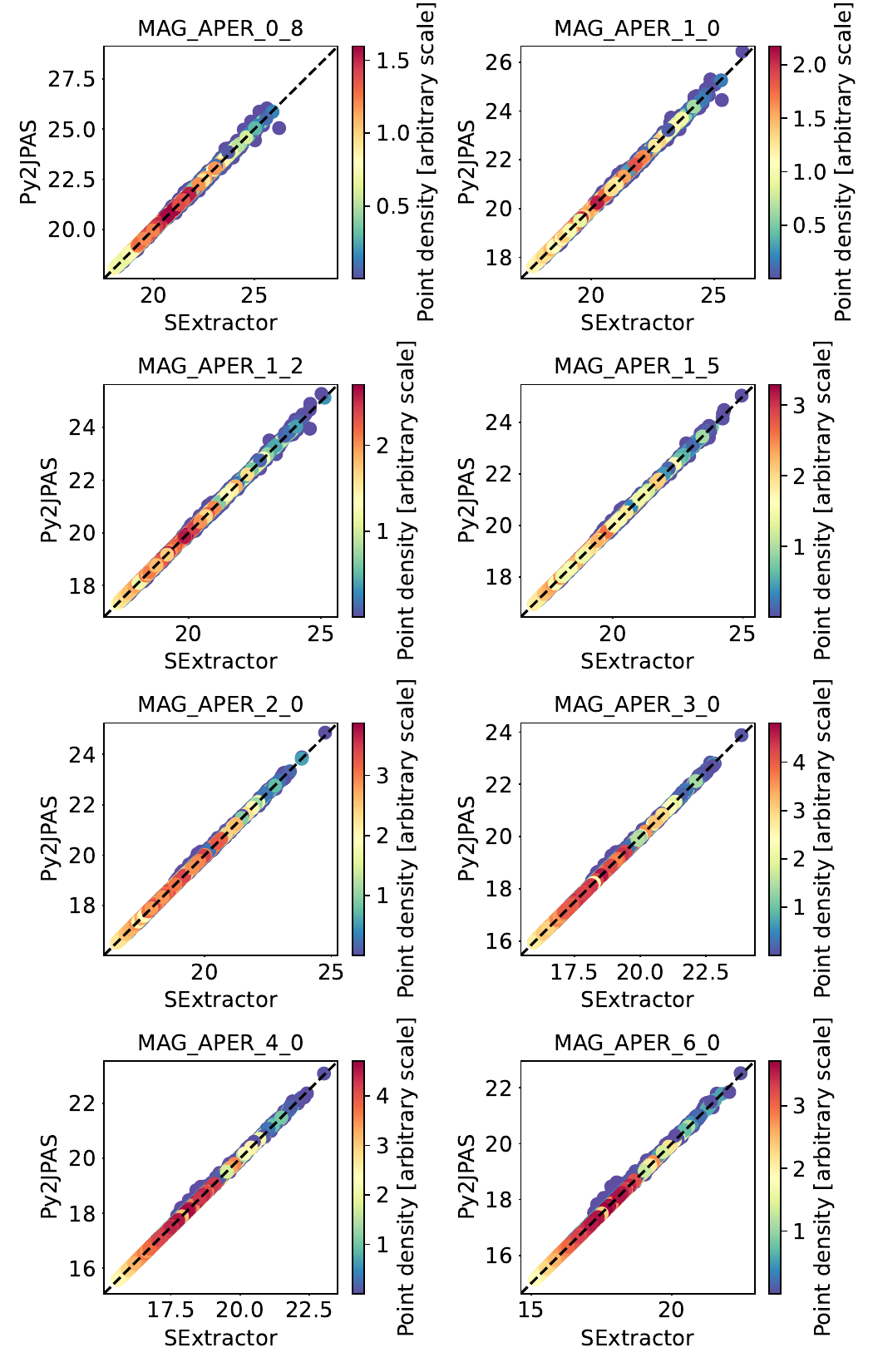}
    \caption[Comparison of the circular photometries of the catalogue and the one obtained with our methodology for the spatially resolved galaxies in \mjp]{Comparison of the circular photometries of the catalogue and the one obtained with our methodology for the spatially resolved galaxies in \mjp. Each point represents one filter of one galaxy. Colour scale represents the density of points. The black dashed line represents the 1:1 relation.}
    \label{fig:circulardensity}
\end{figure}

\begin{figure}
    \centering
    \includegraphics[width=0.85\textwidth]{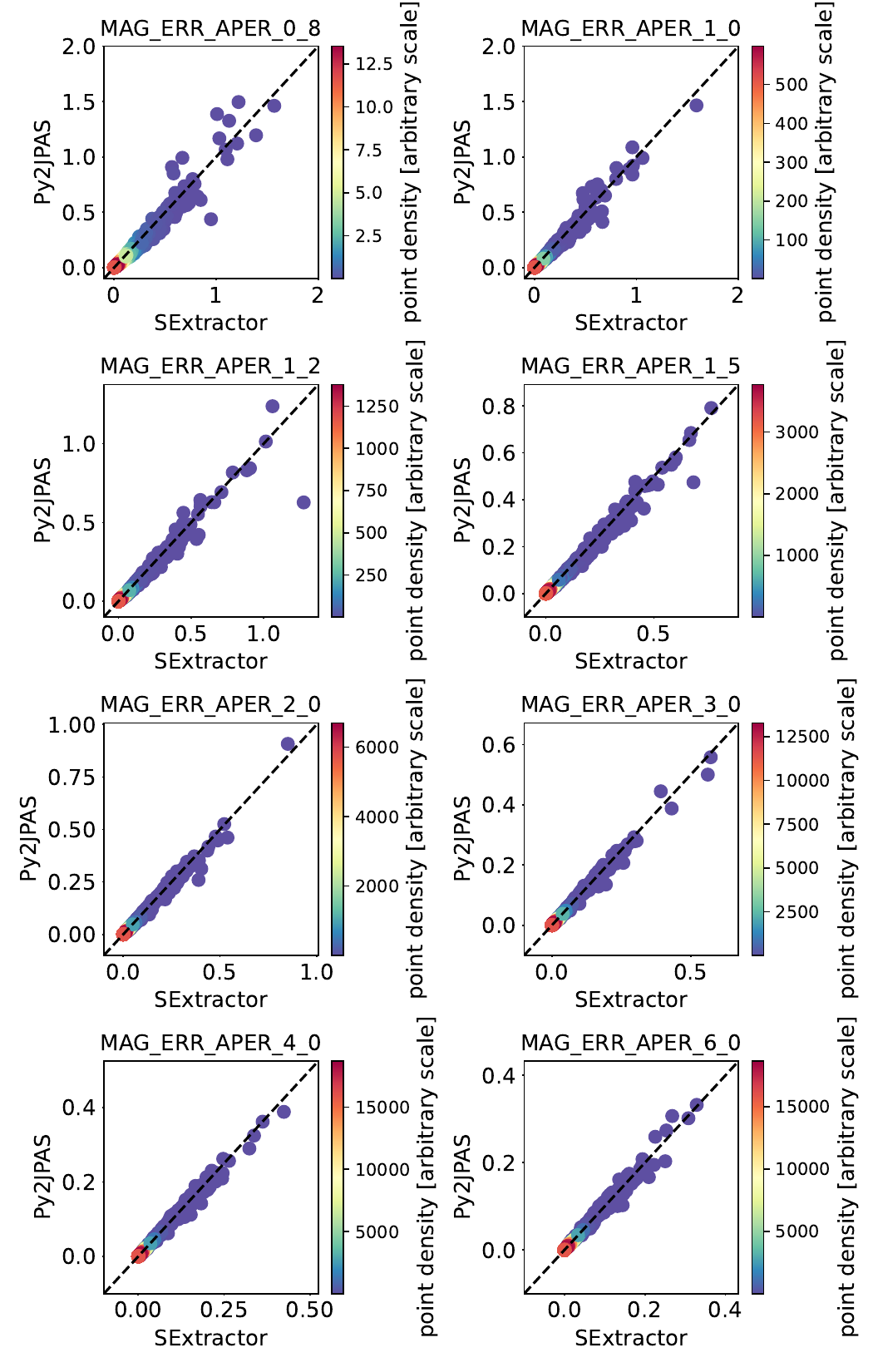}
    \caption[Comparison of errors the circular photometries of the catalogue and the one obtained with our methodology for the spatially resolved galaxies in \mjp]{Comparison of errors the circular photometries of the catalogue and the one obtained with our methodology for the spatially resolved galaxies in \mjp. Each point represents one filter of one galaxy. Colour scale represents the density of points. The black dashed line represents the 1:1 relation. }
    \label{fig:circulardensityerr}
\end{figure}

We finish by showing the comparison with the circular photometries (see Figs.~\ref{fig:circulardensity} and ~\ref{fig:circulardensityerr}). We find an even better agreement than before, and the relation becomes tighter as the aperture increases. This is actually a good result for us, since larger apertures contain more pixels and are more representative test of the calculations (smaller apertures are mainly dominated by the \gls{ZP}). Nonetheless, all the points are concentrated in a relation very close to the 1:1 target relation. We also find that the errors of the magnitudes are recovered with great accuracy, for a large interval of values ($\sim [0.04, 2]$~mag). This is also a good proof of concept, since the sizes of the apertures are unambiguously defined, which guarantees that the integrated region is the same using \sext \ and \PyDJ.

\section{Discussion}
In this chapter we have proven that we are able to accurately reproduce the values of the photometry given in the \mjp \ catalogues. We are aware that this is not a sufficient condition for our final purpose, which is to obtain the properties of the spatially resolved galaxies in \mjp. However, it is a necessary condition, and we have proven that this first requirement is properly fulfilled.

We find that some of the values show large offsets. However, when inspecting every individual galaxy, we find justification for these values. These offset are mostly due to the masks of bright sources or to the local background, which could be estimated using a different aperture for each galaxy. Of course, a dedicated mask and background estimation for each galaxy would provide better results, since it could account for the particularities of each case. However, this will not be feasible once the data from \jp \ arrive. We expect \jp \ data to contain thousands of spatially resolved galaxies, thus our aim to automatise  the process in preparation for this data. The automatising allows to work with larger data samples, at the expense of a worse performance in some particular cases. However, from a statistical point of view, these cases will have a lower weight if the method works well for the vast majority of cases. We find that this is the case of our method, where we accurately reproduce the values of the catalogue of \mjp, which are the publicly available reference values. 

We finish noting that we have used big, elliptical apertures and circular apertures for this chapter. These will not be the geometries used in the next chapters. However, that is the only difference, and the rest of the calculation is the same, so the validity of the results should remain the same. 

\section{Summary and conclusions}
In this chapter we have described and tested \PyDJ, our tool to automatise the analysis of the spatially resolved galaxies of \mjp, in preparation for the future \jp \ releases. We have described and tested our methodology, which can be sumarised in:
\begin{itemize}
    \item Download of scientific images and tables.
    \item Masking of the nearby sources to avoid biases and contamination.
    \item PSF homogenisation, assuming Gaussian models, to provide equivalent apertures thought all the filters.
    \item Magnitudes and flux calculations, using the equations found throughout the text.
\end{itemize}

We have tested the different steps of our methodology using the galaxy 2470--10239, the largest galaxy in the \mjp \ sample. We find that our methodology provides solid masks using the values $\mathrm{rms} = 5, \ \mathrm{npix} = 30$. The PSF homogenisation greatly improves the photometry of the inner regions, eliminating non-physical structures that appear in the \js. We have been able to accurately reproduce the values of the magnitudes in the \magauto, \magpetro, and circular photometries of the publicly available \mjp \ catalogue for the  sample of spatially resolved galaxies in \mjp. We argue that our method providing solid photometric measurements is the first step towards our final goal of studying the properties of this sample of galaxies, which will be done in the next chapters.

\chapter{The case of the galaxy 2470--10239} \label{chapter:MANGA}

\section{Introduction}

There is one galaxy in common in the \mjp \  sample and the \gls{MaNGA} survey \citep{MANGA2015}: the galaxy 2470--10239 ($\alpha =$ $14$~h $15$~min $20.37$~s; $\delta = 52\degree$ $20'$ $45.19''$). This is a bright ($r_{\mathrm{SDSS}} = 14.62$), massive ($M_\star = 10^{11.51}$~$\mathrm{M_\odot}$), red, elliptical galaxy. The redshift of this galaxy is $z=0.074$.  We choose this galaxy because its the largest galaxy in angular size ($\texttt{R\_EFF}\approx 2'$) in \mjp. This way, we can provide the best possible spatially resolved analysis (until the arrival of \jp \ data) as well as using different segmentation approaches to test our tool. In addition, we can use available data from the \gls{MaNGA} survey for these galaxy in order to show a comparison of the performance of our code applied to the \mjp \ data with the results obtained using the \gls{MaNGA} data, in a similar way to the example shown in \cite{Bonoli2020}.

We include this short chapter focused on this galaxy in order to provide a deeper insight in a ``best-case'' scenario for the \mjp \ data. However, we expect to have data from numerous galaxies which are even more suitable for spatially resolved studies in the upcoming data releases from \jp. Therefore, this chapter also aims at providing an example of the science that is to come in a near future.

\section{Data}

\begin{figure}
    \centering
    \includegraphics[width=\textwidth]{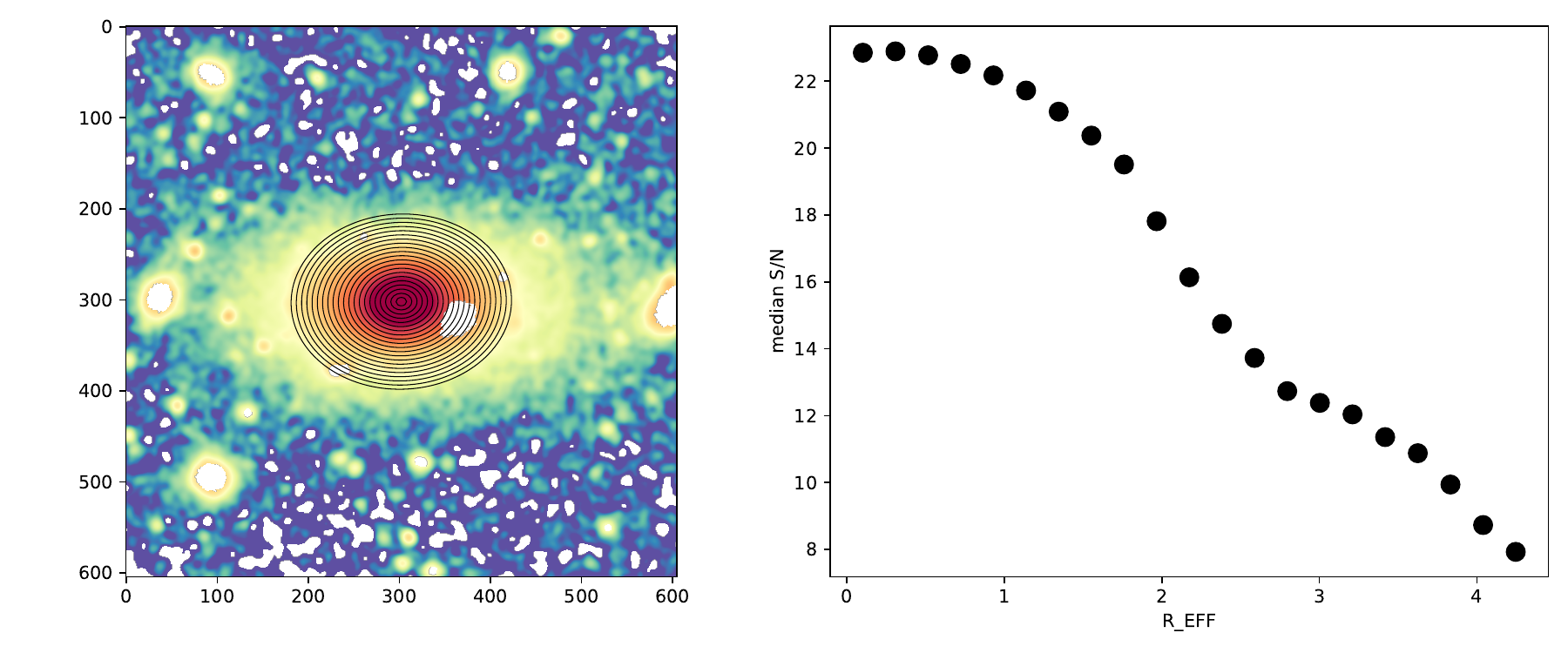}
    \caption[Elliptical apertures and radial variation of the S/N ratio for the galaxy 2470--10239]{Elliptical apertures and radial variation of the S/N ratio for the galaxy 2470--10239. Left panel: Elliptical rings used to obtain the radial profiles of the stellar population properties of the galaxy, over an image of the \rb \ band. Right panel: Median S/N ratio of all the filters as a function of the distance to the centre of the galaxy.  }
    \label{fig:MANGAradial_SN}
\end{figure}

The data used in this chapter comes from two different surveys. First, we will use the photometric data from the \mjp \ survey. The nature of this data has been summarised in Chapter~\ref{chapter:minijp} and further information can be found in the work by \cite{Bonoli2020}. We also use the spectroscopic data of this galaxy from the \gls{MaNGA} survey. Particularly, this galaxy is part of the SDSS DR14 \citep{MANGA2018}. 

In order to show the quality of the \mjp \ data for this galaxy, we show the spatial resolution achievable for this galaxy, using the \gls{FWHM} of the worst \gls{PSF} in all the filters as a minimum size, and how  the S/N ratio varies with the galacto-centric distance (see Fig.~\ref{fig:MANGAradial_SN}). We use the $\mathrm{R\_EFF}$ provided by \sext \ as the normalisation radius. We find that the S/N decreases with distance. It should be expected, since the brightness of the galaxy is expected to follow a Sérsic profile \citep{Sersic1963, Sersic1968}. We note that the S/N ratio is limited by the error of the ZP which, as argued in Chapter~\ref{chapter:code}, is the dominant factor of the error for the brightest regions. Taking into account that for this data release the error of the ZP was conservatively set as $\sigma_\mathrm{ZP} = 0.04$ for all the filters, the maximum S/N achievable is $\mathrm{S/N}\sim 27$, which is close to the S/N reached in inner regions.  This plot also shows that \mjp \ allows us to achieve a median S/N ratio of 10 at distances larger than 3~$\mathrm{R\_EFF}$, allowing us to derive the properties of the galaxy up to large distances from the centre. 

\section{Methodology}

For this chapter, we will mainly use our tool, \PyDJ, to obtain the photometry of the different regions of the galaxy. In order to derive the stellar population properties of these regions, we use \baysea  \ \citep[de Amorim et al. in preparation, see Sect.~\ref{sec:mjp:BaySR} for a summary, and see also][]{Rosa2021}. We shall also use the \gls{ANN} from \cite{Gines2021} to estimate the $\mathrm{H}\alpha$ emission (see Sect.~\ref{sec:mjp:ANN} for a summary). 

Concerning the data from \gls{MaNGA}, we use the \pycasso \ code \citep{Pycasso2017} for the analysis. This code deals with most of the required steps for the analysis, including the calculation of the parameters of the ellipse that better fits the light distribution of the galaxy, the binning of the flux from the the spaxels in the desired regions, and the retrieval of the stellar population properties of the regions using \texttt{STARLIGHT} \citep{cid2005}.

Throughout this chapter, we use the following methods in order to obtain the different regions of the galaxy: 
\begin{itemize}
    \item Elliptical rings with steps of $0.5$~HLR and $0.25$~HLR. The \gls{HLR} and ellipse parameters of these regions were calculated using \pycasso. In order to make sure that the apertures used with \PyDJ \ and \mjp \ data are equivalent to these ones, we used elliptical apertures of the same angular aperture, where $\mathrm{HLR} = 9.127''$ and the same elliptical parameters. These apertures will be used in order to compare the fluxes obtained with both methods.
    \item Elliptical rings using the maximum resolution allowed by the size of the \gls{FWHM} of the worst \gls{PSF}. The parameters of the ellipse are derived using the script included in \pycasso, using the \rb{} as reference. We will only use regions with median S/N ratio above five for the filters with $\lambda_{\mathrm{pivot}} < 5000$ (see Chapter~\ref{chapter:spatiallyresolved} for the discussion of this decision).
    \item The  \gls{BatMAN} binning by \cite{BATMAN}. We choose to only use the \rb{} flux image and its pixel error image as inputs. We restrict the code to find regions in an elliptical aperture with semi-axes $a = 2.5 \times \texttt{PETRO\_RADIUS} \times \texttt{A\_WORLD}$, $b = 2.5 \times \texttt{PETRO\_RADIUS} \times \texttt{B\_WORLD}$.
    \item The Voronoi binning method by \cite{Voronoi}. We use as input the uJPAS band image and a target S/N ratio of 5. We restrict the code to find regions in an elliptical aperture with semi-axes $a = 2.5 \times \texttt{PETRO\_RADIUS} \times \texttt{A\_WORLD}$, $b = 2.5 \times \texttt{PETRO\_RADIUS} \times \texttt{B\_WORLD}$.
\end{itemize}

We use the PSF homogenised images for all the analysis, except for the comparison of the magnitudes with \mjp \ data and \gls{MaNGA} data. For this comparison, we also shift the \gls{MaNGA} data so that all the spectra are into the observed frame.

\begin{figure}
    \centering
    \includegraphics{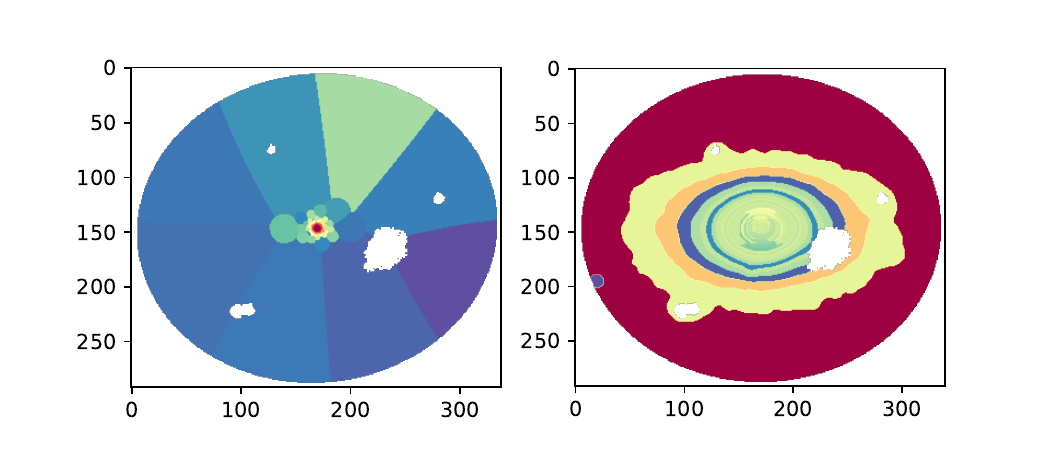}
    \caption[Segmentation maps obtained with the Voronoi and \gls{BatMAN} binning methods for the galaxy 2470--10239.]{Segmentation maps obtained with the Voronoi (left panel) and \gls{BatMAN} binning (right panel) methods for the galaxy 2470--10239. Each colour represents a region.}
    \label{fig:segmentationmaps}
\end{figure}

The segmentation maps obtained with the  \gls{BatMAN} and Voronoi binning methods can be seen in Fig.~\ref{fig:segmentationmaps}. For the  \gls{BatMAN} binning we find that regions follow a pseudo-elliptical rings shape, which could be expected given the distribution of the S/N with the galactocentric distance. On the other hand, the regions obtained with the Voronoi binning are  likely not physical and their distribution does not seem natural. These different maps manifest the approach of each method: the  \gls{BatMAN} binning aims at obtaining physical regions based on the S/N, which in this particular case leads to pseudo-elliptical regions. On the other hand, the Voronoi approach is closer to a purely mathematical method, aiming at reaching a minimum target S/N ratio, regardless of the underlying physics of the regions. After numerous attempts and tries,  the parameters chosen were the ones who gave the best compromise between binning regions and providing a considerable amount of regions. For example, when using the \rb{} as input for the Voronoi binning, most of the pixels were not binned since they already had the required S/N ratio. This is not a desired behaviour, since we know that bluer filters have a lower S/N and therefore performing a pixel by pixel fit would not make sense, because it would be computationally expensive and the results would not be reliable. When using other filters with lower S/N, the binning obtained is similar to the results shown here. Nonetheless, despite our efforts, the regions provided with this choice are not ideal either, since it provides mostly angular sector-like regions, which seem rather unnatural. 

\section{Results}
In order to fully show the capabilities of \PyDJ, we first compare the magnitudes obtained with our tool for the \mjp \ data and the magnitudes obtained with \pycasso \ for the \gls{MaNGA} data using equivalent apertures. We then perform a SED fitting  of the regions obtained with three different methods: the elliptical rings segmentation, in order to obtain the radial profiles of the properties and compare them with the profiles obtained with \pycasso \ and \gls{MaNGA},  the  \gls{BatMAN} binning by \cite{BATMAN} and the Voronoi binning method by \cite{Voronoi}.

\subsection{Flux comparison}

\begin{figure}
    \centering
    \includegraphics[width=\textwidth]{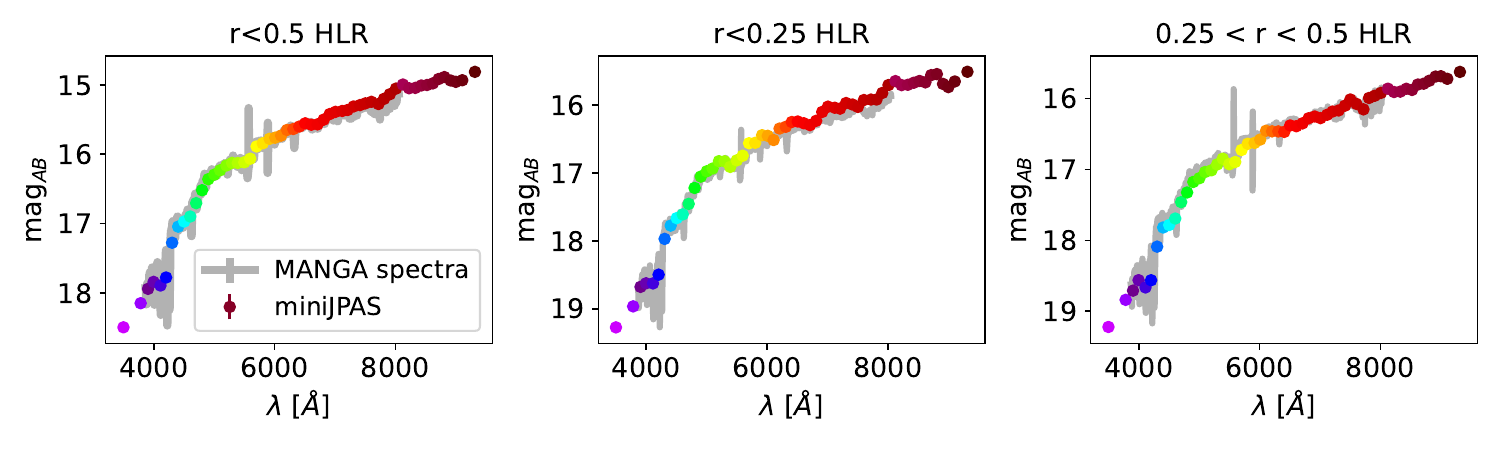}
    \caption[Comparison of the spectra obtained for the galaxy  2470--10239 with \gls{MaNGA} data and \mjp data.]{Comparison of the spectra obtained for the galaxy  2470--10239 with \gls{MaNGA} data and \mjp data. The grey lines represent the spectra obtained from \gls{MaNGA} data. Colour dots represent the \js \ obtained from \mjp data. Left panel shows the spectra for an elliptical aperture of $0.5$~HLR. Middle panel shows the spectra for an elliptical aperture of $0.25$~HLR. Right panel shows the spectral of an elliptical annulus with an inner radius of  $0.25$~HLR and an outer radius of  $0.5$~HLR.}
    \label{fig:MANGA_flux_com}
\end{figure}

We start comparing the magnitudes obtained with the \gls{MaNGA} data and \pycasso \ with the magnitudes obtained with \PyDJ \ for the \mjp \ data for equivalent apertures (see Figure~\ref{fig:MANGA_flux_com}), similarly to the result presented in \cite{Bonoli2020}. We use three different extractions, an elliptical aperture of $r<0.5$~HLR, an elliptical aperture of $r<0.25$~HLR and an elliptical ring of $0.25< r <0.5$~HLR. We can see that the extractions obtained using \PyDJ \ show a very good agreement over all the spectral range with  the spectroscopic data. 

This result acts as a sanity check in several aspects. First, we find that \PyDJ \ is able to retrieve fluxes and magnitudes consistent not only with \sext \ and the \mjp \ catalogues, but also with other surveys and codes. Second, it also shows that these fluxes and magnitudes are consistent not only for integrated values of the galaxy, but also for regions at different distances from the galactic centre. Lastly, it also shows that the calibration is also consistent with other surveys.

\subsection{Radial profiles of the stellar population properties}

\begin{figure}
    \centering
    \includegraphics[width=\textwidth]{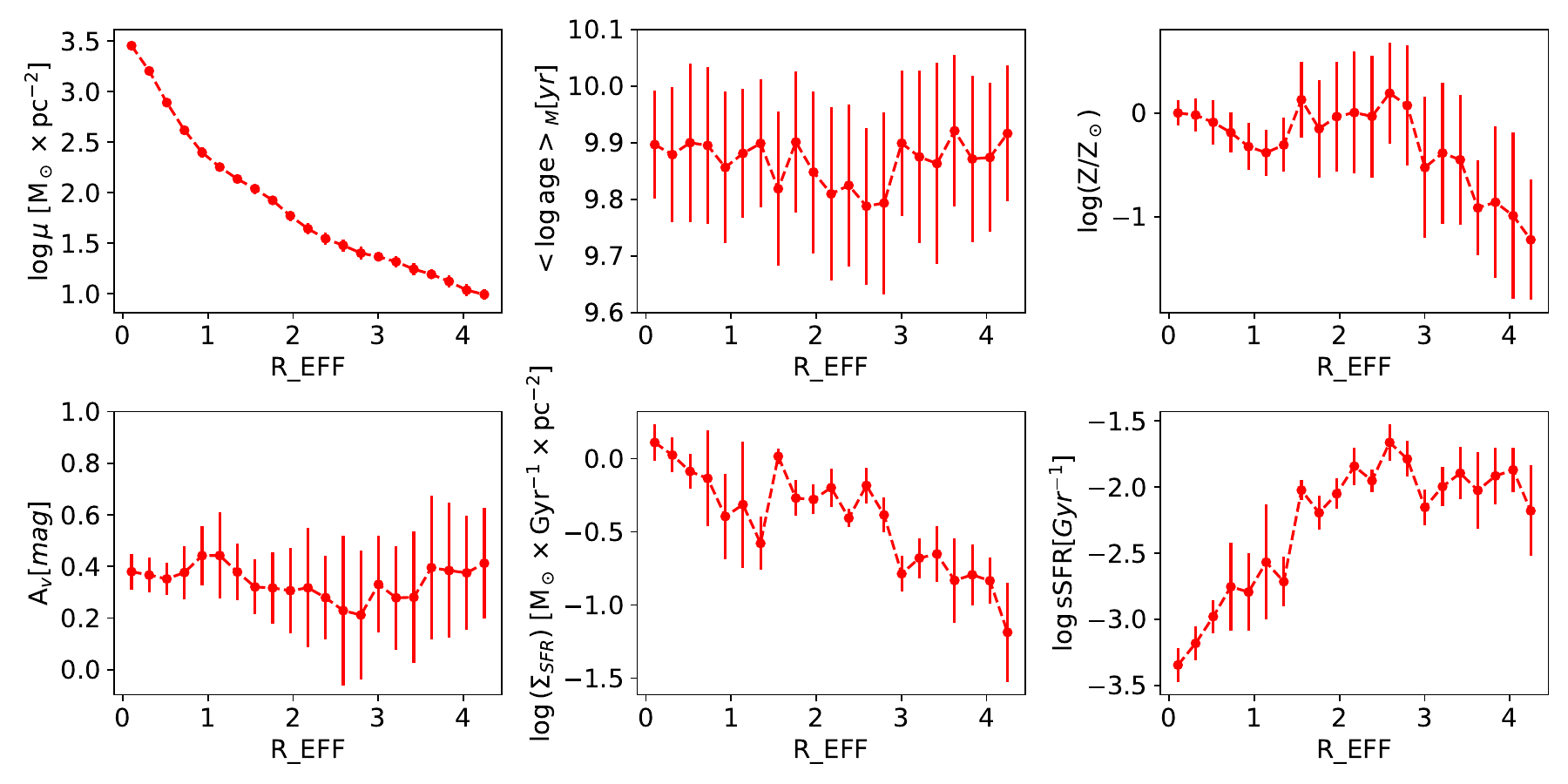}
    \caption{Radial profiles of the stellar population properties of the galaxy $2470-10239$. From left to right, up to bottom: stellar mass density, mass-weigthed age, stellar metallicity, $A_V$ extinction, SFR intensity and sSFR.}
    \label{fig:MANGAradSP}
\end{figure}

We study the radial profiles of several stellar population properties of the galaxy regions. In order to simplify the reading of the text, we specify now which are these properties and the units that we use, and we will not explicitly include the units during the discussion of the results:
\begin{itemize}
    \item The stellar surface mass density, $\mu_\star$, calculated as the total stellar mass of the region divided by its projected area: $\mu_\star = M_\star / A$. We measure $\mu_\star$ in units of $M_\odot \times \mathrm{pc}^{-2}$ and we will generally use the logarithm of this value.
    \item  The mass-weighted stellar age,  $\left <\log \mathrm{age} \right>_M$, calculated as $\left <\log \mathrm{age} \right>_M = \sum_i X_{M,i} \times \log \mathrm{age}_i$, where $X_{M,i}$ is the fraction of mass contributed by the i-th stellar model, with stellar age $ \log \mathrm{age}_i$,  to the total stellar mass of the region, $M_\star$. We measure the age in yr and we calculate the logarithm in this unit. 
    \item  The luminosity-weighted stellar age,  $\left <\log \mathrm{age} \right>_L$, calculated as $\left <\log \mathrm{age} \right>_M = \sum_i X_{L,i} \times \log \mathrm{age}_i$, where $X_{L,i}$ is the fraction of luminosity at $5635$~\AA \ contributed by the i-th stellar model, with stellar age $ \log \mathrm{age}_i$,  to the total stellar mass of the region at said wavelength. We measure the age in yr and we calculate the logarithm in this unit. 
    \item  The stellar metallicity,  $\left < \log \mathrm{Z/Z_\odot} \right >$, calculated as the mean metallicity from the models.
    \item  The extinction $A_V$, using the attenuation law by \citet{calzetti2000} which we added as a foreground screen. This parameter is given in AB magnitudes.
    \item  The intensity of the \gls{SFR}, $\Sigma_\mathrm{SFR}$, calculated as the ratio of the \gls{SFR} of the region divided by the projected area of the region, $\Sigma_\mathrm{SFR} = \mathrm{SFR}/A$. We calculate $\Sigma_\mathrm{SFR}$ in units of $M_\odot \times \mathrm{Gyr}^{-1} \times \mathrm{pc}^{-2}$, and we generally refer to the logarithm of this value.
    \item  The \gls{sSFR}, calculated as the ratio of the \gls{SFR} and the stellar mass of the region, $\mathrm{sSFR} = \mathrm{SFR}/M_\star$.  We calculate \gls{sSFR} in units of $\mathrm{Gyr}^{-1}$, and we generally refer to the logarithm of this value.
\end{itemize}

The radial profiles of these properties can be seen in Fig.~\ref{fig:MANGAradSP}. We obtain the following results:
\begin{itemize}
    \item The stellar mass density shows a very similar profile to that found by \cite{Rosa2014,Rosa2015, Bluck2020} and \cite{Abdurro2023}, this is, the surface mass density decreases as the distance to the centre increases. The profile found resembles a Sérsic profile \citep{Sersic1963} or a de Vaucouleurs profile \citep{deVaucouleurs1948}. Our result is more similar to the profile found by \cite{Rosa2015} than those found by \cite{Bluck2020} and \cite{Abdurro2023}. The values of the stellar mass density go from $\log \mu_\star \approx 3.5$ at the innermost regions up to $\log \mu_\star \approx 1$ at the outermost regions. Values found for this type of galaxies by \cite{Rosa2014,Rosa2015} range from $\log \mu_\star \approx 4$ at the innermost regions up to $\log \mu_\star \approx 2.5$ at 3~R\_EFF. At that distance, we find a value of $\log \mu_\star \approx 1.5$. We note that we are comparing a single galaxy with the average found by \cite{Rosa2015}. Also, results presented in that work use a Salpeter \gls{IMF} which, as also found on that work, on average provides an stellar mass $0.27$~dex larger than the \cite{Chabrier2003} IMF used in our analysis. Therefore, we find that the profiles are compatible, although we find a steeper decrease in $\log \mu_\star$ for this galaxy. If we compare the values of $\log \mu_\star$ that we obtain with those shown by \cite{Bluck2020} for the quiescent galaxies, we find an almost prefect agreement ($\log \mu_\star \approx 3.5$ at the central regions, $\log \mu_\star \approx 2.2$ at 1.4~R\_EFF). The surface mass density profiles found by \cite{Abdurro2023} go from $\log \mu_\star \approx 4$ in central regions up $\log \mu_\star \approx 2$ at 4~R\_EFF. However, they also find a dispersion in these values that is compatible with our results.
    \item The mass weighted age shows a very flat profile at $< \log \mathrm{age}>_M \approx 9.9$. This result matches perfectly the profile shown by \cite{Rosa2014}. However, radial profiles found in a later study by \cite{Rosa2015} for galaxies of similar mass are not so flat and show slightly younger populations in the outer regions of the galaxy, although the variation is lower than $0.2$~dex and the ages found are also compatible with our results. We also note that results from \cite{Rosa2014,Rosa2015} actually correspond to light weighted-ages, but we find no significant difference among the mass and light weighted ages once the errorbars are taken into account. The works by \cite{SanRoman2018,Bluck2020,Parikh2021} and \cite{Abdurro2023} also find flat age profiles. Additionally, table~3 from \cite{SanRoman2018} include several works who also find flat age profiles for early-type galaxies, such as \cite{Davies1993,Mehlert2003,Wu2005,SanchezBlazquez2006b,Reda2007,Rawle2008,Rawle2010,Wilkinson2015, Goddard2017b} or \cite{Zheng2017}.
    \item The stellar metallicity seems to decrease by $\sim 0.2$~dex from the centre up to 1~R\_EFF. Then it seems  to slightly increase up to $1.5$~R\_EFF, flattening up to 3~R\_EFF and then decreasing notably up to 4~R\_EFF. However, the uncertainties grow notably as the distance to the centre increases. Taking them into account, the general profile is compatible with a flat one where $<\log \mathrm{Z/Z_\odot} > \approx -0.2$~dex, although the tendency towards less metal rich values in outer regions seems clear. Metallicity profiles found by \cite{Rosa2015} for this type of galaxies are also rather flat, or slightly negative, but with values indicating that these galaxies are generally more metal rich. The gradient found by \cite{SanRoman2018} is negative, as well as many works summarised in Table~3 from that work, which find even steeper gradients \citep[see e.g.][]{Davies1993,Mehlert2003,Wu2005,SanchezBlazquez2006b, SanchezBlazquez2007,Reda2007,Rawle2008,Rawle2010,Wilkinson2015}. 
    \item The extinction $A_V$ shows a rather flat profile at  $A_V \approx 0.4$ with no significant variation of the properties within the errorbars, which generally increase towards the outer parts of the galaxy (lower S/N). For this parameter, the agreement with the results from \cite{Rosa2015} is not as good as in the previous cases. In that work, they find a steep decrease of the extinction from the central regions up to $\sim 0.5$~R\_EFF, where the profile becomes much flatter. However, this flattening occurs at $A_V \approx 0$. On the other hand, profiles found by \cite{SanRoman2018} are also flat. Nonetheless, we note that we are comparing a single galaxy with the averages for similar galaxies, so a similar behaviour should be found, but discrepancies from the general case can be expected.
    \item The intensity of the SFR also decreases towards the outer regions of the galaxy, from $\log \Sigma_\mathrm{SFR}=0$ down to $\log \Sigma_\mathrm{SFR}=-1$. The profiles found by \cite{Rosa2016} for this type of galaxies decrease slightly up to 2~R\_EFF and then increase slightly, over the same range of values. Nonetheless, since the star formation of these galaxies is usually so low, values themselves only really indicate that the region is quenched. In fact, all the galaxy remains below the quenched limit set by \cite{Bluck2020}, that is, a region that lies from $\log \Sigma_\mathrm{SFR}=0.5$ in central regions down to $\log \Sigma_\mathrm{SFR}=-0.5$ at $\sim 1.4$~\texttt{R\_EFF}.
    \item The \gls{sSFR} shows values increasing from $\log \mathrm{sSFR} \approx -3.5$ up to $\log \mathrm{sSFR} \approx -2$. This behaviour is very similar to the findings by \cite{Rosa2016} and \cite{Abdurro2023}. It could be a indicative of a inside-out quenching scenario, but the most relevant detail is that the galaxy remains well bellow the $\log \mathrm{sSFR}=-1$ limit for quenched galaxies proposed by \cite{Peng2010}.
\end{itemize}

\subsubsection{Comparison with \gls{MaNGA}}

\begin{figure}
    \centering
    \includegraphics[width=0.8\textwidth]{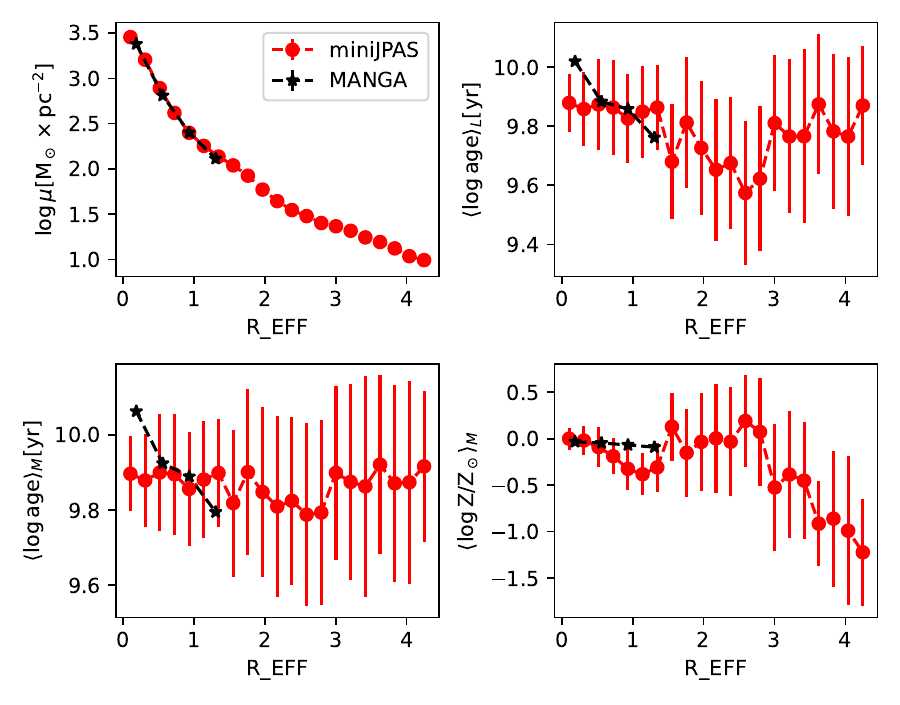}
    \caption{Comparison of the radial profiles of the stellar population porperties of the galaxy 2470--10239 obtained with our methodology and with \gls{MaNGA} data.}
    \label{fig:MANGAradialSPPcomp}
\end{figure}

The comparison of the mass density, mass and light weighted ages and the stellar  metallicity obtained with \PyDJ \ for the \mjp \ data and the results obtained with \pycasso \ for the \gls{MaNGA} data can be seen in Fig.~\ref{fig:MANGAradialSPPcomp}. We find that, even using different bins, codes, and data, the mass density profiles fit perfectly. Even thought the stellar mass density is very related to the light distribution of the galaxy due to the mass-to-light relation of the stellar models, this is still a very good proof of the validity of our methodology.

On the other hand, the age profiles obtained with our code appear to be flatter than the trend that the data from \gls{MaNGA} would suggest, which seem to show a negative gradient towards outer parts of the galaxy. However, we note that the errorbars of the \gls{MaNGA} data are not available, but the points are still compatible with our results, given our own uncertainty intervals. The trend in the metallicity is flat for both data sets, and  the values are compatible within the errorbars. We note that metallicity is usually estimated more precisely with spectroscopy, due to the possibility to use spectral indices. Another possible source of discrepancies is the use of different SED-fitting codes: while \starlight \ is a non-parametric code, \baysea \ uses a parametric \gls{SFH}. Despite these differences in methodology and data, we have shown that, when taking into account the uncertainty intervals, our methodology applied to \mjp \ data provides values of the stellar population properties that are consistent with the results obtained using spectroscopy.

\subsection{Stellar population maps}
In this section, we show the 2D maps of the stellar population properties obtained using the \gls{BatMAN} and the Voronoi binnings. Again, we study stellar mass density, the mass-weighted age, the stellar metallicity, the $A_V$ extinction, the SFR intensity and the sSFR.

\begin{figure}
    \centering
    \includegraphics[width=\textwidth]{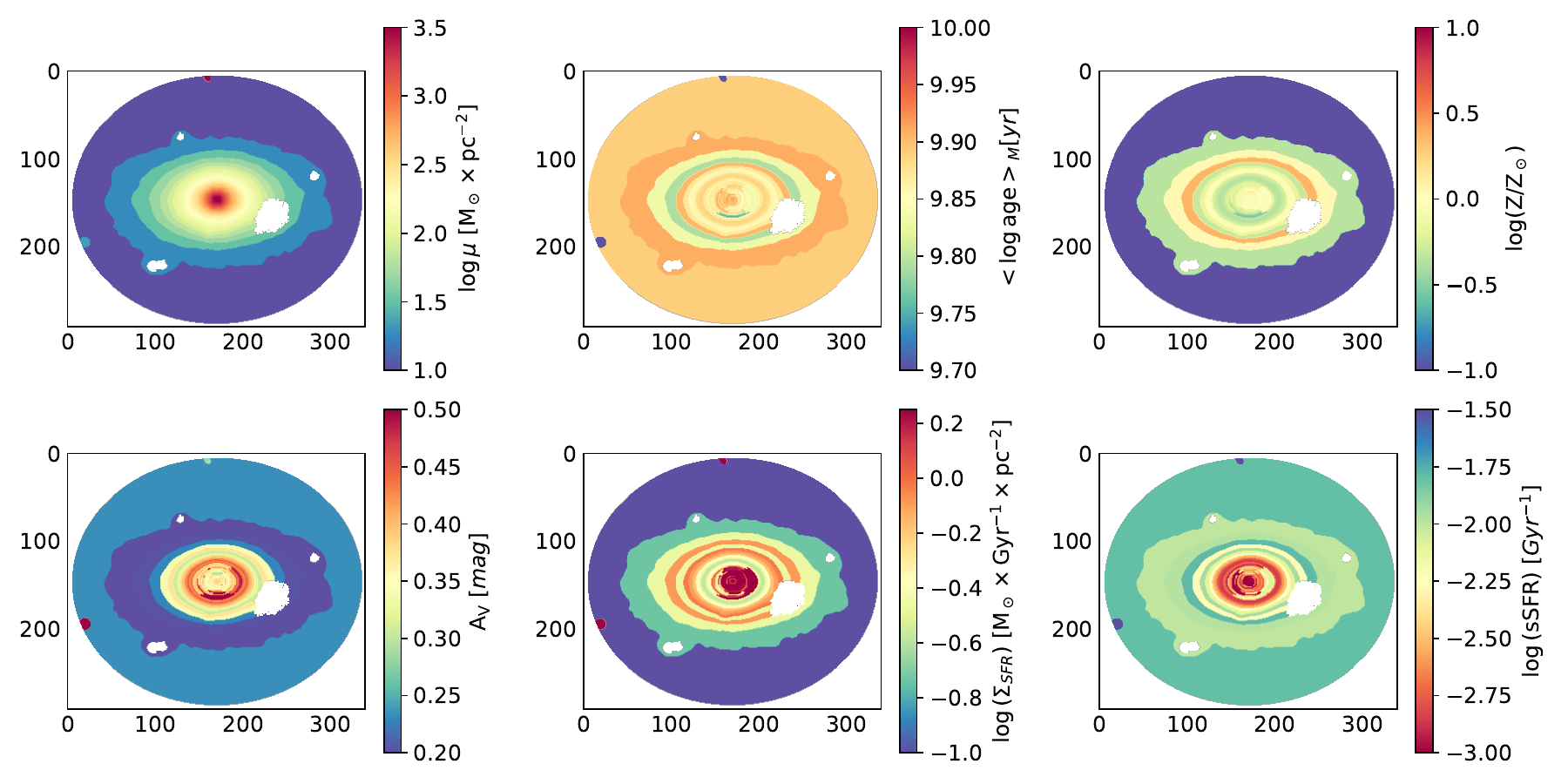}
    \caption[Maps of the stellar population properties of the galaxy 2470--10239 obtained with  \gls{BatMAN}]{Maps of the stellar population properties of the galaxy 2470--10239 obtained with  \gls{BatMAN}. From left to right, up to bottom: stellar mass density, mass-weighted age, stellar metallicity, $A_V$ extinction, SFR intensity and sSFR.}
    \label{fig:MANGABatman}
\end{figure}

The regions obtained with the  \gls{BatMAN} resemble elliptical annulus, with  some irregularities. We show the map of the spatially resolved stellar population properties in Fig.~\ref{fig:MANGABatman}. The behaviour we find is very similar to the one found in the elliptical ring segmentation. The last ring just consists on all the external pixels and its most likely made out of background, so its values should not be taken into account.

The stellar mass surface density shows the smoothest map out of all the properties. The density is clearly higher in the central regions and it decreases steeply towards the outer parts.  The other properties maps are not so smooth and the distribution looks more irregular. However, this is mostly due to the almost flat profile of all these properties, as well as the larger noise. In this sense it is also important to consider that \gls{BatMAN} provides regions of pixels with similar S/N. However, the final S/N of the region is not necessarily high enough. 

We find that the mass weighted age of the regions slightly fluctuates between  the values $< \log \mathrm{age}>_M \approx 9.9$ and $< \log \mathrm{age}>_M \approx 9.85$, with one region slightly younger. This is compatible with a flat profile within the uncertainty found.  The behaviour of the metallicity is similar to the age, showing values around  $<\log \mathrm{Z/Z_\odot} > \approx 0$, but in this case we do not find such steep decrease in the outermost parts. 

On the other hand, the extinction $A_V$ seems to show a not so flat gradient in this map. Values in the central regions are $A_V \approx 0.35$, then increases up to $A_V \approx 0.45$ for some intermediate rings, and then decreases down to $A_V \approx 0.2$. The differences with the radial profile actually are compatible within the errobars, but there might be a metallicity-extinction degeneracy in the outer regions.

Concerning the SFR intensity, we find a clearer distribution. The intensity is notably larger in central regions ($\log \Sigma_\mathrm{SFR} \approx 0.2$) and decreases towards the outer regions down to $\log \Sigma_\mathrm{SFR} \approx -0.7$. Some fluctuations are found in intermediate parts, just like in other properties, but we associate them to the uncertainty of the measurement over a range of values that merely indicate that this regions show no significant star formation. 

The sSFR maps shows a similar behaviour to the radial profile, where all the regions are well below the quench limit, but it increases from inside-out, from $\log \mathrm{sSFR}\approx-3$ up to $\log \mathrm{sSFR}\approx-1.75$. In general, we find that these maps offer a two-dimensional representation of the radial profiles show before, which could be expected due to the geometry of the regions. 

\begin{figure}
    \centering
    \includegraphics[width=\textwidth]{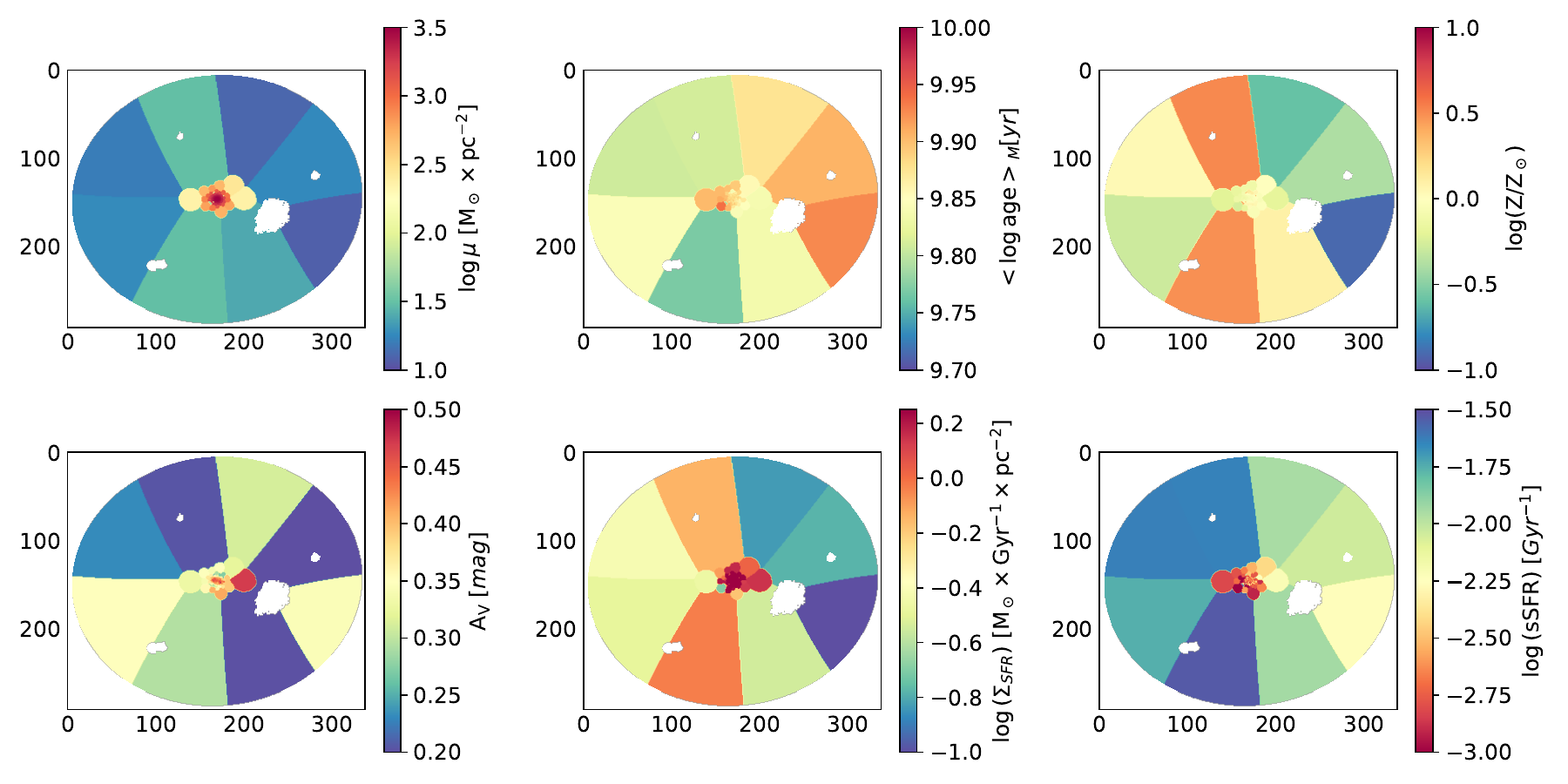}
    \caption[Maps of the stellar population properties of the galaxy 2470--10239 obtained with Voronoi binning.]{Maps of the stellar population properties of the galaxy 2470--10239 obtained with Voronoi binning. From left to right, up to bottom: stellar mass density, mass-weigthed age, stellar metallicity, $A_V$ extinction, SFR instenisty and sSFR.}
    \label{fig:MANGAvoronoi}
\end{figure}

The resulting stellar population properties maps from the Voronoi segmentation can be seen in Fig. ~\ref{fig:MANGAvoronoi}. These maps are hard to interpret, since the outer pixels are binned into angular-sector like regions. These outer regions fluctuate within a range of values that is larger or smaller depending on the property in question. We note that the Voronoi binning does not aim to provide \textit{physical} regions, but rather regions with a high enough S/N ratio. Therefore, the provided regions may be a mix of different stellar populations, which is what likely cause this fluctuations of values in the outer regions. Therefore, this might not be the best segmentation method for this type of galaxy.

The stellar mass density shows a smoother map again. Central regions show a larger mass density ($\log \mu_\star \approx 3.5$) and outer regions show lower values of up to $\log \mu_\star \approx 1$. Fluctuations in the outer regions are not really significant, and values are more similar among them than for the other properties. This is likely due to the fact that this parameter is the one that we retrieve with a larger precision. This map is similar to the one found using \gls{BatMAN} and is compatible with the radial profile using elliptical rings.

The mass weighted stellar age shows very similar values over all the regions, similarly to the \gls{BatMAN} and elliptical ring results. Inner regions values are almost constant, while outer regions show some fluctuations but are lower than $0.1$~dex. Overall, the map is compatible with a constant stellar age over all the regions. 

The stellar metallicity shows an almost constant value of $<\log \mathrm{Z/Z_\odot} > \approx 0$ in the inner regions, but more important fluctuations are found in the outer regions, with values ranging from $<\log \mathrm{Z/Z_\odot} > \approx 0.5$ down to $<\log \mathrm{Z/Z_\odot} > \approx -1$. We associate this fluctuations to the lesser precision in the estimation of this property, which is also seen in the radial profile derived using the elliptical rings. Also, we note that the least metal rich region correspond to the oldest region in the age map, which might be pointing to a degeneracy in the determination of the age and the metallicity. 

The extinction $A_V$ shows a value of $A_V \approx 0.5$ in one of the central regions, but our result for most of the inner regions is $\sim 0.35$. Once again, we find some striking fluctuations in the outer regions, with values in the interval $\sim [0.35,0.2]$. If we look at the three regions with the lowest extinction, we find that some of these values may actually be a consequence of an extinction-metallicity-age degeneracy, as well as the aforementioned mix of different stellar populations. 

The intensity of the SFR is notably higher in the innermost regions ($\log \Sigma_\mathrm{SFR} \approx 0$), and we newly find large fluctuations in the values of the outer regions, from $\log \Sigma_\mathrm{SFR} \approx 0$ down to $\log \Sigma_\mathrm{SFR} \approx -1$. Using the elliptical rings, we found a negative gradient in these property, but it seemed flatter using the regions obtained with \gls{BatMAN}. All the regions are below the quench limit shown by \cite{Bluck2020}.

The \gls{sSFR} is notably lower in the inner regions ($\log \mathrm{sSFR}\approx-3$) than in the outer regions (up to $\log \mathrm{sSFR}\approx-1.75$). This means that the same gradient found with the elliptical rings and with the \gls{BatMAN} segmentation is reproduced, although the fluctuations of the values of the outer regions appear again, like in other properties. All the regions are below the quench limit. 

In summary, we find that the three segmentations provide compatible results. The elliptical rings and \gls{BatMAN} segmentation throw similar results given the geometry of the regions. On the other hand, the Voronoi binning does not perform really well in the outer parts of the galaxy, providing regions that resemble angular sectors and whose results seem to be affected by the uncertainties and the mixture of different stellar populations, as well as the age-metallicity-extinction degeneracy. 

\subsection{Emission-line information}\label{sec:MANGAgal:SED}

\begin{figure}
    \centering
    \includegraphics[width=0.65\textwidth]{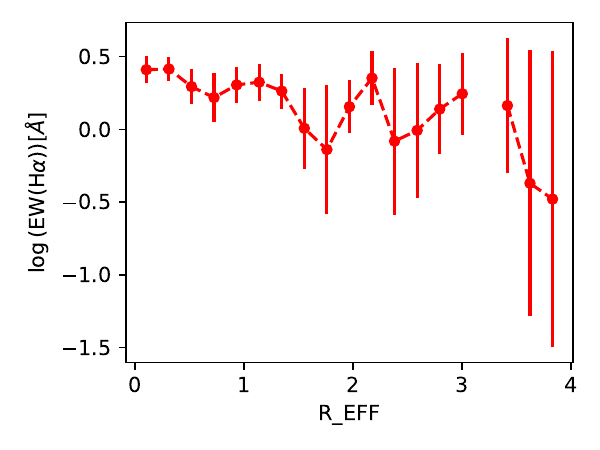}
    \caption{Radial profile of the EW($\mathrm{H}\alpha$)  of the  galaxy 2470--10239.}
    \label{fig:MANGAradEW}
\end{figure}

We conclude the results of these chapter by showing the radial profile of the \gls{EW} of the $\mathrm{H}\alpha$ emission for this galaxy (see Fig.~\ref{fig:MANGAradEW}). We choose to only show this line since this galaxy has no strong emission lines and $\mathrm{H}\alpha$ is the line predicted with a higher precision. Therefore, this is mainly an illustrative example of the capabilities of the code.

The EW($\mathrm{H}\alpha$) predicted with the \gls{ANN} also shows a flat profile around a low value, $ \log \mathrm{EW}(\mathrm{H}\alpha) \approx 0.2$~[\AA], meaning that the galaxy shows no significant emission at any distance of the centre. In fact, this values would belong to the retired region of the WHAN diagram \citep{WHAN_1, WHAN_2}. 

\section{Summary and concussions}

In this chapter we have applied the complete routine of \PyDJ \ to the galaxy 2470--10239. We have also applied the \pycasso \ program to the data available from the \gls{MaNGA} survey. We have found that our tool accurately reproduces the photometry obtained with spectroscopic data using equivalent apertures. Concerning the stellar population properties, we have used three different segmentations: elliptical rings using the maximum size allowed by the \gls{FWHM} of the worst \gls{PSF}, the \gls{BatMAN} binning by \cite{BATMAN} and the Voronoi binning by \cite{Voronoi}. Our main results and conclusions can be summarised as:

\begin{itemize}
    \item The whole workflow of \PyDJ \ provides values of the photometry not only consistent with the \mjp \ data release, but also with other survey (\gls{MaNGA}).
    \item We find radial profiles of the stellar mass density, stellar age, stellar metallicity, extinction $A_V$, intensity of the \gls{SFR}, and the \gls{sSFR} consistent with the literature: the stellar mass density decreases steeply with the distance to the galactic centre; the age, metallicity and extinction profiles are quite flat, the intensity of the \gls{SFR} decreases with the distance to the centre and the \gls{sSFR} increases with galacto-centric distance. Both the intensity of the \gls{SFR} and the \gls{sSFR} are below the quenched limits found in the literature.
    \item When compared to the results obtained with \gls{MaNGA} data and \pycasso, the stellar mass density, the stellar age, and the stellar metallicity profiles are compatible within the uncertainty intervals. The best agreement is found in the stellar mass density, while the age profile might seem to show an steeper gradient with \gls{MaNGA} data, but the values are still within the errorbars of our tool. 
    \item The \gls{BatMAN} binning provides very similar results to the elliptical ring segmentation, but offering a two-dimensional map of the properties instead of a one-dimensional profile. Slight fluctuations are found in some properties as the distance to the centre increases, most likely due to the uncertainty intervals over a small range of values, as well as a possible age-metallicity-extinction degeneracy.
    \item The Voronoi binning provides regions hard to interpret in the outer parts of the galaxy. The mass density profile and, to a lesser extent, the map of the intensity of \gls{SFR} and the \gls{sSFR} are clearly compatible with the radial profile and the \gls{BatMAN} segmentation. On the other hand, results of the stellar age and, more importantly, metallicity and extinction suffer from fluctuations in the outer regions, most likely due to a mixture of stellar populations in the regions, the aforementioned age-metallicity-extinction degeneracy and larger uncertainties. However, the values of the central regions show a great accordance with the other segmentations.
\end{itemize}

\chapter{Spatially resolved properties of the galaxies in \mjp} \label{chapter:spatiallyresolved}

\begin{abstract}
    In this chapter we study the properties of the spatially resolved  galaxies in \mjp, using our tool \PyDJ. We select a total of 51 galaxies that are suitable for this purpose and classify them by their spectral type (red or blue) and their environment (galaxies in groups or in the field), in order to study the effect of the environment. We find that 15 galaxies are red in the field, 9 are red in  groups, 21 are blue in the field, and 6 are blue in  groups. We use elliptical rings of the maximum size allowed by the FWHM of the worst PSF, elliptical rings using steps of $0.7$~\texttt{R\_EFF} to study the properties of the regions of the galaxies, and we use an  inside out segmentation to study the \gls{SFH} of the galaxy. We find that the stellar population properties of galaxies, as well as the emission line properties, are generally distributed in a clear way if we use a mass density--colour diagram. We find that redder, denser regions are usually older, more metal rich, and show lower values of the $\Sigma_\mathrm{SFR}$ and \gls{sSFR} (they are more quiescent) than bluer, less dense regions. These regions also show less intense emission lines. The radial profiles of the properties are compatible with the results from other works, suggesting a inside--out formation scenario, along with the results from the \gls{SFH}. The profiles of red and blue galaxies are clearly different, but we find no significant effect of the environment on them. 
\end{abstract}

\section{Introduction}
The evolution of galaxies is driven by many factors. The total stellar mass is known to correlate with other properties such as the stellar age and metallicity \citep{Gallazzi2005, Peng2015} or the \gls{SFR} \citep{Brinchmann2004,Noeske2007,Salim2007,renzini2015objective}. Processes related with the environment, such as ram-pressure stripping \citep{Gunn1972}, tidal stripping \citep{Malumuth1984}, or harassment \citep{Moore1996} can also play a role in the evolution of galaxies. For example, the fraction of red galaxies is known to correlate with the density of the environment \citep{Balogh2004}, and there exist a well-known density--morphology relation \citep[e.g.][]{Cappellari2011,muzzin2012, fogarty2014}.

However, recent studies show that the evolution of galaxies is mainly driven by local processes \citep[see][for a review]{Sanchez2020, Sanchez2021}. Also, similarly to the integrated case, there is a bimodality in many properties of galaxy regions \citep{Zibetti2017},  and the stellar mass acts both as a local and a global driver of the evolution of galaxies  \citep{Zibetti2022}. Mass density has also been proved to be a  very important \citep[see e.g.][]{Kauffmann2003b, Kauffmann2003a, Kauffmann2006}, although its relevance also depends on the structure of the galaxy \citep{Rosa2014}. However, some relations that are found between the global properties of galaxies do not remain true at smaller scales, such as the correlation between the molecular gas and high-mass star formation \citep{Kruijssen2019}. Therefore, the study of the spatially resolved galaxies is fundamental to better understand the processes that drive the galaxy formation and evolution.

Colour gradients have been observed for many years in different types of galaxies \citep{Peletier1990,Peletier1996,deJong1996,Silva1998,Bell2000,LaBarbera2004,Wu2005,MunozMateos2007,Roche2010,Tortora2010}, which have been usually interpreted as age and metallicity gradients. Thanks to the capabilities of \gls{IFU} surveys, such as CALIFA \citep{CALIFA2012} or MaNGA \citep{MANGA2015}, the PHANGS-MUSE survey \citep{PhangsMuse2022} or the WEAVE-Apertif survey \citep{Hess2020} these gradients have been interpreted more precisely, as well as the properties of galactic regions. In this regard, we remark the works carried out using CALIFA data, which have unveiled the radial structure and properties of the stellar population properties, \gls{SFR}, \gls{SFH}, and mass--to--light ratios of galaxies \citep{Rosa2014,Rosa2015,Rosa2016,Rosa2017,Ruben2017}, as well as the work by \cite{SanRoman2018}, which also studied the stellar population properties of the spatially resolved galaxies using the photometric data from the ALHAMBRA survey, and can be thus considered as the precursor from our work.

The \jp \ survey data  will be excellent to study the properties of the spatially resolved galaxies thanks to its photometric filter system, large \gls{FoV} and footprint, allowing for studying large galaxies without aperture bias or \gls{FoV} limitations, while also performing SED-fitting with an spectral resolution comparable to very low resolution spectroscopy. In this chapter, we aim at studying the spatially resolved properties of the galaxies in \mjp, taking advantage of the capabilities of the \jp \ filter system and the large \gls{FoV} of the survey, in order to unveil the properties of these galaxies and the role that the environment might be playing in their evolution, while also showing the power of the future \jp \ data to perform this kind of studies.

\section{Data}
In this section we describe the nature of our data, all belonging to the \mjp \ survey. We describe the selection of the galaxies we use for our study, as well as their classification into galaxies in field or in groups and inspect their observational properties.

\subsection{The \mjp \ survey}
The data used for this chapter proceeds completely from the \mjp \ survey \citep{Bonoli2020}. The main technical aspects of this survey are summarised in Chapter~\ref{chapter:minijp}, and some scientific results found with this survey that are relevant for our work have been summarised in Chapter~\ref{chapter:minijpresults}.

The data from \mjp \ and \jp \ are ideal to perform IFU-like studies, since its large FoV allows for observing large galaxies in the same pointing, and the filter system allows for retrieving the stellar population properties of the galaxies with great precision \citep[see][]{Rosa2021}. Moreover, the nature of the survey allows for cluster and groups detection, with no selection bias in the targets observed  \citep[see e.g.][]{Doubrawa2023, Maturi2023}. 

\subsection{Sample selection}
The sample used to study the properties of the spatially resolved galaxies in \mjp \ is the same as that in Chapter~\ref{chapter:code} (see Sect.~\ref{sec:code:sample} for a more detailed explanation). In a few words, galaxies have been selected so that: 
\begin{itemize}
    \item Galaxies are at least twice larger than an assumed limiting \gls{FWHM} of the \gls{PSF}, allowing for at least two extractions ($\mathrm{R\_EFF} > 2''$)
    \item There are no edge-on galaxies (ellipticity must be smaller than $0.6$)
    \item There are no artifacts or nearby sources that can bias the photometry (\texttt{MASK FLAGS}=0 for all bands, $\texttt{FLAGS}$ must not contain the flag 1).
    \item Selected objects are galaxies ($\texttt{CLASS STAR} <0.1$)
\end{itemize}

With this criteria we obtain a total number of 51 galaxies. These galaxies are later divided into red and blue galaxies using \cite{Luis2023} selection criterion, which is  an adaptation of the criterion given by \cite{Luis2019}, previously used by \cite{Rosa2021} to segregate the whole galaxy populations in \mjp \  in red and blue galaxies. We consider galaxies to be red if:
\begin{equation}
    (u-r)_{\mathrm{int}}> 0.16 (\log( M_{\star})-10)-0.254(z-0.1)+ 1.689,
\end{equation}
and blue otherwise. With this criterion we obtain a set of 27 blue galaxies and 24 red galaxies.

\subsection{Environmental classification}
To study the effects of the environment, we further classify our sample galaxies into galaxies in the field and galaxies in groups. For such purpose we use the adaptation of the AMICO code \citep[][]{Maturi2005, AMICO} done by \cite{Maturi2023} for the \mjp \ data release. The main aspects of those works are summarised in Sect~\ref{sec:mjp:AMICO}. This code is based in the Optimal Filtering technique. \citep[see ][for more details about Optimal Filtering]{Maturi2005, Bellagamba2011}, and among other parameters, it provides a probabilistic association for each galaxy for each cluster/group detection.

This probabilistic association has been used in previous works in order to classify galaxies and study the effects of  environment in galaxy evolution \citep{Rosa2022, Julio2022}. We shall use the same classification as \cite{Rosa2022}, where we classify galaxies as galaxies in groups if their largest probabilistic association is larger than $0.8$, and we consider them as field galaxies if its probabilistic association is smaller than $0.1$. This gives us a total of 15 red galaxies in the field, 21 blue galaxies in the field, 9 red galaxies in groups and 6 blue galaxies in groups. 

\subsection{Observational properties of the sample}

\begin{figure}
    \centering
    \includegraphics[width=\textwidth]{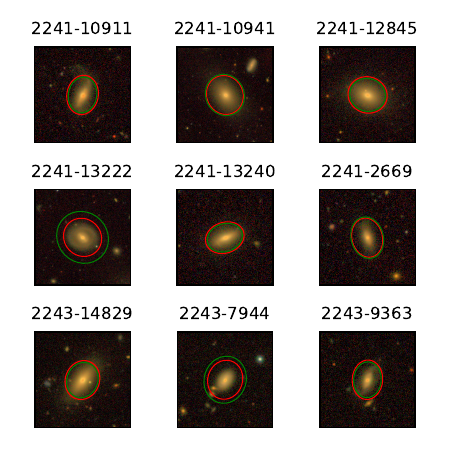}
    \caption[RGB images of the red galaxies in the field from our sample (Part 1)]{RGB images of the red galaxies in the field from our sample (Part 1). The red ellipses shows the \texttt{AUTO} ellipses calculated with Eqs.~\ref{eq:akron} and \ref{eq:bkron}. The green ellipses show the \texttt{PETRO} ellipses calculated with Eqs.~\ref{eq:apetro} and \ref{eq:bpetro}. The band used to make the RGB images are \ib{} (R) \rb{} (G) and \gb{} (B).}
    \label{fig:RedfieldRGB1}
\end{figure}

\begin{figure}
    \centering
    \includegraphics[width=\textwidth]{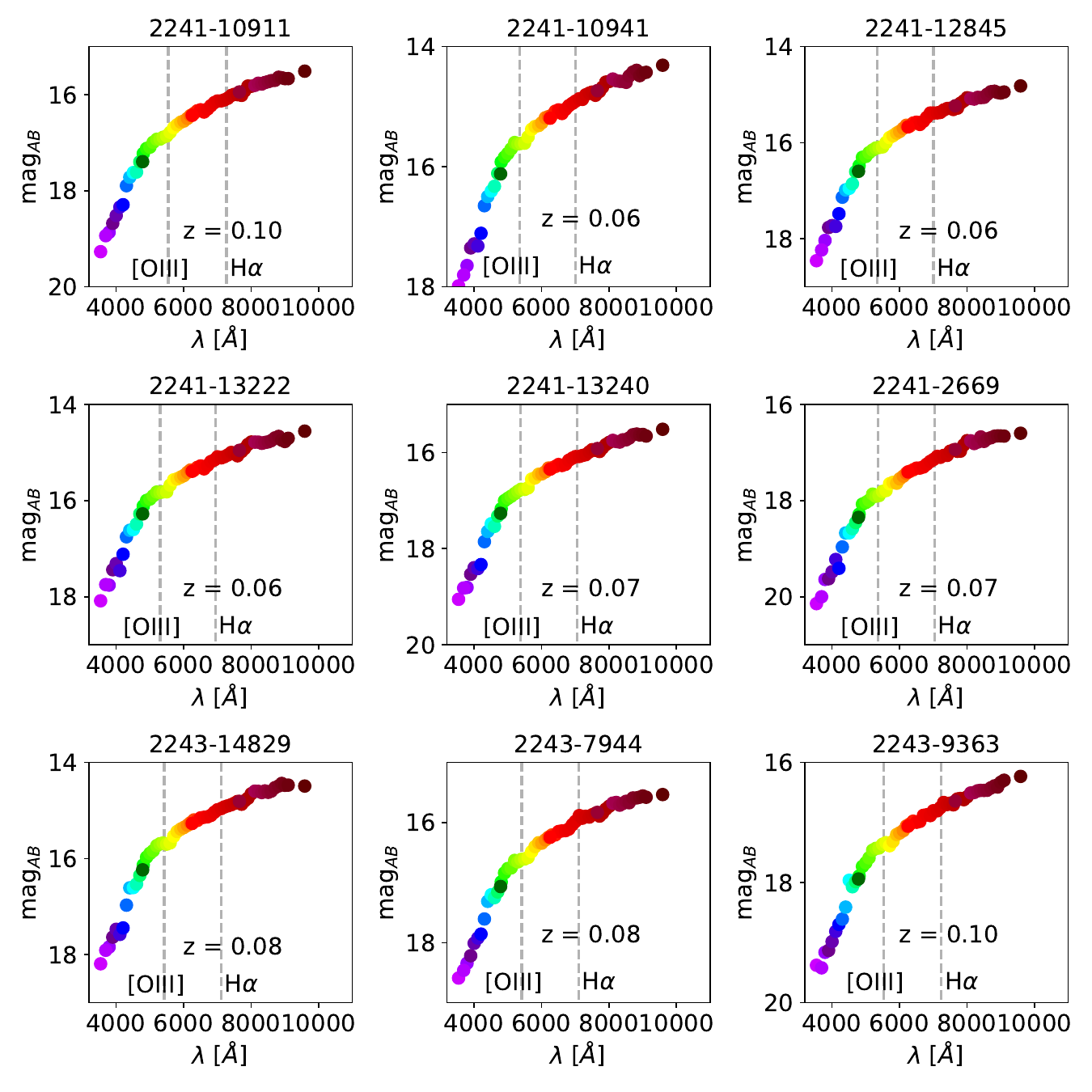}
    \caption[\js \ of  images of the red galaxies in the field from our sample (Part 1)]{\js \ of  images of the red galaxies in the field from our sample (Part 1). Each point is coloured using the internal colour palette generally used by the \jp \ collaboration in order to identify the filters.}
    \label{fig:Redfieldjs1}
\end{figure}

\begin{figure}
    \centering
    \includegraphics[width=\textwidth]{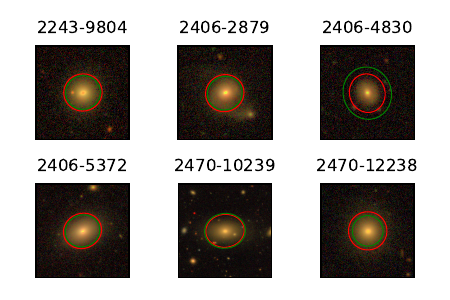}
    \caption[RGB images of the red galaxies in the field from our sample (Part 2)]{RGB images of the red galaxies in the field from our sample (Part 2). The red ellipses shows the \texttt{AUTO} ellipses calculated with Eqs.~\ref{eq:akron} and \ref{eq:bkron}. The green ellipses show the \texttt{PETRO} ellipses calculated with Eqs.~\ref{eq:apetro} and \ref{eq:bpetro}. The band used to make the RGB images are \ib{} (R) \rb{} (G) and \gb{} (B).}
    \label{fig:RedfieldRGB2}
\end{figure}

\begin{figure}
    \centering
    \includegraphics[width=\textwidth]{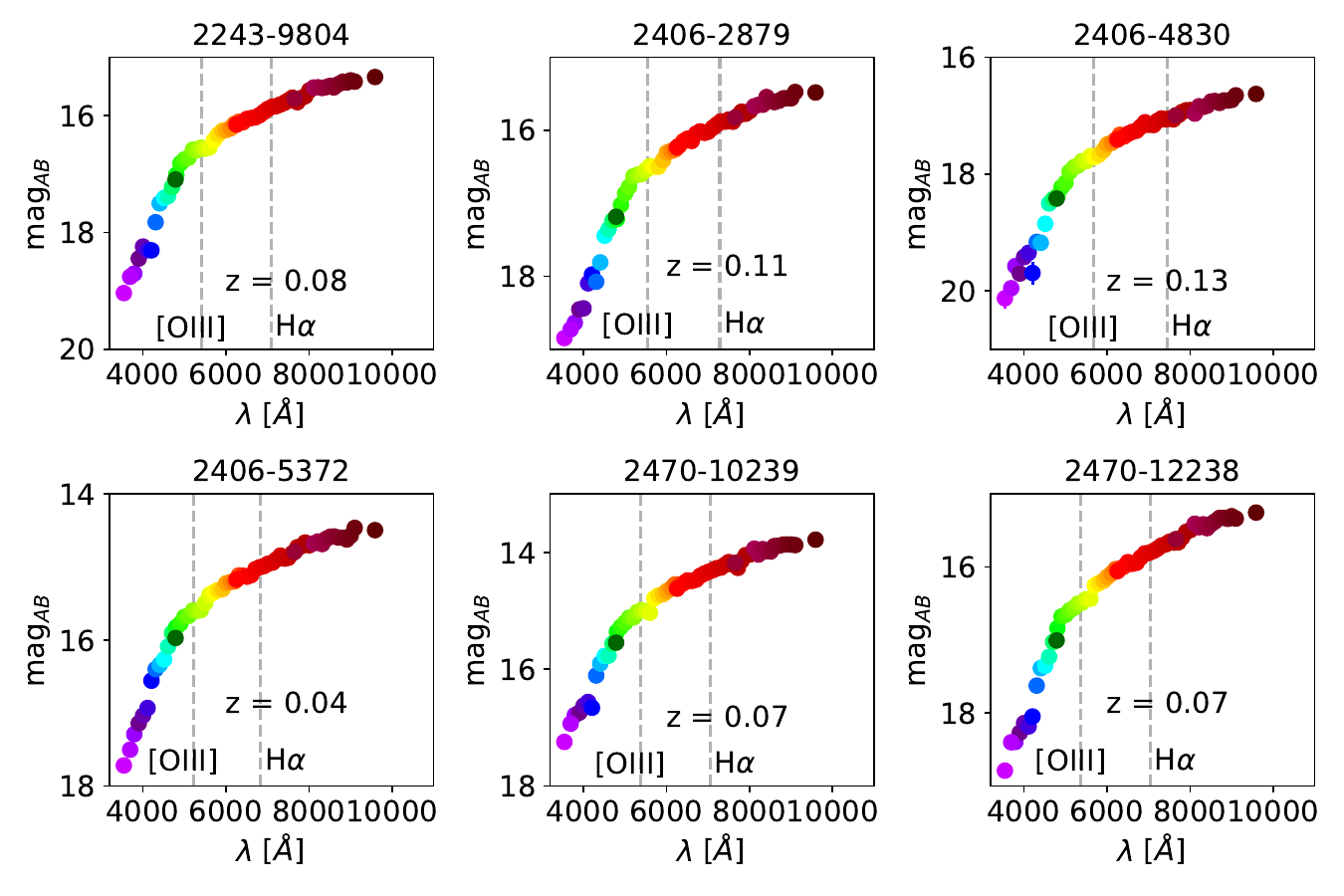}
    \caption[\js \ of  images of the red galaxies in the field from our sample (Part 2)]{\js \ of  images of the red galaxies in the field from our sample (Part 2). Each point is coloured using the internal colour palette generally used by the \jp \ collaboration in order to identify the filters.}
    \label{fig:Redfieldjs2}
\end{figure}

\begin{figure}
    \centering
    \includegraphics[width=\textwidth]{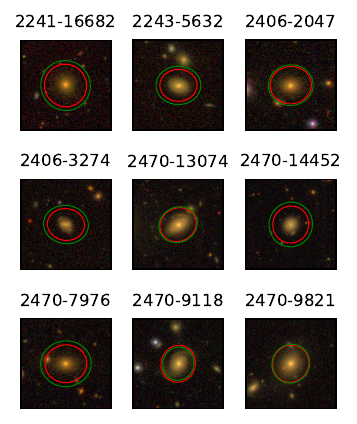}
    \caption[RGB images of the red galaxies in groups from our sample]{RGB images of the red galaxies in groups from our sample. The red ellipses shows the \texttt{AUTO} ellipses calculated with Eqs.~\ref{eq:akron} and \ref{eq:bkron}. The green ellipses show the \texttt{PETRO} ellipses calculated with Eqs.~\ref{eq:apetro} and \ref{eq:bpetro}. The band used to make the RGB images are \ib{} (R) \rb{} (G) and \gb{} (B).}
    \label{fig:RedgroupsRGB}
\end{figure}

\begin{figure}
    \centering
    \includegraphics[width=\textwidth]{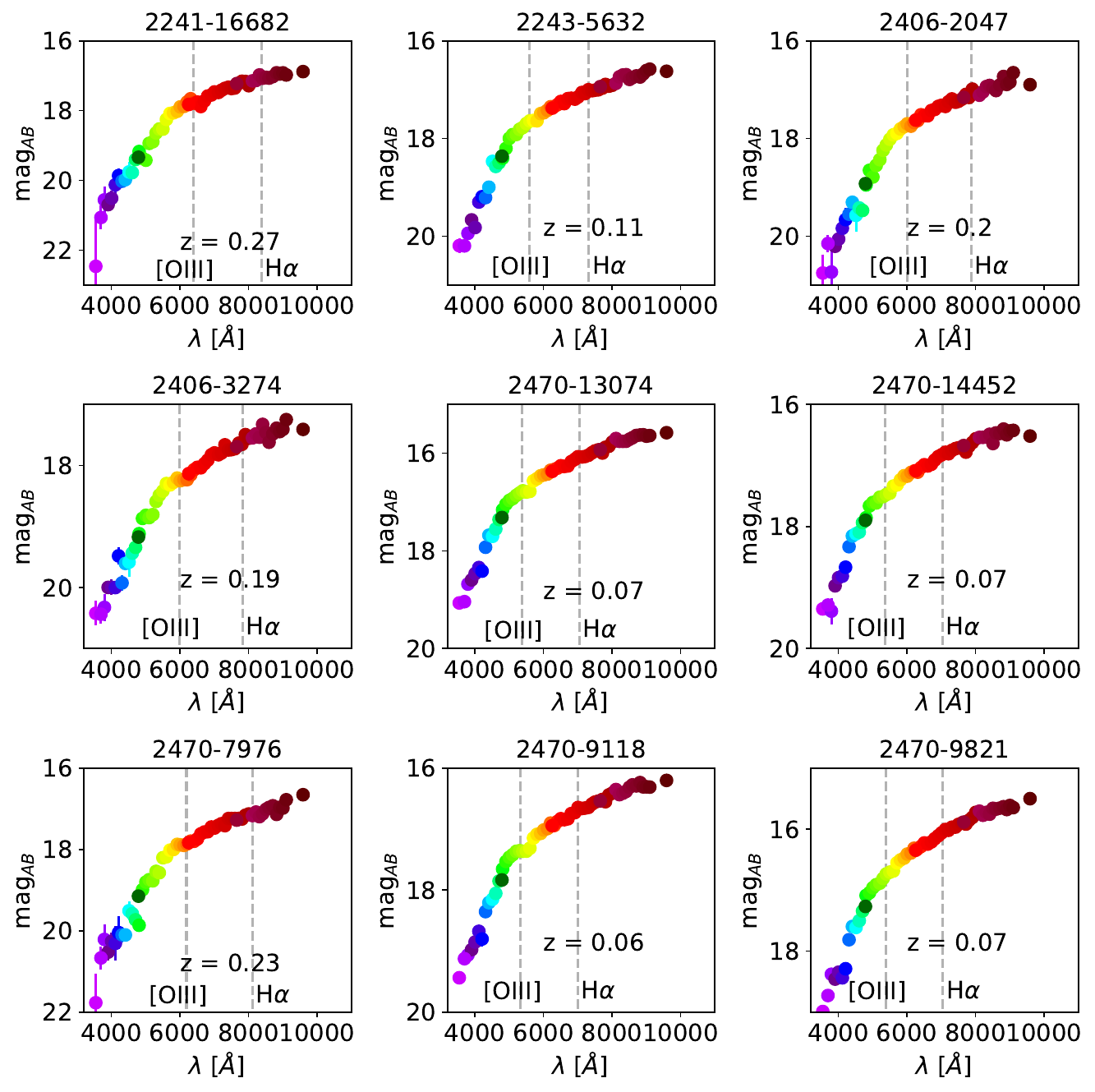}
    \caption[\js \ of  images of the red galaxies in groups from our sample]{\js \ of  images of the red galaxies in groups from our sample. Each point is coloured using the internal colour palette generally used by the \jp \ collaboration in order to identify the filters.}
    \label{fig:Redgroupsjs}
\end{figure}

\begin{figure}
    \centering
    \includegraphics[width=\textwidth]{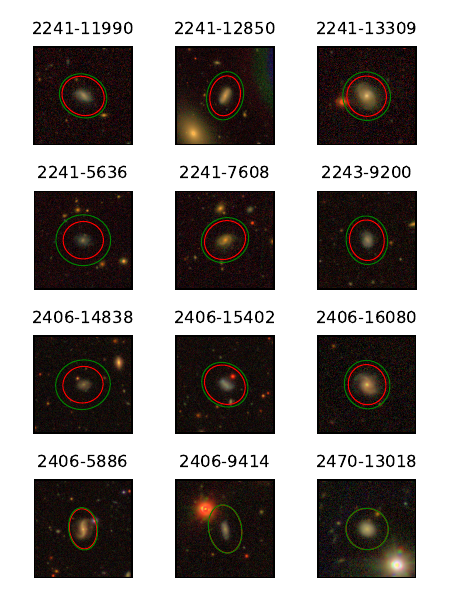}
    \caption[RGB images of the blue galaxies in the field from our sample (Part 1)]{RGB images of the blue galaxies in the field from our sample (Part 1). The red ellipses shows the \texttt{AUTO} ellipses calculated with Eqs.~\ref{eq:akron} and \ref{eq:bkron}. The green ellipses show the \texttt{PETRO} ellipses calculated with Eqs.~\ref{eq:apetro} and \ref{eq:bpetro}. The band used to make the RGB images are \ib{} (R) \rb{} (G) and \gb{} (B).}
    \label{fig:BluefieldRGB1}
\end{figure}

\begin{figure}
    \centering
    \includegraphics[width=\textwidth]{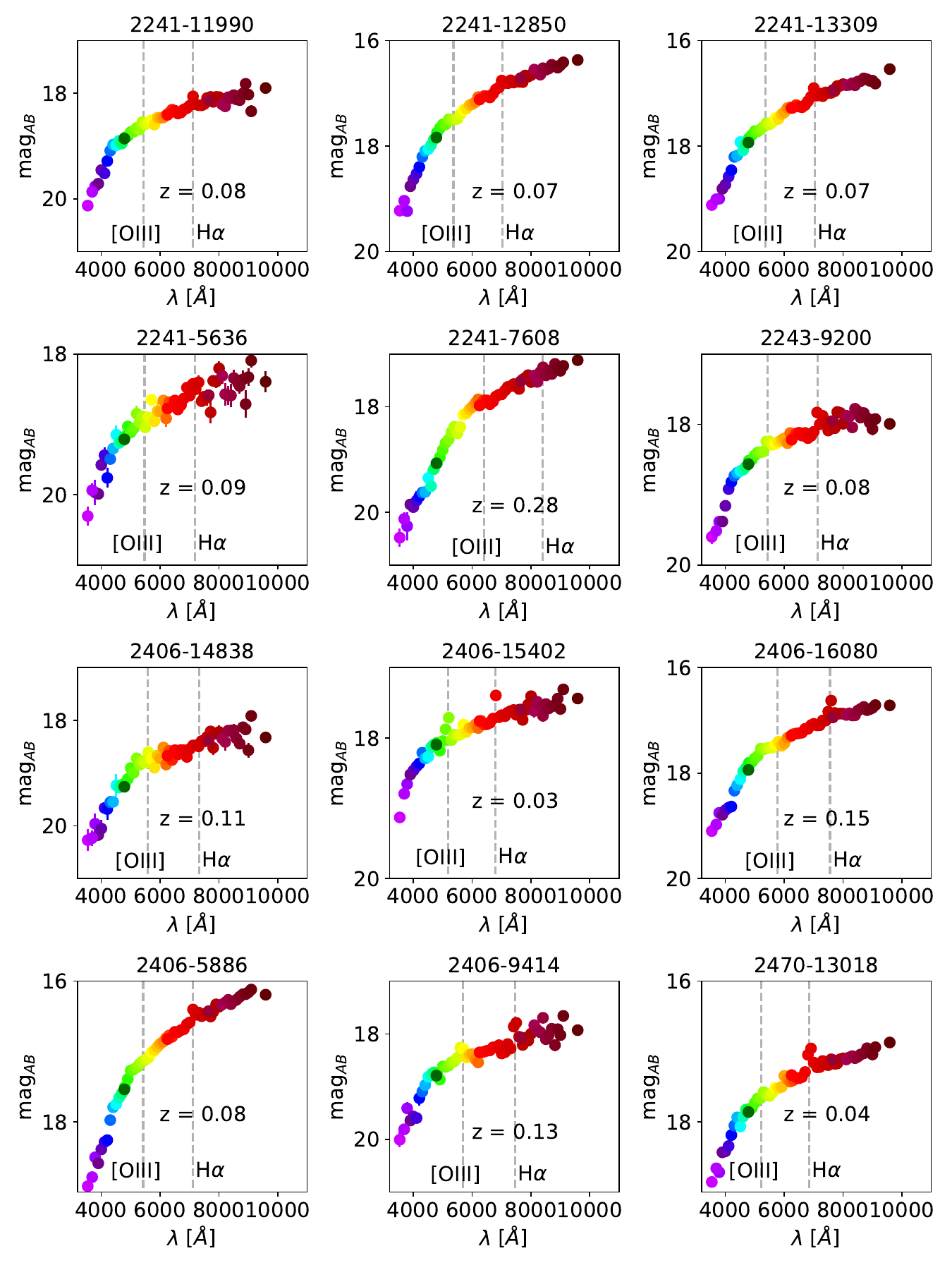}
    \caption[\js \ of  images of the blue galaxies in the field from our sample (Part 1)]{\js \ of  images of the blue galaxies in the field from our sample (Part 1). Each point is coloured using the internal colour palette generally used by the \jp \ collaboration in order to identify the filters.}
    \label{fig:Bluefieldjs1}
\end{figure}

\begin{figure}
    \centering
    \includegraphics[width=\textwidth]{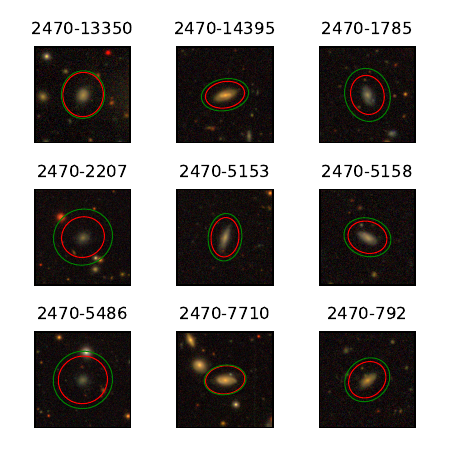}
    \caption[RGB images of the blue galaxies in the field from our sample (Part 2)]{RGB images of the blue galaxies in the field from our sample (Part 2). The red ellipses shows the \texttt{AUTO} ellipses calculated with Eqs.~\ref{eq:akron} and \ref{eq:bkron}. The green ellipses show the \texttt{PETRO} ellipses calculated with Eqs.~\ref{eq:apetro} and \ref{eq:bpetro}. The band used to make the RGB images are \ib{} (R) \rb{} (G) and \gb{} (B).}
    \label{fig:BluefieldRGB2}
\end{figure}

\begin{figure}
    \centering
    \includegraphics[width=\textwidth]{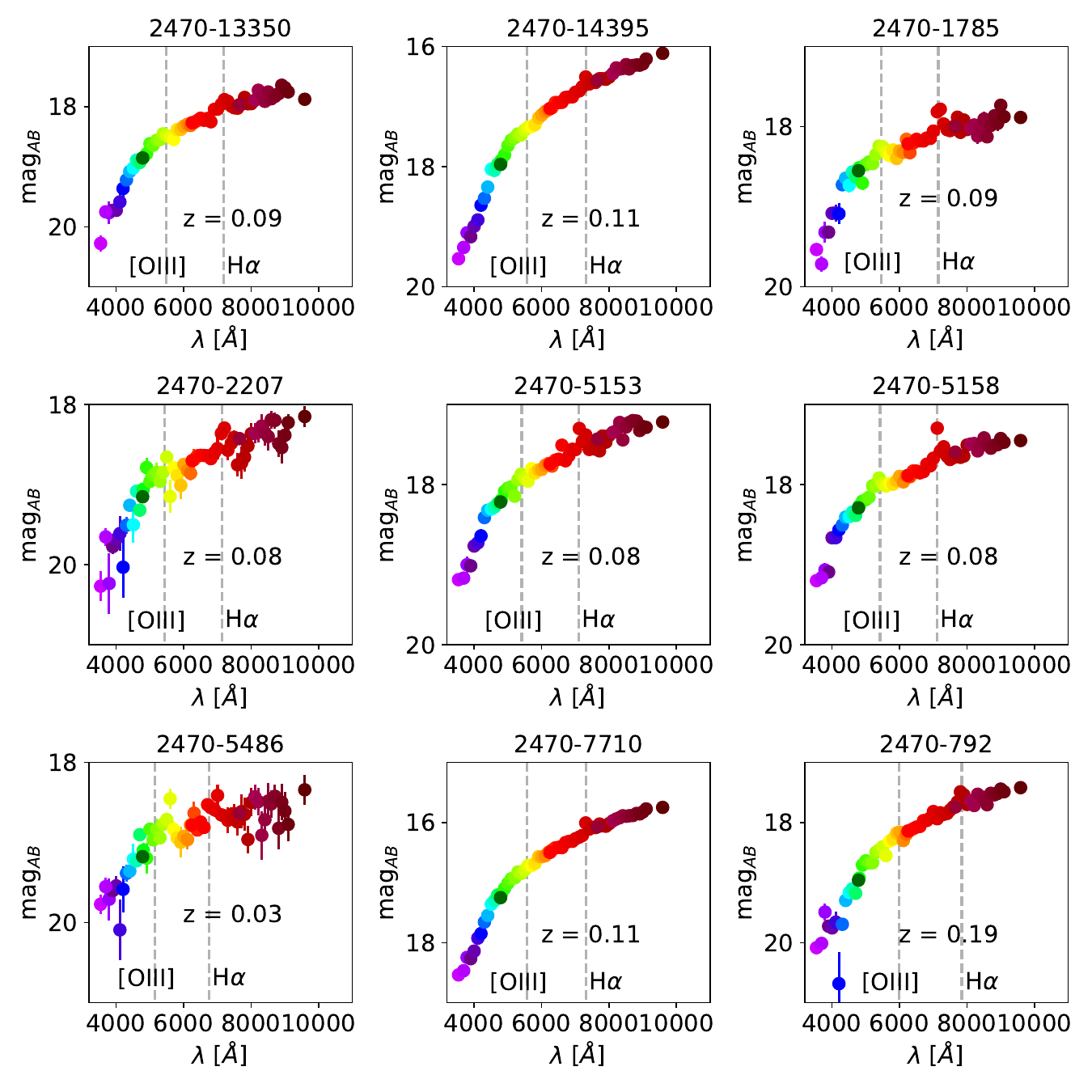}
    \caption[\js \ of  images of the blue galaxies in the field from our sample (Part 2)]{\js \ of  images of the blue galaxies in the field from our sample (Part 2). Each point is coloured using the internal colour palette generally used by the \jp \ collaboration in order to identify the filters.}
    \label{fig:Bluefieldjs2}
\end{figure}

\begin{figure}
    \centering
    \includegraphics[width=\textwidth]{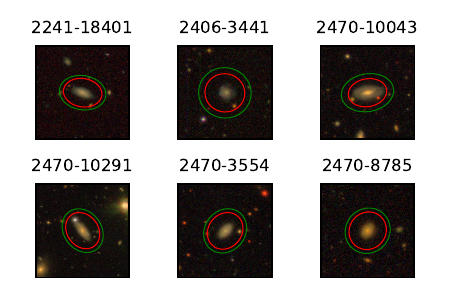}
    \caption[RGB images of the blue galaxies in groups from our sample]{RGB images of the blue galaxies in groups from our sample. The red ellipses shows the \texttt{AUTO} ellipses calculated with Eqs.~\ref{eq:akron} and \ref{eq:bkron}. The green ellipses show the \texttt{PETRO} ellipses calculated with Eqs.~\ref{eq:apetro} and \ref{eq:bpetro}. The band used to make the RGB images are \ib{} (R) \rb{} (G) and \gb{} (B).}
    \label{fig:BluegroupsRGB}
\end{figure}

\begin{figure}
    \centering
    \includegraphics[width=\textwidth]{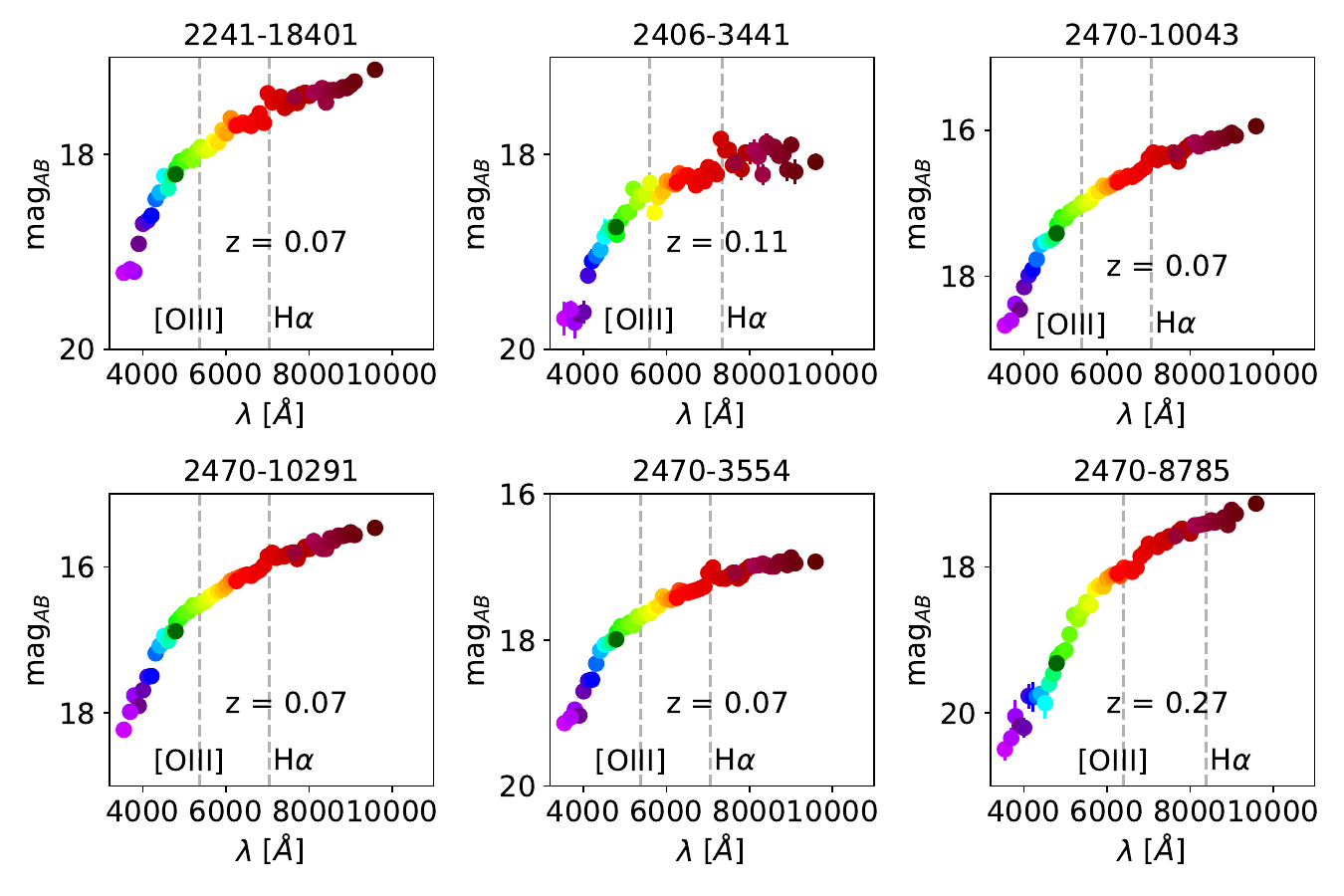}
    \caption[\js \ of  images of the blue galaxies in groups from our sample]{\js \ of  images of the blue galaxies in groups from our sample. Each point is coloured using the internal colour palette generally used by the \jp \ collaboration in order to identify the filters.}
    \label{fig:Bluegroupsjs}
\end{figure}

Before proceeding to the analysis we check the observational properties of the sample and the stamps of the galaxies. We perform a visual inspection of the RGB stamps of all the selected galaxies in order to ensure that all the objects are actual galaxies and not false detections from \sext, or stellar objects miss-classified through the $\texttt{CLASS\_STAR}$ parameter; that the ellipticity criteria did not leave any edge-on galaxy in the sample; and that there are no nearby stars that could bias the photometry. The stamps of the final sample of galaxies, along with their \js,  can be seen in Figs.~\ref{fig:RedfieldRGB1}, \ref{fig:Redfieldjs1}, \ref{fig:RedfieldRGB2}, \ref{fig:Redfieldjs2}, \ref{fig:RedgroupsRGB}, \ref{fig:Redgroupsjs}, \ref{fig:BluefieldRGB1}, \ref{fig:Bluefieldjs1}, \ref{fig:BluefieldRGB2}, \ref{fig:Bluefieldjs2}, \ref{fig:BluegroupsRGB}, \ref{fig:Bluegroupsjs}.

We can see that our final sample fulfils the desired conditions. We also note that, for some galaxies, the ellipticities and orientations of the $\texttt{KRON}$ and $\texttt{PETRO}$ ellipses provided by \sext \ seem to be improvable in order to create apertures that better fit the morphology of the galaxy. However, this effect is generally caused because the light distribution of the galaxy is dominated by a very bright nucleus, which is not so notable in the RGB images. Concerning the \js, red galaxies show a steeper 4000-break than blue galaxies, and some blue galaxies show an excess of flux in the filters where we would expect $\mathrm{H}\alpha$ or [OIII] to be detected but, from these \js \ alone we see no noticeable difference among red and blue galaxies in field and groups.

\begin{figure}
    \centering
    \includegraphics[width=\textwidth]{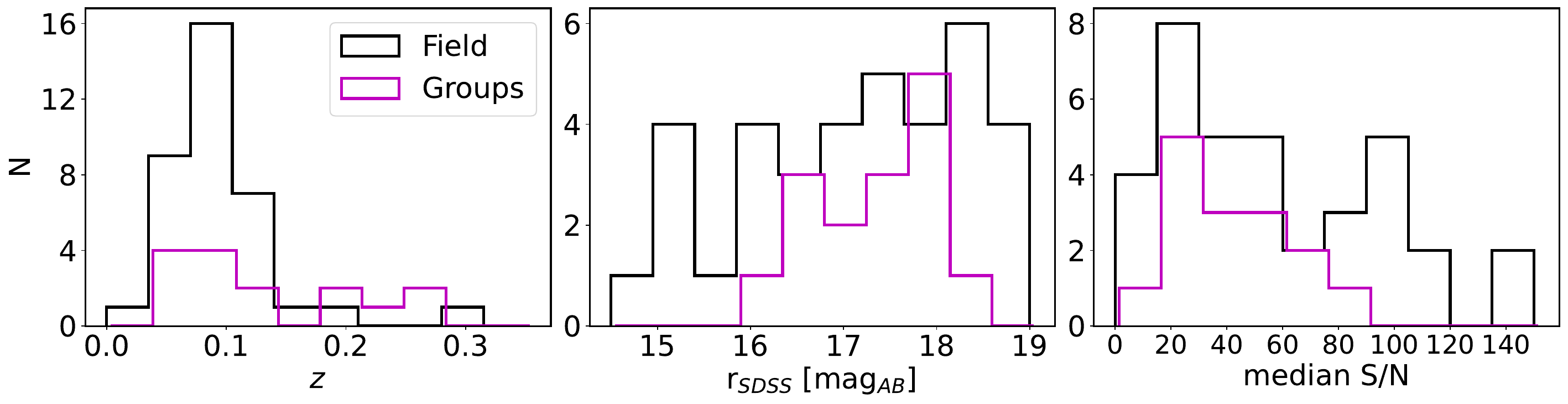}
    \caption[Integrated observational properties of the spatially resolved galaxies in \mjp]{Integrated observational properties of the spatially resolved galaxies in \mjp. First panel shows the histogram of the redshift of the galaxies. Middle panel shows the distribution of the magnitude in the \rb{} of the \magauto \ photometry. Last panel shows the distribution of the median S/N ratio of all the filters in the \magauto \ photometry. Black histograms represent galaxies in the field, and magenta histograms represent galaxies in groups.}
    \label{fig:intobsprop}
\end{figure}

The last check we perform is related to the redshift distribution of the galaxies (see Fig.~\ref{fig:intobsprop}). We find that both field and groups galaxies span a similar redshift range of $z \lessapprox 0.3$, with a peak around $z \sim 0.1$, but with a higher percentage of galaxies at redshifts greater that $0.1$ for galaxies in groups than in the field. The similarity in the redshift distribution and range allows us to make a fair comparison among the properties of galaxies in the two different environments. Additionally, we also find similar distributions in the brightness in the \rb{} and the median S/N ratio of the integrated magnitudes. However, field galaxies also show a significant amount of brighter galaxies with better S/N ratio, but this is due to the larger size of their sample. Once more data from \jp \ are available, this difference might be negligible.

\section{Methodology}
In this section we summarise the methods and codes used for the analysis of our data. These codes are either part of this thesis or have been used in previous works, already published, proving their effectiveness and correct working process.

\subsection{\PyDJ} \label{sec:codeSP}

The main tool used for this analysis is \PyDJ, our code designed to ease and automatise all the process required for the study of the spatially resolved galaxies in \mjp. This tool was described and discussed in detail in Chapter~\ref{chapter:code}, but we offer a brief summary along with the values of the parameters and options selected for each step.

The first step of the process is the download of all the data for each galaxy. This data consists of:

\begin{itemize}
    \item The \infot. This table contains all the information from the \texttt{MagABDualObj} table, the  obtained from running \sext \ on its Dual Mode using the \rb{} band as the reference filter. This table is used to download the images and create the apertures for the segmentation. 
    \item The scientific images. In order to download these images, we use the coordinates of the galaxy, obtained through the \infot, which also provides the semi-major axis of the galaxy, $\texttt{A\_WORLD}$, that we use as a reference for the size of the stamps. We choose to download square stamps with a size of $30 \times \texttt{A\_WORLD}$, since they provide an area large enough to work.
    \item The zero points of the galaxy, required to convert the ADUs into physical units. The calibration of these zero points is described in \cite{Lopez-sanJuan2019}. We also use the coordinates of the galaxy from \infot \ to download this item.
    \item The redhifts of the galaxy. We download both the photo-z table, which contains the $z\mathrm{PDF}$, its median value ($z_{ML}$), its mots likely value ($\mathrm{PHOTOZ}$), and the error, as well as other parameters. We also download the cross-match table with SDSS, selecting the spectroscopic redshift, its error and its flag, in order to use this redshift when available.
    \item The PSF models for all the filters, using the coordinates of the galaxy once again.
\end{itemize}

We then produce a mask in order to avoid including the nearby objects in the extraction of the photometry. We use the \mangle \ mask, provided by the collaboration, as well as the nearby object masks produced through the automatic process previously described, including the deblending option. The values of the parameters used for this step are $\texttt{NPIX} = 10$ and $rms = 5$, because after careful inspection and several tries, we find that it provides in general a good compromise between masking all the nearby objects enough and not masking small variations in the background that could be interpreted as another source, as well as not masking undesired regions of the galaxy.

We degrade the images in the filters in order to homogenise the PSF functions and avoid introducing colour terms that might bias our results  \citep[see e.g.][]{Michard2002, Cyprian2010, Molino2014, GonzalezPerez2011,Er2018,Liao2023}. This procedure is fairly standard and has been used in several other works, such as \cite{Bertin2002,Darnell2009,Desai2012,Desai2016} or \cite{SanRoman2018}.

\begin{figure}
    \centering
    \includegraphics[width=0.5\textwidth]{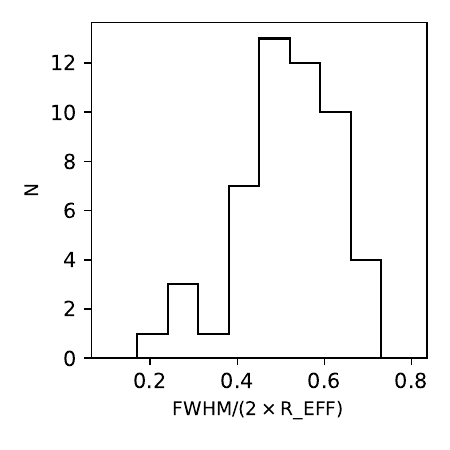}
    \caption[Comparison of the \gls{FWHM} of the worst \gls{PSF} for each galaxy and the double of its effective radius.]{Comparison of the \gls{FWHM} of the worst \gls{PSF} for each galaxy and the double of its effective radius.}
    \label{fig:PSFcheck}
\end{figure}

After all these steps, the images are ready to provide homogeneus apertures along all the filters, and we proceed to divide the galaxy into regions and extract their photometry.  In Fig.~\ref{fig:PSFcheck} we show an histogram of the size of the FWHM of the PSF in comparison to twice the effective radius \texttt{R\_EFF} provided by \sext. We decide to use this radius as a reference since it is publicly available and serves as a good starting point without the need of external or additional codes, and this value is already fixed in the data release. As we can see, our goal of selecting galaxies where we could extract at least two regions within $1$~\texttt{R\_EFF} is generally fulfilled, although there are some smaller galaxies. This may also be a consequence of our approximation of the PSF models as Gaussian distributions, which could provide larger \gls{FWHM} for some cases.

We use three different segmentations in this chapter:
\begin{itemize}
    \item Elliptical rings with major axis of the same size as the \gls{FWHM} of the worst \gls{PSF}. This is done to minimise the effect of the PSF in the photometry. With this option, the code calculates the FWHM of the worst PSF and uses it as the major axis of the ellipses. For each ellipse, it estimates the median S/N ratio of the photometry, and keeps adding generating new ellipses until the median S/N drops below a certain value. Here, we choose the threshold value to be 10. We will refer to this segmentation as the maximum resolution segmentation for short.
    \item Elliptical rings, using the same step of 0.7~\texttt{R\_EFF} for all the galaxies in the sample, in order to plot the radial profiles of the stellar population properties. We choose this step taking Fig.~\ref{fig:PSFcheck} into account, so we can produce radial bins where each galaxy is weighted only once. 
    \item An ``Inside-out'' segmentation, where we define an inner region of $a \leq 0.75$~\texttt{R\_EFF}, where $a$ is the semi-major axis of the ellipse, and an outer region of $0.75 < a \leq 2$~\texttt{R\_EFF}. This will be used in Sect.~\ref{sec:SRSFH} to study the star formation history of the galaxies.    
\end{itemize}

Additionally, we note that in order to generate the ellipses we only use the ellipticity and angular position paramters provided by \sext \ as a first approximation. We generate an object mask with semi-major and semi-minor axes of size $2.5 \times \texttt{A\_WORLD} \times \texttt{PETRO\_RADIUS}$ and $2.5 \times \texttt{B\_WORLD} \times \texttt{PETRO\_RADIUS}$, (following the same notation than the parameters provided in the catalogue). Then, within that aperture, we run the routine from \pycasso \ \citep{Pycasso2017} in order to obtain a better estimation of the parameters.

\section{Results}

In this section, we present the results of our analysis. We start by analysing the integrated properties of our sample, in order to have an idea of the general properties of the galaxies and take them into account in the spatially resolved analysis. We then check the results of the SED fitting, study the radial profiles of the stellar population properties and the gradients of the innermost regions of the galaxy, study the local SFMS, the radial profiles of the line emission and compare the SFH of the inner and outer regions.

\subsection{Integrated properties of the sample}
\begin{figure}
    \centering
    \includegraphics[width=\textwidth]{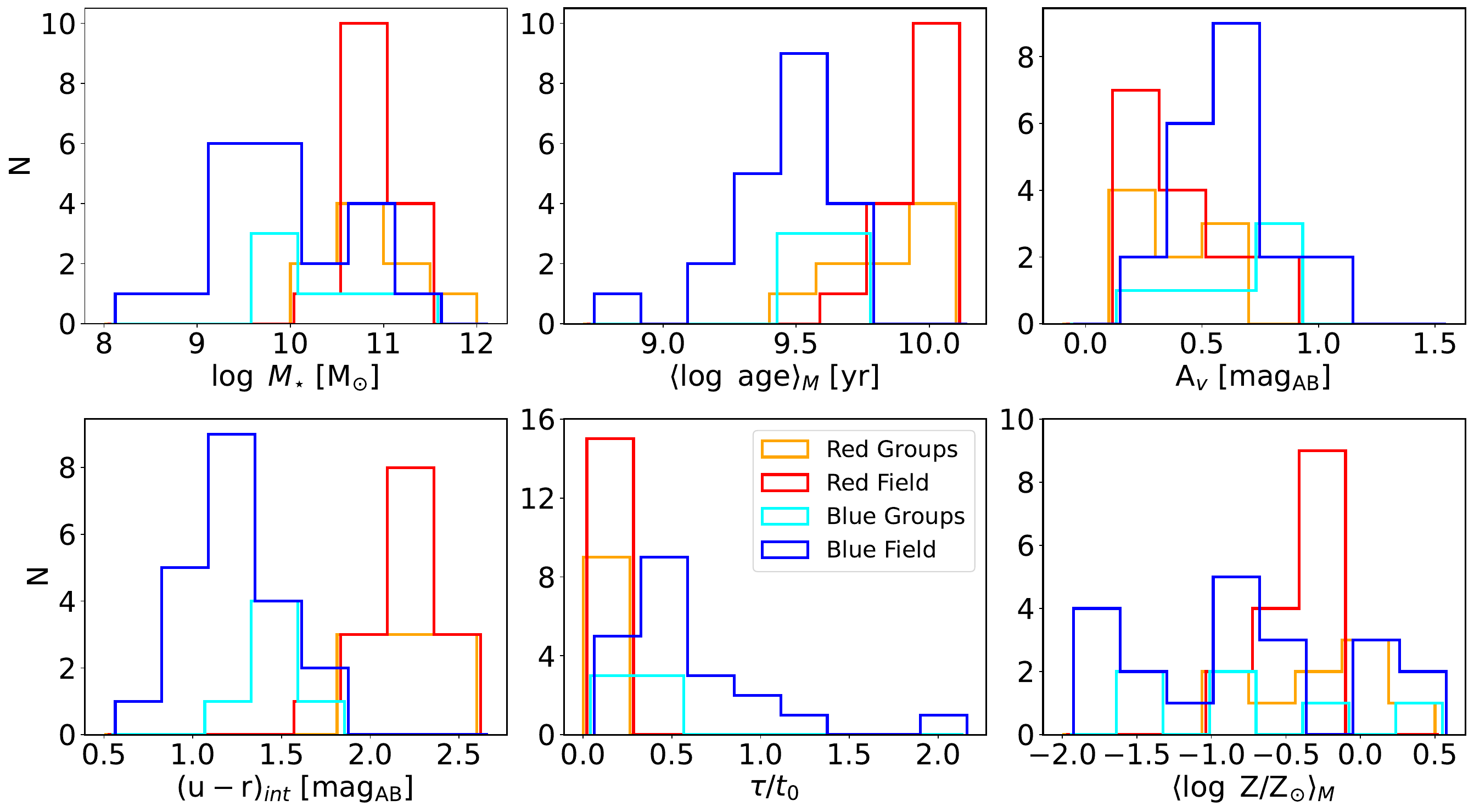}
    \caption[Comparison of the integrated stellar population properties properties of the spatially resolved galaxies by colour and environment]{Comparison of the integrated stellar population properties properties of the spatially resolved galaxies by colour and environment. From left to right, up to bottom: stellar masss, mass-weighted age, extinction $A_V$, $(u-r)_\mathrm{int}$, $\tau/\mathrm{t}_0$, and stellar metallicity. Red histograms represents red galaxies in the field. Orange histogram represents red galaxies in groups. Blue histograms represent blue galaxies in the field. Cyan histograms represent blue galaxies in groups.}
    \label{fig:intSPP}
\end{figure}
It is important to take into account the properties of the integrated galaxies before proceeding to study them in a spatially resolved way, to avoid biases and wrong conclusions in our analysis. Fig.~\ref{fig:intSPP} shows the distribution of the main stellar population properties of our sample. Our field galaxies (blue and red histograms) sample mainly span a range of masses between $10^9$~$M_\odot$ and $10^{11.5}$~$M_\odot$, while galaxies in groups (orange and cyan histograms) masses are limited to the $[10^{9.5},10^{11.5}]$~$M_\odot$ range. Similarly, galaxies in groups show a larger range in the mass-weighted ages, towards younger ages, in comparison to galaxies in groups (the age range is $\sim [10^9,10^{10.1}]$~yr for field galaxies and $\sim [10^{9.5},10^{10.1}]$~yr for group galaxies). The extinction range is similar ($\sim [0,1]$) for galaxies in groups and in the field, with no clear peak, while field galaxies show a tail towards bluer values of $(u-r)_{\mathrm{int}}$, down to $0.5$, while there are no group galaxies with colour below 1~mag. Galaxies in the field also show a tail towards larger values of $\tau / t_0$ (all groups galaxies values are contained in the $[0,0.5]$ range, while field galaxies show values of up to $1.5$), which generally mean star formation episodes more extended in time. The metallicity range is similar, spanning values all over the range of our models. Group galaxies do not show a clear peak, which could be due to the low number sample, while field galaxies show a peak between the values $-1$ and 0. 

If we also take into account the colour of the galaxies, we find that the distribution of the properties of the red galaxies in groups and in the field is very similar, but the distribution of the properties of blue galaxies shows some differences depending on their environment. The stellar masses of red galaxies (red and orange histograms) is in the range between $[10^{10},10^{12}]$~$M_\odot$, with a peak at $\sim 10^{10.8}$~$M_\odot$. The masses of blue galaxies in the field spans a wide range of masses, from  $10^8$~$M_\odot$ up to $10^{11.5}$~$M_\odot$, with a main peak at $\sim 10^{9.5}$~$M_\odot$ and a secondary one at $\sim 10^{10.5}$~$M_\odot$. However, the masses of our sample of blue galaxies in groups are mainly found in a smaller range, from $\sim 10^{9.5}$~$M_\odot$ up to $\sim 10^{11.5}$~$M_\odot$. This means that our sample of blue galaxies in the field contains a larger number of lower mass galaxies, which must be taken into account when comparing the properties of the regions of blue galaxies in the field and in groups. 

The mass-weighted stellar ages of red galaxies are mainly contained in the range $\sim [10^{9.5},10^{10.2}]$~yr, peaking at $\sim 10^{10}$~yr, with no significant difference due to the environment. On the other hand, similarly to the stellar mass, the range of ages found for blue galaxies in the field ($\sim [10^{8.75},10^{9.75}]$~yr) is wider with a younger tail than the range found for blue galaxies in groups ($\sim [10^{9.5},10^{9.75}]$~yr). This distributions of ages and masses reflect the age-mass relation \citep[see e.g.][]{Gallazzi2005,Peng2015} which will also be shown later in the text. 

The distribution of the extinction $A_V$ shows  that all types of galaxies have extinctions in the range $[0,1]$. Red galaxies, both in the field and in groups, peak at lower extinctions, $A_V \approx 0$, while blue galaxies in the field peak at  $A_V \approx 0.5$. However, the distribution of the extinction of the blue galaxies in groups is rather flat, most likely due to the reduced size of the sample. This different peaks can be expected, since blue, star--forming galaxies can show much higher extinctions \citep[see e.g.][]{charlot2000}.

The distribution of the $(u-r)_\mathrm{int}$ colour is mostly bimodal and red and blue galaxies are clearly separated. This can be expected by selection as well as because of the well-known galaxy bi-modality \citep[see e.g.][]{Strateva2001,Bell2004,Baldry2004,Williams2009,Moresco2013,Povic2013,Fritz2014,Luis2019,Rosa2021}. However, it is remarkable that red galaxies show similar colours, in the range $(u-r)_\mathrm{int} \in [2.0,2.5]$~mag, but blue galaxies in groups colours range from $(u-r)_\mathrm{int} \approx 1.0$~mag  up to $(u-r)_\mathrm{int} \approx 2.0$~mag, which clearly lean towards redder colours than those of blue galaxies in the field, whose colours range from $(u-r)_\mathrm{int} \approx 0.5$~mag  up to $(u-r)_\mathrm{int} \approx 2.0$~mag and peaking at   $(u-r)_\mathrm{int} \approx 1$~mag. 

The distribution of $\tau / \mathrm{t}_0 $ clearly separates red and blue galaxies. Values found for red galaxies in the field and in groups are all contained in the lowest bin ($\tau / \mathrm{t}_0 \approx [0,0.25]$), blue galaxies show some larger values, particularly for galaxies in the field  ($\tau / \mathrm{t}_0 \approx [0,0.5]$ for blue galaxies in groups, $\tau / \mathrm{t}_0 \approx [0,1.5]$). This indicates that, according to our $\tau$ models, red galaxies built their masses at earlier epochs and faster than blue galaxies, specially blue galaxies in the field.

The distribution of the metallicity shows that the metallicities of red galaxies are within a narrower range ($\left < \log \mathrm{Z/Z_\odot} \right > \approx [-0.1,0]$ for red galaxies in the field, $\left < \log \mathrm{Z/Z_\odot} \right > \approx [-0.1,0.5]$ for red galaxies in groups) than the values of the metallicity in blue galaxies ($\left < \log \mathrm{Z/Z_\odot} \right > \approx [-2.0,0.5]$ for blue galaxies in the field, $\left < \log \mathrm{Z/Z_\odot} \right > \approx [-1.5,0.5]$ for blue galaxies in groups).

These distributions could partly be a consequence of the different size of the samples, but these biases of more massive, older and redder galaxies in groups are also found in the literature \citep[see e.g.][]{Rosa2022,Julio2022}. We also note that our selection was based on size. Therefore, given the mass-size relation \citep[see e.g.][]{Shen2003,Hyde2009,Poggianti2013,Lange2015,Yoon2023} it is also reasonable that our sample is biased toward more massive galaxies, which are usually older and redder \citep{Kauffmann2003}. In addition, elliptical galaxies are usually redder, older and more massive, while spirals are usually bluer, less massive and younger \citep[see e.g.][]{Hogg2004,Kauffmann2004,Bassett2013}. Since we are removing edge-on galaxies from our sample, which can only be spiral galaxies, we are more likely to be dominated by elliptical galaxies.

\begin{figure}
    \centering
    \includegraphics[width=\textwidth]{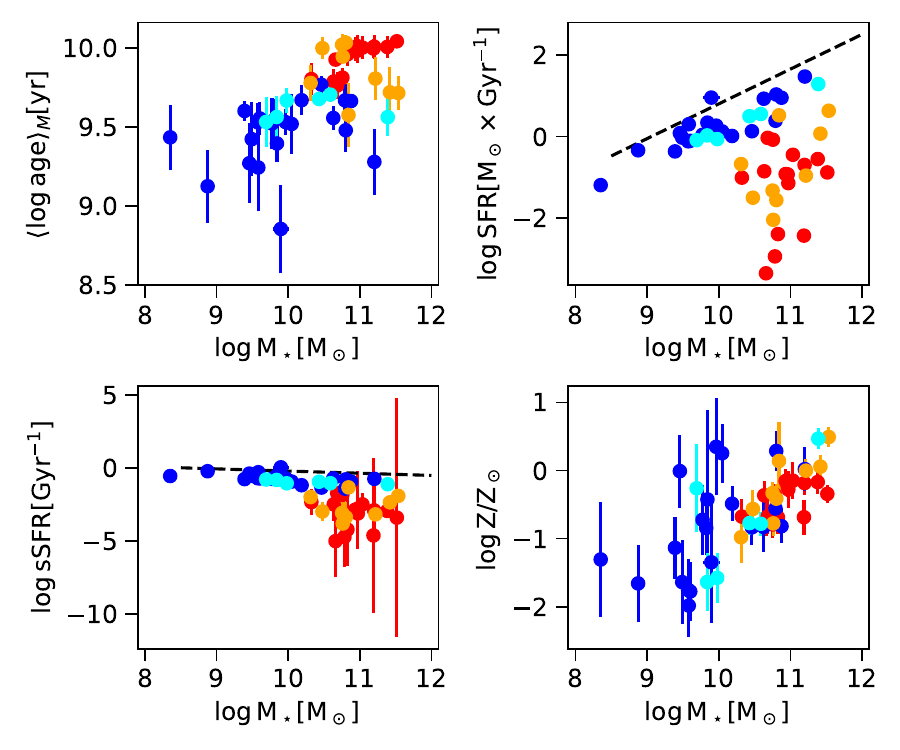}
    \caption[Relations of the integrated properties of the galaxy. From left to right, top to bottom: age--mass relation, SFMS (SFR-mass), SFMS (sSFR-mass) and mass-metallicity relation]{Relations of the integrated properties of the galaxy. From left to right, top to bottom: age--mass relation, SFMS (SFR-mass), SFMS (sSFR-mass) and mass-metallicity relation. Red points represent red galaxies in the field. Orange points represent red galaxies in groups. Blue points represent blue galaxies in the field. Cyan points represent blue galaxies in groups. Black dashed lines represent the fit obtained by \cite{Rosa2022} for the SFMS in the field.}
    \label{fig:intrelations}
\end{figure}

We also explore some other known relations among stellar population properties (see Fig.~\ref{fig:intrelations}). The first panel shows the mass-age relation of galaxies. We find that blue and red galaxies are well distinguished in this diagram, and seem to show different trends. We find no significant differences among red/blue galaxies in groups and field environments. 

The second and third panels show the \gls{SFMS} of galaxies in our sample, in their \gls{SFR} and \gls{sSFR} versions. We use the fit obtained by \cite{Rosa2022} for the whole \mjp \ sample as a reference. We find that blue galaxies lie near this fit, regardless of their environment, while red galaxies are found well below this fit. This behaviour is the same found in the literature \citep[see e.g.][]{Peng2012,Speagle2014}. We note that, even though blue galaxies are close to the fit, they are all below the fit. This means that the star formation rate of most of our blue galaxies have \gls{SFR} that are in the low range of the \mjp \ sample. This could be expected, since bulges are generally quiescent, and disc galaxies are likely to be star--forming \citep{Dimauro2022}, and we have already discussed that, due to our selection criteria, we are likely to loose more disk galaxies and that our sample could be dominated by galaxies with prominent bulges.

The last panel shows the mass-metallicity relation, as described by \cite{Gallazzi2005}. Similarly to the previous cases, we are not able to distinguish the environment of the galaxies, and the mass and the colour of the galaxy seem to play a mayor role in our sample. We do however find a larger range of metallicites for low mass galaxies, while high mass galaxies are concentrated close to a single value ($\sim-0.5$~dex in our sample), as seen in the literature \citep{Gallazzi2005}.

\subsection{SED-fitting check} \label{sec:SEDcheck}

\begin{figure}
    \centering
    \includegraphics[width=0.95\textwidth]{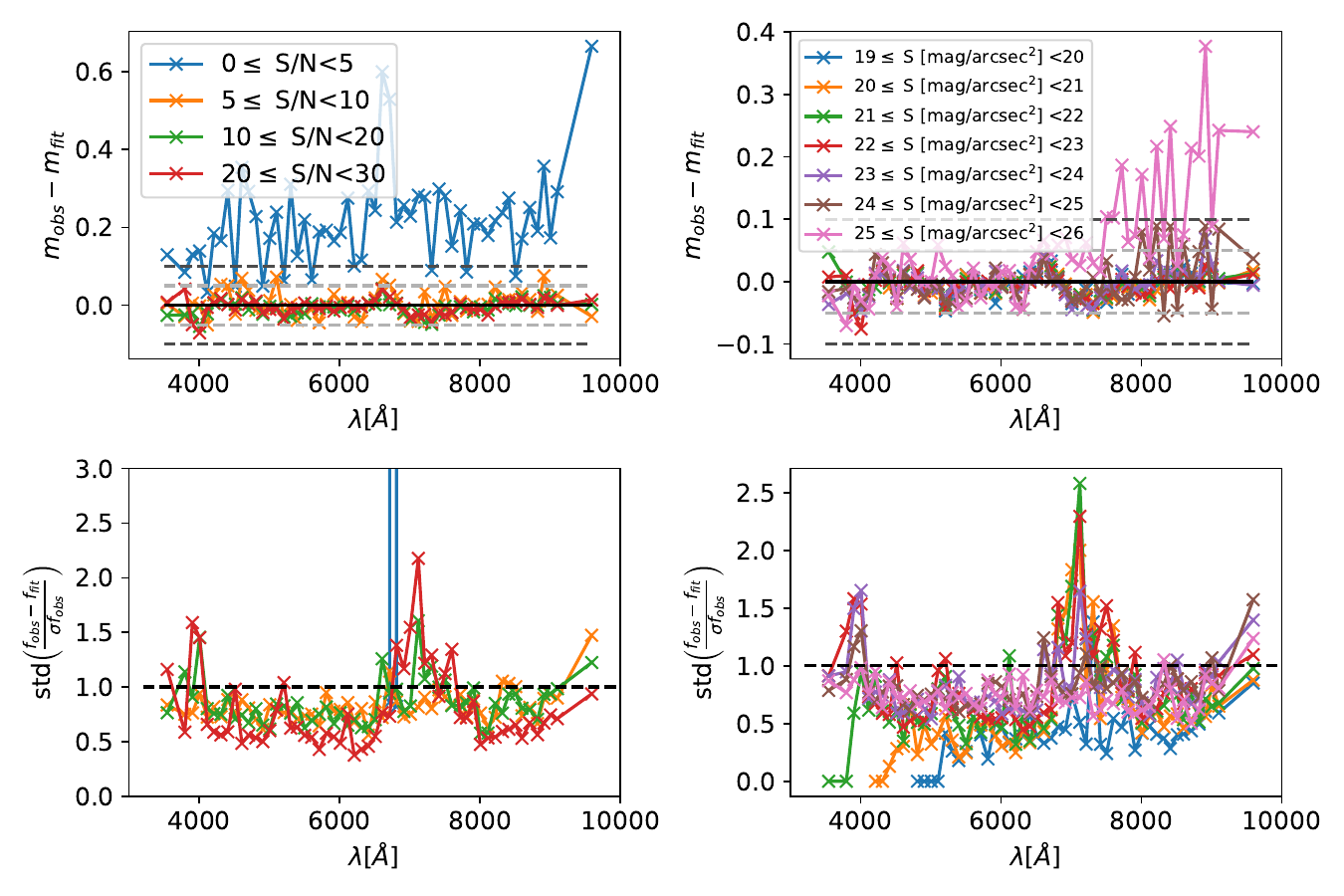}
    \caption[Residuals for the $0.7$~\texttt{R\_EFF} RING segmentation]{Residuals for the $0.7$~\texttt{R\_EFF} RING segmentation. Top panels show the values of the residuals. Bottom panels show the standard deviation of the ratio between the residuals and the observed errors. Left panels show these parameters divided by S/N. Right panels show them divided by surface brightness. Different colours represent different different bins of S/N and surface brightness, respectively. Solid black lines in top panels represent the value 0 of the residuals. Grey dashed lines in top panels represent the limits $[-0.05,0.05]$ of the residuals. Black dashed lines in the top panels represent the limits $[-0.1,0.1]$ of the residuals. Black dashed line of the bottom panels shows the desired value of 1 for ratio between the residuals and the observed errors.}
    \label{fig:residual_rings}
\end{figure}

\begin{figure}
    \centering
    \includegraphics[width=0.95\textwidth]{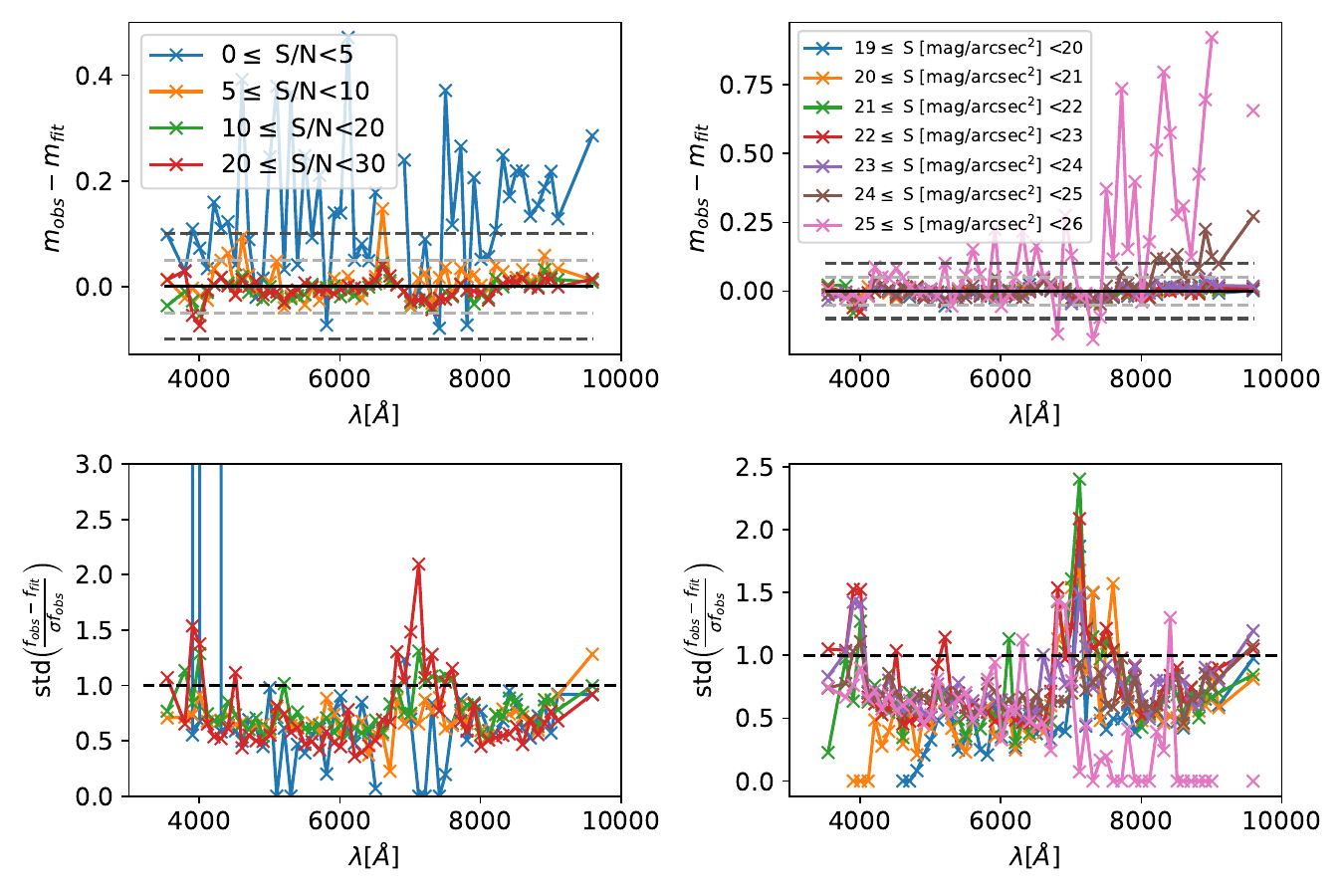}
    \caption[Residuals for the maximum resolution segmentation]{Residuals for the maximum resolution segmentation. Top panels show the values of the residuals. Bottom panels show the standard deviation of the ratio between the residuals and the observed errors. Left panels show these parameters divided by S/N. Right panels show them divided by surface brightness. Different colours represent different different bins of S/N and surface brightness, respectively. Solid black lines in top panels represent the value 0 of the residuals. Grey dashed lines in top panels represent the limits $[-0.05,0.05]$ of the residuals. Black dashed lines in the top panels represent the limits $[-0.1,0.1]$ of the residuals. Black dashed line of the bottom panels shows the desired value of 1 for ratio between the residuals and the observed errors.}
    \label{fig:residual_MAX_RES}
\end{figure}

Before continuing with the analysis of the properties, we perform a check to test the quality of the SED fitting in our data. Our SED-fitting codes have been tested and their efficiency has been proven when applied to \jp \ like data has been proven in previous works \citep{Rosa2021,Rosa2022}. However, their performance has not been tested yet in the regions of spatially resolved galaxies. These tests are  valuable not only for this reason, but also to evaluate the quality of our results and how trustworthy they are.

We start by calculating the residuals of our fit (see Figs.~\ref{fig:residual_rings} and \ref{fig:residual_MAX_RES}). We calculate the residual for each region obtained using the maximum resolution and the $0.7$~\texttt{R\_EFF} rings segmentation, and we calculate the median for each filter, using bins of S/N and surface brightness. Low values indicate that the code was able to retrieve a combination of stellar populations that resembles the extracted spectra. This is a good test for our photometry, since retrieving spectra that are physically plausible indicates that the whole process of the extraction at least provides explainable spectra, given our current knowledge of stellar models. We also calculate the standard deviation of the residuals divided by the calculated error, which is related to the errors. Ideally, we should get values close to one. Lower values would indicate that the errors are overestimated, and larger values that they are underestimated.

We find that, for both photometries (uniform annulus and annulus using the maximum resolution allowed by the size of the PSF), filters with a S/N larger have residuals well within the $[-0.05,0.05]$ range (grey dashed lines in the figures). The same can be said for filters with a surface brightness brighter than 25. This range of residuals in magnitudes means that the relative difference of the measured and fitted fluxes is below 5~\%. Most importantly, we do not find a significant bias with wavelength for these residual. However, we note that, as the S/N ratio or the surface brightness decreases, the residuals of many filters start becoming biased towards larger, positive values of $m_{obs} - m_{fit}$. This means that, for these cases, the code tends to underestimate the flux (or the fitting tends to overestimate the flux of the solution). A possible explanation could be the PSF homogenisation. However, this homogenisation affects more the inner regions of the galaxy, that loose flux in favour of the closer areas, and outer areas are mostly unaffected. Another possible explanation is that the filter close to 4000~\AA \ is, on the contrary, biased to negative values of $m_{obs} - m_{fit}$. This different behaviour on blue and red could be due to the efficiency of the CCDs at different wavelengths, and the SED-fitting code trying to fit all filters with an intermediate solution, or finding that these blue filters are more important for the fit. Note that these filters contain the so-called 4000-break, that is known to be highly correlated with the stellar population properties of galaxies \citep[see e.g.][]{Rosa2005}.

However, when the S/N drops below five, for both photometries, the residuals become noticeably larger, reaching values larger than $0.3$ (a relative difference in fluxes larger than 30~\%) and are biased toward positive values for all filters. This is particularly noticeable in the $0.7$~\texttt{R\_EFF} RING segmentation. In this segmentation, a filter close to the $6500$~\AA \ shows a notably larger residual than the others. This filter is close to the H$\alpha$ emission line rest wavelength. We note that our code does not fit emission lines, which can lead to larger residuals in the emission filters. However, this behaviour is not shown in the maximum resolution segmentation, and the mode of the redshift o our sample is $z\sim0.1$, which would shift the H$\alpha$ emission towards $\sim 7219$~\AA. However, this behaviour is not seen in the maximum resolution segmentation, where filters with wavelengths shorter than $4000$~\AA \ show a bias towards negative residuals, and filters above show positive values. This would support the aforementioned idea that the SED-fitting code might be finding an intermediate solution for both wavelength ranges. We also note that residuals seem to behave better in the maximum resolution segmentation, but we impose a S/N cut in the segmentation itself, which would remove from the analysis the regions with worse S/N and dimmer surface brightness.

When separated by surface brightness bins, we find a that residuals are contained in the $[-0.05,0.05]$ range when the surface brightness is brighter than $\sim 25$~$\mathrm{mag/arcsec^2}$. For dimmer values, the residuals start to increase. Interestingly enough, the red filters are the worst ones, showing a bias toward positive values. This is likely due to fringing effects.

Concerning the standard deviation of  $\mathrm{std} \left ( \frac{f_{obs} - f_{fit}}{\sigma f_{obs}}\right )$, values close to $1$ are desired, since the would show that the errors are correctly estimated. Values for $\mathrm{S/N}<5$ are not shown in the graph for the $0.7$~\texttt{R\_EFF} RING segmentation because they are too large and do not allow for proper interpretation of the graph. In all cases, the interpretation is similar: values fluctuate close to the unity value, some above it, but most below it. This would mean that our error is overestimated. We note that we have included ZP error, which is set to a conservative value \citep[the error provided is $0.04$ for all the filters for all the galaxies, but recent revisions of the calibration method provide much more accurate ZP, see ][]{Lopez-sanJuan2019,LopezSanJuan2023}. However, we decide to keep this value in the equations and provide a more conservative uncertainty. For this parameter, there is no significant dependence with the S/N or the surface brightness. The filter at 7000~\AA \ seems to be the worst one in all cases, with the error being underestimated. These could be due to the H$\alpha$ emission.

\begin{figure}
    \centering
    \includegraphics[width=\textwidth]{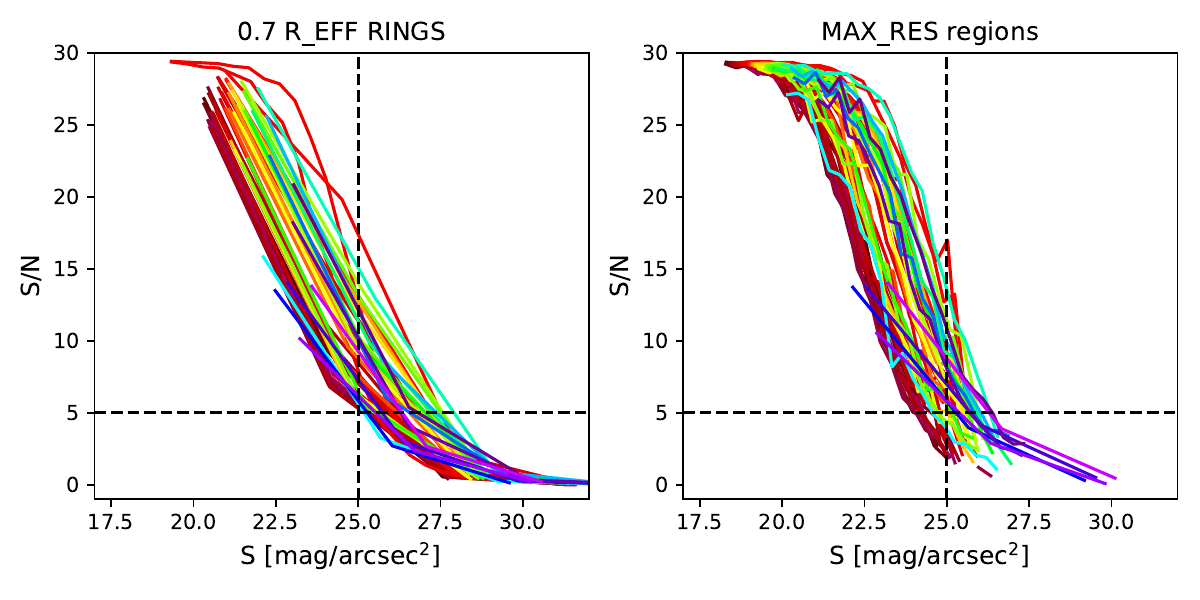}
    \caption{S/N vs surface brightness. Colours represent each filter of \mjp.}
    \label{fig:SN_Surf_brigth_filter}
\end{figure}

\begin{figure}
    \centering
    \includegraphics[width=\textwidth]{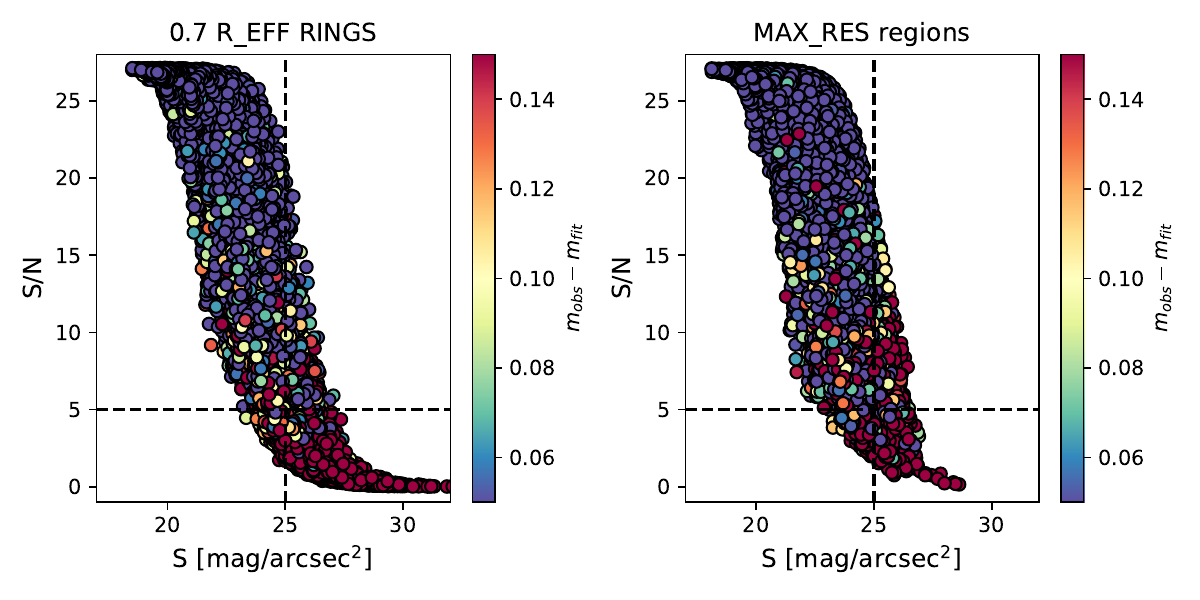}
    \caption{S/N vs surface brightness, color coded with the residuals. Each point represents one filter for one region.}
    \label{fig:SN_Surf_brigth}
\end{figure}

In order to further investigate the relation among the S/N ratio, the surface brightness and the residuals, we plot the first one as a function of the second, coloured by filter (Fig.~\ref{fig:SN_Surf_brigth_filter}) and the residuals (Fig.~\ref{fig:SN_Surf_brigth}). We find that the S/N ratio decays similarly for all the regions, filters and both segmentations as the surface brightness becomes dimmer. We note that this decays seems to be faster  for redder filters. This is in accordance to the previous observation where red filters were the ones to show worse values of the residuals as the surface brightness decreased. Concerning the residuals, the distribution is clear: bright regions with good S/N show the lowest values for the residuals. As soon as the S/N starts to decrease, regions with larger residuals start to appear in both segmentations. For the $0.7$~\texttt{R\_EFF} RINGS segmentation, the cut in S/N seems to be a better filter than the cut in surface brightness, since it removes most of the regions with higher residuals. For the maximum resolution segmentation, however, some regions with higher residuals show a S/N larger than 5, and the threshold $S \gtrsim  25$~$\mathrm{mag/arcsec^2}$ shows some high residual regions. We note here that it is important to consider that these regions are the smallest possible, given the size of the PSF.

\begin{figure}
    \centering
    \includegraphics[width=\textwidth]{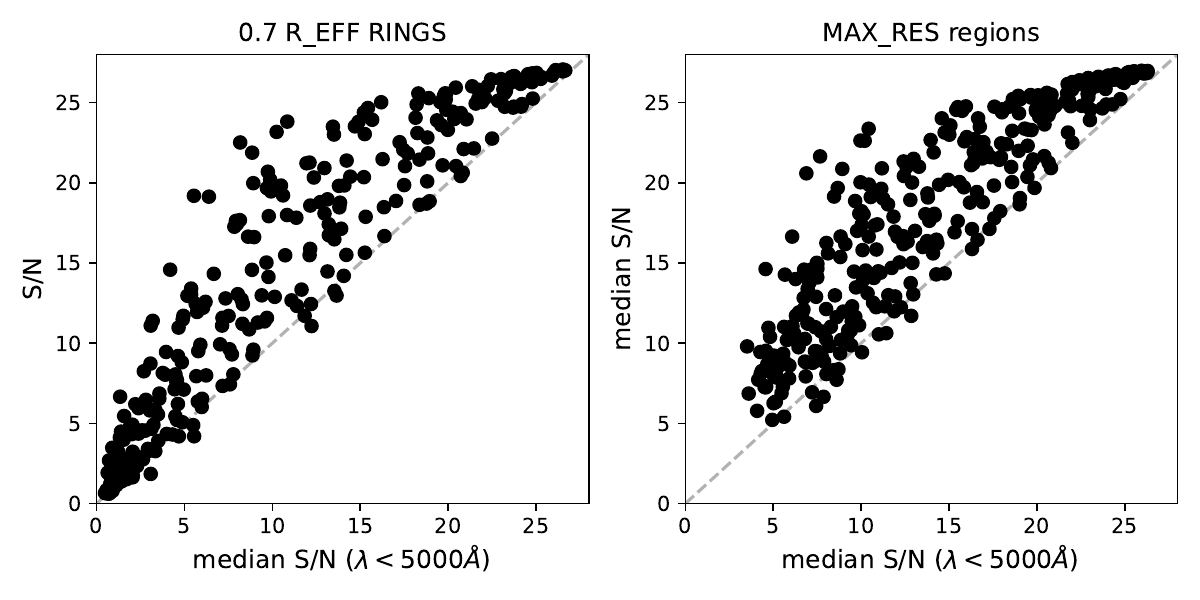}
    \caption[Median S/N  vs median S/N ($\lambda<5000$~\AA)]{Median S/N  vs median S/N ($\lambda<5000$~\AA). Left panel shows the comparison for the $0.7$~\texttt{R\_EFF} RINGS segmentation, and right panel shows the comparison for the maximum resolution segmentation. Grey dashed lines show the identity relation.}
    \label{fig:SN_vs_SN}
\end{figure}

We must also take into account that the median S/N of the \js \ of the region depends on the maximum wavelength considered, since blue filters in general have a lower S/N for the same region. We illustrate this behaviour in Fig.~\ref{fig:SN_vs_SN}. We compare the median S/N of each region for every galaxy with the median S/N of the filters with $\lambda_{\mathrm{pivot}}<5000$~\AA. We find that the general median of the S/N can be up to 10 units larger than the median of these bluer filters. Since these filters are the ones that contain the 4000 break, and due to its correlation with the stellar population properties, we consider that it is better to look at the median S/N in these filters instead of the median of all the filters. Taking into account the results of the previous figures, we find that limiting our analysis to the regions where the median S/N of the filters with $\lambda_{\mathrm{pivot}}<5000$~\AA \ is larger than 5 will provide the most trustworthy results. We know that the surface brightness is also an important factor to consider, but the limiting surface brightness of \jp \ has not been studied yet and its beyond the scope of this work. However, we hope that these results may set a starting point for such purpose.

After all these tests and analysis, we can conclude that our methodology provides solid results in terms of the stellar population properties. We should, however, restrict our analysis to the regions were the median S/N ratio of the bluer filters ($\lambda_{\mathrm{pivot}}<5000$~\AA), since our SED-fitting results show large residual values for filters and regions below this limit.

\subsection{Spatially resolved stellar population properties}

In this section, we study the stellar population properties of the regions of the galaxies in our sample. We start by comparing the properties derived using an elliptical aperture at 1~\texttt{R\_EFF} with those derived using the \magauto \ photometry. We then propose an equivalent diagram to the dust corrected mass-colour diagram, to study how the properties are distributed, and we finish by studying the local main sequence of the star formation. As in Chapter~\ref{chapter:MANGA}, in order to simplify the reading of the text, we specify now which are these properties and the units that we use, and we will not explicitly include the units during the discussion of the results. In this section, we will focus in the following properties:

\begin{itemize}
    \item The stellar surface mass density, $\mu_\star$. We measure $\mu_\star$ in units of $M_\odot \times \mathrm{pc}^{-2}$ and we will generally use the logarithm of this value. 
    \item  The mass-weighted stellar age,  $\left <\log \mathrm{age} \right>_M$.  We measure the age in yr and we calculate the logarithm in this unit. 
    \item  The luminosity-weighted stellar age,  $\left <\log \mathrm{age} \right>_L$.  We measure the age in yr and we calculate the logarithm in this unit. 
    \item  The stellar metallicity,  $\left < \log \mathrm{Z/Z_\odot} \right >$.
    \item  The extinction $A_V$. This parameter is given in AB magnitudes. 
    \item  The intensity of the \gls{SFR}, $\Sigma_\mathrm{SFR}$. We calculate $\Sigma_\mathrm{SFR}$ in units of $M_\odot \times \mathrm{Gyr}^{-1} \times \mathrm{pc}^{-2}$, and we generally refer to the logarithm of this value. 
    \item  The \gls{sSFR}. We calculate \gls{sSFR} in units of $\mathrm{Gyr}^{-1}$, and we generally refer to the logarithm of this value.
\end{itemize}

\subsubsection{Integrated properties vs properties at \texttt{M1R\_EFF}}
\begin{figure}
    \centering
    \includegraphics[width=\textwidth]{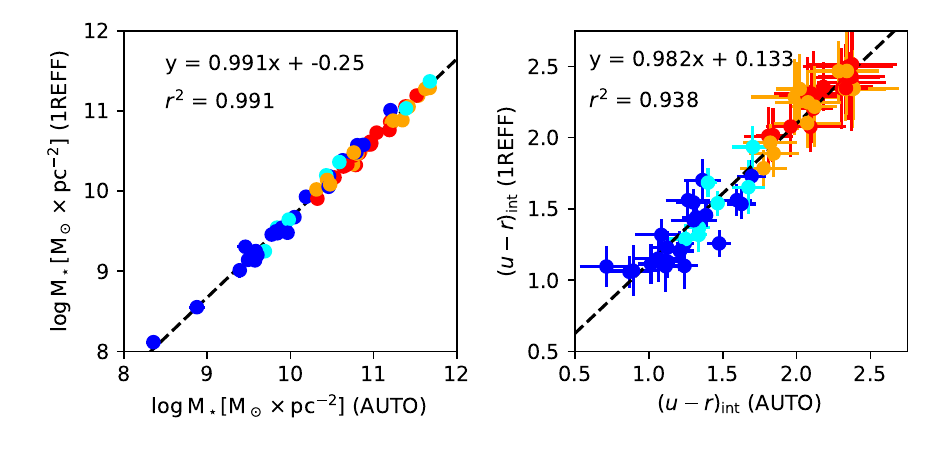}
    \caption[Comparison of the mass and colour obtained at 1~$\mathrm{R\_EFF}$ and with \magauto.]{Comparison of the mass (left panel) and $(u-r)_\mathrm{int}$  colour (right panel) obtained at 1~$\mathrm{R\_EFF}$ and with \magauto. Red points represent red galaxies in the field. Orange points represent red galaxies in groups. Blue points represent blue galaxies in the field. Cyan points represent blue galaxies in groups.}
    \label{fig:AUTOvs1REFF}
\end{figure}

We compare the results obtained for the total mass and the $(u-r)$ colour correction from extinction $(u-r)_\mathrm{int}$ from an extraction performed for an aperture of 1 effective radius and the \magauto \ extraction from the catalogue (which we have shown to be reproduced by our code) in Fig.~\ref{fig:AUTOvs1REFF}. The total mass shows a very linear behaviour, with a slope very close to the unity, but with an offset of $-0.25$. Using the properties of the logarithms, it easy to find that this offset means that the mass contained in an aperture of 1~$\mathrm{R\_EFF}$ is $\sim 56$~\% of the mass contained in the \magauto. This is a good result from several points of view. It is known that there is a strong relation among the luminosity and the mass and of a galaxy, mainly imposed by the stellar models \citep[see e.g.][]{Conroy2013}. We are using the $\mathrm{R\_EFF}$ provided by \sext \ so that our results are easily reproducible, with a value provided in the catalogues. This radius is defined so that an aperture of such size contains half of the light of the galaxy. On the other hand, \magauto \ is defined for the goal of defining an aperture that contains almost all the light of the galaxy, without compromising the S/N ratio. Therefore, it is a solid result that our methodology finds that the mass contained in 1~$\mathrm{R\_EFF}$ is close to half of the mass obtained using \magauto. It also indicates that the estimation of the effective radius provided by \sext \ for our data is good enough for our purposes. We note that there is a relation among the \gls{HMR} and the \gls{HLR}, but they are not exactly equal and this relation changes with the morphological type and the mass of the galaxy \citep{Rosa2015}. Nonetheless, we are not really comparing the \gls{HMR} and the \gls{HLR}, but rather the mass contained at different apertures.

The $(u-r)_\mathrm{int}$ colour shows an offset too, and we find that galaxies show a bluer colour when using the AUTO aperture. This is another good result for our method, since bluer regions are expected to be in the outer parts, while inner regions tend to be redder \citep[see e.g.][]{GonzalezPerez2011,Kennedy2016,Marian2018,Miller2023}. The inclusion of these outer parts of the galaxies should provide a bluer average in consequence.

\subsubsection{Surface mass density-colour diagrams}
\begin{figure}
    \centering
    \includegraphics[width=\textwidth]{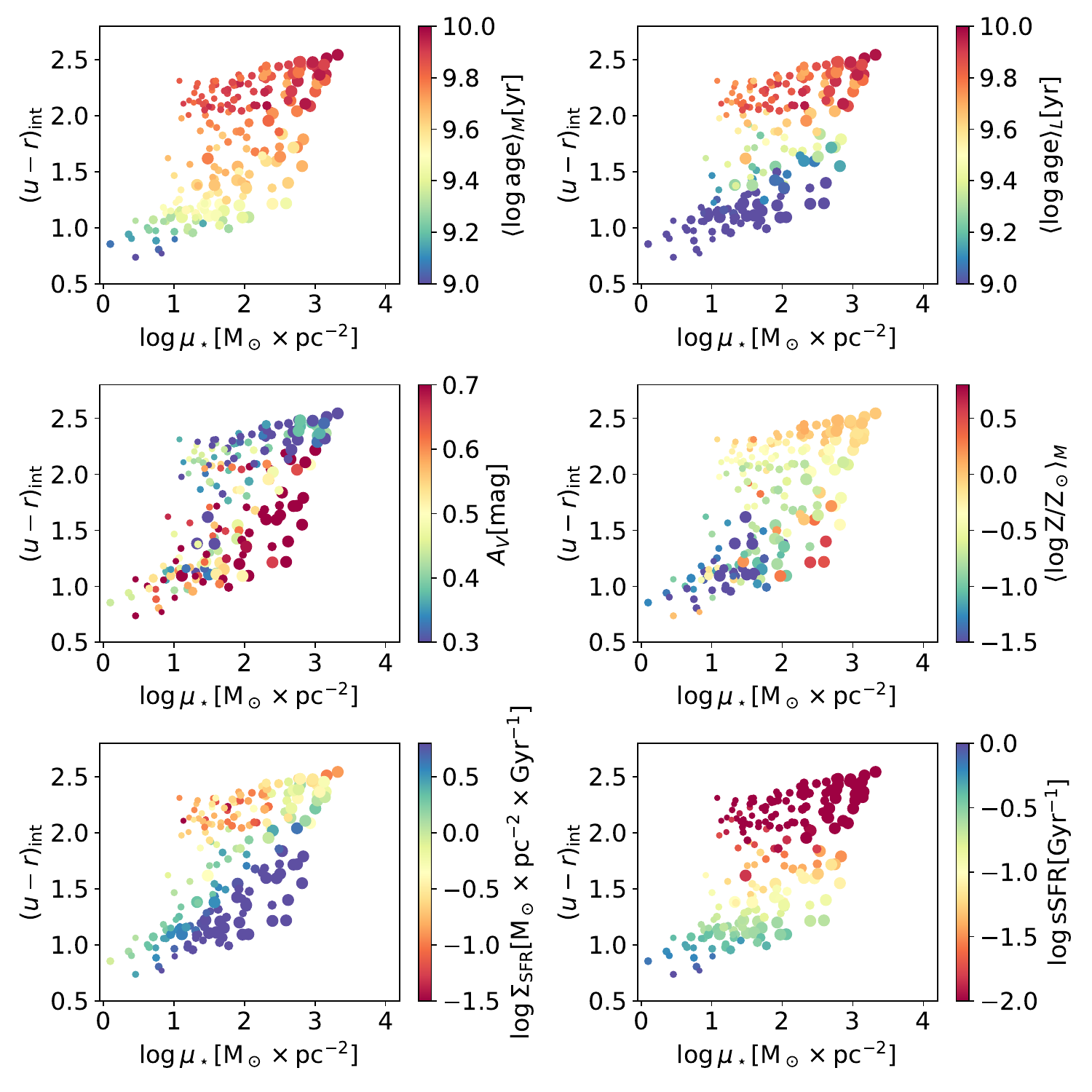}
    \caption[Colour--mass density diagram coloured by the main stellar population properties]{Colour--mass density diagram coloured by the main stellar population properties. From left to right, up to bottom: mass-weighted age, luminosity-weighted age, extinction, stellar metallicity, intensity of the star formation rate and \gls{sSFR}. Point size is inversely proportional to the distance to the galactic centre.}
    \label{fig:colourmassdenSPP}
\end{figure}

In the next step of our analysis, we propose an equivalent diagram to the mass-colour one, but using the mass density instead of the mass. We use again the uniform rings, in order for the galaxies to populate the diagram with a similar amount of points, even though we still use the aforementioned cut in S/N. We find that there is a nice distribution of all the properties in this diagram. The oldest regions, both luminosity and mass-weighted, appear in the redder and denser regions, while the bluer regions are younger, specially those of lower stellar mass density. We find the largest extinctions in the blue regions of larger density. The bluer and less dense regions are the ones with the lowest metallicity, while blue regions with high density are some of the metal richest points in the diagram. However, in general redder and denser regions are metal richer. The intensity of the \gls{SFR} shows a different correlation, but we need to take into account that plotting it against the mass density is the same as the SFMS, which we will study later. Denser and redder regions show the lowest \gls{sSFR}. These regions can be expected to be the most quiescent ones if we look at the radial profiles of the surface mass density and the \gls{sSFR} found by \cite{Rosa2015,Rosa2016}. In general, these diagrams are reproducing the same relations that we found in the colour-mass diagrams used for the integrated properties of galaxies \citep[see e.g.][]{Rosa2021,Rosa2022,Julio2022}, this is, once the mass (density) and the colour of (the region of) the galaxy are known, the other stellar population properties can be easily constrained. We also note that, even there is a correlation with both colour and mass density, the strongest one seems to be with colour. Results by \cite{Luis2019} show that, using $\tau$-delayed models from \cite{Madau1998} to derive the \gls{SFR} of the galaxies from the \gls{ALHAMBRA} survey, the properties are very correlated with the colour of the galaxies. Our findings are similar, since the dependence seems to be stronger with the $(u-r)_\mathrm{int}$ colour that the stellar surface mass density for the stellar ages and the \gls{sSFR}, but the stellar surface mass density also plays a role in the other properties. Nonetheless, this also means that we are recovering  the properties of the models, which is a solid statement for our results.

\begin{figure}
    \centering
    \includegraphics[width=\textwidth]{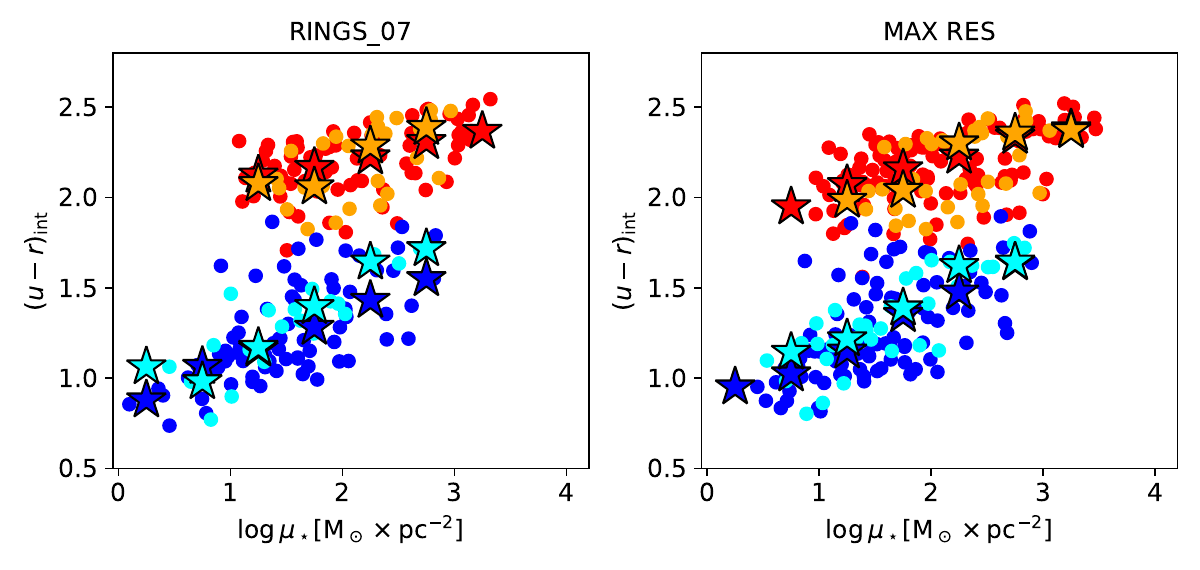}
    \caption[Colour--mass density diagram coloured by the by the environment and colour of the galaxies]{Colour--mass density diagram coloured by the by the environment and colour of the galaxies. Left panel shows the results using the $0.7$~\texttt{R\_EFF} RING segmentation. Right panel shows the results obtained with the maximum resolution segmentation. Red points represent regions belonging to the red galaxies in the field, orange is used for the regions of the red galaxies in groups, blue points are for the regions of blue galaxies in the field and cyan points are for regions of blue galaxies in groups. Point size is proportional to the inverse of the distance to the galaxy centre. Stars represent the median value for each galaxy type in each mass density bin.}
    \label{fig:colourmassdencolorandenv}
\end{figure}

We also use this diagram to study the distributions of the regions of the galaxies according to their colour and environment (see Fig.~\ref{fig:colourmassdencolorandenv}). We also plot the regions of the maximum resolution segmentation. We find that, for both segmentations, regions of red and blue galaxies are very well distinguished. This could be expected since regions of blue galaxies are bluer and regions of red galaxies are redder, as we had seen in the radial profiles. However, it is interesting to see that very few points of the red galaxies go into the redder regions and vice versa for blue galaxies. In addition, the distribution of the points seems to be related not only to colour, but also to stellar mass density, since red and blue points follow a diagonal trend rather than a vertical one. This trend is clearer in the maximum resolution segmentation. On the one hand, we have already mentioned that this diagram might be dominated by higher resolution galaxies. On the other hand, if we take into account the strong relation found between the stellar mass density and the radial distance, we can assume up to a certain point the mass density as a proxy of the distance. Therefore, these diagrams are showing in a different way the colour profiles of the galaxies, and using a better resolution might provide a better idea of the behaviour of red and blue galaxies. 

We do not find any relation of the environment of the galaxy with the position in these diagrams. The points of galaxies in groups are distributed in a similar way to the regions of the galaxies in the field, and the do not seem to cluster in any particular region. This result is in accordance to the results obtained for the radial profiles, where, except for some details in the ages and the \gls{sSFR}, we could not separate galaxies in groups and in the field. This result is similar to the integrated case. In \cite{Julio2022} we found that the distribution of the properties of the galaxies in the cluster mJPC2470-1771 in the mass colour diagram was the same as in the general sample of \mjp. The main difference resided in the fraction of red and blue galaxies and therefore, the clustering of points in the diagram. Here, our results seem to point towards a greater relevance of the mass density and the colour in the spatially resolved properties, rather than the environment, but future data might reveal a different fraction in the distribution of these properties.

\begin{figure}
    \centering
    \includegraphics[width=\textwidth]{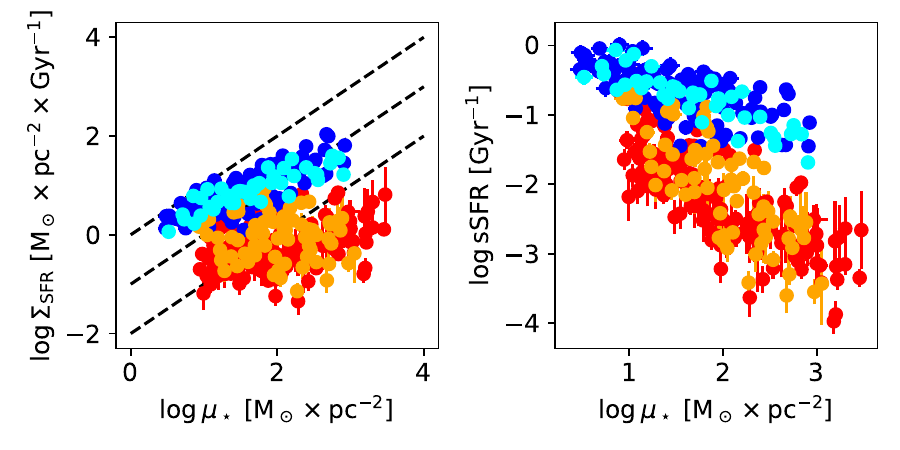}
    \caption[Local star formation main sequence]{Local star formation main sequence. Left panel: intensity of the \gls{SFR} as a function of the stellar surface mass density. Black dashed lines represent loci of constant \gls{sSFR} (from top to bottom, $\mathrm{sSFR} = 1, 0.1, 0.01$~$\mathrm{Gyr}^{-1}$). Right panel: \gls{sSFR} as a function of the stellar surface mass density. Red points represent red galaxies in the field. Orange points represent red galaxies in groups. Blue points represent blue galaxies in the field. Cyan points represent blue galaxies in groups.  }
    \label{fig:localSFMS}
\end{figure}

We finish this part of our analysis of the stellar population properties by reproducing the local SFMS, both in its \gls{SFR} intensity and \gls{sSFR} versions (see Fig.~\ref{fig:localSFMS}). Both versions are similar, just changing the slope from positive to negative. We can not distinguish environments here, but regions of red and blue galaxies are well differentiated. Our results are similar to those by \cite{Rosa2016}, this is, there is a strong relation between the mass density of the region and its \gls{SFR} density or its \gls{sSFR}, very likely a linear fit. Red galaxies are more disperse in the \gls{SFR} density version, but actually it looks like we could fit a linear relation in the \gls{sSFR} version.

\subsection{Radial distribution of the stellar population properties}

We study the radial profiles of the main stellar populations, this is, stellar mass density, mass-weighted and luminosity-weighted stellar age, the extinction parameter, stellar metallicity, extinction corrected colour $\mathrm{(u-r)_\mathrm{int}}$, the intensity of the \gls{SFR} and the \gls{sSFR}. For this analysis, we will use the uniform ring segmentation, every $0.7$~$\mathrm{R\_EFF}$, which is the worst PSF of the galaxy that has the largest PSF in comparison to its effective radius. We choose to do it this way so that every galaxy has the same weight in each bin. Using the maximum resolution segmentation would provide more data points, but for each radius bin, every galaxy would provide a different amount of points. Therefore, the calculated median would by dominated by those galaxies with a better resolution. We also divide our sample in four different groups, in order to study the possible effect of the environment: red galaxies in the field, red galaxies in groups, blue galaxies in the field, and blue galaxies in groups. We also note that in order to compare our results with those of the literature, we will assume that red galaxies are generally quiescent and early-type, while blue galaxies are generally star--forming and late-type \citep[see e.g.][]{Hogg2004,Kauffmann2004,Bassett2013}. We decide to use this proxy since, even thought they are not the same concept, they are known to correlate very well, and it will ease the comparison, because different studies decide to classify their samples according to different criteria.

Additionally, in order to provide a numerical value of the internal gradients for each galaxy, we use the results obtained with the maximum resolution segmentation, and we calculate the gradient of each property, restricted to $r<1.5$~$\mathrm{R\_EFF}$. In consequence, unlike the case of the median profiles, galaxies with a better resolution have the same weight in the plot as those with a worse one. We limit the radius to $r<1.5$~$\mathrm{R\_EFF}$ in order to fit only the central part (we know that, for example, the profile of the mass density is similar to a Sérsic profile and the internal region differs from the external one), and in order to have at least two points for each galaxy.

\subsubsection{Stellar mass surface density}

\begin{figure}
    \centering
    \includegraphics[width=\textwidth]{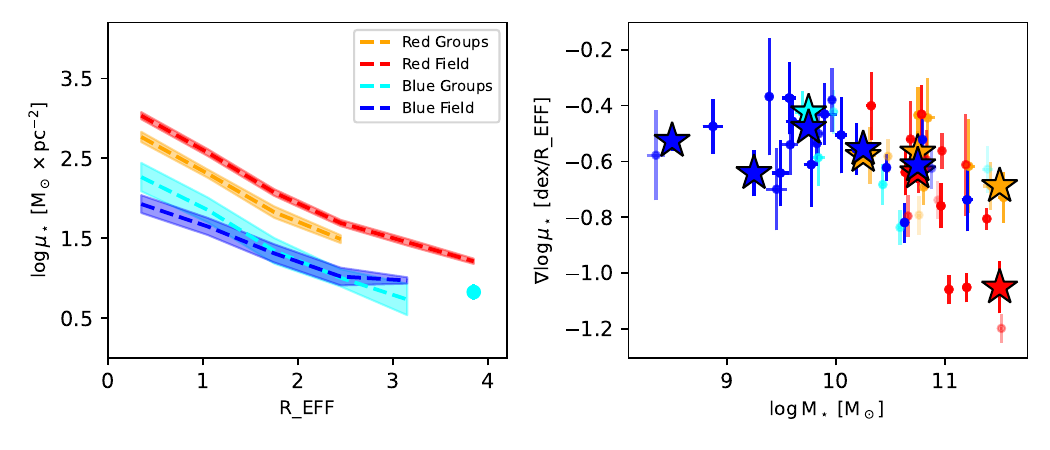}
    \caption[Radial profile and internal gradients of the stellar mass surface density by galaxy colour and environment.]{Radial profile and gradients of the stellar mass surface density by galaxy colour and environment. Red colours indicate red galaxies in the field. Orange colours represent red galaxies in groups. Blue colours represent blue galaxies in the field. Cyan colours represent blue galaxies in groups. Left panel: Radial profile of the mass surface density. Dashed lines represent the median value in the radius bin. Colour shade represent the error of the median. Single points represent bins where only one region was left after the S/N cleaning. Right panel: internal gradients of the mass surface density as a function of the total stellar mass. Circles represent the values of each galaxy. Stars represent the median value for each mass bin.}
    \label{fig:mass_Profile_gradient}
\end{figure}

The stellar mass density clearly decreases as the radius becomes larger for all galaxies colour and environments. Red galaxies in the field show the highest densities among our sample, staying above the rest of the galaxies at all radius. The median value goes from $\log \mu_\star \approx 3$ at $0.7$~\texttt{R\_EFF}, and decreases down to $\log \mu_\star \approx 1.2$ at $\sim 4$~\texttt{R\_EFF}. The shape of the profiles found by \cite{Rosa2014,Rosa2015} are similar for all galaxies regardless of their morphology or mass. However, the values of the density greatly change with these parameters. Most of the red galaxies in the field (and red galaxies in general) are more massive than $\log M_\star = 10$~$[M_\odot]$. For that range of masses, values found by those works at the central regions range from $\log \mu_\star \approx 3.5$ to $\log \mu_\star \approx 4.2$. Our values are lower, but we note that we have applied a PSF homogenisation, which flattens the profile, lowering them at the centre, while this was not done in the CALIFA sample. Also, our radial bins use a lower resolution, given the PSF. A better spatial resolution also provides larger mass densities specially close to the centre (see Fig.~\ref{fig:MANGAradSP} in Chapter~\ref{chapter:MANGA} for an example of a galaxy studied with larger resolution). In fact, if we look at the values found by \cite{Rosa2014,Rosa2015} at $\sim 0.7$~\texttt{R\_EFF} we find that they are $\log \mu_\star \sim3.5$, which is much closer to our results. Finally, results from \cite{Rosa2014,Rosa2015} use a Salpeter \gls{IMF} which on average provides an stellar mass $0.27$ larger than the \cite{Chabrier2003} used our this analysis \citep{Rosa2015}. At larger distances ($\sim3$~\texttt{R\_EFF}) \cite{Rosa2014,Rosa2015} find values in the range $\log \mu_\star \approx [1.5,2.5]$. Our results are also in good agreement with values found by \cite{Bluck2020} for the quiescent galaxies  ($\log \mu_\star \approx 3.5$ at the central regions, $\log \mu_\star \approx 2.2$ at 1.4~\texttt{R\_EFF}),  as well as with the results found by \cite{Abdurro2023} for galaxies in this range of masses ($\log \mu_\star \approx [3,4]$ in central regions up $\log \mu_\star \approx [1,2]$ at 4~\texttt{R\_EFF}). 

Red galaxies in groups densities are close to the red galaxies in field, but slightly less dense (around $\sim 0.1$~dex), but stay above blue galaxies in the field and in groups. These galaxies are also noisier in their outer regions than their counterparts in field, since after applying the S/N cut the profile only reaches $\sim 2.5$~\texttt{R\_EFF}. Galaxies in groups are also more likely to be contaminated by other nearby galaxies. Nonetheless, the shape of the profile and their values greatly resemble the ones of the red galaxies in the field, going from    $\log \mu_\star \approx 2.9$ at $0.7$~\texttt{R\_EFF} down to $\log \mu_\star \approx 1.75$ at $\sim 2.5$~\texttt{R\_EFF}. Since the range of masses covered by red galaxies in the groups and in the field is similar, we also find that the results are compatible with the findings by \cite{Rosa2014,Rosa2015,Bluck2020} and \cite{Abdurro2023}. From these profiles, it might seem that, on average, red galaxies in groups are slightly less dense than red galaxies in the field. A similar result is found by \cite{Bluck2020}, in the sense that quenched satellite galaxies (those that have fallen in a larger dark matter halo) are less dense on average than central galaxies (those that remain in the centre of their dark matter halo). However, the dispersion of the profiles of red galaxies in groups is large, and when looked individually, they also cover the range of values of galaxies in the field. Ideally, we would divide galaxies into bins of different mass and study these two (four, if we consider blue galaxies in the field and in groups) in each bin. However, due to our reduced sample, we shall perform these study for future \jp \ data release and avoid stronger conclusions in the meantime.  Our results are also highly compatible with the findings by \cite{Ana2024}. Their sample of elliptical galaxies and massive spirals ($\log M_\star > 10.5$~$[M_\odot]$), comparable to our sample of red galaxies and blue galaxies in groups, shows mass surface densities that range from $\log \mu_\star \approx 3.5$ at $\sim 0.1$~\texttt{R\_EFF} down to $\log \mu_\star \approx 2$ at $\sim 0.1$~\texttt{R\_EFF}. Our densities might appear to be lower, but we find that this is likely a consequence of the chosen radius bin. In fact, the mass density of these galaxies is $\log \mu_\star \approx 3.1$ at $\sim 0.7$~\texttt{R\_EFF}, more similar to our initial bin.

Blue galaxies, both in the field and in groups show very similar results. Their profile overlap in the region between 1 and 3~\texttt{R\_EFF}. For radii smaller than 1~\texttt{R\_EFF}, the median blue galaxies in groups is slightly higher. However, when taking into account the dispersion of the profiles of the blue galaxies in the field, it can be considered negligible. Moreover, the mass distribution of the blue galaxies in groups is shifted towards higher values, which can lead to higher stellar mass density profiles \citep[see e.g.][]{Rosa2015}. The masses of the blue galaxies is mainly contained in the range $\log M_\star \approx [9,10]$~$[M_\odot]$. Values of the stellar mass density found by \cite{Rosa2014,Rosa2015} in this mass range go from $\log \mu_\star \approx [2.5,3.5]$ in central regions down to  $\log \mu_\star \approx [1,2]$ at $\sim 3$~\texttt{R\_EFF}. Taking into account the aforementioned factors (PSF homogenisation, spatial resolution used, different IMF), our results are compatible: the mass density goes from $\log \mu_\star \approx [1.75,2]$ in central regions down to  $\log \mu_\star \approx [0.75,1]$ at $\sim 3$~\texttt{R\_EFF}. Our results are also compatible with those of \cite{Abdurro2023}, who find $\log \mu_\star \approx [2.5,3.5]$ in central regions and  $\log \mu_\star \approx [1,2]$ at $\sim 3$~\texttt{R\_EFF}, and \cite{Bluck2020}, who find $\log \mu_\star \approx 2.75$ in central regions and  $\log \mu_\star \approx 1.75$ at $\sim 1.4$~\texttt{R\_EFF} for star--forming galaxies and $\log \mu_\star \approx 2.75$ in central regions and  $\log \mu_\star \approx 2$ at $\sim 1.4$~\texttt{R\_EFF} for galaxies in the green-valley, blue galaxies are most likely to be in one of these two categories. Low mass spiral galaxies from \cite{Ana2024}, that is,  spiral galaxies with $\log M_\star < 10.5$~$[M_\odot]$, show values compatible with our sample of blue galaxies in the field once the bin size is taken into account: $\log \mu_\star \approx 2.5$ at $\sim 0.1$~\texttt{R\_EFF} down to $\log \mu_\star \approx 1.2$ at $\sim 2$~\texttt{R\_EFF}, with $\log \mu_\star \approx 2.2$ at $\sim 0.7$~\texttt{R\_EFF}.

Regarding the gradients and their relation with the total stellar mass (see right panel of Fig.~\ref{fig:mass_Profile_gradient}), we find that there is clearly a negative gradient for all galaxies (the mass density decreases with the distance to the centre). If we look at the dependence of the values of the gradients, we find that gradients become more negative as total  stellar mass increases. This trend is more obvious for stellar masses $\log M_\star >9.5$~$[M_\odot]$. Results for masses lower than that limit do not seem to follow such trend, but they might be affected by the low number of points. The values of the gradients seems to be more affected by the stellar mass than the colour of the galaxy or its environment, since the median value of the gradient in each mass bin is very similar, regardless of these two classification criteria. There is one exception, which is the difference in the gradient of the highest mass red galaxies, where the median gradient of the red galaxies in field is notably more negative that the median gradient of red galaxies in groups. This difference however can be caused by the low number of points in the red galaxies in groups in that bin. However, the median gradient of the red galaxies in field supposes a significant break from the tendency found in previous points. The gradients go from $\nabla \log \mu_\star \approx -0.4$~$\mathrm{dex/R\_EFF}$ at $\log M_\star \approx 9.5$~$[M_\odot]$ down to  $\nabla \log \mu_\star \approx [-1,-0.8]$~$\mathrm{dex/R\_EFF}$ at $\log M_\star > 11$~$[M_\odot]$. In the work by \cite{Rosa2015} a strong correlation between this gradient and the total stellar mass was also found but the gradients are more negative ($\nabla \log \mu_\star \sim -0.75$~$\mathrm{dex/R\_EFF}$ for $\log M_\star \approx 9.5$~$[M_\odot]$, $\nabla \log \mu_\star \sim -1.2$~$\mathrm{dex/R\_EFF}$ for $\log M_\star > 11$~$[M_\odot]$). The steepness of the relation is similar (a difference of $\sim 0.5$~$\mathrm{dex/R\_EFF}$ over the mass range) but values are notably flatter for our sample. This could be due to the PSF homogenisation, since we are measuring the inner gradient and central regions are more affected by the homogenisation, as well as by the size of the radius bin. Even thought we are using the maximum size allowed by the PSF, some galaxies are still very small.

\subsubsection{Stellar ages}

\begin{figure}
    \centering
    \includegraphics[width=\textwidth]{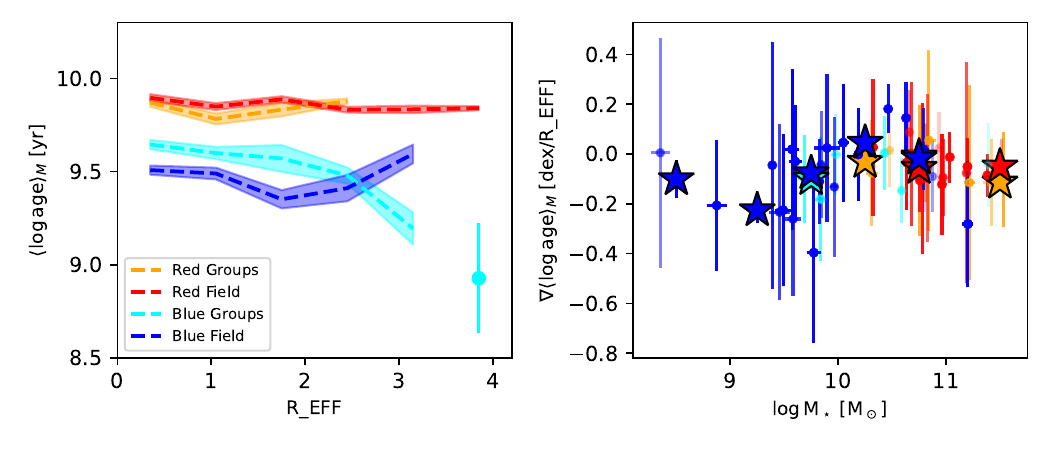}
    \caption[Radial profile and internal gradients of the mass-weighted stellar age by galaxy colour and environment.]{Radial profile and internal gradients of the mass-weighted stellar age by galaxy colour and environment. Red colours indicate red galaxies in the field. Orange colours represent red galaxies in groups. Blue colours represent blue galaxies in the field. Cyan colours represent blue galaxies in groups. Left panel: Radial profile of the mass surface density. Dashed lines represent the median value in the radius bin. Colour shade represent the error of the median. Single points represent bins where only one region was left after the S/N cleaning. Right panel: internal gradients of the mass surface density as a function of the total stellar mass. Circles represent the values of each galaxy. Stars represent the median value for each mass bin.}
    \label{fig:ageM_Profile_gradient}
\end{figure}

\begin{figure}
    \centering
    \includegraphics[width=\textwidth]{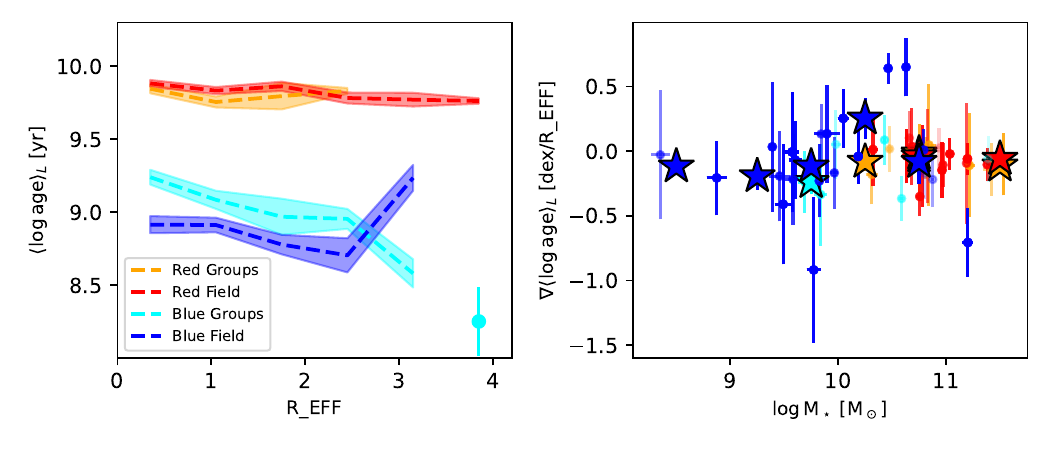}
    \caption[Radial profile and internal gradients of the luminosity-weighted stellar age by galaxy colour and environment.]{Radial profile and internal gradients of the luminosity-weighted stellar age by galaxy colour and environment. Red colours indicate red galaxies in the field. Orange colours represent red galaxies in groups. Blue colours represent blue galaxies in the field. Cyan colours represent blue galaxies in groups. Left panel: Radial profile of the mass surface density. Dashed lines represent the median value in the radius bin. Colour shade represent the error of the median. Single points represent bins where only one region was left after the S/N cleaning. Right panel: internal gradients of the mass surface density as a function of the total stellar mass. Circles represent the values of each galaxy. Stars represent the median value for each mass bin.}
    \label{fig:ageL_Profile_gradient}
\end{figure}

Both mass-weighted (see Fig.~\ref{fig:ageM_Profile_gradient}) and luminosity-weighted stellar (see Fig.~\ref{fig:ageL_Profile_gradient}) ages show a similar profile for red galaxies, both in the field and in groups. These profiles are almost flat at a value of $\left <\log \mathrm{age} \right>_M \approx 9.9$ and $\left <\log \mathrm{age} \right>_L \approx 9.9$, respectively. Red galaxies in the field seem to show an slight increase in the age up to 1~\texttt{R\_EFF}, and then an slight decrease, while ages of red galaxies seem to show the opposite behaviour (it decreases and then increases). However, given the usual errors in the determination of the errors, this is compatible with a flat profile for both types of galaxies at a similar age, with no significant differences due to the environment. We might be limited by the ages of the model and the SED-fitting code could be trying to provide the oldest age possible. Age profiles found by \cite{Rosa2014, Rosa2015} for the luminosity-weighted age depend on the mass and morphology of the galaxy. In general terms, profiles are similar: the age decreases with a steeper gradient from the centre up to $\sim 1$~\texttt{R\_EFF}, point from where the stellar age stills decreases but at a slower rate. The initial steeper decrease is more evident for galaxies with a total stellar mass in the range of $\log M_\star = [10.1,11.2]$~$[M_\odot]$, particularly for galaxies in the mass range $\log M_\star = [10.1,10.6]$~$[M_\odot]$. Our sample of galaxies is within this mass range, but generally in the more massive range. The size of the radius binning and the PSF homogenisation might be at play again. The range of ages covered by galaxies in that range of masses in the works by \cite{Rosa2014,Rosa2015} is $\left <\log \mathrm{age} \right>_L \approx [9.5,10]$, with outer parts of galaxies in the lower mass end reaching values down to $\left <\log \mathrm{age} \right>_L \approx 9$. Our results are compatible with the findings for the high-mass end of the range of stellar masses. However, luminosity-weighted age profiles found by \cite{SanRoman2018} for this type of galaxies are shown to be mostly flat in an age range around $\left <\log \mathrm{age} \right>_L \approx 9.9$. Results found by \cite{Bluck2020} for quiescent galaxies are similar, with almost flat profiles at $\left <\log \mathrm{age} \right>_L \approx 9.9$, regardless o whether they are central o satellites galaxies. Mass-weighted age profiles found by \cite{Abdurro2023} are also mostly flat for all galaxies with $\log M_\star >9.5$~$[M_\odot]$, and become even flatter as the mass increases. In particular, for quiescent galaxies, ages are $\left <\log \mathrm{age} \right>_M \approx 9.9$. Therefore, our results are in great agreement with those by \cite{SanRoman2018,Bluck2020,Abdurro2023}. The luminosity-weighted stellar ages found by \cite{Ana2024} for their sample of elliptical galaxies are compatible with our results, showing values of the ages $\left <\log \mathrm{age} \right>_L \approx 9.9$, but their profiles show a steeper decrease, with ages down to $\left <\log \mathrm{age} \right>_L \approx 9.5$ at 2~\texttt{R\_EFF}.

Blue galaxies exhibit more noticeable differences between their mass and luminosity-weighted age profiles. The most most significant one is an offset of $\sim 0.5$~$\mathrm{dex}$ between the mass and luminosity-weighted ages (luminosity ages are younger). The profile is similar in general term for both ages and for both types of galaxies. Blue galaxies in groups ages decrease until $\sim 2.5$~$\mathrm{R\_EFF}$ and then the age decrease more steeply. The initial decrease seems slightly flatter for the mass-weighted-age (from $\left <\log \mathrm{age} \right>_M \approx 9.6$ down to $\left <\log \mathrm{age} \right>_M \approx 9.5$) than for the luminosity-weighted age (from $\left <\log \mathrm{age} \right>_M \approx 9.25$ down to $\left <\log \mathrm{age} \right>_M \approx 8.9$). However the last steep decrease might be caused by a single galaxy, whose age can also have been measured with a great uncertainty (the S/N decreases with distance too). The mass-weighted age of blue galaxies in field decreases from from $\left <\log \mathrm{age} \right>_M \approx 9.5$ down to $\left <\log \mathrm{age} \right>_M \approx 9.4$ at 2~\texttt{R\_EFF} and then increases up to $\left <\log \mathrm{age} \right>_M \approx 9.7$ at 3~\texttt{R\_EFF}. Similarly, the luminosity-weighted age decreases from $\left <\log \mathrm{age} \right>_L \approx 8.9$ down to $\left <\log \mathrm{age} \right>_L \approx 8.7$ at $2.5$~\texttt{R\_EFF} and then increases up to $\left <\log \mathrm{age} \right>_L \approx 9.4$ at 3~\texttt{R\_EFF}. This final increase in both ages is, similarly to the steep decrease found in blue galaxies in groups, due to a single point in that radius bin. Another possible explanation is a degeneracy found in both types of galaxies among the age, the metallicity and the extinction. The age-metallicity has been known for long \cite{Worthey1994,Worthey1999}. Indeed, if we compare our results with other works, these older ages do not appear in the outer parts of galaxies, neither the steeper decrease in outer regions \citep[see e.g.][]{Rosa2014, Rosa2015, Bluck2020, Abdurro2023}. Therefore, it is likely an effect of the number of points in each bin and their mass. Aside from these outer parts, which can also be affected by a lower S/N ratio, results for the ages of blue galaxies is in great agreement with the literature: the works by \cite{Rosa2014,Rosa2015} find luminosity-weighted age profiles rather flat, but with ages slightly younger populations at larger radii for masses $\log M_\star <10.1$~$[M_\odot]$, which is the mass interval where most of the blue galaxies are found. Ages for these galaxies are around $\left <\log \mathrm{age} \right>_L \approx 9$. This is also the case for the star--forming galaxies studied by \cite{Bluck2020} and \cite{Abdurro2023}. We note that blue galaxies in groups seem to show slightly older ages than blue galaxies in the field, but taking into account the uncertainty intervals and the mass distribution (blue galaxies in groups in our sample are generally more massive than blue galaxies in the field) it is hard to associate this difference to the environment. In this regard, the sample of spiral galaxies studied by \cite{Ana2024} shows an offset between low mass ($\left <\log \mathrm{age} \right>_L \approx 9.2$ at $0.1$~\texttt{R\_EFF} down to $\left <\log \mathrm{age} \right>_L \approx 8.5$ at 2~\texttt{R\_EFF}) and high mass galaxies ($\left <\log \mathrm{age} \right>_L \approx 9.5$ at $0.1$~\texttt{R\_EFF} down to $\left <\log \mathrm{age} \right>_L \approx 9.0$ at 2~\texttt{R\_EFF}). In this sense, our results are similar, although we find generally flatter profiles and slightly younger ages.

Concerning the relation of the gradients with the total stellar mass (see right panels of Figs.~\ref{fig:ageM_Profile_gradient} and \ref{fig:ageL_Profile_gradient}), we find that the gradients of both mass and luminosity-weighted ages are negative for low mass galaxies ($\nabla \left <\log \mathrm{age} \right> \sim -0.2$ for galaxies with mass $\log M_\star \sim 9$~$[M_\odot]$, and their values keep increasing with mass, regardless of the colour and environment. For galaxies with $\log M_\star >10$~$[M_\odot]$, the gradient of both ages is  $\nabla \left <\log \mathrm{age} \right> \sim 0$. The work by \cite{Rosa2015} finds that the gradient becomes more negative up to $\log M_\star \sim 11$~$[M_\odot]$ and then starts becoming flatter, over a range of values of $\nabla \left <\log \mathrm{age} \right> \approx [-0.5,0]$. The range of values that we find is similar for luminosity-weighted ages, although we find that the tendency changes at much lower masses. The work by \cite{Breda2020} finds the same relation of the gradients with the mass that \cite{Rosa2015}, but they find that the gradients start to increase at $\log M_\star \sim 11$~$[M_\odot]$, which is closer to our results. A strong relation between the mass an the gradient of stellar age is found by \cite{Parikh2021}, in the same way that our results: the gradient is more negative for low-mass galaxies and it becomes flatter as the mass increases. Regarding the values of the gradients, massive early types galaxies are commonly found to have rather flat age gradients \citep[see e.g.][and references therein]{SanRoman2018, Parikh2021}, which is in accordance with our results. However, \cite{Rosa2015, Bluck2020} still find negative gradients for this type of galaxy ($\nabla \left <\log \mathrm{age} \right> \sim -0.1$). For blue/star--forming/early type galaxies, negative age gradients are found by works such as \cite{Rosa2015,Breda2020,Bluck2020,Parikh2021}. The values of the gradients found by these works are compatible with our results, although gradients found for late-type galaxies by \cite{Parikh2021} are more negative (steeper gradient). Overall, we find that our results are in good agreement with the literature.

\subsubsection{Extinction}

\begin{figure}
    \centering
    \includegraphics[width=\textwidth]{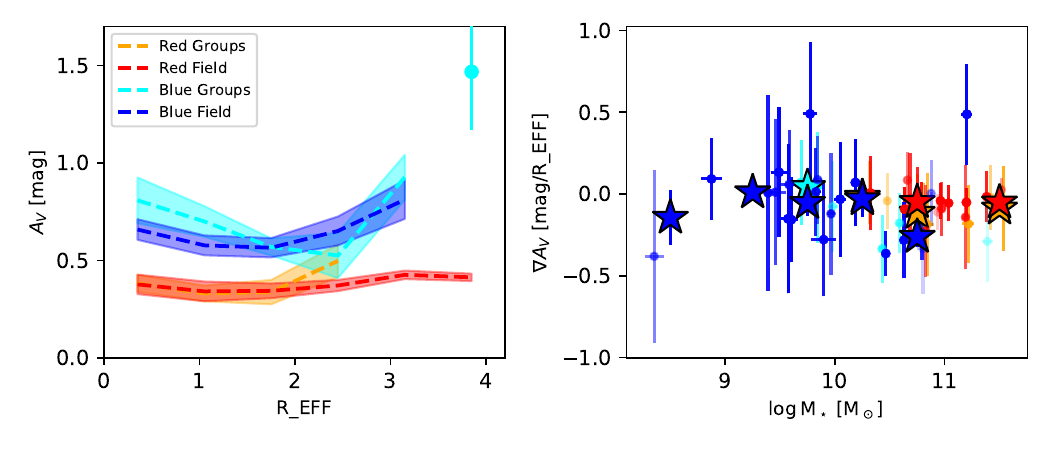}
    \caption[Radial profile and internal gradients of the extinction $A_V$ by galaxy colour and environment.]{Radial profile and internal gradients of the $A_V$ by galaxy colour and environment. Red colours indicate red galaxies in the field. Orange colours represent red galaxies in groups. Blue colours represent blue galaxies in the field. Cyan colours represent blue galaxies in groups. Left panel: Radial profile of the mass surface density. Dashed lines represent the median value in the radius bin. Colour shade represent the error of the median. Single points represent bins where only one region was left after the S/N cleaning. Right panel: internal gradients of the mass surface density as a function of the total stellar mass. Circles represent the values of each galaxy. Stars represent the median value for each mass bin.}
    \label{fig:AV_Profile_gradient}
\end{figure}

The extinction $A_V$ (see Fig.~\ref{fig:AV_Profile_gradient}) shows a rather flat profile for red galaxies, both in groups and in the field, at $A_V \approx 0.4$. Red galaxies in the field seem to increase their $A_V$ very slightly from $\sim 1$~\texttt{R\_EFF}, and even though red galaxies in groups might show a steeper increase in the extinction from $\sim 2$~\texttt{R\_EFF}, but the lack of points at larger distances does not allow for confirming this tendency. Nonetheless, we can not distinguish any effect of the environment in this parameter given the uncertainty intervals. Radial profiles of the extinction $A_V$ found by \cite{Rosa2015} find an steep decrease from the central region up to $\sim 0.5$~\texttt{R\_EFF}, regardless of the mass of the galaxy, and then keeps decreasing, except for the most massive galaxies. Because of the chosen radius bin we would be unable to see the initial steep decrease, but we do not find it either when using the maximum resolution rings. Values found for $A_V$ by \cite{Rosa2015} are in the range $A_V \approx [0,0.2]$ once the gradient becomes flatter at radii larger than $\sim 0.5$~\texttt{R\_EFF}. Our results are more or less compatible outside the initial steeper decrease, although we find generally larger extinctions. However, our profiles and their values are very similar to the findings by \cite{SanRoman2018}. In this work, they find very small positive gradients for this type of galaxies, for a range of values of $A_V \approx [0.2,0.5]$

On the other hand, the profile decreases with the distance to the galactic centre for blue galaxies in groups, from $A_V \approx 0.9$ in central regions down to $A_V \approx 0.5$ at $\sim 2.5$~\texttt{R\_EFF} and then slightly increases steeply (up to $A_V \approx 1.5$ at $\sim 4$~\texttt{R\_EFF}). This steep increase is complementary to the steep decrease in the luminosity-weighted age of blue galaxies in groups. Note that the error of the median becomes zero in that point unlike in the other groups of galaxies. This hints that this behaviour might be caused by a degeneracy between both parameters in a single galaxy (the standard deviation of a single point is zero). For blue galaxies in the field, the behaviour is similar, this is the  extinction decreases from $A_V \approx 0.75$ in central regions down to $A_V \approx 0.5$ at $\sim 2$~\texttt{R\_EFF} and then slightly increases up to $A_V \approx 0.9$ at $\sim 3$~\texttt{R\_EFF}. This increase does not coincide with a steep decrease in the stellar age, so in this case this degeneration does not seem to be playing a role. Nonetheless, the change in the gradient is not large and it can be caused by more massive galaxies, which might reach a higher S/N ratio. 
We can not separate field and group galaxies either, except for a slightly larger median value of the extinction of blue galaxies in groups, but taking into account results from \cite{Rosa2015}, this might also be a consequence of the total mass distribution. The comparison with our results with those of \cite{Rosa2015} shows that both works find a decrease in the internal regions of this type of galaxies. Values of the extinction found by \cite{Rosa2015} are within the range $A_V \approx [0.4,0.6]$ for galaxies in the mass range $\log M_\star=[9.1,11.2]$~$[M_\odot]$, which is compatible with our range of values, although we find galaxies reaching higher extinctions.

The gradients of the extinction $A_V$ (see right panel of Fig.~\ref{fig:AV_Profile_gradient}) show a large dispersion. Most of them are within the range $ \nabla A_V \approx [-0.2,0.2]$~mag/R\_EFF, and the median gradient is generally negative ($\sim -0.1$). There seems to be no strong relation with the mass. The median gradient of blue galaxies in the field increases with mass for galaxies with mass $\log M_\star < 9.5$~$[M_\odot]$, but it could be an effect of the low number of galaxies. Then it flattens at $ \nabla A_V \approx -0.1$~mag/R\_EFF and decreases down to $ \nabla A_V \approx 0.1$~mag/R\_EFF, but this last decrease could be again an effect of the low number of points in the last mass bin. This is also the case for the difference in the median values found at $\log M_\star \sim 9.75$~$[M_\odot]$ of the gradient of blue galaxies in the groups and in the field. Red galaxies in groups and in the field might show no significantly different values of the median gradient, and both seem to faintly become flatter with total stellar mass. The comparison with the results from \cite{Rosa2015} shows that the dispersion in the values of the internal gradients is also very large. However, the range of values, although compatible, is not the same ($ \nabla A_V \approx [-0.6,0.2]$~mag/R\_EFF). We may not be able to find steeper gradients due to the PSF homogenisation, the radius bin or the size of the sample. They also find that the gradient becomes stronger (more negative) with masses up to $\log M_\star \approx 11$~$[M_\odot]$, and then becomes flatter. Our results might be compatible with this phenomena, but the mass where the tendency changes would be much lower ($\log M_\star \approx 10$~$[M_\odot]$).  More data is required to further investigate this relation. However, we note that our results for the massive red galaxies are highly compatible with the findings by \cite{SanRoman2018} ($ \nabla A_V \approx -0.03$~mag/R\_EFF).

\subsubsection{Metallicity}

\begin{figure}
    \centering
    \includegraphics[width=\textwidth]{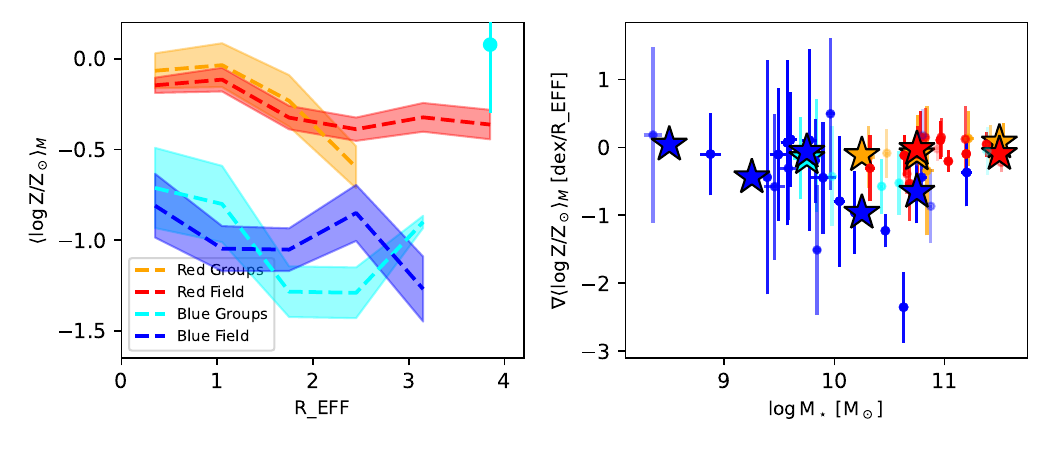}
    \caption[Radial profile and internal gradients of the stellar metallicity by galaxy colour and environment.]{Radial profile and internal gradients of the stellar metallicity  by galaxy colour and environment. Red colours indicate red galaxies in the field. Orange colours represent red galaxies in groups. Blue colours represent blue galaxies in the field. Cyan colours represent blue galaxies in groups. Left panel: Radial profile of the mass surface density. Dashed lines represent the median value in the radius bin. Colour shade represent the error of the median. Single points represent bins where only one region was left after the S/N cleaning. Right panel: internal gradients of the mass surface density as a function of the total stellar mass. Circles represent the values of each galaxy. Stars represent the median value for each mass bin.}
    \label{fig:logZ_Profile_gradient}
\end{figure}

We find that the profiles of the stellar metallicites of all the types of galaxies decrease with the distance to the centre (see Fig.~\ref{fig:logZ_Profile_gradient}). Values of red galaxies in the field decrease from $\left < \log \mathrm{Z/Z_\odot} \right > \approx -0.25$ in central regions town to $\left < \log \mathrm{Z/Z_\odot} \right > \approx -0.5$ at $\sim 2$~\texttt{R\_EFF}. Then it increases back to  $\left < \log \mathrm{Z/Z_\odot} \right > \approx -0.3$ and remains rather flat. Taking into account the usual error bars of the metallicity, and that only the most extended galaxies contribute to the last bins, this profile can be compatible with a flat one, but also with a decreasing profile up to $\sim 2$~\texttt{R\_EFF}, and then constant for further distances. Red galaxies in groups show a flatter profile with values $\left < \log \mathrm{Z/Z_\odot} \right > \approx -0.25$ up to $\sim 2$~\texttt{R\_EFF} and then start decreasing more steeply, down to $\left < \log \mathrm{Z/Z_\odot} \right > \approx -0.5$ at $\sim 2.5$~\texttt{R\_EFF}. Even though the gradients seem different, given the errors of the median it is not possible to solidly affirm that the environment is playing a key role in the profiles of the metallicity of red galaxies. Radial profiles of the metallicity found by \cite{Rosa2014, Rosa2015} show a modest decrease of the metallicity with the distance to the centre, by a factor lower than $0.2$~dex over a distance of 3~\texttt{R\_EFF}, for galaxies in the mass range $\log M_\star = [10.6,11.8]$~$[M_\odot]$. Values of the metallicity in that range of mass found by \cite{Rosa2014,Rosa2015} are within the interval $\left < \log \mathrm{Z/Z_\odot} \right > \approx [-0.2,0.1]$. Our results are generally more metal poor. This is also true when compared with finding by \cite{SanRoman2018} for this type of galaxies, which are summarised in a decreasing profile from $\left < \log \mathrm{Z/Z_\odot} \right > \approx 0.1$ in central regions town to $\left < \log \mathrm{Z/Z_\odot} \right > \approx -0.3$ at $\sim 3$~\texttt{R\_EFF}. This difference is large, but the uncertainties in the metallicity are generally large (see e.g. Fig~\ref{fig:MANGAradSP} in Chapter~\ref{chapter:MANGA}) and values of all works can be compatible within this uncertainty intervals. Additionally, \cite{SanRoman2019} showed that different methodologies can lead to systematic differences in IFU-like studies. 

The radial profile of blue galaxies in groups and in the field is very similar, overlapping for the most part of it. The profile of both types of galaxies decreases from $\left < \log \mathrm{Z/Z_\odot} \right > \approx -0.5$ in central regions down to $\left < \log \mathrm{Z/Z_\odot} \right > \approx -1.25$ at $\sim 3$~\texttt{R\_EFF}. The profile of blue galaxies in groups then shows much more metal rich values at the external parts, from $\sim 2.5$~\texttt{R\_EFF}. However, as in other cases, we find that this is caused by a single point which might be also suffering from a degeneracy with stellar age, since this is the same breaking point of the extinction and luminosity-weighted age of blue galaxies in groups. At this point, ages become noticeably younger, and the extinction and metallicity become much larger. The SED-fitting seems to have run into this degeneracy for that point.  Additionally, blue galaxies in the field show a break in their gradient which becomes steeper at $\sim 2.5$~\texttt{R\_EFF}. This could be caused by a lesser number of points contributing to this part of the profile, but we also note that this break coincides with an opposite behaviour in the luminosity-weighted age, which becomes unexpectedly older for these regions. This would support the idea of the degeneracy among age, extinction, and metallicity. If we compare our results with those by \cite{Rosa2014,Rosa2015}, focusing in the mass range $\log M_\star = [9.1,10.6]$~$[M_\odot]$, we find that the metallicity decreases with distance too, although at a slower rate. In fact, for galaxies with masses $\log M_\star < 10.1$~$[M_\odot]$, the profile looks rather flat. This also indicates that some of the aforementioned discrepancies found when comparing the different stellar population properties could also be a consequence of not separating the galaxies by mass, which is not possible in this study due to the limited size of our sample.

The gradients of  the stellar metallicity (see right panel of Fig.~\ref{fig:logZ_Profile_gradient}) are generally negative. The dispersion is large, as well as the errorbars. The relation with the mass is not clear and in general terms seems rather flat. Given the typical uncertainties in the gradients, there is no significant difference among their median values in a same mass bin, regardless of the colour and environment, with the exception of high mass blue galaxies in the field. This difference, however, is most likely due to the lesser number of that type of galaxies in those bins. Overall, most gradients are within the range $ \nabla \left < \log \mathrm{Z/Z_\odot} \right > \approx [-1,0.2]$. Gradients found by \cite{Rosa2015} are within the range $ \nabla \left < \log \mathrm{Z/Z_\odot} \right > \approx [-4,0.2]$, which is compatible with our results, although we find even steeper ones. They also find that the relation with the total stellar mass is not clear and that it shows a large dispersion. In general terms, our results are compatible with those by \cite{Rosa2015}, but we estimate the metallicity more poorly than other parameters. Our results are also in agreement with those by \cite{Parikh2021}, who find flat metalicity gradients in elliptical galaxies and negative gradients for late type galaxies, as well as with many works in the literature that, in general find negative gradients of the metallicity \citep[see e.g.][]{Davies1993,Mehlert2003,SanchezBlazquez2006b,SanchezBlazquez2007,Reda2007,Zheng2017}.

\subsubsection{Colour}
\begin{figure}
    \centering
    \includegraphics[width=\textwidth]{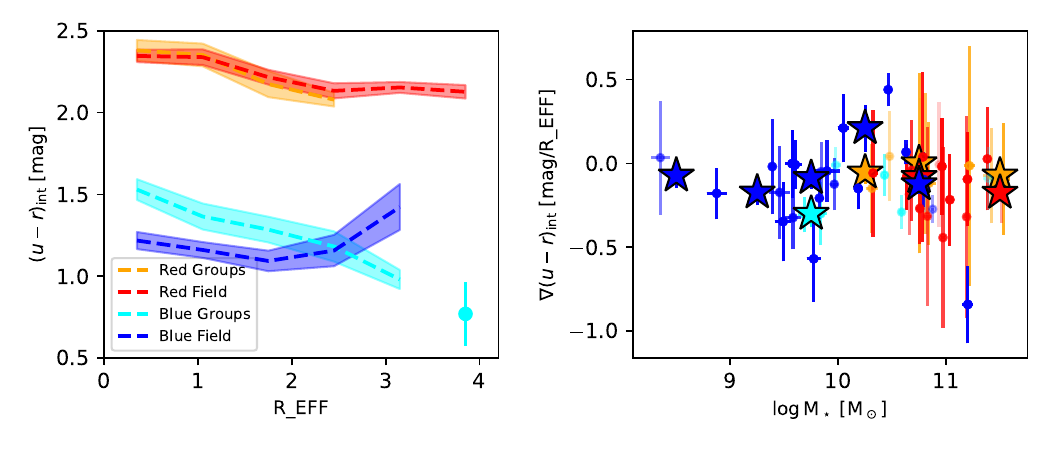}
    \caption[Radial profile and internal gradients of the extinction $A_V$ by galaxy colour and environment.]{Radial profile and internal gradients of the $A_V$ by galaxy colour and environment. Red colours indicate red galaxies in the field. Orange colours represent red galaxies in groups. Blue colours represent blue galaxies in the field. Cyan colours represent blue galaxies in groups. Left panel: Radial profile of the mass surface density. Dashed lines represent the median value in the radius bin. Colour shade represent the error of the median. Single points represent bins where only one region was left after the S/N cleaning. Right panel: internal gradients of the mass surface density as a function of the total stellar mass. Circles represent the values of each galaxy. Stars represent the median value for each mass bin.}
    \label{fig:colour_Profile_gradient}
\end{figure}

The $(u-r)_\mathrm{int}$ colour profiles clearly show the difference between red and blue galaxies (see Fig.~\ref{fig:colour_Profile_gradient}). The profiles of red galaxies in the field and in groups are very similar, with no significant difference due to the environment. In central regions of red galaxies, we find that $(u-r)_\mathrm{int} \approx 2.4$, and it decreases down to $(u-r)_\mathrm{int} \approx 2.1$ at $\sim 1.5$~\texttt{R\_EFF}. Then, the profile flattens and remains at that value. The most noticeable difference is that the colour of red galaxies in groups remains approximately constant up to $\sim 1.2$~\texttt{R\_EFF} and then starts decreasing, while the colour of red galaxies in the field starts decreasing until a distance of up to $\sim 1.2$~\texttt{R\_EFF} and then remains constant. Given the error of the median and the different number of red galaxies in the field and in groups, along with the small difference, it is difficult to suggest that this is actually an effect of the environment. 

Blue galaxies in the field and in groups show much more different profiles. The $(u-r)_\mathrm{int}$ colour of blue galaxies in groups decreases from $(u-r)_\mathrm{int} \approx 1.55$~mag in central regions, down to  $(u-r)_\mathrm{int} \approx 0.75$~mag at $\sim 4$~\texttt{R\_EFF}. There is a break point at $\sim 2.5$~\texttt{R\_EFF}, where the colour becomes bluer more rapidly. As discussed with previous properties which showed a breaking point at the same distance, this could be caused by a single galaxy. However, it also shows that the breaks seen in the luminosity-weighted ages, extinction, and metallicity, as well as their degeneracy, is likely caused by this break. On the other hand, blue galaxies in the field show a much flatter profile. The median value of the colour is  $(u-r)_\mathrm{int} \approx 1.25$~mag in central regions and decreases less than $0.1$~mag up to $\sim 2$~\texttt{R\_EFF}, and then it starts increasing up to $(u-r)_\mathrm{int} \approx 1.5$~mag at $\sim 3$~\texttt{R\_EFF}. This reddening of the outer parts could be an effect of more massive galaxies dominating this regions, since they are more likely to be more luminous and have a better S/N ratio at larger distances. Similarly, the difference in the values of the colour found in blue galaxies in the field and in the groups is not large enough to be solidly considered an effect of the environment, particularly if we take into account that the mass distribution of blue galaxies in the field leans toward more massive galaxies.

The gradient in the colour (see right panel of Fig.~\ref{fig:colour_Profile_gradient}) is negative or almost flat for most galaxies, indicating that they become bluer towards their outskirts. This result is found in the literature, as well as a general consensus that colour gradients tend to be flatter in spheroidal galaxies \citep[see e.g.][]{GonzalezPerez2011,Kennedy2016,Marian2018,Miller2023}. We do not find a  strong relation with the mass, but the gradient seems to become steeper as the mass increases up to $\log M_\star \approx 10$~$[M_\odot]$, where it starts to increase (becoming flatter). For blue galaxies in the field, the median value of the gradient even becomes positive at this mass, but it can be due to the low number of points in that bin.

\subsubsection{Intensity of the \gls{SFR}}
\begin{figure}
    \centering
    \includegraphics[width=\textwidth]{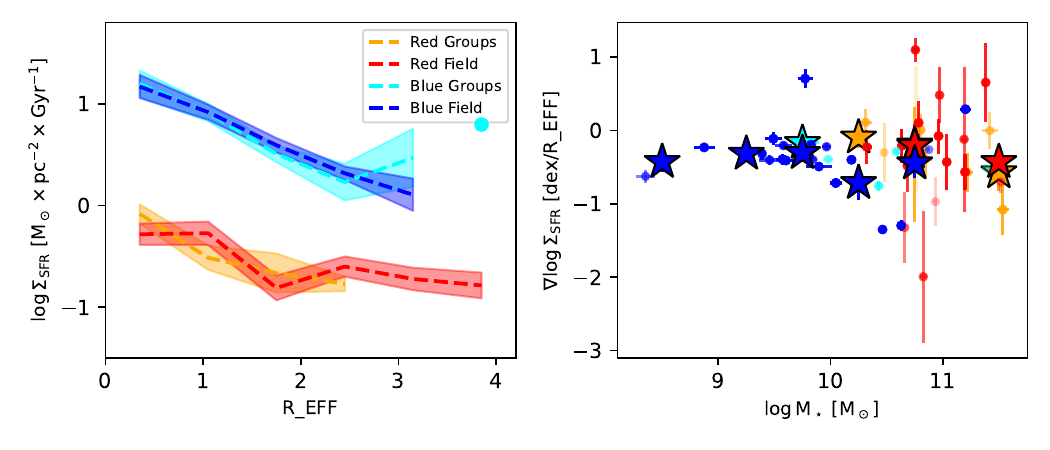}
    \caption[Radial profile and internal gradients of the intensity of the \gls{SFR} by galaxy colour and environment.]{Radial profile and internal gradients of the intensity of the \gls{SFR} by galaxy colour and environment. Red colours indicate red galaxies in the field. Orange colours represent red galaxies in groups. Blue colours represent blue galaxies in the field. Cyan colours represent blue galaxies in groups. Left panel: Radial profile of the mass surface density. Dashed lines represent the median value in the radius bin. Colour shade represent the error of the median. Single points represent bins where only one region was left after the S/N cleaning. Right panel: internal gradients of the mass surface density as a function of the total stellar mass. Circles represent the values of each galaxy. Stars represent the median value for each mass bin.}
    \label{fig:SFR_Profile_gradient}
\end{figure}

The profiles of the intensity of the \gls{SFR}, $\Sigma_\mathrm{SFR}$, (see Fig.~\ref{fig:SFR_Profile_gradient}) show a clear bi-modality, with red and blue galaxies clearly divided. We find that the intensity of the \gls{SFR} decreases with the distance to the centre of the galaxy for all the types of galaxies. Red galaxies in groups and in the field overlap for the greatest part of the profile. The profile decreases from $\log \Sigma_{\mathrm{SFR}} \approx 0$ in central regions down to $\log \Sigma_{\mathrm{SFR}} \approx -1$ at $\sim 2$~\texttt{R\_EFF}, and it the profile flattens. 

Similarly, the profile of blue galaxies also decreases with distance, and blue galaxies in groups and in the field overlap almost perfectly. The profile decreases from $\log \Sigma_{\mathrm{SFR}} \approx 1.2$ in central regions down to $\log \Sigma_{\mathrm{SFR}} \approx 0$ at $\sim 3$~\texttt{R\_EFF}. Blue galaxies in groups show once more a break in the tendency at $\sim 2.5$~\texttt{R\_EFF}, where the intensity of the \gls{SFR} suddenly starts to increase. As argued before, this is likely caused by a single point with a higher mass. 

If we compare our results with the literature, we find and overall good agreement. Results by \cite{Rosa2016} suggest that these profiles depend more on the morphology of the galaxy than on its mass. In that work, the intensity of the \gls{SFR} of spiral galaxies decreases from $\log \Sigma_{\mathrm{SFR}} \approx 2$ in central regions down to $\log \Sigma_{\mathrm{SFR}} \approx 0.5$ at $\sim 3$~\texttt{R\_EFF}. Meanwhile, elliptical and lenticular galaxies show rather flat profiles around values of $\log \Sigma_{\mathrm{SFR}} \approx 0$. This results are similar to our findings, although there are two main differences: in general, we find lower intensities in blue galaxies (compared to elliptical galaxies), and we do find an initial decrease in the intensity of red galaxies that is only shown in the profiles of the most massive elliptical in the work by \cite{Rosa2016}, starting at $\log \Sigma_{\mathrm{SFR}} \approx 1$, which is a higher intensity that what we find in our results. This difference might be caused by the differences in the radial bin, this is, since our ellipses are larger, they include more pixels with a lower intensity, which lowers the median value. This is particularly relevant taking into account that values found by \cite{Rosa2016} at $\sim 0.7$~\texttt{R\_EFF} are closer to $\log \Sigma_{\mathrm{SFR}} \approx 1.5$ for spiral galaxies, and that the profile is steep: the difference along 1~\texttt{R\_EFF} are of the order of 1~dex. If we compare our results with those found by \cite{Bluck2020}, we find that red galaxies are below their quiescent limit, and that blue galaxies are compatible with the range of values found for star--forming galaxies ($\log \Sigma_{\mathrm{SFR}} \approx 1$ in central regions and $\log \Sigma_{\mathrm{SFR}} \approx 0$ at $1.4$~\texttt{R\_EFF}. Results by \cite{Ana2024} are highly compatible with ours: their sample of spiral galaxies show values from $\log \Sigma_{\mathrm{SFR}} \approx 1$ at $\sim 0.1$~\texttt{R\_EFF} down to $\log \Sigma_{\mathrm{SFR}} \approx 0.5$ at $\sim 2$~\texttt{R\_EFF}, an their sample of elliptical galaxies shows values in the range $\log \Sigma_{\mathrm{SFR}} \approx [0,0.5]$, although they seem flatter.

The gradients of the intensity of the \gls{SFR} (see right panel of Fig.~\ref{fig:SFR_Profile_gradient}) seem to be related to the total mass. They increase with the total stellar mass from $\nabla \log \Sigma_{\mathrm{SFR}} \approx -0.5$~dex/R\_EFF for masses $\log M_\star <9 $~$[M_\odot]$ up to $\nabla \Sigma_{\mathrm{SFR}} \approx -0.25$~dex/R\_EFF for masses $\log M_\star \approx 10$~$[M_\odot]$, and then it decreases more steeply (down to $\nabla \log \Sigma_{\mathrm{SFR}} \approx -1.25$~dex/R\_EFF for red galaxies in the field with  $\log M_\star >11 $~$[M_\odot]$. Here, we find a larger difference in the median gradient of red galaxies in the field and in groups, with flatter profiles for galaxies in the groups. However, taking into account the uncertainties in the gradients and that these galaxies do not show any significant star formation and that they are in the quiescent regime, this result is not clear to interpret. It is also interesting to note that for galaxies with masses $\log M_\star > 10$~$[M_\odot]$ the dispersion in the values of the gradients greatly increases. This can be interpreted as a consequence of galaxies starting to enter in the quiescent regime, so the profile is less significant, since the values of the intensity of the \gls{SFR} only show that there is no significant star formation taking place.

We remark that the values of the gradients found for blue galaxies are compatible with those found by \cite{Bluck2020} for central, star--forming galaxies ($\nabla \log \Sigma_{\mathrm{SFR}} = -0.38$~dex/R\_EFF) as well as the average for all star--forming galaxies ($\nabla \log \Sigma_{\mathrm{SFR}} = -0.44$~dex/R\_EFF). Our results are also compatible with the gradients found for galaxies in the green valley, all for centrals ($\nabla \log \Sigma_{\mathrm{SFR}} \approx -0.24$~dex/R\_EFF) satellites ($\nabla \log \Sigma_{\mathrm{SFR}} \approx -0.43$~dex/R\_EFF) and the average ($\nabla \log \Sigma_{\mathrm{SFR}} \approx -0.28$~dex/R\_EFF), but the gradient found for star--forming satellites ($\nabla \log \Sigma_{\mathrm{SFR}} \approx -0.65$~dex/R\_EFF) is only compatible with our results for galaxies with mass $\log M_\star > 10$~$[M_\odot]$.

\subsubsection{Specific star formation rate}
\begin{figure}
    \centering
    \includegraphics[width=\textwidth]{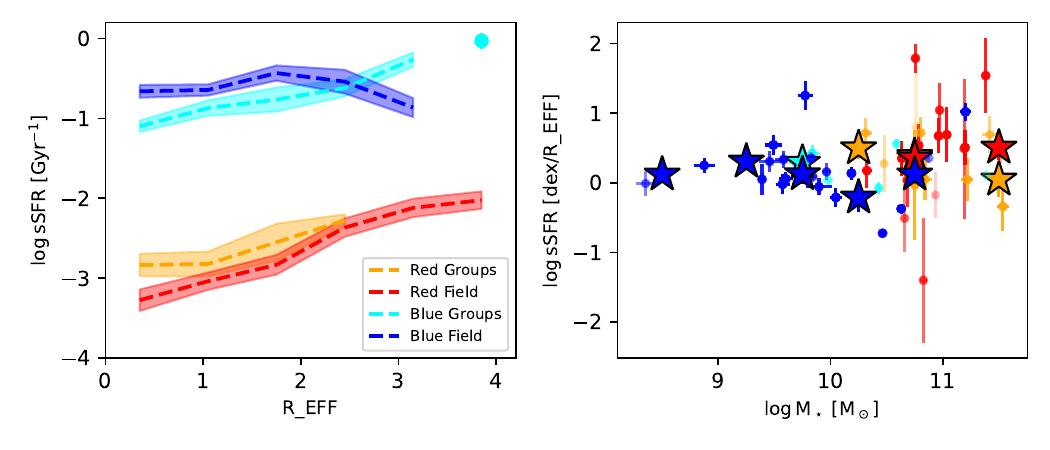}
    \caption[Radial profile and internal gradients of the \gls{sSFR} by galaxy colour and environment.]{Radial profile and internal gradients of the \gls{sSFR} by galaxy colour and environment. Red colours indicate red galaxies in the field. Orange colours represent red galaxies in groups. Blue colours represent blue galaxies in the field. Cyan colours represent blue galaxies in groups. Left panel: Radial profile of the mass surface density. Dashed lines represent the median value in the radius bin. Colour shade represent the error of the median. Single points represent bins where only one region was left after the S/N cleaning. Right panel: internal gradients of the mass surface density as a function of the total stellar mass. Circles represent the values of each galaxy. Stars represent the median value for each mass bin.}
    \label{fig:sSFR_Profile_gradient}
\end{figure}

The radial profiles of the \gls{sSFR} (see Fig.~\ref{fig:sSFR_Profile_gradient}) are analogous to those found for the intensity of the \gls{SFR}. We find a clear bimodality that divides galaxies into red and blue. The \gls{sSFR} slightly increases with the distance to the galactic centre for red galaxies in the groups and in the field. Both types of galaxies overlap within the uncertainties of the median along all the profile. Values increase from $\log \mathrm{sSFR} \approx -3$ in the central regions up to  $\log \mathrm{sSFR} \approx -2$ at 4~\texttt{R\_EFF} for red galaxies in the field and $\log \mathrm{sSFR} \approx -2.5$ at $2.5$~\texttt{R\_EFF} for red galaxies in groups. This is the same profile found by \cite{Rosa2016} for elliptical an lenticular galaxies, over the same range of values. Profiles found by \cite{Abdurro2023} for quiescent galaxies are also very similar to ours, particularly those for masses $\log M_\star >10.5$~$[M_\odot]$. These galaxies remain below the $\log \mathrm{sSFR} = -1$ threshold established by \cite{Peng2010} to segregate galaxies into active and quiescent.  The differences between environments are again negligible for red galaxies. The range of values found by \cite{Ana2024} for their sample of elliptical galaxies is similar to our results, $\log \mathrm{sSFR} = [-3,2]$, although their gradients seems steeper. 

Blue galaxies in the field show rather flat profiles at $\log \mathrm{sSFR} \approx -0.85$, while the profile of blue galaxies in groups increase from $\log \mathrm{sSFR} \approx -1$ in central regions up to $\log \mathrm{sSFR} \approx 0$ at 4~\texttt{R\_EFF}. However, taking into account the effects seen in other properties caused by the single point in the last radial bin, this increase could in reality be much flatter. Blue galaxies both in the field and in groups are active according to the $\log \mathrm{sSFR} = -1$ threshold established by \cite{Peng2010} The \gls{sSFR} of blue galaxies in the field seems to be slightly larger than the \gls{sSFR} of blue galaxies in groups. This could be an indicator of the quenching of blue galaxies in the groups. However, we must consider that due to the limited size of our sample, the statistical significance of this difference needs to be improved, particularly given the small difference shown once we take into account the uncertainty intervals. Also, the mass distribution of the mass blue galaxies in groups and field is not exactly the same (field galaxies show galaxies with a lower mass, see Fig.~\ref{fig:intrelations}), which can also be a consequence of our limited sample. Therefore, more data is required to confirm this hint about the possible effect of the environment. This also holds truth for the differences observed in the ages and the colour profiles. Profiles found for the \gls{sSFR} by \cite{Rosa2016} depend on the morphology of the galaxy, and some spirals are shown to be below the quiescent limit. However, our profiles for the blue galaxies are almost identical to those found by \cite{Rosa2016} for Sc and Sd galaxies. Radial profiles found for star--forming galaxies and galaxies in the green valley by \cite{Abdurro2023} are also compatible with our results, with faintly increasing profiles over a similar range of values. However the sample of spiral galaxies from \cite{Ana2024} shows quenched values ($\log \mathrm{sSFR} < -1$) in the inner regions, but become star forming at $0.7$~\texttt{R\_EFF} for low mass spirals and at $1$~\texttt{R\_EFF} for high mass spirals. Points where their profiles become flatter and compatible with our results.

The \gls{sSFR} gradients are positive for almost all galaxies, but show a similar relation with the mass to the one found for the intensity of the \gls{SFR}, this is, the gradients seem to increase with the total stellar mass from $\nabla \log \mathrm{sSFR} \approx 0$~dex/R\_EFF for masses $\log M_\star <9 $~$[M_\odot]$ up to $\log \mathrm{sSFR} \approx 0.4$~dex/R\_EFF for masses $\log M_\star \approx 10$~$[M_\odot]$, and then it decreases more steeply down to $\nabla \log \mathrm{sSFR} \approx -0.1$~dex/R\_EFF for red galaxies in the field with  $\log M_\star \approx 11 $~$[M_\odot]$. Similarly to the gradients of the intensity of the \gls{SFR}, the dispersion increases notably for masses $\log M_\star > 10.5 $~$[M_\odot]$, where quiescent galaxies begin to dominate and the profile becomes less significant by itself. We mostly find positive gradients, which favour an inside-out quenching scenario.

\subsection{Emission lines}

In this section, we study the main predictions of the ANN for the regions we have obtained with our methodology. We use the homogeneous rings segmentation for the same reasons as in the previous section, this is, in order to prevent higher resolution galaxies from dominating the radial profiles and mass density-colour diagrams. The radial profiles of the predicted equivalent width of H$\alpha$, H$\beta$, $\mathrm{[NII]}$, and $\mathrm{[OIII]}$, as well as the ratios $\mathrm{[NII]/H}\alpha$ and $\mathrm{[OIII]/H}\beta$, which are the ones used to build the BPT diagrams and separate star--forming galaxies and galaxies hosting and \gls{AGN}.

\subsubsection{Line emission of the regions}

\begin{figure}
    \centering
    \includegraphics[width=\textwidth]{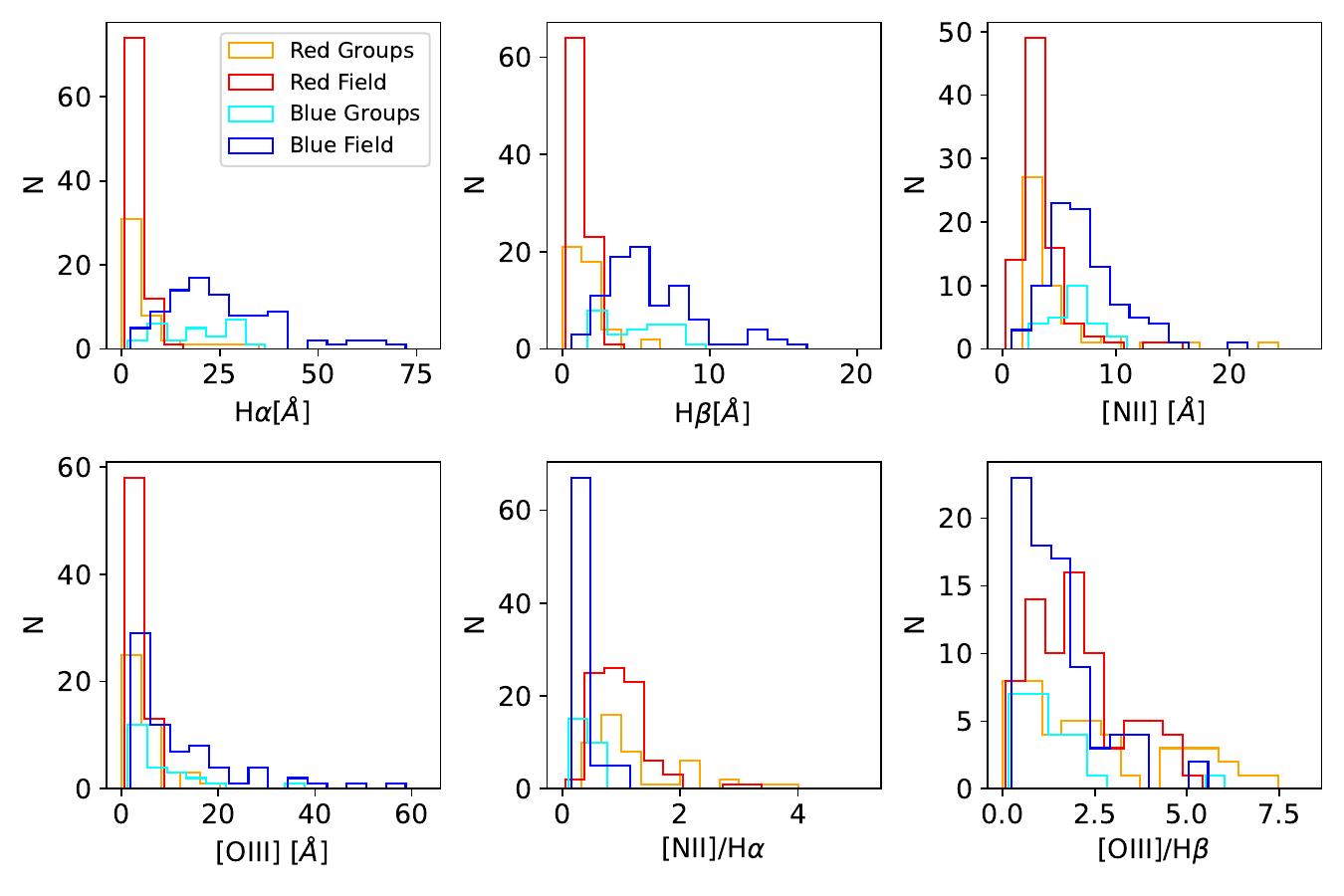}
    \caption[Histograms of the EW and ratios of the line emission of the regions of our sample of galaxies, by colour and environment]{Histograms of the EW and ratios of the line emission of the regions of our sample of galaxies, by colour and environment. From left to right, up to bottom: $\mathrm{EW(H\alpha)}$, $\mathrm{EW(H\beta)}$, $\mathrm{EW([NII])}$, $\mathrm{EW([OIII])}$, $\mathrm{[NII]/H\alpha}$, $\mathrm{[OIII]/H\beta}$. Red histograms represents red galaxies in the field. Orange histogram represents red galaxies in groups. Blue histograms represent blue galaxies in the field. Cyan histograms represent blue galaxies in groups}
    \label{fig:EWhist}
\end{figure}

We start our analysis by studying the distribution of the values of the EW and ratios of the line emission of the regions of the galaxies in our sample, divided by their colour and their environment (see Fig.\ref{fig:EWhist}). We find that the $\mathrm{EW(H\alpha)}$ of red galaxies, in the field and in groups is very low ($<10$~\AA). However, blue galaxies show a larger range of values, and their emission is generally much higher than that of red galaxies: the $\mathrm{EW(H\alpha)}$ of blue galaxies in groups ranges from a few \AA \ up to $\sim 30$~\AA, and the emission of blue galaxies in the field ranges from a few \AA \ up to almost $\sim 75$~\AA, with a peak of the distribution at $\sim 20$~\AA. We find no notable effect of the environment in red galaxies, and the differences in the distributions of blue galaxies can be due to the different size of the samples. This distribution of the properties can be expected when the relation between the mass and the $\mathrm{EW(H\alpha)}$ is taken into account \citep[see e. g.][]{Fumagalli2012, Sobral2014, Khostovan2021}. Additionally, $\mathrm{EW(H\alpha)}$ can be used as a tracer of the \gls{SFR} \citep{Kennicutt1998,Kennicutt2012,Marmolqueralto2016,Khostovan2021}, so it can be expected that the regions of blue galaxies show higher values of $\mathrm{EW(H\alpha)}$, since these galaxies are usually star--forming \citep[see e.g.][]{Peng2010,Bluck2014} and we have already shown with our analysis that their regions indeed have higher intensities of the \gls{SFR}.

The distribution of the $\mathrm{EW(H\beta)}$ shows similar results: red galaxies in groups and in the field do not show strong emission ($\mathrm{EW(H\beta)} \approx [0,3]$~\AA) but blue galaxies do. The values of blue galaxies in groups range from $\mathrm{EW(H\beta)} = 0$~\AA \ up to $\mathrm{EW(H\beta)} \approx 10$~\AA, while blue galaxies in the field show values in the range  $\mathrm{EW(H\beta)} = [0,17]$~\AA, peaking at $\mathrm{EW(H\beta)} = 5$~\AA.

Results for the $\mathrm{EW([NII])}$ are slightly different. Red galaxies still show low values of the emission, but there is a greater number of regions with $\mathrm{EW([NII])} > 0$~\AA. Values of $\mathrm{EW([NII])}$ range from 0~\AA \ up to $\sim 15$~\AA, peaking at $\sim 15$~\AA. We note that, due to the width of the filters, the observed emission of $\mathrm{H\alpha}$ and [NII] always falls in the same filter, which can lead to some  degeneracies in the estimation of the EW of both lines \citep{Gines2021} and some of these values could be compensating small variations in the relative flux of the filters, with a prediction of low $\mathrm{H}\alpha$. Values of $\mathrm{EW([NII])}$ for blue galaxies are generally higher, spanning a range from $\mathrm{EW([NII])} \approx 1$~\AA \ up to $\mathrm{EW([NII])} \approx 10$~\AA \ for blue galaxies in groups and from $\mathrm{EW([NII])} \approx 0$~\AA \ up to $\mathrm{EW([NII])} \approx 15$~\AA \ for blue galaxies in the field, both peaking at  $\mathrm{EW([NII])} \approx 5$~\AA.

The values of the $\mathrm{EW([OIII])}$ are also notably lower for red galaxies ($\mathrm{EW([OIII])} \approx [0,10]$~\AA \ for red galaxies in the field and in groups) than for blue galaxies ($\mathrm{EW([OIII])} \approx [0,20]$~\AA \ for blue galaxies in groups, $\mathrm{EW([OIII])} \approx [0,60]$~\AA \ for blue galaxies in the field).

Results for the $\mathrm{[NII]/H\alpha}$ ratio is higher for red galaxies ($\mathrm{[NII]/H\alpha}\approx [0,4]$ \ for both red galaxies in the field and in groups, peaking at $\mathrm{[NII]/H\alpha}\approx1$ ) than for blue galaxies ($\mathrm{[NII]/H\alpha}\approx [0,1]$ for both galaxies in groups and in the field). Even though these ratios can also be unprecise when the estimation of the EWs are low. However, we note that in the WHAN diagram \citep{WHAN_1,WHAN_2}, star--forming galaxies are found in the areas of lower $\mathrm{[NII]/H\alpha}$, as we find here for blue galaxies.

The distribution of the $\mathrm{[OIII]/H\beta}$ ratio is not very different for blue and red galaxies, may they be in groups or in the field. Most values are within the range $\mathrm{[OIII]/H\beta} \approx [0,5]$, with a slight tail for red galaxies in groups with values of up to $\mathrm{[OIII]/H\beta}=7.5$. This ratio can also suffer from inaccuracies due to almost null estimations of the values of the EW. However, we note that in the BPT diagram \citep{BPT}, galaxies with higher ratios are more likely to be classified as \gls{AGN} hosts.

\begin{figure}
    \centering
    \includegraphics[width=\textwidth]{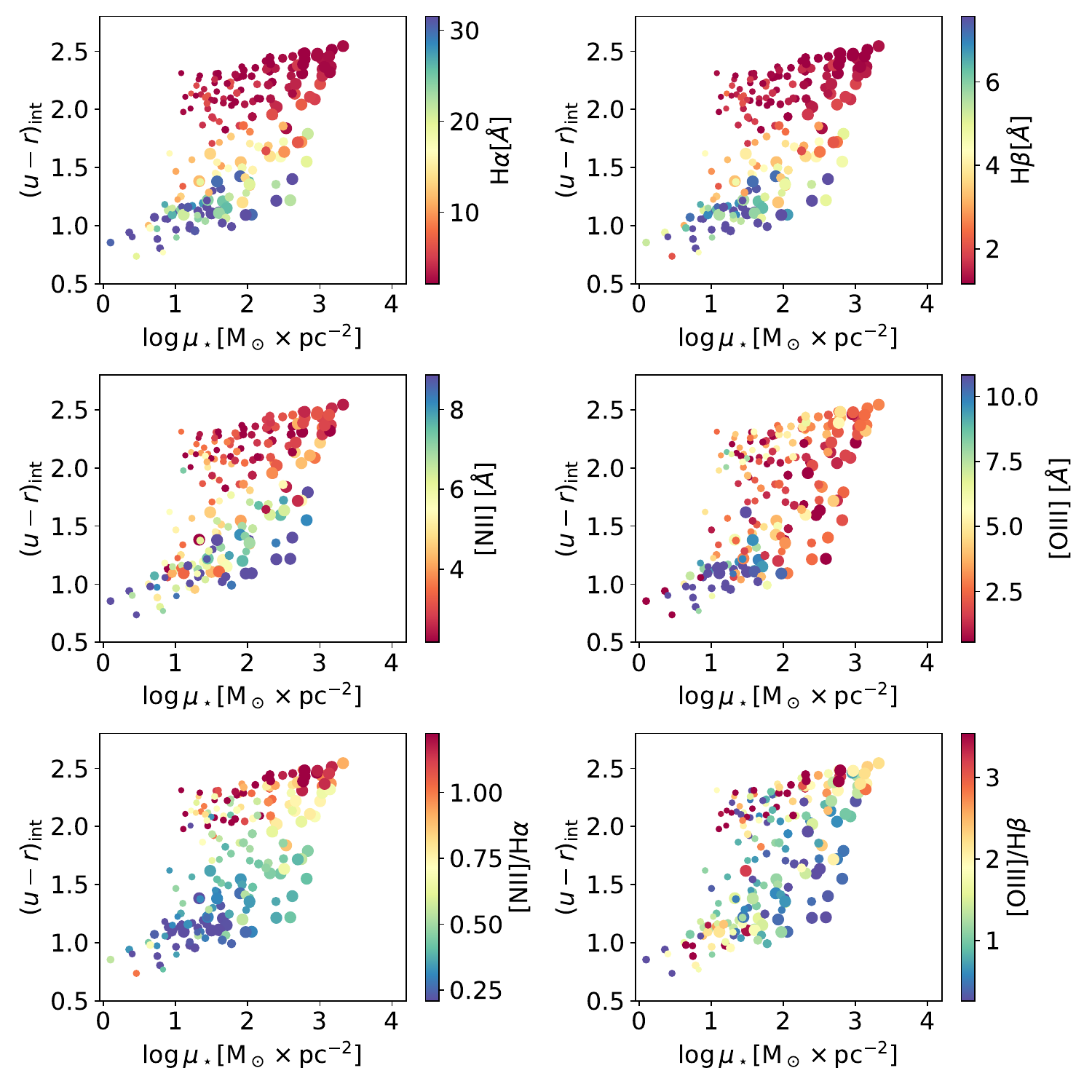}
    \caption[Colour--mass density diagram coloured by the main stellar population properties.]{Colour--mass density diagram coloured by the main stellar population properties.  From left to right, up to bottom: $\mathrm{EW(H\alpha)}$, $\mathrm{EW(H\beta)}$, $\mathrm{EW([NII])}$, $\mathrm{EW([OIII])}$, $\mathrm{[NII]/H\alpha}$, $\mathrm{[OIII]/H\beta}$. Red histograms represents red galaxies in the field. Orange histogram represents red galaxies in groups. Blue histograms represent blue galaxies in the field. Cyan histograms represent blue galaxies in groups}
    \label{fig:colourmassdenEW}
\end{figure}

The mass density-colour diagram (see Fig.~\ref{fig:colourmassdenEW}) offers a clear interpretation, since the equivalent widths and the line ratios show a clear distribution in the diagram. There is a strong correlation with the colour, particularly for the EW of H$\alpha$ and H$\beta$. This can be partly due to the fact that the ANN are trained with colours, but there is also evidence of relation of emission and blue/red galaxies \citep[see e.g.][and references therein]{Gines2022}. However, the correlation with the stellar mass density must not be depreciated.

We find the equivalent width of the emission increases in blue, low stellar mass surface density regions. However, in these regions we also find the largest variability of the values. Using the same arguments about the mass density as proxy of the distance to the centre as before, these regions are in general the outermost parts of blue galaxies. The variability can be expected, since not all galaxies show the same emission, and extreme emission line galaxies exist \citep[see e.g.][and references therein]{Iglesias2022,Gines2022}. On the other hand, as regions become redder, particularly those of higher density, the equivalent widths of the lines decrease notably. These regions are generally the innermost parts of red galaxies, which are known to be generally quiescent and show very low values of the EW. The equivalent width of $\mathrm{[OIII]}$ however, is only noticeable in the lowest density blue regions, and becomes negligible quickly. However, we have already pointed that our estimation of the $\mathrm{[OIII]}$ is the most uncertain one.  

The ratios of the lines present quite the opposite behaviour in comparison to equivalent widths. We find the lowest values of the $\mathrm{[NII]/H}\alpha$ are clearly found on the blue low-mass density , while higher values of the ratio are found in the redder and mass-denser regions. It is also in the redder and mass-denser regions where we find the largest dispersion of values. This makes sense, since we would expect blue regions to be in the star--forming part of the WHAN and BPT diagrams, while red regions should be more likely to be in the LINER or retired parts of these diagrams. In the WHAN diagram, the star--forming regions are those with lower $\mathrm{[NII]/H}\alpha$ ratio, while in the BPT diagram galaxies with a lower $\mathrm{[NII]/H}\alpha$ ratio require a higher $\mathrm{[OIII]/H}\beta$ in order to be classified as \gls{AGN} hosts instead of star--forming regions. Also, denser regions are expected to be the nucleus of the galaxy (where the \gls{AGN} effect should be more prominent), and it is in said regions where we find the largest values of the ratio. A similar argumentation can be made with the BPT diagram and the $\mathrm{[OIII]/H}\beta$ ratio. This ratio shows a larger dispersion in our mass-density-colour regions. On the one hand, this value is predicted more poorly. On the other hand, its relation with red and blue galaxies is more flexible in the BPT diagram, due to its relation with the other ratio.

\subsubsection{WHAN and BPT diagrams}

\begin{figure}
    \centering
    \includegraphics[width=\textwidth]{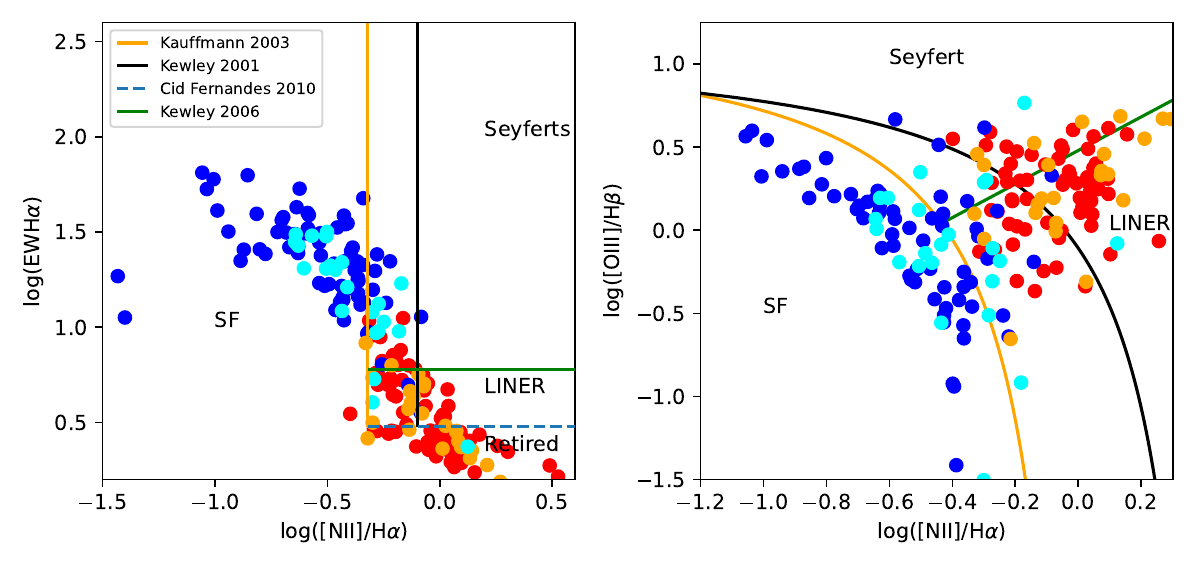}
    \caption[WHAN and BPT diagrams of the regions of the spatially resolved galaxies in \mjp]{WHAN (left panel) and BPT (right panel) diagrams of the regions of the spatially resolved galaxies in \mjp. Red points represent red galaxies in the field. Orange points represent red galaxies in groups. Blue points represent blue galaxies in the field. Cyan points represent blue galaxies in groups. }
    \label{fig:WHANspatially}
\end{figure}

In order to check if these arguments are seen in the actual WHAN and BPT diagrams, we also plot them using our regions obtained with the homogeneous ring segmentation (see Fig.~\ref{fig:WHANspatially}). Here we can confirm our previous speculations. The regions of blue galaxies are generally placed in the star--forming region of blue diagrams, with some of them showing a ``composite'' behaviour in the sense that they are found between the \cite{Kauffmann2003} and the \cite{Kewley2001} lines. Very few of these regions appear in the LINER region of the diagrams, and we find only one point in the WHAN diagram and three in the BPT that might be classified as Seyfert. On the other hand, regions of red galaxies are mainly found in the LiNER and retired regions of the WHAN diagram, while in the BPT diagram these regions are mostly found in the "composite" region, with a higher number of regions classified as Seyferts. Here, we note the poor precision of our estimation of the $\mathrm{[OIII]/H}\beta$ ratio. We can not separate galaxies in groups from galaxies in the field. In summary, we found that our results regarding the emission lines are self consistent.

\subsubsection{Radial profiles of the emission lines}
\begin{figure}
    \centering
    \includegraphics[width=\textwidth]{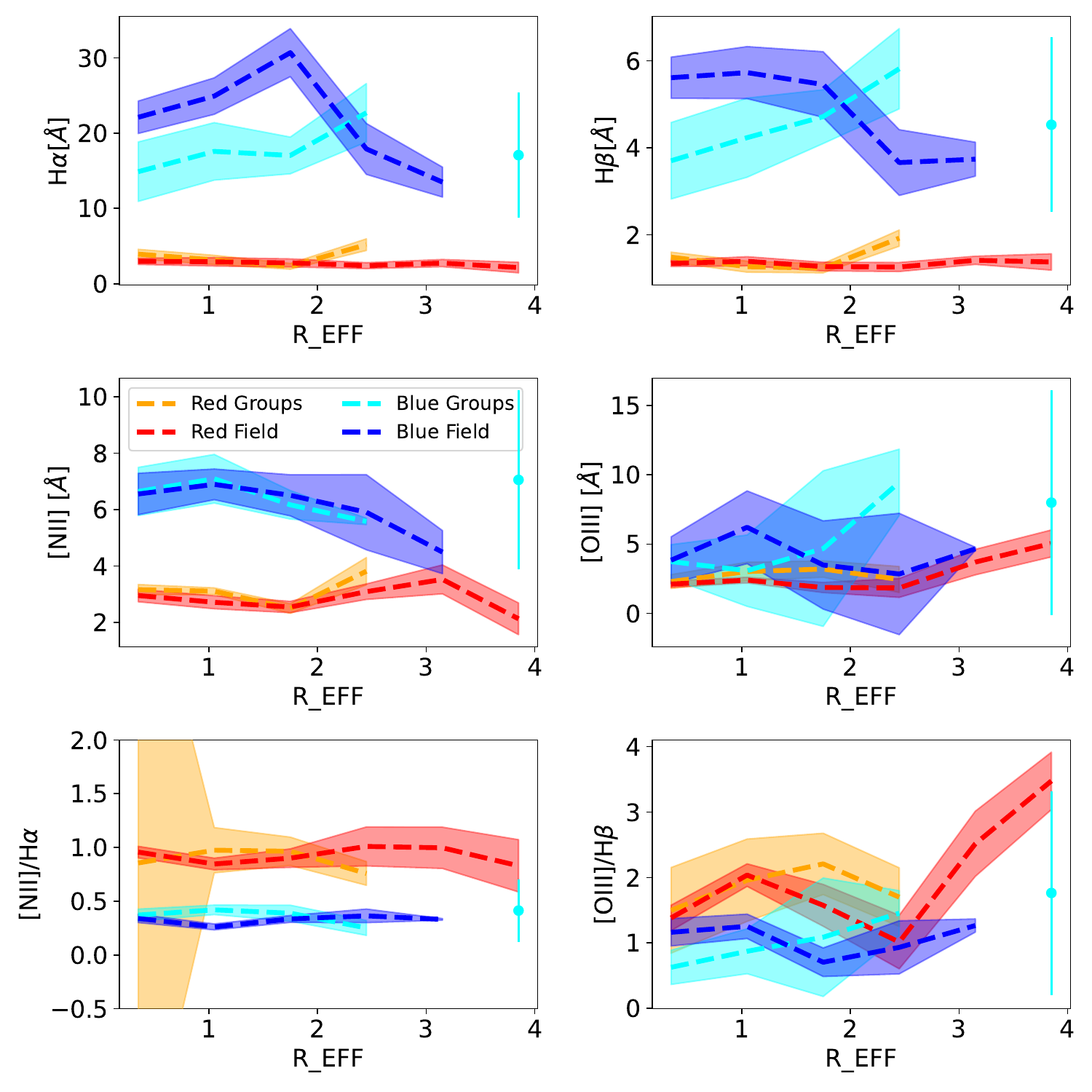}
    \caption[Radial profiles of EW and emission line ratios]{Radial profiles of EW and emission line ratios. From left to right, up to bottom: $\mathrm{EW(H\alpha)}$, $\mathrm{EW(H\beta)}$, $\mathrm{EW([NII])}$, $\mathrm{EW([OIII])}$, $\mathrm{[NII]/H\alpha}$, $\mathrm{[OIII]/H\beta}$. Red colours indicate red galaxies in the field. Orange colours represent red galaxies in groups. Blue colours represent blue galaxies in the field. Cyan colours represent blue galaxies in groups. Left panel: Radial profile of the mass surface density. Dashed lines represent the median value in the radius bin. Colour shade represent the error of the median. Single points represent bins where only one region was left after the S/N cleaning}
    \label{fig:medianEW}
\end{figure}

The radial profiles of the equivalent widths and ratios (see Fig.~\ref{fig:medianEW}) are hard to interpret. We find a similar behaviour of the profiles for the equivalent widths of H$\alpha$, H$\beta$, and $\mathrm{[NII]}$. Red galaxies, both in groups and in the field, show flat profiles compatible with the absence of emission. On the other hand, blue galaxies profiles show emission at most distances to the centre. This is accordance with the known integrated properties of red and blue galaxies, this is, red galaxy are usually quiescent and show no intense emission lines, while blue galaxies are usually star--forming and can show strong emission lines, particularly H$\alpha$, which is very related to the star formation \citep{Kennicutt1998,Kennicutt2012,Marmolqueralto2016,Khostovan2021}.  There might be a difference among blue galaxies in the field and blue galaxies in groups, since the profile of blue galaxies in groups appears to be flatter, while the emission seems to grow towards the outer regions of the galaxy for blue galaxies in the field, up to $\sim 2$~$\mathrm{R\_EFF}$, where the emission drops drastically. This drop could be real, due to the galaxy becoming very dim and reaching the limits of the galaxy, or it could be actually be a consequence of the S/N ratio of this outer regions. The growth in emission is steeper for $\mathrm{H}\alpha$, then for $\mathrm{H}\beta$ and is flatter for $\mathrm{[NIII]}$. A possible explanation is that $\mathrm{H}\alpha$ is usually on the most intense emission lines in the optical regime in galaxies (we do indeed find larger values of its equivalent width than for any other emission line). This allows for a larger variation along the galaxy.  

However, the radial profiles of $\mathrm{[OIII]}$ are very hard to interpret. The uncertainty interval of the blue galaxies in the field is really wide, giving a rather flat profile. The emissions of blue galaxies in groups seem to grow steeply up to $\sim 2$~$\mathrm{R\_EFF}$, and then decreases fast. The same doubt as in the previous emission lines arises for this phenomena, as well as the fact that we know that the last point comes from a single galaxy region. The values of red galaxies for this emission line appear to grow slightly toward outer areas, but given the low values of the equivalent widths and the uncertainty intervals, it is not possible to confidently affirm so. We also note that the ANN find it harder to predict the values of $\mathrm{[OIII]}$ when there is little emission \citep{Gines2021}.

The profiles of the ratios of the lines show the opposite behaviour than the emissions lines: they are higher for red galaxies than for blue galaxies. This makes sense if we take into account that both the BPT diagram and the WHAN diagram define their regions for star--forming as where the values of the $\mathrm{[NII]/H}\alpha$ ratio (and $\mathrm{[OIII]/H}\beta$ too in the BPT) are lower. Red and blue galaxies are well differentiated (although the values are more similar for the $\mathrm{[OIII]/H}\beta$ ratio), and it is not possible to see any effect of the environment. The profiles of $\mathrm{[NII]/H}\alpha$ are rather flat, while the profiles of $\mathrm{[OIII]/H}\beta$ seem to grow towards outer regions of the galaxy for red galaxies and blue galaxies in groups, while it seems to decrease for blue galaxies in the field. This seems to be consequence of the EW of $\mathrm{[OIII]}$, so it may be unreal.

\subsection{SFH}\label{sec:SRSFH}

\begin{figure}
    \centering
    \includegraphics[width=\textwidth]{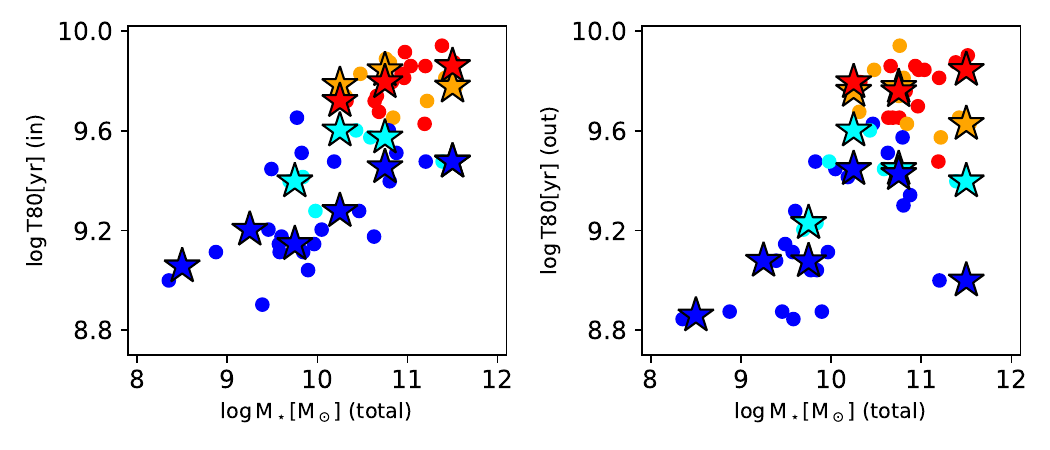}
    \caption[$\mathrm{T80}$ vs galaxy total mass]{$\mathrm{T80}$ vs galaxy total mass. Left panel shows values for the inner region. Right panel shows the values for the outer region. Red points represent red galaxies in the field. Orange points represent red galaxies in groups. Blue points represent blue galaxies in the field. Cyan points represent blue galaxies in groups. Stars represent the median value for each type of galaxy in each stellar mass bin.}
    \label{fig:T80}
\end{figure}

\begin{figure}
    \centering
    \includegraphics[width=\textwidth]{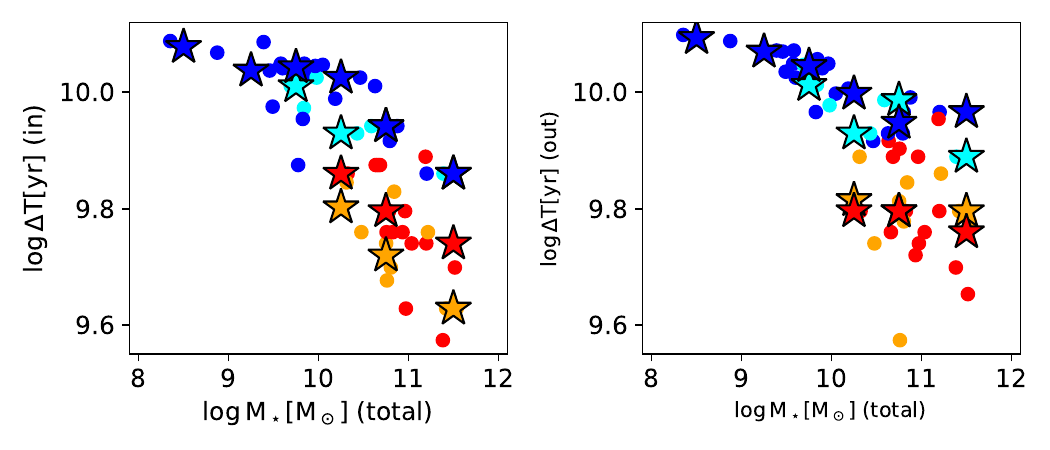}
    \caption[$\Delta \mathrm{T}$ vs galaxy total mass.]{$\Delta \mathrm{T}$ vs galaxy total mass. Left panel shows values for the inner region. Right panel shows the values for the outer region. Red points represent red galaxies in the field. Orange points represent red galaxies in groups. Blue points represent blue galaxies in the field. Cyan points represent blue galaxies in groups.  Stars represent the median value for each type of galaxy in each stellar mass bin.}
    \label{fig:deltaT}
\end{figure}

We finish our section of results by studying the SFH of the galaxies. For this section, we shall use the inside-out segmentation, in order to compare the SFH of the inner and outer regions. We study two parameters: $\mathrm{T80}$, which is the loockback time at which the galaxy has formed 80~\% of its stellar mass (taking into account the mass loss due to stars reaching the end of their lifetime); and $\Delta \mathrm{T}$ which is how long it took for the galaxy to form 80~\% of its total mass. This second parameter takes into account when the galaxy did start forming stars. We show this parameters as a function of the total stellar mass, colour coded by the colour of the galaxy and its environment, in Figs.~\ref{fig:T80} and \ref{fig:deltaT}, respectively.

We find that $\mathrm{T80}$ has a strong correlation with the mass. Red, more massive galaxies show higher values of these parameter, both inside and outside the galaxy, which indicates that these galaxies formed their stars in earlier cosmological epochs. The values are similar for both inner and outer regions. This means that both regions were formed at similar times, given the precision of our stellar model base. On the other hand, blue galaxies  show lower values, which would indicate a formation time closer to the present. For these galaxies, we find a larger dispersion in the values of the inner regions, which also show values generally larger. The interpretation of this result would suggest that blue galaxies have formed their inner parts, possibly bulges, earlier than their outer parts, this is, and inside-out growth model, as observed in many works \citep[see e.g.][]{MunozMateos2007,Perez2013,SanchezBlazquez2014,IbarraMedel2016,Zheng2017}. 

The $\Delta \mathrm{T}$ parameter also shows a strong correlation with the mass (see Fig.~\ref{fig:deltaT}) in the inner and outer parts of the galaxy. Red galaxies show smaller values, indicating a faster formation process, both in inner and outer regions. It is known that red galaxies are required to follow a faster evolutionary track in order to be able to become red, according to the current galaxy formation and evolution models \citep[see e.g.][]{Bell2004,Faber2007,Muzzin2013}. On the other hand blue galaxies shows longer formation times, which are consistent with these galaxies still showing blue stars that contribute to their colour. The formation time of both regions is generally similar, although inner parts seem to show a larger dispersion towards smaller values. This would indicate, as the previous case, that inner parts formed slightly faster than outer parts.

Our results are, in summary, consistent with those of the literature concerning the formation and evolution of red and blue galaxies and with an inside-out formation model. However, we find no significant difference between galaxies in groups and in the field. The environment could have been expected to have an impact in the formation and evolution time of the galaxy through processes such as ram pressure stripping, \citep{Gunn1972}, tidal stripping \citep{Malumuth1984}, or harassment \citep{Moore1996}. Nonetheless, the size of our sample is limited and a larger amount of data might show some differences with a greater statistical significance.

\section{Discussion}

\begin{figure}
    \centering
    \includegraphics[width=\textwidth]{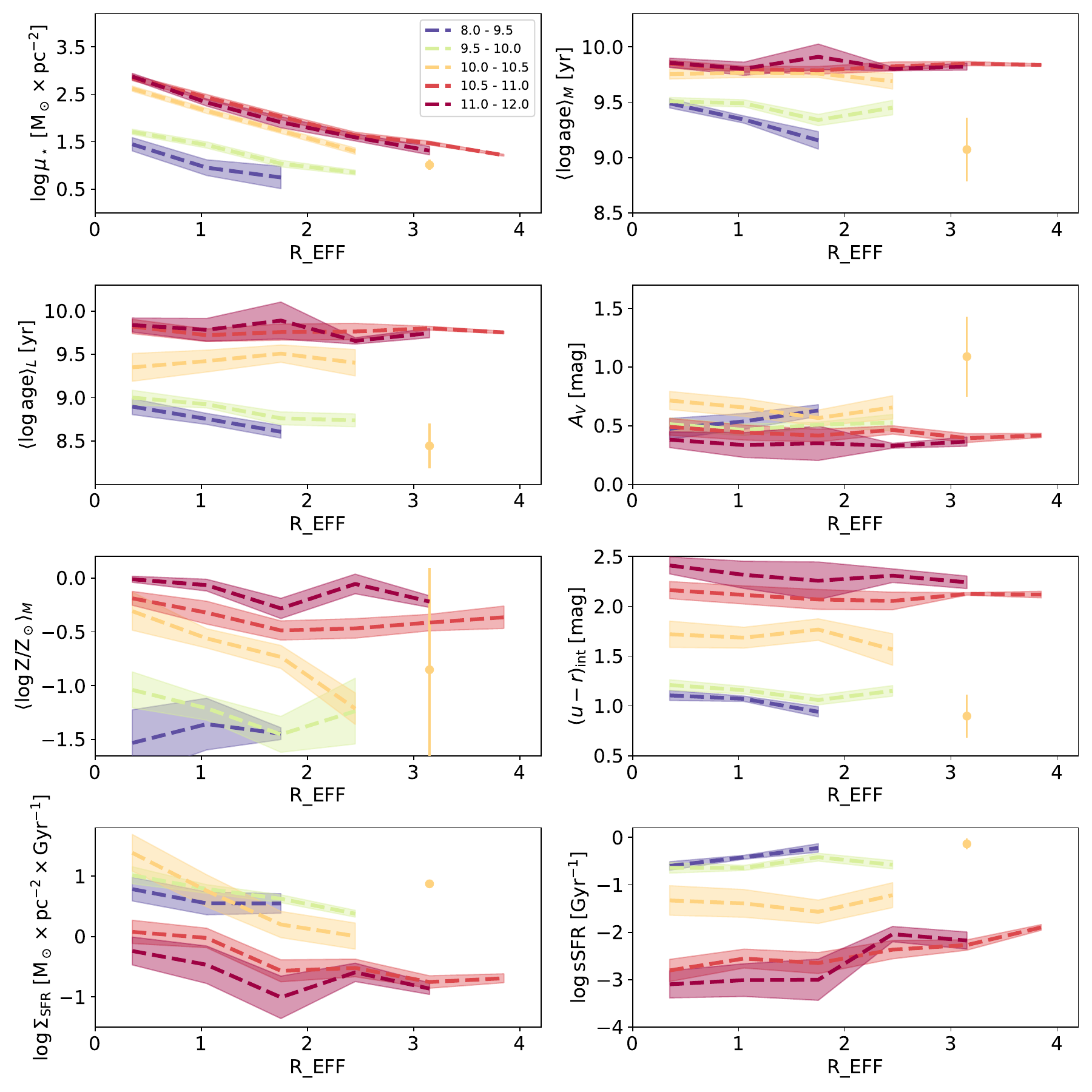}
    \caption[Radial profiles of the stellar population properties, divided by mass bins.]{Radial profiles of the stellar population properties, divided by mass bins. From left to right, up to bottom: stellar mass surface density, mass-weighted age, luminosity-weighted age, extinction, stellar metallicity, $(u-r)_\mathrm{int}$ colour, intensity of the star formation rate and \gls{sSFR}. Different colours represent mass bins. Bluer colours represent lower mass bins and redder colours represent higher mass bins. The mass bins used are $M_\star = [10^{8},10^{9.5}]$~$M_\odot$, $[10^{9.5},10^{10}]$~$M_\odot$, $[10^{10},10^{10.5}]$~$M_\odot$, $[10^{10.5},10^{11}]$~$M_\odot$ and $[10^{11},10^{12}]$~$M_\odot$. Dashed lines represent the median value in each mass bin, shades represent the error of the median and single points represent radius bins with only one region. }
    \label{fig:radialprofmass}
\end{figure}

Throughout this chapter, we have found radial profiles and gradients of the stellar population properties that are in agreement with the literature, particularly (but not restricted to) the works by \cite{Rosa2014,Rosa2015,Rosa2017,SanRoman2018,Bluck2020,Parikh2021} and \cite{Abdurro2023}, and we are somehow reproducing the integrated properties of red and blue galaxies (red galaxies are more massive, redder, metal richer,  and have a lower intensity of the \gls{SFR} and \gls{sSFR}) in an smaller scale. The profiles show clear differences between red and blue galaxies, but no significant difference between galaxies in the field and in groups. We note here that we are limited by several factors. The first one and probably the most important one, our sample of galaxies is quite small. The results themselves for each galaxy are trustworthy because of our selection and because of the proofs performed in Sect.~\ref{sec:SEDcheck}, but their statistical significance will be improved once more data are available. Another factor is that our selection of galaxies is limited to groups, which usually have lower masses than clusters. This is important because the efficiency of several environment related process increases with the mass of group or cluster \citep{Alonso2012, Raj2019}. Related to this point, in \cite{Rosa2022} we found differences between the integrated properties of galaxies in groups and in the field, but many of them due to the fraction of red and blue galaxies, and the properties of galaxies with the same mass and colour also showed similar properties, regardless of their environment, similarly to the results found here. Lastly, the resolution of the binning of the galaxy affects the profile: a worse spatial resolution provides a flatter profile. Therefore, we may be able to provide profiles with better resolution showing more significant differences in the future, when data of galaxies with larger apparent sizes is available.

In general, we find that, because of the mass, red and blue galaxies are mostly well separated. However, similarly to the case of the radial profiles, we can't see any differences between galaxies in groups and field. Regarding the results related to the gradients of the stellar population properties, the two galaxies in our sample with mass below $10^9$~$M_\odot$ are separated from the rest and, taking them as particular cases, we can find some tendencies in the gradients with the mass.


Concerning the role of environment on galaxy evolution, it is common to compare it to the role of the stellar mass, as reflected by the division of the quenching process into mass quenching and environmental quenching \citep[see e.g.][]{Peng2010,Ilbert2013}. In fact, the radial profiles of the stellar population properties studied by \cite{Rosa2015} show a dependence not only on the morphological type of the galaxy, but also on its total stellar mass, as well as the galaxies studied by \cite{Ana2024}. Furthermore, the work by \cite{Zibetti2022} points to the stellar mass both as local and global driver of the evolution of galaxies. For this reason we decide to include the radial profiles of the stellar population properties previously studied, but divided in mass bins (see Fig.~\ref{fig:radialprofmass}). 

We find that, for most properties, their values are well differentiated by mass at all distances. In particular, the stellar mass surface density is always higher for massive galaxies than for low mass galaxies, and massive galaxies are also older, more metal rich, redder at all distances than low mass galaxies. Additionally, the profiles of the \gls{sSFR} also show clear offsets with mass, where low mass galaxies have significantly higher \gls{sSFR} than massive galaxies. The most notable exception is the extinction, which shows similar profiles for all masses. On the other hand, the radial profiles of the intensity of the \gls{SFR} are almost bimodal, with galaxies in the range $M_\star = [10^{8},10^{10.5}]$~$M_\odot$ showing compatible profiles within the uncertainty intervals, but still clearly differentiated from galaxies with masses larger than $10^{10.5}$~$M_\odot$. These results show that the mass does indeed play a very relevant role in the determination of the local properties of galaxies, as found in the aforementioned works.

However, we also point that the mass of our groups may be too low for the environment to produce any significant effect in our spatially resolved galaxies. Indeed, the mass of the groups in \mjp \ is very low in comparison to massive clusters, as shown by \cite{Rosa2022}. In that work, we showed how the quenched fraction excess of galaxies was significantly higher for the cluster mJPC2470-1771 than for low mass groups. Therefore, we may see more significant effects in the properties of the spatially resolved galaxies in groups and clusters with higher mass in the future data releases of \mjp.

\section{Summary and conclusions}
We have applied our tool, \PyDJ, to the spatially resolved galaxies from the \mjp \ survey in order to study the effect of the environment in the properties of the spatially resolved galaxies. We have selected a total of 51 galaxies, from which 15 are red and in the field, 9 are red and are in a groups, 21 are blue and in the field, and 6 are blue in a group. We have checked the observational and the integrated properties  of the sample, checking that red galaxies in the field and in groups are comparable between them, as well as blue galaxies in the field and in groups, but we must take into account that the sample of blue galaxies in the groups is biased towards higher masses. We have also check that this sample of galaxies verifies well known integrated relations, such as the mass--age and mass--metallicity relations and the star-forming main sequence. 

In the spatially resolved analysis, we have mainly used elliptical rings of semi-major axis the size of the FWHM of the worst PSF for each galaxy, and elliptical rings of $0.7$~\texttt{R\_EFF}. We have checked the residuals of our SED-fitting, finding that they are below the $5$~\% relative error in flux, with no significant bias for the filters, as long as the S/N ratio is higher than 5. We also find that errors are generally well estimated in this regime, or slightly overestimated. After studying the relation of the S/N and the surface brightness of the regions, we decide to remove from our analysis those regions with a median $\mathrm{S/N}<5$ in the filters with $\lambda_{pivot} < 5000$~\AA. We have then studied the properties of these regions (first, stellar population properties, then, line emission), using a mass density--colour diagram and their radial profiles and gradients. Lastly, we compare the \gls{SFH} inside-out of galaxies, using elliptical rings defined from the centre up to $0.7$~\texttt{R\_EFF} for inner regions, and from $0.7$~\texttt{R\_EFF} to $2.5$~\texttt{R\_EFF}. Our main conclusions are:

\begin{itemize}
    \item Our tool, \PyDJ, provides solid magnitude measurements that offer reliable galaxy properties.
    \item The  properties of the regions are distributed clearly in the mass density-colour diagrams, similarly to the integrated mass-colour diagram. We find that redder, denser regions are usually older, more metal rich, and show lower values of the $\Sigma_\mathrm{SFR}$ and \gls{sSFR} (they are more quiescent) than bluer, less dense regions. The highest value of the extinction $A_V$ are found in blue, dense regions, as well as some of the most metal rich regions. The regions of red and blue galaxies remain clearly separated in this diagrams.
    \item We are able to reproduce the local star formation main sequence found in other works, such as \cite{Rosa2016}, which implies a thight relation between the $\Sigma_\mathrm{SFR}$ or \gls{sSFR} and the mass surface density. 
    \item Radial profiles of the stellar population properties. The radial profiles of the properties of the galaxies that we obtain are compatible with an extensive literature, such as \cite{Rosa2014,Rosa2015,Rosa2016,Bluck2020,Abdurro2023,Ana2024}. The profiles of the red and blue galaxies are clearly different, but we do not find any remarkable effect of the environment. The gradients of these properties do, in fact, depended more on the total stellar mass of the galaxy.    
    \item Emission lines. We find that the EW of the emission lines predicted using the ANN from \cite{Gines2021} also are clearly distributed in the mass density--colour diagrams, with the highest EW generally found in the lowest density, bluest regions. The ration of $\mathrm{[NII]/H\alpha}$ is also clearly distributed in this diagram, with the highest ratio found in the reddest and densest regions. radial profiles hard to determine. Diagrams are useful. The distribution of the regions in the WHAN and BPT diagrams show that regions of blue galaxies are generally star-forming, and regions of red galaxies are generally classified as LINER or retired. The radial profiles of $\mathrm{H}\alpha$ and $\mathrm{H}\beta$ seem to increase towards outer regions of blue galaxies, but are flat for red galaxies. The profiles of [NII] and [OIII], as well as the profiles of the ratios, are mostly flat for all galaxies.
    \item The comparison of the \gls{SFH} of the inner and outer regions suggests and inside-outside formation scenario.
    \item We find that in general, the properties of the regions of red and blue galaxies well distinguished, but that there is no significant effect of the environment in the properties of these regions. 
\end{itemize}

\chapter{The galaxy populations of mJPC2470--1771, the largest cluster in \mjp} \label{chapter:cluster}

Chapter based on the article published in Astronomy \& Astrophysics by J. E. Rodríguez-Martín et al. 2022, volume 666, id.A160, 24 pp. DOI: 10.1051/0004-6361/202243245

\begin{abstract}
The Javalambre-Physics of the Accelerating Universe Astrophysical Survey (J-PAS) is a photometric survey that is poised to scan several thousands of square degrees of the sky. It will use 54 narrow-band filters, combining the benefits of low-resolution spectra and photometry. Its offshoot, \mjp, \ is a 1 deg$^2$ survey that uses \jp \ filter system with the Pathfinder camera. In this work, we study mJPC2470-1771, the most massive cluster detected in \mjp. We survey the stellar population properties of the members, their star formation rates (SFR), star formation histories (SFH), the emission line galaxy (ELG) population, spatial distribution of these properties, and the ensuing effects of the environment. This work shows the power of  \jp \ to study the role of environment in galaxy evolution.  
 We used a spectral energy distribution (SED) fitting code to derive the stellar population properties of the galaxy members: stellar mass, extinction, metallicity, $(u-r)_{\mathrm{res}}$ and $(u-r)_{\mathrm{int}}$ colours, mass-weighted age, the SFH that is parametrised by a delayed-$\tau$ model ($\tau$, $t_0$), and SFRs. We used artificial neural networks for the identification of the ELG population via the detection of the H$\alpha$, [NII], H$\beta$, and [OIII] nebular emission. We used the Ew(H$\alpha$)-[NII] (WHAN) and [OIII]/H$\beta$-[NII]/H$\alpha$ (BPT) diagrams to separate them into individual star-forming galaxies and AGNs. 
We find that the fraction of red galaxies increases with the cluster-centric radius; and at $0.5$~R$_{200}$ the red and blue fractions are both equal. The redder, more metallic, and more massive galaxies tend to be inside the central part of the cluster, 
whereas blue, less metallic, and less massive galaxies are mainly located outside of the inner $0.5$~R$_{200}$. We selected 49 ELG, with $65.3$~\% of the them likely to be star-forming galaxies, dominated by blue galaxies, and $24$~\% likely to have an AGN (Seyfert or LINER galaxies). The rest are difficult to classify and are most likely composite galaxies. 
These latter galaxies are red, and their abundance decreases with the cluster-centric radius; in contrast, the fraction of star-forming galaxies increases outwards up to $R_{200}$. 
Our results are compatible with an scenario in which galaxy  members  were formed roughly at the same epoch, but blue galaxies  have had more recent star formation episodes, and they are quenching out from  within the cluster centre. The spatial distribution of red galaxies and their properties suggest that they were quenched prior to the cluster accretion or an earlier cluster accretion epoch. AGN feedback or mass might also stand as an obstacle in the quenching of these galaxies.
\end{abstract}



\section{Introduction}
\label{sec:Introduction}
Galaxies in clusters interact with each other as well as with the intracluster medium through processes such as ram-pressure stripping \citep{Gunn1972}, tidal stripping \citep{Malumuth1984}, or harassment \citep{Moore1996}. These processes affect the galaxies' star formation and evolutionary processes and can lead to a greater presence of massive galaxies in dense environments and lower star formation rates (SFRs) in such regions \citep[e.g.][]{Lewis2002, Gomez2003, Baldry2006}. Galaxies in clusters are therefore a great laboratory for studying the role of environment in galaxy evolution.

Certainly, interactions within galaxy clusters play a relevant role in the transformation of galaxies (\citealt{boselli2006}). Since the pioneering work by  \cite{Dressler1980} it is well-known that there is a morphology-density relation that may imply a connection between dense environments and the transformation and evolution of galaxies. This relation shows that as local galaxy density increases, so does the fraction of early-type galaxies, and the fraction of spirals decreases. \citet{Dressler1980} explains this relation as a reflection of the time scale of the formation of the disc of galaxies. This morphology-density relation has been confirmed by many works both at the levels of the nearby universe \citep[e.g.][]{Cappellari2011, fogarty2014} and at higher redshift \citep[e.g.][]{muzzin2012}. This relation could be the result of galaxy-galaxy merging processes and ram-pressure stripping, since these activities can lead to the formation of a spheroidal component, resulting in the morphological transformation of late- to early-type galaxies \citep{Boselli2008, Rijcke2010, Joshi2020, Peschken2020, Janz2021}.

The effect of dense environments can be also seen in the properties of galaxy populations.  Density strongly affects the stellar mass distribution and, at fixed stellar mass, the star formation rate and nuclear activity depend on the density as well, however, the structural parameters are independent of the environment \citep{Kauffmann2004}. \cite{Balogh2004} showed that, at fixed luminosity, the mean $(u-r)$ colour of red and blue galaxies is almost independent of the environment, but the fraction of red galaxies increases with density. These authors propose that the transformations from blue to red must occur very rapidly (in this case, the process is known as `quenching') or at high redshift. Along this line, \cite{Bower1990} results also point out that galaxies in denser environments are older on average, meaning that galaxies in denser environments have had their star formation truncated at earlier epochs, as opposed to galaxies in less dense environments. This age-density relation is also seen in red sequence galaxies in the work by \cite{Cooper2010MNRAS}, showing also a weak correlation between metal-rich galaxies and denser environments. This age-density relation is supported by several other works \citep[e.g. ][]{Trager2000, Thomas2005,Clemens2006, Bernardi2006, Smith2008}

Clusters, in particular those formed at more recent times, are also dynamically in-mature structures that have doubled their mass since $z\sim 0.5$ \citep{boylan2009, gao2012}.  The accretion times for $z = 0$ cluster members are quite extended, with $\sim 20$~\% of them incorporated into the cluster halo more than 7~Gyr ago and  $\sim 20$~\% within the last 2 Gyr  (\citealt{berrier2009}). Thus, the galaxy populations in clusters have evolved rapidly since $z\sim$0.5, with the accretion of star-forming galaxies into the cluster and their transformation into early-type red galaxies.

The so-called Butcher-Oemler effect \citep{ButcherOemler1978, ButcherOemler1984} also reflects the evolutionary nature of clusters. It shows that the fraction of blue galaxies is larger for clusters at higher redshift  (e.g. \citealt{balogh2000,Ellingson2001,diaferio2001}). Moreover, these studies  have found that blue galaxies are mostly located outside the cluster cores and that the effect is not significant for distances larger than $0.5$~R$_{200}$. In fact, passive galaxies are mainly located in the virialised regions while the emission line galaxies are more common in the outskirts of the clusters \citep{Haines2012, Haines2015, Noble2013, Noble2016, Mercurio2021}. In contrast,  the Faint Infrared Grism Survey \citep[FIGS, ][]{FIGS} revealed that [OIII] emitters are more common close to groups, but there is no evidence of a relation between SFR and local galaxy density \citep{Pharo2020}.

Quenching is an important effect related not only to the environment, but also to the galaxy mass. \cite{Peng2010} separated the effects of the mass and the environment in halting star formation and they considered quenching as a combination of both effects (mass-quenching and environment-quenching). Other works have also shown that certain processes that are in some way related to the galaxy stellar mass can suppress the star formation in galaxies independently of the environment \citep[see e.g.][]{ Peng2012, ArcilaOsejo2019, Contini2020, Guo2021}. In fact, AGN feedback can play a relevant role by heating the infalling gas, thus preventing further star formation in the galaxy  \citep{dimatteo2005, Fabian2012, Mcnamara2012}, although this approach remains open for debate \citep{Esposito2022, Wang2022}. The central velocity dispersion is  correlated with the mass of the central black hole of galaxies, so it is  connected to the AGN feedback and it has been shown to play a crucial role in quenching \citep[see e.g.][]{Bluck2020, Brownson2022}.

Some environmental processes can temporarily enhance the star formation due to the inflow of gas toward the central part (e.g. galaxy-galaxy interactions) or the compression of the gas \citep[e.g. ram pressure stripping][]{Joseph1985, Park2009, Ellison2013, RuizLara2020, Boselli2021, Mazzi2021, Lizee2021}. Nevertheless, these environment mechanisms eventually shut down the process of star formation by heating or removing the gas from galaxies  \citep{Altalo2015, Davies2015,Lisenfeld2017,Joshi2019}.

Alternatively, the halo mass is proposed as the main property that is causally linked to the rapid shut down of the star formation  \citep[see e.g.][]{Bluck2014, Woo2015, MonteroDorta2021} because the fraction of quenched galaxies is more correlated with the group or cluster halo mass  at a fixed M$_\star$ than with M$_\star$ at a fixed halo mass \citep{woo2013}. A bimodality exists in the specific star formation rates (sSFR, i.e. the ratio of the SFR to the stellar mass) of satellite galaxies (those falling into denser haloes) and they are more likely to be quenched than field galaxies \citep{Wetzel2012}. In addition, mass and environment quenching could be connected to massive halos that heat the cold-accreted gas, thereby preventing further star formation \citep{dekel2006}.  Furthermore, the quenching of satellites is correlated not only to halo-mass, but it is also anticorrelated with regard to the group and cluster radial distance  \citep{woo2013, Woo2017}. Nonetheless, the IllustrisTNG simulations show that the dependence of the quenched galaxy fraction with the cluster-centric radius is also a function of the mass of halos. Thus,  although the fraction of quenched low-mass satellites in less massive halos  is higher closer to the centre, it is independent of the distance in massive halos \citep{Donnari2021}. However, after several decades of debate, the relevance of the different processes that can lead to the quenching of star formation and the transformation of galaxies in clusters is not yet clear. 

Photometric surveys with broad band filters, such as the Local Cluster Substructure Survey \citep[LoCuSS, ][]{Haines2015}, the Advance Large Homogeneous Area Medium Band Redshift Astronomical \citep[ALHAMBRA][]{Ascaso2015}, the VIMOS Public Extragalactic Redshift Survey \citep[VIPERS][]{Haines2017},  the Subaru Strategic Program with the Hyper Suprime-Cam \citep[HSC-SSP, ][]{Lin2017, Jian2018},  and the Gemini Observations of Galaxies in Rich Environments \citep[GOGREEN][]{McNab2021}, have been essential in the detection and study of galaxy clusters. Surveys with narrow band filters at specific wavelengths aimed at selecting emission lines \citep{Lin2017, Koyama2018, Hayashi2020} and identifying infalling galaxies in the outskirt of clusters \citep{kodama2004} have been very useful for identifying the galaxy populations in clusters. However, these surveys could suffer from several problems related to contamination from: 1) field interlopers due to the lack of precise redshifts for the galaxy cluster membership; 2) dusty star-forming galaxies in the cluster red galaxy population due to the non-correction of colours by extinction; 3) AGNs in the star-forming galaxy population. Furthermore, the small FoV of some instruments, \citep[e.g. with Tunable Filters, as in][]{SanchezPortal2015, RodriguezdelPino2017} hinders the observation of the whole cluster or beyond the cluster centre. 

The Javalambre-Physics of the Accelerating Universe Astrophysical Survey (J-PAS; \citealt{Benitez2009,Benitez2014}) is poised to overcome the problems associated with broad and narrow band photometric surveys. \jp{} will be a very powerful tool in detecting galaxy clusters and providing new clues for the understanding of the role of dense environment in quenching star formation in galaxies.  
\jp{} is a photometric survey that will  scan thousands of square degrees of the sky. With its 54 narrow-band filters (FWHM$\sim 145$~\AA, with a difference of $\sim 100~$\AA\ between the central wavelength of each one), plus two medium and four broadband filters, it will provide data of a scope that is comparable to very-low-resolution spectroscopy ($R\sim 60, \Delta \lambda \sim 100$~\AA). 

\jp{} is ideal for studies focused on the role of environment in galaxy evolution thanks to its capability to detect galaxy clusters and groups \citep[see][]{Rosa2022}. It will be able to provide robust cluster or group detection based on accurate photometric redshifts \citep{HC2021}.The sensitivity of the survey allows us to easily observe the whole galaxy cluster memberships brighter than 22.5 in $r$ band and  to study the quenching as a function of cluster-centric radius. \jp \ is ideal for SED fitting and for identifying and characterise the blue and red galaxy populations \citep{Rosa2021}. Given its spectral coverage and resolution, it is capable of identifying emission line objects and also measuring the lines H$\alpha$, [NII]$\lambda$6584, H$\beta$, and [OIII]$\lambda$5007 in clusters at $z<0.35$ \citep{Gines2021, Gines2022}. These lines are relevant to discriminate between the AGN and star-forming (SF) populations and to study their spatial distribution within the cluster. 

At present, there is  data available using the \jp{} photometric system: miniJPAS \citep{Bonoli2020}. In this chapter, we identify the galaxy populations to study the variation of galaxy properties as a function of the cluster-centric radius in the largest cluster detected in \mjp{}, mJPC2470-1771. The ultimate goal is to demonstrate the capability and the power of \jp \ for investigating the characterisation of galaxy populations in galaxy clusters, as well as the role of environment in quenching the star formation. This will allow us to shed light on the processes responsible for transforming blue and star-forming galaxies into red galaxies in this dense environments.

This chapter is structured as follows.  In Sect.~\ref{sec:Data}, we briefly summarise the  \mjp \ observations and calibrations as well as the selection of the cluster members. In Sect.~\ref{sec:SP}, we describe the methods used to identify and study the stellar population properties of the galaxies and we compare the results obtained with different photometries. In Sect.~\ref{sec:charact}, we present the stellar population properties of these galaxy populations, we divide our ELG into star-forming (SF) galaxies and galaxies with an active galactic nuclei (AGN), and we study the SFR of the cluster galaxies. In Sect.~\ref{sec:discussion}, we discuss our results in terms of their spatial and radial distributions and in Sect.~\ref{sec:conclus}, we summarise our results and present our conclusions.

Throughout this chapter, we assume a Lambda cold dark matter ($\Lambda$CDM) cosmology with $h = 0.674$, $\Omega _{\mathrm{M}} = 0.315$, $\Omega _\Lambda = 0.685$, based on the latest results from \cite{Planck2020}. This is the same cosmology used by \citet{Bonoli2020}. We use the AB magnitude system \cite{Oke1983}. We use the standard notation $M_{\Delta}$ for the mass enclosed within a sphere of radius $R_{\Delta}$, within which the mean overdensity equals $\Delta \times \rho_c(z)$ at a particular redshift $z$; that is, $M_{\Delta} = (4\pi \Delta/3)\rho_c(z)R_{\Delta}^{3}$.

\section{Data: \mjp \ }
\label{sec:Data}

\begin{figure}
\centering
\includegraphics[width=0.75\textwidth]{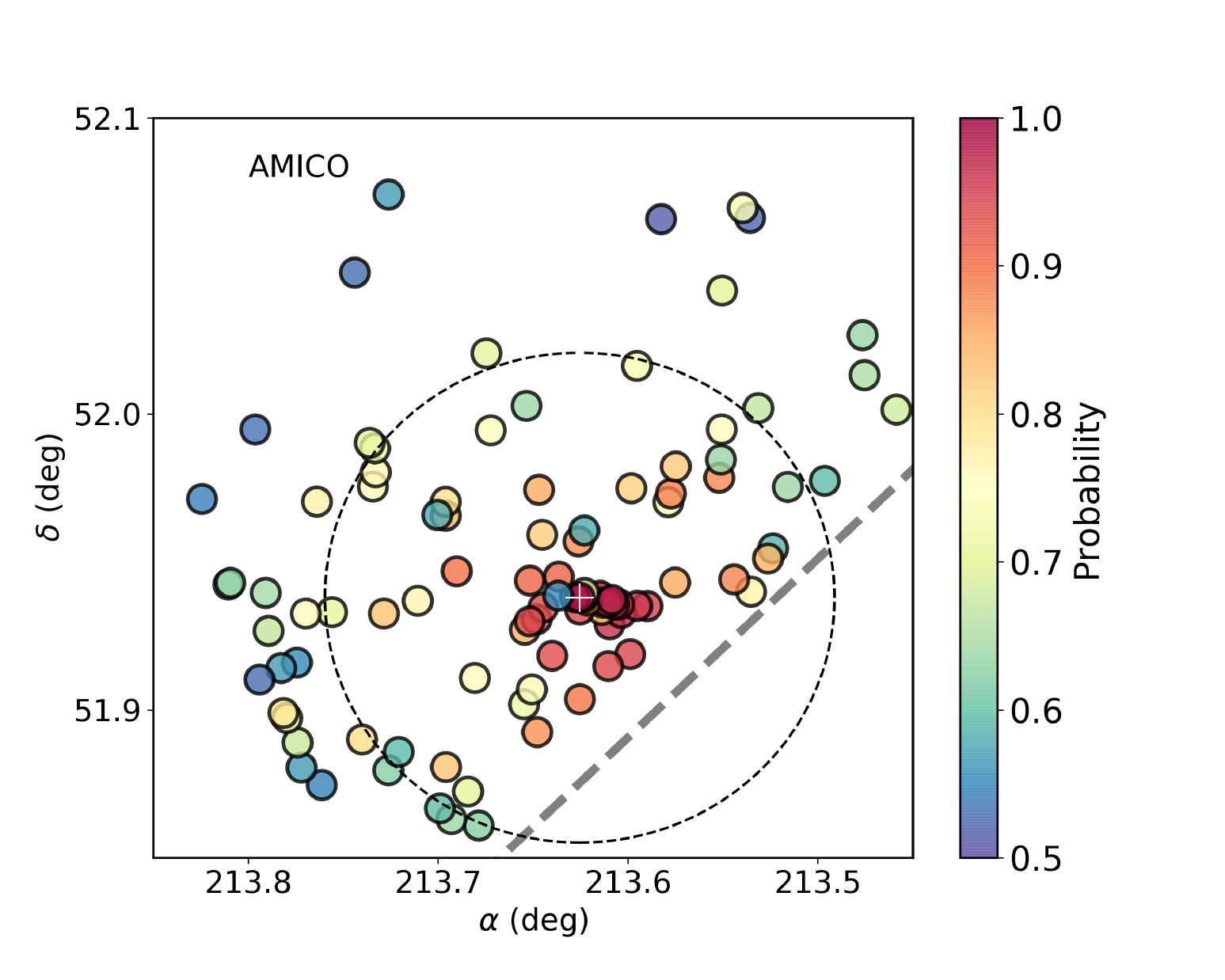}
\caption[Galaxy members of mJPC2470-1771 in the sky plane]{Galaxy members of mJPC2470-1771 in the sky plane. The colour bar indicates the AMICO probability of being a member galaxy. The grey dashed line indicates the edge of the field of view of \mjp. The black dashed circle indicates the value of $R_{200}$. The white cross represents the position of the BCG.
}
\label{fig:mapProb}
\end{figure}

\begin{figure*}
\centering
\includegraphics[width=\textwidth]{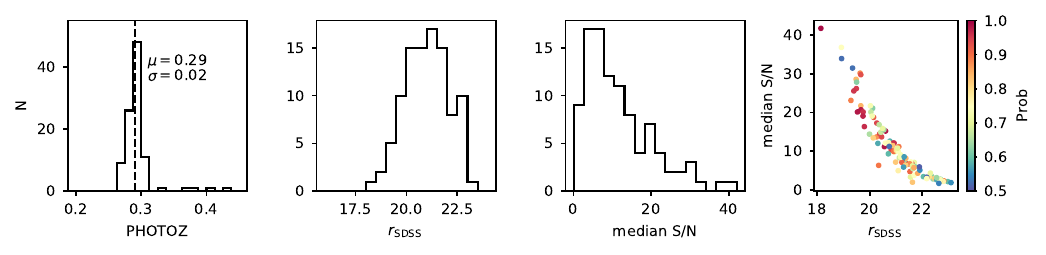}
\caption[Observational properties of the galaxies in the cluster]{Observational properties of the galaxies in the cluster. First panel: Redshift (\photozbest{} ) distribution. The median ($\mu$) and standard deviation ($\sigma$) of \photozbest{} are written in the figure. The black dashed line represents the median of \photozbest{}. Second and third panels: Distributions of the \rb{} magnitude, and median S/N of the narrow-band filters. Fourth panel: Median \magpsfcor \ S/N in the narrow-band filters as a function of \rb{} magnitude. Colour coding indicates the probabilistic association given by AMICO.
}
\label{fig:histobs}
\end{figure*}

\subsection{Observations and calibration}
The \mjp \ survey \citep{Bonoli2020} is a $1~\mathrm{deg}^2$ imaging survey performed at the  Observatorio Astrofísico de Javalambre, \citep[OAJ, ][]{OAJ} using the 2.5m Javalambre Survey Telescope (JST/T250, \citealt{T250}), which provides a good image quality along the optical spectral range (3300--11000 \AA). The instrument used for the data acquisition is the  JPAS-Pathfinder camera. It has a single charge-coupled device (CCD)  with 9.2k$\times$9.2k pixel. The resulting field of view (FoV) is $0.27 ~\mathrm{deg}^2$ and the pixel scale is $0.23"~\mathrm{pixel}^{-1}$. The survey consists of four pointings along the AEGIS stripe \citep{AEGIS2007}. 

One of the greatest strengths of J-PAS  resides in its photometric system. It consists of 54 narrow-band (NB) filters with a full width at half maximum (FWHM)   of 145~\AA \ spaced by 100~\AA, covering the spectral range from 3780~\AA \ to 9100~\AA. There are two broader filters complementing these NB ones: \uja{}, a medium band filter with FWHM of 495~\AA \ and centred at 3497~\AA \ and J1007, a high-pass filter centred at 9316~\AA.  This system provides low-resolution spectra ($R\sim60$), referred to as \js, \ and allows us to detect, identify, and characterise the stellar population properties of galaxies up to $z \sim 1$ \citep{Rosa2021}. The filter system was originally optimised to accurately measure photometric redshifts (photo-$z$) for cosmological studies \citep{Benitez2009, Benitez2014, Bonoli2020}. In addition, four SSDS-like broadband filters are included: \ujp{}, \gb{}, \rb{}, and \ib{}. In particular, \rb{} is used as the reference detection band for the miniJPAS `dual-mode' catalogues. More information about the filter system can be found in \cite{Martin-Franch2012} and \cite{Bonoli2020}.

The area observed in \mjp \ overlaps with the AEGIS field, which is located in the north galactic hemisphere with coordinates: ($\alpha$, $\delta$) = ($215.00^{\circ}$, $+53.00^{\circ}$). It is composed of four pointings covering a total area of 1~deg$^2$. The depth is deeper than 22~mag for filters with $\lambda<7500$~\AA \ and is $\sim 22$~mag for longer wavelengths. The data was processed by the Data Processing and Archiving Unit (UPAD, \citealt{Cristobal-Hornillos2014}) at Centro de Estudios de F\'isica del Cosmos de Arag\'on (CEFCA). Further details on the different processes involved (the processing of single images, the final coadded images, the PSF treatment, the photometry and its calibration, and the masks) can be found in \cite{Bonoli2020}. Nonetheless, the data used in this work  were obtained with SExtractor dual-mode \citep{Bertin1996}. The photometric calibration is an adaptation of the methodology presented in \cite{Lopez-sanJuan2019}. All the images and catalogues are available through the CEFCA Web portal\footnote{\url{https://archive.cefca.es/catalogues}}, which also offers advanced tools for data searches, visualisations, and data queries \citep[see ][ and a future paper is also forthcoming; Civera et al., in prep.]{Civera2020}

\subsection{Identification of galaxy members}
\label{sec:galaxymembers}

\begin{figure*}
\centering
\includegraphics[width=0.8\textwidth]{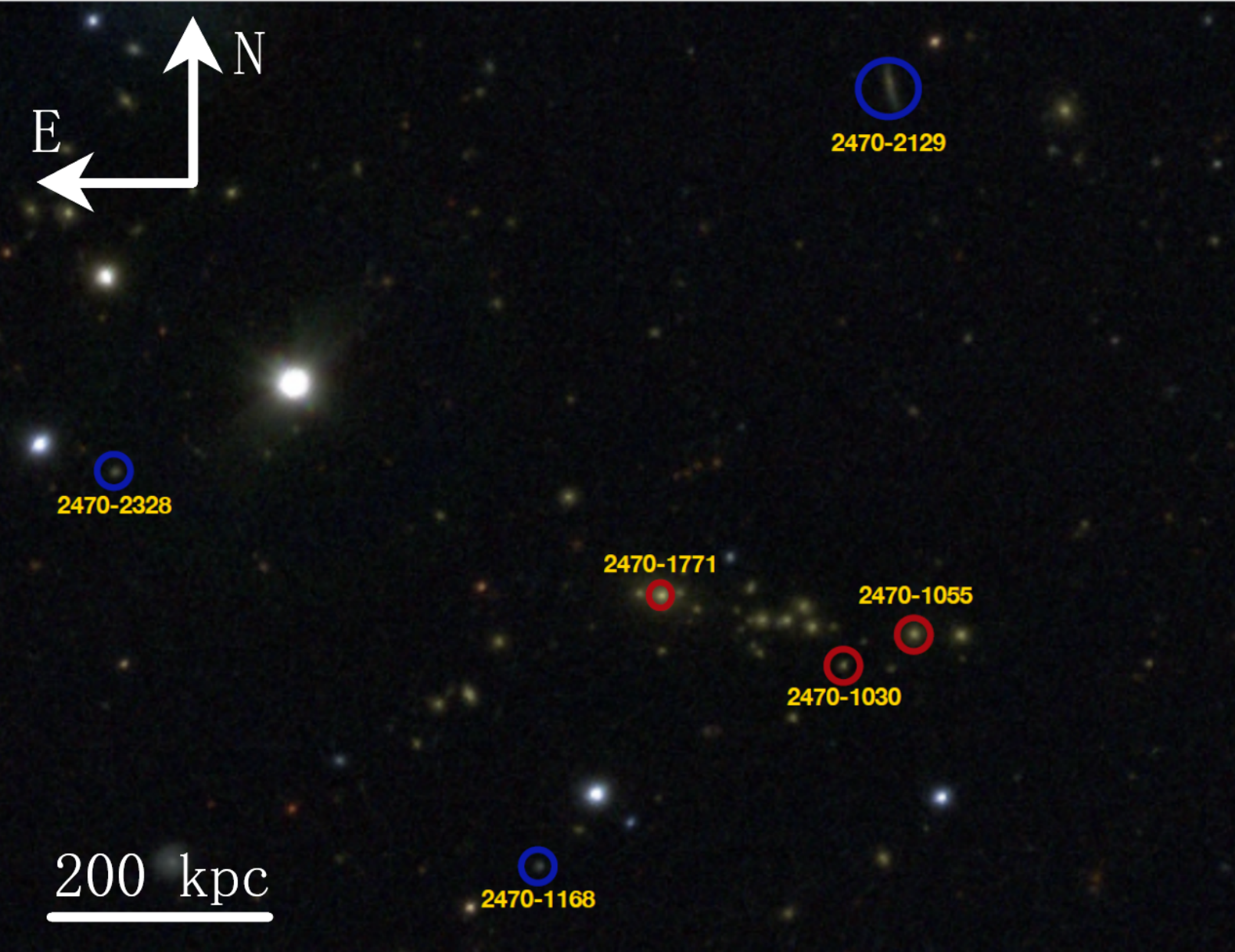}
\includegraphics[width=0.98\textwidth]{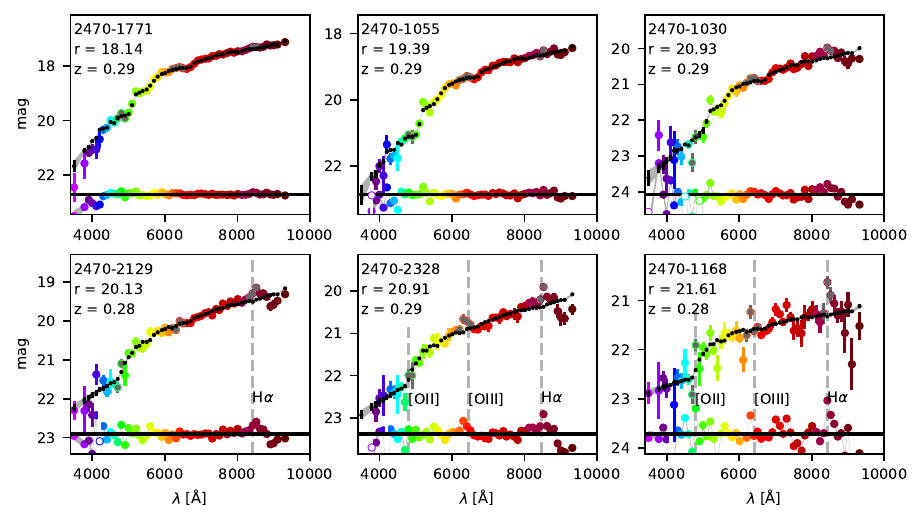}
\caption[mJPC2470-1771 and \js{} examples]{mJPC2470-1771 and \js{} examples. Top panel: \mjp \ view of mJPC2470-1771. Prominent red and blue galaxies in the cluster are marked with red and blue circles, respectively. The BCG corresponds to 2470-1771, with an spectroscopic redshift of $z=0.289$. Bottom panel: \magpsfcor \ \js \ of three red galaxies (top row) and three blue galaxies (bottom row) that are marked with circles in the top panel. 
The mean model fitted by \baysea \ is plotted as black points, and the grey band shows the magnitudes of the mean model $\pm$ one $\sigma$ uncertainty level.
The difference between the model and the best model fitted magnitudes are plotted as a small coloured points around the black bottom line. Masked filter (white coloured circles) and filters overlapping with the emission lines H$\alpha$, [NII], [OIII], H$\beta,$ and [OII] (darker grey coloured circles) are not used in the fit. Grey vertical dashed lines show the wavelengths corresponding to detectable emission lines.
}
\label{fig:jspectra}
\end{figure*}

The reference code for the detection of galaxy clusters in this work is Adaptive  Matched  Identifier of  Clustered  Objects (AMICO,  \citealt{2005A&A...436...37M, AMICO}), which is an algorithm based on the optimal filtering technique \citep[see e.g.][]{Postman1996, Bellagamba2011, AMICO}. It uses a statistical description of the background noise and a template to characterise the signal of the clusters. The signal is defined as the product of the template and an amplitude, plus the noise component. It uses several inputs, mainly the galaxies' sky positions, their magnitudes, and their redshifts, to compute this amplitude and other parameters. Our parameter of interest is the association probability assigned to each galaxy, which represents the probability of the galaxy of being a member of the cluster. All the details and AMICO inputs used for making this catalogue of galaxy clusters  in miniJPAS are detailed in the work by \cite{Maturi2023}, but a summary was provided in Sect.\ref{sec:mjp:AMICO}. 

For this work, we use the results from AMICO when using the \photozbest{} (the redshift corresponding to the maximum of the redshift probability density function, $z\mathrm{PDF}$ see \citealt{HC2021}). The choice of the redshift affects the galaxy members identified for the cluster, but for our purpose (the identification and characterisation of the galaxy populations in  mJPC2470-1771), there are no significant differences (see Appendix \ref{appendix:AMICOversions}). We used this catalogue since it is based on the same redshift as the one used for the SED-fitting analysis carried out by \cite{Rosa2021} using \baysea \ (see Sec. \ref{sec:method}).

The cluster, mJPC2470--1771, was identified as the most massive in \mjp \ by \cite{Bonoli2020}. The redshift of the cluster is $z=0.29$. In total, there are 99 objects (see Fig.~\ref{fig:mapProb}) with  AMICO association probability higher than $0.5$ and brighter than 22.5 in the $r$ band. We identify the brightest cluster galaxy (BCG) as the galaxy with the highest luminosity in the \rb{} and the highest stellar mass. Its ID in the \mjp \ catalogue is 2470-1771 and its coordinates are $(\alpha, \delta)=(213.6254^{\circ}$, $+51.9379^{\circ}$). 

The catalogue of this cluster has been tested with the follow-up on Gemini/GMOS observation of 38 galaxies with probabilistic association larger than $0.5$ (Carrasco et al., in prep.). Two galaxies failed to be classified by AMICO as members of the cluster, which have probabilities $0.58$ and $0.62$ in  the catalogue. Therefore, we estimated that AMICO classification only fails in 5\%  of cases.

Using GMOS spectroscopy, we estimated that $R_{200}$ \footnote{$R_{200}$ is the radius where the mean mass overdensity is 200 times the critical density at the cluster's redshift.} is 1304 kpc, and the halo mass is  $M_{200}=3.3\times 10^{14} ~ M_\odot$. These estimates are based on the measurement of the velocity dispersion. The measurement took all the observed members and applied the Clean routine from \cite{CLEAN} on it. This routine iteratively estimates the velocity dispersion and removes the outliers based on the caustic profile. Velocity dispersion is estimated using MAD \citep{MAD}. 
Using this value of $R_{200}$, we see that some members of the cluster may be outside of our observing FoV (see  Fig. \ref{fig:mapProb}).

In fact, assuming that galaxies are symmetrically distributed up to $R_{200}$, we estimated that nine galaxies that are between 0.5~$R_{200}$ and $R_{200}$ may not be included in our observing FoV. These galaxies represent only 12$\%$ of the sample; thus, any conclusion within $R_{200}$ is robust. Outside $R_{200}$, the incompleteness could be higher, but it is difficult to evaluate, and could be up to $\sim$20-30$\%$ if the galaxy members show a circular symmetry. Thus, conclusions outside $R_{200}$ must be taken with caution. In any case, extensive properties, such as the stellar mass surface density, are corrected for this incompleteness.

The cluster has also been detected with other IDs, such as MaxBCG J213.62543+51.93786 \citep{Koester2007}, who found a detected richness of 19 (scaled richness 17); WHL J141430.1+515616 \citep{Wen2012}, who found a $R_{200}$ of $1.2$~Mpc, 30 objects inside $R_{200}$, and a richness of 34, or RM J141430.1+515616.5 \citep{Rozo2015a, Rozo2015b}. None of these works are dedicated to a specific study of the properties of the galaxy members of this cluster; moreover, they are incomplete in their detection memberships. Thus, 
our work is the first and almost complete ($\sim$ 10$\%$ outside of the FOV) study inside $R_{200}$ for cluster memberships brighter than 22.5 (AB) in the $r$-band.


\subsection{Observational properties of galaxy members}
\label{sec:observationalproperties}
We have two different available photometries, \magauto \ and \magpsfcor. The first one is provided by  \sext{} and estimates the total flux of the galaxy using an adaptive scaled aperture \citep[see ][and \sext{ manual for further details.}]{Bertin1996}. Using the same approach as \cite{Molino2019}, \magpsfcor \ is aimed at correcting for the differences in PSF among different bands. It uses an aperture with the same shape as the Kron radius (smaller than the one used by \magauto) to provide robust colours determination \citep[see ][for further details]{Bonoli2020}. Due to their different apertures and extraction procedures \citep{Bonoli2020}, results between both may vary from one galaxy to another. In particular, \magauto \ uses a larger aperture, so it may include outer regions of the galaxy in the integration process, which tend to contain younger, blue stars. In fact, \cite{Rosa2021} compared the values of the stellar population properties obtained fitting the data from \magauto \ and \magpsfcor, finding that the main difference is that, on average, masses are $0.2$~dex larger in \magauto \ and rest frame colours are also bluer by $-0.09$~mag. Therefore, for our analysis, we used \magpsfcor, given its better signal-to-noise ratio (S/N). Its smaller aperture also allows for an improved detection of the emission lines in the centre of the galaxies.

We first looked at the observational properties of the galaxies (see Fig. \ref{fig:histobs}). The median measured redshift of the cluster's galaxies is $z=0.29$, with a standard deviation of $\sigma = 0.02$. There are four galaxies with redshift greater than 0.35. We looked at the $z\mathrm{PDF}$ of the galaxies in the galaxies in the cluster. These four galaxies are the only ones that show a multimodal distribution with peaks of similar amplitude (more than $\sim 50$~\% of the amplitude of the maximum peak). Due to their $z\mathrm{PDF}$ and their \photozbest{}, we decided to remove these four galaxies from our analysis.

The distribution of \rb \ peaks at around $\sim 21$~mag and most galaxies are brighter than $22.5$~mag. The number of galaxies in each bin steeply decreases with increasing median S/N, but the peak of the distribution is close to 10.  We see that brighter galaxies have a better S/N. Although there are galaxies with different probability and brightness, most of the galaxies with probability higher than $\sim 0.8$ have a magnitude brighter than 20 and a S/N higher than $\sim 11$. On the other hand, galaxies with probability lower than $\sim 0.6$ have a S/N higher than $\sim 10$. 

\section{Identification of the galaxy populations} 
\label{sec:SP}
The purpose of this section is to identify the red (RG), blue (BG), and emission line (ELG) galaxies in the cluster. First, we explain the method for retrieving the stellar population properties of the galaxies based on the \js{} fits. Then we describe the methods to identify ELGs. 


\subsection{\js{} fits}
\label{sec:method}

We used \baysea\ (de Amorim et al., in prep.), a parametric SED fitting code, to obtain the stellar population properties from \js.
\baysea \ is an adaptation of the method developed by \cite{lopez-fernandez2018} in order to use \jp \ magnitudes as input.
The code generates synthetic \js\ from parametric SFH models. For a given observed \js\ we performed a Markov Chain Monte Carlo (MCMC) exploration of the parameter space, thus obtaining a sample of parameters that approximates the probability density function (PDF) of the model. In this work we assumed a delayed-$\tau$ model given by:

\begin{equation}
    \psi(t)=\frac {M_{ini}} {\tau^2 \left [ 1- e^{-\frac{t_0}{\tau}} \left ( \frac{t_0}{\tau}+ 1 \right )   \right ]}(t_0-t)e^{-\frac{t_0-t}{\tau}},
    \label{eq:SFH}
\end{equation}
where $t$ is the look-back time, $t_0$ is the (look-back) time when the star formation began, $\tau$ is a measurement of how extended in time the star formation was, and  ${M_{ini}}$ is the total mass of formed stars.
This model also includes stellar metallicity ($Z$) and dust attenuation ($A_V$) which, combined with stellar population model spectra and a foreground dust screen extinction curve, results in a set of model \js \footnote{The stellar population spectra are preprocessed and converted to observed-frame magnitudes for a grid of redshifts, using \jp \ filter curves. }.
The complete set of parameters is $(t_0, \tau, A_V, Z)$.
We also obtained the stellar mass ($M_\star$) from the scaling factor of the model with relation to the observed \js.
From these parameters, we can calculate the mass-weighted and light-weighted ages and rest-frame colours.

We let $100$ chains walk the parameter space for 2200 steps.
The autocorrelation time\footnote{The assessment of autocorrelation time and convergence of the chains was performed in a small sample of \js. This is a fairly manual process, as with any MCMC convergence study. We consider the burn-in phase to be over at around $5 \times$ the autocorrelation time, which we assume is conservative enough.} of the chains for this model is around 120 steps, we discarded the first 1200 steps as a burn-in phase.
In the end, we got a total of 100,000 samples of the parameter space.
For each galaxy, we took the mean and standard deviation of the parameters and properties of the samples as an estimate of their expected value and uncertainties.

Emission lines are not included in the models. Because some of the NBs can be affected by strong contributions from the Balmer (H$\alpha$, H$\beta$) and optical collisionally-excited ([OIII]$\lambda$5007, 4959, [OII]$\lambda$3727, [NII]$\lambda$6589, 6548) emission lines, we removed those bands where these lines could be  at the redshift of each galaxy  from the fits. In this way, we ensure that the fit is done only over the stellar continuum since the nebular continuum is negligible in most of the SF galaxies. Only objects with HII regions with very extreme emission lines \citep[e.g. \ewha{} > 1000 \AA][]{Rosa1994} would be affected by this assumption. These galaxies are not present in this sample.

A more detailed explanation of the method with a global study of the galaxies in the AEGIS field can be found in \cite{Rosa2021}. Since the data used in this chapter are a subsample from \cite{Rosa2021}, the models are computed using the initial mass function (IMF) by \cite{Chabrier2003} and the latest versions of the \cite{C&B2003} stellar population synthesis models \citep{Plat2019}. We chose the attenuation law by \citet{calzetti2000} which we added as a foreground screen. We also note that, unless stated otherwise, the term 'mass' refers to the stellar mass (M$_\star$) derived by \baysea.

In Fig. \ref{fig:jspectra}, we show the \js \ of six galaxies (three red and three blue) along with the fits obtained with \baysea. These galaxies are identified with red and blue circles in the top panel. This serves as an example of the aspect of \jp \ data, the effectiveness of \baysea \ and it also manifests the capability of \jp \ to detect line emission (which will be exploited in Sect.\ref{sec:ELG}).

\subsection{Identification of red and blue galaxies in the cluster}

\label{sec:Masscolour}

Throughout this chapter, we use two different colours: $(u-r)_{\mathrm{res}}$ and $(u-r)_{\mathrm{int}}$. Both of them are rest-frame colours derived from the star formation history obtained from the SED fitting (see Sect. \ref{sec:method}), but the first one is not corrected from extinction while the second one is calculated including the reddening correction in the synthetic SED.

The bimodal distribution of the AEGIS galaxy population is shown in the galaxy stellar mass-colour diagram \citep{Rosa2021}. In this work, we show that $(u-r)_{\mathrm{int}}$ is more useful than $(u-r)_{\mathrm{res}}$ to discriminate between the red star-forming and quiescent galaxies because it accounts for the fraction of red star-forming galaxy population of the sample. We use an adaptation of the criterion given by \cite{Luis2019}, previously been used by \cite{Rosa2021} to segregate the whole galaxy populations in miniJPAS in red and blue galaxies. We consider galaxies to be red if:
\begin{equation}
    (u-r)_{\mathrm{int}}> 0.16 (\log( M_{\star})-10)-0.254(z-0.1)+ 1.792,
    \label{eq:red-blue}
\end{equation}
and blue otherwise.

\subsection{ELG identification}

\begin{figure}
    \centering
    \includegraphics[width=0.65\textwidth]{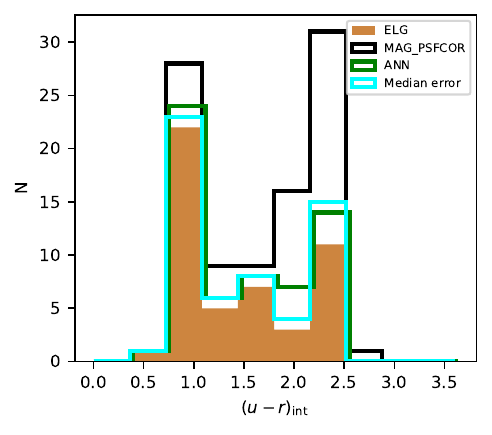}
    \caption[$(u-r)_{\mathrm{int}}$ histogram comparing the emission line galaxies selection criteria]{$(u-r)_{\mathrm{int}}$ histogram comparing the emission line galaxies selection criteria. The black line shows the distribution of all the galaxies in the AMICO catalogue. Cyan histogram shows the distribution for the median error. Same for green but with the ANN method. Orange solid histogram shows the distribution for the common galaxies for both methods. The different distributions have been shifted a little bit in the plot to show more clearly the similarities and differences between them.  }
    \label{fig:method_comparisson}
\end{figure}

\label{sec:ELG}
As seen in Fig. \ref{fig:jspectra}, the \js \ are capable of showing the line emission as excess flux in a given filter. In this section, we describe two methods for classifying the galaxies as emitters or non-emitters, with respect solely to the H$\alpha$ emission. We applied the methods to the cluster catalogue to characterise the  emission line galaxy populations.

\subsubsection{Median error method}
This method is based on our \jp{} but also uses a prior based on the result from our SED fitting code, since we distinguish between red and blue galaxies. The base idea is that when looking at the filter that is sensible to the observed line wavelength, a galaxy that presents H$\alpha$ (or H$\alpha$ +[NII], as in this method we cannot separate the emission of both lines)  emission will show a lower magnitude value in the observed \js \ ($\mathrm{m}_{\mathrm{obs}}$) than in the stellar continuum fit ($\mathrm{m}_{\mathrm{fit}}$). However, it is not enough to simply consider that $\mathrm{m}_{\mathrm{fit}}-\mathrm{m}_{\mathrm{obs}}>0$. We must establish a threshold value. A first consideration to make is that the difference must be greater than the observed error. In order to estimate the observed error of the fitted stellar continuum, we consider the median error in the five filters closer to the band where H$\alpha$ is, symmetrically distributed. If we only choose three filters, our estimation could be contaminated with other lines close to H$\alpha$, such as [SII]. Choosing seven filters only changes the final set in one galaxy. Choosing more filters would mean estimating the continuum too far from H$\alpha$, given the width of \jp \ filters. Besides, the blue galaxies in our sample are noisier than red galaxies. The median S/N in the five closest filters to H$\alpha$ at the cluster redshift is almost three times better for red galaxies than for blue galaxies. We find that the larger magnitude difference of the blue galaxies is not enough to compensate their worse S/N. This implies that if the same threshold (three sigma) is applied, there will be a bias towards the detection of less blue emission line galaxies. Lastly, we must take into account that due to the uncertainties in the \photozbest{} determination, the filter with the closest central wavelength to the calculated line wavelength might not be the one showing the line emission. This is a consequence of our method being fine tuned to fit a set of galaxies observed with spectroscopy in the cluster area (see Appendix~\ref{appendix:GMOS}). Therefore, our method proceeds as follows. First, we find the closest filter to the observed line wavelength. Then, we calculate the median error of that filter and the four adjacent ones (symmetrically distributed) $\eta ( \epsilon_{\mathrm{m}_{\mathrm{obs}}})$. Finally, we look at the closest filter, the previous and the next one, and we classify the galaxy as an emission line galaxy if one of those filters satisfies the following condition:
\begin{equation}
    \mathrm{m}_{\mathrm{fit}}-\mathrm{m}_{\mathrm{obs}}> \theta \cdot \eta (\epsilon_{\mathrm{m}_{\mathrm{obs}}}),
\end{equation}
\noindent where $\theta=1$ for blue galaxies and $\theta=3$ for red galaxies in order to account for their better S/N in the filters closer to the H$\alpha$ wavelength at the cluster redshift. 
These values of $\theta$ were chosen in order to account for this differences in the S/N ratio. They have been tested with the data from GMOS spectroscopic observations of 13 galaxies with clear H$\alpha$ emission in the spectra.

This method allows us to identify emission line galaxies in the cluster, but we do not use it to estimate the fluxes of the lines; in this respect, the ANN method is more useful.  

\subsubsection{ANN}
This method uses the equivalent width of H$\alpha$, EW(H$\alpha$) predictions made by \cite{Gines2021} ANN. In that work, two different ANNs are trained using synthetic photometry (in the sense that real spectra are processed to obtain \jp \ magnitudes) obtained from the Calar Alto Legacy Integral Field Area survey \citep[CALIFA,][]{CALIFA2012} and the Mapping nearby Galaxies at Apache Point Observatory survey \citep[MANGA,][]{MANGA2015}. One of the ANN is trained to calculate the EW of H$\alpha$, H$\beta$, [OIII], and [NII]. The other ANN is trained in order to classify the galaxies into ELG and quiescent galaxies. 
The CALIFA and MANGA galaxies contain millions of spaxels with different astrophysical conditions, which include regions with high and low star-formation activity as well as variations in the gas-phase metallicity or the dust distribution. Furthermore, CALIFA and MANGA survey contain galaxies in different environments (clusters, groups and field) since they were selected to avoid environmental bias \citep{Walcher2014,Wake2017}. Therefore, we do not expect our prediction to be unlikely in the cluster under study. Nevertheless, by construction, our training set includes, on average, a smaller amount of spaxels ionized by the presence of AGN or shocks waves. In \cite{Gines2021} we showed that we do not miss a fraction of AGN larger than $3$~\% over the whole sample of galaxies used from SDSS. This confirms that the transfer from the training sample to our current data is trustworthy.

As explained in \cite{Gines2021}, there is a minimum measurable EW for a photometric filter. Therefore, the criteria we use is simply to consider the galaxy as an emission line galaxy if the EW given by the ANN is greater than the minimum measurable EW, taking the error bars into account. This is: 
\begin{equation*}
    \frac{\Delta '}{\mathrm{S/N}-1} <  \mathrm{EW_{H\alpha_{\mathrm{ANN}}}}+ \epsilon \mathrm{EW_{H\alpha_{\mathrm{ANN}}}} \ ; \  \mathrm{EW_{H\alpha_{\mathrm{ANN}}}} > \epsilon \mathrm{EW_{H\alpha_{\mathrm{ANN}}}} ,
\end{equation*}
\noindent where $\Delta '$ is the equivalent width of the filter and can be calculated as: 

\begin{equation}
    \Delta ' = \frac{\int \lambda T (\lambda ) \mathrm{d}\lambda}{\lambda_z T (\lambda_z )} 
,\end{equation}

\noindent where $T$ is the normalised transmittance of the filter and $\lambda_z$ is the observed emission line wavelength. To compute the minimum measurable EW, we find the J-PAS filter with a central wavelength closest to the observed H$\alpha$ wavelength given the galaxy's redshift. Since the minimum EW is associated with a certain filter, the $\Delta '$ and S/N parameters must be considered in the same filter and it does not make any sense to consider a median value for any of them. Also, since we are using the ANN predictions, here we can separate the H$\alpha$ emission from the [NII] emission.

This method is very useful not only for identifying emission line galaxies, but also to predict the EW of H$\alpha$, H$\beta$, [NII], and [OIII] lines, along with their respective ratios ([NII]/H$\alpha$, [OIII]/H$\beta$). We are able to reach a precision in the $\log$([NII]/H$\alpha$) of $0.09$~dex for SF galaxies and average S/N $\sim$10 in the \js{}. This is independent of the redshift of the galaxies as we prove in \citep{Gines2021}, where we tested our results with a sample of SDSS galaxies within the redhisft range of $0 < z < 0.35$.
 
\subsubsection{ELG final set}
We then applied both methods to all the galaxies in the cluster. The median error method selects 57 galaxies as galaxies with emission lines, while the ANN method selects 50 galaxies in total. We decided to be more conservative and consider as ELG population the intersection of both groups, since this defines a more robust subset. A total of 49 galaxies remains. A comparison of the three sets with the whole cluster colour distribution can be seen in Fig. \ref{fig:method_comparisson}. The $(u-r)_\mathrm{int}$ distribution is very similar in both methods, peaking around $(u-r)_\mathrm{int}\approx 1$ and with a lower peak at $(u-r)_\mathrm{int}\approx 2.2$. This peak is easily understood when looking at the general distribution, since there is also a peak at this value, even greater than the bluer one. When defining the median error method and establishing the values of the multipliers, there is a risk of generating a greater bias towards blue galaxies greater than desired in order to account for the larger errors. Comparing its histogram with the ANN, we can see that the proportion of blue and red galaxies remains very similar, so we can trust this method. As a final comment, these results are coherent with what we would expect, since most of the ELG are blue.

\section{Characterisation of the galaxy populations}
\label{sec:charact}

\begin{table*}[]
    \centering
    \resizebox{\textwidth}{!}{%
     
    \begin{tabular}{c c c c c c c c c}
    \hline
    \hline
         {\small Property} & {\small Galaxies} & {\small RG}  & {\small BG}  & {\small ELG} &  {\small ELG-R} &  {\small ELG-B} &  {\small SF} &  {\small AGN}\\
         \hline
$\log M_\star$ & $10.0 \pm 0.65$ & $10.4 \pm 0.32$ & $9.63 \pm 0.64$ & $9.89 \pm 0.71$ & $10.5 \pm 0.36$ & $9.62 \pm 0.65$ & $9.56 \pm 0.60$ & $10.5 \pm 0.42$\\
$ A_V$ & $0.57 \pm 0.43$ & $0.32 \pm 0.19$ & $0.84 \pm 0.44$ & $0.63 \pm 0.42$ & $0.32 \pm 0.25$ & $0.76 \pm 0.40$ & $0.65 \pm 0.41$ & $0.49 \pm 0.35$\\
$ \log Z_\star>$ & $0.09 \pm 0.45$ & $0.29 \pm 0.20$ & $-0.1 \pm 0.53$ & $0.05 \pm 0.50$ & $0.35 \pm 0.19$ & $-0.0 \pm 0.54$ & $-0.0 \pm 0.53$ & $0.35 \pm 0.25$\\
$ (u-r)_\mathrm{res}$ & $2.04 \pm 0.51$ & $2.42 \pm 0.10$ & $1.64 \pm 0.48$ & $1.85 \pm 0.55$ & $2.43 \pm 0.08$ & $1.59 \pm 0.47$ & $1.60 \pm 0.50$ & $2.31 \pm 0.28$\\
$ (u-r)_\mathrm{int}$ & $1.67 \pm 0.60$ & $2.21 \pm 0.16$ & $1.10 \pm 0.30$ & $1.44 \pm 0.58$ & $2.22 \pm 0.16$ & $1.09 \pm 0.30$ & $1.18 \pm 0.46$ & $1.98 \pm 0.43$\\
$ <\log age>_\mathrm{M}$ & $9.44 \pm 0.24$ & $9.61 \pm 0.13$ & $9.26 \pm 0.21$ & $9.34 \pm 0.23$ & $9.53 \pm 0.17$ & $9.26 \pm 0.21$ & $9.27 \pm 0.23$ & $9.47 \pm 0.22$\\
$\tau / t_0 $ & $0.52 \pm 0.60$ & $0.12 \pm 0.02$ & $0.94 \pm 0.63$ & $0.71 \pm 0.66$ & $0.11 \pm 0.02$ & $0.97 \pm 0.64$ & $0.95 \pm 0.68$ & $0.23 \pm 0.27$\\
         \hline
    \end{tabular}}
\caption[Mean and standard deviation values of the stellar population properties of the galaxies in the cluster]{Mean and standard deviation values of the stellar population properties of the galaxies in the cluster.    The properties are for red and blue galaxies (RG, BG), emission lines galaxies (ELG), ELG with red (ELG-R) or blue (ELG-B) colours, star-forming galaxies (SF), and  galaxies with an AGN. }
    \label{tab:sp_auto_psfcor}
\end{table*}

In this section, we analyse the stellar population properties of the galaxies belonging to the cluster. First, the sample is divided in red and blue galaxies. Then, ELG are characterised by their stellar populations by dividing also the galaxies in star-forming (SF) and galaxies with an active galactic nucleae (AGN). The average and dispersion values of the stellar population properties are summarised in Table~\ref{tab:sp_auto_psfcor}.

\subsection{The red and blue galaxy populations}

\begin{figure*}
\centering
\includegraphics[width=\textwidth]{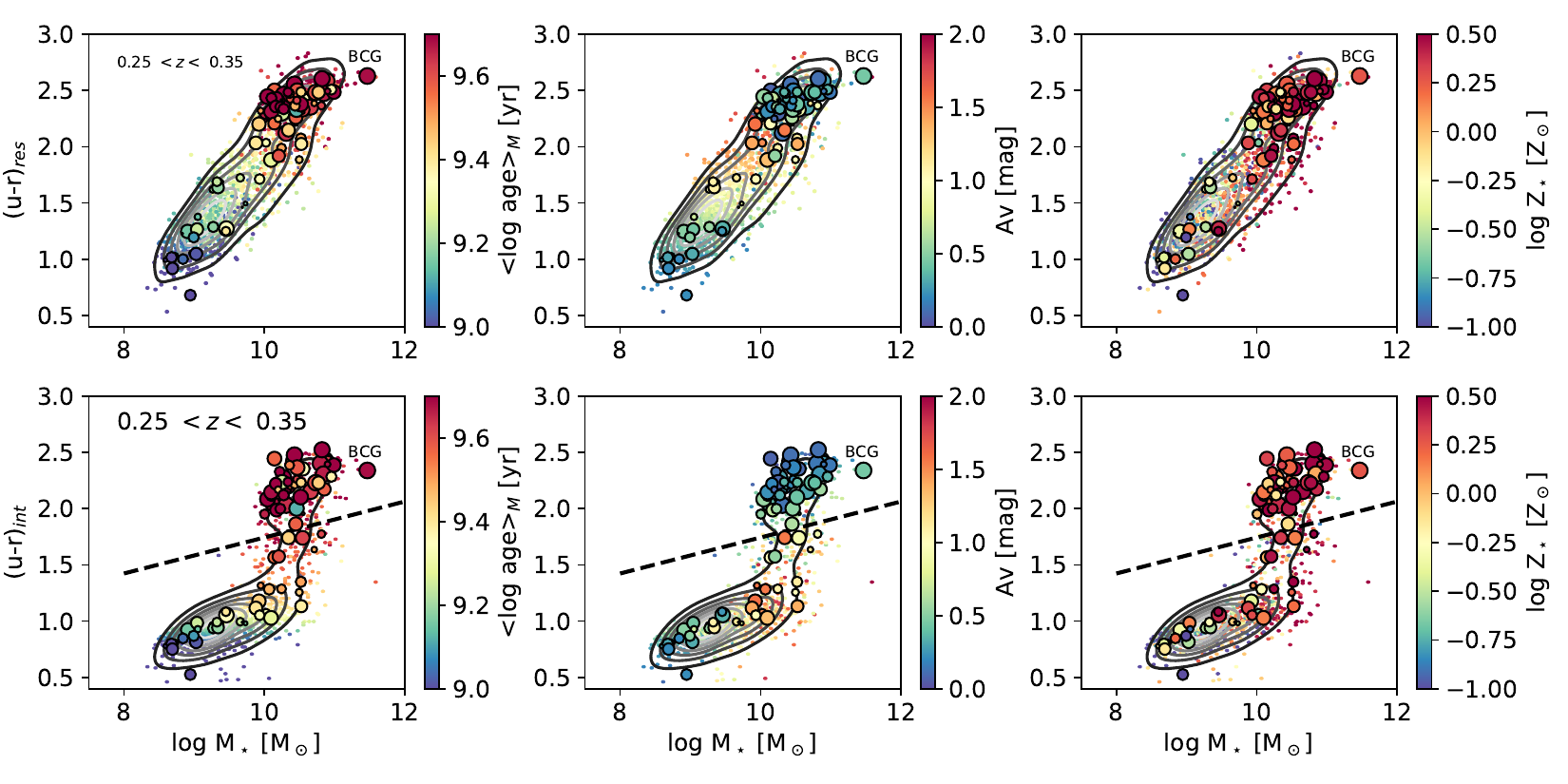}
\caption[$(u-r)_{\mathrm{res}}$ (top panels) and $(u-r)_{\mathrm{int}}$ (bottom panels) colour vs. stellar mass for the  redshift bin 0.25$< z <$0.35 derived by \baysea \ from the AEGIS galaxy populations (contour) and galaxy cluster members (circles)]{$(u-r)_{\mathrm{res}}$ (top panels) and $(u-r)_{\mathrm{int}}$ (bottom panels) colour vs. stellar mass for the  redshift bin 0.25$< z <$0.35 derived by \baysea \ from the AEGIS galaxy populations (contour) and galaxy cluster members (circles). The coloured bar shows the distribution of the stellar population properties age, extinction, and metallicity (from left to right). The size of the circles indicates the probability of the galaxy to be member of the cluster. The position of the brightest galaxy in the cluster (BCG) in each panel is marked. The dashed line in the $(u-r)_{\mathrm{int}}$ divides blue galaxies (below the line) and red galaxies (above the line).
}
\label{fig:masscolour}
\end{figure*}

\begin{figure*}
\centering
\includegraphics[width=\textwidth]{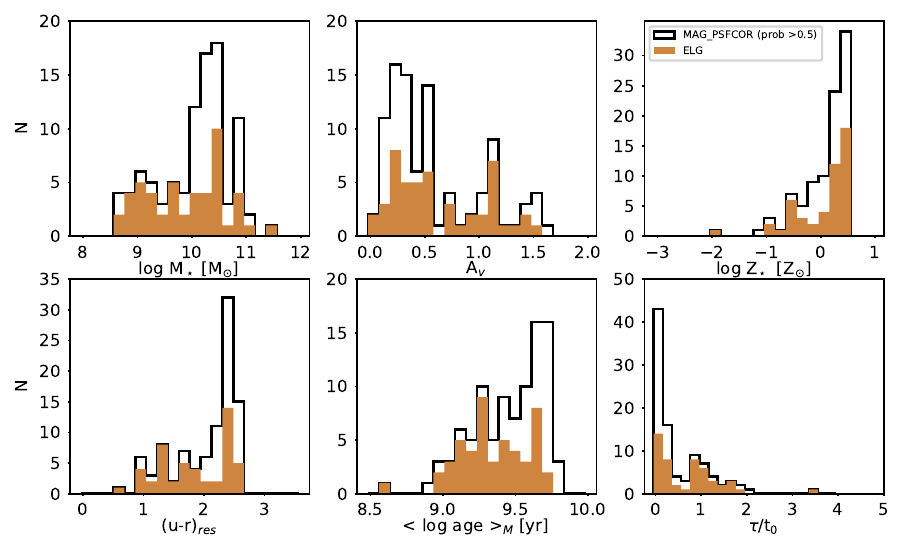}
\caption[Stellar population properties distribution for the emission line galaxy population]{Stellar population properties distribution for the emission line galaxy population. Black histogram shows the distribution for all the galaxy members in the \magpsfcor \ photometry. Brown solid histogram shows the distribution for the objects selected as emission line galaxies. From left to right and from upper to bottom: Stellar mass, extinction, stellar metallicity, $(u-r)_{\mathrm{res}}$ colour, mean mass-weighted age, and ratio between the SFR parameters $\tau$ and t0. 
}
\label{fig:em_line_hist}
\end{figure*}

\begin{figure}
    \centering
    \includegraphics[width=0.65\textwidth]{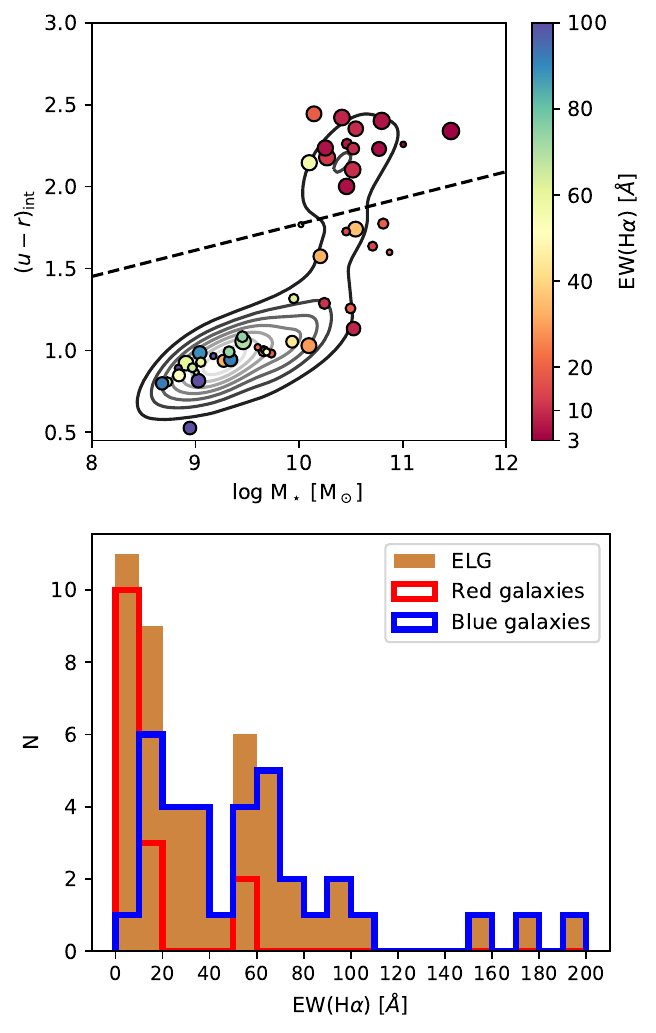}
    \caption[EW(H$\alpha$) of the ELG population. Top panel: Colour-mass diagram for the ELG galaxy population]{ EW(H$\alpha$) of the ELG population. Top panel: Colour-mass diagram for the ELG galaxy population. The colour bar shows the H$\alpha$ equivalent width. Galaxies above the black dashed line are considered to be red. Galaxies below are considered to be blue. Bottom panel: H$\alpha$ equivalent width distribution histogram. Green histogram shows the distribution of the galaxies selected with the ANN method. Orange solid histogram shows the distribution for the selected galaxies. Red histogram shows the distribution of the selected red galaxies. The same holds true for blue histogram and blue galaxies.}
    \label{fig:EW_Ha}
\end{figure}

The bimodal distribution of the red and blue galaxies in the cluster can be seen in the colour-mass diagram (see Fig.~\ref{fig:masscolour}) The comparison between the $ (u-r)_\mathrm{res}$ and the $(u-r)_\mathrm{int}$ shows how the extinction correction moves a significant number of galaxies from the redder (upper) regions of the diagram to the bluer regions (below the black dashed line). The colour code also shows how, on average, redder galaxies are older and more metal-rich. Galaxies with larger extinctions are located in the middle region of the diagram, which could be considered as an equivalent of the green valley. The comparison with \cite{Rosa2021} results for the whole AEGIS catalogue shows that the distribution of these properties in the colour-mass diagrams remains the same. This would indicate that, for fixed values of the colour and mass, the effect of the environment on these properties is negligible.
We find that, on average, red galaxies are more massive than blue galaxies by $\sim 0.8$~dex. Blue galaxies also show a larger variance in mass. The extinction $A_V$ is significantly larger on average ($\sim 0.5$~mag) for blue galaxies. This is expected because most of the blue galaxies are star-forming, and the extinction that young stars experience is almost  double than that for the old stellar population \citep{charlot2000}. In contrast, blue galaxies are less metal rich than red galaxies by $\sim 0.1$~dex.  On average, red galaxies are older by $0.4$~dex. The value of  $\tau/t_0$  is nine times larger for blue galaxies than for red galaxies; thus, the star formation lasts longer in the blue galaxy population. 

The total fraction of red galaxies is $0.52$ ($0.48$ for blue galaxies). The fraction of red galaxies in the whole catalogue of \mjp \ at the cluster's redshift, obtained by \cite{Rosa2021},  using \baysea, is around $0.2$ or even lower. This is supported by works in the literature such as \cite{Balogh2004}. If we assume a symmetric distribution within $R_{200}$, and if all the missing galaxies were blue (worst case scenario), the fraction of red galaxies inside $R_{200}$ would be 0.55 (compared to the current observed fraction of $0.62$ inside $R_{200}$), which is still higher than the fraction of red galaxies in the field. Instead, if we assumed a symmetrical distribution keeping the same amount of blue and red galaxies in the missing area, we would find the fraction of red galaxies to be even larger ($0.64$).

\subsection{ELG population}

In Fig.~\ref{fig:em_line_hist} we compare the distribution of the stellar properties of the ELG with the whole sample and we summarise them in Table~\ref{tab:sp_auto_psfcor}. We find that their values span the same ranges than the properties of the whole catalogue. However, the distribution themselves are different. The stellar mass still peaks at $\log M_\star \approx 10.5$~[$M_\odot$], but the contribution of galaxies with $\log M_\star < 10$~[$M_\odot$] becomes more significant. In fact, most of the galaxies in such range are classified as ELG. On average, ELGs are less massive than the whole sample by 0.1 dex.
The distribution of $A_V$ shows that most of the galaxies that are not selected as ELG exhibit values lower than 0.5, but the distribution remains similar (the average only becomes $0.06$~mag lower). A similar behaviour is found for the metallicity, where most of the galaxies with $ \log Z_\star\lesssim$-0.5 are ELG, but the peak of the distribution is still the same as the whole set. The average only becomes lower by $0.04$~dex.
Nonetheless, the distribution of $ (u-r)_\mathrm{res}$ changes significantly. The peak of the distribution is still found at $ (u-r)_\mathrm{res}\approx 2.5$~mag, but most of the galaxies with $ (u-r)_\mathrm{res}< 2$ are ELG, and only a few galaxies with $ (u-r)_\mathrm{res}> 2$ are ELG. Moreover, the peak of the stellar ages is now found at  $ <\log age>_\mathrm{M} \approx 9.25$, with most of the young galaxies being ELG and only a few of the old galaxies showing emission lines (ELG are younger by $0.1$~dex on average). Furthermore, most of the galaxies  with $\tau /t_0 \lesssim 0.8$ are not ELG, and almost all the galaxies $\tau /t_0 \gtrsim 0.8$ are ELG, and the average value of ELGs is almost 50~\% larger than the average of the whole sample.

These results show that ELGs have properties similar to the BG population. However, they display differences that suggest that ELG is a mix of red and blue populations, where RG are significantly less abundant than the BG population. To further investigate  this point, we explore the colour-mass diagram and the distribution of the inferred EW(H$\alpha$), dividing the ELG into red (ELG-R) and blue (ELG-B) galaxies (Fig. \ref{fig:EW_Ha}). We see that galaxies with the lowest predicted EW ($<10$~\AA \ approximately) are all red galaxies, while galaxies  above this value are all blue -- except for three of them. Two of them are particularly notable, having a predicted EW above 50 \AA. However, when looking at the spectra, we find that the inferred emission may be a result from incomplete background subtraction due to fringing effect that suffer some of the red filters, which is translated into a variation of the measured magnitude that could be interpreted as an emission line by the ANN, due to the magnitude difference among one filter and its adjacent ones. With that exception, we can conclude that ELG-R are characterised by low estimated values of EW(H$\alpha$), while ELG-B have EW(H$\alpha$) $>$ 10 \AA. Taking into account that our ELG-R are more massive than our ELG-B, we find that our results are coherent with the EW(H$\alpha$)-mass relation found in the literature \citep[see e. g.][]{Fumagalli2012, Sobral2014, Khostovan2021}.

\subsection{Star-forming galaxies and AGN populations}
\label{sec:AGN}

\begin{figure*}
    \centering
    \includegraphics[width=\textwidth]{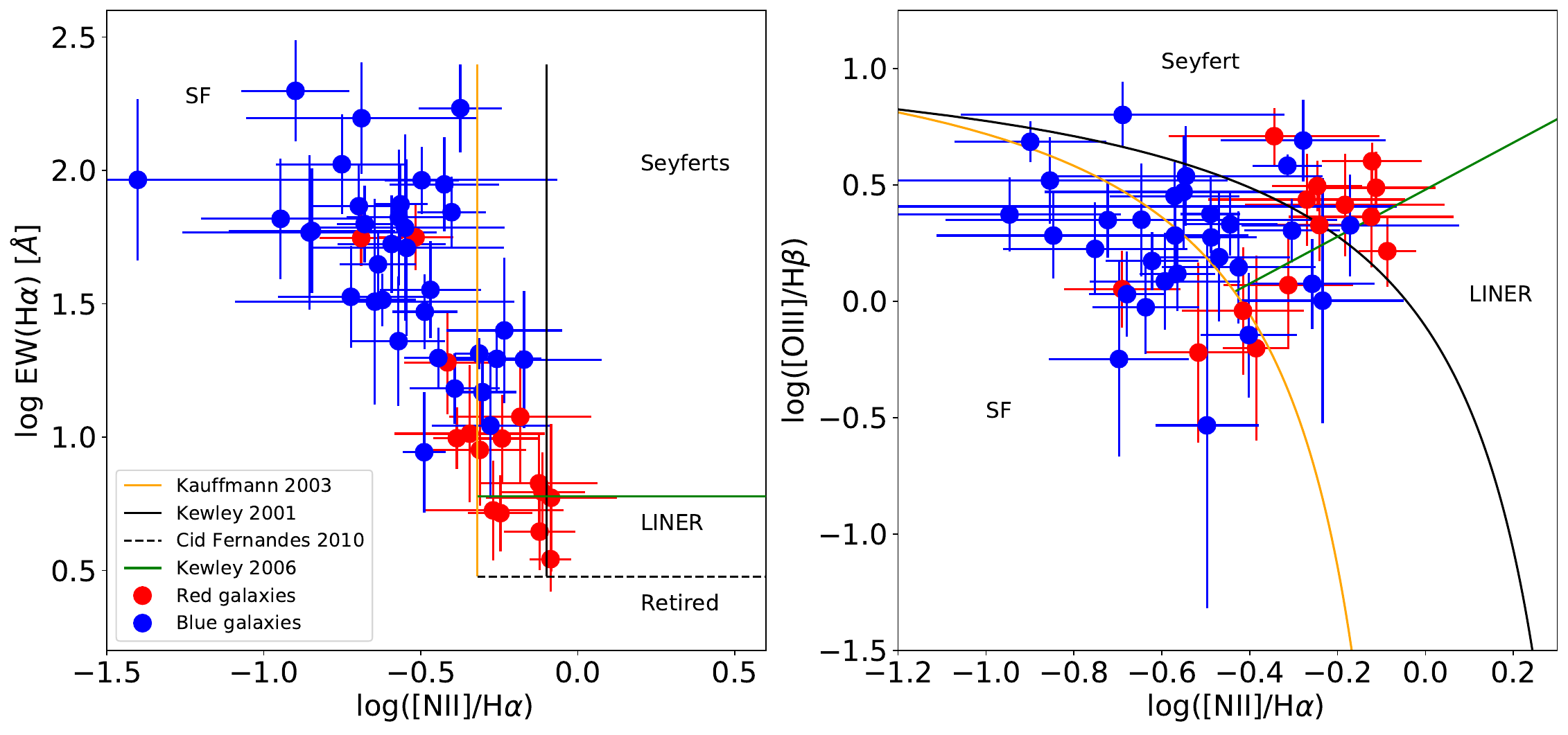}
    \caption[ELG classification diagrams. Left panel: WHAN diagram with the galaxies classified as emission line galaxies]{ELG classification diagrams. Left panel: WHAN diagram with the galaxies classified as emission line galaxies. Red points represent red galaxies and blue points, blue galaxies. The solid orange and black lines represent the \cite{WHAN_1} transposition of the \cite{Kauffmann2003} and \cite{Kewley2001} SF-AGN distiction criteria, and the green solid line represents the transposition of the \cite{Kewley2006} made by \cite{WHAN_1}. The dashed black line represents the distinction between retired galaxies and LINERs \citep{WHAN_2}. Right panel: BPT diagram for the emission line galaxy population. The colour coding is the same as the left panel.}
    \label{fig:WHAN+BPT}
\end{figure*}

The ELG population can be a mix of star-forming galaxies (SF) and AGN galaxies. To find the abundance of these two classes,  we use two different diagrams: the WHAN diagram, introduced by \cite{WHAN_1} (CF10), and the BPT diagram \citep{BPT}. There are a number of works that present their own criteria to separate the SF, Seyferts and LINERs in the BPT diagram, but here we use the results from three in particular: \cite{Kauffmann2003} (K03) and \cite{Kewley2001} (K01) to distinct SF galaxies from galaxies with a potential AGN, and we use the transposition to this diagram made by \cite{WHAN_1} of the separation criteria between Seyferts and LINERs found by \cite{Kewley2006} (K06). In the case of the WHAN diagram, we use the criteria from  CF10 and \cite{WHAN_2} (CF11). In this work, several criteria are presented, but for consistency we choose the transpositions made in these works (CF10 and CF11) of the same criteria used in the BPT (K03 and K01) in addition to the criteria of studies where galaxies with EW(H$\alpha$)$<3$~\AA  \ are considered to be retired galaxies \footnote{This is the name given by CF10 to red-quiescent galaxies with weak H$\alpha$ emission, that is probably produced by post-AGB stars.}.  
An example of each type of galaxy, with its \js \ and its position in the WHAN and BPT diagrams, respectively, can be seen in Fig \ref{fig:Examples_AGN}. This figure is useful to explain more clearly how we interpret the position of galaxies in these diagrams.

Figure~\ref{fig:WHAN+BPT} shows the WHAN and BPT diagrams with all the ELG population. Table \ref{tab:AGN} shows the classification for each of these galaxies in both diagrams.
Since a different classification can be derived from each diagram and due to the error bars obtained for many galaxies, it is not trivial to assign a label to each ELG. Therefore, we use a probabilistic approach in the following way: we calculate the area of the error box in the WHAN diagram and we calculate the fraction of this area that falls in each of the diagram regions. We define this fraction as the probability representing how likely it is for that galaxy to be a SF, Seyfert, LINER, retired, or composite (SF-Seyfert or SF-LINER) galaxy. The error bars plotted in this figure take into account the correlation among the emission lines through the calculation of the Pearson coefficient and its inclusion in the error budget.

We find that 32 galaxies ($65.3$~\% of the ELG) have a probability greater than $0.7$ of being associated with the SF region (33 above $0.5$, representing $67.3$~\%). We selected these galaxies as the SF population. Only one galaxy (which represents 2~\% of the ELG) has a combined probability in the Seyfert and LINER region greater than 0.5 in the WHAN diagram. The rest of the galaxies are difficult to uniquely identify as AGN using only the WHAN diagram. Due to this restriction,  we selected as AGNs those galaxies that are above the K01 curve in the BPT diagram if the probability of being SF in the WHAN diagram is less than 0.5. With these criteria, only 2470-3670 need to be excluded from the AGN sample (see Appendix~\ref{sec:ELG_class} and Fig \ref{fig:Examples_AGN}). Galaxies between the K01 and K03 lines likely have contributions from AGN, but we cannot resolve whether they are Syferts, LINERs, or composite galaxies. Thus, we do not include them as part of the AGN sample, nor as SF if they are not classified as SF in the WHAN diagram. 

In Table~\ref{tab:sp_auto_psfcor}, we summarise the stellar population properties of these galaxies. If we compare the SF galaxies with the blue ELG, we find that the differences in the average and standard deviation values are negligible except for the extinction $A_V$, which are lower for SF galaxies, and $(u-r)_{int}$ colours, which are slightly redder (but with a greater standard deviation) for SF galaxies. This indicates that most of the blue ELG are SF galaxies.

The comparison between the values of the red ELG and AGN populations shows that the main difference between them resides in a larger extinction on average for the AGNs, slightly bluer $(u-r)_{res}$ and $(u-r)_{int}$ colours, younger ages and higher values of $\tau/t_0$. This indicates that the sample selected as AGN is a mixture of blue and red ELGs and that we are not able to fully separate the contributions of pure AGNs from star formation in galaxies, or that most of these galaxies are actually composite. 

We focus here on the H$\alpha$ emission in order to select our ELG sample. This line  can be used as a tracer of the star formation \citep[see e.g.][]{Kennicutt1998, Kennicutt2012, Kewley2002, Garn2010, Oteo201, CT2015} or the presence of AGNs \ \citep[see e.g.][]{Osterbrock1985,Veilleux1987,Osterbrock1989, Kewley2001, WHAN_2}.  Therefore, taking into account the relation between galaxies in the blue cloud and a higher star formation than galaxies in the red sequence, \citep[see e.g.][]{Kauffmann2003a, Kauffmann2003b, Baldry2004, Brinchmann2004, Gallazzi2005, Mateus2006, Mateus2007}, it is reasonable to assume that an important fraction of our selected ELG are blue galaxies, which are generally classified as star-forming galaxies according to the WHAN and BPT diagrams, and that red galaxies are generally classified as LINERs or retired galaxies. A similar discussion with compatible results can be found in the works of \cite{Chies2015, RodriguezdelPino2017}. 

\subsection{The star formation rate}
\label{sec:SFR}

\begin{figure}
\centering
\includegraphics[width=0.65\textwidth]{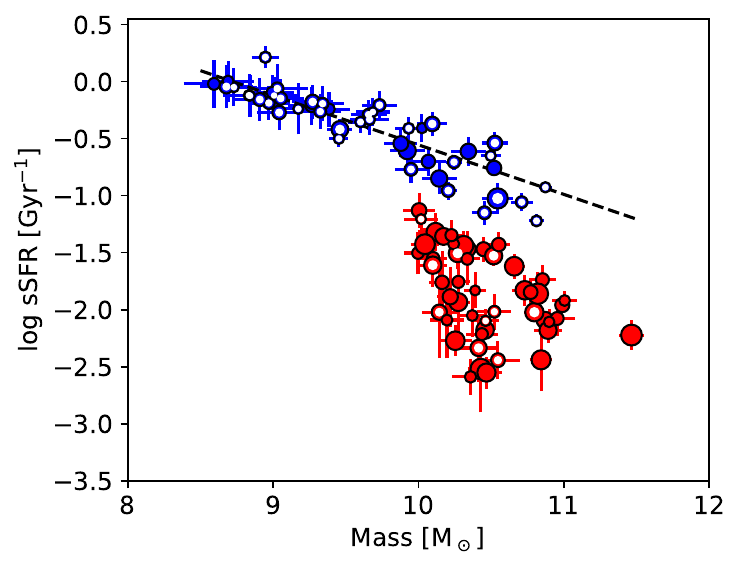}
\caption[Star formation main sequence of the galaxies in the cluster mJPC2470--1771]{Star formation main sequence. Red dots represent the red galaxies. Blue dots represent  blue galaxies. White dots represent the ELG population (selected in Sect.\ref{sec:ELG}) with inferred EW(H$\alpha)>6$~\AA. Dot size is proportional to the inverse distance to the BCG} 
\label{fig:main_sequence}
\end{figure}

In order to calculate the star formation rate (SFR) and the specific SFR (sSFR = SFR/M$_\star$), we use the star formation history (SFH), derived from the SED-fits to add up the stellar mass formed in the last 20 Myr and divide it by 20 Myr.\footnote{In this text, sSFR is expressed in units of Gyr$^{-1}$; and log sSFR is the decimal logarithm  of the sSFR}. The mean and standard deviation values we obtain for the sSFR for each set of galaxies are $0.25 \pm 0.32$~Gyr$^{-1}$ for the whole sample, then $0.020 \pm 0.016$~Gyr$^{-1}$ for red galaxies, $0.49 \pm 0.32$~Gyr$^{-1}$ for blue galaxies, $0.35 \pm 0.35$~Gyr$^{-1}$ for the ELGs, $0.016 \pm 0.015$~Gyr$^{-1}$ for the red ELGs, $0.50 \pm 0.32$~Gyr$^{-1}$r for the blue ELGs, and $0.48 \pm 0.34$~Gyr$^{-1}$ for SF galaxies. 

We find that the mean value of the sSFR of the blue galaxies is $\sim 25$ times larger than the mean value of the red galaxies. This is accordance with the literature \citep[see e.g.][]{Kauffmann2003a, Kauffmann2003b, Baldry2004, Brinchmann2004, Gallazzi2005, Mateus2006, Mateus2007}. The difference in the values obtained for the red galaxies and the red ELGs is negligible, as well as the difference between blue galaxies and blue ELGs. 

The star-forming main sequence \citep{Noeske2007} is a relation between the SFR and the stellar mass of galaxies in the form of a power law \citep[see e.g.][]{Elbaz2007,Speagle2014,Sparre2015,cano2016spatially, vilella2021j}. The work by \cite{Nantais2020} supports that the relation remains constant with density, and \cite{Speagle2014} and \cite{Santini2017} works find no variation with redshift in the slope, but \cite{Noeske2007} find variations with redshift.

We study the main sequence of the star formation in Fig. \ref{fig:main_sequence}.We find that blue galaxies are well placed in the main sequence. The low mass-and-high sSFR end of the main sequence is dominated by blue ELGs. This is compatible with young stars as the main mechanism of H$\alpha$ emission for blue galaxies (as seen in Sect.\ref{sec:AGN}). Meanwhile, red ELGs are mainly found in the low sSFR region, so their H$\alpha$ emission is probably due to other mechanism, such as the presence of an AGN. We fit a main sequence of the star formation using the SF galaxies. 
The obtained fit (see black dashed line in Fig. \ref{fig:main_sequence}) is: 
\begin{equation}
    \log \mathrm{sSFR}= 
    (-0.43 \pm 0.07)\log M_\star + (3.78\pm 0.64).  
\end{equation}

Translating this fit to SFR instead of sSFR, the obtained slope is $0.57$. These results are lower than the ones obtained by \cite{Gines2022} analysing the whole AEGIS field. In that work, they calculate the SFR through the H$\alpha$ flux and the SFH provided by \baysea. The values of the slope (in the SFR versus mass fit) are both higher than our results. This means that sSFR decreases more rapidly with mass for the galaxies in this cluster than for galaxies in lower density environments, which is the dominant population in AEGIS \citep{Rosa2022}. However, the SF galaxies with 
$\log M_\star <9.8$~$M_\odot$ shows a flatter slope that  would suggest that the sSFR is almost independent of the mass. This also holds true for the results from other works, such as \cite{boogaard2018muse}, \cite{vilella2021j}, \cite{puertas2017aperture}, \cite{renzini2015objective}, \cite{zahid2012census}, \cite{shin2021metal}, \cite{belfiore2016sdss}, \cite{cano2016spatially}, \cite{cano2019sdss}, and \cite{sanchez2018sdss}.

\section{Discussion}
\label{sec:discussion}

\begin{figure*}
\centering
\includegraphics[width=\textwidth]{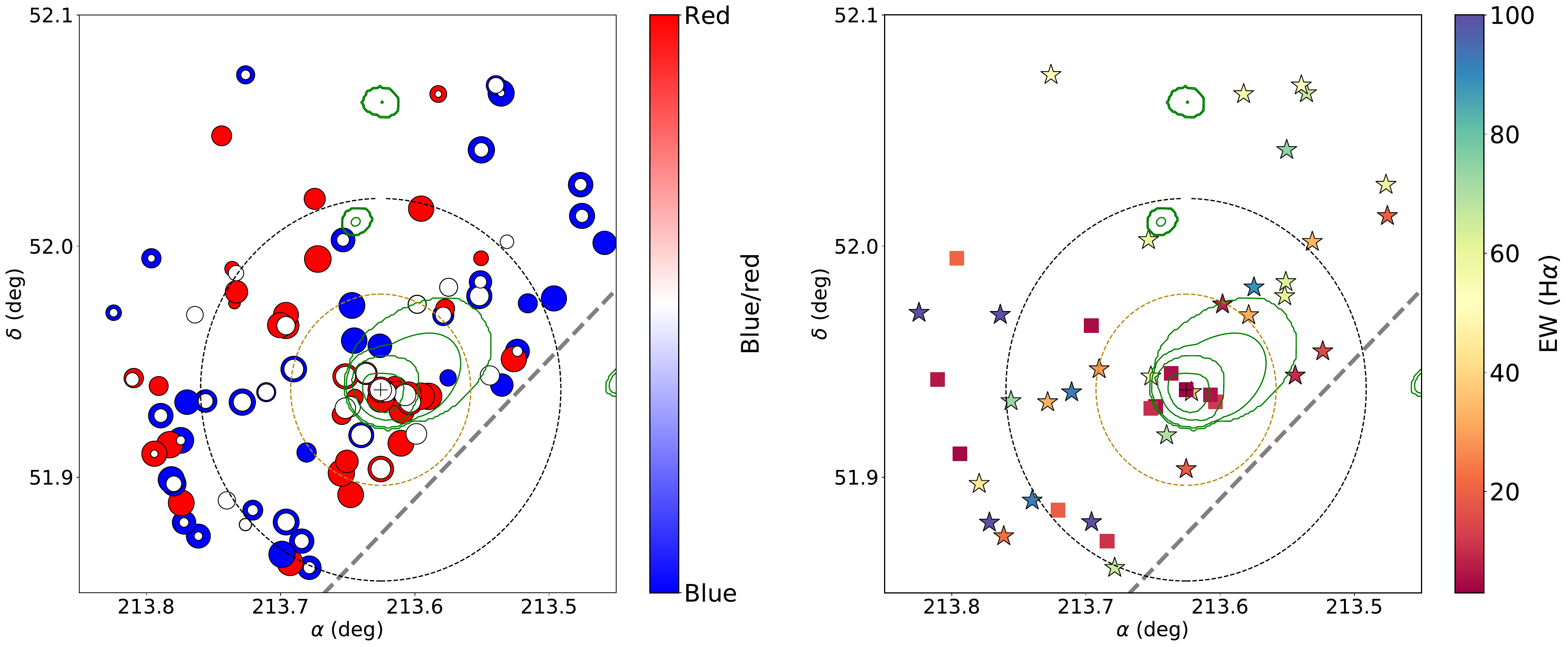}
\caption[Spatial distribution of the galaxy populations in the cluster]{Spatial distribution of the galaxy populations in the cluster. Left panel: Spatial distribution of the red, blue, and  emission line galaxy populations. Blue dots represent blue galaxies. Red galaxies are represented with red dots. White dots over red and blue dots represent the ELG. Dot size is proportional to the AMICO association probability. The dashed golden and black circles represents the $0.5$~R$_{200}$ and R$_{200}$ distances to the BCG, respectively. The grey dashed line represents the limit of the FoV of \mjp. The black cross represents the position of the BCG. Green contours represent the X-ray emission in the 0.5-2 keV energy band from XMM data \citep{Bonoli2020}. Energy levels are $3.654 \times 10^{-16}$, $1.218 \times 10^{ -15}$, $3.654 \times 10^{ -15}$ and $1.218 \times 10^{-14}$~ergs~s$^{-1}$~cm$^{-2}$ arcmin$^{-2}$. Rigt panel: spatial distribution of the SF and AGN. Stars represent SF galaxies and squares the AGNs. The color code represent the inferred} EW(H$\alpha$). The rest of the symbols are the same as in the left panel.
\label{fig:spatial_distribution_with_contours}
\end{figure*}

The spatial distribution of the galaxy populations and the variation of galaxy properties with the cluster-centric radius is a key piece of information in improving our understanding of the role of  environment for quenching the star formation in galaxies  \citep{Donnari2021,Dacunha2022, Niemiec2022}, Galaxy-galaxy or galaxy-ICM interactions are more efficient at the cluster centre, where the density of galaxies and the density of the gas are higher. Therefore, it is expected that environmental processes are more efficient inside the virial radius  \citep{Alonso2012, Raj2019}. However, other processes, such as galaxy harassment, ram-pressure stripping, and starvation can act outside the virialised region, being also effective at the cluster periphery  \citep{Bahe2013, Zinger2018, Lacerna2022}. The analysis of the galaxy populations, SFR and SFH have been proven to be very useful to study quenching and cluster formation scenarios \citep[see e.g.][]{vonderLinden2010}.

In this section we discuss the distribution of the galaxy populations within the cluster. The fraction of red, blue, and star-forming galaxies as well as AGN as a function of the radial distance to the cluster centre provide clues about the relevance of the environment and AGN feedback in the quenching of star formation. We also study the variation of the SFH parameters of galaxies that are in the central part ($r\leq 0.5$~R$_{200}$), with respect to outskirt regions ($r >0.5$~R$_{200}$); thus, the SFH-distance relation provides information about the accretion history and the differential quenching time scales. We finish the discussion with the variation of stellar population properties with cluster-centric radius; in particular,  the sSFR-distance relation traces how the environment-quenching process proceeds.

\subsection{Spatial distribution of the galaxy populations}

The 2D map distribution of the galaxy populations of the cluster is shown in  Fig.~\ref{fig:spatial_distribution_with_contours}.  Most of the red galaxies are located inside the inner region (d$<$ $0.5~R_{200}$ from the BCG), while half of the blue galaxies are located around $0.9~R_{200}$. This visual inspection is corroborated by the mean distance of the RG, which is $0.60$~R$_{200}$, while for BG is around $0.98$~R$_{200}$, which is almost the same as the mean distance of ELGs ($0.90$~R$_{200}$, $0.98$~R$_{200}$ for the SF galaxies, and $0.64$~R$_{200}$ for the AGNs). Moreover, the distribution of the ELGs is very similar to that of BGs, because most of the ELGs are BGs. This indicates that RGs are more prominent in denser environments than BGs, as seen in previous works \citep[see e.g][]{Balogh2004}.  Also, ELGs appear to show a more uniform distance distribution, which  is in accordance with the results of the literature, such as \cite{Haines2012, Haines2015, Noble2013, Noble2016, Mercurio2021}. However, our ELG population is not composed exclusively of SF galaxies, as there are also AGN candidates; this is justified by the presence of ELGs in denser environments. In fact, most of the AGNs are located in the central region, while the number of SF galaxies increases with distance.

\begin{figure*}
\centering
\includegraphics[width=\textwidth]{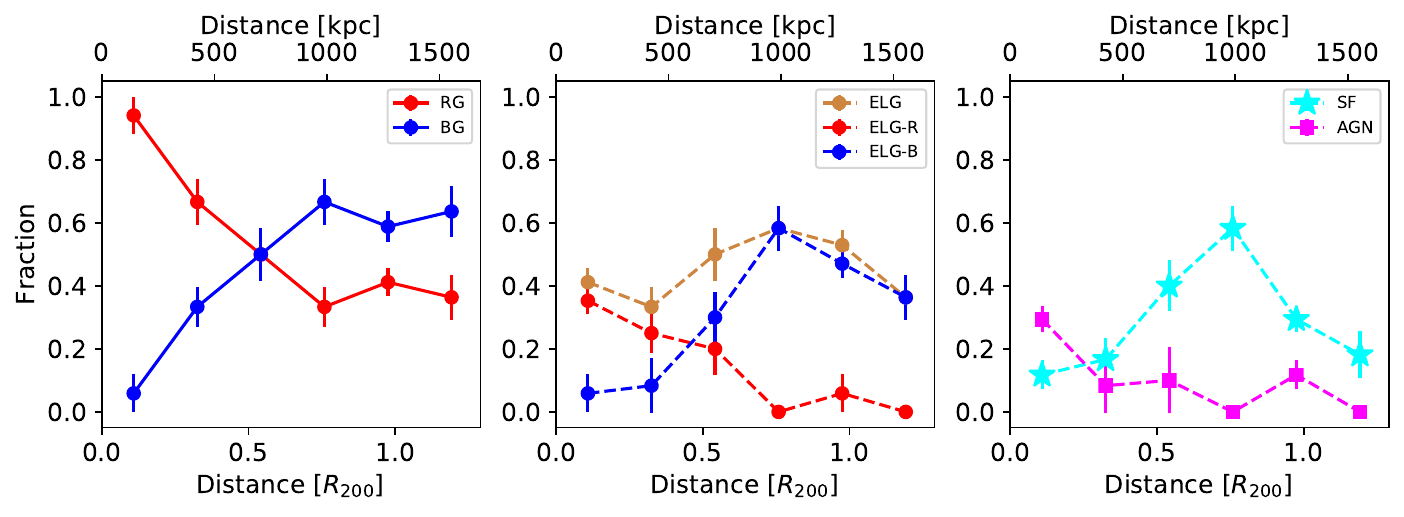}
\caption[Radial variation of the galaxy populations in the cluster]{Radial variation of the galaxy populations in the cluster. First panel: Radial variation of the fraction of red and blue galaxies. Red points represent the fraction of red galaxies, blue points represent the fraction of blue galaxies. Second panel: Fraction of ELG, red ELG, and blue ELG as a function of distance. Peru points are the total fraction of ELG, red points are the fraction of red ELG, and blue points are the fraction of blue ELG. Third panel: Fraction of SF galaxies and AGNs as a function of the distance to the BCG. Cyan stars represent the fraction of star-forming galaxies. Magenta squares represent the fraction of AGN.}
\label{fig:fraction}
\end{figure*}

To quantify the spatial variation of the galaxy populations, we discuss how the fraction of the different galaxy populations changes with the distance to the BCG (see Fig.~\ref{fig:fraction}). Red galaxies clearly dominate over blue galaxies in the inner regions (up to $d\sim 0.5$~R$_{200}$). At this point, we are not affected by the incompleteness of the observations. The fraction of red galaxies steeply decreases, as the fraction of blue galaxies increases with the distance, but the fraction of red galaxies remains above the value obtained by \cite{Rosa2021} even at distances larger than R$_{200}$. It is also interesting to note that the fraction of red galaxies is equal to the fraction of blue galaxies at $d\approx0.5$~R$_{200}$. Thus, we can conclude that inside the virialised region, the red galaxy population dominates over the blue one. 

The fraction of ELG slightly increases with distance, up to $d\approx$~R$_{200}$ and then decreases again. This decrease of the ELG fraction for $d>$~R$_{200}$ could be a consequence of the possible observational incompleteness of our sample at larger distances. The fraction remains below $0.6$ at all distances. This fraction is  below the $\sim$80$\%$ star-forming field AEGIS population at $z=0.2-0.3$ \citep{Gines2022}.
If we separate blue and red ELG, we find that most of the ELG are red in the inner areas, but their fraction rapidly decreases and is negligible at distances larger than $d\approx0.5$~R$_{200}$. In contrast, the fraction of blue ELG is negligible inside $d\approx0.3$~R$_{200}$, and then increases steeply. An almost equal behaviour is found for AGN and SF galaxy populations, respectively. As it happens for the blue ELG, the fraction of SF galaxies increases up to $0.6$ as the distance increases. Although it decreases to lower values at distances higher than $d\approx0.8$~R$_{200}$. This may be due to our incompleteness in the observations or to the presence of composite blue galaxies that cannot be clearly classified neither as SF nor as AGN.

Thus,  blue and star-forming galaxies are more common beyond the cluster virialised region, in agreement with other works, such as \cite{Haines2012, Haines2015, Noble2013, Noble2016, Mercurio2021}. As reported by \citet{Olave-Rojas2018}, we find that the fraction of red galaxies remains higher than in the field at distances larger than R$_{200}$. However, in contrast to \citet{Guglielmo2019}, we do not find a larger fraction of SF galaxies than blue galaxies, except for the most central region, where a reactivation of the SF may be taking place, due to the aforementioned mechanisms. The radial profile of the AGN fraction could compatible with the results by \citet{Peluso2022}, who find a significant abundance of AGNs in galaxies that have suffered ram-pressure stripping, taking into account the relation between ram-pressure stripping and the ICM density AGN feedback might play a role at the centre due to the increase of AGN fraction toward it; but this study could not explain the large fraction ($>$ 50$\%$) of red galaxies within 0.5~$R_{200}$. Other processes related to the environment, along with possible previous mass-quenching, may be acting in the cluster.

\subsection{Stellar population properties: Radial variation of the colours, mass, and ages}

\begin{figure*}
\centering
\includegraphics[width=\textwidth]{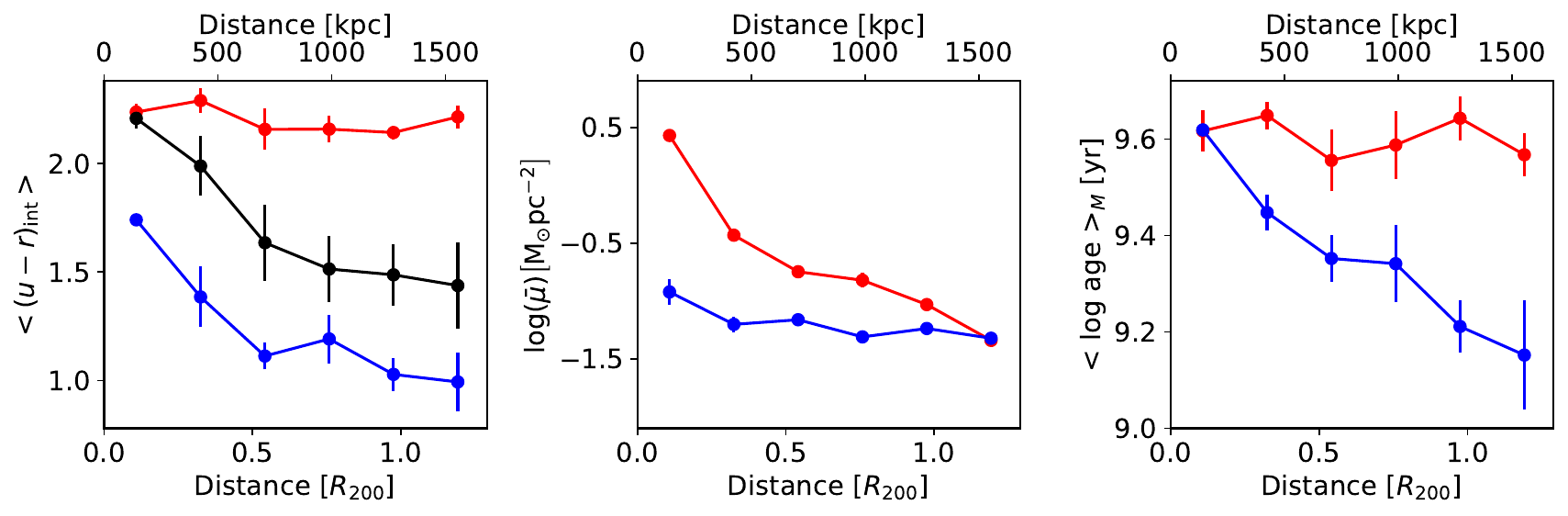}
\caption[Radial distribution of the stellar mass surface density, the mean mass-weighted age and the mean $(u-r)_\mathrm{int}$ colour]{Radial distribution of the stellar mass surface density, the mean mass-weighted age and the mean $(u-r)_\mathrm{int}$ colour. The grey dashed lines represent the limit of the FoV of \mjp. Red dots represent the values for the red galaxies and blue dots represent the values for the blue galaxies. Black dots represent the values of $(u-r)_\mathrm{int}$ for all the galaxies.
}
\label{fig:mass_age_radial}
\end{figure*}

The variation of the abundance of red and blue galaxies has an important effect on the radial variation of the galaxy properties, particularly on the colours and ages of the stellar populations. 
 
We studied the radial profiles of the $(u-r)_\mathrm{int}$ colour, the mass density (with the area corrected from mask an incompleteness effects), and the $< \log \mathrm{age} >_M$ (see Fig.~\ref{fig:mass_age_radial}). The colour, $(u-r)_\mathrm{int}$, decreases with the radial distance; however, this is in part due to the radial variation of the fraction of red and blue galaxies in the cluster. When the galaxy population is segregated in red and blue galaxies, we find that most of the red galaxies have very similar $(u-r)_{int}$, while blue galaxies become bluer going from the cluster centre and further out towards the edges. This is probably associated with a change in the age of the blue galaxy population. Red galaxies ages stay approximately constant at $< \log \mathrm{age} >_M\approx 9.6$. \ \footnote{This is the logarithm of the stellar ages expressed in yr}. Meanwhile, the mean age of BGs decreases as the distance to the BCG increases by about $0.3$~dex in the inner $0.5$~R$_{200}$. The stellar mass surface density of the red galaxies decreases with the distance to the BCG, reflecting that the most massive galaxies are sited in the inner $0.5$~R$_{200}$. Blue galaxies show a lower mass density and a smoother slope than RGs, but they show similar stellar mass density as the red galaxies beyond $0.5$~R$_{200}$.

Therefore, we find that red, more massive, older galaxies, are found in the inner areas while we find, blue, younger, and less massive galaxies in the outskirts. These properties are generally associated with galaxies in the red sequence (which also, show low values of star formation) and the blue cloud with usually higher levels of star formation than most galaxies in the red sequence \citep[see e.g.][]{Kauffmann2003a, Kauffmann2003b, Baldry2004, Brinchmann2004, Gallazzi2005, Mateus2006, Mateus2007}.

The stellar ages of the blue galaxies show a clear gradient with cluster-centric distance. However, the mean ages of the red galaxies is almost constant with the radial distance. It suggests that these galaxies were probably quenched earlier than their accretion to the cluster or during the first epoch of the accretion.

\subsection{SFH: Spatial variation}
\label{sec:SFH}

\begin{table}
    \resizebox{\textwidth}{!}{%
    \centering
    \begin{tabular}{c c c c c c c c c}
    \hline
    \hline
        {\small SP} & {\small RG} &  {\small BG}  &  {\small RG ($d<0.5$)}& {\small BG ($d<0.5$)} & {\small RG ($0.5<d<1$)} &  {\small BG ($0.5<d<1$)} & {\small RG ($d>1$) } &   {\small BG ($d>1$)} \\
     \hline
$t_0$ & $6.44 \pm 1.76$ & $6.00 \pm 1.72$ & $6.64 \pm 1.75$ & $7.69 \pm 0.17$ & $6.30 \pm 1.86$ & $6.37 \pm 1.59$ & $6.18 \pm 1.65$ & $5.33 \pm 1.66$\\
$\tau/t_0$ & $0.12 \pm 0.02$ & $0.94 \pm 0.63$ & $0.12 \pm 0.03$ & $0.46 \pm 0.21$ & $0.12 \pm 0.02$ & $0.83 \pm 0.49$ & $0.11 \pm 0.02$ & $1.14 \pm 0.70$\\
        \hline
    \end{tabular}}
    \caption[Averages and dispersions for the SFH parameters]{ Averages and dispersions for the SFH parameters.  Values are the mean and standard deviation of the red and blue galaxy members, and for galaxies inside $0.5$~R$_{200}$, between $0.5$~R$_{200}$ and R$_{200}$ and outside R$_{200}$.}
    \label{tab:SFHparam}
\end{table}

\begin{figure*}
\centering
\includegraphics[width=\textwidth]{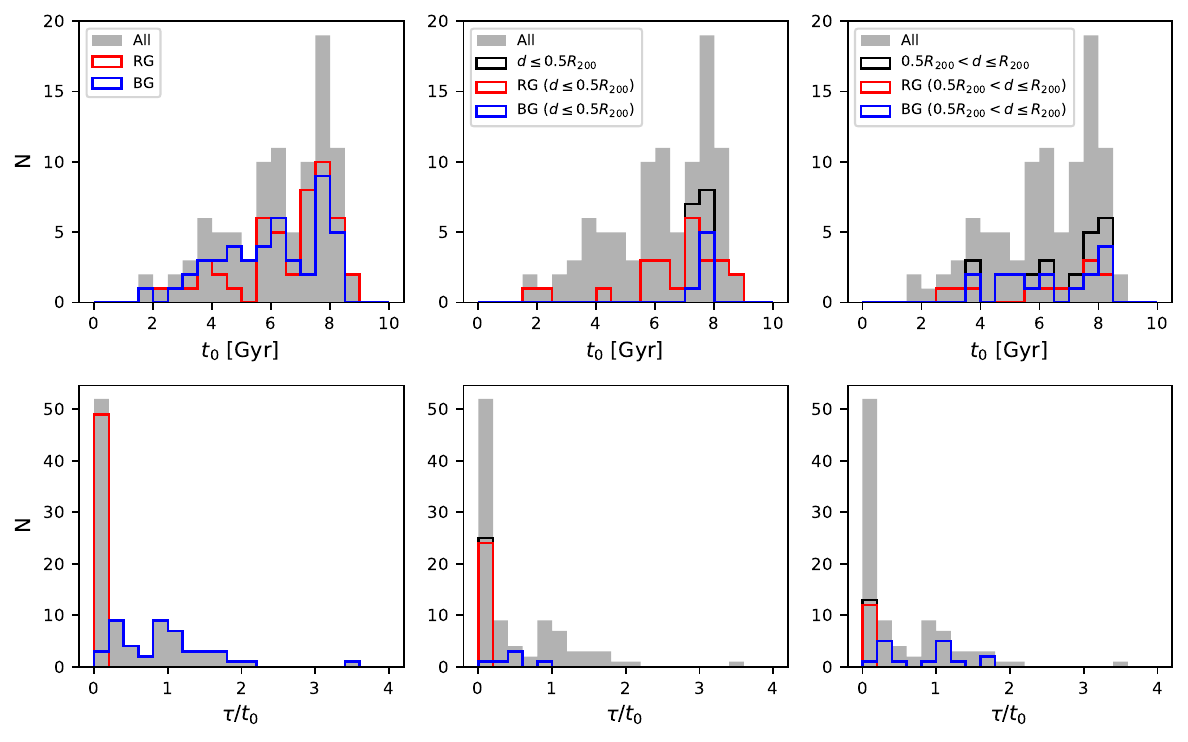}
\caption[Distributions of SFH parameters t0 (top row) and  $\tau$/t0 (bottom row), shown from left to right]{Distributions of SFH parameters t0 (top row) and  $\tau$/t0 (bottom row), shown from left to right. Grey histograms represent the distribution of all the galaxies in the cluster in all panels. Red, blue, and black histograms represent the distribution of red, blue and all galaxies at different cluster-centric distances: All the galaxies in the cluster (first column), galaxies within $0.5$~R$_{200}$ (second column), and galaxies within $0.5$~R$_{200}$ and R$_{200}$ (third column)
}
\label{fig:hist_t0_tau}
\end{figure*}

We go on to investigate the spatial variation of the SFH of the cluster galaxies. Before doing that, however, we comment here on the uncertainties involved.

The reliability of our methodology has been previously assessed in \citet{Rosa2021}, where we have shown that the SFH of a complete sub-sample of \mjp{} galaxies selected at $z \sim 0.1$ constrains the cosmic evolution of the star formation rate density up to $z \sim 3$, producing results in good agreement with those derived from cosmological surveys. We have further shown that the galaxy properties (stellar mass, ages, and metallicity) can be inferred by fitting the \js{} with the non-parametric codes \muff{}, \alstar{}, and \tgas{}, and the results are similar to those obtained by \baysea{} using the same delayed-$\tau$ model used in this work (see Table 1 in \citealt{Rosa2021}).

Reassuring as these statistical results may be, we should not lose sight of the inherent difficulties in estimating SFHs of individual galaxies \citep[e.g.][]{ocvirk2006}. Our Bayesian analysis, based on an analytical prescription for the SFH (Eq.\ \ref{eq:SFH}), is just one out of a vast spectrum of alternative approaches. Parametric models such as our delayed-$\tau$ model are known to lack flexibility to emulate the diversity of SFHs in galaxies \citep[e.g.][]{dressler2016,pacifici2016} and to induce considerable biases in some cases \citep[e.g]{lower2020}. Non-parametric models alleviate these problems, but the higher dimensionality associated with the added flexibility requires extra care when specifying the priors, which may have a significant impact in the estimated properties \citep{leja2019}. Moreover, it has been previously shown that only a few characteristic episodes in the SFH can be retrieved from the SED-fitting \citep{cid2005, ocvirk2006, asari2007,  tojeiro2017}. 

Despite all the caveats involved, our previous analysis of $\sim$8000 \mjp{} galaxies, where we detected only small differences between properties derived though parametric and non-parametric codes (significantly below the 0.4 dex in \logM{}, inferred for the SED-fitting of broad band photometry of mock data of cosmological galaxy formation simulations by \citealt{lower2020}) gives us confidence that we can use our results to investigate the general trend of the spatial variation of the SFH among the galaxy cluster members. The comparative and statistical nature of this analysis further alleviates worries associated with the SFH parameters derived for each galaxy; however, as discussed above, this approach should be treated with caution.

We focus on two parameters for this study: $t_0$, the lookback time when the star formation began, and $\tau/$t$_0$, a measure of the extent of the star formation that is better constrained than $t_0$ or $\tau$. Table \ref{tab:SFHparam} and Fig.\ \ref{fig:hist_t0_tau} summarise our results. We divide the galaxies into blue and red ones once again. We further divide galaxies by their (projected) distances to the BCG into smaller than $0.5$~R$_{200}$, between $0.5$~R$_{200}$ and $1$~R$_{200}$, and larger than R$_{200}$ bins. This allows us to distinguish the effect of the environment in the SFH for the central virialised area and the outer regions.

The parameter $t_0$ shows similar values for red and blue galaxies at all  distances. This would suggest that most galaxies started forming stars roughly at the same epoch (around $\sim 6.5$~Gyr). The main differences appear in the blue galaxies within and outside of 0.5$~R_{200}$. Blue galaxies in the inner region show the highest mean value of $t_0$, but they are very few and this value is compatible with the one obtained for red galaxies at similar distances. Blue galaxies in the outer region ($d >$~R$_{200}$) show a lower value for $t_0$. This could be a consequence of these galaxies being in the cluster infall region \citep{Rines2006}.

Since values of $t_0$ are very similar for most galaxies all over the cluster, we interpret the low values of $\tau/\mathrm{t}_0$ as short episodes of star formation and large values of this parameter as star formation processes that are more extended over time. Red galaxy values of $\tau/\mathrm{t}_0 \sim 0.12,$ no matter their distance to the BCG,  suggest that their star formation was shut down very fast. On the contrary, blue galaxies show larger values of this fraction than red galaxies at all distances and there is a clear increase in $\tau/\mathrm{t}_0$ as the distance to the BCG increases.

Thus, these results suggest a faster quenching process for blue galaxies in the dense (inner) regions;  while red galaxies might be quenched earlier on and independently of the distance to the cluster centre in an earlier cluster accretion epoch. Moreover, the quenching of the star formation of red galaxies might be linked to the AGN or galaxy stellar mass, rather than to the environment because at the smaller cluster-centric distance is where we find the most massive galaxies and the fraction of AGN is larger.

\subsection{sSFR: Radial variation}

\begin{figure}
\centering
\includegraphics[width=0.55\textwidth]{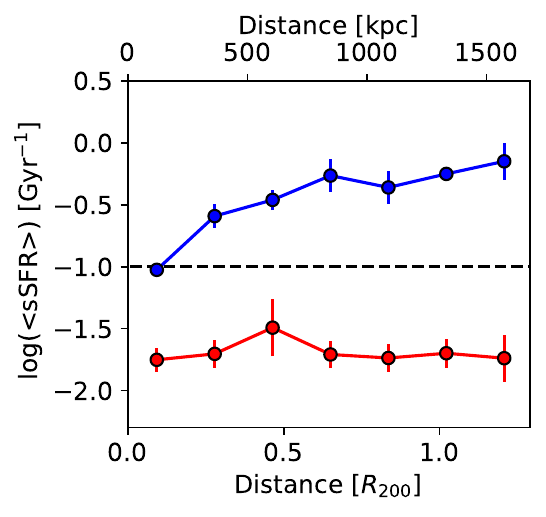}
\caption[Radial distribution of the mean sSFR. The gray dashed lines shows the limit of the FoV of \mjp]{Radial distribution of the mean sSFR. The gray dashed lines shows the limit of the FoV of \mjp. The black dashed line shows \cite{Peng2010} criteria to distinguish among star-forming and quiescent galaxies. Red dots represent red galaxies. Blue dots represent  blue galaxies. }
\label{fig:sSFR}
\end{figure}

From the SFH parameters, we can conclude that red galaxies have already been quenched independently of their position in the cluster; whereas for blue galaxies, the quenching process proceeds from the inner to the outer regions of the cluster. To confirm this conclusion, we study the variation of the sSFR with the distance (see Fig.~\ref{fig:sSFR}). It is worth noticing that blue galaxies have a sSFR at all the cluster-centric radius that are above $sSFR=0.1$~Gyr$^{-1}$, which is the threshold adopted by \cite{Peng2010} to differentiate star-forming galaxies from quenched galaxies. The mean sSFR of blue galaxies also clearly increases from inside-out of the cluster; in contrast, the mean sSFR of the red galaxies remains constant within the error bars. This  suggests  that red galaxies have quenched before their accretion to the cluster or were quenched within it in an earlier accretion epoch, while blue galaxies are still in the process of quenching. This could be related to the pre-processing effects during the infalling processes \citep[assuming galaxies are incorporated in substructures already evolving; see e.g.][]{Gavazzi2003, Aguerri2017, Donnari2021}. Our results differ from those of \citet{Knowles2022}, who find no dependence of the star formation with the cluster-centric distance -- but the distinction between red and blue galaxies is key for this result. In particular, the results of \cite{Balogh1999}  (in a redshift range similar to that of our cluster)  show that the last episode of star formation is more recent for galaxies in the outskirts than in inner regions.

\subsection{On the pre-processing scenario}

To sum up, all the galaxies were formed around the same epoch (with the exception of some outer blue ones). Red galaxies had shorter star formation periods and have a similar SFH, independently of their position in the cluster. Meanwhile, blue galaxies are still forming stars or have been forming them until very recently, and galaxies in the inner regions are quenching faster than in the outer ones. However, red galaxies were quenched earlier on, independently of their position on the cluster. These  results suggest different evolutionary paths and accretion histories for red and blue galaxies.

Illustris cosmological simulations have shown that pre-processing plays a relevant role in quenching galaxies \citep{Donnari2021}. They find that satellites can be quenched before infalling in dense environment, or after being accreted into any host; or while being members of pre-processing hosts other than the actual one where they are found today. AGN feedback and mass-quenching may be acting in the pre-processing host phases. This is a possible scenario for explaining the spatial variation of the SFH of red and blue galaxies, their abundance, AGN fraction, and variation of the galaxies properties with the radial distance to the cluster centre.

Another scenario is the 'slow-then-rapid' quenching \citep[see e.g.][]{Maier2019, Roberts2019, Kipper2021}, whereby galaxies undergo slow quenching processes and, once they enter a dense environment, start a faster quenching phase. Results from \cite{Pallero2022}, using the \texttt{C-EAGLE} simulation support this scenario. They show that these processes usually become relevant at $\sim \mathrm{R}_{200}$, where the ICM reaches a density high enough for ram-pressure stripping to  become relevant. They also find that the fraction of galaxies quenched in situ, in comparison the fraction of galaxies quenched because of pre-processing, decreases as $M_{200}$ increases. 

A combination of both scenarios may serve us to interpret the spatial and radial distributions of the stellar population properties that are found. We identified red galaxies with very similar properties along the whole cluster. Red galaxies may quenched within smaller structures that were later accreted to the cluster.  According to the results from \citet{Pallero2022} , for a cluster of this mass ($M_{200}=3.3\times 10^{14} ~ \mathrm{M}_\odot$), we would expect a similar fraction of galaxies quenched inside the cluster and quenched via pre-processing. However, we find that $\sim 73$~\% of the red galaxies are within R$_{200}$, so in order to be compatible with these results, some of the inner red galaxies would have to be part of a different halo. On the other hand, the results from \citet{Donnari2021}  show that the pre-processing scenario is relevant for low mass galaxies and that massive galaxies quench on their own.  

If we assume that some of the blue galaxies belong to the original halo and some have been accreted later, we could explain the behaviour of blue galaxies and the greater dispersion of their properties, as well as the greater amount of quenching among inner ones. However, \citet{Pallero2022} estimate that the quenching timescale for galaxies once the in-fall beyond R$_{200}$ is $\sim 1$~Gyr, but our estimations of $\Delta t_q$ are larger for blue galaxies, and only some of the red galaxies are compatible with these values, regardless of their distance to the cluster centre. These suggest that the accretion and evolution scenario may be more complex and a different model is required.

\subsection{Quenched fraction excesss}
\begin{figure}
    \centering
    \includegraphics[width=0.55\textwidth]{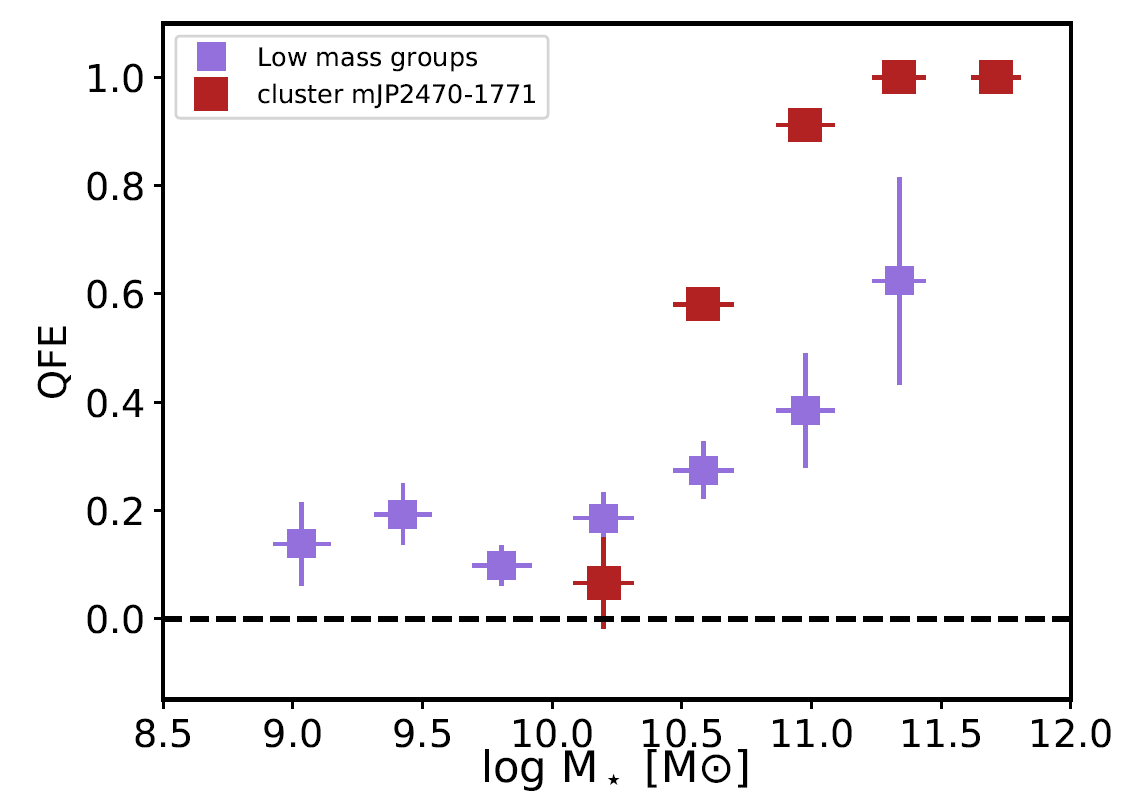}
    \caption[Comparison of the quenched fraction excess derived from \mjp \ groups with $M_{\mathrm{group}}<5 \times 10^{11}$~$\mathrm{M_\odot}$ and for the cluster mJPC2470--1771. Picture taken from \cite{Rosa2022}]{Comparison of the quenched fraction excess derived from \mjp \ groups with $M_{\mathrm{group}}<5 \times 10^{11}$~$\mathrm{M_\odot}$ and for the cluster mJPC2470--1771. Picture taken from \cite{Rosa2022}.}
    \label{fig:QFE}
\end{figure}

We finish the discussion of this chapter by recovering a result from \cite{Rosa2022}, work co-led by the author of this thesis, that is also related to the cluster studied in this chapter. Following the definition by \cite{McNab2021}, we use the quenched fraction excess:

\begin{equation}
    \mathrm{QFE} = (f^{\mathrm{F}}_{SF} - f^{\mathrm{G}}_{SF})/f^{\mathrm{F}}_{SF},
\end{equation}

\noindent where QFE is the quenched fraction excess, $f^{\mathrm{F}}_{SF}$ is the fraction of star--forming galaxies in the field and $f^{\mathrm{G}}_{SF}$ is the fraction of star--forming galaxies in groups. Others works have used equivalent parameters, such as the environmental quenching efficiency \citep{Peng2010,Wetzel2015,Nantais2017,vanderburg2018}, the transition fraction \citep{Bosch2008}, or the conversion fraction \citep{Balogh2016,Fossati2017}. These parameters illustrate the role that the environment has in quenching the star-forming galaxies, turning them into quiescent ones. We compare the QFE obtained for the cluster and for the low mass groups ($M_{\mathrm{group}}<5 \times 10^{11}$~$\mathrm{M_\odot}$) of the \mjp \ survey in Fig.~\ref{fig:QFE}. The comparison shows that, for galaxies with masses $\mathrm{M}_\star > 10^{10.25}$~$M_\odot$, the QFE found for the cluster is around twice larger than for groups with lower masses. This is in great agreement with the results by \cite{Donnari2021,Donnaris2021compare} who, using the IllustrisTNG simulations, find that the fraction of quenched galaxies with mass $\mathrm{M}_\star > 10^{10.25}$~$M_\odot$ increases with the halo mass of the group, from $0.2$ up to $0.7$, but remains approximately constant for massive haloes ($M>10^{14}$). Along with the results from \cite{Rosa2022}, we interpret this findings as a consequence of the deeper gravitational potential well of clusters, which favours galaxy--galaxy interactions and other interactions, such as ram pressure stripping, which has indeed been shown to be more efficient in clusters \citep[see e.g.][and references therein]{Singh2019} or harassment, also more efficient in clusters \citep[see e.g.][and references therein]{Bialas2015}. These processes along with the pre-processing scenario already discussed, lead to an increased quenching effciency in clusters.

\section{Summary and conclusions}
\label{sec:conclus}
In this chapter, we study the stellar population properties of the \mjp \ cluster mJPC2470-1771 using the \jp \ photometric filter system. Its redshift is $z=0.29$, its mass $M_{200}=3.3\times 10^{14} ~ \mathrm{M}_\odot$, and its radius $R_{200}=1.304 ~ \mathrm{kpc}$.  We used the fossil record method for stellar populations and we analysed the SEDs (\js) of the galaxy members of the cluster. The  cluster was detected and its members were selected using the AMICO implementation for \mjp \ \citep{Maturi2023} based on  \photozbest{}, with an ultimate selection of 99 objects.

We used the \baysea \ code to fit the stellar continuum and constrain the stellar population properties by assuming a delayed-$\tau$ model for the SFH. 
The parameters obtained with \baysea \ are the stellar  mass,  the  metallicity,   and the  extinction $A_V$,  $t_0$,  and $\tau$. We used these parameters and fittings to calculate the mass- and light-weighted ages and the extinction-corrected rest frame $(u-r)$ colours. 
We established a set of criteria to select the ELG population, using the median error of the closest filters to H$\alpha$ wavelength and the predictions for the EW(H$\alpha$), EW([NII]), EW(H$\beta$), and EW([OIII]) made with \cite{Gines2021} ANN. We used the WHAN and BPT diagrams to separate SF, AGNs, and quiescent-retired galaxies.
We studied the spatial distribution of the stellar population properties in the cluster as well as the radial distribution of the abundances of red, blue, SF galaxies, and AGN hosts. The main conclusions of our analysis are as follows:

\begin{itemize}
    \item{We observe a fraction of red galaxies (52~\%) that is larger than that in the whole AEGIS field set of galaxies with redshift $0.25<z<0.35$, which is $\sim 20\%$. The distribution of the stellar population properties in the mass-colour diagrams is the same as the whole set.}
    
    \item{We selected a total of 48 ELG. These are dominated by young galaxies and most of the blue, less massive galaxies have been selected as ELGs. There are red galaxies in this set, showing the lowest inferred values of EW(H$\alpha$), being the median value equal to 8.96 \AA. 65.3\% of these galaxies are probably star-forming galaxies, while 24.4\% could be AGNs and the rest could be SF, AGNs, or composite galaxies.}
    
    \item{The red, older, more massive galaxies are mainly located in the inner part ($d< 0.5$~R$_{200}$) of the cluster, where the density is higher. The blue, and SF galaxies are more numerous at ($d> 0.5$~R$_{200}$), and their abundance increases with radial distance, being equal to the red galaxy fraction at $d\sim 0.5$~R$_{200}$. The abundance of the AGNs population decreases with the radial distance and it is higher at the cluster centre.}
    
    \item{In analysing the SFH, we find that galaxy members were formed roughly at the same epoch, but blue galaxies have experienced more recent star formation periods. Our results are compatible with a scenario where red galaxies are quenched prior to the cluster accretion or an earlier cluster accretion epoch; whereas blue galaxies may be in the transition to be quenched. This is also supported by the radial distribution of the red and blue galaxy populations, because the mean stellar age remains constant for red galaxies, but decreases for blue galaxies along with the distance to the BCG.} 
       
    \item{The sSFR of the red galaxies is almost constant with radial distance at sSFR $\sim 0.02$ Gyr$^{-1}$. The sSFR of blue galaxies decreases with the cluster-centric radius from sSFR $\sim 0.1$~Gyr$^{-1}$ to above $\sim 0.7$~Gyr$^{-1}$ beyond $0.5$~R$_{200}$.  This suggests that the quenching of blue galaxies is progressing from the inside-out of the cluster.
    AGN feedback or mass (or both) might also be intervening in the quenching of red galaxies. }

\end{itemize}

Our results show that the environment plays a relevant role in galaxy evolution, mainly manifested through a larger fraction of red, more massive galaxies in denser regions, as well as larger fraction of AGN hosts, lower, SFRs, and shorter star formation episodes, compared to less dense regions. This also shows the power of \jp{} in studies of the role of the environment in galaxy evolution. We developed a methodology that will be transferable to \jp \ data, providing more solid results regarding the relation between galaxy evolution and the environment.   









\chapter{Future work} \label{c5}

\epigraph{\textit{It isn't where you came from, it's where you're going that counts.}
}{Ella Fitzgerald}

The work carried out along this thesis has shed light into the question on what is the role played by the environment on the evolution of galaxies. However, with new answers, new questions arise, paving the way for new works and new studies. The future work related to this thesis can be divided into more technical works, and scientific works.

Most of the technical future work is related to \PyDJ. Some improvements and functions can be implemented in order the add more flexibility into certain aspects, such as the desired stamps of the galaxies, which could take into account the orientation of the galaxy, or limiting the download to certain set of specified filters, as well as improving the compatibility with external masks files defined by the user, like the \texttt{ds9} region files, for example. It could also be useful to include further information concerning the images of each filter in the header of the cubes, and we might need to explore ways to optimise the size of the final files in the hard drive, in order to be prepared for the large amount of data that is to come from the \jp \ data releases. 

However, the most interesting improvement would be without doubt making a routine that deconvolves the \gls{PSF} of the stamps, instead of degrading them to match the worst PSF. There are already published works in this regard, for example those by \cite{Starck2002,Prato2012,Farrens2017,Sreejith2023,Wang2023} that are worth of studying and testing, although additional modifications or considerations might be required to fit our data and needs. If this objective was achieved, we would be able to improve the spatial resolution of our analysis and the size of the regions would be limited by the S/N ratio rather than the PSF. This way, we could take the make the most of \jp \ pixel scale and, for example, obtain more precise radial profiles, which we have shown that can be smothered because of larger bins.  

The tools and methodology developed in this thesis have been proven to be precise and solid. Since the \jp \ data acquisition will be done using the JPCam instead of the Pathfinder camera, more tests and calibrations will be required in order to provide reliable catalogues. Our tools can be helpful during this data validation processes. 

Another technical work, but with strong scientific relation and implications, is to develop an efficient segmentation code that divided galaxies into regions based on the underlying physics. We showed in Chapter~\ref{chapter:MANGA} that Voronoi binning code by \cite{Voronoi} provides non-physical regions in the outer parts of elliptical galaxies \citep[see also ][]{SanRoman2018}. On the other hand, the \gls{BatMAN} code by \cite{BATMAN} provides regions that resemble more the expected structures of a galaxy, but its computing time is too long for large galaxies or large data samples. 

From the scientific point of view, the most important caveat in our analysis is the small size of our samples, both in the study of the spatially resolved galaxies and in the study of the galaxy populations in galaxy clusters. However, we have taken advantage of this limitation by carrying out a more detailed analysis that has allowed us to develop a methodology that takes into account other details of the data that might have been overlooked otherwise. More data would also allow us to divide galaxies into mass bins, which would let us to better distinguish between the effects of the mass and the environment.

In Chapter~\ref{chapter:cluster} we discussed some caveats that might be caused by the use of a parametric SED-fitting code. Our tool now allows us to further study the possible consequences of using a parametric code, like \baysea, and a non parametric code, like \muff \ \citep{Luis2015,Luis2023}, and constrain the actual impact of that different assumptions on the \gls{SFH} can have on the results on a smaller scale. 

Lastly, the properties derived from the regions of the spatially resolved galaxies could be used as a training input for machine learning and artificial intelligence algorithms. In particular, it might be possible to build a foundation model that would be able to predict different parameters and perform several tasks, even helping to disentangle the nature of the physical processes that take place at smaller scales in galaxies.  

\chapter{Summary and conclusions} \label{chapter:conclussion}
\epigraph{\textit{La satisfacción del deber cumplido}}{Julio Rodríguez Soto \\ My grandfather}

Throughout this thesis we have used the available data from the \mjp \ survey to study the role of  environment on galaxy evolution. For this purpose, we have used two different approaches. Firstly, we have taken advantage of its large \gls{FoV} and its photometric filter system, which provide excellent data for unbiased IFU--like studies in different environments. On the other hand, we studied the galaxy population in the most massive galaxy cluster detected in the miniJPAS footprint, the cluster mJPC2470--1771.

The first step to deal with the requirements for the IFU-like study of the \mjp \ galaxies was to develop a tool that automatised all the required processes for the analysis of the data. These mainly include the download of scientific tables and images, the masking of nearby sources, the PSF homogenisation of the images in the different filters, the segmentation of the galaxy into regions, the computation of the values of the fluxes and magnitudes of the different regions, and the estimation of the stellar population properties and line emission through external codes. These steps, explained and tested in Chapters~\ref{chapter:code} and \ref{chapter:MANGA}, are essential for the study of the properties of the spatially resolved galaxies.

After testing our methodology, in Chapter~\ref{chapter:spatiallyresolved} we run our tool on the sample of spatially resolved galaxies in \mjp, selecting those galaxies without biased photometry due to nearby sources in any of the bands, and that are also large enough in comparison to the \gls{FWHM} of the worst \gls{PSF}, amounting to a total of 51 galaxies. Additionally, using results from the SED-fitting of the integrated photometry, as well as the AMICO catalogue of galaxy groups by \cite{Maturi2023}, we further classified our sample into 15 galaxies that are red in the field, 9 that are red and are in  groups, 21 that are blue and in the field, and 6 that are blue in  groups. With this classification we studied the stellar population properties and the emission lines of the regions of these spatially resolved galaxies, as well as the effect of  environment, by comparing the properties of the four categories of galaxies, using stellar surface mass density--colour diagrams, radial profiles of the properties, stellar population gradients, as well as studying the relation between some of these properties and comparing the \gls{SFH} of inner and outer parts of these galaxies. 

Lastly, we study the effect of  environment from the point of view of the integrated galaxies in Chapter~\ref{chapter:cluster}. For such purpose, we study the properties of the galaxies in the cluster mJPC2470--1771, also selected using the probabilistic association provided by AMICO. We select a total of 99 galaxies. Using once again the SED--fitting code \baysea, as well as the \gls{ANN} trained by \cite{Gines2021}, we are able to study the stellar population properties of these galaxies, as well as their \gls{SFR}, \gls{SFH}, end emission lines (H$\alpha$, H$\beta$, $\mathrm{[NII]}$, and $\mathrm{[OIII]}$), as well as classifying them into star-forming galaxies and AGN-host candidates via the WHAN and BPT diagrams.

The main conclusions of this thesis are:

 \begin{itemize}
    \item Our tool, \PyDJ, effectively automatises the analysis of the properties of the spatially resolved galaxies in \mjp, providing solid measurements of the fluxes and magnitudes of the regions of spatially resolved galaxies. In particular, the values obtained with our tool for different apertures are consistent with the magnitudes of the \mjp \ catalogues, which were obtained using \sext, with median differences below $0.1$~mag, that is, a flux discrepancy below 10\%.
    \item Using the largest galaxy in the \mjp \ data release, 2470--10239, our tool also produces flux and magnitude measurements that are consistent with those obtained using \pycasso \ and spectroscopy from the \gls{MaNGA} survey. Furthermore, the estimated stellar population properties estimated with \pycasso \ and \starlight \ are also compatible with our results obtained using \PyDJ \ and \baysea \ within the uncertainty intervals. 
    
    \item The homogenisation of the \gls{PSF} greatly improves the obtained \js \ of the central regions, by removing aperture effects that appear as non-physical variations from band to band. This also improves the accuracy of the SED-fitting of these regions and reduces the SED-fitting residuals. 
    
    \item The SED-fitting of the \js \ of the regions of the spatially resolved galaxies yield  residuals without significant bias in any band. For regions with $\mathrm{S/N}>5$, residuals are generally below $0.05$~mag, which translate into a relative difference lower than 5~\%. The errors are generally slightly overestimated ($\sim 20$~\%), most likely due to the assumption of an error of $0.04$ in the \gls{ZP}. For S/N ratios lower than 5, the residuals increase notably and are generally biased towards values that indicate an underestimation of the flux, and the errors become underestimated too. 
    \item The stellar population properties of the regions, as well as the \gls{EW} and ratios of the emission lines are properly defined by their loci on a mass density--colour diagrams, in a equivalent way to their integrated counterparts. We find that redder, denser regions are usually older, more metal rich, and show lower values of the $\Sigma_\mathrm{SFR}$ and \gls{sSFR}  than bluer and less dense regions. The higher extinction $A_V$ values are found in blue, dense regions, and metal-rich regions. Additionally, the regions of red and blue galaxies remain clearly separated in these diagrams, but there is no evidence of environment effects.
    \item The radial profiles of the stellar population properties are compatible with those found in the literature. In particular, we find that the stellar mass surface density clearly decreases towards outer parts, as well as the $(u-r)_\mathrm{int}$ colour, the intensity of the \gls{SFR}, and the stellar ages of blue galaxies. On the other hand, the stellar ages of red galaxies remain approximately constant, and the \gls{sSFR} increases toward the outer regions of the galaxies. The radial profiles of red and blue galaxies are distinguished, but we do not find significant effect of the environment. 
    \item The gradients of most properties show a weak dependency on the stellar mass, but not on the galaxy spectral type or environment. 
    \item The results found for the \gls{SFH} of inner and outer parts of the galaxy, along with the radial profiles of the ages, support an inside--out formation scenario, as well as inside--out quenching mechanism.
    
    \item The  fraction of red galaxies in the mJPC2470--1771 cluster is 52~\%, which is more than the double of the fraction of red galaxies in the whole AEGIS field set of galaxies with redshift $0.25<z<0.35$, which is $\sim 20\%$. 
    
    \item The properties of the galaxies in the cluster are properly defined by their loci in the mass--colour diagrams. For the same colour and stellar mass, the properties of the galaxies are similar to their counterparts in the field at a similar redshift.
    
    \item We  are able to detect a total of 48 ELG in the cluster, which are mainly young and blue galaxies. Using the WHAN and BPT diagrams, we find that a 65.3\% fraction of these galaxies are likely star-forming galaxies, while 24.4\% may be AGNs and the rest may be SF, AGNs, or composite galaxies.
    
    \item The fraction of red galaxies increases towards the cluster centre, as well as the abundance of AGN hosts. On the other hand, the fraction of blue and SF galaxies increases towards outer regions of the cluster. 

    \item The median \gls{sSFR} of red galaxies is almost constant ($ \log \mathrm{sSFR} \sim 0.02$~[Gyr]$^{-1}$). However, the \gls{sSFR} of blue galaxies shows a dependency on the distance to the centre of the cluster, rising from sSFR $\sim 0.1$~Gyr$^{-1}$ at inner regions, up to  $\sim 0.7$~Gyr$^{-1}$ at larger distances than $0.5$~R$_{200}$. These results suggest an inside--out quenching scenario in terms of the cluster-centric distance.
    
    \item The \gls{SFH} of the galaxies in the mJPC2470-1771 cluster show that they were formed roughly at a similar epoch, where blue galaxies have remained blue due to  recent star formation episodes. Our results are compatible with a scenario where red galaxies are quenched prior to the cluster accretion or an earlier cluster accretion epoch; whereas blue galaxies may be in the transition to be quenched. This is also supported by the radial distribution of the red and blue galaxy populations, because the mean stellar age remains constant for red galaxies, but decreases for blue galaxies along with the distance to the BCG.   

 \end{itemize}

 The general picture that can be drawn from these conclusions is that the environment does  play a role on galaxy evolution. However, this role is mostly reflected in the galaxy populations found in more dense regions, such as clusters and groups. A larger fraction of red, quiescent galaxies is found in this type of environment, but the properties of these galaxies are similar to their counterparts in the field, for a given stellar mass and colour. Regarding  blue galaxies, this is also true, although the distribution of their properties is shifted towards more massive and redder values. Similarly, the properties of the regions of the spatially resolved galaxies are well determined by their colour and stellar mass density, but the mass of the groups and the galaxy number density is not high enough as to trigger any significant difference on the properties of the regions of galaxies in groups when compared to galaxies in the field. Additionally, processes such as the ram pressure stripping depend on the density of the gas of the ICM, which is lower in groups than in clusters. This may be a consequence of the typical mass of the groups in our spatially resolved sample, since it might not be high enough to produce a significant effect on the properties of the galaxies, as opposed to the case of massive galaxy clusters. The  importance of the mass of the group or the cluster is observed for example in the quenched fraction excess, which is significantly larger in the cluster than in the low mass groups found in miniJPAS.

\addcontentsline{toc}{section}{\bf ~~~~~~~~~~~REFERENCES}
\bibliographystyle{references/aa}
\bibliography{main}

\begin{thebibliography}{521}
\expandafter\ifx\csname natexlab\endcsname\relax\def\natexlab#1{#1}\fi

\bibitem[{Abadi {et~al.}(2015)Abadi, Agarwal, Barham, Brevdo, Chen, Citro,
  Corrado, Davis, Dean, Devin, Ghemawat, Goodfellow, Harp, Irving, Isard, Jia,
  Jozefowicz, Kaiser, Kudlur, Levenberg, Mané, Monga, Moore, Murray, Olah,
  Schuster, Shlens, Steiner, Sutskever, Talwar, Tucker, Vanhoucke, Vasudevan,
  Viégas, Vinyals, Warden, Wattenberg, Wicke, Yu, \& Zheng}]{Abadi2015}
Abadi, M., Agarwal, A., Barham, P., {et~al.} 2015, TensorFlow: Large-Scale
  Machine Learning on Heterogeneous Distributed Systems

\bibitem[{{Abadi} {et~al.}(1999){Abadi}, {Moore}, \& {Bower}}]{Abadi1999}
{Abadi}, M.~G., {Moore}, B., \& {Bower}, R.~G. 1999,
  \href{https://doi.org/10.1046/j.1365-8711.1999.02715.x}{\mnras}, 308, 947,
  [\href{https://arxiv.org/abs/astro-ph/9903436}{astro-ph/9903436}].

\bibitem[{{Abdurro'uf} {et~al.}(2022){Abdurro'uf}, {Accetta}, {Aerts}, {Silva
  Aguirre}, {Ahumada}, {Ajgaonkar}, {Filiz Ak}, {Alam}, {Allende Prieto},
  {Almeida}, {Anders}, {Anderson}, {Andrews}, {Anguiano}, {Aquino-Ort{\'\i}z},
  {Arag{\'o}n-Salamanca}, {Argudo-Fern{\'a}ndez}, {Ata}, {Aubert},
  {Avila-Reese}, {Badenes}, {Barb{\'a}}, {Barger}, {Barrera-Ballesteros},
  {Beaton}, {Beers}, {Belfiore}, {Bender}, {Bernardi}, {Bershady}, {Beutler},
  {Bidin}, {Bird}, {Bizyaev}, {Blanc}, {Blanton}, {Boardman}, {Bolton},
  {Boquien}, {Borissova}, {Bovy}, {Brandt}, {Brown}, {Brownstein}, {Brusa},
  {Buchner}, {Bundy}, {Burchett}, {Bureau}, {Burgasser}, {Cabang}, {Campbell},
  {Cappellari}, {Carlberg}, {Wanderley}, {Carrera}, {Cash}, {Chen}, {Chen},
  {Cherinka}, {Chiappini}, {Choi}, {Chojnowski}, {Chung}, {Clerc}, {Cohen},
  {Comerford}, {Comparat}, {da Costa}, {Covey}, {Crane}, {Cruz-Gonzalez},
  {Culhane}, {Cunha}, {Dai}, {Damke}, {Darling}, {Davidson}, {Davies},
  {Dawson}, {De Lee}, {Diamond-Stanic}, {Cano-D{\'\i}az}, {S{\'a}nchez},
  {Donor}, {Duckworth}, {Dwelly}, {Eisenstein}, {Elsworth}, {Emsellem},
  {Eracleous}, {Escoffier}, {Fan}, {Farr}, {Feng}, {Fern{\'a}ndez-Trincado},
  {Feuillet}, {Filipp}, {Fillingham}, {Frinchaboy}, {Fromenteau}, {Galbany},
  {Garc{\'\i}a}, {Garc{\'\i}a-Hern{\'a}ndez}, {Ge}, {Geisler}, {Gelfand},
  {G{\'e}ron}, {Gibson}, {Goddy}, {Godoy-Rivera}, {Grabowski}, {Green},
  {Greener}, {Grier}, {Griffith}, {Guo}, {Guy}, {Hadjara}, {Harding},
  {Hasselquist}, {Hayes}, {Hearty}, {Hern{\'a}ndez}, {Hill}, {Hogg},
  {Holtzman}, {Horta}, {Hsieh}, {Hsu}, {Hsu}, {Huber}, {Huertas-Company},
  {Hutchinson}, {Hwang}, {Ibarra-Medel}, {Chitham}, {Ilha}, {Imig}, {Jaekle},
  {Jayasinghe}, {Ji}, {Johnson}, {Jones}, {J{\"o}nsson}, {Katkov}, {Khalatyan},
  {Kinemuchi}, {Kisku}, {Knapen}, {Kneib}, {Kollmeier}, {Kong}, {Kounkel},
  {Kreckel}, {Krishnarao}, {Lacerna}, {Lane}, {Langgin}, {Lavender}, {Law},
  {Lazarz}, {Leung}, {Leung}, {Lewis}, {Li}, {Li}, {Lian}, {Liang}, {Lin},
  {Lin}, {Lin}, {Lintott}, {Long}, {Longa-Pe{\~n}a}, {L{\'o}pez-Cob{\'a}},
  {Lu}, {Lundgren}, {Luo}, {Mackereth}, {de la Macorra}, {Mahadevan},
  {Majewski}, {Manchado}, {Mandeville}, {Maraston}, {Margalef-Bentabol},
  {Masseron}, {Masters}, {Mathur}, {McDermid}, {Mckay}, {Merloni},
  {Merrifield}, {Meszaros}, {Miglio}, {Di Mille}, {Minniti}, {Minsley},
  {Monachesi}, {Moon}, {Mosser}, {Mulchaey}, {Muna}, {Mu{\~n}oz}, {Myers},
  {Myers}, {Nadathur}, {Nair}, {Nandra}, {Neumann}, {Newman}, {Nidever},
  {Nikakhtar}, {Nitschelm}, {O'Connell}, {Garma-Oehmichen}, {Luan Souza de
  Oliveira}, {Olney}, {Oravetz}, {Ortigoza-Urdaneta}, {Osorio}, {Otter},
  {Pace}, {Padilla}, {Pan}, {Pan}, {Parikh}, {Parker}, {Peirani}, {Pe{\~n}a
  Ram{\'\i}rez}, {Penny}, {Percival}, {Perez-Fournon}, {Pinsonneault},
  {Poidevin}, {Poovelil}, {Price-Whelan}, {B{\'a}rbara de Andrade Queiroz},
  {Raddick}, {Ray}, {Rembold}, {Riddle}, {Riffel}, {Riffel}, {Rix}, {Robin},
  {Rodr{\'\i}guez-Puebla}, {Roman-Lopes}, {Rom{\'a}n-Z{\'u}{\~n}iga}, {Rose},
  {Ross}, {Rossi}, {Rubin}, {Salvato}, {S{\'a}nchez}, {S{\'a}nchez-Gallego},
  {Sanderson}, {Santana Rojas}, {Sarceno}, {Sarmiento}, {Sayres}, {Sazonova},
  {Schaefer}, {Schiavon}, {Schlegel}, {Schneider}, {Schultheis}, {Schwope},
  {Serenelli}, {Serna}, {Shao}, {Shapiro}, {Sharma}, {Shen}, {Shetrone}, {Shu},
  {Simon}, {Skrutskie}, {Smethurst}, {Smith}, {Sobeck}, {Spoo}, {Sprague},
  {Stark}, {Stassun}, {Steinmetz}, {Stello}, {Stone-Martinez},
  {Storchi-Bergmann}, {Stringfellow}, {Stutz}, {Su}, {Taghizadeh-Popp},
  {Talbot}, {Tayar}, {Telles}, {Teske}, {Thakar}, {Theissen}, {Tkachenko},
  {Thomas}, {Tojeiro}, {Hernandez Toledo}, {Troup}, {Trump}, {Trussler},
  {Turner}, {Tuttle}, {Unda-Sanzana}, {V{\'a}zquez-Mata}, {Valentini},
  {Valenzuela}, {Vargas-Gonz{\'a}lez}, {Vargas-Maga{\~n}a}, {Alfaro},
  {Villanova}, {Vincenzo}, {Wake}, {Warfield}, {Washington}, {Weaver},
  {Weijmans}, {Weinberg}, {Weiss}, {Westfall}, {Wild}, {Wilde}, {Wilson},
  {Wilson}, {Wilson}, {Wolf}, {Wood-Vasey}, {Yan}, {Zamora}, {Zasowski},
  {Zhang}, {Zhao}, {Zheng}, {Zheng}, \& {Zhu}}]{MANGA2022}
{Abdurro'uf}, {Accetta}, K., {Aerts}, C., {et~al.} 2022,
  \href{https://doi.org/10.3847/1538-4365/ac4414}{\apjs}, 259, 35,
  [\href{https://arxiv.org/abs/2112.02026}{2112.02026}].

\bibitem[{{Abdurro'uf} {et~al.}(2023){Abdurro'uf}, {Coe}, {Jung}, {Ferguson},
  {Brammer}, {Iyer}, {Bradley}, {Dayal}, {Windhorst}, {Zitrin}, {Meena},
  {Oguri}, {Diego}, {Kokorev}, {Dimauro}, {Adamo}, {Conselice}, {Welch},
  {Vanzella}, {Hsiao}, {Xu}, {Roy}, \& {Mulcahey}}]{Abdurro2023}
{Abdurro'uf}, {Coe}, D., {Jung}, I., {et~al.} 2023,
  \href{https://doi.org/10.3847/1538-4357/acba06}{\apj}, 945, 117,
  [\href{https://arxiv.org/abs/2301.02209}{2301.02209}].

\bibitem[{{Abolfathi} {et~al.}(2018){Abolfathi}, {Aguado}, {Aguilar}, {Allende
  Prieto}, {Almeida}, {Ananna}, {Anders}, {Anderson}, {Andrews}, {Anguiano},
  {Arag{\'o}n-Salamanca}, {Argudo-Fern{\'a}ndez}, {Armengaud}, {Ata},
  {Aubourg}, {Avila-Reese}, {Badenes}, {Bailey}, {Balland}, {Barger},
  {Barrera-Ballesteros}, {Bartosz}, {Bastien}, {Bates}, {Baumgarten},
  {Bautista}, {Beaton}, {Beers}, {Belfiore}, {Bender}, {Bernardi}, {Bershady},
  {Beutler}, {Bird}, {Bizyaev}, {Blanc}, {Blanton}, {Blomqvist}, {Bolton},
  {Boquien}, {Borissova}, {Bovy}, {Bradna Diaz}, {Brandt}, {Brinkmann},
  {Brownstein}, {Bundy}, {Burgasser}, {Burtin}, {Busca}, {Ca{\~n}as},
  {Cano-D{\'\i}az}, {Cappellari}, {Carrera}, {Casey}, {Cervantes Sodi}, {Chen},
  {Cherinka}, {Chiappini}, {Choi}, {Chojnowski}, {Chuang}, {Chung}, {Clerc},
  {Cohen}, {Comerford}, {Comparat}, {Correa do Nascimento}, {da Costa},
  {Cousinou}, {Covey}, {Crane}, {Cruz-Gonzalez}, {Cunha}, {da Silva Ilha},
  {Damke}, {Darling}, {Davidson}, {Dawson}, {de Icaza Lizaola}, {de la
  Macorra}, {de la Torre}, {De Lee}, {de Sainte Agathe}, {Deconto Machado},
  {Dell'Agli}, {Delubac}, {Diamond-Stanic}, {Donor}, {Downes}, {Drory}, {du Mas
  des Bourboux}, {Duckworth}, {Dwelly}, {Dyer}, {Ebelke}, {Davis Eigenbrot},
  {Eisenstein}, {Elsworth}, {Emsellem}, {Eracleous}, {Erfanianfar},
  {Escoffier}, {Fan}, {Fern{\'a}ndez Alvar}, {Fernandez-Trincado}, {Fernando
  Cirolini}, {Feuillet}, {Finoguenov}, {Fleming}, {Font-Ribera}, {Freischlad},
  {Frinchaboy}, {Fu}, {G{\'o}mez Maqueo Chew}, {Galbany}, {Garc{\'\i}a
  P{\'e}rez}, {Garcia-Dias}, {Garc{\'\i}a-Hern{\'a}ndez}, {Garma Oehmichen},
  {Gaulme}, {Gelfand}, {Gil-Mar{\'\i}n}, {Gillespie}, {Goddard}, {Gonz{\'a}lez
  Hern{\'a}ndez}, {Gonzalez-Perez}, {Grabowski}, {Green}, {Grier}, {Gueguen},
  {Guo}, {Guy}, {Hagen}, {Hall}, {Harding}, {Hasselquist}, {Hawley}, {Hayes},
  {Hearty}, {Hekker}, {Hernandez}, {Hernandez Toledo}, {Hogg},
  {Holley-Bockelmann}, {Holtzman}, {Hou}, {Hsieh}, {Hunt}, {Hutchinson},
  {Hwang}, {Jimenez Angel}, {Johnson}, {Jones}, {J{\"o}nsson}, {Jullo}, {Khan},
  {Kinemuchi}, {Kirkby}, {Kirkpatrick}, {Kitaura}, {Knapp}, {Kneib},
  {Kollmeier}, {Lacerna}, {Lane}, {Lang}, {Law}, {Le Goff}, {Lee}, {Li}, {Li},
  {Lian}, {Liang}, {Lima}, {Lin}, {Long}, {Lucatello}, {Lundgren}, {Mackereth},
  {MacLeod}, {Mahadevan}, {Maia}, {Majewski}, {Manchado}, {Maraston},
  {Mariappan}, {Marques-Chaves}, {Masseron}, {Masters}, {McDermid}, {McGreer},
  {Melendez}, {Meneses-Goytia}, {Merloni}, {Merrifield}, {Meszaros}, {Meza},
  {Minchev}, {Minniti}, {Mueller}, {Muller-Sanchez}, {Muna}, {Mu{\~n}oz},
  {Myers}, {Nair}, {Nandra}, {Ness}, {Newman}, {Nichol}, {Nidever},
  {Nitschelm}, {Noterdaeme}, {O'Connell}, {Oelkers}, {Oravetz}, {Oravetz},
  {Ort{\'\i}z}, {Osorio}, {Pace}, {Padilla}, {Palanque-Delabrouille},
  {Palicio}, {Pan}, {Pan}, {Parikh}, {P{\^a}ris}, {Park}, {Peirani},
  {Pellejero-Ibanez}, {Penny}, {Percival}, {Perez-Fournon}, {Petitjean},
  {Pieri}, {Pinsonneault}, {Pisani}, {Prada}, {Prakash}, {Queiroz}, {Raddick},
  {Raichoor}, {Barboza Rembold}, {Richstein}, {Riffel}, {Riffel}, {Rix},
  {Robin}, {Rodr{\'\i}guez Torres}, {Rom{\'a}n-Z{\'u}{\~n}iga}, {Ross},
  {Rossi}, {Ruan}, {Ruggeri}, {Ruiz}, {Salvato}, {S{\'a}nchez}, {S{\'a}nchez},
  {Sanchez Almeida}, {S{\'a}nchez-Gallego}, {Santana Rojas}, {Santiago},
  {Schiavon}, {Schimoia}, {Schlafly}, {Schlegel}, {Schneider}, {Schuster},
  {Schwope}, {Seo}, {Serenelli}, {Shen}, {Shen}, {Shetrone}, {Shull}, {Silva
  Aguirre}, {Simon}, {Skrutskie}, {Slosar}, {Smethurst}, {Smith}, {Sobeck},
  {Somers}, {Souter}, {Souto}, {Spindler}, {Stark}, {Stassun}, {Steinmetz},
  {Stello}, {Storchi-Bergmann}, {Streblyanska}, {Stringfellow}, {Su{\'a}rez},
  {Sun}, {Szigeti}, {Taghizadeh-Popp}, {Talbot}, {Tang}, {Tao}, {Tayar},
  {Tembe}, {Teske}, {Thakar}, {Thomas}, {Tissera}, {Tojeiro}, {Tremonti},
  {Troup}, {Urry}, {Valenzuela}, {van den Bosch}, {Vargas-Gonz{\'a}lez},
  {Vargas-Maga{\~n}a}, {Vazquez}, {Villanova}, {Vogt}, {Wake}, {Wang},
  {Weaver}, {Weijmans}, {Weinberg}, {Westfall}, {Whelan}, {Wilcots}, {Wild},
  {Williams}, {Wilson}, {Wood-Vasey}, {Wylezalek}, {Xiao}, {Yan}, {Yang},
  {Ybarra}, {Y{\`e}che}, {Zakamska}, {Zamora}, {Zarrouk}, {Zasowski}, {Zhang},
  {Zhao}, {Zhao}, {Zheng}, {Zheng}, {Zhou}, {Zhu}, {Zinn}, \&
  {Zou}}]{MANGA2018}
{Abolfathi}, B., {Aguado}, D.~S., {Aguilar}, G., {et~al.} 2018,
  \href{https://doi.org/10.3847/1538-4365/aa9e8a}{\apjs}, 235, 42,
  [\href{https://arxiv.org/abs/1707.09322}{1707.09322}].

\bibitem[{{Aguado} {et~al.}(2019){Aguado}, {Ahumada}, {Almeida}, {Anderson},
  {Andrews}, {Anguiano}, {Aquino Ort{\'\i}z}, {Arag{\'o}n-Salamanca},
  {Argudo-Fern{\'a}ndez}, {Aubert}, {Avila-Reese}, {Badenes}, {Barboza
  Rembold}, {Barger}, {Barrera-Ballesteros}, {Bates}, {Bautista}, {Beaton},
  {Beers}, {Belfiore}, {Bernardi}, {Bershady}, {Beutler}, {Bird}, {Bizyaev},
  {Blanc}, {Blanton}, {Blomqvist}, {Bolton}, {Boquien}, {Borissova}, {Bovy},
  {Brandt}, {Brinkmann}, {Brownstein}, {Bundy}, {Burgasser}, {Byler}, {Cano
  Diaz}, {Cappellari}, {Carrera}, {Cervantes Sodi}, {Chen}, {Cherinka}, {Choi},
  {Chung}, {Coffey}, {Comerford}, {Comparat}, {Covey}, {da Silva Ilha}, {da
  Costa}, {Dai}, {Damke}, {Darling}, {Davies}, {Dawson}, {de Sainte Agathe},
  {Deconto Machado}, {Del Moro}, {De Lee}, {Diamond-Stanic}, {Dom{\'\i}nguez
  S{\'a}nchez}, {Donor}, {Drory}, {du Mas des Bourboux}, {Duckworth}, {Dwelly},
  {Ebelke}, {Emsellem}, {Escoffier}, {Fern{\'a}ndez-Trincado}, {Feuillet},
  {Fischer}, {Fleming}, {Fraser-McKelvie}, {Freischlad}, {Frinchaboy}, {Fu},
  {Galbany}, {Garcia-Dias}, {Garc{\'\i}a-Hern{\'a}ndez}, {Garma Oehmichen},
  {Geimba Maia}, {Gil-Mar{\'\i}n}, {Grabowski}, {Gu}, {Guo}, {Ha},
  {Harrington}, {Hasselquist}, {Hayes}, {Hearty}, {Hernandez Toledo}, {Hicks},
  {Hogg}, {Holley-Bockelmann}, {Holtzman}, {Hsieh}, {Hunt}, {Hwang},
  {Ibarra-Medel}, {Jimenez Angel}, {Johnson}, {Jones}, {J{\"o}nsson},
  {Kinemuchi}, {Kollmeier}, {Krawczyk}, {Kreckel}, {Kruk}, {Lacerna}, {Lan},
  {Lane}, {Law}, {Lee}, {Li}, {Lian}, {Lin}, {Lin}, {Lintott}, {Long},
  {Longa-Pe{\~n}a}, {Mackereth}, {de la Macorra}, {Majewski}, {Malanushenko},
  {Manchado}, {Maraston}, {Mariappan}, {Marinelli}, {Marques-Chaves},
  {Masseron}, {Masters}, {McDermid}, {Medina Pe{\~n}a}, {Meneses-Goytia},
  {Merloni}, {Merrifield}, {Meszaros}, {Minniti}, {Minsley}, {Muna}, {Myers},
  {Nair}, {Correa do Nascimento}, {Newman}, {Nitschelm}, {Olmstead}, {Oravetz},
  {Oravetz}, {Ortega Minakata}, {Pace}, {Padilla}, {Palicio}, {Pan}, {Pan},
  {Parikh}, {Parker}, {Peirani}, {Penny}, {Percival}, {Perez-Fournon},
  {Peterken}, {Pinsonneault}, {Prakash}, {Raddick}, {Raichoor}, {Riffel},
  {Riffel}, {Rix}, {Robin}, {Roman-Lopes}, {Rose}, {Ross}, {Rossi}, {Rowlands},
  {Rubin}, {S{\'a}nchez}, {S{\'a}nchez-Gallego}, {Sayres}, {Schaefer},
  {Schiavon}, {Schimoia}, {Schlafly}, {Schlegel}, {Schneider}, {Schultheis},
  {Seo}, {Shamsi}, {Shao}, {Shen}, {Shetty}, {Simonian}, {Smethurst}, {Sobeck},
  {Souter}, {Spindler}, {Stark}, {Stassun}, {Steinmetz}, {Storchi-Bergmann},
  {Stringfellow}, {Su{\'a}rez}, {Sun}, {Taghizadeh-Popp}, {Talbot}, {Tayar},
  {Thakar}, {Thomas}, {Tissera}, {Tojeiro}, {Troup}, {Unda-Sanzana},
  {Valenzuela}, {Vargas-Maga{\~n}a}, {V{\'a}zquez-Mata}, {Wake}, {Weaver},
  {Weijmans}, {Westfall}, {Wild}, {Wilson}, {Woods}, {Yan}, {Yang}, {Zamora},
  {Zasowski}, {Zhang}, {Zheng}, {Zheng}, {Zhu}, {Zinn}, \& {Zou}}]{MANGA2019}
{Aguado}, D.~S., {Ahumada}, R., {Almeida}, A., {et~al.} 2019,
  \href{https://doi.org/10.3847/1538-4365/aaf651}{\apjs}, 240, 23,
  [\href{https://arxiv.org/abs/1812.02759}{1812.02759}].

\bibitem[{{Aguerri} {et~al.}(2017){Aguerri}, {Agulli}, {Diaferio}, \& {Dalla
  Vecchia}}]{Aguerri2017}
{Aguerri}, J.~A.~L., {Agulli}, I., {Diaferio}, A., \& {Dalla Vecchia}, C. 2017,
  \href{https://doi.org/10.1093/mnras/stx457}{\mnras}, 468, 364,
  [\href{https://arxiv.org/abs/1703.00740}{1703.00740}].

\bibitem[{{Aihara} {et~al.}(2018){Aihara}, {Arimoto}, {Armstrong}, {Arnouts},
  {Bahcall}, {Bickerton}, {Bosch}, {Bundy}, {Capak}, {Chan}, {Chiba}, {Coupon},
  {Egami}, {Enoki}, {Finet}, {Fujimori}, {Fujimoto}, {Furusawa}, {Furusawa},
  {Goto}, {Goulding}, {Greco}, {Greene}, {Gunn}, {Hamana}, {Harikane},
  {Hashimoto}, {Hattori}, {Hayashi}, {Hayashi}, {He{\l}miniak}, {Higuchi},
  {Hikage}, {Ho}, {Hsieh}, {Huang}, {Huang}, {Ikeda}, {Imanishi}, {Inoue},
  {Iwasawa}, {Iwata}, {Jaelani}, {Jian}, {Kamata}, {Karoji}, {Kashikawa},
  {Katayama}, {Kawanomoto}, {Kayo}, {Koda}, {Koike}, {Kojima}, {Komiyama},
  {Konno}, {Koshida}, {Koyama}, {Kusakabe}, {Leauthaud}, {Lee}, {Lin}, {Lin},
  {Lupton}, {Mandelbaum}, {Matsuoka}, {Medezinski}, {Mineo}, {Miyama},
  {Miyatake}, {Miyazaki}, {Momose}, {More}, {More}, {Moritani}, {Moriya},
  {Morokuma}, {Mukae}, {Murata}, {Murayama}, {Nagao}, {Nakata}, {Niida},
  {Niikura}, {Nishizawa}, {Obuchi}, {Oguri}, {Oishi}, {Okabe}, {Okamoto},
  {Okura}, {Ono}, {Onodera}, {Onoue}, {Osato}, {Ouchi}, {Price}, {Pyo}, {Sako},
  {Sawicki}, {Shibuya}, {Shimasaku}, {Shimono}, {Shirasaki}, {Silverman},
  {Simet}, {Speagle}, {Spergel}, {Strauss}, {Sugahara}, {Sugiyama}, {Suto},
  {Suyu}, {Suzuki}, {Tait}, {Takada}, {Takata}, {Tamura}, {Tanaka}, {Tanaka},
  {Tanaka}, {Tanaka}, {Terai}, {Terashima}, {Toba}, {Tominaga}, {Toshikawa},
  {Turner}, {Uchida}, {Uchiyama}, {Umetsu}, {Uraguchi}, {Urata}, {Usuda},
  {Utsumi}, {Wang}, {Wang}, {Wong}, {Yabe}, {Yamada}, {Yamanoi}, {Yasuda},
  {Yeh}, {Yonehara}, \& {Yuma}}]{Aihara2018Presentation}
{Aihara}, H., {Arimoto}, N., {Armstrong}, R., {et~al.} 2018,
  \href{https://doi.org/10.1093/pasj/psx066}{\pasj}, 70, S4,
  [\href{https://arxiv.org/abs/1704.05858}{1704.05858}].

\bibitem[{{Alatalo} {et~al.}(2015){Alatalo}, {Appleton}, {Lisenfeld},
  {Bitsakis}, {Lanz}, {Lacy}, {Charmandaris}, {Cluver}, {Dopita}, {Guillard},
  {Jarrett}, {Kewley}, {Nyland}, {Ogle}, {Rasmussen}, {Rich},
  {Verdes-Montenegro}, {Xu}, \& {Yun}}]{Altalo2015}
{Alatalo}, K., {Appleton}, P.~N., {Lisenfeld}, U., {et~al.} 2015,
  \href{https://doi.org/10.1088/0004-637X/812/2/117}{\apj}, 812, 117,
  [\href{https://arxiv.org/abs/1509.05779}{1509.05779}].

\bibitem[{{Alonso} {et~al.}(2012){Alonso}, {Mesa}, {Padilla}, \&
  {Lambas}}]{Alonso2012}
{Alonso}, S., {Mesa}, V., {Padilla}, N., \& {Lambas}, D.~G. 2012,
  \href{https://doi.org/10.1051/0004-6361/201117901}{\aap}, 539, A46,
  [\href{https://arxiv.org/abs/1111.2292}{1111.2292}].

\bibitem[{{Andersson} \& {Davies}(2019)}]{Andersson2019}
{Andersson}, E.~P. \& {Davies}, M.~B. 2019,
  \href{https://doi.org/10.1093/mnras/stz709}{\mnras}, 485, 4134,
  [\href{https://arxiv.org/abs/1808.08087}{1808.08087}].

\bibitem[{{Ando} {et~al.}(2022){Ando}, {Shimasaku}, {Momose}, {Ito}, {Sawicki},
  \& {Shimakawa}}]{Ando2022}
{Ando}, M., {Shimasaku}, K., {Momose}, R., {et~al.} 2022,
  \href{https://doi.org/10.1093/mnras/stac1049}{\mnras}, 513, 3252,
  [\href{https://arxiv.org/abs/2201.05185}{2201.05185}].

\bibitem[{{Arcila-Osejo} {et~al.}(2019){Arcila-Osejo}, {Sawicki}, {Arnouts},
  {Golob}, {Moutard}, \& {Sorba}}]{ArcilaOsejo2019}
{Arcila-Osejo}, L., {Sawicki}, M., {Arnouts}, S., {et~al.} 2019,
  \href{https://doi.org/10.1093/mnras/stz1169}{\mnras}, 486, 4880,
  [\href{https://arxiv.org/abs/1904.11654}{1904.11654}].

\bibitem[{{Arnouts} \& {Ilbert}(2011)}]{Lephare2011}
{Arnouts}, S. \& {Ilbert}, O. 2011, {LePHARE: Photometric Analysis for Redshift
  Estimate}, Astrophysics Source Code Library, record ascl:1108.009

\bibitem[{{Arnouts} {et~al.}(2013){Arnouts}, {Le Floc'h}, {Chevallard},
  {Johnson}, {Ilbert}, {Treyer}, {Aussel}, {Capak}, {Sanders}, {Scoville},
  {McCracken}, {Milliard}, {Pozzetti}, \& {Salvato}}]{Arnouts2013}
{Arnouts}, S., {Le Floc'h}, E., {Chevallard}, J., {et~al.} 2013,
  \href{https://doi.org/10.1051/0004-6361/201321768}{\aap}, 558, A67,
  [\href{https://arxiv.org/abs/1309.0008}{1309.0008}].

\bibitem[{{Asari} {et~al.}(2007){Asari}, {Cid Fernandes}, {Stasi{\'n}ska},
  {Torres-Papaqui}, {Mateus}, {Sodr{\'e}}, {Schoenell}, \& {Gomes}}]{asari2007}
{Asari}, N.~V., {Cid Fernandes}, R., {Stasi{\'n}ska}, G., {et~al.} 2007,
  \href{https://doi.org/10.1111/j.1365-2966.2007.12255.x}{\mnras}, 381, 263,
  [\href{https://arxiv.org/abs/0707.3578}{0707.3578}].

\bibitem[{{Ascaso} {et~al.}(2015){Ascaso}, {Ben{\'\i}tez},
  {Fern{\'a}ndez-Soto}, {Arnalte-Mur}, {L{\'o}pez-Sanjuan}, {Molino},
  {Schoenell}, {Jim{\'e}nez-Teja}, {Merson}, {Huertas-Company},
  {D{\'\i}az-Garc{\'\i}a}, {Mart{\'\i}nez}, {Cenarro}, {Dupke}, {M{\'a}rquez},
  {Masegosa}, {Nieves-Seoane}, {Povi{\'c}}, {Varela}, {Viironen}, {Aguerri},
  {Olmo}, {Moles}, {Perea}, {Alfaro}, {Aparicio-Villegas}, {Broadhurst},
  {Cabrera-Ca{\~n}o}, {Castander}, {Cepa}, {Cervi{\~n}o}, {Delgado},
  {Crist{\'o}bal-Hornillos}, {Hurtado-Gil}, {Husillos}, {Infante}, {Prada}, \&
  {Quintana}}]{Ascaso2015}
{Ascaso}, B., {Ben{\'\i}tez}, N., {Fern{\'a}ndez-Soto}, A., {et~al.} 2015,
  \href{https://doi.org/10.1093/mnras/stv1317}{\mnras}, 452, 549,
  [\href{https://arxiv.org/abs/1506.03823}{1506.03823}].

\bibitem[{{Bacon} {et~al.}(2001){Bacon}, {Copin}, {Monnet}, {Miller},
  {Allington-Smith}, {Bureau}, {Carollo}, {Davies}, {Emsellem}, {Kuntschner},
  {Peletier}, {Verolme}, \& {de Zeeuw}}]{SAURON2001}
{Bacon}, R., {Copin}, Y., {Monnet}, G., {et~al.} 2001,
  \href{https://doi.org/10.1046/j.1365-8711.2001.04612.x}{\mnras}, 326, 23,
  [\href{https://arxiv.org/abs/astro-ph/0103451}{astro-ph/0103451}].

\bibitem[{{Bah{\'e}} {et~al.}(2013){Bah{\'e}}, {McCarthy}, {Balogh}, \&
  {Font}}]{Bahe2013}
{Bah{\'e}}, Y.~M., {McCarthy}, I.~G., {Balogh}, M.~L., \& {Font}, A.~S. 2013,
  \href{https://doi.org/10.1093/mnras/stt109}{\mnras}, 430, 3017,
  [\href{https://arxiv.org/abs/1210.8407}{1210.8407}].

\bibitem[{{Bakos} {et~al.}(2008){Bakos}, {Trujillo}, \& {Pohlen}}]{Bakos2008}
{Bakos}, J., {Trujillo}, I., \& {Pohlen}, M. 2008,
  \href{https://doi.org/10.1086/591671}{\apjl}, 683, L103,
  [\href{https://arxiv.org/abs/0807.2776}{0807.2776}].

\bibitem[{{Baldry} {et~al.}(2006){Baldry}, {Balogh}, {Bower}, {Glazebrook},
  {Nichol}, {Bamford}, \& {Budavari}}]{Baldry2006}
{Baldry}, I.~K., {Balogh}, M.~L., {Bower}, R.~G., {et~al.} 2006,
  \href{https://doi.org/10.1111/j.1365-2966.2006.11081.x}{\mnras}, 373, 469,
  [\href{https://arxiv.org/abs/astro-ph/0607648}{astro-ph/0607648}].

\bibitem[{{Baldry} {et~al.}(2004){Baldry}, {Glazebrook}, {Brinkmann},
  {Ivezi{\'c}}, {Lupton}, {Nichol}, \& {Szalay}}]{Baldry2004}
{Baldry}, I.~K., {Glazebrook}, K., {Brinkmann}, J., {et~al.} 2004,
  \href{https://doi.org/10.1086/380092}{\apj}, 600, 681,
  [\href{https://arxiv.org/abs/astro-ph/0309710}{astro-ph/0309710}].

\bibitem[{{Baldwin} {et~al.}(1981){Baldwin}, {Phillips}, \& {Terlevich}}]{BPT}
{Baldwin}, J.~A., {Phillips}, M.~M., \& {Terlevich}, R. 1981,
  \href{https://doi.org/10.1086/130766}{\pasp}, 93, 5.

\bibitem[{{Balogh} {et~al.}(2004){Balogh}, {Baldry}, {Nichol}, {Miller},
  {Bower}, \& {Glazebrook}}]{Balogh2004}
{Balogh}, M.~L., {Baldry}, I.~K., {Nichol}, R., {et~al.} 2004,
  \href{https://doi.org/10.1086/426079}{\apjl}, 615, L101,
  [\href{https://arxiv.org/abs/astro-ph/0406266}{astro-ph/0406266}].

\bibitem[{{Balogh} {et~al.}(2016){Balogh}, {McGee}, {Mok}, {Muzzin}, {van der
  Burg}, {Bower}, {Finoguenov}, {Hoekstra}, {Lidman}, {Mulchaey}, {Noble},
  {Parker}, {Tanaka}, {Wilman}, {Webb}, {Wilson}, \& {Yee}}]{Balogh2016}
{Balogh}, M.~L., {McGee}, S.~L., {Mok}, A., {et~al.} 2016,
  \href{https://doi.org/10.1093/mnras/stv2949}{\mnras}, 456, 4364,
  [\href{https://arxiv.org/abs/1511.07344}{1511.07344}].

\bibitem[{{Balogh} {et~al.}(1999){Balogh}, {Morris}, {Yee}, {Carlberg}, \&
  {Ellingson}}]{Balogh1999}
{Balogh}, M.~L., {Morris}, S.~L., {Yee}, H.~K.~C., {Carlberg}, R.~G., \&
  {Ellingson}, E. 1999, \href{https://doi.org/10.1086/308056}{\apj}, 527, 54,
  [\href{https://arxiv.org/abs/astro-ph/9906470}{astro-ph/9906470}].

\bibitem[{{Balogh} {et~al.}(2000){Balogh}, {Navarro}, \& {Morris}}]{balogh2000}
{Balogh}, M.~L., {Navarro}, J.~F., \& {Morris}, S.~L. 2000,
  \href{https://doi.org/10.1086/309323}{\apj}, 540, 113,
  [\href{https://arxiv.org/abs/astro-ph/0004078}{astro-ph/0004078}].

\bibitem[{{Bassett} {et~al.}(2013){Bassett}, {Papovich}, {Lotz}, {Bell},
  {Finkelstein}, {Newman}, {Tran}, {Almaini}, {Lani}, {Cooper}, {Croton},
  {Dekel}, {Ferguson}, {Kocevski}, {Koekemoer}, {Koo}, {McGrath}, {McIntosh},
  \& {Wechsler}}]{Bassett2013}
{Bassett}, R., {Papovich}, C., {Lotz}, J.~M., {et~al.} 2013,
  \href{https://doi.org/10.1088/0004-637X/770/1/58}{\apj}, 770, 58,
  [\href{https://arxiv.org/abs/1305.0607}{1305.0607}].

\bibitem[{{Beers} {et~al.}(1990){Beers}, {Flynn}, \& {Gebhardt}}]{MAD}
{Beers}, T.~C., {Flynn}, K., \& {Gebhardt}, K. 1990,
  \href{https://doi.org/10.1086/115487}{\aj}, 100, 32.

\bibitem[{{Bekki}(2009)}]{Bekki2009}
{Bekki}, K. 2009,
  \href{https://doi.org/10.1111/j.1365-2966.2009.15431.x}{\mnras}, 399, 2221,
  [\href{https://arxiv.org/abs/0907.4409}{0907.4409}].

\bibitem[{{Belfiore} {et~al.}(2016){Belfiore}, {Maiolino}, {Maraston},
  {Emsellem}, {Bershady}, {Masters}, {Yan}, {Bizyaev}, {Boquien}, {Brownstein},
  {Bundy}, {Drory}, {Heckman}, {Law}, {Roman-Lopes}, {Pan}, {Stanghellini},
  {Thomas}, {Weijmans}, \& {Westfall}}]{belfiore2016sdss}
{Belfiore}, F., {Maiolino}, R., {Maraston}, C., {et~al.} 2016,
  \href{https://doi.org/10.1093/mnras/stw1234}{\mnras}, 461, 3111,
  [\href{https://arxiv.org/abs/1605.07189}{1605.07189}].

\bibitem[{{Bell} \& {de Jong}(2000)}]{Bell2000}
{Bell}, E.~F. \& {de Jong}, R.~S. 2000,
  \href{https://doi.org/10.1046/j.1365-8711.2000.03138.x}{\mnras}, 312, 497,
  [\href{https://arxiv.org/abs/astro-ph/9909402}{astro-ph/9909402}].

\bibitem[{{Bell} \& {de Jong}(2001)}]{Bell2001}
{Bell}, E.~F. \& {de Jong}, R.~S. 2001,
  \href{https://doi.org/10.1086/319728}{\apj}, 550, 212,
  [\href{https://arxiv.org/abs/astro-ph/0011493}{astro-ph/0011493}].

\bibitem[{{Bell} {et~al.}(2003){Bell}, {McIntosh}, {Katz}, \&
  {Weinberg}}]{Bell2003}
{Bell}, E.~F., {McIntosh}, D.~H., {Katz}, N., \& {Weinberg}, M.~D. 2003,
  \href{https://doi.org/10.1086/378847}{\apjs}, 149, 289,
  [\href{https://arxiv.org/abs/astro-ph/0302543}{astro-ph/0302543}].

\bibitem[{{Bell} {et~al.}(2004){Bell}, {Wolf}, {Meisenheimer}, {Rix}, {Borch},
  {Dye}, {Kleinheinrich}, {Wisotzki}, \& {McIntosh}}]{Bell2004}
{Bell}, E.~F., {Wolf}, C., {Meisenheimer}, K., {et~al.} 2004,
  \href{https://doi.org/10.1086/420778}{\apj}, 608, 752,
  [\href{https://arxiv.org/abs/astro-ph/0303394}{astro-ph/0303394}].

\bibitem[{{Bellagamba} {et~al.}(2011){Bellagamba}, {Maturi}, {Hamana},
  {Meneghetti}, {Miyazaki}, \& {Moscardini}}]{Bellagamba2011}
{Bellagamba}, F., {Maturi}, M., {Hamana}, T., {et~al.} 2011,
  \href{https://doi.org/10.1111/j.1365-2966.2011.18202.x}{\mnras}, 413, 1145,
  [\href{https://arxiv.org/abs/1006.0610}{1006.0610}].

\bibitem[{{Bellagamba} {et~al.}(2018){Bellagamba}, {Roncarelli}, {Maturi}, \&
  {Moscardini}}]{AMICO}
{Bellagamba}, F., {Roncarelli}, M., {Maturi}, M., \& {Moscardini}, L. 2018,
  \href{https://doi.org/10.1093/mnras/stx2701}{\mnras}, 473, 5221,
  [\href{https://arxiv.org/abs/1705.03029}{1705.03029}].

\bibitem[{{Benitez} {et~al.}(2014){Benitez}, {Dupke}, {Moles}, {Sodre},
  {Cenarro}, {Marin-Franch}, {Taylor}, {Cristobal}, {Fernandez-Soto}, {Mendes
  de Oliveira}, {Cepa-Nogue}, {Abramo}, {Alcaniz}, {Overzier},
  {Hernandez-Monteagudo}, {Alfaro}, {Kanaan}, {Carvano}, {Reis}, {Martinez
  Gonzalez}, {Ascaso}, {Ballesteros}, {Xavier}, {Varela}, {Ederoclite},
  {Vazquez Ramio}, {Broadhurst}, {Cypriano}, {Angulo}, {Diego}, {Zandivarez},
  {Diaz}, {Melchior}, {Umetsu}, {Spinelli}, {Zitrin}, {Coe}, {Yepes}, {Vielva},
  {Sahni}, {Marcos-Caballero}, {Shu Kitaura}, {Maroto}, {Masip}, {Tsujikawa},
  {Carneiro}, {Gonzalez Nuevo}, {Carvalho}, {Reboucas}, {Carvalho}, {Abdalla},
  {Bernui}, {Pigozzo}, {Ferreira}, {Chandrachani Devi}, {Bengaly}, {Campista},
  {Amorim}, {Asari}, {Bongiovanni}, {Bonoli}, {Bruzual}, {Cardiel}, {Cava},
  {Cid Fernandes}, {Coelho}, {Cortesi}, {Delgado}, {Diaz Garcia}, {Espinosa},
  {Galliano}, {Gonzalez-Serrano}, {Falcon-Barroso}, {Fritz}, {Fernandes},
  {Gorgas}, {Hoyos}, {Jimenez-Teja}, {Lopez-Aguerri}, {Lopez-San Juan},
  {Mateus}, {Molino}, {Novais}, {OMill}, {Oteo}, {Perez-Gonzalez}, {Poggianti},
  {Proctor}, {Ricciardelli}, {Sanchez-Blazquez}, {Storchi-Bergmann}, {Telles},
  {Schoennell}, {Trujillo}, {Vazdekis}, {Viironen}, {Daflon},
  {Aparicio-Villegas}, {Rocha}, {Ribeiro}, {Borges}, {Martins}, {Marcolino},
  {Martinez-Delgado}, {Perez-Torres}, {Siffert}, {Calvao}, {Sako}, {Kessler},
  {Alvarez-Candal}, {De Pra}, {Roig}, {Lazzaro}, {Gorosabel}, {Lopes de
  Oliveira}, {Lima-Neto}, {Irwin}, {Liu}, {Alvarez}, {Balmes}, {Chueca},
  {Costa-Duarte}, {da Costa}, {Dantas}, {Diaz}, {Fabregat}, {Ferrari},
  {Gavela}, {Gracia}, {Gruel}, {Gutierrez}, {Guzman}, {Hernandez-Fernandez},
  {Herranz}, {Hurtado-Gil}, {Jablonsky}, {Laporte}, {Le Tiran}, {Licandro},
  {Lima}, {Martin}, {Martinez}, {Montero}, {Penteado}, {Pereira}, {Peris},
  {Quilis}, {Sanchez-Portal}, {Soja}, {Solano}, {Torra}, \&
  {Valdivielso}}]{Benitez2014}
{Benitez}, N., {Dupke}, R., {Moles}, M., {et~al.} 2014, arXiv e-prints,
  arXiv:1403.5237, [\href{https://arxiv.org/abs/1403.5237}{1403.5237}].

\bibitem[{{Ben{\'\i}tez} {et~al.}(2009){Ben{\'\i}tez}, {Gazta{\~n}aga},
  {Miquel}, {Castander}, {Moles}, {Crocce}, {Fern{\'a}ndez-Soto}, {Fosalba},
  {Ballesteros}, {Campa}, {Cardiel-Sas}, {Castilla}, {Crist{\'o}bal-Hornillos},
  {Delfino}, {Fern{\'a}ndez}, {Fern{\'a}ndez-Sopuerta}, {Garc{\'\i}a-Bellido},
  {Lobo}, {Mart{\'\i}nez}, {Ortiz}, {Pacheco}, {Paredes}, {Pons-Border{\'\i}a},
  {S{\'a}nchez}, {S{\'a}nchez}, {Varela}, \& {de Vicente}}]{Benitez2009}
{Ben{\'\i}tez}, N., {Gazta{\~n}aga}, E., {Miquel}, R., {et~al.} 2009,
  \href{https://doi.org/10.1088/0004-637X/691/1/241}{\apj}, 691, 241,
  [\href{https://arxiv.org/abs/0807.0535}{0807.0535}].

\bibitem[{{Bernardi} {et~al.}(2006){Bernardi}, {Nichol}, {Sheth}, {Miller}, \&
  {Brinkmann}}]{Bernardi2006}
{Bernardi}, M., {Nichol}, R.~C., {Sheth}, R.~K., {Miller}, C.~J., \&
  {Brinkmann}, J. 2006, \href{https://doi.org/10.1086/499522}{\aj}, 131, 1288,
  [\href{https://arxiv.org/abs/astro-ph/0509360}{astro-ph/0509360}].

\bibitem[{{Berrier} {et~al.}(2009){Berrier}, {Stewart}, {Bullock}, {Purcell},
  {Barton}, \& {Wechsler}}]{berrier2009}
{Berrier}, J.~C., {Stewart}, K.~R., {Bullock}, J.~S., {et~al.} 2009,
  \href{https://doi.org/10.1088/0004-637X/690/2/1292}{\apj}, 690, 1292,
  [\href{https://arxiv.org/abs/0804.0426}{0804.0426}].

\bibitem[{{Bertin}(2010)}]{Bertin2010Swarp}
{Bertin}, E. 2010, {SWarp: Resampling and Co-adding FITS Images Together},
  Astrophysics Source Code Library, record ascl:1010.068

\bibitem[{{Bertin}(2011)}]{Bertin2011}
{Bertin}, E. 2011, in Astronomical Society of the Pacific Conference Series,
  Vol. 442, Astronomical Data Analysis Software and Systems XX, ed. I.~N.
  {Evans}, A.~{Accomazzi}, D.~J. {Mink}, \& A.~H. {Rots}, 435

\bibitem[{{Bertin} \& {Arnouts}(1996)}]{Bertin1996}
{Bertin}, E. \& {Arnouts}, S. 1996,
  \href{https://doi.org/10.1051/aas:1996164}{\aaps}, 117, 393.

\bibitem[{{Bertin} {et~al.}(2002){Bertin}, {Mellier}, {Radovich}, {Missonnier},
  {Didelon}, \& {Morin}}]{Bertin2002}
{Bertin}, E., {Mellier}, Y., {Radovich}, M., {et~al.} 2002, in Astronomical
  Society of the Pacific Conference Series, Vol. 281, Astronomical Data
  Analysis Software and Systems XI, ed. D.~A. {Bohlender}, D.~{Durand}, \&
  T.~H. {Handley}, 228

\bibitem[{{Bessell}(2005)}]{Bessel2005}
{Bessell}, M.~S. 2005,
  \href{https://doi.org/10.1146/annurev.astro.41.082801.100251}{\araa}, 43,
  293.

\bibitem[{{Bialas} {et~al.}(2015){Bialas}, {Lisker}, {Olczak}, {Spurzem}, \&
  {Kotulla}}]{Bialas2015}
{Bialas}, D., {Lisker}, T., {Olczak}, C., {Spurzem}, R., \& {Kotulla}, R. 2015,
  \href{https://doi.org/10.1051/0004-6361/201425235}{\aap}, 576, A103,
  [\href{https://arxiv.org/abs/1503.01965}{1503.01965}].

\bibitem[{{Blanton} {et~al.}(2001){Blanton}, {Dalcanton}, {Eisenstein},
  {Loveday}, {Strauss}, {SubbaRao}, {Weinberg}, {Anderson}, {Annis}, {Bahcall},
  {Bernardi}, {Brinkmann}, {Brunner}, {Burles}, {Carey}, {Castander},
  {Connolly}, {Csabai}, {Doi}, {Finkbeiner}, {Friedman}, {Frieman}, {Fukugita},
  {Gunn}, {Hennessy}, {Hindsley}, {Hogg}, {Ichikawa}, {Ivezi{\'c}}, {Kent},
  {Knapp}, {Lamb}, {Leger}, {Long}, {Lupton}, {McKay}, {Meiksin}, {Merelli},
  {Munn}, {Narayanan}, {Newcomb}, {Nichol}, {Okamura}, {Owen}, {Pier}, {Pope},
  {Postman}, {Quinn}, {Rockosi}, {Schlegel}, {Schneider}, {Shimasaku},
  {Siegmund}, {Smee}, {Snir}, {Stoughton}, {Stubbs}, {Szalay}, {Szokoly},
  {Thakar}, {Tremonti}, {Tucker}, {Uomoto}, {Vanden Berk}, {Vogeley},
  {Waddell}, {Yanny}, {Yasuda}, \& {York}}]{Blanton2001}
{Blanton}, M.~R., {Dalcanton}, J., {Eisenstein}, D., {et~al.} 2001,
  \href{https://doi.org/10.1086/320405}{\aj}, 121, 2358,
  [\href{https://arxiv.org/abs/astro-ph/0012085}{astro-ph/0012085}].

\bibitem[{{Blanton} {et~al.}(2005){Blanton}, {Eisenstein}, {Hogg}, {Schlegel},
  \& {Brinkmann}}]{Blanton2005}
{Blanton}, M.~R., {Eisenstein}, D., {Hogg}, D.~W., {Schlegel}, D.~J., \&
  {Brinkmann}, J. 2005, \href{https://doi.org/10.1086/422897}{\apj}, 629, 143,
  [\href{https://arxiv.org/abs/astro-ph/0310453}{astro-ph/0310453}].

\bibitem[{{Blanton} \& {Moustakas}(2009)}]{Blanton2009}
{Blanton}, M.~R. \& {Moustakas}, J. 2009,
  \href{https://doi.org/10.1146/annurev-astro-082708-101734}{\araa}, 47, 159,
  [\href{https://arxiv.org/abs/0908.3017}{0908.3017}].

\bibitem[{{Bluck} {et~al.}(2011){Bluck}, {Conselice}, {Almaini}, {Laird},
  {Nandra}, \& {Gr{\"u}tzbauch}}]{Bluck2011}
{Bluck}, A. F.~L., {Conselice}, C.~J., {Almaini}, O., {et~al.} 2011,
  \href{https://doi.org/10.1111/j.1365-2966.2010.17521.x}{\mnras}, 410, 1174,
  [\href{https://arxiv.org/abs/1008.2162}{1008.2162}].

\bibitem[{{Bluck} {et~al.}(2020){Bluck}, {Maiolino}, {Piotrowska}, {Trussler},
  {Ellison}, {S{\'a}nchez}, {Thorp}, {Teimoorinia}, {Moreno}, \&
  {Conselice}}]{Bluck2020}
{Bluck}, A. F.~L., {Maiolino}, R., {Piotrowska}, J.~M., {et~al.} 2020,
  \href{https://doi.org/10.1093/mnras/staa2806}{\mnras}, 499, 230,
  [\href{https://arxiv.org/abs/2009.05341}{2009.05341}].

\bibitem[{{Bluck} {et~al.}(2014){Bluck}, {Mendel}, {Ellison}, {Moreno},
  {Simard}, {Patton}, \& {Starkenburg}}]{Bluck2014}
{Bluck}, A. F.~L., {Mendel}, J.~T., {Ellison}, S.~L., {et~al.} 2014,
  \href{https://doi.org/10.1093/mnras/stu594}{\mnras}, 441, 599,
  [\href{https://arxiv.org/abs/1403.5269}{1403.5269}].

\bibitem[{{Bluck} {et~al.}(2023){Bluck}, {Piotrowska}, \&
  {Maiolino}}]{Bluck2023}
{Bluck}, A. F.~L., {Piotrowska}, J.~M., \& {Maiolino}, R. 2023,
  \href{https://doi.org/10.3847/1538-4357/acac7c}{\apj}, 944, 108,
  [\href{https://arxiv.org/abs/2301.03677}{2301.03677}].

\bibitem[{{Bonoli} {et~al.}(2021){Bonoli}, {Mar{\'\i}n-Franch}, {Varela},
  {V{\'a}zquez Rami{\'o}}, {Abramo}, {Cenarro}, {Dupke}, {V{\'\i}lchez},
  {Crist{\'o}bal-Hornillos}, {Gonz{\'a}lez Delgado},
  {Hern{\'a}ndez-Monteagudo}, {L{\'o}pez-Sanjuan}, {Muniesa}, {Civera},
  {Ederoclite}, {Hern{\'a}n-Caballero}, {Marra}, {Baqui}, {Cortesi},
  {Cypriano}, {Daflon}, {de Amorim}, {D{\'\i}az-Garc{\'\i}a}, {Diego},
  {Mart{\'\i}nez-Solaeche}, {P{\'e}rez}, {Placco}, {Prada}, {Queiroz},
  {Alcaniz}, {Alvarez-Candal}, {Cepa}, {Maroto}, {Roig}, {Siffert}, {Taylor},
  {Benitez}, {Moles}, {Sodr{\'e}}, {Carneiro}, {Mendes de Oliveira}, {Abdalla},
  {Angulo}, {Aparicio Resco}, {Balaguera-Antol{\'\i}nez}, {Ballesteros},
  {Brito-Silva}, {Broadhurst}, {Carrasco}, {Castro}, {Cid Fernandes}, {Coelho},
  {de Melo}, {Doubrawa}, {Fernandez-Soto}, {Ferrari}, {Finoguenov},
  {Garc{\'\i}a-Benito}, {Iglesias-P{\'a}ramo}, {Jim{\'e}nez-Teja}, {Kitaura},
  {Laur}, {Lopes}, {Lucatelli}, {Mart{\'\i}nez}, {Maturi}, {Overzier},
  {Pigozzo}, {Quartin}, {Rodr{\'\i}guez-Mart{\'\i}n}, {Salzano}, {Tamm},
  {Tempel}, {Umetsu}, {Valdivielso}, {von Marttens}, {Zitrin},
  {D{\'\i}az-Mart{\'\i}n}, {L{\'o}pez-Alegre}, {L{\'o}pez-Sainz},
  {Yanes-D{\'\i}az}, {Rueda-Teruel}, {Rueda-Teruel}, {Abril Iba{\~n}ez}, {L
  Ant{\'o}n Bravo}, {Bello Ferrer}, {Bielsa}, {Casino}, {Castillo}, {Chueca},
  {Cuesta}, {Garzar{\'a}n Calderaro}, {Iglesias-Marzoa}, {{\'I}niguez},
  {Lamadrid Gutierrez}, {Lopez-Martinez}, {Lozano-P{\'e}rez}, {Ma{\'\i}cas
  Sacrist{\'a}n}, {Molina-Ib{\'a}{\~n}ez}, {Moreno-Signes}, {Rodr{\'\i}guez
  Llano}, {Royo Navarro}, {Tilve Rua}, {Andrade}, {Alfaro}, {Akras},
  {Arnalte-Mur}, {Ascaso}, {Barbosa}, {Beltr{\'a}n Jim{\'e}nez}, {Benetti},
  {Bengaly}, {Bernui}, {Blanco-Pillado}, {Borges Fernandes}, {Bregman},
  {Bruzual}, {Calderone}, {Carvano}, {Casarini}, {Chaves-Montero},
  {Chies-Santos}, {Coutinho de Carvalho}, {Dimauro}, {Duarte Puertas},
  {Figueruelo}, {Gonz{\'a}lez-Serrano}, {Guerrero}, {Gurung-L{\'o}pez},
  {Herranz}, {Huertas-Company}, {Irwin}, {Izquierdo-Villalba}, {Kanaan},
  {Kehrig}, {Kirkpatrick}, {Lim}, {Lopes}, {Lopes de Oliveira},
  {Marcos-Caballero}, {Mart{\'\i}nez-Delgado}, {Mart{\'\i}nez-Gonz{\'a}lez},
  {Mart{\'\i}nez-Somonte}, {Oliveira}, {Orsi}, {Penna-Lima}, {Reis}, {Spinoso},
  {Tsujikawa}, {Vielva}, {Vitorelli}, {Xia}, {Yuan}, {Arroyo-Polonio},
  {Dantas}, {Galarza}, {Gon{\c{c}}alves}, {Gon{\c{c}}alves}, {Gonzalez},
  {Gonzalez}, {Greisel}, {Jim{\'e}nez-Esteban}, {Landim}, {Lazzaro}, {Magris},
  {Monteiro-Oliveira}, {Pereira}, {Rebou{\c{c}}as}, {Rodriguez-Espinosa},
  {Santos da Costa}, \& {Telles}}]{Bonoli2020}
{Bonoli}, S., {Mar{\'\i}n-Franch}, A., {Varela}, J., {et~al.} 2021,
  \href{https://doi.org/10.1051/0004-6361/202038841}{\aap}, 653, A31,
  [\href{https://arxiv.org/abs/2007.01910}{2007.01910}].

\bibitem[{{Boogaard} {et~al.}(2018){Boogaard}, {Brinchmann}, {Bouch{\'e}},
  {Paalvast}, {Bacon}, {Bouwens}, {Contini}, {Gunawardhana}, {Inami}, {Marino},
  {Maseda}, {Mitchell}, {Nanayakkara}, {Richard}, {Schaye}, {Schreiber},
  {Tacchella}, {Wisotzki}, \& {Zabl}}]{boogaard2018muse}
{Boogaard}, L.~A., {Brinchmann}, J., {Bouch{\'e}}, N., {et~al.} 2018,
  \href{https://doi.org/10.1051/0004-6361/201833136}{\aap}, 619, A27,
  [\href{https://arxiv.org/abs/1808.04900}{1808.04900}].

\bibitem[{{Book} \& {Benson}(2010)}]{Book2010}
{Book}, L.~G. \& {Benson}, A.~J. 2010,
  \href{https://doi.org/10.1088/0004-637X/716/1/810}{\apj}, 716, 810,
  [\href{https://arxiv.org/abs/1001.2305}{1001.2305}].

\bibitem[{{Boquien} {et~al.}(2019){Boquien}, {Burgarella}, {Roehlly}, {Buat},
  {Ciesla}, {Corre}, {Inoue}, \& {Salas}}]{Cigale2019}
{Boquien}, M., {Burgarella}, D., {Roehlly}, Y., {et~al.} 2019,
  \href{https://doi.org/10.1051/0004-6361/201834156}{\aap}, 622, A103,
  [\href{https://arxiv.org/abs/1811.03094}{1811.03094}].

\bibitem[{{Boselli} {et~al.}(2008){Boselli}, {Boissier}, {Cortese}, \&
  {Gavazzi}}]{Boselli2008}
{Boselli}, A., {Boissier}, S., {Cortese}, L., \& {Gavazzi}, G. 2008,
  \href{https://doi.org/10.1086/525513}{\apj}, 674, 742,
  [\href{https://arxiv.org/abs/0801.2113}{0801.2113}].

\bibitem[{{Boselli} \& {Gavazzi}(2006)}]{boselli2006}
{Boselli}, A. \& {Gavazzi}, G. 2006,
  \href{https://doi.org/10.1086/500691}{\pasp}, 118, 517,
  [\href{https://arxiv.org/abs/astro-ph/0601108}{astro-ph/0601108}].

\bibitem[{{Boselli} {et~al.}(2021){Boselli}, {Lupi}, {Epinat}, {Amram},
  {Fossati}, {Anderson}, {Boissier}, {Boquien}, {Consolandi}, {C{\^o}t{\'e}},
  {Cuillandre}, {Ferrarese}, {Galbany}, {Gavazzi}, {G{\'o}mez-L{\'o}pez},
  {Gwyn}, {Hensler}, {Hutchings}, {Kuncarayakti}, {Longobardi}, {Peng},
  {Plana}, {Postma}, {Roediger}, {Roehlly}, {Schimd}, {Trinchieri}, \&
  {Vollmer}}]{Boselli2021}
{Boselli}, A., {Lupi}, A., {Epinat}, B., {et~al.} 2021,
  \href{https://doi.org/10.1051/0004-6361/202039046}{\aap}, 646, A139,
  [\href{https://arxiv.org/abs/2012.07377}{2012.07377}].

\bibitem[{{Boselli} {et~al.}(2016){Boselli}, {Roehlly}, {Fossati}, {Buat},
  {Boissier}, {Boquien}, {Burgarella}, {Ciesla}, {Gavazzi}, \&
  {Serra}}]{Boselli2016}
{Boselli}, A., {Roehlly}, Y., {Fossati}, M., {et~al.} 2016,
  \href{https://doi.org/10.1051/0004-6361/201629221}{\aap}, 596, A11,
  [\href{https://arxiv.org/abs/1609.00545}{1609.00545}].

\bibitem[{{Bower} {et~al.}(2006){Bower}, {Benson}, {Malbon}, {Helly}, {Frenk},
  {Baugh}, {Cole}, \& {Lacey}}]{Bower2006}
{Bower}, R.~G., {Benson}, A.~J., {Malbon}, R., {et~al.} 2006,
  \href{https://doi.org/10.1111/j.1365-2966.2006.10519.x}{\mnras}, 370, 645,
  [\href{https://arxiv.org/abs/astro-ph/0511338}{astro-ph/0511338}].

\bibitem[{{Bower} {et~al.}(1990){Bower}, {Ellis}, {Rose}, \&
  {Sharples}}]{Bower1990}
{Bower}, R.~G., {Ellis}, R.~S., {Rose}, J.~A., \& {Sharples}, R.~M. 1990,
  \href{https://doi.org/10.1086/115347}{\aj}, 99, 530.

\bibitem[{{Bower} {et~al.}(2008){Bower}, {McCarthy}, \& {Benson}}]{Bower2008}
{Bower}, R.~G., {McCarthy}, I.~G., \& {Benson}, A.~J. 2008,
  \href{https://doi.org/10.1111/j.1365-2966.2008.13869.x}{\mnras}, 390, 1399,
  [\href{https://arxiv.org/abs/0808.2994}{0808.2994}].

\bibitem[{{Boylan-Kolchin} {et~al.}(2009){Boylan-Kolchin}, {Springel}, {White},
  {Jenkins}, \& {Lemson}}]{boylan2009}
{Boylan-Kolchin}, M., {Springel}, V., {White}, S. D.~M., {Jenkins}, A., \&
  {Lemson}, G. 2009,
  \href{https://doi.org/10.1111/j.1365-2966.2009.15191.x}{\mnras}, 398, 1150,
  [\href{https://arxiv.org/abs/0903.3041}{0903.3041}].

\bibitem[{{Bravo-Alfaro} {et~al.}(2011){Bravo-Alfaro}, {Scott}, {Brinks},
  {Cortese}, {Granados}, {Navarro-Poupard}, {Mayya}, \&
  {Durret}}]{BravoAlfaro2011}
{Bravo-Alfaro}, H., {Scott}, T.~C., {Brinks}, E., {et~al.} 2011, in Tracing the
  Ancestry of Galaxies, ed. C.~{Carignan}, F.~{Combes}, \& K.~C. {Freeman},
  Vol. 277, 296--299

\bibitem[{{Breda} {et~al.}(2020){Breda}, {Papaderos}, {Gomes}, {V{\'\i}lchez},
  {Ziegler}, {Hirschmann}, {Cardoso}, {Lagos}, \& {Buitrago}}]{Breda2020}
{Breda}, I., {Papaderos}, P., {Gomes}, J.~M., {et~al.} 2020,
  \href{https://doi.org/10.1051/0004-6361/201937193}{\aap}, 635, A177,
  [\href{https://arxiv.org/abs/2001.05738}{2001.05738}].

\bibitem[{{Brinchmann} {et~al.}(2004){Brinchmann}, {Charlot}, {White},
  {Tremonti}, {Kauffmann}, {Heckman}, \& {Brinkmann}}]{Brinchmann2004}
{Brinchmann}, J., {Charlot}, S., {White}, S.~D.~M., {et~al.} 2004,
  \href{https://doi.org/10.1111/j.1365-2966.2004.07881.x}{\mnras}, 351, 1151,
  [\href{https://arxiv.org/abs/astro-ph/0311060}{astro-ph/0311060}].

\bibitem[{{Brough} {et~al.}(2013){Brough}, {Croom}, {Sharp}, {Hopkins},
  {Taylor}, {Baldry}, {Gunawardhana}, {Liske}, {Norberg}, {Robotham}, {Bauer},
  {Bland-Hawthorn}, {Colless}, {Foster}, {Kelvin}, {Lara-Lopez},
  {L{\'o}pez-S{\'a}nchez}, {Loveday}, {Owers}, {Pimbblet}, \&
  {Prescott}}]{Brough2013}
{Brough}, S., {Croom}, S., {Sharp}, R., {et~al.} 2013,
  \href{https://doi.org/10.1093/mnras/stt1489}{\mnras}, 435, 2903,
  [\href{https://arxiv.org/abs/1308.2985}{1308.2985}].

\bibitem[{{Brownson} {et~al.}(2022){Brownson}, {Bluck}, {Maiolino}, \&
  {Jones}}]{Brownson2022}
{Brownson}, S., {Bluck}, A. F.~L., {Maiolino}, R., \& {Jones}, G.~C. 2022,
  \href{https://doi.org/10.1093/mnras/stab3749}{\mnras}, 511, 1913,
  [\href{https://arxiv.org/abs/2201.02484}{2201.02484}].

\bibitem[{{Bruzual} \& {Charlot}(2003)}]{C&B2003}
{Bruzual}, G. \& {Charlot}, S. 2003,
  \href{https://doi.org/10.1046/j.1365-8711.2003.06897.x}{\mnras}, 344, 1000,
  [\href{https://arxiv.org/abs/astro-ph/0309134}{astro-ph/0309134}].

\bibitem[{{Bundy} {et~al.}(2015){Bundy}, {Bershady}, {Law}, {Yan}, {Drory},
  {MacDonald}, {Wake}, {Cherinka}, {S{\'a}nchez-Gallego}, {Weijmans}, {Thomas},
  {Tremonti}, {Masters}, {Coccato}, {Diamond-Stanic}, {Arag{\'o}n-Salamanca},
  {Avila-Reese}, {Badenes}, {Falc{\'o}n-Barroso}, {Belfiore}, {Bizyaev},
  {Blanc}, {Bland-Hawthorn}, {Blanton}, {Brownstein}, {Byler}, {Cappellari},
  {Conroy}, {Dutton}, {Emsellem}, {Etherington}, {Frinchaboy}, {Fu}, {Gunn},
  {Harding}, {Johnston}, {Kauffmann}, {Kinemuchi}, {Klaene}, {Knapen},
  {Leauthaud}, {Li}, {Lin}, {Maiolino}, {Malanushenko}, {Malanushenko}, {Mao},
  {Maraston}, {McDermid}, {Merrifield}, {Nichol}, {Oravetz}, {Pan}, {Parejko},
  {Sanchez}, {Schlegel}, {Simmons}, {Steele}, {Steinmetz}, {Thanjavur},
  {Thompson}, {Tinker}, {van den Bosch}, {Westfall}, {Wilkinson}, {Wright},
  {Xiao}, \& {Zhang}}]{MANGA2015}
{Bundy}, K., {Bershady}, M.~A., {Law}, D.~R., {et~al.} 2015,
  \href{https://doi.org/10.1088/0004-637X/798/1/7}{\apj}, 798, 7,
  [\href{https://arxiv.org/abs/1412.1482}{1412.1482}].

\bibitem[{{Butcher} \& {Oemler}(1978)}]{ButcherOemler1978}
{Butcher}, H. \& {Oemler}, A., J. 1978,
  \href{https://doi.org/10.1086/155751}{\apj}, 219, 18.

\bibitem[{{Butcher} \& {Oemler}(1984)}]{ButcherOemler1984}
{Butcher}, H. \& {Oemler}, A., J. 1984,
  \href{https://doi.org/10.1086/162519}{\apj}, 285, 426.

\bibitem[{{Calabretta} \& {Greisen}(2002)}]{FitsWCS2002II}
{Calabretta}, M.~R. \& {Greisen}, E.~W. 2002,
  \href{https://doi.org/10.1051/0004-6361:20021327}{\aap}, 395, 1077,
  [\href{https://arxiv.org/abs/astro-ph/0207413}{astro-ph/0207413}].

\bibitem[{{Calzetti} {et~al.}(2000){Calzetti}, {Armus}, {Bohlin}, {Kinney},
  {Koornneef}, \& {Storchi-Bergmann}}]{calzetti2000}
{Calzetti}, D., {Armus}, L., {Bohlin}, R.~C., {et~al.} 2000,
  \href{https://doi.org/10.1086/308692}{\apj}, 533, 682,
  [\href{https://arxiv.org/abs/astro-ph/9911459}{astro-ph/9911459}].

\bibitem[{{Cano-D{\'\i}az} {et~al.}(2019){Cano-D{\'\i}az}, {{\'A}vila-Reese},
  {S{\'a}nchez}, {Hern{\'a}ndez-Toledo}, {Rodr{\'\i}guez-Puebla}, {Boquien}, \&
  {Ibarra-Medel}}]{cano2019sdss}
{Cano-D{\'\i}az}, M., {{\'A}vila-Reese}, V., {S{\'a}nchez}, S.~F., {et~al.}
  2019, \href{https://doi.org/10.1093/mnras/stz1894}{\mnras}, 488, 3929,
  [\href{https://arxiv.org/abs/1907.04386}{1907.04386}].

\bibitem[{{Cano-D{\'\i}az} {et~al.}(2016){Cano-D{\'\i}az}, {S{\'a}nchez},
  {Zibetti}, {Ascasibar}, {Bland-Hawthorn}, {Ziegler}, {Gonz{\'a}lez Delgado},
  {Walcher}, {Garc{\'\i}a-Benito}, {Mast}, {Mendoza-P{\'e}rez},
  {Falc{\'o}n-Barroso}, {Galbany}, {Husemann}, {Kehrig}, {Marino},
  {S{\'a}nchez-Bl{\'a}zquez}, {L{\'o}pez-Cob{\'a}}, {L{\'o}pez-S{\'a}nchez}, \&
  {Vilchez}}]{cano2016spatially}
{Cano-D{\'\i}az}, M., {S{\'a}nchez}, S.~F., {Zibetti}, S., {et~al.} 2016,
  \href{https://doi.org/10.3847/2041-8205/821/2/L26}{\apjl}, 821, L26,
  [\href{https://arxiv.org/abs/1602.02770}{1602.02770}].

\bibitem[{{Cantalupo}(2010)}]{Cantalupo2010}
{Cantalupo}, S. 2010,
  \href{https://doi.org/10.1111/j.1745-3933.2010.00806.x}{\mnras}, 403, L16,
  [\href{https://arxiv.org/abs/0912.4149}{0912.4149}].

\bibitem[{{Capak} {et~al.}(2007){Capak}, {Aussel}, {Ajiki}, {McCracken},
  {Mobasher}, {Scoville}, {Shopbell}, {Taniguchi}, {Thompson}, {Tribiano},
  {Sasaki}, {Blain}, {Brusa}, {Carilli}, {Comastri}, {Carollo}, {Cassata},
  {Colbert}, {Ellis}, {Elvis}, {Giavalisco}, {Green}, {Guzzo}, {Hasinger},
  {Ilbert}, {Impey}, {Jahnke}, {Kartaltepe}, {Kneib}, {Koda}, {Koekemoer},
  {Komiyama}, {Leauthaud}, {Le Fevre}, {Lilly}, {Liu}, {Massey}, {Miyazaki},
  {Murayama}, {Nagao}, {Peacock}, {Pickles}, {Porciani}, {Renzini}, {Rhodes},
  {Rich}, {Salvato}, {Sanders}, {Scarlata}, {Schiminovich}, {Schinnerer},
  {Scodeggio}, {Sheth}, {Shioya}, {Tasca}, {Taylor}, {Yan}, \&
  {Zamorani}}]{Capak2007}
{Capak}, P., {Aussel}, H., {Ajiki}, M., {et~al.} 2007,
  \href{https://doi.org/10.1086/519081}{\apjs}, 172, 99,
  [\href{https://arxiv.org/abs/0704.2430}{0704.2430}].

\bibitem[{{Cappellari}(2016)}]{Cappellari2016}
{Cappellari}, M. 2016,
  \href{https://doi.org/10.1146/annurev-astro-082214-122432}{\araa}, 54, 597,
  [\href{https://arxiv.org/abs/1602.04267}{1602.04267}].

\bibitem[{{Cappellari} \& {Copin}(2003)}]{Voronoi}
{Cappellari}, M. \& {Copin}, Y. 2003,
  \href{https://doi.org/10.1046/j.1365-8711.2003.06541.x}{\mnras}, 342, 345,
  [\href{https://arxiv.org/abs/astro-ph/0302262}{astro-ph/0302262}].

\bibitem[{{Cappellari} {et~al.}(2011){Cappellari}, {Emsellem}, {Krajnovi{\'c}},
  {McDermid}, {Scott}, {Verdoes Kleijn}, {Young}, {Alatalo}, {Bacon}, {Blitz},
  {Bois}, {Bournaud}, {Bureau}, {Davies}, {Davis}, {de Zeeuw}, {Duc},
  {Khochfar}, {Kuntschner}, {Lablanche}, {Morganti}, {Naab}, {Oosterloo},
  {Sarzi}, {Serra}, \& {Weijmans}}]{Cappellari2011}
{Cappellari}, M., {Emsellem}, E., {Krajnovi{\'c}}, D., {et~al.} 2011,
  \href{https://doi.org/10.1111/j.1365-2966.2010.18174.x}{\mnras}, 413, 813,
  [\href{https://arxiv.org/abs/1012.1551}{1012.1551}].

\bibitem[{{Casado} {et~al.}(2017){Casado}, {Ascasibar}, {Garc{\'\i}a-Benito},
  {Guidi}, {Choudhury}, {Bellocchi}, {S{\'a}nchez}, \& {D{\'\i}az}}]{BATMAN}
{Casado}, J., {Ascasibar}, Y., {Garc{\'\i}a-Benito}, R., {et~al.} 2017,
  \href{https://doi.org/10.1093/mnras/stw3362}{\mnras}, 466, 3989,
  [\href{https://arxiv.org/abs/1607.07299}{1607.07299}].

\bibitem[{{Catal{\'a}n-Torrecilla} {et~al.}(2015){Catal{\'a}n-Torrecilla}, {Gil
  de Paz}, {Castillo-Morales}, {Iglesias-P{\'a}ramo}, {S{\'a}nchez},
  {Kennicutt}, {P{\'e}rez-Gonz{\'a}lez}, {Marino}, {Walcher}, {Husemann},
  {Garc{\'\i}a-Benito}, {Mast}, {Gonz{\'a}lez Delgado}, {Mu{\~n}oz-Mateos},
  {Bland-Hawthorn}, {Bomans}, {Del Olmo}, {Galbany}, {Gomes}, {Kehrig},
  {L{\'o}pez-S{\'a}nchez}, {Mendoza}, {Monreal-Ibero}, {P{\'e}rez-Torres},
  {S{\'a}nchez-Bl{\'a}zquez}, {Vilchez}, \& {Califa Collaboration}}]{CT2015}
{Catal{\'a}n-Torrecilla}, C., {Gil de Paz}, A., {Castillo-Morales}, A.,
  {et~al.} 2015, \href{https://doi.org/10.1051/0004-6361/201526023}{\aap}, 584,
  A87, [\href{https://arxiv.org/abs/1507.03801}{1507.03801}].

\bibitem[{{Cenarro} {et~al.}(2018{\natexlab{a}}){Cenarro}, {Ederoclite},
  {{\'I}{\~n}iguez}, {Mar{\'\i}n-Franch}, {V{\'a}zquez Rami{\'o}},
  {Yanes-D{\'\i}az}, {Chueca}, {Lasso-Cabrera}, {Rueda-Teruel}, {Rueda-Teruel},
  {Crist{\'o}bal-Hornillos}, {Varela}, {L{\'o}pez-Alegre}, {Bello},
  {Ant{\'o}n-Bravo}, {Bielsa de Toledo}, {Dom{\'\i}nguez-Mart{\'\i}nez},
  {Moreno-Signes}, {Iglesias-Marzoa}, {D{\'\i}az-Mart{\'\i}n}, {Civera},
  {Hern{\'a}ndez-Fuertes}, {Muniesa-Gallardo}, {Castillo},
  {L{\'o}pez-S{\'a}inz}, {Moles}, {Lousberg}, {Bastin}, \&
  {Pirnay}}]{Cenarro2018}
{Cenarro}, A.~J., {Ederoclite}, A., {{\'I}{\~n}iguez}, C., {et~al.}
  2018{\natexlab{a}}, in Society of Photo-Optical Instrumentation Engineers
  (SPIE) Conference Series, Vol. 10700, Ground-based and Airborne Telescopes
  VII, ed. H.~K. {Marshall} \& J.~{Spyromilio}, 107000D

\bibitem[{{Cenarro} {et~al.}(2018{\natexlab{b}}){Cenarro}, {Ederoclite},
  {{\'I}{\~n}iguez}, {Mar{\'\i}n-Franch}, {V{\'a}zquez Rami{\'o}},
  {Yanes-D{\'\i}az}, {Chueca}, {Lasso-Cabrera}, {Rueda-Teruel}, {Rueda-Teruel},
  {Crist{\'o}bal-Hornillos}, {Varela}, {L{\'o}pez-Alegre}, {Bello},
  {Ant{\'o}n-Bravo}, {Bielsa de Toledo}, {Dom{\'\i}nguez-Mart{\'\i}nez},
  {Moreno-Signes}, {Iglesias-Marzoa}, {D{\'\i}az-Mart{\'\i}n}, {Civera},
  {Hern{\'a}ndez-Fuertes}, {Muniesa-Gallardo}, {Castillo},
  {L{\'o}pez-S{\'a}inz}, {Moles}, {Lousberg}, {Bastin}, \& {Pirnay}}]{T250}
{Cenarro}, A.~J., {Ederoclite}, A., {{\'I}{\~n}iguez}, C., {et~al.}
  2018{\natexlab{b}}, in Society of Photo-Optical Instrumentation Engineers
  (SPIE) Conference Series, Vol. 10700, Ground-based and Airborne Telescopes
  VII, ed. H.~K. {Marshall} \& J.~{Spyromilio}, 107000D

\bibitem[{{Cenarro} {et~al.}(2019){Cenarro}, {Moles},
  {Crist{\'o}bal-Hornillos}, {Mar{\'\i}n-Franch}, {Ederoclite}, {Varela},
  {L{\'o}pez-Sanjuan}, {Hern{\'a}ndez-Monteagudo}, {Angulo}, {V{\'a}zquez
  Rami{\'o}}, {Viironen}, {Bonoli}, {Orsi}, {Hurier}, {San Roman}, {Greisel},
  {Vilella-Rojo}, {D{\'\i}az-Garc{\'\i}a}, {Logro{\~n}o-Garc{\'\i}a},
  {Gurung-L{\'o}pez}, {Spinoso}, {Izquierdo-Villalba}, {Aguerri}, {Allende
  Prieto}, {Bonatto}, {Carvano}, {Chies-Santos}, {Daflon}, {Dupke},
  {Falc{\'o}n-Barroso}, {Gon{\c{c}}alves}, {Jim{\'e}nez-Teja}, {Molino},
  {Placco}, {Solano}, {Whitten}, {Abril}, {Ant{\'o}n}, {Bello}, {Bielsa de
  Toledo}, {Castillo-Ram{\'\i}rez}, {Chueca}, {Civera},
  {D{\'\i}az-Mart{\'\i}n}, {Dom{\'\i}nguez-Mart{\'\i}nez},
  {Garzar{\'a}n-Calderaro}, {Hern{\'a}ndez-Fuertes}, {Iglesias-Marzoa},
  {I{\~n}iguez}, {Jim{\'e}nez Ruiz}, {Kruuse}, {Lamadrid}, {Lasso-Cabrera},
  {L{\'o}pez-Alegre}, {L{\'o}pez-Sainz}, {Ma{\'\i}cas}, {Moreno-Signes},
  {Muniesa}, {Rodr{\'\i}guez-Llano}, {Rueda-Teruel}, {Rueda-Teruel},
  {Soriano-Lagu{\'\i}a}, {Tilve}, {Valdivielso}, {Yanes-D{\'\i}az}, {Alcaniz},
  {Mendes de Oliveira}, {Sodr{\'e}}, {Coelho}, {Lopes de Oliveira}, {Tamm},
  {Xavier}, {Abramo}, {Akras}, {Alfaro}, {Alvarez-Candal}, {Ascaso}, {Beasley},
  {Beers}, {Borges Fernandes}, {Bruzual}, {Buzzo}, {Carrasco}, {Cepa},
  {Cortesi}, {Costa-Duarte}, {De Pr{\'a}}, {Favole}, {Galarza}, {Galbany},
  {Garcia}, {Gonz{\'a}lez Delgado}, {Gonz{\'a}lez-Serrano},
  {Guti{\'e}rrez-Soto}, {Hernandez-Jimenez}, {Kanaan}, {Kuncarayakti},
  {Landim}, {Laur}, {Licandro}, {Lima Neto}, {Lyman}, {Ma{\'\i}z
  Apell{\'a}niz}, {Miralda-Escud{\'e}}, {Morate}, {Nogueira-Cavalcante},
  {Novais}, {Oncins}, {Oteo}, {Overzier}, {Pereira}, {Rebassa-Mansergas},
  {Reis}, {Roig}, {Sako}, {Salvador-Rusi{\~n}ol}, {Sampedro},
  {S{\'a}nchez-Bl{\'a}zquez}, {Santos}, {Schmidtobreick}, {Siffert}, {Telles},
  \& {Vilchez}}]{Cenarro2019}
{Cenarro}, A.~J., {Moles}, M., {Crist{\'o}bal-Hornillos}, D., {et~al.} 2019,
  \href{https://doi.org/10.1051/0004-6361/201833036}{\aap}, 622, A176,
  [\href{https://arxiv.org/abs/1804.02667}{1804.02667}].

\bibitem[{{Cenarro} {et~al.}(2014{\natexlab{a}}){Cenarro}, {Moles},
  {Mar{\'\i}n-Franch}, {Crist{\'o}bal-Hornillos}, {Yanes D{\'\i}az},
  {Ederoclite}, {Varela}, {V{\'a}zquez Rami{\'o}}, {Valdivielso},
  {Ben{\'\i}tez}, {Cepa}, {Dupke}, {Fern{\'a}ndez-Soto}, {Mendes de Oliveira},
  {Sodr{\'e}}, {Taylor}, {Rueda-Teruel}, {Rueda-Teruel}, {Luis-Simoes},
  {Chueca}, {Ant{\'o}n}, {Bello}, {D{\'\i}az-Mart{\'\i}n},
  {Guill{\'e}n-Civera}, {Hern{\'a}ndez-Fuertes}, {Iglesias-Marzoa},
  {Jim{\'e}nez-Mej{\'\i}as}, {Lasso-Cabrera}, {L{\'o}pez-Alegre},
  {L{\'o}pez-Sainz}, {Rodr{\'\i}guez-Hern{\'a}ndez}, {Su{\'a}rez}, {Lamadrid},
  {Ma{\'\i}cas}, {Abril-Iba{\~n}ez}, {Tilve}, \&
  {Rodr{\'\i}guez-Llano}}]{Cenarro2014}
{Cenarro}, A.~J., {Moles}, M., {Mar{\'\i}n-Franch}, A., {et~al.}
  2014{\natexlab{a}}, in Society of Photo-Optical Instrumentation Engineers
  (SPIE) Conference Series, Vol. 9149, Observatory Operations: Strategies,
  Processes, and Systems V, ed. A.~B. {Peck}, C.~R. {Benn}, \& R.~L. {Seaman},
  91491I

\bibitem[{{Cenarro} {et~al.}(2014{\natexlab{b}}){Cenarro}, {Moles},
  {Mar{\'\i}n-Franch}, {Crist{\'o}bal-Hornillos}, {Yanes D{\'\i}az},
  {Ederoclite}, {Varela}, {V{\'a}zquez Rami{\'o}}, {Valdivielso},
  {Ben{\'\i}tez}, {Cepa}, {Dupke}, {Fern{\'a}ndez-Soto}, {Mendes de Oliveira},
  {Sodr{\'e}}, {Taylor}, {Rueda-Teruel}, {Rueda-Teruel}, {Luis-Simoes},
  {Chueca}, {Ant{\'o}n}, {Bello}, {D{\'\i}az-Mart{\'\i}n},
  {Guill{\'e}n-Civera}, {Hern{\'a}ndez-Fuertes}, {Iglesias-Marzoa},
  {Jim{\'e}nez-Mej{\'\i}as}, {Lasso-Cabrera}, {L{\'o}pez-Alegre},
  {L{\'o}pez-Sainz}, {Rodr{\'\i}guez-Hern{\'a}ndez}, {Su{\'a}rez}, {Lamadrid},
  {Ma{\'\i}cas}, {Abril-Iba{\~n}ez}, {Tilve}, \& {Rodr{\'\i}guez-Llano}}]{OAJ}
{Cenarro}, A.~J., {Moles}, M., {Mar{\'\i}n-Franch}, A., {et~al.}
  2014{\natexlab{b}}, in Society of Photo-Optical Instrumentation Engineers
  (SPIE) Conference Series, Vol. 9149, Observatory Operations: Strategies,
  Processes, and Systems V, ed. A.~B. {Peck}, C.~R. {Benn}, \& R.~L. {Seaman},
  91491I

\bibitem[{{Chabrier}(2003)}]{Chabrier2003}
{Chabrier}, G. 2003, \href{https://doi.org/10.1086/376392}{\pasp}, 115, 763,
  [\href{https://arxiv.org/abs/astro-ph/0304382}{astro-ph/0304382}].

\bibitem[{{Chambers} {et~al.}(2016){Chambers}, {Magnier}, {Metcalfe},
  {Flewelling}, {Huber}, {Waters}, {Denneau}, {Draper}, {Farrow}, {Finkbeiner},
  {Holmberg}, {Koppenhoefer}, {Price}, {Rest}, {Saglia}, {Schlafly}, {Smartt},
  {Sweeney}, {Wainscoat}, {Burgett}, {Chastel}, {Grav}, {Heasley}, {Hodapp},
  {Jedicke}, {Kaiser}, {Kudritzki}, {Luppino}, {Lupton}, {Monet}, {Morgan},
  {Onaka}, {Shiao}, {Stubbs}, {Tonry}, {White}, {Ba{\~n}ados}, {Bell},
  {Bender}, {Bernard}, {Boegner}, {Boffi}, {Botticella}, {Calamida},
  {Casertano}, {Chen}, {Chen}, {Cole}, {Deacon}, {Frenk}, {Fitzsimmons},
  {Gezari}, {Gibbs}, {Goessl}, {Goggia}, {Gourgue}, {Goldman}, {Grant},
  {Grebel}, {Hambly}, {Hasinger}, {Heavens}, {Heckman}, {Henderson}, {Henning},
  {Holman}, {Hopp}, {Ip}, {Isani}, {Jackson}, {Keyes}, {Koekemoer}, {Kotak},
  {Le}, {Liska}, {Long}, {Lucey}, {Liu}, {Martin}, {Masci}, {McLean}, {Mindel},
  {Misra}, {Morganson}, {Murphy}, {Obaika}, {Narayan}, {Nieto-Santisteban},
  {Norberg}, {Peacock}, {Pier}, {Postman}, {Primak}, {Rae}, {Rai}, {Riess},
  {Riffeser}, {Rix}, {R{\"o}ser}, {Russel}, {Rutz}, {Schilbach}, {Schultz},
  {Scolnic}, {Strolger}, {Szalay}, {Seitz}, {Small}, {Smith}, {Soderblom},
  {Taylor}, {Thomson}, {Taylor}, {Thakar}, {Thiel}, {Thilker}, {Unger},
  {Urata}, {Valenti}, {Wagner}, {Walder}, {Walter}, {Watters}, {Werner},
  {Wood-Vasey}, \& {Wyse}}]{Chambers2016}
{Chambers}, K.~C., {Magnier}, E.~A., {Metcalfe}, N., {et~al.} 2016, arXiv
  e-prints, arXiv:1612.05560,
  [\href{https://arxiv.org/abs/1612.05560}{1612.05560}].

\bibitem[{{Charlot} \& {Fall}(2000)}]{charlot2000}
{Charlot}, S. \& {Fall}, S.~M. 2000,
  \href{https://doi.org/10.1086/309250}{\apj}, 539, 718,
  [\href{https://arxiv.org/abs/astro-ph/0003128}{astro-ph/0003128}].

\bibitem[{{Chies-Santos} {et~al.}(2015){Chies-Santos}, {Rodr{\'\i}guez del
  Pino}, {Arag{\'o}n-Salamanca}, {Bamford}, {Gray}, {Wolf}, {B{\"o}hm},
  {Maltby}, {Pintos-Castro}, {Sanch{\'e}z-Portal}, \& {Weinzirl}}]{Chies2015}
{Chies-Santos}, A.~L., {Rodr{\'\i}guez del Pino}, B., {Arag{\'o}n-Salamanca},
  A., {et~al.} 2015, \href{https://doi.org/10.1093/mnras/stv779}{\mnras}, 450,
  4458, [\href{https://arxiv.org/abs/1504.02053}{1504.02053}].

\bibitem[{Chollet {et~al.}(2015)}]{chollet2015keras}
Chollet, F. {et~al.} 2015, Keras

\bibitem[{{Cid Fernandes} {et~al.}(2005){Cid Fernandes}, {Mateus}, {Sodr{\'e}},
  {Stasi{\'n}ska}, \& {Gomes}}]{cid2005}
{Cid Fernandes}, R., {Mateus}, A., {Sodr{\'e}}, L., {Stasi{\'n}ska}, G., \&
  {Gomes}, J.~M. 2005,
  \href{https://doi.org/10.1111/j.1365-2966.2005.08752.x}{\mnras}, 358, 363,
  [\href{https://arxiv.org/abs/astro-ph/0412481}{astro-ph/0412481}].

\bibitem[{{Cid Fernandes} {et~al.}(2011){Cid Fernandes}, {Stasi{\'n}ska},
  {Mateus}, \& {Vale Asari}}]{WHAN_2}
{Cid Fernandes}, R., {Stasi{\'n}ska}, G., {Mateus}, A., \& {Vale Asari}, N.
  2011, \href{https://doi.org/10.1111/j.1365-2966.2011.18244.x}{\mnras}, 413,
  1687, [\href{https://arxiv.org/abs/1012.4426}{1012.4426}].

\bibitem[{{Cid Fernandes} {et~al.}(2010){Cid Fernandes}, {Stasi{\'n}ska},
  {Schlickmann}, {Mateus}, {Vale Asari}, {Schoenell}, \& {Sodr{\'e}}}]{WHAN_1}
{Cid Fernandes}, R., {Stasi{\'n}ska}, G., {Schlickmann}, M.~S., {et~al.} 2010,
  \href{https://doi.org/10.1111/j.1365-2966.2009.16185.x}{\mnras}, 403, 1036,
  [\href{https://arxiv.org/abs/0912.1643}{0912.1643}].

\bibitem[{{Civera} \& {Hern{\'a}ndez}(2020)}]{Civera2020}
{Civera}, T. \& {Hern{\'a}ndez}, J. 2020, in Astronomical Society of the
  Pacific Conference Series, Vol. 527, Astronomical Data Analysis Software and
  Systems XXIX, ed. R.~{Pizzo}, E.~R. {Deul}, J.~D. {Mol}, J.~{de Plaa}, \&
  H.~{Verkouter}, 101

\bibitem[{{Cleland} \& {McGee}(2021)}]{Cleland2021}
{Cleland}, C. \& {McGee}, S.~L. 2021,
  \href{https://doi.org/10.1093/mnras/staa3267}{\mnras}, 500, 590,
  [\href{https://arxiv.org/abs/2006.16307}{2006.16307}].

\bibitem[{{Clemens} {et~al.}(2006){Clemens}, {Bressan}, {Nikolic}, {Alexander},
  {Annibali}, \& {Rampazzo}}]{Clemens2006}
{Clemens}, M.~S., {Bressan}, A., {Nikolic}, B., {et~al.} 2006,
  \href{https://doi.org/10.1111/j.1365-2966.2006.10530.x}{\mnras}, 370, 702,
  [\href{https://arxiv.org/abs/astro-ph/0603714}{astro-ph/0603714}].

\bibitem[{{Coe} {et~al.}(2006){Coe}, {Ben{\'\i}tez}, {S{\'a}nchez}, {Jee},
  {Bouwens}, \& {Ford}}]{Coe2006}
{Coe}, D., {Ben{\'\i}tez}, N., {S{\'a}nchez}, S.~F., {et~al.} 2006,
  \href{https://doi.org/10.1086/505530}{\aj}, 132, 926,
  [\href{https://arxiv.org/abs/astro-ph/0605262}{astro-ph/0605262}].

\bibitem[{{Combes}(2017)}]{Combes2017}
{Combes}, F. 2017, \href{https://doi.org/10.3389/fspas.2017.00010}{Frontiers in
  Astronomy and Space Sciences}, 4, 10,
  [\href{https://arxiv.org/abs/1707.09621}{1707.09621}].

\bibitem[{{Conrado} {et~al.}(2024){Conrado}, {Gonz{\'a}lez Delgado},
  {Garc{\'\i}a-Benito}, {P{\'e}rez}, {Verley}, {Ruiz-Lara},
  {S{\'a}nchez-Menguiano}, {Duarte Puertas}, {Jim{\'e}nez},
  {Dom{\'\i}nguez-G{\'o}mez}, {Espada}, {Argudo-Fern{\'a}ndez},
  {Alc{\'a}zar-Laynez}, {Bl{\'a}zquez-Calero}, {Bidaran}, {Zurita}, {Peletier},
  {Torres-R{\'\i}os}, {Florido}, {Rodr{\'\i}guez Mart{\'\i}nez}, {del
  Moral-Castro}, {van de Weygaert}, {Falc{\'o}n-Barroso}, {Lugo-Aranda},
  {S{\'a}nchez}, {van der Hulst}, {Courtois}, {Ferr{\'e}-Mateu},
  {S{\'a}nchez-Bl{\'a}zquez}, {Rom{\'a}n}, \& {Aceituno}}]{Ana2024}
{Conrado}, A.~M., {Gonz{\'a}lez Delgado}, R.~M., {Garc{\'\i}a-Benito}, R.,
  {et~al.} 2024, \href{https://doi.org/10.1051/0004-6361/202449414}{\aap}, 687,
  A98, [\href{https://arxiv.org/abs/2404.10823}{2404.10823}].

\bibitem[{{Conroy}(2013)}]{Conroy2013}
{Conroy}, C. 2013,
  \href{https://doi.org/10.1146/annurev-astro-082812-141017}{\araa}, 51, 393,
  [\href{https://arxiv.org/abs/1301.7095}{1301.7095}].

\bibitem[{{Contini} {et~al.}(2020){Contini}, {Gu}, {Ge}, {Rhee}, {Yi}, \&
  {Kang}}]{Contini2020}
{Contini}, E., {Gu}, Q., {Ge}, X., {et~al.} 2020,
  \href{https://doi.org/10.3847/1538-4357/ab6730}{\apj}, 889, 156,
  [\href{https://arxiv.org/abs/2001.01369}{2001.01369}].

\bibitem[{{Cooper} {et~al.}(2010){Cooper}, {Gallazzi}, {Newman}, \&
  {Yan}}]{Cooper2010MNRAS}
{Cooper}, M.~C., {Gallazzi}, A., {Newman}, J.~A., \& {Yan}, R. 2010,
  \href{https://doi.org/10.1111/j.1365-2966.2009.16020.x}{\mnras}, 402, 1942,
  [\href{https://arxiv.org/abs/0910.0245}{0910.0245}].

\bibitem[{{Cooper} {et~al.}(2012){Cooper}, {Griffith}, {Newman}, {Coil},
  {Davis}, {Dutton}, {Faber}, {Guhathakurta}, {Koo}, {Lotz}, {Weiner},
  {Willmer}, \& {Yan}}]{Cooper2012}
{Cooper}, M.~C., {Griffith}, R.~L., {Newman}, J.~A., {et~al.} 2012,
  \href{https://doi.org/10.1111/j.1365-2966.2011.19938.x}{\mnras}, 419, 3018,
  [\href{https://arxiv.org/abs/1109.5698}{1109.5698}].

\bibitem[{{Cortijo-Ferrero} {et~al.}(2017{\natexlab{a}}){Cortijo-Ferrero},
  {Gonz{\'a}lez Delgado}, {P{\'e}rez}, {Cid Fernandes}, {Garc{\'\i}a-Benito},
  {Di Matteo}, {S{\'a}nchez}, {de Amorim}, {Lacerda}, {L{\'o}pez
  Fern{\'a}ndez}, \& {Tadhunter}}]{CortijoFerrero2017a}
{Cortijo-Ferrero}, C., {Gonz{\'a}lez Delgado}, R.~M., {P{\'e}rez}, E., {et~al.}
  2017{\natexlab{a}}, \href{https://doi.org/10.1051/0004-6361/201731217}{\aap},
  607, A70, [\href{https://arxiv.org/abs/1707.05324}{1707.05324}].

\bibitem[{{Cortijo-Ferrero} {et~al.}(2017{\natexlab{b}}){Cortijo-Ferrero},
  {Gonz{\'a}lez Delgado}, {P{\'e}rez}, {Cid Fernandes}, {S{\'a}nchez}, {de
  Amorim}, {Di Matteo}, {Garc{\'\i}a-Benito}, {Lacerda}, {L{\'o}pez
  Fern{\'a}ndez}, \& {Tadhunter}}]{CortijoFerrero2017b}
{Cortijo-Ferrero}, C., {Gonz{\'a}lez Delgado}, R.~M., {P{\'e}rez}, E., {et~al.}
  2017{\natexlab{b}}, \href{https://doi.org/10.1093/mnras/stx383}{\mnras}, 467,
  3898, [\href{https://arxiv.org/abs/1702.06544}{1702.06544}].

\bibitem[{{Cortijo-Ferrero} {et~al.}(2017{\natexlab{c}}){Cortijo-Ferrero},
  {Gonz{\'a}lez Delgado}, {P{\'e}rez}, {S{\'a}nchez}, {Cid Fernandes}, {de
  Amorim}, {Di Matteo}, {Garc{\'\i}a-Benito}, {Lacerda}, {L{\'o}pez
  Fern{\'a}ndez}, {Tadhunter}, {Villar-Mart{\'\i}n}, \&
  {Roth}}]{CortijoFerrero2017c}
{Cortijo-Ferrero}, C., {Gonz{\'a}lez Delgado}, R.~M., {P{\'e}rez}, E., {et~al.}
  2017{\natexlab{c}}, \href{https://doi.org/10.1051/0004-6361/201730669}{\aap},
  606, A95, [\href{https://arxiv.org/abs/1706.01896}{1706.01896}].

\bibitem[{{Costa-Souza} {et~al.}(2024){Costa-Souza}, {Riffel}, {Dors},
  {Riffel}, \& {da Rocha-Poppe}}]{Costa-Souza2024}
{Costa-Souza}, J.~H., {Riffel}, R.~A., {Dors}, O.~L., {Riffel}, R., \& {da
  Rocha-Poppe}, P.~C. 2024,
  \href{https://doi.org/10.1093/mnras/stad3809}{\mnras}, 527, 9192,
  [\href{https://arxiv.org/abs/2310.15842}{2310.15842}].

\bibitem[{{Courtes}(1982)}]{Courtes1982}
{Courtes}, G. 1982, in Astrophysics and Space Science Library, Vol.~92, IAU
  Colloq. 67: Instrumentation for Astronomy with Large Optical Telescopes, ed.
  C.~M. {Humphries}, 123

\bibitem[{{Covey} {et~al.}(2007){Covey}, {Ivezi{\'c}}, {Schlegel},
  {Finkbeiner}, {Padmanabhan}, {Lupton}, {Ag{\"u}eros}, {Bochanski}, {Hawley},
  {West}, {Seth}, {Kimball}, {Gogarten}, {Claire}, {Haggard}, {Kaib},
  {Schneider}, \& {Sesar}}]{Covey2007}
{Covey}, K.~R., {Ivezi{\'c}}, {\v{Z}}., {Schlegel}, D., {et~al.} 2007,
  \href{https://doi.org/10.1086/522052}{\aj}, 134, 2398,
  [\href{https://arxiv.org/abs/0707.4473}{0707.4473}].

\bibitem[{{Cowie} \& {Songaila}(1977)}]{Cowie1977}
{Cowie}, L.~L. \& {Songaila}, A. 1977,
  \href{https://doi.org/10.1038/266501a0}{\nat}, 266, 501.

\bibitem[{{Crist{\'o}bal-Hornillos} {et~al.}(2014){Crist{\'o}bal-Hornillos},
  {Varela}, {Ederoclite}, {V{\'a}zquez Rami{\'o}}, {L{\'o}pez-Sainz},
  {Hern{\'a}ndez-Fuertes}, {Civera}, {Muniesa}, {Moles}, {Cenarro},
  {Mar{\'\i}n-Franch}, \& {Yanes-D{\'\i}az}}]{Cristobal-Hornillos2014}
{Crist{\'o}bal-Hornillos}, D., {Varela}, J., {Ederoclite}, A., {et~al.} 2014,
  in Society of Photo-Optical Instrumentation Engineers (SPIE) Conference
  Series, Vol. 9152, Software and Cyberinfrastructure for Astronomy III, ed.
  G.~{Chiozzi} \& N.~M. {Radziwill}, 91520O

\bibitem[{{Croom} {et~al.}(2021){Croom}, {Owers}, {Scott}, {Poetrodjojo},
  {Groves}, {van de Sande}, {Barone}, {Cortese}, {D'Eugenio}, {Bland-Hawthorn},
  {Bryant}, {Oh}, {Brough}, {Agostino}, {Casura}, {Catinella}, {Colless},
  {Cecil}, {Davies}, {Drinkwater}, {Driver}, {Ferreras}, {Foster},
  {Fraser-McKelvie}, {Lawrence}, {Leslie}, {Liske}, {L{\'o}pez-S{\'a}nchez},
  {Lorente}, {McElroy}, {Medling}, {Obreschkow}, {Richards}, {Sharp}, {Sweet},
  {Taranu}, {Taylor}, {Tescari}, {Thomas}, {Tocknell}, \&
  {Vaughan}}]{Croom2021}
{Croom}, S.~M., {Owers}, M.~S., {Scott}, N., {et~al.} 2021,
  \href{https://doi.org/10.1093/mnras/stab229}{\mnras}, 505, 991,
  [\href{https://arxiv.org/abs/2101.12224}{2101.12224}].

\bibitem[{{Croton} {et~al.}(2006){Croton}, {Springel}, {White}, {De Lucia},
  {Frenk}, {Gao}, {Jenkins}, {Kauffmann}, {Navarro}, \& {Yoshida}}]{Croton2006}
{Croton}, D.~J., {Springel}, V., {White}, S. D.~M., {et~al.} 2006,
  \href{https://doi.org/10.1111/j.1365-2966.2005.09675.x}{\mnras}, 365, 11,
  [\href{https://arxiv.org/abs/astro-ph/0508046}{astro-ph/0508046}].

\bibitem[{{Cypriano} {et~al.}(2010){Cypriano}, {Amara}, {Voigt}, {Bridle},
  {Abdalla}, {R{\'e}fr{\'e}gier}, {Seiffert}, \& {Rhodes}}]{Cyprian2010}
{Cypriano}, E.~S., {Amara}, A., {Voigt}, L.~M., {et~al.} 2010,
  \href{https://doi.org/10.1111/j.1365-2966.2010.16461.x}{\mnras}, 405, 494,
  [\href{https://arxiv.org/abs/1001.0759}{1001.0759}].

\bibitem[{{Dacunha} {et~al.}(2022){Dacunha}, {Belyakov}, {Adhikari}, {Shin},
  {Goldstein}, \& {Jain}}]{Dacunha2022}
{Dacunha}, T., {Belyakov}, M., {Adhikari}, S., {et~al.} 2022,
  \href{https://doi.org/10.1093/mnras/stac392}{\mnras}, 512, 4378,
  [\href{https://arxiv.org/abs/2111.06499}{2111.06499}].

\bibitem[{{Darnell} {et~al.}(2009){Darnell}, {Bertin}, {Gower}, {Ngeow},
  {Desai}, {Mohr}, {Adams}, {Daues}, {Gower}, {Ngeow}, {Desai}, {Beldica},
  {Freemon}, {Lin}, {Neilsen}, {Tucker}, {da Costa}, {Martelli}, {Ogando},
  {Jarvis}, \& {Sheldon}}]{Darnell2009}
{Darnell}, T., {Bertin}, E., {Gower}, M., {et~al.} 2009, in Astronomical
  Society of the Pacific Conference Series, Vol. 411, Astronomical Data
  Analysis Software and Systems XVIII, ed. D.~A. {Bohlender}, D.~{Durand}, \&
  P.~{Dowler}, 18

\bibitem[{{Davies} {et~al.}(2015){Davies}, {Robotham}, {Driver}, {Alpaslan},
  {Baldry}, {Bland-Hawthorn}, {Brough}, {Brown}, {Cluver}, {Drinkwater},
  {Foster}, {Grootes}, {Konstantopoulos}, {Lara-L{\'o}pez},
  {L{\'o}pez-S{\'a}nchez}, {Loveday}, {Meyer}, {Moffett}, {Norberg}, {Owers},
  {Popescu}, {De Propris}, {Sharp}, {Tuffs}, {Wang}, {Wilkins}, {Dunne},
  {Bourne}, \& {Smith}}]{Davies2015}
{Davies}, L.~J.~M., {Robotham}, A.~S.~G., {Driver}, S.~P., {et~al.} 2015,
  \href{https://doi.org/10.1093/mnras/stv1241}{\mnras}, 452, 616,
  [\href{https://arxiv.org/abs/1507.04447}{1507.04447}].

\bibitem[{{Davies} {et~al.}(1993){Davies}, {Sadler}, \&
  {Peletier}}]{Davies1993}
{Davies}, R.~L., {Sadler}, E.~M., \& {Peletier}, R.~F. 1993,
  \href{https://doi.org/10.1093/mnras/262.3.650}{\mnras}, 262, 650.

\bibitem[{{Davis} {et~al.}(1985){Davis}, {Efstathiou}, {Frenk}, \&
  {White}}]{Davis1985}
{Davis}, M., {Efstathiou}, G., {Frenk}, C.~S., \& {White}, S.~D.~M. 1985,
  \href{https://doi.org/10.1086/163168}{\apj}, 292, 371.

\bibitem[{{Davis} {et~al.}(2007){Davis}, {Guhathakurta}, {Konidaris}, {Newman},
  {Ashby}, {Biggs}, {Barmby}, {Bundy}, {Chapman}, {Coil}, {Conselice},
  {Cooper}, {Croton}, {Eisenhardt}, {Ellis}, {Faber}, {Fang}, {Fazio},
  {Georgakakis}, {Gerke}, {Goss}, {Gwyn}, {Harker}, {Hopkins}, {Huang},
  {Ivison}, {Kassin}, {Kirby}, {Koekemoer}, {Koo}, {Laird}, {Le Floc'h}, {Lin},
  {Lotz}, {Marshall}, {Martin}, {Metevier}, {Moustakas}, {Nandra}, {Noeske},
  {Papovich}, {Phillips}, {Rich}, {Rieke}, {Rigopoulou}, {Salim},
  {Schiminovich}, {Simard}, {Smail}, {Small}, {Weiner}, {Willmer}, {Willner},
  {Wilson}, {Wright}, \& {Yan}}]{AEGIS2007}
{Davis}, M., {Guhathakurta}, P., {Konidaris}, N.~P., {et~al.} 2007,
  \href{https://doi.org/10.1086/517931}{\apjl}, 660, L1,
  [\href{https://arxiv.org/abs/astro-ph/0607355}{astro-ph/0607355}].

\bibitem[{{de Amorim} {et~al.}(2017){de Amorim}, {Garc{\'\i}a-Benito}, {Cid
  Fernandes}, {Cortijo-Ferrero}, {Gonz{\'a}lez Delgado}, {Lacerda}, {L{\'o}pez
  Fern{\'a}ndez}, {P{\'e}rez}, \& {Vale Asari}}]{Pycasso2017}
{de Amorim}, A.~L., {Garc{\'\i}a-Benito}, R., {Cid Fernandes}, R., {et~al.}
  2017, \href{https://doi.org/10.1093/mnras/stx1805}{\mnras}, 471, 3727,
  [\href{https://arxiv.org/abs/1707.05293}{1707.05293}].

\bibitem[{{de Jong}(1996)}]{deJong1996}
{de Jong}, R.~S. 1996,
  \href{https://doi.org/10.48550/arXiv.astro-ph/9604010}{\aap}, 313, 377,
  [\href{https://arxiv.org/abs/astro-ph/9604010}{astro-ph/9604010}].

\bibitem[{{De Lucia} {et~al.}(2006){De Lucia}, {Springel}, {White}, {Croton},
  \& {Kauffmann}}]{deLucia2006}
{De Lucia}, G., {Springel}, V., {White}, S. D.~M., {Croton}, D., \&
  {Kauffmann}, G. 2006,
  \href{https://doi.org/10.1111/j.1365-2966.2005.09879.x}{\mnras}, 366, 499,
  [\href{https://arxiv.org/abs/astro-ph/0509725}{astro-ph/0509725}].

\bibitem[{{De Lucia} {et~al.}(2012){De Lucia}, {Weinmann}, {Poggianti},
  {Arag{\'o}n-Salamanca}, \& {Zaritsky}}]{deLucia2012}
{De Lucia}, G., {Weinmann}, S., {Poggianti}, B.~M., {Arag{\'o}n-Salamanca}, A.,
  \& {Zaritsky}, D. 2012,
  \href{https://doi.org/10.1111/j.1365-2966.2012.20983.x}{\mnras}, 423, 1277,
  [\href{https://arxiv.org/abs/1111.6590}{1111.6590}].

\bibitem[{{De Rijcke} {et~al.}(2010){De Rijcke}, {Van Hese}, \&
  {Buyle}}]{Rijcke2010}
{De Rijcke}, S., {Van Hese}, E., \& {Buyle}, P. 2010,
  \href{https://doi.org/10.1088/2041-8205/724/2/L171}{\apjl}, 724, L171.

\bibitem[{{de Vaucouleurs}(1948)}]{deVaucouleurs1948}
{de Vaucouleurs}, G. 1948, Annales d'Astrophysique, 11, 247.

\bibitem[{{de Zeeuw} {et~al.}(2002){de Zeeuw}, {Bureau}, {Emsellem}, {Bacon},
  {Carollo}, {Copin}, {Davies}, {Kuntschner}, {Miller}, {Monnet}, {Peletier},
  \& {Verolme}}]{Sauron2002}
{de Zeeuw}, P.~T., {Bureau}, M., {Emsellem}, E., {et~al.} 2002,
  \href{https://doi.org/10.1046/j.1365-8711.2002.05059.x}{\mnras}, 329, 513,
  [\href{https://arxiv.org/abs/astro-ph/0109511}{astro-ph/0109511}].

\bibitem[{{Dekel} \& {Birnboim}(2006)}]{dekel2006}
{Dekel}, A. \& {Birnboim}, Y. 2006,
  \href{https://doi.org/10.1111/j.1365-2966.2006.10145.x}{\mnras}, 368, 2,
  [\href{https://arxiv.org/abs/astro-ph/0412300}{astro-ph/0412300}].

\bibitem[{{Dekel} \& {Silk}(1986)}]{Dekel1986}
{Dekel}, A. \& {Silk}, J. 1986, \href{https://doi.org/10.1086/164050}{\apj},
  303, 39.

\bibitem[{{Desai} {et~al.}(2012){Desai}, {Armstrong}, {Mohr}, {Semler}, {Liu},
  {Bertin}, {Allam}, {Barkhouse}, {Bazin}, {Buckley-Geer}, {Cooper}, {Hansen},
  {High}, {Lin}, {Lin}, {Ngeow}, {Rest}, {Song}, {Tucker}, \&
  {Zenteno}}]{Desai2012}
{Desai}, S., {Armstrong}, R., {Mohr}, J.~J., {et~al.} 2012,
  \href{https://doi.org/10.1088/0004-637X/757/1/83}{\apj}, 757, 83,
  [\href{https://arxiv.org/abs/1204.1210}{1204.1210}].

\bibitem[{{Desai} {et~al.}(2016){Desai}, {Mohr}, {Bertin}, {K{\"u}mmel}, \&
  {Wetzstein}}]{Desai2016}
{Desai}, S., {Mohr}, J.~J., {Bertin}, E., {K{\"u}mmel}, M., \& {Wetzstein}, M.
  2016, \href{https://doi.org/10.1016/j.ascom.2016.04.002}{Astronomy and
  Computing}, 16, 67, [\href{https://arxiv.org/abs/1601.07182}{1601.07182}].

\bibitem[{{Di Matteo} {et~al.}(2005){Di Matteo}, {Springel}, \&
  {Hernquist}}]{dimatteo2005}
{Di Matteo}, T., {Springel}, V., \& {Hernquist}, L. 2005,
  \href{https://doi.org/10.1038/nature03335}{\nat}, 433, 604,
  [\href{https://arxiv.org/abs/astro-ph/0502199}{astro-ph/0502199}].

\bibitem[{{Diaferio} {et~al.}(2001){Diaferio}, {Kauffmann}, {Balogh}, {White},
  {Schade}, \& {Ellingson}}]{diaferio2001}
{Diaferio}, A., {Kauffmann}, G., {Balogh}, M.~L., {et~al.} 2001,
  \href{https://doi.org/10.1046/j.1365-8711.2001.04303.x}{\mnras}, 323, 999,
  [\href{https://arxiv.org/abs/astro-ph/0005485}{astro-ph/0005485}].

\bibitem[{{D{\'\i}az-Garc{\'\i}a}
  {et~al.}(2019{\natexlab{a}}){D{\'\i}az-Garc{\'\i}a}, {Cenarro},
  {L{\'o}pez-Sanjuan}, {Ferreras}, {Cervi{\~n}o}, {Fern{\'a}ndez-Soto},
  {Gonz{\'a}lez Delgado}, {M{\'a}rquez}, {Povi{\'c}}, {San Roman}, {Viironen},
  {Moles}, {Crist{\'o}bal-Hornillos}, {L{\'o}pez-Comazzi}, {Alfaro},
  {Aparicio-Villegas}, {Ben{\'\i}tez}, {Broadhurst}, {Cabrera-Ca{\~n}o},
  {Castander}, {Cepa}, {Husillos}, {Infante}, {Aguerri}, {Mart{\'\i}nez},
  {Masegosa}, {Molino}, {del Olmo}, {Perea}, {Prada}, \& {Quintana}}]{Luis2019}
{D{\'\i}az-Garc{\'\i}a}, L.~A., {Cenarro}, A.~J., {L{\'o}pez-Sanjuan}, C.,
  {et~al.} 2019{\natexlab{a}},
  \href{https://doi.org/10.1051/0004-6361/201832788}{\aap}, 631, A156,
  [\href{https://arxiv.org/abs/1711.10590}{1711.10590}].

\bibitem[{{D{\'\i}az-Garc{\'\i}a}
  {et~al.}(2019{\natexlab{b}}){D{\'\i}az-Garc{\'\i}a}, {Cenarro},
  {L{\'o}pez-Sanjuan}, {Ferreras}, {Fern{\'a}ndez-Soto}, {Gonz{\'a}lez
  Delgado}, {M{\'a}rquez}, {Masegosa}, {San Roman}, {Viironen}, {Bonoli},
  {Cervi{\~n}o}, {Moles}, {Crist{\'o}bal-Hornillos}, {Alfaro},
  {Aparicio-Villegas}, {Ben{\'\i}tez}, {Broadhurst}, {Cabrera-Ca{\~n}o},
  {Castander}, {Cepa}, {Husillos}, {Infante}, {Aguerri}, {Mart{\'\i}nez},
  {Molino}, {del Olmo}, {Perea}, {Prada}, \& {Quintana}}]{Luis2019b}
{D{\'\i}az-Garc{\'\i}a}, L.~A., {Cenarro}, A.~J., {L{\'o}pez-Sanjuan}, C.,
  {et~al.} 2019{\natexlab{b}},
  \href{https://doi.org/10.1051/0004-6361/201832882}{\aap}, 631, A157,
  [\href{https://arxiv.org/abs/1802.06813}{1802.06813}].

\bibitem[{{D{\'\i}az-Garc{\'\i}a} {et~al.}(2015){D{\'\i}az-Garc{\'\i}a},
  {Cenarro}, {L{\'o}pez-Sanjuan}, {Ferreras}, {Varela}, {Viironen},
  {Crist{\'o}bal-Hornillos}, {Moles}, {Mar{\'\i}n-Franch}, {Arnalte-Mur},
  {Ascaso}, {Cervi{\~n}o}, {Gonz{\'a}lez Delgado}, {M{\'a}rquez}, {Masegosa},
  {Molino}, {Povi{\'c}}, {Alfaro}, {Aparicio-Villegas}, {Ben{\'\i}tez},
  {Broadhurst}, {Cabrera-Ca{\~n}o}, {Castander}, {Cepa}, {Fern{\'a}ndez-Soto},
  {Husillos}, {Infante}, {Aguerri}, {Mart{\'\i}nez}, {del Olmo}, {Perea},
  {Prada}, {Quintana}, \& {Gruel}}]{Luis2015}
{D{\'\i}az-Garc{\'\i}a}, L.~A., {Cenarro}, A.~J., {L{\'o}pez-Sanjuan}, C.,
  {et~al.} 2015, \href{https://doi.org/10.1051/0004-6361/201425582}{\aap}, 582,
  A14, [\href{https://arxiv.org/abs/1505.07555}{1505.07555}].

\bibitem[{{D{\'\i}az-Garc{\'\i}a}
  {et~al.}(2019{\natexlab{c}}){D{\'\i}az-Garc{\'\i}a}, {Cenarro},
  {L{\'o}pez-Sanjuan}, {Peralta de Arriba}, {Ferreras}, {Cervi{\~n}o},
  {M{\'a}rquez}, {Masegosa}, {del Olmo}, \& {Perea}}]{Luis2019c}
{D{\'\i}az-Garc{\'\i}a}, L.~A., {Cenarro}, A.~J., {L{\'o}pez-Sanjuan}, C.,
  {et~al.} 2019{\natexlab{c}},
  \href{https://doi.org/10.1051/0004-6361/201935257}{\aap}, 631, A158,
  [\href{https://arxiv.org/abs/1901.05983}{1901.05983}].

\bibitem[{{Dimauro} {et~al.}(2022){Dimauro}, {Daddi}, {Shankar}, {Cattaneo},
  {Huertas-Company}, {Bernardi}, {Caro}, {Dupke}, {H{\"a}u{\ss}ler},
  {Johnston}, {Cortesi}, {Mei}, \& {Peletier}}]{Dimauro2022}
{Dimauro}, P., {Daddi}, E., {Shankar}, F., {et~al.} 2022,
  \href{https://doi.org/10.1093/mnras/stac884}{\mnras}, 513, 256,
  [\href{https://arxiv.org/abs/2203.15819}{2203.15819}].

\bibitem[{{Donnari} {et~al.}(2021{\natexlab{a}}){Donnari}, {Pillepich},
  {Joshi}, {Nelson}, {Genel}, {Marinacci}, {Rodriguez-Gomez}, {Pakmor},
  {Torrey}, {Vogelsberger}, \& {Hernquist}}]{Donnari2021}
{Donnari}, M., {Pillepich}, A., {Joshi}, G.~D., {et~al.} 2021{\natexlab{a}},
  \href{https://doi.org/10.1093/mnras/staa3006}{\mnras}, 500, 4004,
  [\href{https://arxiv.org/abs/2008.00005}{2008.00005}].

\bibitem[{{Donnari} {et~al.}(2021{\natexlab{b}}){Donnari}, {Pillepich},
  {Nelson}, {Marinacci}, {Vogelsberger}, \& {Hernquist}}]{Donnaris2021compare}
{Donnari}, M., {Pillepich}, A., {Nelson}, D., {et~al.} 2021{\natexlab{b}},
  \href{https://doi.org/10.1093/mnras/stab1950}{\mnras}, 506, 4760,
  [\href{https://arxiv.org/abs/2008.00004}{2008.00004}].

\bibitem[{{Dooley} {et~al.}(2016){Dooley}, {Peter}, {Vogelsberger}, {Zavala},
  \& {Frebel}}]{Dooley2016}
{Dooley}, G.~A., {Peter}, A. H.~G., {Vogelsberger}, M., {Zavala}, J., \&
  {Frebel}, A. 2016, \href{https://doi.org/10.1093/mnras/stw1309}{\mnras}, 461,
  710, [\href{https://arxiv.org/abs/1603.08919}{1603.08919}].

\bibitem[{{Doubrawa} {et~al.}(2023){Doubrawa}, {Cypriano}, {Finoguenov},
  {Lopes}, {Maturi}, {Gonzalez}, \& {Dupke}}]{Doubrawa2023}
{Doubrawa}, L., {Cypriano}, E.~S., {Finoguenov}, A., {et~al.} 2023,
  \href{https://doi.org/10.1093/mnras/stad3024}{\mnras}, 526, 4285,
  [\href{https://arxiv.org/abs/2312.10564}{2312.10564}].

\bibitem[{{Dressler}(1980)}]{Dressler1980}
{Dressler}, A. 1980, \href{https://doi.org/10.1086/157753}{\apj}, 236, 351.

\bibitem[{{Dressler} {et~al.}(2016){Dressler}, {Kelson}, {Abramson},
  {Gladders}, {Oemler}, {Poggianti}, {Mulchaey}, {Vulcani}, {Shectman},
  {Williams}, \& {McCarthy}}]{dressler2016}
{Dressler}, A., {Kelson}, D.~D., {Abramson}, L.~E., {et~al.} 2016,
  \href{https://doi.org/10.3847/1538-4357/833/2/251}{\apj}, 833, 251,
  [\href{https://arxiv.org/abs/1607.02143}{1607.02143}].

\bibitem[{{Duarte Puertas} {et~al.}(2017){Duarte Puertas}, {Vilchez},
  {Iglesias-P{\'a}ramo}, {Kehrig}, {P{\'e}rez-Montero}, \&
  {Rosales-Ortega}}]{puertas2017aperture}
{Duarte Puertas}, S., {Vilchez}, J.~M., {Iglesias-P{\'a}ramo}, J., {et~al.}
  2017, \href{https://doi.org/10.1051/0004-6361/201629044}{\aap}, 599, A71,
  [\href{https://arxiv.org/abs/1611.07935}{1611.07935}].

\bibitem[{{Duarte Puertas} {et~al.}(2022){Duarte Puertas}, {Vilchez},
  {Iglesias-P{\'a}ramo}, {Moll{\'a}}, {P{\'e}rez-Montero}, {Kehrig},
  {Pilyugin}, \& {Zinchenko}}]{DuartePuertas2022}
{Duarte Puertas}, S., {Vilchez}, J.~M., {Iglesias-P{\'a}ramo}, J., {et~al.}
  2022, \href{https://doi.org/10.1051/0004-6361/202141571}{\aap}, 666, A186,
  [\href{https://arxiv.org/abs/2205.01203}{2205.01203}].

\bibitem[{{Ebeling} {et~al.}(2014){Ebeling}, {Stephenson}, \&
  {Edge}}]{Ebeling2014}
{Ebeling}, H., {Stephenson}, L.~N., \& {Edge}, A.~C. 2014,
  \href{https://doi.org/10.1088/2041-8205/781/2/L40}{\apjl}, 781, L40,
  [\href{https://arxiv.org/abs/1312.6135}{1312.6135}].

\bibitem[{{Efstathiou}(2000)}]{Efstathiou2000}
{Efstathiou}, G. 2000,
  \href{https://doi.org/10.1046/j.1365-8711.2000.03665.x}{\mnras}, 317, 697,
  [\href{https://arxiv.org/abs/astro-ph/0002245}{astro-ph/0002245}].

\bibitem[{{Elbaz} {et~al.}(2007){Elbaz}, {Daddi}, {Le Borgne}, {Dickinson},
  {Alexander}, {Chary}, {Starck}, {Brandt}, {Kitzbichler}, {MacDonald},
  {Nonino}, {Popesso}, {Stern}, \& {Vanzella}}]{Elbaz2007}
{Elbaz}, D., {Daddi}, E., {Le Borgne}, D., {et~al.} 2007,
  \href{https://doi.org/10.1051/0004-6361:20077525}{\aap}, 468, 33,
  [\href{https://arxiv.org/abs/astro-ph/0703653}{astro-ph/0703653}].

\bibitem[{{Ellingson} {et~al.}(2001){Ellingson}, {Lin}, {Yee}, \&
  {Carlberg}}]{Ellingson2001}
{Ellingson}, E., {Lin}, H., {Yee}, H.~K.~C., \& {Carlberg}, R.~G. 2001,
  \href{https://doi.org/10.1086/318423}{\apj}, 547, 609,
  [\href{https://arxiv.org/abs/astro-ph/0010141}{astro-ph/0010141}].

\bibitem[{{Ellison} {et~al.}(2013){Ellison}, {Mendel}, {Patton}, \&
  {Scudder}}]{Ellison2013}
{Ellison}, S.~L., {Mendel}, J.~T., {Patton}, D.~R., \& {Scudder}, J.~M. 2013,
  \href{https://doi.org/10.1093/mnras/stt1562}{\mnras}, 435, 3627,
  [\href{https://arxiv.org/abs/1308.3707}{1308.3707}].

\bibitem[{{Ellison} {et~al.}(2018){Ellison}, {S{\'a}nchez}, {Ibarra-Medel},
  {Antonio}, {Mendel}, \& {Barrera-Ballesteros}}]{Ellison2018}
{Ellison}, S.~L., {S{\'a}nchez}, S.~F., {Ibarra-Medel}, H., {et~al.} 2018,
  \href{https://doi.org/10.1093/mnras/stx2882}{\mnras}, 474, 2039,
  [\href{https://arxiv.org/abs/1711.00915}{1711.00915}].

\bibitem[{{Emsellem} {et~al.}(2022){Emsellem}, {Schinnerer}, {Santoro},
  {Belfiore}, {Pessa}, {McElroy}, {Blanc}, {Congiu}, {Groves}, {Ho}, {Kreckel},
  {Razza}, {Sanchez-Blazquez}, {Egorov}, {Faesi}, {Klessen}, {Leroy}, {Meidt},
  {Querejeta}, {Rosolowsky}, {Scheuermann}, {Anand}, {Barnes},
  {Be{\v{s}}li{\'c}}, {Bigiel}, {Boquien}, {Cao}, {Chevance}, {Dale},
  {Eibensteiner}, {Glover}, {Grasha}, {Henshaw}, {Hughes}, {Koch}, {Kruijssen},
  {Lee}, {Liu}, {Pan}, {Pety}, {Saito}, {Sandstrom}, {Schruba}, {Sun},
  {Thilker}, {Usero}, {Watkins}, \& {Williams}}]{PhangsMuse2022}
{Emsellem}, E., {Schinnerer}, E., {Santoro}, F., {et~al.} 2022,
  \href{https://doi.org/10.1051/0004-6361/202141727}{\aap}, 659, A191,
  [\href{https://arxiv.org/abs/2110.03708}{2110.03708}].

\bibitem[{{Enia} {et~al.}(2020){Enia}, {Rodighiero}, {Morselli}, {Casasola},
  {Bianchi}, {Rodriguez-Mu{\~n}oz}, {Mancini}, {Renzini}, {Popesso}, {Cassata},
  {Negrello}, \& {Franceschini}}]{Enia2020}
{Enia}, A., {Rodighiero}, G., {Morselli}, L., {et~al.} 2020,
  \href{https://doi.org/10.1093/mnras/staa433}{\mnras}, 493, 4107,
  [\href{https://arxiv.org/abs/2001.04479}{2001.04479}].

\bibitem[{{Epinat} {et~al.}(2024){Epinat}, {Contini}, {Mercier}, {Ciesla},
  {Lemaux}, {Johnson}, {Richard}, {Brinchmann}, {Boogaard}, {Carton},
  {Michel-Dansac}, {Bacon}, {Krajnovi{\'c}}, {Finley}, {Schroetter}, {Ventou},
  {Abril-Melgarejo}, {Boselli}, {Bouch{\'e}}, {Kollatschny}, {Kova{\v{c}}},
  {Paalvast}, {Soucail}, {Urrutia}, \& {Weilbacher}}]{Epinat2024}
{Epinat}, B., {Contini}, T., {Mercier}, W., {et~al.} 2024,
  \href{https://doi.org/10.1051/0004-6361/202348038}{\aap}, 683, A205,
  [\href{https://arxiv.org/abs/2312.00924}{2312.00924}].

\bibitem[{{Er} {et~al.}(2018){Er}, {Hoekstra}, {Schrabback}, {Cardone},
  {Scaramella}, {Maoli}, {Vicinanza}, {Gillis}, \& {Rhodes}}]{Er2018}
{Er}, X., {Hoekstra}, H., {Schrabback}, T., {et~al.} 2018,
  \href{https://doi.org/10.1093/mnras/sty685}{\mnras}, 476, 5645,
  [\href{https://arxiv.org/abs/1708.06085}{1708.06085}].

\bibitem[{{Esposito} {et~al.}(2022){Esposito}, {Vallini}, {Pozzi}, {Casasola},
  {Mingozzi}, {Vignali}, {Gruppioni}, \& {Salvestrini}}]{Esposito2022}
{Esposito}, F., {Vallini}, L., {Pozzi}, F., {et~al.} 2022,
  \href{https://doi.org/10.1093/mnras/stac313}{\mnras}, 512, 686,
  [\href{https://arxiv.org/abs/2202.00697}{2202.00697}].

\bibitem[{{Faber} {et~al.}(2007){Faber}, {Willmer}, {Wolf}, {Koo}, {Weiner},
  {Newman}, {Im}, {Coil}, {Conroy}, {Cooper}, {Davis}, {Finkbeiner}, {Gerke},
  {Gebhardt}, {Groth}, {Guhathakurta}, {Harker}, {Kaiser}, {Kassin},
  {Kleinheinrich}, {Konidaris}, {Kron}, {Lin}, {Luppino}, {Madgwick},
  {Meisenheimer}, {Noeske}, {Phillips}, {Sarajedini}, {Schiavon}, {Simard},
  {Szalay}, {Vogt}, \& {Yan}}]{Faber2007}
{Faber}, S.~M., {Willmer}, C.~N.~A., {Wolf}, C., {et~al.} 2007,
  \href{https://doi.org/10.1086/519294}{\apj}, 665, 265,
  [\href{https://arxiv.org/abs/astro-ph/0506044}{astro-ph/0506044}].

\bibitem[{{Fabian}(2012)}]{Fabian2012}
{Fabian}, A.~C. 2012,
  \href{https://doi.org/10.1146/annurev-astro-081811-125521}{\araa}, 50, 455,
  [\href{https://arxiv.org/abs/1204.4114}{1204.4114}].

\bibitem[{{Fang} {et~al.}(2018){Fang}, {Faber}, {Koo}, {Rodr{\'\i}guez-Puebla},
  {Guo}, {Barro}, {Behroozi}, {Brammer}, {Chen}, {Dekel}, {Ferguson},
  {Gawiser}, {Giavalisco}, {Kartaltepe}, {Kocevski}, {Koekemoer}, {McGrath},
  {McIntosh}, {Newman}, {Pacifici}, {Pandya}, {P{\'e}rez-Gonz{\'a}lez},
  {Primack}, {Salmon}, {Trump}, {Weiner}, {Willner}, {Acquaviva}, {Dahlen},
  {Finkelstein}, {Finlator}, {Fontana}, {Galametz}, {Grogin}, {Gruetzbauch},
  {Johnson}, {Mobasher}, {Papovich}, {Pforr}, {Salvato}, {Santini}, {van der
  Wel}, {Wiklind}, \& {Wuyts}}]{Fang2018}
{Fang}, J.~J., {Faber}, S.~M., {Koo}, D.~C., {et~al.} 2018,
  \href{https://doi.org/10.3847/1538-4357/aabcba}{\apj}, 858, 100,
  [\href{https://arxiv.org/abs/1710.05489}{1710.05489}].

\bibitem[{{Fang} {et~al.}(2012){Fang}, {Faber}, {Salim}, {Graves}, \&
  {Rich}}]{Fang2012}
{Fang}, J.~J., {Faber}, S.~M., {Salim}, S., {Graves}, G.~J., \& {Rich}, R.~M.
  2012, \href{https://doi.org/10.1088/0004-637X/761/1/23}{\apj}, 761, 23,
  [\href{https://arxiv.org/abs/1210.6062}{1210.6062}].

\bibitem[{{Farrens} {et~al.}(2017){Farrens}, {Ngol{\`e} Mboula}, \&
  {Starck}}]{Farrens2017}
{Farrens}, S., {Ngol{\`e} Mboula}, F.~M., \& {Starck}, J.~L. 2017,
  \href{https://doi.org/10.1051/0004-6361/201629709}{\aap}, 601, A66,
  [\href{https://arxiv.org/abs/1703.02305}{1703.02305}].

\bibitem[{{Fogarty} {et~al.}(2014){Fogarty}, {Scott}, {Owers}, {Brough},
  {Croom}, {Pracy}, {Houghton}, {Bland-Hawthorn}, {Colless}, {Davies}, {Jones},
  {Allen}, {Bryant}, {Goodwin}, {Green}, {Konstantopoulos}, {Lawrence},
  {Richards}, {Cortese}, \& {Sharp}}]{fogarty2014}
{Fogarty}, L.~M.~R., {Scott}, N., {Owers}, M.~S., {et~al.} 2014,
  \href{https://doi.org/10.1093/mnras/stu1165}{\mnras}, 443, 485,
  [\href{https://arxiv.org/abs/1406.3899}{1406.3899}].

\bibitem[{{Folkes} {et~al.}(1999){Folkes}, {Ronen}, {Price}, {Lahav},
  {Colless}, {Maddox}, {Deeley}, {Glazebrook}, {Bland-Hawthorn}, {Cannon},
  {Cole}, {Collins}, {Couch}, {Driver}, {Dalton}, {Efstathiou}, {Ellis},
  {Frenk}, {Kaiser}, {Lewis}, {Lumsden}, {Peacock}, {Peterson}, {Sutherland},
  \& {Taylor}}]{Folkes1999}
{Folkes}, S., {Ronen}, S., {Price}, I., {et~al.} 1999,
  \href{https://doi.org/10.1046/j.1365-8711.1999.02721.x}{\mnras}, 308, 459,
  [\href{https://arxiv.org/abs/astro-ph/9903456}{astro-ph/9903456}].

\bibitem[{{Fossati} {et~al.}(2017){Fossati}, {Wilman}, {Mendel}, {Saglia},
  {Galametz}, {Beifiori}, {Bender}, {Chan}, {Fabricius}, {Bandara}, {Brammer},
  {Davies}, {F{\"o}rster Schreiber}, {Genzel}, {Hartley}, {Kulkarni}, {Lang},
  {Momcheva}, {Nelson}, {Skelton}, {Tacconi}, {Tadaki}, {{\"U}bler}, {van
  Dokkum}, {Wisnioski}, {Whitaker}, {Wuyts}, \& {Wuyts}}]{Fossati2017}
{Fossati}, M., {Wilman}, D.~J., {Mendel}, J.~T., {et~al.} 2017,
  \href{https://doi.org/10.3847/1538-4357/835/2/153}{\apj}, 835, 153,
  [\href{https://arxiv.org/abs/1611.07524}{1611.07524}].

\bibitem[{{Foster} {et~al.}(2012){Foster}, {Hopkins}, {Gunawardhana},
  {Lara-L{\'o}pez}, {Sharp}, {Steele}, {Taylor}, {Driver}, {Baldry}, {Bamford},
  {Liske}, {Loveday}, {Norberg}, {Peacock}, {Alpaslan}, {Bauer},
  {Bland-Hawthorn}, {Brough}, {Cameron}, {Colless}, {Conselice}, {Croom},
  {Frenk}, {Hill}, {Jones}, {Kelvin}, {Kuijken}, {Nichol}, {Owers},
  {Parkinson}, {Pimbblet}, {Popescu}, {Prescott}, {Robotham}, {Lopez-Sanchez},
  {Sutherland}, {Thomas}, {Tuffs}, {van Kampen}, \& {Wijesinghe}}]{Foster2012}
{Foster}, C., {Hopkins}, A.~M., {Gunawardhana}, M., {et~al.} 2012,
  \href{https://doi.org/10.1051/0004-6361/201220050}{\aap}, 547, A79,
  [\href{https://arxiv.org/abs/1209.1636}{1209.1636}].

\bibitem[{{Fritz} {et~al.}(2014){Fritz}, {Scodeggio}, {Ilbert}, {Bolzonella},
  {Davidzon}, {Coupon}, {Garilli}, {Guzzo}, {Zamorani}, {Abbas}, {Adami},
  {Arnouts}, {Bel}, {Bottini}, {Branchini}, {Cappi}, {Cucciati}, {De Lucia},
  {de la Torre}, {Franzetti}, {Fumana}, {Granett}, {Iovino}, {Krywult}, {Le
  Brun}, {Le F{\`e}vre}, {Maccagni}, {Ma{\l}ek}, {Marulli}, {McCracken},
  {Paioro}, {Polletta}, {Pollo}, {Schlagenhaufer}, {Tasca}, {Tojeiro},
  {Vergani}, {Zanichelli}, {Burden}, {Di Porto}, {Marchetti}, {Marinoni},
  {Mellier}, {Moscardini}, {Nichol}, {Peacock}, {Percival}, {Phleps}, \&
  {Wolk}}]{Fritz2014}
{Fritz}, A., {Scodeggio}, M., {Ilbert}, O., {et~al.} 2014,
  \href{https://doi.org/10.1051/0004-6361/201322379}{\aap}, 563, A92,
  [\href{https://arxiv.org/abs/1401.6137}{1401.6137}].

\bibitem[{{Fujita}(2004)}]{Fujita2004}
{Fujita}, Y. 2004, \href{https://doi.org/10.1093/pasj/56.1.29}{\pasj}, 56, 29,
  [\href{https://arxiv.org/abs/astro-ph/0311193}{astro-ph/0311193}].

\bibitem[{{Fujita} \& {Nagashima}(1999)}]{Fujita1999}
{Fujita}, Y. \& {Nagashima}, M. 1999,
  \href{https://doi.org/10.1086/307139}{\apj}, 516, 619,
  [\href{https://arxiv.org/abs/astro-ph/9812378}{astro-ph/9812378}].

\bibitem[{{Fumagalli} {et~al.}(2012){Fumagalli}, {Patel}, {Franx}, {Brammer},
  {van Dokkum}, {da Cunha}, {Kriek}, {Lundgren}, {Momcheva}, {Rix}, {Schmidt},
  {Skelton}, {Whitaker}, {Labbe}, \& {Nelson}}]{Fumagalli2012}
{Fumagalli}, M., {Patel}, S.~G., {Franx}, M., {et~al.} 2012,
  \href{https://doi.org/10.1088/2041-8205/757/2/L22}{\apjl}, 757, L22,
  [\href{https://arxiv.org/abs/1206.2645}{1206.2645}].

\bibitem[{{Gaia Collaboration} {et~al.}(2018){Gaia Collaboration}, {Brown},
  {Vallenari}, {Prusti}, {de Bruijne}, {Babusiaux}, {Bailer-Jones}, {Biermann},
  {Evans}, {Eyer}, {Jansen}, {Jordi}, {Klioner}, {Lammers}, {Lindegren},
  {Luri}, {Mignard}, {Panem}, {Pourbaix}, {Randich}, {Sartoretti}, {Siddiqui},
  {Soubiran}, {van Leeuwen}, {Walton}, {Arenou}, {Bastian}, {Cropper},
  {Drimmel}, {Katz}, {Lattanzi}, {Bakker}, {Cacciari}, {Casta{\~n}eda},
  {Chaoul}, {Cheek}, {De Angeli}, {Fabricius}, {Guerra}, {Holl}, {Masana},
  {Messineo}, {Mowlavi}, {Nienartowicz}, {Panuzzo}, {Portell}, {Riello},
  {Seabroke}, {Tanga}, {Th{\'e}venin}, {Gracia-Abril}, {Comoretto},
  {Garcia-Reinaldos}, {Teyssier}, {Altmann}, {Andrae}, {Audard},
  {Bellas-Velidis}, {Benson}, {Berthier}, {Blomme}, {Burgess}, {Busso},
  {Carry}, {Cellino}, {Clementini}, {Clotet}, {Creevey}, {Davidson}, {De
  Ridder}, {Delchambre}, {Dell'Oro}, {Ducourant},
  {Fern{\'a}ndez-Hern{\'a}ndez}, {Fouesneau}, {Fr{\'e}mat}, {Galluccio},
  {Garc{\'\i}a-Torres}, {Gonz{\'a}lez-N{\'u}{\~n}ez}, {Gonz{\'a}lez-Vidal},
  {Gosset}, {Guy}, {Halbwachs}, {Hambly}, {Harrison}, {Hern{\'a}ndez},
  {Hestroffer}, {Hodgkin}, {Hutton}, {Jasniewicz}, {Jean-Antoine-Piccolo},
  {Jordan}, {Korn}, {Krone-Martins}, {Lanzafame}, {Lebzelter}, {L{\"o}ffler},
  {Manteiga}, {Marrese}, {Mart{\'\i}n-Fleitas}, {Moitinho}, {Mora}, {Muinonen},
  {Osinde}, {Pancino}, {Pauwels}, {Petit}, {Recio-Blanco}, {Richards},
  {Rimoldini}, {Robin}, {Sarro}, {Siopis}, {Smith}, {Sozzetti}, {S{\"u}veges},
  {Torra}, {van Reeven}, {Abbas}, {Abreu Aramburu}, {Accart}, {Aerts},
  {Altavilla}, {{\'A}lvarez}, {Alvarez}, {Alves}, {Anderson}, {Andrei},
  {Anglada Varela}, {Antiche}, {Antoja}, {Arcay}, {Astraatmadja}, {Bach},
  {Baker}, {Balaguer-N{\'u}{\~n}ez}, {Balm}, {Barache}, {Barata}, {Barbato},
  {Barblan}, {Barklem}, {Barrado}, {Barros}, {Barstow}, {Bartholom{\'e}
  Mu{\~n}oz}, {Bassilana}, {Becciani}, {Bellazzini}, {Berihuete}, {Bertone},
  {Bianchi}, {Bienaym{\'e}}, {Blanco-Cuaresma}, {Boch}, {Boeche}, {Bombrun},
  {Borrachero}, {Bossini}, {Bouquillon}, {Bourda}, {Bragaglia}, {Bramante},
  {Breddels}, {Bressan}, {Brouillet}, {Br{\"u}semeister}, {Brugaletta},
  {Bucciarelli}, {Burlacu}, {Busonero}, {Butkevich}, {Buzzi}, {Caffau},
  {Cancelliere}, {Cannizzaro}, {Cantat-Gaudin}, {Carballo}, {Carlucci},
  {Carrasco}, {Casamiquela}, {Castellani}, {Castro-Ginard}, {Charlot},
  {Chemin}, {Chiavassa}, {Cocozza}, {Costigan}, {Cowell}, {Crifo}, {Crosta},
  {Crowley}, {Cuypers}, {Dafonte}, {Damerdji}, {Dapergolas}, {David}, {David},
  {de Laverny}, {De Luise}, {De March}, {de Martino}, {de Souza}, {de Torres},
  {Debosscher}, {del Pozo}, {Delbo}, {Delgado}, {Delgado}, {Di Matteo},
  {Diakite}, {Diener}, {Distefano}, {Dolding}, {Drazinos}, {Dur{\'a}n},
  {Edvardsson}, {Enke}, {Eriksson}, {Esquej}, {Eynard Bontemps}, {Fabre},
  {Fabrizio}, {Faigler}, {Falc{\~a}o}, {Farr{\`a}s Casas}, {Federici},
  {Fedorets}, {Fernique}, {Figueras}, {Filippi}, {Findeisen}, {Fonti},
  {Fraile}, {Fraser}, {Fr{\'e}zouls}, {Gai}, {Galleti}, {Garabato},
  {Garc{\'\i}a-Sedano}, {Garofalo}, {Garralda}, {Gavel}, {Gavras}, {Gerssen},
  {Geyer}, {Giacobbe}, {Gilmore}, {Girona}, {Giuffrida}, {Glass}, {Gomes},
  {Granvik}, {Gueguen}, {Guerrier}, {Guiraud}, {Guti{\'e}rrez-S{\'a}nchez},
  {Haigron}, {Hatzidimitriou}, {Hauser}, {Haywood}, {Heiter}, {Helmi}, {Heu},
  {Hilger}, {Hobbs}, {Hofmann}, {Holland}, {Huckle}, {Hypki}, {Icardi},
  {Jan{\ss}en}, {Jevardat de Fombelle}, {Jonker}, {Juh{\'a}sz}, {Julbe},
  {Karampelas}, {Kewley}, {Klar}, {Kochoska}, {Kohley}, {Kolenberg},
  {Kontizas}, {Kontizas}, {Koposov}, {Kordopatis}, {Kostrzewa-Rutkowska},
  {Koubsky}, {Lambert}, {Lanza}, {Lasne}, {Lavigne}, {Le Fustec}, {Le
  Poncin-Lafitte}, {Lebreton}, {Leccia}, {Leclerc}, {Lecoeur-Taibi},
  {Lenhardt}, {Leroux}, {Liao}, {Licata}, {Lindstr{\o}m}, {Lister}, {Livanou},
  {Lobel}, {L{\'o}pez}, {Managau}, {Mann}, {Mantelet}, {Marchal}, {Marchant},
  {Marconi}, {Marinoni}, {Marschalk{\'o}}, {Marshall}, {Martino}, {Marton},
  {Mary}, {Massari}, {Matijevi{\v{c}}}, {Mazeh}, {McMillan}, {Messina},
  {Michalik}, {Millar}, {Molina}, {Molinaro}, {Moln{\'a}r}, {Montegriffo},
  {Mor}, {Morbidelli}, {Morel}, {Morris}, {Mulone}, {Muraveva}, {Musella},
  {Nelemans}, {Nicastro}, {Noval}, {O'Mullane}, {Ord{\'e}novic},
  {Ord{\'o}{\~n}ez-Blanco}, {Osborne}, {Pagani}, {Pagano}, {Pailler},
  {Palacin}, {Palaversa}, {Panahi}, {Pawlak}, {Piersimoni}, {Pineau}, {Plachy},
  {Plum}, {Poggio}, {Poujoulet}, {Pr{\v{s}}a}, {Pulone}, {Racero}, {Ragaini},
  {Rambaux}, {Ramos-Lerate}, {Regibo}, {Reyl{\'e}}, {Riclet}, {Ripepi}, {Riva},
  {Rivard}, {Rixon}, {Roegiers}, {Roelens}, {Romero-G{\'o}mez}, {Rowell},
  {Royer}, {Ruiz-Dern}, {Sadowski}, {Sagrist{\`a} Sell{\'e}s}, {Sahlmann},
  {Salgado}, {Salguero}, {Sanna}, {Santana-Ros}, {Sarasso}, {Savietto},
  {Schultheis}, {Sciacca}, {Segol}, {Segovia}, {S{\'e}gransan}, {Shih},
  {Siltala}, {Silva}, {Smart}, {Smith}, {Solano}, {Solitro}, {Sordo}, {Soria
  Nieto}, {Souchay}, {Spagna}, {Spoto}, {Stampa}, {Steele},
  {Steidelm{\"u}ller}, {Stephenson}, {Stoev}, {Suess}, {Surdej}, {Szabados},
  {Szegedi-Elek}, {Tapiador}, {Taris}, {Tauran}, {Taylor}, {Teixeira},
  {Terrett}, {Teyssandier}, {Thuillot}, {Titarenko}, {Torra Clotet}, {Turon},
  {Ulla}, {Utrilla}, {Uzzi}, {Vaillant}, {Valentini}, {Valette}, {van Elteren},
  {Van Hemelryck}, {van Leeuwen}, {Vaschetto}, {Vecchiato}, {Veljanoski},
  {Viala}, {Vicente}, {Vogt}, {von Essen}, {Voss}, {Votruba}, {Voutsinas},
  {Walmsley}, {Weiler}, {Wertz}, {Wevers}, {Wyrzykowski}, {Yoldas},
  {{\v{Z}}erjal}, {Ziaeepour}, {Zorec}, {Zschocke}, {Zucker}, {Zurbach}, \&
  {Zwitter}}]{Gaia2018}
{Gaia Collaboration}, {Brown}, A.~G.~A., {Vallenari}, A., {et~al.} 2018,
  \href{https://doi.org/10.1051/0004-6361/201833051}{\aap}, 616, A1,
  [\href{https://arxiv.org/abs/1804.09365}{1804.09365}].

\bibitem[{{Gallazzi} \& {Bell}(2009)}]{Gallazzi2009}
{Gallazzi}, A. \& {Bell}, E.~F. 2009,
  \href{https://doi.org/10.1088/0067-0049/185/2/253}{\apjs}, 185, 253,
  [\href{https://arxiv.org/abs/0910.1591}{0910.1591}].

\bibitem[{{Gallazzi} {et~al.}(2005){Gallazzi}, {Charlot}, {Brinchmann},
  {White}, \& {Tremonti}}]{Gallazzi2005}
{Gallazzi}, A., {Charlot}, S., {Brinchmann}, J., {White}, S. D.~M., \&
  {Tremonti}, C.~A. 2005,
  \href{https://doi.org/10.1111/j.1365-2966.2005.09321.x}{\mnras}, 362, 41,
  [\href{https://arxiv.org/abs/astro-ph/0506539}{astro-ph/0506539}].

\bibitem[{{Gallazzi} {et~al.}(2021){Gallazzi}, {Pasquali}, {Zibetti}, \&
  {Barbera}}]{Gallazzi2021}
{Gallazzi}, A.~R., {Pasquali}, A., {Zibetti}, S., \& {Barbera}, F.~L. 2021,
  \href{https://doi.org/10.1093/mnras/stab265}{\mnras}, 502, 4457,
  [\href{https://arxiv.org/abs/2010.04733}{2010.04733}].

\bibitem[{{Gao} {et~al.}(2012){Gao}, {Navarro}, {Frenk}, {Jenkins}, {Springel},
  \& {White}}]{gao2012}
{Gao}, L., {Navarro}, J.~F., {Frenk}, C.~S., {et~al.} 2012,
  \href{https://doi.org/10.1111/j.1365-2966.2012.21564.x}{\mnras}, 425, 2169,
  [\href{https://arxiv.org/abs/1201.1940}{1201.1940}].

\bibitem[{{Gao} {et~al.}(2018){Gao}, {Bao}, {Yuan}, {Kong}, {Zou}, {Zhou},
  {Gu}, {Lin}, {Liang}, \& {Huang}}]{Gao2018}
{Gao}, Y., {Bao}, M., {Yuan}, Q., {et~al.} 2018,
  \href{https://doi.org/10.3847/1538-4357/aae9ef}{\apj}, 869, 15,
  [\href{https://arxiv.org/abs/1810.08928}{1810.08928}].

\bibitem[{{Garc{\'\i}a-Benito} {et~al.}(2017){Garc{\'\i}a-Benito},
  {Gonz{\'a}lez Delgado}, {P{\'e}rez}, {Cid Fernandes}, {Cortijo-Ferrero},
  {L{\'o}pez Fern{\'a}ndez}, {de Amorim}, {Lacerda}, {Vale Asari}, \&
  {S{\'a}nchez}}]{Ruben2017}
{Garc{\'\i}a-Benito}, R., {Gonz{\'a}lez Delgado}, R.~M., {P{\'e}rez}, E.,
  {et~al.} 2017, \href{https://doi.org/10.1051/0004-6361/201731357}{\aap}, 608,
  A27, [\href{https://arxiv.org/abs/1709.00413}{1709.00413}].

\bibitem[{{Garc{\'\i}a-Benito} {et~al.}(2015){Garc{\'\i}a-Benito}, {Zibetti},
  {S{\'a}nchez}, {Husemann}, {de Amorim}, {Castillo-Morales}, {Cid Fernandes},
  {Ellis}, {Falc{\'o}n-Barroso}, {Galbany}, {Gil de Paz}, {Gonz{\'a}lez
  Delgado}, {Lacerda}, {L{\'o}pez-Fernandez}, {de Lorenzo-C{\'a}ceres},
  {Lyubenova}, {Marino}, {Mast}, {Mendoza}, {P{\'e}rez}, {Vale Asari},
  {Aguerri}, {Ascasibar}, {Bekerait{\.{e}}}, {Bland-Hawthorn},
  {Barrera-Ballesteros}, {Bomans}, {Cano-D{\'\i}az}, {Catal{\'a}n-Torrecilla},
  {Cortijo}, {Delgado-Inglada}, {Demleitner}, {Dettmar}, {D{\'\i}az},
  {Florido}, {Gallazzi}, {Garc{\'\i}a-Lorenzo}, {Gomes}, {Holmes},
  {Iglesias-P{\'a}ramo}, {Jahnke}, {Kalinova}, {Kehrig}, {Kennicutt},
  {L{\'o}pez-S{\'a}nchez}, {M{\'a}rquez}, {Masegosa}, {Meidt}, {Mendez-Abreu},
  {Moll{\'a}}, {Monreal-Ibero}, {Morisset}, {del Olmo}, {Papaderos},
  {P{\'e}rez}, {Quirrenbach}, {Rosales-Ortega}, {Roth}, {Ruiz-Lara},
  {S{\'a}nchez-Bl{\'a}zquez}, {S{\'a}nchez-Menguiano}, {Singh}, {Spekkens},
  {Stanishev}, {Torres-Papaqui}, {van de Ven}, {Vilchez}, {Walcher}, {Wild},
  {Wisotzki}, {Ziegler}, {Alves}, {Barrado}, {Quintana}, \&
  {Aceituno}}]{CALIFA2015}
{Garc{\'\i}a-Benito}, R., {Zibetti}, S., {S{\'a}nchez}, S.~F., {et~al.} 2015,
  \href{https://doi.org/10.1051/0004-6361/201425080}{\aap}, 576, A135,
  [\href{https://arxiv.org/abs/1409.8302}{1409.8302}].

\bibitem[{{Garn} {et~al.}(2010){Garn}, {Sobral}, {Best}, {Geach}, {Smail},
  {Cirasuolo}, {Dalton}, {Dunlop}, {McLure}, \& {Farrah}}]{Garn2010}
{Garn}, T., {Sobral}, D., {Best}, P.~N., {et~al.} 2010,
  \href{https://doi.org/10.1111/j.1365-2966.2009.16042.x}{\mnras}, 402, 2017,
  [\href{https://arxiv.org/abs/0911.2511}{0911.2511}].

\bibitem[{{Gavazzi} {et~al.}(2003){Gavazzi}, {Cortese}, {Boselli},
  {Iglesias-Paramo}, {V{\'\i}lchez}, \& {Carrasco}}]{Gavazzi2003}
{Gavazzi}, G., {Cortese}, L., {Boselli}, A., {et~al.} 2003,
  \href{https://doi.org/10.1086/378264}{\apj}, 597, 210,
  [\href{https://arxiv.org/abs/astro-ph/0307075}{astro-ph/0307075}].

\bibitem[{{Ge} {et~al.}(2024){Ge}, {Wang}, {Lei}, {Guo}, {Jiang}, \&
  {Cao}}]{Ge2024}
{Ge}, X., {Wang}, H.-T., {Lei}, C.-L., {et~al.} 2024,
  \href{https://doi.org/10.1088/1674-4527/ad1c77}{Research in Astronomy and
  Astrophysics}, 24, 035006.

\bibitem[{{Girichidis} {et~al.}(2012){Girichidis}, {Federrath}, {Banerjee}, \&
  {Klessen}}]{Girichidis2012}
{Girichidis}, P., {Federrath}, C., {Banerjee}, R., \& {Klessen}, R.~S. 2012,
  \href{https://doi.org/10.1111/j.1365-2966.2011.20073.x}{\mnras}, 420, 613,
  [\href{https://arxiv.org/abs/1110.1924}{1110.1924}].

\bibitem[{{Goddard} {et~al.}(2017){Goddard}, {Thomas}, {Maraston}, {Westfall},
  {Etherington}, {Riffel}, {Mallmann}, {Zheng}, {Argudo-Fern{\'a}ndez}, {Lian},
  {Bershady}, {Bundy}, {Drory}, {Law}, {Yan}, {Wake}, {Weijmans}, {Bizyaev},
  {Brownstein}, {Lane}, {Maiolino}, {Masters}, {Merrifield}, {Nitschelm},
  {Pan}, {Roman-Lopes}, {Storchi-Bergmann}, \& {Schneider}}]{Goddard2017b}
{Goddard}, D., {Thomas}, D., {Maraston}, C., {et~al.} 2017,
  \href{https://doi.org/10.1093/mnras/stw3371}{\mnras}, 466, 4731,
  [\href{https://arxiv.org/abs/1612.01546}{1612.01546}].

\bibitem[{{G{\'o}mez} {et~al.}(2003){G{\'o}mez}, {Nichol}, {Miller}, {Balogh},
  {Goto}, {Zabludoff}, {Romer}, {Bernardi}, {Sheth}, {Hopkins}, {Castander},
  {Connolly}, {Schneider}, {Brinkmann}, {Lamb}, {SubbaRao}, \&
  {York}}]{Gomez2003}
{G{\'o}mez}, P.~L., {Nichol}, R.~C., {Miller}, C.~J., {et~al.} 2003,
  \href{https://doi.org/10.1086/345593}{\apj}, 584, 210,
  [\href{https://arxiv.org/abs/astro-ph/0210193}{astro-ph/0210193}].

\bibitem[{{Gondhalekar} {et~al.}(2022){Gondhalekar}, {de Souza}, \&
  {Chies-Santos}}]{galmask2022}
{Gondhalekar}, Y., {de Souza}, R.~S., \& {Chies-Santos}, A.~L. 2022,
  \href{https://doi.org/10.3847/2515-5172/ac780b}{Research Notes of the
  American Astronomical Society}, 6, 128,
  [\href{https://arxiv.org/abs/2206.06787}{2206.06787}].

\bibitem[{{Gonz{\'a}lez Delgado} {et~al.}(2005){Gonz{\'a}lez Delgado},
  {Cervi{\~n}o}, {Martins}, {Leitherer}, \& {Hauschildt}}]{Rosa2005}
{Gonz{\'a}lez Delgado}, R.~M., {Cervi{\~n}o}, M., {Martins}, L.~P.,
  {Leitherer}, C., \& {Hauschildt}, P.~H. 2005,
  \href{https://doi.org/10.1111/j.1365-2966.2005.08692.x}{\mnras}, 357, 945,
  [\href{https://arxiv.org/abs/astro-ph/0501204}{astro-ph/0501204}].

\bibitem[{{Gonz{\'a}lez Delgado} {et~al.}(2016){Gonz{\'a}lez Delgado}, {Cid
  Fernandes}, {P{\'e}rez}, {Garc{\'\i}a-Benito}, {L{\'o}pez Fern{\'a}ndez},
  {Lacerda}, {Cortijo-Ferrero}, {de Amorim}, {Vale Asari}, {S{\'a}nchez},
  {Walcher}, {Wisotzki}, {Mast}, {Alves}, {Ascasibar}, {Bland-Hawthorn},
  {Galbany}, {Kennicutt}, {M{\'a}rquez}, {Masegosa}, {Moll{\'a}},
  {S{\'a}nchez-Bl{\'a}zquez}, \& {V{\'\i}lchez}}]{Rosa2016}
{Gonz{\'a}lez Delgado}, R.~M., {Cid Fernandes}, R., {P{\'e}rez}, E., {et~al.}
  2016, \href{https://doi.org/10.1051/0004-6361/201628174}{\aap}, 590, A44,
  [\href{https://arxiv.org/abs/1603.00874}{1603.00874}].

\bibitem[{{Gonz{\'a}lez Delgado} {et~al.}(2021){Gonz{\'a}lez Delgado},
  {D{\'\i}az-Garc{\'\i}a}, {de Amorim}, {Bruzual}, {Cid Fernandes},
  {P{\'e}rez}, {Bonoli}, {Cenarro}, {Coelho}, {Cortesi}, {Garc{\'\i}a-Benito},
  {L{\'o}pez Fern{\'a}ndez}, {Mart{\'\i}nez-Solaeche},
  {Rodr{\'\i}guez-Mart{\'\i}n}, {Magris}, {Mej{\'\i}a-Narvaez}, {Brito-Silva},
  {Abramo}, {Diego}, {Dupke}, {Hern{\'a}n-Caballero},
  {Hern{\'a}ndez-Monteagudo}, {L{\'o}pez-Sanjuan}, {Mar{\'\i}n-Franch},
  {Marra}, {Moles}, {Montero-Dorta}, {Queiroz}, {Sodr{\'e}}, {Varela},
  {V{\'a}zquez Rami{\'o}}, {V{\'\i}lchez}, {Baqui}, {Ben{\'\i}tez},
  {Crist{\'o}bal-Hornillos}, {Ederoclite}, {Mendes de Oliveira}, {Civera},
  {Muniesa}, {Taylor}, {Tempel}, \& {J-PAS Collaboration}}]{Rosa2021}
{Gonz{\'a}lez Delgado}, R.~M., {D{\'\i}az-Garc{\'\i}a}, L.~A., {de Amorim}, A.,
  {et~al.} 2021, \href{https://doi.org/10.1051/0004-6361/202039849}{\aap}, 649,
  A79, [\href{https://arxiv.org/abs/2102.13121}{2102.13121}].

\bibitem[{{Gonz{\'a}lez Delgado} {et~al.}(2015){Gonz{\'a}lez Delgado},
  {Garc{\'\i}a-Benito}, {P{\'e}rez}, {Cid Fernandes}, {de Amorim},
  {Cortijo-Ferrero}, {Lacerda}, {L{\'o}pez Fern{\'a}ndez}, {Vale-Asari},
  {S{\'a}nchez}, {Moll{\'a}}, {Ruiz-Lara}, {S{\'a}nchez-Bl{\'a}zquez},
  {Walcher}, {Alves}, {Aguerri}, {Bekerait{\'e}}, {Bland-Hawthorn}, {Galbany},
  {Gallazzi}, {Husemann}, {Iglesias-P{\'a}ramo}, {Kalinova},
  {L{\'o}pez-S{\'a}nchez}, {Marino}, {M{\'a}rquez}, {Masegosa}, {Mast},
  {M{\'e}ndez-Abreu}, {Mendoza}, {del Olmo}, {P{\'e}rez}, {Quirrenbach}, \&
  {Zibetti}}]{Rosa2015}
{Gonz{\'a}lez Delgado}, R.~M., {Garc{\'\i}a-Benito}, R., {P{\'e}rez}, E.,
  {et~al.} 2015, \href{https://doi.org/10.1051/0004-6361/201525938}{\aap}, 581,
  A103, [\href{https://arxiv.org/abs/1506.04157}{1506.04157}].

\bibitem[{{Gonz{\'a}lez Delgado} {et~al.}(2014){Gonz{\'a}lez Delgado},
  {P{\'e}rez}, {Cid Fernandes}, {Garc{\'\i}a-Benito}, {de Amorim},
  {S{\'a}nchez}, {Husemann}, {Cortijo-Ferrero}, {L{\'o}pez Fern{\'a}ndez},
  {S{\'a}nchez-Bl{\'a}zquez}, {Bekeraite}, {Walcher}, {Falc{\'o}n-Barroso},
  {Gallazzi}, {van de Ven}, {Alves}, {Bland-Hawthorn}, {Kennicutt}, {Kupko},
  {Lyubenova}, {Mast}, {Moll{\'a}}, {Marino}, {Quirrenbach}, {V{\'\i}lchez}, \&
  {Wisotzki}}]{Rosa2014}
{Gonz{\'a}lez Delgado}, R.~M., {P{\'e}rez}, E., {Cid Fernandes}, R., {et~al.}
  2014, \href{https://doi.org/10.1051/0004-6361/201322011}{\aap}, 562, A47,
  [\href{https://arxiv.org/abs/1310.5517}{1310.5517}].

\bibitem[{{Gonz{\'a}lez Delgado} {et~al.}(2017){Gonz{\'a}lez Delgado},
  {P{\'e}rez}, {Cid Fernandes}, {Garc{\'\i}a-Benito}, {L{\'o}pez
  Fern{\'a}ndez}, {Vale Asari}, {Cortijo-Ferrero}, {de Amorim}, {Lacerda},
  {S{\'a}nchez}, {Lehnert}, \& {Walcher}}]{Rosa2017}
{Gonz{\'a}lez Delgado}, R.~M., {P{\'e}rez}, E., {Cid Fernandes}, R., {et~al.}
  2017, \href{https://doi.org/10.1051/0004-6361/201730883}{\aap}, 607, A128,
  [\href{https://arxiv.org/abs/1706.06119}{1706.06119}].

\bibitem[{{Gonzalez-Delgado} {et~al.}(1994){Gonzalez-Delgado}, {Perez},
  {Tenorio-Tagle}, {Vilchez}, {Terlevich}, {Terlevich}, {Telles},
  {Rodriguez-Espinosa}, {Mas-Hesse}, {Garcia-Vargas}, {Diaz}, {Cepa}, \&
  {Castaneda}}]{Rosa1994}
{Gonzalez-Delgado}, R.~M., {Perez}, E., {Tenorio-Tagle}, G., {et~al.} 1994,
  \href{https://doi.org/10.1086/174992}{\apj}, 437, 239.

\bibitem[{{Gonz{\'a}lez Delgado} {et~al.}(2022){Gonz{\'a}lez Delgado},
  {Rodr{\'\i}guez-Mart{\'\i}n}, {D{\'\i}az-Garc{\'\i}a}, {de Amorim},
  {Garc{\'\i}a-Benito}, {Mart{\'\i}nez-Solaeche}, {Lopes}, {Maturi},
  {P{\'e}rez}, {Cid Fernandes}, {Cortesi}, {Finoguenov}, {Carrasco},
  {Hern{\'a}n-Caballero}, {Abramo}, {Alcaniz}, {Ben{\'\i}tez}, {Bonoli},
  {Cenarro}, {Crist{\'o}bal-Hornillos}, {Diego}, {Dupke}, {Ederoclite},
  {Fern{\'a}ndez-Ontiveros}, {L{\'o}pez-Sanjuan}, {Mar{\'\i}n-Franch},
  {M{\'a}rquez}, {Mendes de Oliveira}, {Moles}, {Pintos}, {Sodr{\'e}},
  {Taylor}, {Varela}, {V{\'a}zquez Rami{\'o}}, \& {V{\'\i}lchez}}]{Rosa2022}
{Gonz{\'a}lez Delgado}, R.~M., {Rodr{\'\i}guez-Mart{\'\i}n}, J.~E.,
  {D{\'\i}az-Garc{\'\i}a}, L.~A., {et~al.} 2022,
  \href{https://doi.org/10.1051/0004-6361/202244030}{\aap}, 666, A84,
  [\href{https://arxiv.org/abs/2207.05770}{2207.05770}].

\bibitem[{{Gonzalez-Perez} {et~al.}(2011){Gonzalez-Perez}, {Castander}, \&
  {Kauffmann}}]{GonzalezPerez2011}
{Gonzalez-Perez}, V., {Castander}, F.~J., \& {Kauffmann}, G. 2011,
  \href{https://doi.org/10.1111/j.1365-2966.2010.17744.x}{\mnras}, 411, 1151,
  [\href{https://arxiv.org/abs/1008.2354}{1008.2354}].

\bibitem[{{Goodman} \& {Weare}(2010)}]{Emcee}
{Goodman}, J. \& {Weare}, J. 2010,
  \href{https://doi.org/10.2140/camcos.2010.5.65}{Communications in Applied
  Mathematics and Computational Science}, 5, 65.

\bibitem[{{Green} {et~al.}(2018{\natexlab{a}}){Green}, {Croom}, {Scott},
  {Cortese}, {Medling}, {D'Eugenio}, {Bryant}, {Bland-Hawthorn}, {Allen},
  {Sharp}, {Ho}, {Groves}, {Drinkwater}, {Mannering}, {Harischandra}, {van de
  Sande}, {Thomas}, {O'Toole}, {McDermid}, {Vuong}, {Sealey}, {Bauer},
  {Brough}, {Catinella}, {Cecil}, {Colless}, {Couch}, {Driver}, {Federrath},
  {Foster}, {Goodwin}, {Hampton}, {Hopkins}, {Jones}, {Konstantopoulos},
  {Lawrence}, {Leon-Saval}, {Liske}, {L{\'o}pez-S{\'a}nchez}, {Lorente},
  {Mould}, {Obreschkow}, {Owers}, {Richards}, {Robotham}, {Schaefer}, {Sweet},
  {Taranu}, {Tescari}, {Tonini}, \& {Zafar}}]{Green2018SAMI}
{Green}, A.~W., {Croom}, S.~M., {Scott}, N., {et~al.} 2018{\natexlab{a}},
  \href{https://doi.org/10.1093/mnras/stx3135}{\mnras}, 475, 716,
  [\href{https://arxiv.org/abs/1707.08402}{1707.08402}].

\bibitem[{{Green} {et~al.}(2018{\natexlab{b}}){Green}, {Schlafly},
  {Finkbeiner}, {Rix}, {Martin}, {Burgett}, {Draper}, {Flewelling}, {Hodapp},
  {Kaiser}, {Kudritzki}, {Magnier}, {Metcalfe}, {Tonry}, {Wainscoat}, \&
  {Waters}}]{Green2018}
{Green}, G.~M., {Schlafly}, E.~F., {Finkbeiner}, D., {et~al.}
  2018{\natexlab{b}}, \href{https://doi.org/10.1093/mnras/sty1008}{\mnras},
  478, 651, [\href{https://arxiv.org/abs/1801.03555}{1801.03555}].

\bibitem[{{Greisen} \& {Calabretta}(2002)}]{FitsWCS2002I}
{Greisen}, E.~W. \& {Calabretta}, M.~R. 2002,
  \href{https://doi.org/10.1051/0004-6361:20021326}{\aap}, 395, 1061,
  [\href{https://arxiv.org/abs/astro-ph/0207407}{astro-ph/0207407}].

\bibitem[{{Greisen} {et~al.}(2006){Greisen}, {Calabretta}, {Valdes}, \&
  {Allen}}]{FitsWCS2006}
{Greisen}, E.~W., {Calabretta}, M.~R., {Valdes}, F.~G., \& {Allen}, S.~L. 2006,
  \href{https://doi.org/10.1051/0004-6361:20053818}{\aap}, 446, 747,
  [\href{https://arxiv.org/abs/astro-ph/0507293}{astro-ph/0507293}].

\bibitem[{{Guglielmo} {et~al.}(2019){Guglielmo}, {Poggianti}, {Vulcani},
  {Maurogordato}, {Fritz}, {Bolzonella}, {Fotopoulou}, {Adami}, \&
  {Pierre}}]{Guglielmo2019}
{Guglielmo}, V., {Poggianti}, B.~M., {Vulcani}, B., {et~al.} 2019,
  \href{https://doi.org/10.1051/0004-6361/201834970}{\aap}, 625, A112,
  [\href{https://arxiv.org/abs/1903.12293}{1903.12293}].

\bibitem[{{Gunn} \& {Gott}(1972)}]{Gunn1972}
{Gunn}, J.~E. \& {Gott}, J.~Richard, I. 1972,
  \href{https://doi.org/10.1086/151605}{\apj}, 176, 1.

\bibitem[{{Guo} {et~al.}(2021){Guo}, {Carleton}, {Bell}, {Chen}, {Dekel},
  {Faber}, {Giavalisco}, {Kocevski}, {Koekemoer}, {Koo}, {Kurczynski}, {Lee},
  {Liu}, {Papovich}, \& {P{\'e}rez-Gonz{\'a}lez}}]{Guo2021}
{Guo}, Y., {Carleton}, T., {Bell}, E.~F., {et~al.} 2021,
  \href{https://doi.org/10.3847/1538-4357/abf115}{\apj}, 914, 7,
  [\href{https://arxiv.org/abs/2105.12144}{2105.12144}].

\bibitem[{{Gutcke} {et~al.}(2017){Gutcke}, {Macci{\`o}}, {Dutton}, \&
  {Stinson}}]{Gutcke2017}
{Gutcke}, T.~A., {Macci{\`o}}, A.~V., {Dutton}, A.~A., \& {Stinson}, G.~S.
  2017, \href{https://doi.org/10.1093/mnras/stx005}{\mnras}, 466, 4614,
  [\href{https://arxiv.org/abs/1701.01130}{1701.01130}].

\bibitem[{{Haines} {et~al.}(2017){Haines}, {Iovino}, {Krywult}, {Guzzo},
  {Davidzon}, {Bolzonella}, {Garilli}, {Scodeggio}, {Granett}, {de la Torre},
  {De Lucia}, {Abbas}, {Adami}, {Arnouts}, {Bottini}, {Cappi}, {Cucciati},
  {Franzetti}, {Fritz}, {Gargiulo}, {Le Brun}, {Le F{\`e}vre}, {Maccagni},
  {Ma{\l}ek}, {Marulli}, {Moutard}, {Polletta}, {Pollo}, {Tasca}, {Tojeiro},
  {Vergani}, {Zanichelli}, {Zamorani}, {Bel}, {Branchini}, {Coupon}, {Ilbert},
  {Moscardini}, {Peacock}, \& {Siudek}}]{Haines2017}
{Haines}, C.~P., {Iovino}, A., {Krywult}, J., {et~al.} 2017,
  \href{https://doi.org/10.1051/0004-6361/201630118}{\aap}, 605, A4,
  [\href{https://arxiv.org/abs/1611.07050}{1611.07050}].

\bibitem[{{Haines} {et~al.}(2012){Haines}, {Pereira}, {Sanderson}, {Smith},
  {Egami}, {Babul}, {Edge}, {Finoguenov}, {Moran}, \& {Okabe}}]{Haines2012}
{Haines}, C.~P., {Pereira}, M.~J., {Sanderson}, A.~J.~R., {et~al.} 2012,
  \href{https://doi.org/10.1088/0004-637X/754/2/97}{\apj}, 754, 97,
  [\href{https://arxiv.org/abs/1205.6818}{1205.6818}].

\bibitem[{{Haines} {et~al.}(2015){Haines}, {Pereira}, {Smith}, {Egami},
  {Babul}, {Finoguenov}, {Ziparo}, {McGee}, {Rawle}, {Okabe}, \&
  {Moran}}]{Haines2015}
{Haines}, C.~P., {Pereira}, M.~J., {Smith}, G.~P., {et~al.} 2015,
  \href{https://doi.org/10.1088/0004-637X/806/1/101}{\apj}, 806, 101,
  [\href{https://arxiv.org/abs/1504.05604}{1504.05604}].

\bibitem[{{Hamilton}(1993)}]{Mangle1993}
{Hamilton}, A.~J.~S. 1993, \href{https://doi.org/10.1086/173288}{\apj}, 417,
  19.

\bibitem[{{Hamilton} \& {Tegmark}(2004)}]{Mangle2004}
{Hamilton}, A.~J.~S. \& {Tegmark}, M. 2004,
  \href{https://doi.org/10.1111/j.1365-2966.2004.07490.x}{\mnras}, 349, 115,
  [\href{https://arxiv.org/abs/astro-ph/0306324}{astro-ph/0306324}].

\bibitem[{{Hayashi} {et~al.}(2020){Hayashi}, {Shimakawa}, {Tanaka}, {Onodera},
  {Koyama}, {Inoue}, {Komiyama}, {Lee}, {Lin}, \& {Yabe}}]{Hayashi2020}
{Hayashi}, M., {Shimakawa}, R., {Tanaka}, M., {et~al.} 2020,
  \href{https://doi.org/10.1093/pasj/psaa076}{\pasj}, 72, 86,
  [\href{https://arxiv.org/abs/2007.07413}{2007.07413}].

\bibitem[{{Heasley}(1999)}]{Heasley1999}
{Heasley}, J.~N. 1999, in Astronomical Society of the Pacific Conference
  Series, Vol. 189, Precision CCD Photometry, ed. E.~R. {Craine}, D.~L.
  {Crawford}, \& R.~A. {Tucker}, 56

\bibitem[{{Henriques} \& {Thomas}(2010)}]{Henriques2010}
{Henriques}, B. M.~B. \& {Thomas}, P.~A. 2010,
  \href{https://doi.org/10.1111/j.1365-2966.2009.16151.x}{\mnras}, 403, 768,
  [\href{https://arxiv.org/abs/0909.2150}{0909.2150}].

\bibitem[{{Hern{\'a}n-Caballero} {et~al.}(2021){Hern{\'a}n-Caballero},
  {Varela}, {L{\'o}pez-Sanjuan}, {Muniesa}, {Civera}, {Chaves-Montero},
  {D{\'\i}az-Garc{\'\i}a}, {Laur}, {Hern{\'a}ndez-Monteagudo}, {Abramo},
  {Angulo}, {Crist{\'o}bal-Hornillos}, {Gonz{\'a}lez Delgado}, {Greisel},
  {Orsi}, {Queiroz}, {Sobral}, {Tamm}, {Tempel}, {V{\'a}zquez-Rami{\'o}},
  {Alcaniz}, {Ben{\'\i}tez}, {Bonoli}, {Carneiro}, {Cenarro}, {Dupke},
  {Ederoclite}, {Mar{\'\i}n-Franch}, {Mendes de Oliveira}, {Moles},
  {Sodr{\'e}}, {Taylor}, {Cypriano}, \& {Mart{\'\i}nez-Solaeche}}]{HC2021}
{Hern{\'a}n-Caballero}, A., {Varela}, J., {L{\'o}pez-Sanjuan}, C., {et~al.}
  2021, \href{https://doi.org/10.1051/0004-6361/202141236}{\aap}, 654, A101,
  [\href{https://arxiv.org/abs/2108.03271}{2108.03271}].

\bibitem[{{Hess} {et~al.}(2020){Hess}, {Falcon-Barroso}, {Ascasibar},
  {Perez-Martin}, {Serra}, {Weijmans}, \& {Weave-Apertif Team}}]{Hess2020}
{Hess}, K.~M., {Falcon-Barroso}, J., {Ascasibar}, Y., {et~al.} 2020, in
  American Astronomical Society Meeting Abstracts, Vol. 235, American
  Astronomical Society Meeting Abstracts \#235, 459.03

\bibitem[{{High} {et~al.}(2009){High}, {Stubbs}, {Rest}, {Stalder}, \&
  {Challis}}]{High2009}
{High}, F.~W., {Stubbs}, C.~W., {Rest}, A., {Stalder}, B., \& {Challis}, P.
  2009, \href{https://doi.org/10.1088/0004-6256/138/1/110}{\aj}, 138, 110,
  [\href{https://arxiv.org/abs/0903.5302}{0903.5302}].

\bibitem[{{Hogg} {et~al.}(2004){Hogg}, {Blanton}, {Brinchmann}, {Eisenstein},
  {Schlegel}, {Gunn}, {McKay}, {Rix}, {Bahcall}, {Brinkmann}, \&
  {Meiksin}}]{Hogg2004}
{Hogg}, D.~W., {Blanton}, M.~R., {Brinchmann}, J., {et~al.} 2004,
  \href{https://doi.org/10.1086/381749}{\apjl}, 601, L29,
  [\href{https://arxiv.org/abs/astro-ph/0307336}{astro-ph/0307336}].

\bibitem[{{Hook} {et~al.}(2004){Hook}, {J{\o}rgensen}, {Allington-Smith},
  {Davies}, {Metcalfe}, {Murowinski}, \& {Crampton}}]{Hook2004}
{Hook}, I.~M., {J{\o}rgensen}, I., {Allington-Smith}, J.~R., {et~al.} 2004,
  \href{https://doi.org/10.1086/383624}{\pasp}, 116, 425.

\bibitem[{{Hsieh} {et~al.}(2017){Hsieh}, {Lin}, {Lin}, {Pan}, {Hsu},
  {S{\'a}nchez}, {Cano-D{\'\i}az}, {Zhang}, {Yan}, {Barrera-Ballesteros},
  {Boquien}, {Riffel}, {Brownstein}, {Cruz-Gonz{\'a}lez}, {Hagen}, {Ibarra},
  {Pan}, {Bizyaev}, {Oravetz}, \& {Simmons}}]{Hsieh2017}
{Hsieh}, B.~C., {Lin}, L., {Lin}, J.~H., {et~al.} 2017,
  \href{https://doi.org/10.3847/2041-8213/aa9d80}{\apjl}, 851, L24,
  [\href{https://arxiv.org/abs/1711.09162}{1711.09162}].

\bibitem[{{Hyde} \& {Bernardi}(2009)}]{Hyde2009}
{Hyde}, J.~B. \& {Bernardi}, M. 2009,
  \href{https://doi.org/10.1111/j.1365-2966.2009.14445.x}{\mnras}, 394, 1978,
  [\href{https://arxiv.org/abs/0810.4922}{0810.4922}].

\bibitem[{{Ibarra-Medel} {et~al.}(2016){Ibarra-Medel}, {S{\'a}nchez},
  {Avila-Reese}, {Hern{\'a}ndez-Toledo}, {Gonz{\'a}lez}, {Drory}, {Bundy},
  {Bizyaev}, {Cano-D{\'\i}az}, {Malanushenko}, {Pan}, {Roman-Lopes}, \&
  {Thomas}}]{IbarraMedel2016}
{Ibarra-Medel}, H.~J., {S{\'a}nchez}, S.~F., {Avila-Reese}, V., {et~al.} 2016,
  \href{https://doi.org/10.1093/mnras/stw2126}{\mnras}, 463, 2799,
  [\href{https://arxiv.org/abs/1609.01304}{1609.01304}].

\bibitem[{{Iglesias-P{\'a}ramo} {et~al.}(2022){Iglesias-P{\'a}ramo}, {Arroyo},
  {Kehrig}, {V{\'\i}lchez}, {Duarte Puertas}, {P{\'e}rez-Montero}, {Breda},
  {Jim{\'e}nez-Teja}, {L{\'o}pez Sanjuan}, {Lumbreras-Calle}, {Coelho},
  {Gurung-L{\'o}pez}, {Queiroz}, {M{\'a}rquez}, {Povi{\'c}}, {Gonz{\'a}lez
  Delgado}, {Chaves-Montero}, {Sobral}, {Hern{\'a}n-Caballero},
  {Fern{\'a}ndez-Ontiveros}, {D{\'\i}az-Garc{\'\i}a}, {Alvarez-Candal},
  {Abramo}, {Alcaniz}, {Ben{\'\i}tez}, {Bonoli}, {Cenarro},
  {Crist{\'o}bal-Hornillos}, {Dupke}, {Ederoclite}, {Mar{\'\i}n-Franch},
  {Mendes de Oliveira}, {Moles}, {Sodr{\'e}}, {Taylor}, {Varela}, {V{\'a}zquez
  Rami{\'o}}, \& {J-PAS Team}}]{Iglesias2022}
{Iglesias-P{\'a}ramo}, J., {Arroyo}, A., {Kehrig}, C., {et~al.} 2022,
  \href{https://doi.org/10.1051/0004-6361/202243931}{\aap}, 665, A95.

\bibitem[{{Ilbert} {et~al.}(2013){Ilbert}, {McCracken}, {Le F{\`e}vre},
  {Capak}, {Dunlop}, {Karim}, {Renzini}, {Caputi}, {Boissier}, {Arnouts},
  {Aussel}, {Comparat}, {Guo}, {Hudelot}, {Kartaltepe}, {Kneib}, {Krogager},
  {Le Floc'h}, {Lilly}, {Mellier}, {Milvang-Jensen}, {Moutard}, {Onodera},
  {Richard}, {Salvato}, {Sanders}, {Scoville}, {Silverman}, {Taniguchi},
  {Tasca}, {Thomas}, {Toft}, {Tresse}, {Vergani}, {Wolk}, \&
  {Zirm}}]{Ilbert2013}
{Ilbert}, O., {McCracken}, H.~J., {Le F{\`e}vre}, O., {et~al.} 2013,
  \href{https://doi.org/10.1051/0004-6361/201321100}{\aap}, 556, A55,
  [\href{https://arxiv.org/abs/1301.3157}{1301.3157}].

\bibitem[{{Infante-Sainz} {et~al.}(2020){Infante-Sainz}, {Trujillo}, \&
  {Rom{\'a}n}}]{Infante-sainz2020}
{Infante-Sainz}, R., {Trujillo}, I., \& {Rom{\'a}n}, J. 2020,
  \href{https://doi.org/10.1093/mnras/stz3111}{\mnras}, 491, 5317,
  [\href{https://arxiv.org/abs/1911.01430}{1911.01430}].

\bibitem[{{Jaff{\'e}} {et~al.}(2018){Jaff{\'e}}, {Poggianti}, {Moretti},
  {Gullieuszik}, {Smith}, {Vulcani}, {Fasano}, {Fritz}, {Tonnesen}, {Bettoni},
  {Hau}, {Biviano}, {Bellhouse}, \& {McGee}}]{Jaffe2018}
{Jaff{\'e}}, Y.~L., {Poggianti}, B.~M., {Moretti}, A., {et~al.} 2018,
  \href{https://doi.org/10.1093/mnras/sty500}{\mnras}, 476, 4753,
  [\href{https://arxiv.org/abs/1802.07297}{1802.07297}].

\bibitem[{{Janz} {et~al.}(2021){Janz}, {Salo}, {Su}, \& {Venhola}}]{Janz2021}
{Janz}, J., {Salo}, H., {Su}, A.~H., \& {Venhola}, A. 2021,
  \href{https://doi.org/10.1051/0004-6361/202039408}{\aap}, 647, A80,
  [\href{https://arxiv.org/abs/2101.10728}{2101.10728}].

\bibitem[{{Jian} {et~al.}(2018){Jian}, {Lin}, {Oguri}, {Nishizawa}, {Takada},
  {More}, {Koyama}, {Tanaka}, \& {Komiyama}}]{Jian2018}
{Jian}, H.-Y., {Lin}, L., {Oguri}, M., {et~al.} 2018,
  \href{https://doi.org/10.1093/pasj/psx096}{\pasj}, 70, S23,
  [\href{https://arxiv.org/abs/1704.06219}{1704.06219}].

\bibitem[{{J{\o}rgensen}(1999)}]{Jorgensen1999}
{J{\o}rgensen}, I. 1999,
  \href{https://doi.org/10.1046/j.1365-8711.1999.02555.x}{\mnras}, 306, 607,
  [\href{https://arxiv.org/abs/astro-ph/9902250}{astro-ph/9902250}].

\bibitem[{{Joseph} \& {Wright}(1985)}]{Joseph1985}
{Joseph}, R.~D. \& {Wright}, G.~S. 1985,
  \href{https://doi.org/10.1093/mnras/214.2.87}{\mnras}, 214, 87.

\bibitem[{{Joshi} {et~al.}(2019){Joshi}, {Appleton}, {Blanc}, {Guillard},
  {Rich}, {Struck}, {Freeland}, {Peterson}, {Helou}, \& {Alatalo}}]{Joshi2019}
{Joshi}, B.~A., {Appleton}, P.~N., {Blanc}, G.~A., {et~al.} 2019,
  \href{https://doi.org/10.3847/1538-4357/ab2124}{\apj}, 878, 161,
  [\href{https://arxiv.org/abs/1812.07743}{1812.07743}].

\bibitem[{{Joshi} {et~al.}(2020){Joshi}, {Pillepich}, {Nelson}, {Marinacci},
  {Springel}, {Rodriguez-Gomez}, {Vogelsberger}, \& {Hernquist}}]{Joshi2020}
{Joshi}, G.~D., {Pillepich}, A., {Nelson}, D., {et~al.} 2020,
  \href{https://doi.org/10.1093/mnras/staa1668}{\mnras}, 496, 2673,
  [\href{https://arxiv.org/abs/2004.01191}{2004.01191}].

\bibitem[{{Kampakoglou} \& {Benson}(2007)}]{Kampakoglou2007}
{Kampakoglou}, M. \& {Benson}, A.~J. 2007,
  \href{https://doi.org/10.1111/j.1365-2966.2006.11223.x}{\mnras}, 374, 775,
  [\href{https://arxiv.org/abs/astro-ph/0607024}{astro-ph/0607024}].

\bibitem[{{Kauffmann} {et~al.}(2006){Kauffmann}, {Heckman}, {De Lucia},
  {Brinchmann}, {Charlot}, {Tremonti}, {White}, \& {Brinkmann}}]{Kauffmann2006}
{Kauffmann}, G., {Heckman}, T.~M., {De Lucia}, G., {et~al.} 2006,
  \href{https://doi.org/10.1111/j.1365-2966.2006.10061.x}{\mnras}, 367, 1394,
  [\href{https://arxiv.org/abs/astro-ph/0510405}{astro-ph/0510405}].

\bibitem[{{Kauffmann} {et~al.}(2003{\natexlab{a}}){Kauffmann}, {Heckman},
  {Tremonti}, {Brinchmann}, {Charlot}, {White}, {Ridgway}, {Brinkmann},
  {Fukugita}, {Hall}, {Ivezi{\'c}}, {Richards}, \& {Schneider}}]{Kauffmann2003}
{Kauffmann}, G., {Heckman}, T.~M., {Tremonti}, C., {et~al.} 2003{\natexlab{a}},
  \href{https://doi.org/10.1111/j.1365-2966.2003.07154.x}{\mnras}, 346, 1055,
  [\href{https://arxiv.org/abs/astro-ph/0304239}{astro-ph/0304239}].

\bibitem[{{Kauffmann} {et~al.}(2003{\natexlab{b}}){Kauffmann}, {Heckman},
  {White}, {Charlot}, {Tremonti}, {Brinchmann}, {Bruzual}, {Peng}, {Seibert},
  {Bernardi}, {Blanton}, {Brinkmann}, {Castander}, {Cs{\'a}bai}, {Fukugita},
  {Ivezic}, {Munn}, {Nichol}, {Padmanabhan}, {Thakar}, {Weinberg}, \&
  {York}}]{Kauffmann2003b}
{Kauffmann}, G., {Heckman}, T.~M., {White}, S. D.~M., {et~al.}
  2003{\natexlab{b}},
  \href{https://doi.org/10.1046/j.1365-8711.2003.06291.x}{\mnras}, 341, 33,
  [\href{https://arxiv.org/abs/astro-ph/0204055}{astro-ph/0204055}].

\bibitem[{{Kauffmann} {et~al.}(2003{\natexlab{c}}){Kauffmann}, {Heckman},
  {White}, {Charlot}, {Tremonti}, {Peng}, {Seibert}, {Brinkmann}, {Nichol},
  {SubbaRao}, \& {York}}]{Kauffmann2003a}
{Kauffmann}, G., {Heckman}, T.~M., {White}, S. D.~M., {et~al.}
  2003{\natexlab{c}},
  \href{https://doi.org/10.1046/j.1365-8711.2003.06292.x}{\mnras}, 341, 54,
  [\href{https://arxiv.org/abs/astro-ph/0205070}{astro-ph/0205070}].

\bibitem[{{Kauffmann} {et~al.}(1993){Kauffmann}, {White}, \&
  {Guiderdoni}}]{Kauffmann1993}
{Kauffmann}, G., {White}, S.~D.~M., \& {Guiderdoni}, B. 1993,
  \href{https://doi.org/10.1093/mnras/264.1.201}{\mnras}, 264, 201.

\bibitem[{{Kauffmann} {et~al.}(2004){Kauffmann}, {White}, {Heckman},
  {M{\'e}nard}, {Brinchmann}, {Charlot}, {Tremonti}, \&
  {Brinkmann}}]{Kauffmann2004}
{Kauffmann}, G., {White}, S. D.~M., {Heckman}, T.~M., {et~al.} 2004,
  \href{https://doi.org/10.1111/j.1365-2966.2004.08117.x}{\mnras}, 353, 713,
  [\href{https://arxiv.org/abs/astro-ph/0402030}{astro-ph/0402030}].

\bibitem[{{Keim} {et~al.}(2022){Keim}, {van Dokkum}, {Danieli}, {Lokhorst},
  {Li}, {Shen}, {Abraham}, {Chen}, {Gilhuly}, {Liu}, {Merritt}, {Miller},
  {Pasha}, \& {Polzin}}]{Keim2022}
{Keim}, M.~A., {van Dokkum}, P., {Danieli}, S., {et~al.} 2022,
  \href{https://doi.org/10.3847/1538-4357/ac7dab}{\apj}, 935, 160,
  [\href{https://arxiv.org/abs/2109.09778}{2109.09778}].

\bibitem[{{Kelly} {et~al.}(2014){Kelly}, {von der Linden}, {Applegate},
  {Allen}, {Allen}, {Burchat}, {Burke}, {Ebeling}, {Capak}, {Czoske},
  {Donovan}, {Mantz}, \& {Morris}}]{Kelly2014}
{Kelly}, P.~L., {von der Linden}, A., {Applegate}, D.~E., {et~al.} 2014,
  \href{https://doi.org/10.1093/mnras/stt1946}{\mnras}, 439, 28,
  [\href{https://arxiv.org/abs/1208.0602}{1208.0602}].

\bibitem[{{Kennedy} {et~al.}(2016){Kennedy}, {Bamford}, {H{\"a}u{\ss}ler},
  {Brough}, {Holwerda}, {Hopkins}, {Vika}, \& {Vulcani}}]{Kennedy2016}
{Kennedy}, R., {Bamford}, S.~P., {H{\"a}u{\ss}ler}, B., {et~al.} 2016,
  \href{https://doi.org/10.1051/0004-6361/201628715}{\aap}, 593, A84,
  [\href{https://arxiv.org/abs/1608.03495}{1608.03495}].

\bibitem[{{Kennicutt}(1998)}]{Kennicutt1998}
{Kennicutt}, Robert~C., J. 1998,
  \href{https://doi.org/10.1146/annurev.astro.36.1.189}{\araa}, 36, 189,
  [\href{https://arxiv.org/abs/astro-ph/9807187}{astro-ph/9807187}].

\bibitem[{{Kennicutt} \& {Evans}(2012)}]{Kennicutt2012}
{Kennicutt}, R.~C. \& {Evans}, N.~J. 2012,
  \href{https://doi.org/10.1146/annurev-astro-081811-125610}{\araa}, 50, 531,
  [\href{https://arxiv.org/abs/1204.3552}{1204.3552}].

\bibitem[{{Kewley} {et~al.}(2001){Kewley}, {Dopita}, {Sutherland}, {Heisler},
  \& {Trevena}}]{Kewley2001}
{Kewley}, L.~J., {Dopita}, M.~A., {Sutherland}, R.~S., {Heisler}, C.~A., \&
  {Trevena}, J. 2001, \href{https://doi.org/10.1086/321545}{\apj}, 556, 121,
  [\href{https://arxiv.org/abs/astro-ph/0106324}{astro-ph/0106324}].

\bibitem[{{Kewley} {et~al.}(2002){Kewley}, {Geller}, {Jansen}, \&
  {Dopita}}]{Kewley2002}
{Kewley}, L.~J., {Geller}, M.~J., {Jansen}, R.~A., \& {Dopita}, M.~A. 2002,
  \href{https://doi.org/10.1086/344487}{\aj}, 124, 3135,
  [\href{https://arxiv.org/abs/astro-ph/0208508}{astro-ph/0208508}].

\bibitem[{{Kewley} {et~al.}(2006){Kewley}, {Groves}, {Kauffmann}, \&
  {Heckman}}]{Kewley2006}
{Kewley}, L.~J., {Groves}, B., {Kauffmann}, G., \& {Heckman}, T. 2006,
  \href{https://doi.org/10.1111/j.1365-2966.2006.10859.x}{\mnras}, 372, 961,
  [\href{https://arxiv.org/abs/astro-ph/0605681}{astro-ph/0605681}].

\bibitem[{{Khostovan} {et~al.}(2021){Khostovan}, {Malhotra}, {Rhoads},
  {Harish}, {Jiang}, {Wang}, {Wold}, {Zheng}, {Barrientos}, {Coughlin}, {Hu},
  {Infante}, {Perez}, {Pharo}, {Valdes}, \& {Walker}}]{Khostovan2021}
{Khostovan}, A.~A., {Malhotra}, S., {Rhoads}, J.~E., {et~al.} 2021,
  \href{https://doi.org/10.1093/mnras/stab778}{\mnras}, 503, 5115,
  [\href{https://arxiv.org/abs/2103.10959}{2103.10959}].

\bibitem[{{Kiiveri} {et~al.}(2021){Kiiveri}, {Gruen}, {Finoguenov}, {Erben},
  {van Waerbeke}, {Rykoff}, {Miller}, {Hagstotz}, {Dupke}, {Patrick Henry},
  {Kneib}, {Gozaliasl}, {Kirkpatrick}, {Cibirka}, {Clerc}, {Costanzi},
  {Cypriano}, {Rozo}, {Shan}, {Spinelli}, {Valiviita}, \&
  {Weller}}]{Kiiveri2021}
{Kiiveri}, K., {Gruen}, D., {Finoguenov}, A., {et~al.} 2021,
  \href{https://doi.org/10.1093/mnras/staa3936}{\mnras}, 502, 1494,
  [\href{https://arxiv.org/abs/2101.02257}{2101.02257}].

\bibitem[{{Kipper} {et~al.}(2021){Kipper}, {Tamm}, {Tempel}, {de Propris}, \&
  {Ganeshaiah Veena}}]{Kipper2021}
{Kipper}, R., {Tamm}, A., {Tempel}, E., {de Propris}, R., \& {Ganeshaiah
  Veena}, P. 2021, \href{https://doi.org/10.1051/0004-6361/202039648}{\aap},
  647, A32, [\href{https://arxiv.org/abs/2101.08549}{2101.08549}].

\bibitem[{{Klimentowski} {et~al.}(2009){Klimentowski}, {{\L}okas},
  {Kazantzidis}, {Mayer}, \& {Mamon}}]{Klimentowski2009}
{Klimentowski}, J., {{\L}okas}, E.~L., {Kazantzidis}, S., {Mayer}, L., \&
  {Mamon}, G.~A. 2009,
  \href{https://doi.org/10.1111/j.1365-2966.2009.15046.x}{\mnras}, 397, 2015,
  [\href{https://arxiv.org/abs/0803.2464}{0803.2464}].

\bibitem[{{Knowles} {et~al.}(2022){Knowles}, {Cotton}, {Rudnick}, {Camilo},
  {Goedhart}, {Deane}, {Ramatsoku}, {Bietenholz}, {Br{\"u}ggen}, {Button},
  {Chen}, {Chibueze}, {Clarke}, {de Gasperin}, {Ianjamasimanana}, {J{\'o}zsa},
  {Hilton}, {Kesebonye}, {Kolokythas}, {Kraan-Korteweg}, {Lawrie}, {Lochner},
  {Loubser}, {Marchegiani}, {Mhlahlo}, {Moodley}, {Murphy}, {Namumba},
  {Oozeer}, {Parekh}, {Pillay}, {Passmoor}, {Ramaila}, {Ranchod},
  {Retana-Montenegro}, {Sebokolodi}, {Sikhosana}, {Smirnov}, {Thorat},
  {Venturi}, {Abbott}, {Adam}, {Adams}, {Aldera}, {Bauermeister}, {Bennett},
  {Bode}, {Botha}, {Botha}, {Brederode}, {Buchner}, {Burger}, {Cheetham}, {de
  Villiers}, {Dikgale-Mahlakoana}, {du Toit}, {Esterhuyse}, {Fadana},
  {Fanaroff}, {Fataar}, {Foley}, {Fourie}, {Frank}, {Gamatham}, {Gatsi},
  {Geyer}, {Gouws}, {Gumede}, {Heywood}, {Hlakola}, {Hokwana}, {Hoosen},
  {Horn}, {Horrell}, {Hugo}, {Isaacson}, {Jonas}, {Jordaan}, {Joubert},
  {Julie}, {Kapp}, {Kasper}, {Kenyon}, {Kotz{\'e}}, {Kotze}, {Kriek}, {Kriel},
  {Krishnan}, {Kusel}, {Legodi}, {Lehmensiek}, {Liebenberg}, {Lord}, {Lunsky},
  {Madisa}, {Magnus}, {Main}, {Makhaba}, {Makhathini}, {Malan}, {Manley},
  {Marais}, {Maree}, {Martens}, {Mauch}, {McAlpine}, {Merry}, {Millenaar},
  {Mokone}, {Monama}, {Mphego}, {New}, {Ngcebetsha}, {Ngoasheng}, {Ockards},
  {Otto}, {Patel}, {Peens-Hough}, {Perkins}, {Ramanujam}, {Ramudzuli},
  {Ratcliffe}, {Renil}, {Robyntjies}, {Rust}, {Salie}, {Sambu}, {Schollar},
  {Schwardt}, {Schwartz}, {Serylak}, {Siebrits}, {Sirothia}, {Slabber},
  {Sofeya}, {Taljaard}, {Tasse}, {Tiplady}, {Toruvanda}, {Twum}, {van Balla},
  {van der Byl}, {van der Merwe}, {van Dyk}, {Van Tonder}, {Van Wyk}, {Venter},
  {Venter}, {Welz}, {Williams}, \& {Xaia}}]{Knowles2022}
{Knowles}, K., {Cotton}, W.~D., {Rudnick}, L., {et~al.} 2022,
  \href{https://doi.org/10.1051/0004-6361/202141488}{\aap}, 657, A56,
  [\href{https://arxiv.org/abs/2111.05673}{2111.05673}].

\bibitem[{{Kodama} {et~al.}(2004){Kodama}, {Balogh}, {Smail}, {Bower}, \&
  {Nakata}}]{kodama2004}
{Kodama}, T., {Balogh}, M.~L., {Smail}, I., {Bower}, R.~G., \& {Nakata}, F.
  2004, \href{https://doi.org/10.1111/j.1365-2966.2004.08271.x}{\mnras}, 354,
  1103, [\href{https://arxiv.org/abs/astro-ph/0408037}{astro-ph/0408037}].

\bibitem[{{Koester} {et~al.}(2007){Koester}, {McKay}, {Annis}, {Wechsler},
  {Evrard}, {Bleem}, {Becker}, {Johnston}, {Sheldon}, {Nichol}, {Miller},
  {Scranton}, {Bahcall}, {Barentine}, {Brewington}, {Brinkmann}, {Harvanek},
  {Kleinman}, {Krzesinski}, {Long}, {Nitta}, {Schneider}, {Sneddin}, {Voges},
  \& {York}}]{Koester2007}
{Koester}, B.~P., {McKay}, T.~A., {Annis}, J., {et~al.} 2007,
  \href{https://doi.org/10.1086/509599}{\apj}, 660, 239,
  [\href{https://arxiv.org/abs/astro-ph/0701265}{astro-ph/0701265}].

\bibitem[{{Kova{\v{c}}} {et~al.}(2010){Kova{\v{c}}}, {Lilly}, {Knobel},
  {Bolzonella}, {Iovino}, {Carollo}, {Scarlata}, {Sargent}, {Cucciati},
  {Zamorani}, {Pozzetti}, {Tasca}, {Scodeggio}, {Kampczyk}, {Peng}, {Oesch},
  {Zucca}, {Finoguenov}, {Contini}, {Kneib}, {Le F{\`e}vre}, {Mainieri},
  {Renzini}, {Bardelli}, {Bongiorno}, {Caputi}, {Coppa}, {de la Torre}, {de
  Ravel}, {Franzetti}, {Garilli}, {Lamareille}, {Le Borgne}, {Le Brun},
  {Maier}, {Mignoli}, {Pello}, {Perez Montero}, {Ricciardelli}, {Silverman},
  {Tanaka}, {Tresse}, {Vergani}, {Abbas}, {Bottini}, {Cappi}, {Cassata},
  {Cimatti}, {Fumana}, {Guzzo}, {Koekemoer}, {Leauthaud}, {Maccagni},
  {Marinoni}, {McCracken}, {Memeo}, {Meneux}, {Porciani}, {Scaramella}, \&
  {Scoville}}]{Kovac2010}
{Kova{\v{c}}}, K., {Lilly}, S.~J., {Knobel}, C., {et~al.} 2010,
  \href{https://doi.org/10.1088/0004-637X/718/1/86}{\apj}, 718, 86,
  [\href{https://arxiv.org/abs/0909.2032}{0909.2032}].

\bibitem[{{Koyama} {et~al.}(2018){Koyama}, {Hayashi}, {Tanaka}, {Kodama},
  {Shimakawa}, {Yamamoto}, {Nakata}, {Tanaka}, {Suzuki}, {Tadaki}, {Nishizawa},
  {Yabe}, {Toba}, {Lin}, {Jian}, \& {Komiyama}}]{Koyama2018}
{Koyama}, Y., {Hayashi}, M., {Tanaka}, M., {et~al.} 2018,
  \href{https://doi.org/10.1093/pasj/psx078}{\pasj}, 70, S21,
  [\href{https://arxiv.org/abs/1704.05979}{1704.05979}].

\bibitem[{{Koyama} {et~al.}(2014){Koyama}, {Kodama}, {Tadaki}, {Hayashi},
  {Tanaka}, \& {Shimakawa}}]{Koyama2014}
{Koyama}, Y., {Kodama}, T., {Tadaki}, K.-i., {et~al.} 2014,
  \href{https://doi.org/10.1088/0004-637X/789/1/18}{\apj}, 789, 18,
  [\href{https://arxiv.org/abs/1405.4165}{1405.4165}].

\bibitem[{{Kron}(1980)}]{Kron1980}
{Kron}, R.~G. 1980, \href{https://doi.org/10.1086/190669}{\apjs}, 43, 305.

\bibitem[{{Kruijssen} {et~al.}(2019){Kruijssen}, {Schruba}, {Chevance},
  {Longmore}, {Hygate}, {Haydon}, {McLeod}, {Dalcanton}, {Tacconi}, \& {van
  Dishoeck}}]{Kruijssen2019}
{Kruijssen}, J.~M.~D., {Schruba}, A., {Chevance}, M., {et~al.} 2019,
  \href{https://doi.org/10.1038/s41586-019-1194-3}{\nat}, 569, 519,
  [\href{https://arxiv.org/abs/1905.08801}{1905.08801}].

\bibitem[{{Kuijken} {et~al.}(2019){Kuijken}, {Heymans}, {Dvornik},
  {Hildebrandt}, {de Jong}, {Wright}, {Erben}, {Bilicki}, {Giblin}, {Shan},
  {Getman}, {Grado}, {Hoekstra}, {Miller}, {Napolitano}, {Paolilo}, {Radovich},
  {Schneider}, {Sutherland}, {Tewes}, {Tortora}, {Valentijn}, \& {Verdoes
  Kleijn}}]{Kujiken2019}
{Kuijken}, K., {Heymans}, C., {Dvornik}, A., {et~al.} 2019,
  \href{https://doi.org/10.1051/0004-6361/201834918}{\aap}, 625, A2,
  [\href{https://arxiv.org/abs/1902.11265}{1902.11265}].

\bibitem[{{Kumari} {et~al.}(2021){Kumari}, {Maiolino}, {Trussler}, {Mannucci},
  {Cresci}, {Curti}, {Marconi}, \& {Belfiore}}]{Kumari2021}
{Kumari}, N., {Maiolino}, R., {Trussler}, J., {et~al.} 2021,
  \href{https://doi.org/10.1051/0004-6361/202140757}{\aap}, 656, A140,
  [\href{https://arxiv.org/abs/2108.12437}{2108.12437}].

\bibitem[{{Kuntschner} {et~al.}(2001){Kuntschner}, {Lucey}, {Smith}, {Hudson},
  \& {Davies}}]{Kuntschner2001}
{Kuntschner}, H., {Lucey}, J.~R., {Smith}, R.~J., {Hudson}, M.~J., \& {Davies},
  R.~L. 2001, \href{https://doi.org/10.1046/j.1365-8711.2001.04263.x}{\mnras},
  323, 615, [\href{https://arxiv.org/abs/astro-ph/0011234}{astro-ph/0011234}].

\bibitem[{{La Barbera} {et~al.}(2012){La Barbera}, {Ferreras}, {de Carvalho},
  {Bruzual}, {Charlot}, {Pasquali}, \& {Merlin}}]{LaBarbera2012}
{La Barbera}, F., {Ferreras}, I., {de Carvalho}, R.~R., {et~al.} 2012,
  \href{https://doi.org/10.1111/j.1365-2966.2012.21848.x}{\mnras}, 426, 2300,
  [\href{https://arxiv.org/abs/1208.0587}{1208.0587}].

\bibitem[{{La Barbera} {et~al.}(2004){La Barbera}, {Merluzzi}, {Busarello},
  {Massarotti}, \& {Mercurio}}]{LaBarbera2004}
{La Barbera}, F., {Merluzzi}, P., {Busarello}, G., {Massarotti}, M., \&
  {Mercurio}, A. 2004, \href{https://doi.org/10.1051/0004-6361:20047157}{\aap},
  425, 797, [\href{https://arxiv.org/abs/astro-ph/0307482}{astro-ph/0307482}].

\bibitem[{{Labb{\'e}} {et~al.}(2003){Labb{\'e}}, {Franx}, {Rudnick},
  {F{\"o}rster Schreiber}, {Rix}, {Moorwood}, {van Dokkum}, {van der Werf},
  {R{\"o}ttgering}, {van Starkenburg}, {van der Wel}, {Kuijken}, \&
  {Daddi}}]{Labbe2003}
{Labb{\'e}}, I., {Franx}, M., {Rudnick}, G., {et~al.} 2003,
  \href{https://doi.org/10.1086/346140}{\aj}, 125, 1107,
  [\href{https://arxiv.org/abs/astro-ph/0212236}{astro-ph/0212236}].

\bibitem[{{Lacerda} {et~al.}(2022){Lacerda}, {S{\'a}nchez},
  {Mej{\'\i}a-Narv{\'a}ez}, {Camps-Fari{\~n}a}, {Espinosa-Ponce},
  {Barrera-Ballesteros}, {Ibarra-Medel}, \& {Lugo-Aranda}}]{Lacerda2022}
{Lacerda}, E. A.~D., {S{\'a}nchez}, S.~F., {Mej{\'\i}a-Narv{\'a}ez}, A.,
  {et~al.} 2022, \href{https://doi.org/10.1016/j.newast.2022.101895}{\na}, 97,
  101895, [\href{https://arxiv.org/abs/2202.08027}{2202.08027}].

\bibitem[{{Lacerna} {et~al.}(2022){Lacerna}, {Rodriguez}, {Montero-Dorta},
  {O'Mill}, {Cora}, {Artale}, {Ruiz}, {Hough}, \&
  {Vega-Mart{\'\i}nez}}]{Lacerna2022}
{Lacerna}, I., {Rodriguez}, F., {Montero-Dorta}, A.~D., {et~al.} 2022,
  \href{https://doi.org/10.1093/mnras/stac1020}{\mnras}, 513, 2271,
  [\href{https://arxiv.org/abs/2110.09536}{2110.09536}].

\bibitem[{{Lange} {et~al.}(2015){Lange}, {Driver}, {Robotham}, {Kelvin},
  {Graham}, {Alpaslan}, {Andrews}, {Baldry}, {Bamford}, {Bland-Hawthorn},
  {Brough}, {Cluver}, {Conselice}, {Davies}, {Haeussler}, {Konstantopoulos},
  {Loveday}, {Moffett}, {Norberg}, {Phillipps}, {Taylor},
  {L{\'o}pez-S{\'a}nchez}, \& {Wilkins}}]{Lange2015}
{Lange}, R., {Driver}, S.~P., {Robotham}, A. S.~G., {et~al.} 2015,
  \href{https://doi.org/10.1093/mnras/stu2467}{\mnras}, 447, 2603,
  [\href{https://arxiv.org/abs/1411.6355}{1411.6355}].

\bibitem[{{Larson}(1974)}]{Larson1974}
{Larson}, R.~B. 1974, \href{https://doi.org/10.1093/mnras/169.2.229}{\mnras},
  169, 229.

\bibitem[{{Larson} {et~al.}(1980){Larson}, {Tinsley}, \&
  {Caldwell}}]{Larson1980}
{Larson}, R.~B., {Tinsley}, B.~M., \& {Caldwell}, C.~N. 1980,
  \href{https://doi.org/10.1086/157917}{\apj}, 237, 692.

\bibitem[{{Le F{\`e}vre} {et~al.}(2005){Le F{\`e}vre}, {Vettolani}, {Garilli},
  {Tresse}, {Bottini}, {Le Brun}, {Maccagni}, {Picat}, {Scaramella},
  {Scodeggio}, {Zanichelli}, {Adami}, {Arnaboldi}, {Arnouts}, {Bardelli},
  {Bolzonella}, {Cappi}, {Charlot}, {Ciliegi}, {Contini}, {Foucaud},
  {Franzetti}, {Gavignaud}, {Guzzo}, {Ilbert}, {Iovino}, {McCracken}, {Marano},
  {Marinoni}, {Mathez}, {Mazure}, {Meneux}, {Merighi}, {Paltani}, {Pell{\`o}},
  {Pollo}, {Pozzetti}, {Radovich}, {Zamorani}, {Zucca}, {Bondi}, {Bongiorno},
  {Busarello}, {Lamareille}, {Mellier}, {Merluzzi}, {Ripepi}, \&
  {Rizzo}}]{VVDS2005}
{Le F{\`e}vre}, O., {Vettolani}, G., {Garilli}, B., {et~al.} 2005,
  \href{https://doi.org/10.1051/0004-6361:20041960}{\aap}, 439, 845,
  [\href{https://arxiv.org/abs/astro-ph/0409133}{astro-ph/0409133}].

\bibitem[{{Leja} {et~al.}(2019){Leja}, {Carnall}, {Johnson}, {Conroy}, \&
  {Speagle}}]{leja2019}
{Leja}, J., {Carnall}, A.~C., {Johnson}, B.~D., {Conroy}, C., \& {Speagle},
  J.~S. 2019, \href{https://doi.org/10.3847/1538-4357/ab133c}{\apj}, 876, 3,
  [\href{https://arxiv.org/abs/1811.03637}{1811.03637}].

\bibitem[{{Lewis}(2009)}]{Lewis2009}
{Lewis}, A. 2009,
  \href{https://doi.org/10.1111/j.1365-2966.2009.15161.x}{\mnras}, 398, 471,
  [\href{https://arxiv.org/abs/0901.0649}{0901.0649}].

\bibitem[{{Lewis} {et~al.}(2002){Lewis}, {Balogh}, {De Propris}, {Couch},
  {Bower}, {Offer}, {Bland-Hawthorn}, {Baldry}, {Baugh}, {Bridges}, {Cannon},
  {Cole}, {Colless}, {Collins}, {Cross}, {Dalton}, {Driver}, {Efstathiou},
  {Ellis}, {Frenk}, {Glazebrook}, {Hawkins}, {Jackson}, {Lahav}, {Lumsden},
  {Maddox}, {Madgwick}, {Norberg}, {Peacock}, {Percival}, {Peterson},
  {Sutherland}, \& {Taylor}}]{Lewis2002}
{Lewis}, I., {Balogh}, M., {De Propris}, R., {et~al.} 2002,
  \href{https://doi.org/10.1046/j.1365-8711.2002.05558.x}{\mnras}, 334, 673,
  [\href{https://arxiv.org/abs/astro-ph/0203336}{astro-ph/0203336}].

\bibitem[{{Liao} \& {Cooper}(2023)}]{Liao2023}
{Liao}, L.-W. \& {Cooper}, A.~P. 2023,
  \href{https://doi.org/10.1093/mnras/stac3327}{\mnras}, 518, 3999,
  [\href{https://arxiv.org/abs/2209.14166}{2209.14166}].

\bibitem[{{Lin} {et~al.}(2019{\natexlab{a}}){Lin}, {Hsieh}, {Pan}, {Rembold},
  {S{\'a}nchez}, {Argudo-Fern{\'a}ndez}, {Rowlands}, {Belfiore}, {Bizyaev},
  {Lacerna}, {Riffel}, {Rong}, {Yuan}, {Drory}, {Maiolino}, \&
  {Wilcots}}]{Lin2019b}
{Lin}, L., {Hsieh}, B.-C., {Pan}, H.-A., {et~al.} 2019{\natexlab{a}},
  \href{https://doi.org/10.3847/1538-4357/aafa84}{\apj}, 872, 50,
  [\href{https://arxiv.org/abs/1901.05126}{1901.05126}].

\bibitem[{{Lin} {et~al.}(2019{\natexlab{b}}){Lin}, {Pan}, {Ellison},
  {Belfiore}, {Shi}, {S{\'a}nchez}, {Hsieh}, {Rowlands}, {Ramya}, {Thorp},
  {Li}, \& {Maiolino}}]{Lin2019a}
{Lin}, L., {Pan}, H.-A., {Ellison}, S.~L., {et~al.} 2019{\natexlab{b}},
  \href{https://doi.org/10.3847/2041-8213/ab4815}{\apjl}, 884, L33,
  [\href{https://arxiv.org/abs/1909.11243}{1909.11243}].

\bibitem[{{Lin} {et~al.}(2017){Lin}, {Hsieh}, {Lin}, {Oguri}, {Chen}, {Tanaka},
  {Chiu}, {Huang}, {Kodama}, {Leauthaud}, {More}, {Nishizawa}, {Bundy}, {Lin},
  \& {Miyazaki}}]{Lin2017}
{Lin}, Y.-T., {Hsieh}, B.-C., {Lin}, S.-C., {et~al.} 2017,
  \href{https://doi.org/10.3847/1538-4357/aa9bf5}{\apj}, 851, 139,
  [\href{https://arxiv.org/abs/1709.04484}{1709.04484}].

\bibitem[{{Lisenfeld} {et~al.}(2017){Lisenfeld}, {Alatalo}, {Zucker},
  {Appleton}, {Gallagher}, {Guillard}, \& {Johnson}}]{Lisenfeld2017}
{Lisenfeld}, U., {Alatalo}, K., {Zucker}, C., {et~al.} 2017,
  \href{https://doi.org/10.1051/0004-6361/201730898}{\aap}, 607, A110,
  [\href{https://arxiv.org/abs/1708.09159}{1708.09159}].

\bibitem[{{Liu} {et~al.}(2015){Liu}, {Pan}, {Hao}, {Hoyle}, {Constantin}, \&
  {Vogeley}}]{Liu2015}
{Liu}, C.-X., {Pan}, D.~C., {Hao}, L., {et~al.} 2015,
  \href{https://doi.org/10.1088/0004-637X/810/2/165}{\apj}, 810, 165,
  [\href{https://arxiv.org/abs/1509.04430}{1509.04430}].

\bibitem[{{Liu} {et~al.}(2021){Liu}, {Yee}, {Drissen}, {Sivanandam},
  {Pintos-Castro}, {Alcorn}, {Hsieh}, {Lin}, {Lin}, {Muzzin}, {Noble}, \&
  {Old}}]{Liu2021}
{Liu}, Q., {Yee}, H.~K.~C., {Drissen}, L., {et~al.} 2021,
  \href{https://doi.org/10.3847/1538-4357/abd71e}{\apj}, 908, 228,
  [\href{https://arxiv.org/abs/2101.01887}{2101.01887}].

\bibitem[{{Liz{\'e}e} {et~al.}(2021){Liz{\'e}e}, {Vollmer}, {Braine}, \&
  {Nehlig}}]{Lizee2021}
{Liz{\'e}e}, T., {Vollmer}, B., {Braine}, J., \& {Nehlig}, F. 2021,
  \href{https://doi.org/10.1051/0004-6361/202038910}{\aap}, 645, A111,
  [\href{https://arxiv.org/abs/2011.10531}{2011.10531}].

\bibitem[{{Logro{\~n}o-Garc{\'\i}a} {et~al.}(2019){Logro{\~n}o-Garc{\'\i}a},
  {Vilella-Rojo}, {L{\'o}pez-Sanjuan}, {Varela}, {Viironen}, {Muniesa},
  {Cenarro}, {Crist{\'o}bal-Hornillos}, {Ederoclite}, {Mar{\'\i}n-Franch},
  {Moles}, {V{\'a}zquez Rami{\'o}}, {Bonoli}, {D{\'\i}az-Garc{\'\i}a}, {Orsi},
  {San Roman}, {Akras}, {Chies-Santos}, {Coelho}, {Daflon}, {Costa-Duarte},
  {Dupke}, {Galbany}, {Gonz{\'a}lez Delgado}, {Hernandez-Jimenez}, {Lopes de
  Oliveira}, {Mendes de Oliveira}, {Oteo}, {Gon{\c{c}}alves},
  {S{\'a}nchez-Portal}, {Schmidtobreick}, \& {Sodr{\'e}}}]{Logrono2019}
{Logro{\~n}o-Garc{\'\i}a}, R., {Vilella-Rojo}, G., {L{\'o}pez-Sanjuan}, C.,
  {et~al.} 2019, \href{https://doi.org/10.1051/0004-6361/201732487}{\aap}, 622,
  A180, [\href{https://arxiv.org/abs/1804.04039}{1804.04039}].

\bibitem[{Logroño~García {et~al.}(2023)Logroño~García, López San~Juan, \&
  Varela~López}]{TesisLogrono}
Logroño~García, R., López San~Juan, C., \& Varela~López, J.~M. 2023, PhD
  thesis, University of Zaragoza, Spain, presentado: 12 06 2023

\bibitem[{{Loh} \& {Spillar}(1986)}]{Loh1986}
{Loh}, E.~D. \& {Spillar}, E.~J. 1986,
  \href{https://doi.org/10.1086/164062}{\apj}, 303, 154.

\bibitem[{{{\L}okas}(2023)}]{Lokas2023}
{{\L}okas}, E.~L. 2023,
  \href{https://doi.org/10.1051/0004-6361/202347735}{\aap}, 678, A147,
  [\href{https://arxiv.org/abs/2309.07494}{2309.07494}].

\bibitem[{{Lopes} {et~al.}(2024){Lopes}, {Ribeiro}, \& {Brambila}}]{Lopes2024}
{Lopes}, P. A.~A., {Ribeiro}, A. L.~B., \& {Brambila}, D. 2024,
  \href{https://doi.org/10.1093/mnrasl/slad134}{\mnras}, 527, L19,
  [\href{https://arxiv.org/abs/2309.11578}{2309.11578}].

\bibitem[{{Lopes} {et~al.}(2014){Lopes}, {Ribeiro}, \& {Rembold}}]{Lopes2014}
{Lopes}, P.~A.~A., {Ribeiro}, A.~L.~B., \& {Rembold}, S.~B. 2014,
  \href{https://doi.org/10.1093/mnras/stt2064}{\mnras}, 437, 2430,
  [\href{https://arxiv.org/abs/1310.6309}{1310.6309}].

\bibitem[{{L{\'o}pez Fern{\'a}ndez}(2017)}]{RafaTesis}
{L{\'o}pez Fern{\'a}ndez}, R. 2017, PhD thesis, University of Granada, Spain

\bibitem[{{L{\'o}pez Fern{\'a}ndez} {et~al.}(2016){L{\'o}pez Fern{\'a}ndez},
  {Cid Fernandes}, {Gonz{\'a}lez Delgado}, {Vale Asari}, {P{\'e}rez},
  {Garc{\'\i}a-Benito}, {de Amorim}, {Lacerda}, {Cortijo-Ferrero}, \&
  {S{\'a}nchez}}]{Rafa2016}
{L{\'o}pez Fern{\'a}ndez}, R., {Cid Fernandes}, R., {Gonz{\'a}lez Delgado},
  R.~M., {et~al.} 2016, \href{https://doi.org/10.1093/mnras/stw260}{\mnras},
  458, 184, [\href{https://arxiv.org/abs/1602.01123}{1602.01123}].

\bibitem[{{L{\'o}pez Fern{\'a}ndez} {et~al.}(2018){L{\'o}pez Fern{\'a}ndez},
  {Gonz{\'a}lez Delgado}, {P{\'e}rez}, {Garc{\'\i}a-Benito}, {Cid Fernandes},
  {Schoenell}, {S{\'a}nchez}, {Gallazzi}, {S{\'a}nchez-Bl{\'a}zquez}, {Vale
  Asari}, \& {Walcher}}]{lopez-fernandez2018}
{L{\'o}pez Fern{\'a}ndez}, R., {Gonz{\'a}lez Delgado}, R.~M., {P{\'e}rez}, E.,
  {et~al.} 2018, \href{https://doi.org/10.1051/0004-6361/201732358}{\aap}, 615,
  A27, [\href{https://arxiv.org/abs/1802.10118}{1802.10118}].

\bibitem[{{L{\'o}pez-Sanjuan} {et~al.}(2019{\natexlab{a}}){L{\'o}pez-Sanjuan},
  {D{\'\i}az-Garc{\'\i}a}, {Cenarro}, {Fern{\'a}ndez-Soto}, {Viironen},
  {Molino}, {Ben{\'\i}tez}, {Crist{\'o}bal-Hornillos}, {Moles}, {Varela},
  {Arnalte-Mur}, {Ascaso}, {Castander}, {Cervi{\~n}o}, {Gonz{\'a}lez Delgado},
  {Husillos}, {M{\'a}rquez}, {Masegosa}, {Del Olmo}, {Povi{\'c}}, \&
  {Perea}}]{LopezSanJuan2019ML}
{L{\'o}pez-Sanjuan}, C., {D{\'\i}az-Garc{\'\i}a}, L.~A., {Cenarro}, A.~J.,
  {et~al.} 2019{\natexlab{a}},
  \href{https://doi.org/10.1051/0004-6361/201833402}{\aap}, 622, A51,
  [\href{https://arxiv.org/abs/1805.03609}{1805.03609}].

\bibitem[{{L{\'o}pez-Sanjuan} {et~al.}(2019{\natexlab{b}}){L{\'o}pez-Sanjuan},
  {Varela}, {Crist{\'o}bal-Hornillos}, {V{\'a}zquez Rami{\'o}}, {Carrasco},
  {Tremblay}, {Whitten}, {Placco}, {Mar{\'\i}n-Franch}, {Cenarro},
  {Ederoclite}, {Alfaro}, {Coelho}, {Civera}, {Hern{\'a}ndez-Fuertes},
  {Jim{\'e}nez-Esteban}, {Jim{\'e}nez-Teja}, {Ma{\'\i}z Apell{\'a}niz},
  {Sobral}, {V{\'\i}lchez}, {Alcaniz}, {Angulo}, {Dupke},
  {Hern{\'a}ndez-Monteagudo}, {Mendes de Oliveira}, {Moles}, \&
  {Sodr{\'e}}}]{Lopez-sanJuan2019}
{L{\'o}pez-Sanjuan}, C., {Varela}, J., {Crist{\'o}bal-Hornillos}, D., {et~al.}
  2019{\natexlab{b}}, \href{https://doi.org/10.1051/0004-6361/201936405}{\aap},
  631, A119, [\href{https://arxiv.org/abs/1907.12939}{1907.12939}].

\bibitem[{{L{\'o}pez-Sanjuan} {et~al.}(2024){L{\'o}pez-Sanjuan}, {V{\'a}zquez
  Rami{\'o}}, {Xiao}, {Yuan}, {Carrasco}, {Varela}, {Crist{\'o}bal-Hornillos},
  {Tremblay}, {Ederoclite}, {Mar{\'\i}n-Franch}, {Cenarro}, {Coelho}, {Daflon},
  {del Pino}, {Dom{\'\i}nguez S{\'a}nchez}, {Fern{\'a}ndez-Ontiveros},
  {Hern{\'a}n-Caballero}, {Jim{\'e}nez-Esteban}, {Alcaniz}, {Angulo}, {Dupke},
  {Hern{\'a}ndez-Monteagudo}, {Moles}, \& {Sodr{\'e}}}]{LopezSanJuan2023}
{L{\'o}pez-Sanjuan}, C., {V{\'a}zquez Rami{\'o}}, H., {Xiao}, K., {et~al.}
  2024, \href{https://doi.org/10.1051/0004-6361/202346012}{\aap}, 683, A29,
  [\href{https://arxiv.org/abs/2301.12395}{2301.12395}].

\bibitem[{{Lower} {et~al.}(2020){Lower}, {Narayanan}, {Leja}, {Johnson},
  {Conroy}, \& {Dav{\'e}}}]{lower2020}
{Lower}, S., {Narayanan}, D., {Leja}, J., {et~al.} 2020,
  \href{https://doi.org/10.3847/1538-4357/abbfa7}{\apj}, 904, 33,
  [\href{https://arxiv.org/abs/2006.03599}{2006.03599}].

\bibitem[{{Ly} {et~al.}(2007){Ly}, {Malkan}, {Kashikawa}, {Shimasaku}, {Doi},
  {Nagao}, {Iye}, {Kodama}, {Morokuma}, \& {Motohara}}]{Ly2007}
{Ly}, C., {Malkan}, M.~A., {Kashikawa}, N., {et~al.} 2007,
  \href{https://doi.org/10.1086/510828}{\apj}, 657, 738,
  [\href{https://arxiv.org/abs/astro-ph/0610846}{astro-ph/0610846}].

\bibitem[{{Ma} {et~al.}(2016){Ma}, {Hopkins}, {Faucher-Gigu{\`e}re}, {Zolman},
  {Muratov}, {Kere{\v{s}}}, \& {Quataert}}]{Xiangcheng2016}
{Ma}, X., {Hopkins}, P.~F., {Faucher-Gigu{\`e}re}, C.-A., {et~al.} 2016,
  \href{https://doi.org/10.1093/mnras/stv2659}{\mnras}, 456, 2140,
  [\href{https://arxiv.org/abs/1504.02097}{1504.02097}].

\bibitem[{{MacArthur} {et~al.}(2004){MacArthur}, {Courteau}, {Bell}, \&
  {Holtzman}}]{McArthur2004}
{MacArthur}, L.~A., {Courteau}, S., {Bell}, E., \& {Holtzman}, J.~A. 2004,
  \href{https://doi.org/10.1086/383525}{\apjs}, 152, 175,
  [\href{https://arxiv.org/abs/astro-ph/0401437}{astro-ph/0401437}].

\bibitem[{{Madau} {et~al.}(1998){Madau}, {Pozzetti}, \&
  {Dickinson}}]{Madau1998}
{Madau}, P., {Pozzetti}, L., \& {Dickinson}, M. 1998,
  \href{https://doi.org/10.1086/305523}{\apj}, 498, 106,
  [\href{https://arxiv.org/abs/astro-ph/9708220}{astro-ph/9708220}].

\bibitem[{{Mahajan}(2013)}]{Mahajan2013}
{Mahajan}, S. 2013, \href{https://doi.org/10.1093/mnrasl/slt021}{\mnras}, 431,
  L117, [\href{https://arxiv.org/abs/1302.2905}{1302.2905}].

\bibitem[{{Maier} {et~al.}(2019){Maier}, {Ziegler}, {Haines}, \&
  {Smith}}]{Maier2019}
{Maier}, C., {Ziegler}, B.~L., {Haines}, C.~P., \& {Smith}, G.~P. 2019,
  \href{https://doi.org/10.1051/0004-6361/201834290}{\aap}, 621, A131,
  [\href{https://arxiv.org/abs/1809.07675}{1809.07675}].

\bibitem[{{Malumuth} \& {Richstone}(1984)}]{Malumuth1984}
{Malumuth}, E.~M. \& {Richstone}, D.~O. 1984,
  \href{https://doi.org/10.1086/161626}{\apj}, 276, 413.

\bibitem[{{Mamon} {et~al.}(2013){Mamon}, {Biviano}, \& {Bou{\'e}}}]{CLEAN}
{Mamon}, G.~A., {Biviano}, A., \& {Bou{\'e}}, G. 2013,
  \href{https://doi.org/10.1093/mnras/sts565}{\mnras}, 429, 3079,
  [\href{https://arxiv.org/abs/1212.1455}{1212.1455}].

\bibitem[{{Marian} {et~al.}(2018){Marian}, {Ziegler}, {Kuchner}, \&
  {Verdugo}}]{Marian2018}
{Marian}, V., {Ziegler}, B., {Kuchner}, U., \& {Verdugo}, M. 2018,
  \href{https://doi.org/10.1051/0004-6361/201832750}{\aap}, 617, A34,
  [\href{https://arxiv.org/abs/1806.10625}{1806.10625}].

\bibitem[{{Mar{\'\i}n-Franch} {et~al.}(2012){Mar{\'\i}n-Franch}, {Chueca},
  {Moles}, {Benitez}, {Taylor}, {Cepa}, {Cenarro}, {Cristobal-Hornillos},
  {Ederoclite}, {Gruel}, {Hern{\'a}ndez-Fuertes}, {L{\'o}pez-Sainz},
  {Luis-Simoes}, {Rueda-Teruel}, {Rueda-Teruel}, {Varela}, {Yanes-D{\'\i}az},
  {Brauneck}, {Danielou}, {Dupke}, {Fern{\'a}ndez-Soto}, {Mendes de Oliveira},
  \& {Sodr{\'e}}}]{Martin-Franch2012}
{Mar{\'\i}n-Franch}, A., {Chueca}, S., {Moles}, M., {et~al.} 2012, in Society
  of Photo-Optical Instrumentation Engineers (SPIE) Conference Series, Vol.
  8450, Modern Technologies in Space- and Ground-based Telescopes and
  Instrumentation II, ed. R.~{Navarro}, C.~R. {Cunningham}, \& E.~{Prieto},
  84503S

\bibitem[{{Mar{\'\i}n-Franch} {et~al.}(2017){Mar{\'\i}n-Franch}, {Taylor},
  {Santoro}, {Laporte}, {Cepa}, {Lasso-Cabrera}, {Yanes-D{\'\i}az}, {Cenarro},
  {Cristobal-Hornillos}, {Ederoclite}, {Moles}, {Varela}, {V{\'a}zquez
  Rami{\'o}}, {Ant{\'o}n}, {Bello}, {Civera}, {Castillo}, {Chueca},
  {Guill{\'e}n-Civera}, {Hern{\'a}ndez-Fuertes}, {Igual}, {{\'I}niguez},
  {L{\'o}pez-Alegre}, {L{\'o}pez-Sainz}, {Nevot}, {Milla}, {Muniesa}, {Pe},
  {Rueda-Teruel}, {Rueda-Teruel}, {S{\'a}nchez}, {Benitez}, {Dupke},
  {Fern{\'a}ndez-Soto}, {Mendes de Oliveira}, \& {Sodr{\'e}}}]{MarinFranch2017}
{Mar{\'\i}n-Franch}, A., {Taylor}, K., {Santoro}, F.~G., {et~al.} 2017, in
  Highlights on Spanish Astrophysics IX, ed. S.~{Arribas}, A.~{Alonso-Herrero},
  F.~{Figueras}, C.~{Hern{\'a}ndez-Monteagudo}, A.~{S{\'a}nchez-Lavega}, \&
  S.~{P{\'e}rez-Hoyos}, 670--675

\bibitem[{{M{\'a}rmol-Queralt{\'o}} {et~al.}(2016){M{\'a}rmol-Queralt{\'o}},
  {McLure}, {Cullen}, {Dunlop}, {Fontana}, \& {McLeod}}]{Marmolqueralto2016}
{M{\'a}rmol-Queralt{\'o}}, E., {McLure}, R.~J., {Cullen}, F., {et~al.} 2016,
  \href{https://doi.org/10.1093/mnras/stw1212}{\mnras}, 460, 3587,
  [\href{https://arxiv.org/abs/1511.01911}{1511.01911}].

\bibitem[{{Mart{\'\i}nez-Solaeche} {et~al.}(2021){Mart{\'\i}nez-Solaeche},
  {Gonz{\'a}lez Delgado}, {Garc{\'\i}a-Benito}, {de Amorim}, {P{\'e}rez},
  {Rodr{\'\i}guez-Mart{\'\i}n}, {D{\'\i}az-Garc{\'\i}a}, {Cid Fernandes},
  {L{\'o}pez-Sanjuan}, {Bonoli}, {Cenarro}, {Dupke}, {Mar{\'\i}n-Franch},
  {Varela}, {V{\'a}zquez Rami{\'o}}, {Abramo}, {Crist{\'o}bal-Hornillos},
  {Moles}, {Alcaniz}, {Baqui}, {Benitez}, {Carneiro}, {Cortesi}, {Ederoclite},
  {Marra}, {Mendes de Oliveira}, {Sodr{\'e}}, {V{\'\i}lchez}, \&
  {Taylor}}]{Gines2021}
{Mart{\'\i}nez-Solaeche}, G., {Gonz{\'a}lez Delgado}, R.~M.,
  {Garc{\'\i}a-Benito}, R., {et~al.} 2021,
  \href{https://doi.org/10.1051/0004-6361/202039146}{\aap}, 647, A158,
  [\href{https://arxiv.org/abs/2008.04287}{2008.04287}].

\bibitem[{{Mart{\'\i}nez-Solaeche} {et~al.}(2022){Mart{\'\i}nez-Solaeche},
  {Gonz{\'a}lez Delgado}, {Garc{\'\i}a-Benito}, {D{\'\i}az-Garc{\'\i}a},
  {Rodr{\'\i}guez-Mart{\'\i}n}, {P{\'e}rez}, {de Amorim}, {Duarte Puertas},
  {Sodr{\'e}}, {Sobral}, {Chaves-Montero}, {V{\'\i}lchez},
  {Hern{\'a}n-Caballero}, {L{\'o}pez-Sanjuan}, {Cortesi}, {Bonoli}, {Cenarro},
  {Dupke}, {Mar{\'\i}n-Franch}, {Varela}, {V{\'a}zquez Rami{\'o}}, {Abramo},
  {Crist{\'o}bal-Hornillos}, {Moles}, {Alcaniz}, {Benitez}, {Ederoclite},
  {Marra}, {Mendes de Oliveira}, {Taylor}, \&
  {Fern{\'a}ndez-Ontiveros}}]{Gines2022}
{Mart{\'\i}nez-Solaeche}, G., {Gonz{\'a}lez Delgado}, R.~M.,
  {Garc{\'\i}a-Benito}, R., {et~al.} 2022,
  \href{https://doi.org/10.1051/0004-6361/202142812}{\aap}, 661, A99,
  [\href{https://arxiv.org/abs/2204.01698}{2204.01698}].

\bibitem[{{Massari} {et~al.}(2020){Massari}, {Marasco}, {Beltramo-Martin},
  {Milli}, {Fiorentino}, {Tolstoy}, \& {Kerber}}]{Massari2020}
{Massari}, D., {Marasco}, A., {Beltramo-Martin}, O., {et~al.} 2020,
  \href{https://doi.org/10.1051/0004-6361/201937359}{\aap}, 634, L5,
  [\href{https://arxiv.org/abs/2001.08134}{2001.08134}].

\bibitem[{{Mateus} {et~al.}(2007){Mateus}, {Sodr{\'e}}, {Cid Fernandes}, \&
  {Stasi{\'n}ska}}]{Mateus2007}
{Mateus}, A., {Sodr{\'e}}, L., {Cid Fernandes}, R., \& {Stasi{\'n}ska}, G.
  2007, \href{https://doi.org/10.1111/j.1365-2966.2006.11290.x}{\mnras}, 374,
  1457, [\href{https://arxiv.org/abs/astro-ph/0604063}{astro-ph/0604063}].

\bibitem[{{Mateus} {et~al.}(2006){Mateus}, {Sodr{\'e}}, {Cid Fernandes},
  {Stasi{\'n}ska}, {Schoenell}, \& {Gomes}}]{Mateus2006}
{Mateus}, A., {Sodr{\'e}}, L., {Cid Fernandes}, R., {et~al.} 2006,
  \href{https://doi.org/10.1111/j.1365-2966.2006.10565.x}{\mnras}, 370, 721,
  [\href{https://arxiv.org/abs/astro-ph/0511578}{astro-ph/0511578}].

\bibitem[{{Mathis} {et~al.}(2006){Mathis}, {Charlot}, \&
  {Brinchmann}}]{Mathis2006}
{Mathis}, H., {Charlot}, S., \& {Brinchmann}, J. 2006,
  \href{https://doi.org/10.1111/j.1365-2966.2005.09790.x}{\mnras}, 365, 385,
  [\href{https://arxiv.org/abs/astro-ph/0511203}{astro-ph/0511203}].

\bibitem[{{Maturi} {et~al.}(2005{\natexlab{a}}){Maturi}, {Bartelmann},
  {Meneghetti}, \& {Moscardini}}]{2005A&A...436...37M}
{Maturi}, M., {Bartelmann}, M., {Meneghetti}, M., \& {Moscardini}, L.
  2005{\natexlab{a}}, \href{https://doi.org/10.1051/0004-6361:20041785}{\aap},
  436, 37,
  [\href{https://arxiv.org/abs/arXiv:astro-ph/0408064}{arXiv:astro-ph/0408064}].

\bibitem[{{Maturi} {et~al.}(2019){Maturi}, {Bellagamba}, {Radovich},
  {Roncarelli}, {Sereno}, {Moscardini}, {Bardelli}, \&
  {Puddu}}]{Bellagamba2019}
{Maturi}, M., {Bellagamba}, F., {Radovich}, M., {et~al.} 2019,
  \href{https://doi.org/10.1093/mnras/stz294}{\mnras}, 485, 498,
  [\href{https://arxiv.org/abs/1810.02811}{1810.02811}].

\bibitem[{{Maturi} {et~al.}(2023){Maturi}, {Finoguenov}, {Lopes}, {Gonz{\'a}lez
  Delgado}, {Dupke}, {Cypriano}, {Carrasco}, {Diego}, {Penna-Lima}, {Doubrawa},
  {V{\'\i}lchez}, {Moscardini}, {Marra}, {Bonoli},
  {Rodr{\'\i}guez-Mart{\'\i}n}, {Zitrin}, {M{\'a}rquez},
  {Hern{\'a}n-Caballero}, {Jim{\'e}nez-Teja}, {Abramo}, {Alcaniz}, {Benitez},
  {Carneiro}, {Cenarro}, {Crist{\'o}bal-Hornillos}, {Ederoclite},
  {L{\'o}pez-Sanjuan}, {Mar{\'\i}n-Franch}, {Mendes de Oliveira}, {Moles},
  {Sodr{\'e}}, {Taylor}, {Varela}, {V{\'a}zquez Rami{\'o}}, \&
  {Fern{\'a}ndez-Ontiveros}}]{Maturi2023}
{Maturi}, M., {Finoguenov}, A., {Lopes}, P.~A.~A., {et~al.} 2023,
  \href{https://doi.org/10.1051/0004-6361/202245323}{\aap}, 678, A145,
  [\href{https://arxiv.org/abs/2307.06412}{2307.06412}].

\bibitem[{{Maturi} {et~al.}(2005{\natexlab{b}}){Maturi}, {Meneghetti},
  {Bartelmann}, {Dolag}, \& {Moscardini}}]{Maturi2005}
{Maturi}, M., {Meneghetti}, M., {Bartelmann}, M., {Dolag}, K., \& {Moscardini},
  L. 2005{\natexlab{b}},
  \href{https://doi.org/10.1051/0004-6361:20042600}{\aap}, 442, 851,
  [\href{https://arxiv.org/abs/astro-ph/0412604}{astro-ph/0412604}].

\bibitem[{{Mazzi} {et~al.}(2021){Mazzi}, {Girardi}, {Zaggia}, {Pastorelli},
  {Rubele}, {Bressan}, {Cioni}, {Clementini}, {Cusano}, {Rocha}, {Gullieuszik},
  {Kerber}, {Marigo}, {Ripepi}, {Bekki}, {Bell}, {de Grijs}, {Groenewegen},
  {Ivanov}, {Oliveira}, {Sun}, \& {van Loon}}]{Mazzi2021}
{Mazzi}, A., {Girardi}, L., {Zaggia}, S., {et~al.} 2021,
  \href{https://doi.org/10.1093/mnras/stab2399}{\mnras}, 508, 245,
  [\href{https://arxiv.org/abs/2108.07225}{2108.07225}].

\bibitem[{{McNab} {et~al.}(2021){McNab}, {Balogh}, {van der Burg}, {Forestell},
  {Webb}, {Vulcani}, {Rudnick}, {Muzzin}, {Cooper}, {McGee}, {Biviano},
  {Cerulo}, {Chan}, {De Lucia}, {Demarco}, {Finoguenov}, {Forrest}, {Golledge},
  {Jablonka}, {Lidman}, {Nantais}, {Old}, {Pintos-Castro}, {Poggianti},
  {Reeves}, {Wilson}, {Yee}, \& {Zaritsky}}]{McNab2021}
{McNab}, K., {Balogh}, M.~L., {van der Burg}, R. F.~J., {et~al.} 2021,
  \href{https://doi.org/10.1093/mnras/stab2558}{\mnras}, 508, 157,
  [\href{https://arxiv.org/abs/2109.03105}{2109.03105}].

\bibitem[{{McNamara} \& {Nulsen}(2007)}]{Mcnamara2007}
{McNamara}, B.~R. \& {Nulsen}, P.~E.~J. 2007,
  \href{https://doi.org/10.1146/annurev.astro.45.051806.110625}{\araa}, 45,
  117, [\href{https://arxiv.org/abs/0709.2152}{0709.2152}].

\bibitem[{{McNamara} \& {Nulsen}(2012)}]{Mcnamara2012}
{McNamara}, B.~R. \& {Nulsen}, P.~E.~J. 2012,
  \href{https://doi.org/10.1088/1367-2630/14/5/055023}{New Journal of Physics},
  14, 055023, [\href{https://arxiv.org/abs/1204.0006}{1204.0006}].

\bibitem[{{Medling} {et~al.}(2018){Medling}, {Cortese}, {Croom}, {Green},
  {Groves}, {Hampton}, {Ho}, {Davies}, {Kewley}, {Moffett}, {Schaefer},
  {Taylor}, {Zafar}, {Bekki}, {Bland-Hawthorn}, {Bloom}, {Brough}, {Bryant},
  {Catinella}, {Cecil}, {Colless}, {Couch}, {Drinkwater}, {Driver},
  {Federrath}, {Foster}, {Goldstein}, {Goodwin}, {Hopkins}, {Lawrence},
  {Leslie}, {Lewis}, {Lorente}, {Owers}, {McDermid}, {Richards}, {Sharp},
  {Scott}, {Sweet}, {Taranu}, {Tescari}, {Tonini}, {van de Sande}, {Walcher},
  \& {Wright}}]{Medling2018}
{Medling}, A.~M., {Cortese}, L., {Croom}, S.~M., {et~al.} 2018,
  \href{https://doi.org/10.1093/mnras/sty127}{\mnras}, 475, 5194,
  [\href{https://arxiv.org/abs/1801.04283}{1801.04283}].

\bibitem[{{Mehlert} {et~al.}(2003){Mehlert}, {Thomas}, {Saglia}, {Bender}, \&
  {Wegner}}]{Mehlert2003}
{Mehlert}, D., {Thomas}, D., {Saglia}, R.~P., {Bender}, R., \& {Wegner}, G.
  2003, \href{https://doi.org/10.1051/0004-6361:20030886}{\aap}, 407, 423,
  [\href{https://arxiv.org/abs/astro-ph/0306219}{astro-ph/0306219}].

\bibitem[{{Menanteau} {et~al.}(2004){Menanteau}, {Ford}, {Illingworth},
  {Sirianni}, {Blakeslee}, {Meurer}, {Martel}, {Ben{\'\i}tez}, {Postman},
  {Franx}, {Ardila}, {Bartko}, {Bouwens}, {Broadhurst}, {Brown}, {Burrows},
  {Cheng}, {Clampin}, {Cross}, {Feldman}, {Golimowski}, {Gronwall}, {Hartig},
  {Infante}, {Kimble}, {Krist}, {Lesser}, {Miley}, {Rosati}, {Sparks}, {Tran},
  {Tsvetanov}, {White}, \& {Zheng}}]{Menanteau2004}
{Menanteau}, F., {Ford}, H.~C., {Illingworth}, G.~D., {et~al.} 2004,
  \href{https://doi.org/10.1086/422568}{\apj}, 612, 202,
  [\href{https://arxiv.org/abs/astro-ph/0405326}{astro-ph/0405326}].

\bibitem[{{Mendel} {et~al.}(2009){Mendel}, {Proctor}, {Rasmussen}, {Brough}, \&
  {Forbes}}]{Mendel2009}
{Mendel}, J.~T., {Proctor}, R.~N., {Rasmussen}, J., {Brough}, S., \& {Forbes},
  D.~A. 2009, \href{https://doi.org/10.1111/j.1365-2966.2009.14689.x}{\mnras},
  396, 2103, [\href{https://arxiv.org/abs/0903.1092}{0903.1092}].

\bibitem[{{Mendes de Oliveira} {et~al.}(2019){Mendes de Oliveira}, {Ribeiro},
  {Schoenell}, {Kanaan}, {Overzier}, {Molino}, {Sampedro}, {Coelho}, {Barbosa},
  {Cortesi}, {Costa-Duarte}, {Herpich}, {Hernandez-Jimenez}, {Placco},
  {Xavier}, {Abramo}, {Saito}, {Chies-Santos}, {Ederoclite}, {Lopes de
  Oliveira}, {Gon{\c{c}}alves}, {Akras}, {Almeida}, {Almeida-Fernandes},
  {Beers}, {Bonatto}, {Bonoli}, {Cypriano}, {Vinicius-Lima}, {de Souza},
  {Fabiano de Souza}, {Ferrari}, {Gon{\c{c}}alves}, {Gonzalez},
  {Guti{\'e}rrez-Soto}, {Hartmann}, {Jaffe}, {Kerber}, {Lima-Dias}, {Lopes},
  {Menendez-Delmestre}, {Nakazono}, {Novais}, {Ortega-Minakata}, {Pereira},
  {Perottoni}, {Queiroz}, {Reis}, {Santos}, {Santos-Silva}, {Santucci},
  {Barbosa}, {Siffert}, {Sodr{\'e}}, {Torres-Flores}, {Westera}, {Whitten},
  {Alcaniz}, {Alonso-Garc{\'\i}a}, {Alencar}, {Alvarez-Candal}, {Amram},
  {Azanha}, {Barb{\'a}}, {Bernardinelli}, {Borges Fernandes}, {Branco},
  {Brito-Silva}, {Buzzo}, {Caffer}, {Campillay}, {Cano}, {Carvano}, {Castejon},
  {Cid Fernandes}, {Dantas}, {Daflon}, {Damke}, {de la Reza}, {de Melo de
  Azevedo}, {De Paula}, {Diem}, {Donnerstein}, {Dors}, {Dupke}, {Eikenberry},
  {Escudero}, {Faifer}, {Far{\'\i}as}, {Fernandes}, {Fernandes}, {Fontes},
  {Galarza}, {Hirata}, {Katena}, {Gregorio-Hetem},
  {Hern{\'a}ndez-Fern{\'a}ndez}, {Izzo}, {Jaque Arancibia}, {Jatenco-Pereira},
  {Jim{\'e}nez-Teja}, {Kann}, {Krabbe}, {Labayru}, {Lazzaro}, {Lima Neto},
  {Lopes}, {Magalh{\~a}es}, {Makler}, {de Menezes}, {Miralda-Escud{\'e}},
  {Monteiro-Oliveira}, {Montero-Dorta}, {Mu{\~n}oz-Elgueta}, {Nemmen}, {Nilo
  Castell{\'o}n}, {Oliveira}, {Ort{\'\i}z}, {Pattaro}, {Pereira}, {Quint},
  {Riguccini}, {Rocha Pinto}, {Rodrigues}, {Roig}, {Rossi}, {Saha}, {Santos},
  {Schnorr M{\"u}ller}, {Sesto}, {Silva}, {Smith Castelli}, {Teixeira},
  {Telles}, {Thom de Souza}, {Th{\"o}ne}, {Trevisan}, {de Ugarte Postigo},
  {Urrutia-Viscarra}, {Veiga}, {Vika}, {Vitorelli}, {Werle}, {Werner}, \&
  {Zaritsky}}]{SPLUS2019}
{Mendes de Oliveira}, C., {Ribeiro}, T., {Schoenell}, W., {et~al.} 2019,
  \href{https://doi.org/10.1093/mnras/stz1985}{\mnras}, 489, 241,
  [\href{https://arxiv.org/abs/1907.01567}{1907.01567}].

\bibitem[{{Mercurio} {et~al.}(2021){Mercurio}, {Rosati}, {Biviano},
  {Annunziatella}, {Girardi}, {Sartoris}, {Nonino}, {Brescia}, {Riccio},
  {Grillo}, {Balestra}, {Caminha}, {De Lucia}, {Gobat}, {Seitz}, {Tozzi},
  {Scodeggio}, {Vanzella}, {Angora}, {Bergamini}, {Borgani}, {Demarco},
  {Meneghetti}, {Strazzullo}, {Tortorelli}, {Umetsu}, {Fritz}, {Gruen},
  {Kelson}, {Lombardi}, {Maier}, {Postman}, {Rodighiero}, \&
  {Ziegler}}]{Mercurio2021}
{Mercurio}, A., {Rosati}, P., {Biviano}, A., {et~al.} 2021,
  \href{https://doi.org/10.1051/0004-6361/202142168}{\aap}, 656, A147,
  [\href{https://arxiv.org/abs/2109.03305}{2109.03305}].

\bibitem[{{Michard}(2002)}]{Michard2002}
{Michard}, R. 2002, \href{https://doi.org/10.1051/0004-6361:20011813}{\aap},
  384, 763.

\bibitem[{{Miller} {et~al.}(2023){Miller}, {van Dokkum}, \&
  {Mowla}}]{Miller2023}
{Miller}, T.~B., {van Dokkum}, P., \& {Mowla}, L. 2023,
  \href{https://doi.org/10.3847/1538-4357/acbc74}{\apj}, 945, 155,
  [\href{https://arxiv.org/abs/2207.05895}{2207.05895}].

\bibitem[{{Moles} {et~al.}(2008){Moles}, {Aguerri}, {Alfaro}, {Ben{\'\i}tez},
  {Broadhurst}, {Cabrera-Ca{\~n}o}, {Castander}, {Cepa}, {Cervi{\~n}o},
  {Fern{\'a}ndez Soto}, {Gonz{\'a}lez Delgado}, {Infante}, {Mart{\'\i}nez},
  {Masegosa}, {M{\'a}rquez}, {Del Olmo}, {Perea}, {Prada}, {Quintana}, \&
  {S{\'a}nchez}}]{Moles2008}
{Moles}, M., {Aguerri}, J.~A.~L., {Alfaro}, E.~J., {et~al.} 2008, in
  Astronomical Society of the Pacific Conference Series, Vol. 390, Pathways
  Through an Eclectic Universe, ed. J.~H. {Knapen}, T.~J. {Mahoney}, \&
  A.~{Vazdekis}, 495

\bibitem[{{Moles} {et~al.}(2010){Moles}, {S{\'a}nchez}, {Lamadrid}, {Cenarro},
  {Crist{\'o}bal-Hornillos}, {Maicas}, \& {Aceituno}}]{Moles2010}
{Moles}, M., {S{\'a}nchez}, S.~F., {Lamadrid}, J.~L., {et~al.} 2010,
  \href{https://doi.org/10.1086/651084}{\pasp}, 122, 363,
  [\href{https://arxiv.org/abs/0912.3762}{0912.3762}].

\bibitem[{{Molino} {et~al.}(2014){Molino}, {Ben{\'\i}tez}, {Moles},
  {Fern{\'a}ndez-Soto}, {Crist{\'o}bal-Hornillos}, {Ascaso},
  {Jim{\'e}nez-Teja}, {Schoenell}, {Arnalte-Mur}, {Povi{\'c}}, {Coe},
  {L{\'o}pez-Sanjuan}, {D{\'\i}az-Garc{\'\i}a}, {Varela}, {Stefanon},
  {Cenarro}, {Matute}, {Masegosa}, {M{\'a}rquez}, {Perea}, {Del Olmo},
  {Husillos}, {Alfaro}, {Aparicio-Villegas}, {Cervi{\~n}o}, {Huertas-Company},
  {Aguerri}, {Broadhurst}, {Cabrera-Ca{\~n}o}, {Cepa}, {Gonz{\'a}lez},
  {Infante}, {Mart{\'\i}nez}, {Prada}, \& {Quintana}}]{Molino2014}
{Molino}, A., {Ben{\'\i}tez}, N., {Moles}, M., {et~al.} 2014,
  \href{https://doi.org/10.1093/mnras/stu387}{\mnras}, 441, 2891,
  [\href{https://arxiv.org/abs/1306.4968}{1306.4968}].

\bibitem[{{Molino} {et~al.}(2019){Molino}, {Costa-Duarte}, {Mendes de
  Oliveira}, {Cenarro}, {Lima Neto}, {Cypriano}, {Sodr{\'e}}, {Coelho},
  {Chow-Mart{\'\i}nez}, {Monteiro-Oliveira}, {Sampedro}, {Cristobal-Hornillos},
  {Varela}, {Ederoclite}, {Chies-Santos}, {Schoenell}, {Ribeiro},
  {Mar{\'\i}n-Franch}, {L{\'o}pez-Sanjuan}, {Hern{\'a}ndez-Fern{\'a}ndez},
  {Cortesi}, {V{\'a}zquez Rami{\'o}}, {Santos}, {Cibirka}, {Novais}, {Pereira},
  {Hern{\'a}ndez-Jimenez}, {Jimenez-Teja}, {Moles}, {Ben{\'\i}tez}, \&
  {Dupke}}]{Molino2019}
{Molino}, A., {Costa-Duarte}, M.~V., {Mendes de Oliveira}, C., {et~al.} 2019,
  \href{https://doi.org/10.1051/0004-6361/201731348}{\aap}, 622, A178,
  [\href{https://arxiv.org/abs/1804.03640}{1804.03640}].

\bibitem[{{Montero-Dorta} {et~al.}(2021){Montero-Dorta}, {Chaves-Montero},
  {Artale}, \& {Favole}}]{MonteroDorta2021}
{Montero-Dorta}, A.~D., {Chaves-Montero}, J., {Artale}, M.~C., \& {Favole}, G.
  2021, \href{https://doi.org/10.1093/mnras/stab2556}{\mnras}, 508, 940,
  [\href{https://arxiv.org/abs/2105.05274}{2105.05274}].

\bibitem[{{Moore} {et~al.}(1996){Moore}, {Katz}, {Lake}, {Dressler}, \&
  {Oemler}}]{Moore1996}
{Moore}, B., {Katz}, N., {Lake}, G., {Dressler}, A., \& {Oemler}, A. 1996,
  \href{https://doi.org/10.1038/379613a0}{\nat}, 379, 613,
  [\href{https://arxiv.org/abs/astro-ph/9510034}{astro-ph/9510034}].

\bibitem[{{Moore} {et~al.}(1998){Moore}, {Lake}, \& {Katz}}]{Moore1998}
{Moore}, B., {Lake}, G., \& {Katz}, N. 1998,
  \href{https://doi.org/10.1086/305264}{\apj}, 495, 139,
  [\href{https://arxiv.org/abs/astro-ph/9701211}{astro-ph/9701211}].

\bibitem[{{Moore} {et~al.}(1999){Moore}, {Lake}, {Quinn}, \&
  {Stadel}}]{Moore1999}
{Moore}, B., {Lake}, G., {Quinn}, T., \& {Stadel}, J. 1999,
  \href{https://doi.org/10.1046/j.1365-8711.1999.02345.x}{\mnras}, 304, 465,
  [\href{https://arxiv.org/abs/astro-ph/9811127}{astro-ph/9811127}].

\bibitem[{{Moorman} {et~al.}(2016){Moorman}, {Moreno}, {White}, {Vogeley},
  {Hoyle}, {Giovanelli}, \& {Haynes}}]{Moorman2016}
{Moorman}, C.~M., {Moreno}, J., {White}, A., {et~al.} 2016,
  \href{https://doi.org/10.3847/0004-637X/831/2/118}{\apj}, 831, 118,
  [\href{https://arxiv.org/abs/1601.04092}{1601.04092}].

\bibitem[{{Moorthy} \& {Holtzman}(2006)}]{Moorthy2006}
{Moorthy}, B.~K. \& {Holtzman}, J.~A. 2006,
  \href{https://doi.org/10.1111/j.1365-2966.2006.10722.x}{\mnras}, 371, 583,
  [\href{https://arxiv.org/abs/astro-ph/0512346}{astro-ph/0512346}].

\bibitem[{{Moresco} {et~al.}(2013){Moresco}, {Pozzetti}, {Cimatti}, {Zamorani},
  {Bolzonella}, {Lamareille}, {Mignoli}, {Zucca}, {Lilly}, {Carollo},
  {Contini}, {Kneib}, {Le F{\`e}vre}, {Mainieri}, {Renzini}, {Scodeggio},
  {Bardelli}, {Bongiorno}, {Caputi}, {Cucciati}, {de la Torre}, {de Ravel},
  {Franzetti}, {Garilli}, {Iovino}, {Kampczyk}, {Knobel}, {Kova{\v{c}}}, {Le
  Borgne}, {Le Brun}, {Maier}, {Pell{\'o}}, {Peng}, {Perez-Montero},
  {Presotto}, {Silverman}, {Tanaka}, {Tasca}, {Tresse}, {Vergani}, {Barnes},
  {Bordoloi}, {Cappi}, {Diener}, {Koekemoer}, {Le Floc'h}, {L{\'o}pez-Sanjuan},
  {McCracken}, {Nair}, {Oesch}, {Scarlata}, {Scoville}, \&
  {Welikala}}]{Moresco2013}
{Moresco}, M., {Pozzetti}, L., {Cimatti}, A., {et~al.} 2013,
  \href{https://doi.org/10.1051/0004-6361/201321797}{\aap}, 558, A61,
  [\href{https://arxiv.org/abs/1305.1308}{1305.1308}].

\bibitem[{{Mu{\~n}oz-Mateos} {et~al.}(2007){Mu{\~n}oz-Mateos}, {Gil de Paz},
  {Boissier}, {Zamorano}, {Jarrett}, {Gallego}, \& {Madore}}]{MunozMateos2007}
{Mu{\~n}oz-Mateos}, J.~C., {Gil de Paz}, A., {Boissier}, S., {et~al.} 2007,
  \href{https://doi.org/10.1086/511812}{\apj}, 658, 1006,
  [\href{https://arxiv.org/abs/astro-ph/0612017}{astro-ph/0612017}].

\bibitem[{{Muzzin} {et~al.}(2013){Muzzin}, {Marchesini}, {Stefanon}, {Franx},
  {McCracken}, {Milvang-Jensen}, {Dunlop}, {Fynbo}, {Brammer}, {Labb{\'e}}, \&
  {van Dokkum}}]{Muzzin2013}
{Muzzin}, A., {Marchesini}, D., {Stefanon}, M., {et~al.} 2013,
  \href{https://doi.org/10.1088/0004-637X/777/1/18}{\apj}, 777, 18,
  [\href{https://arxiv.org/abs/1303.4409}{1303.4409}].

\bibitem[{{Muzzin} {et~al.}(2014){Muzzin}, {van der Burg}, {McGee}, {Balogh},
  {Franx}, {Hoekstra}, {Hudson}, {Noble}, {Taranu}, {Webb}, {Wilson}, \&
  {Yee}}]{Muzzin2014}
{Muzzin}, A., {van der Burg}, R.~F.~J., {McGee}, S.~L., {et~al.} 2014,
  \href{https://doi.org/10.1088/0004-637X/796/1/65}{\apj}, 796, 65,
  [\href{https://arxiv.org/abs/1402.7077}{1402.7077}].

\bibitem[{{Muzzin} {et~al.}(2012){Muzzin}, {Wilson}, {Yee}, {Gilbank},
  {Hoekstra}, {Demarco}, {Balogh}, {van Dokkum}, {Franx}, {Ellingson}, {Hicks},
  {Nantais}, {Noble}, {Lacy}, {Lidman}, {Rettura}, {Surace}, \&
  {Webb}}]{muzzin2012}
{Muzzin}, A., {Wilson}, G., {Yee}, H.~K.~C., {et~al.} 2012,
  \href{https://doi.org/10.1088/0004-637X/746/2/188}{\apj}, 746, 188,
  [\href{https://arxiv.org/abs/1112.3655}{1112.3655}].

\bibitem[{Nair \& Hinton(2010)}]{ReLU}
Nair, V. \& Hinton, G. 2010, in , 807--814

\bibitem[{{Nantais} {et~al.}(2020){Nantais}, {Wilson}, {Muzzin}, {Old},
  {Demarco}, {Cerulo}, {Balogh}, {Rudnick}, {Chan}, {Cooper}, {Forrest},
  {Hayden}, {Lidman}, {Noble}, {Perlmutter}, {Rhea}, {Surace}, {van der Burg},
  \& {van Kampen}}]{Nantais2020}
{Nantais}, J., {Wilson}, G., {Muzzin}, A., {et~al.} 2020,
  \href{https://doi.org/10.1093/mnras/staa2872}{\mnras}, 499, 3061,
  [\href{https://arxiv.org/abs/2009.07861}{2009.07861}].

\bibitem[{{Nantais} {et~al.}(2017){Nantais}, {Muzzin}, {van der Burg},
  {Wilson}, {Lidman}, {Foltz}, {DeGroot}, {Noble}, {Cooper}, \&
  {Demarco}}]{Nantais2017}
{Nantais}, J.~B., {Muzzin}, A., {van der Burg}, R. F.~J., {et~al.} 2017,
  \href{https://doi.org/10.1093/mnrasl/slw224}{\mnras}, 465, L104,
  [\href{https://arxiv.org/abs/1610.08058}{1610.08058}].

\bibitem[{{Niemiec} {et~al.}(2022){Niemiec}, {Giocoli}, {Cohen}, {Jauzac},
  {Jullo}, \& {Limousin}}]{Niemiec2022}
{Niemiec}, A., {Giocoli}, C., {Cohen}, E., {et~al.} 2022,
  \href{https://doi.org/10.1093/mnras/stac832}{\mnras}, 512, 6021,
  [\href{https://arxiv.org/abs/2201.07817}{2201.07817}].

\bibitem[{{Nipoti} \& {Binney}(2007)}]{Nipoti2007}
{Nipoti}, C. \& {Binney}, J. 2007,
  \href{https://doi.org/10.1111/j.1365-2966.2007.12505.x}{\mnras}, 382, 1481,
  [\href{https://arxiv.org/abs/0707.4147}{0707.4147}].

\bibitem[{{Noble} {et~al.}(2013){Noble}, {Webb}, {Muzzin}, {Wilson}, {Yee}, \&
  {van der Burg}}]{Noble2013}
{Noble}, A.~G., {Webb}, T.~M.~A., {Muzzin}, A., {et~al.} 2013,
  \href{https://doi.org/10.1088/0004-637X/768/2/118}{\apj}, 768, 118,
  [\href{https://arxiv.org/abs/1303.4997}{1303.4997}].

\bibitem[{{Noble} {et~al.}(2016){Noble}, {Webb}, {Yee}, {Muzzin}, {Wilson},
  {van der Burg}, {Balogh}, \& {Shupe}}]{Noble2016}
{Noble}, A.~G., {Webb}, T.~M.~A., {Yee}, H.~K.~C., {et~al.} 2016,
  \href{https://doi.org/10.3847/0004-637X/816/2/48}{\apj}, 816, 48,
  [\href{https://arxiv.org/abs/1511.00584}{1511.00584}].

\bibitem[{{Noeske} {et~al.}(2007){Noeske}, {Weiner}, {Faber}, {Papovich},
  {Koo}, {Somerville}, {Bundy}, {Conselice}, {Newman}, {Schiminovich}, {Le
  Floc'h}, {Coil}, {Rieke}, {Lotz}, {Primack}, {Barmby}, {Cooper}, {Davis},
  {Ellis}, {Fazio}, {Guhathakurta}, {Huang}, {Kassin}, {Martin}, {Phillips},
  {Rich}, {Small}, {Willmer}, \& {Wilson}}]{Noeske2007}
{Noeske}, K.~G., {Weiner}, B.~J., {Faber}, S.~M., {et~al.} 2007,
  \href{https://doi.org/10.1086/517926}{\apjl}, 660, L43,
  [\href{https://arxiv.org/abs/astro-ph/0701924}{astro-ph/0701924}].

\bibitem[{{Noirot} {et~al.}(2022){Noirot}, {Sawicki}, {Abraham},
  {Brada{\v{c}}}, {Iyer}, {Moutard}, {Pacifici}, {Ravindranath}, \&
  {Willott}}]{Noirot2022}
{Noirot}, G., {Sawicki}, M., {Abraham}, R., {et~al.} 2022,
  \href{https://doi.org/10.1093/mnras/stac668}{\mnras}, 512, 3566,
  [\href{https://arxiv.org/abs/2203.06185}{2203.06185}].

\bibitem[{{Ocvirk} {et~al.}(2006){Ocvirk}, {Pichon}, {Lan{\c{c}}on}, \&
  {Thi{\'e}baut}}]{ocvirk2006}
{Ocvirk}, P., {Pichon}, C., {Lan{\c{c}}on}, A., \& {Thi{\'e}baut}, E. 2006,
  \href{https://doi.org/10.1111/j.1365-2966.2005.09182.x}{\mnras}, 365, 46,
  [\href{https://arxiv.org/abs/astro-ph/0505209}{astro-ph/0505209}].

\bibitem[{{Oke} \& {Gunn}(1983)}]{Oke1983}
{Oke}, J.~B. \& {Gunn}, J.~E. 1983,
  \href{https://doi.org/10.1086/160817}{\apj}, 266, 713.

\bibitem[{{Olave-Rojas} {et~al.}(2018){Olave-Rojas}, {Cerulo}, {Demarco},
  {Jaff{\'e}}, {Mercurio}, {Rosati}, {Balestra}, \& {Nonino}}]{Olave-Rojas2018}
{Olave-Rojas}, D., {Cerulo}, P., {Demarco}, R., {et~al.} 2018,
  \href{https://doi.org/10.1093/mnras/sty1669}{\mnras}, 479, 2328,
  [\href{https://arxiv.org/abs/1806.08435}{1806.08435}].

\bibitem[{{Osterbrock}(1989)}]{Osterbrock1989}
{Osterbrock}, D.~E. 1989, {Astrophysics of gaseous nebulae and active galactic
  nuclei}

\bibitem[{{Osterbrock} \& {De Robertis}(1985)}]{Osterbrock1985}
{Osterbrock}, D.~E. \& {De Robertis}, M.~M. 1985,
  \href{https://doi.org/10.1086/131676}{\pasp}, 97, 1129.

\bibitem[{{Osuna} {et~al.}(2008){Osuna}, {Ortiz}, {Lusted}, {Dowler}, {Szalay},
  {Shirasaki}, {Nieto-Santisteban}, {Ohishi}, {O'Mullane}, {VOQL-TEG Group}, \&
  {VOQL Working Group.}}]{ADQL2008}
{Osuna}, P., {Ortiz}, I., {Lusted}, J., {et~al.} 2008, {IVOA Astronomical Data
  Query Language Version 2.00}, IVOA Recommendation 30 October 2008

\bibitem[{{Oteo} {et~al.}(2015){Oteo}, {Sobral}, {Ivison}, {Smail}, {Best},
  {Cepa}, \& {P{\'e}rez-Garc{\'\i}a}}]{Oteo201}
{Oteo}, I., {Sobral}, D., {Ivison}, R.~J., {et~al.} 2015,
  \href{https://doi.org/10.1093/mnras/stv1284}{\mnras}, 452, 2018,
  [\href{https://arxiv.org/abs/1506.02670}{1506.02670}].

\bibitem[{{Paccagnella} {et~al.}(2019){Paccagnella}, {Vulcani}, {Poggianti},
  {Moretti}, {Fritz}, {Gullieuszik}, \& {Fasano}}]{Paccagnella2019}
{Paccagnella}, A., {Vulcani}, B., {Poggianti}, B.~M., {et~al.} 2019,
  \href{https://doi.org/10.1093/mnras/sty2728}{\mnras}, 482, 881,
  [\href{https://arxiv.org/abs/1805.11475}{1805.11475}].

\bibitem[{{Pacifici} {et~al.}(2016){Pacifici}, {Kassin}, {Weiner}, {Holden},
  {Gardner}, {Faber}, {Ferguson}, {Koo}, {Primack}, {Bell}, {Dekel}, {Gawiser},
  {Giavalisco}, {Rafelski}, {Simons}, {Barro}, {Croton}, {Dav{\'e}}, {Fontana},
  {Grogin}, {Koekemoer}, {Lee}, {Salmon}, {Somerville}, \&
  {Behroozi}}]{pacifici2016}
{Pacifici}, C., {Kassin}, S.~A., {Weiner}, B.~J., {et~al.} 2016,
  \href{https://doi.org/10.3847/0004-637X/832/1/79}{\apj}, 832, 79,
  [\href{https://arxiv.org/abs/1609.03572}{1609.03572}].

\bibitem[{{Pacucci} {et~al.}(2023){Pacucci}, {Ni}, \& {Loeb}}]{Pacucci2023}
{Pacucci}, F., {Ni}, Y., \& {Loeb}, A. 2023,
  \href{https://doi.org/10.3847/2041-8213/acff5e}{\apjl}, 956, L37,
  [\href{https://arxiv.org/abs/2309.02487}{2309.02487}].

\bibitem[{{Pallero} {et~al.}(2022){Pallero}, {G{\'o}mez}, {Padilla},
  {Bah{\'e}}, {Vega-Mart{\'\i}nez}, \& {Torres-Flores}}]{Pallero2022}
{Pallero}, D., {G{\'o}mez}, F.~A., {Padilla}, N.~D., {et~al.} 2022,
  \href{https://doi.org/10.1093/mnras/stab3318}{\mnras}, 511, 3210,
  [\href{https://arxiv.org/abs/2012.08593}{2012.08593}].

\bibitem[{{Parikh} {et~al.}(2021){Parikh}, {Thomas}, {Maraston}, {Westfall},
  {Andrews}, {Boardman}, {Drory}, \& {Oyarzun}}]{Parikh2021}
{Parikh}, T., {Thomas}, D., {Maraston}, C., {et~al.} 2021,
  \href{https://doi.org/10.1093/mnras/stab449}{\mnras}, 502, 5508,
  [\href{https://arxiv.org/abs/2102.06703}{2102.06703}].

\bibitem[{{Park} \& {Hwang}(2009)}]{Park2009}
{Park}, C. \& {Hwang}, H.~S. 2009,
  \href{https://doi.org/10.1088/0004-637X/699/2/1595}{\apj}, 699, 1595,
  [\href{https://arxiv.org/abs/0812.2088}{0812.2088}].

\bibitem[{{Pascual} {et~al.}(2007){Pascual}, {Gallego}, \&
  {Zamorano}}]{Pascual2007}
{Pascual}, S., {Gallego}, J., \& {Zamorano}, J. 2007,
  \href{https://doi.org/10.1086/510600}{\pasp}, 119, 30,
  [\href{https://arxiv.org/abs/astro-ph/0611121}{astro-ph/0611121}].

\bibitem[{{Pasquali} {et~al.}(2010){Pasquali}, {Gallazzi}, {Fontanot}, {van den
  Bosch}, {De Lucia}, {Mo}, \& {Yang}}]{Pasquali2010}
{Pasquali}, A., {Gallazzi}, A., {Fontanot}, F., {et~al.} 2010,
  \href{https://doi.org/10.1111/j.1365-2966.2010.17074.x}{\mnras}, 407, 937,
  [\href{https://arxiv.org/abs/0912.1853}{0912.1853}].

\bibitem[{{Pasquali} {et~al.}(2019){Pasquali}, {Smith}, {Gallazzi}, {De Lucia},
  {Zibetti}, {Hirschmann}, \& {Yi}}]{Pasquali2019}
{Pasquali}, A., {Smith}, R., {Gallazzi}, A., {et~al.} 2019,
  \href{https://doi.org/10.1093/mnras/sty3530}{\mnras}, 484, 1702,
  [\href{https://arxiv.org/abs/1901.04238}{1901.04238}].

\bibitem[{{Paulin-Henriksson} {et~al.}(2008){Paulin-Henriksson}, {Amara},
  {Voigt}, {Refregier}, \& {Bridle}}]{PaulinHenriksson2008}
{Paulin-Henriksson}, S., {Amara}, A., {Voigt}, L., {Refregier}, A., \&
  {Bridle}, S.~L. 2008,
  \href{https://doi.org/10.1051/0004-6361:20079150}{\aap}, 484, 67,
  [\href{https://arxiv.org/abs/0711.4886}{0711.4886}].

\bibitem[{{Peebles}(1982)}]{Peebles1982}
{Peebles}, P.~J.~E. 1982, \href{https://doi.org/10.1086/183911}{\apjl}, 263,
  L1.

\bibitem[{{Peletier} {et~al.}(1990){Peletier}, {Davies}, \&
  {Illingworth}}]{Peletier1990}
{Peletier}, R., {Davies}, R.~L., \& {Illingworth}, G. 1990, in Dynamics and
  Interactions of Galaxies, ed. R.~{Wielen}, 267--269

\bibitem[{{Peletier} \& {Balcells}(1996)}]{Peletier1996}
{Peletier}, R.~F. \& {Balcells}, M. 1996,
  \href{https://doi.org/10.1086/117958}{\aj}, 111, 2238,
  [\href{https://arxiv.org/abs/astro-ph/9602088}{astro-ph/9602088}].

\bibitem[{{Peletier} \& {de Grijs}(1998)}]{Peletier1998}
{Peletier}, R.~F. \& {de Grijs}, R. 1998,
  \href{https://doi.org/10.1046/j.1365-8711.1998.02071.x}{\mnras}, 300, L3,
  [\href{https://arxiv.org/abs/astro-ph/9808232}{astro-ph/9808232}].

\bibitem[{{Peluso} {et~al.}(2022){Peluso}, {Vulcani}, {Poggianti}, {Moretti},
  {Radovich}, {Smith}, {Jaff{\'e}}, {Crossett}, {Gullieuszik}, {Fritz}, \&
  {Ignesti}}]{Peluso2022}
{Peluso}, G., {Vulcani}, B., {Poggianti}, B.~M., {et~al.} 2022,
  \href{https://doi.org/10.3847/1538-4357/ac4225}{\apj}, 927, 130,
  [\href{https://arxiv.org/abs/2111.02538}{2111.02538}].

\bibitem[{{Peng} {et~al.}(2015){Peng}, {Maiolino}, \& {Cochrane}}]{Peng2015}
{Peng}, Y., {Maiolino}, R., \& {Cochrane}, R. 2015,
  \href{https://doi.org/10.1038/nature14439}{\nat}, 521, 192,
  [\href{https://arxiv.org/abs/1505.03143}{1505.03143}].

\bibitem[{{Peng} {et~al.}(2010){Peng}, {Lilly}, {Kova{\v{c}}}, {Bolzonella},
  {Pozzetti}, {Renzini}, {Zamorani}, {Ilbert}, {Knobel}, {Iovino}, {Maier},
  {Cucciati}, {Tasca}, {Carollo}, {Silverman}, {Kampczyk}, {de Ravel},
  {Sanders}, {Scoville}, {Contini}, {Mainieri}, {Scodeggio}, {Kneib}, {Le
  F{\`e}vre}, {Bardelli}, {Bongiorno}, {Caputi}, {Coppa}, {de la Torre},
  {Franzetti}, {Garilli}, {Lamareille}, {Le Borgne}, {Le Brun}, {Mignoli},
  {Perez Montero}, {Pello}, {Ricciardelli}, {Tanaka}, {Tresse}, {Vergani},
  {Welikala}, {Zucca}, {Oesch}, {Abbas}, {Barnes}, {Bordoloi}, {Bottini},
  {Cappi}, {Cassata}, {Cimatti}, {Fumana}, {Hasinger}, {Koekemoer},
  {Leauthaud}, {Maccagni}, {Marinoni}, {McCracken}, {Memeo}, {Meneux}, {Nair},
  {Porciani}, {Presotto}, \& {Scaramella}}]{Peng2010}
{Peng}, Y.-j., {Lilly}, S.~J., {Kova{\v{c}}}, K., {et~al.} 2010,
  \href{https://doi.org/10.1088/0004-637X/721/1/193}{\apj}, 721, 193,
  [\href{https://arxiv.org/abs/1003.4747}{1003.4747}].

\bibitem[{{Peng} {et~al.}(2012){Peng}, {Lilly}, {Renzini}, \&
  {Carollo}}]{Peng2012}
{Peng}, Y.-j., {Lilly}, S.~J., {Renzini}, A., \& {Carollo}, M. 2012,
  \href{https://doi.org/10.1088/0004-637X/757/1/4}{\apj}, 757, 4,
  [\href{https://arxiv.org/abs/1106.2546}{1106.2546}].

\bibitem[{{P{\'e}rez} {et~al.}(2013){P{\'e}rez}, {Cid Fernandes}, {Gonz{\'a}lez
  Delgado}, {Garc{\'\i}a-Benito}, {S{\'a}nchez}, {Husemann}, {Mast},
  {Rod{\'o}n}, {Kupko}, {Backsmann}, {de Amorim}, {van de Ven}, {Walcher},
  {Wisotzki}, {Cortijo-Ferrero}, \& {CALIFA Collaboration}}]{Perez2013}
{P{\'e}rez}, E., {Cid Fernandes}, R., {Gonz{\'a}lez Delgado}, R.~M., {et~al.}
  2013, \href{https://doi.org/10.1088/2041-8205/764/1/L1}{\apjl}, 764, L1,
  [\href{https://arxiv.org/abs/1301.1679}{1301.1679}].

\bibitem[{{Peschken} {et~al.}(2020){Peschken}, {{\L}okas}, \&
  {Athanassoula}}]{Peschken2020}
{Peschken}, N., {{\L}okas}, E.~L., \& {Athanassoula}, E. 2020,
  \href{https://doi.org/10.1093/mnras/staa299}{\mnras}, 493, 1375,
  [\href{https://arxiv.org/abs/1909.01033}{1909.01033}].

\bibitem[{{Petrosian}(1976)}]{Petrosian1976}
{Petrosian}, V. 1976,
  \href{https://doi.org/10.1086/18230110.1086/182253}{\apjl}, 210, L53.

\bibitem[{{Pharo} {et~al.}(2020){Pharo}, {Malhotra}, {Rhoads}, {Pirzkal},
  {Finkelstein}, {Ryan}, {Cimatti}, {Christensen}, {Hathi}, {Koekemoer},
  {Harish}, {Smith}, {Straughn}, {Windhorst}, {Ferreras}, {Gronwall}, {Hibon},
  {Larson}, {O'Connell}, {Pasquali}, \& {Tilvi}}]{Pharo2020}
{Pharo}, J., {Malhotra}, S., {Rhoads}, J.~E., {et~al.} 2020,
  \href{https://doi.org/10.3847/1538-4357/ab5f5c}{\apj}, 888, 79,
  [\href{https://arxiv.org/abs/1912.02261}{1912.02261}].

\bibitem[{{Pickles} \& {Depagne}(2010)}]{Pickles2010}
{Pickles}, A. \& {Depagne}, {\'E}. 2010,
  \href{https://doi.org/10.1086/657947}{\pasp}, 122, 1437,
  [\href{https://arxiv.org/abs/1011.2020}{1011.2020}].

\bibitem[{{Piotrowska} {et~al.}(2022){Piotrowska}, {Bluck}, {Maiolino}, \&
  {Peng}}]{Piotrowska2022}
{Piotrowska}, J.~M., {Bluck}, A. F.~L., {Maiolino}, R., \& {Peng}, Y. 2022,
  \href{https://doi.org/10.1093/mnras/stab3673}{\mnras}, 512, 1052,
  [\href{https://arxiv.org/abs/2112.07672}{2112.07672}].

\bibitem[{{Piraino-Cerda} {et~al.}(2024){Piraino-Cerda}, {Jaff{\'e}},
  {Louren{\c{c}}o}, {Crossett}, {Salinas}, {Kim}, {Sheen}, {Kelkar}, {Pallero},
  \& {Bravo-Alfaro}}]{PirainoCerda2024}
{Piraino-Cerda}, F., {Jaff{\'e}}, Y.~L., {Louren{\c{c}}o}, A.~C., {et~al.}
  2024, \href{https://doi.org/10.1093/mnras/stad3957}{\mnras}, 528, 919,
  [\href{https://arxiv.org/abs/2401.06973}{2401.06973}].

\bibitem[{{Pirzkal} {et~al.}(2017){Pirzkal}, {Malhotra}, {Ryan}, {Rothberg},
  {Grogin}, {Finkelstein}, {Koekemoer}, {Rhoads}, {Larson}, {Christensen},
  {Cimatti}, {Ferreras}, {Gardner}, {Gronwall}, {Hathi}, {Hibon}, {Joshi},
  {Kuntschner}, {Meurer}, {O'Connell}, {Oestlin}, {Pasquali}, {Pharo},
  {Straughn}, {Walsh}, {Watson}, {Windhorst}, {Zakamska}, \& {Zirm}}]{FIGS}
{Pirzkal}, N., {Malhotra}, S., {Ryan}, R.~E., {et~al.} 2017,
  \href{https://doi.org/10.3847/1538-4357/aa81cc}{\apj}, 846, 84,
  [\href{https://arxiv.org/abs/1706.02669}{1706.02669}].

\bibitem[{{Planck Collaboration} {et~al.}(2020){Planck Collaboration},
  {Aghanim}, {Akrami}, {Ashdown}, {Aumont}, {Baccigalupi}, {Ballardini},
  {Banday}, {Barreiro}, {Bartolo}, {Basak}, {Battye}, {Benabed}, {Bernard},
  {Bersanelli}, {Bielewicz}, {Bock}, {Bond}, {Borrill}, {Bouchet}, {Boulanger},
  {Bucher}, {Burigana}, {Butler}, {Calabrese}, {Cardoso}, {Carron},
  {Challinor}, {Chiang}, {Chluba}, {Colombo}, {Combet}, {Contreras}, {Crill},
  {Cuttaia}, {de Bernardis}, {de Zotti}, {Delabrouille}, {Delouis}, {Di
  Valentino}, {Diego}, {Dor{\'e}}, {Douspis}, {Ducout}, {Dupac}, {Dusini},
  {Efstathiou}, {Elsner}, {En{\ss}lin}, {Eriksen}, {Fantaye}, {Farhang},
  {Fergusson}, {Fernandez-Cobos}, {Finelli}, {Forastieri}, {Frailis},
  {Fraisse}, {Franceschi}, {Frolov}, {Galeotta}, {Galli}, {Ganga},
  {G{\'e}nova-Santos}, {Gerbino}, {Ghosh}, {Gonz{\'a}lez-Nuevo}, {G{\'o}rski},
  {Gratton}, {Gruppuso}, {Gudmundsson}, {Hamann}, {Handley}, {Hansen},
  {Herranz}, {Hildebrandt}, {Hivon}, {Huang}, {Jaffe}, {Jones}, {Karakci},
  {Keih{\"a}nen}, {Keskitalo}, {Kiiveri}, {Kim}, {Kisner}, {Knox},
  {Krachmalnicoff}, {Kunz}, {Kurki-Suonio}, {Lagache}, {Lamarre}, {Lasenby},
  {Lattanzi}, {Lawrence}, {Le Jeune}, {Lemos}, {Lesgourgues}, {Levrier},
  {Lewis}, {Liguori}, {Lilje}, {Lilley}, {Lindholm}, {L{\'o}pez-Caniego},
  {Lubin}, {Ma}, {Mac{\'\i}as-P{\'e}rez}, {Maggio}, {Maino}, {Mandolesi},
  {Mangilli}, {Marcos-Caballero}, {Maris}, {Martin}, {Martinelli},
  {Mart{\'\i}nez-Gonz{\'a}lez}, {Matarrese}, {Mauri}, {McEwen}, {Meinhold},
  {Melchiorri}, {Mennella}, {Migliaccio}, {Millea}, {Mitra},
  {Miville-Desch{\^e}nes}, {Molinari}, {Montier}, {Morgante}, {Moss}, {Natoli},
  {N{\o}rgaard-Nielsen}, {Pagano}, {Paoletti}, {Partridge}, {Patanchon},
  {Peiris}, {Perrotta}, {Pettorino}, {Piacentini}, {Polastri}, {Polenta},
  {Puget}, {Rachen}, {Reinecke}, {Remazeilles}, {Renzi}, {Rocha}, {Rosset},
  {Roudier}, {Rubi{\~n}o-Mart{\'\i}n}, {Ruiz-Granados}, {Salvati}, {Sandri},
  {Savelainen}, {Scott}, {Shellard}, {Sirignano}, {Sirri}, {Spencer},
  {Sunyaev}, {Suur-Uski}, {Tauber}, {Tavagnacco}, {Tenti}, {Toffolatti},
  {Tomasi}, {Trombetti}, {Valenziano}, {Valiviita}, {Van Tent}, {Vibert},
  {Vielva}, {Villa}, {Vittorio}, {Wandelt}, {Wehus}, {White}, {White},
  {Zacchei}, \& {Zonca}}]{Planck2020}
{Planck Collaboration}, {Aghanim}, N., {Akrami}, Y., {et~al.} 2020,
  \href{https://doi.org/10.1051/0004-6361/201833910}{\aap}, 641, A6,
  [\href{https://arxiv.org/abs/1807.06209}{1807.06209}].

\bibitem[{{Plat} {et~al.}(2019){Plat}, {Charlot}, {Bruzual}, {Feltre},
  {Vidal-Garc{\'\i}a}, {Morisset}, {Chevallard}, \& {Todt}}]{Plat2019}
{Plat}, A., {Charlot}, S., {Bruzual}, G., {et~al.} 2019,
  \href{https://doi.org/10.1093/mnras/stz2616}{\mnras}, 490, 978,
  [\href{https://arxiv.org/abs/1909.07386}{1909.07386}].

\bibitem[{{Poggianti} {et~al.}(2009){Poggianti}, {Arag{\'o}n-Salamanca},
  {Zaritsky}, {De Lucia}, {Milvang-Jensen}, {Desai}, {Jablonka}, {Halliday},
  {Rudnick}, {Varela}, {Bamford}, {Best}, {Clowe}, {Noll}, {Saglia},
  {Pell{\'o}}, {Simard}, {von der Linden}, \& {White}}]{Poggianti2009}
{Poggianti}, B.~M., {Arag{\'o}n-Salamanca}, A., {Zaritsky}, D., {et~al.} 2009,
  \href{https://doi.org/10.1088/0004-637X/693/1/112}{\apj}, 693, 112,
  [\href{https://arxiv.org/abs/0811.0252}{0811.0252}].

\bibitem[{{Poggianti} {et~al.}(2013){Poggianti}, {Calvi}, {Bindoni},
  {D'Onofrio}, {Moretti}, {Valentinuzzi}, {Fasano}, {Fritz}, {De Lucia},
  {Vulcani}, {Bettoni}, {Gullieuszik}, \& {Omizzolo}}]{Poggianti2013}
{Poggianti}, B.~M., {Calvi}, R., {Bindoni}, D., {et~al.} 2013,
  \href{https://doi.org/10.1088/0004-637X/762/2/77}{\apj}, 762, 77,
  [\href{https://arxiv.org/abs/1211.1005}{1211.1005}].

\bibitem[{{Poggianti} {et~al.}(2017){Poggianti}, {Moretti}, {Gullieuszik},
  {Fritz}, {Jaff{\'e}}, {Bettoni}, {Fasano}, {Bellhouse}, {Hau}, {Vulcani},
  {Biviano}, {Omizzolo}, {Paccagnella}, {D'Onofrio}, {Cava}, {Sheen}, {Couch},
  \& {Owers}}]{Poggianti2017}
{Poggianti}, B.~M., {Moretti}, A., {Gullieuszik}, M., {et~al.} 2017,
  \href{https://doi.org/10.3847/1538-4357/aa78ed}{\apj}, 844, 48,
  [\href{https://arxiv.org/abs/1704.05086}{1704.05086}].

\bibitem[{{Postman} {et~al.}(1996){Postman}, {Lubin}, {Gunn}, {Oke}, {Hoessel},
  {Schneider}, \& {Christensen}}]{Postman1996}
{Postman}, M., {Lubin}, L.~M., {Gunn}, J.~E., {et~al.} 1996,
  \href{https://doi.org/10.1086/117811}{\aj}, 111, 615,
  [\href{https://arxiv.org/abs/astro-ph/9511011}{astro-ph/9511011}].

\bibitem[{{Povi{\'c}} {et~al.}(2013){Povi{\'c}}, {Huertas-Company}, {Aguerri},
  {M{\'a}rquez}, {Masegosa}, {Husillos}, {Molino}, {Crist{\'o}bal-Hornillos},
  {Perea}, {Ben{\'\i}tez}, {Olmo}, {Fern{\'a}ndez-Soto}, {Jim{\'e}nez-Teja},
  {Moles}, {Alfaro}, {Aparicio-Villegas}, {Ascaso}, {Broadhurst},
  {Cabrera-Ca{\~n}o}, {Castander}, {Cepa}, {Fernandez Lorenzo}, {Cervi{\~n}o},
  {Delgado}, {Infante}, {L{\'o}pez-Sanjuan}, {Mart{\'\i}nez}, {Matute}, {Oteo},
  {P{\'e}rez-Garc{\'\i}a}, {Prada}, \& {Quintana}}]{Povic2013}
{Povi{\'c}}, M., {Huertas-Company}, M., {Aguerri}, J.~A.~L., {et~al.} 2013,
  \href{https://doi.org/10.1093/mnras/stt1538}{\mnras}, 435, 3444,
  [\href{https://arxiv.org/abs/1308.3146}{1308.3146}].

\bibitem[{{Prato} {et~al.}(2012){Prato}, {Cavicchioli}, {Zanni}, {Boccacci}, \&
  {Bertero}}]{Prato2012}
{Prato}, M., {Cavicchioli}, R., {Zanni}, L., {Boccacci}, P., \& {Bertero}, M.
  2012, \href{https://doi.org/10.1051/0004-6361/201118681}{\aap}, 539, A133,
  [\href{https://arxiv.org/abs/1210.2258}{1210.2258}].

\bibitem[{{Raj} {et~al.}(2019){Raj}, {Iodice}, {Napolitano}, {Spavone}, {Su},
  {Peletier}, {Davis}, {Zabel}, {Hilker}, {Mieske}, {Falcon Barroso},
  {Cantiello}, {van de Ven}, {Watkins}, {Salo}, {Schipani}, {Capaccioli}, \&
  {Venhola}}]{Raj2019}
{Raj}, M.~A., {Iodice}, E., {Napolitano}, N.~R., {et~al.} 2019,
  \href{https://doi.org/10.1051/0004-6361/201935433}{\aap}, 628, A4,
  [\href{https://arxiv.org/abs/1906.08704}{1906.08704}].

\bibitem[{{Rawle} {et~al.}(2010){Rawle}, {Smith}, \& {Lucey}}]{Rawle2010}
{Rawle}, T.~D., {Smith}, R.~J., \& {Lucey}, J.~R. 2010,
  \href{https://doi.org/10.1111/j.1365-2966.2009.15722.x}{\mnras}, 401, 852,
  [\href{https://arxiv.org/abs/0909.3844}{0909.3844}].

\bibitem[{{Rawle} {et~al.}(2008){Rawle}, {Smith}, {Lucey}, \&
  {Swinbank}}]{Rawle2008}
{Rawle}, T.~D., {Smith}, R.~J., {Lucey}, J.~R., \& {Swinbank}, A.~M. 2008,
  \href{https://doi.org/10.1111/j.1365-2966.2008.13720.x}{\mnras}, 389, 1891,
  [\href{https://arxiv.org/abs/0807.3545}{0807.3545}].

\bibitem[{{Read} {et~al.}(2006{\natexlab{a}}){Read}, {Wilkinson}, {Evans},
  {Gilmore}, \& {Kleyna}}]{Read2006b}
{Read}, J.~I., {Wilkinson}, M.~I., {Evans}, N.~W., {Gilmore}, G., \& {Kleyna},
  J.~T. 2006{\natexlab{a}},
  \href{https://doi.org/10.1111/j.1365-2966.2005.09959.x}{\mnras}, 367, 387,
  [\href{https://arxiv.org/abs/astro-ph/0511759}{astro-ph/0511759}].

\bibitem[{{Read} {et~al.}(2006{\natexlab{b}}){Read}, {Wilkinson}, {Evans},
  {Gilmore}, \& {Kleyna}}]{Read2006}
{Read}, J.~I., {Wilkinson}, M.~I., {Evans}, N.~W., {Gilmore}, G., \& {Kleyna},
  J.~T. 2006{\natexlab{b}},
  \href{https://doi.org/10.1111/j.1365-2966.2005.09861.x}{\mnras}, 366, 429,
  [\href{https://arxiv.org/abs/astro-ph/0506687}{astro-ph/0506687}].

\bibitem[{{Reda} {et~al.}(2007){Reda}, {Proctor}, {Forbes}, {Hau}, \&
  {Larsen}}]{Reda2007}
{Reda}, F.~M., {Proctor}, R.~N., {Forbes}, D.~A., {Hau}, G. K.~T., \& {Larsen},
  S.~S. 2007, \href{https://doi.org/10.1111/j.1365-2966.2007.11755.x}{\mnras},
  377, 1772, [\href{https://arxiv.org/abs/astro-ph/0703545}{astro-ph/0703545}].

\bibitem[{{Renzini}(2013)}]{Renzini2013}
{Renzini}, A. 2013, in The Intriguing Life of Massive Galaxies, ed.
  D.~{Thomas}, A.~{Pasquali}, \& I.~{Ferreras}, Vol. 295, 377--382

\bibitem[{{Renzini} \& {Peng}(2015)}]{renzini2015objective}
{Renzini}, A. \& {Peng}, Y.-j. 2015,
  \href{https://doi.org/10.1088/2041-8205/801/2/L29}{\apjl}, 801, L29,
  [\href{https://arxiv.org/abs/1502.01027}{1502.01027}].

\bibitem[{{Rinaldi} {et~al.}(2022){Rinaldi}, {Caputi}, {van Mierlo}, {Ashby},
  {Caminha}, \& {Iani}}]{Rinaldi2022}
{Rinaldi}, P., {Caputi}, K.~I., {van Mierlo}, S.~E., {et~al.} 2022,
  \href{https://doi.org/10.3847/1538-4357/ac5d39}{\apj}, 930, 128,
  [\href{https://arxiv.org/abs/2112.03935}{2112.03935}].

\bibitem[{{Rines} \& {Diaferio}(2006)}]{Rines2006}
{Rines}, K. \& {Diaferio}, A. 2006, \href{https://doi.org/10.1086/506017}{\aj},
  132, 1275, [\href{https://arxiv.org/abs/astro-ph/0602032}{astro-ph/0602032}].

\bibitem[{{Rix} {et~al.}(2004){Rix}, {Barden}, {Beckwith}, {Bell}, {Borch},
  {Caldwell}, {H{\"a}ussler}, {Jahnke}, {Jogee}, {McIntosh}, {Meisenheimer},
  {Peng}, {Sanchez}, {Somerville}, {Wisotzki}, \& {Wolf}}]{Rix2004}
{Rix}, H.-W., {Barden}, M., {Beckwith}, S. V.~W., {et~al.} 2004,
  \href{https://doi.org/10.1086/420885}{\apjs}, 152, 163,
  [\href{https://arxiv.org/abs/astro-ph/0401427}{astro-ph/0401427}].

\bibitem[{{Roberts} {et~al.}(2019){Roberts}, {Parker}, {Brown}, {Joshi},
  {Hlavacek-Larrondo}, \& {Wadsley}}]{Roberts2019}
{Roberts}, I.~D., {Parker}, L.~C., {Brown}, T., {et~al.} 2019,
  \href{https://doi.org/10.3847/1538-4357/ab04f7}{\apj}, 873, 42,
  [\href{https://arxiv.org/abs/1902.02820}{1902.02820}].

\bibitem[{{Roche} {et~al.}(2010){Roche}, {Bernardi}, \& {Hyde}}]{Roche2010}
{Roche}, N., {Bernardi}, M., \& {Hyde}, J. 2010,
  \href{https://doi.org/10.1111/j.1365-2966.2010.16976.x}{\mnras}, 407, 1231,
  [\href{https://arxiv.org/abs/0911.0044}{0911.0044}].

\bibitem[{{Rodr{\'\i}guez del Pino} {et~al.}(2017){Rodr{\'\i}guez del Pino},
  {Arag{\'o}n-Salamanca}, {Chies-Santos}, {Weinzirl}, {Bamford}, {Gray},
  {B{\"o}hm}, {Wolf}, \& {Maltby}}]{RodriguezdelPino2017}
{Rodr{\'\i}guez del Pino}, B., {Arag{\'o}n-Salamanca}, A., {Chies-Santos},
  A.~L., {et~al.} 2017, \href{https://doi.org/10.1093/mnras/stx228}{\mnras},
  467, 4200, [\href{https://arxiv.org/abs/1701.06483}{1701.06483}].

\bibitem[{{Rodr{\'\i}guez-Mart{\'\i}n}
  {et~al.}(2022){Rodr{\'\i}guez-Mart{\'\i}n}, {Gonz{\'a}lez Delgado},
  {Mart{\'\i}nez-Solaeche}, {D{\'\i}az-Garc{\'\i}a}, {de Amorim},
  {Garc{\'\i}a-Benito}, {P{\'e}rez}, {Cid Fernandes}, {Carrasco}, {Maturi},
  {Finoguenov}, {Lopes}, {Cortesi}, {Lucatelli}, {Diego}, {Chies-Santos},
  {Dupke}, {Jim{\'e}nez-Teja}, {V{\'\i}lchez}, {Abramo}, {Alcaniz},
  {Ben{\'\i}tez}, {Bonoli}, {Cenarro}, {Crist{\'o}bal-Hornillos}, {Ederoclite},
  {Hern{\'a}n-Caballero}, {L{\'o}pez-Sanjuan}, {Mar{\'\i}n-Franch}, {Mendes de
  Oliveira}, {Moles}, {Sodr{\'e}}, {Taylor}, {Varela}, {V{\'a}zquez Rami{\'o}},
  \& {M{\'a}rquez}}]{Julio2022}
{Rodr{\'\i}guez-Mart{\'\i}n}, J.~E., {Gonz{\'a}lez Delgado}, R.~M.,
  {Mart{\'\i}nez-Solaeche}, G., {et~al.} 2022,
  \href{https://doi.org/10.1051/0004-6361/202243245}{\aap}, 666, A160,
  [\href{https://arxiv.org/abs/2207.10101}{2207.10101}].

\bibitem[{{Rohr} {et~al.}(2023){Rohr}, {Pillepich}, {Nelson}, {Zinger},
  {Joshi}, \& {Ayromlou}}]{Rohr2023}
{Rohr}, E., {Pillepich}, A., {Nelson}, D., {et~al.} 2023,
  \href{https://doi.org/10.1093/mnras/stad2101}{\mnras}, 524, 3502,
  [\href{https://arxiv.org/abs/2304.09196}{2304.09196}].

\bibitem[{{Rojas} {et~al.}(2005){Rojas}, {Vogeley}, {Hoyle}, \&
  {Brinkmann}}]{Rojas2005}
{Rojas}, R.~R., {Vogeley}, M.~S., {Hoyle}, F., \& {Brinkmann}, J. 2005,
  \href{https://doi.org/10.1086/428476}{\apj}, 624, 571,
  [\href{https://arxiv.org/abs/astro-ph/0409074}{astro-ph/0409074}].

\bibitem[{{Roth} {et~al.}(2005){Roth}, {Kelz}, {Fechner}, {Hahn}, {Bauer},
  {Becker}, {B{\"o}hm}, {Christensen}, {Dionies}, {Paschke}, {Popow}, {Wolter},
  {Schmoll}, {Laux}, \& {Altmann}}]{Roth2005}
{Roth}, M.~M., {Kelz}, A., {Fechner}, T., {et~al.} 2005,
  \href{https://doi.org/10.1086/429877}{\pasp}, 117, 620,
  [\href{https://arxiv.org/abs/astro-ph/0502581}{astro-ph/0502581}].

\bibitem[{{Roth} {et~al.}(1997){Roth}, {Seydack}, {Bauer}, \&
  {Laux}}]{Roth1997}
{Roth}, M.~M., {Seydack}, M., {Bauer}, S.-M., \& {Laux}, U. 1997, in Society of
  Photo-Optical Instrumentation Engineers (SPIE) Conference Series, Vol. 2871,
  Optical Telescopes of Today and Tomorrow, ed. A.~L. {Ardeberg}, 1235--1245

\bibitem[{{Rozo} {et~al.}(2015{\natexlab{a}}){Rozo}, {Rykoff}, {Bartlett}, \&
  {Melin}}]{Rozo2015a}
{Rozo}, E., {Rykoff}, E.~S., {Bartlett}, J.~G., \& {Melin}, J.-B.
  2015{\natexlab{a}}, \href{https://doi.org/10.1093/mnras/stv605}{\mnras}, 450,
  592, [\href{https://arxiv.org/abs/1401.7716}{1401.7716}].

\bibitem[{{Rozo} {et~al.}(2015{\natexlab{b}}){Rozo}, {Rykoff}, {Becker},
  {Reddick}, \& {Wechsler}}]{Rozo2015b}
{Rozo}, E., {Rykoff}, E.~S., {Becker}, M., {Reddick}, R.~M., \& {Wechsler},
  R.~H. 2015{\natexlab{b}},
  \href{https://doi.org/10.1093/mnras/stv1560}{\mnras}, 453, 38,
  [\href{https://arxiv.org/abs/1410.1193}{1410.1193}].

\bibitem[{{Ruiz-Lara} {et~al.}(2020){Ruiz-Lara}, {Gallart}, {Monelli},
  {Nidever}, {Dorta}, {Choi}, {Olsen}, {Besla}, {Bernard}, {Cassisi},
  {Massana}, {No{\"e}l}, {P{\'e}rez}, {Rusakov}, {Cioni}, {Majewski}, {van der
  Marel}, {Mart{\'\i}nez-Delgado}, {Monachesi}, {Monteagudo}, {Mu{\~n}oz},
  {Stringfellow}, {Surot}, {Vivas}, {Walker}, \& {Zaritsky}}]{RuizLara2020}
{Ruiz-Lara}, T., {Gallart}, C., {Monelli}, M., {et~al.} 2020,
  \href{https://doi.org/10.1051/0004-6361/202038392}{\aap}, 639, L3,
  [\href{https://arxiv.org/abs/2006.10759}{2006.10759}].

\bibitem[{{Ruszkowski} \& {Begelman}(2002)}]{Ruszkowski2002}
{Ruszkowski}, M. \& {Begelman}, M.~C. 2002,
  \href{https://doi.org/10.1086/344170}{\apj}, 581, 223,
  [\href{https://arxiv.org/abs/astro-ph/0207471}{astro-ph/0207471}].

\bibitem[{{Salim} {et~al.}(2007){Salim}, {Rich}, {Charlot}, {Brinchmann},
  {Johnson}, {Schiminovich}, {Seibert}, {Mallery}, {Heckman}, {Forster},
  {Friedman}, {Martin}, {Morrissey}, {Neff}, {Small}, {Wyder}, {Bianchi},
  {Donas}, {Lee}, {Madore}, {Milliard}, {Szalay}, {Welsh}, \& {Yi}}]{Salim2007}
{Salim}, S., {Rich}, R.~M., {Charlot}, S., {et~al.} 2007,
  \href{https://doi.org/10.1086/519218}{\apjs}, 173, 267,
  [\href{https://arxiv.org/abs/0704.3611}{0704.3611}].

\bibitem[{{Salvador-Sol{\'e}} {et~al.}(2022){Salvador-Sol{\'e}}, {Manrique}, \&
  {Botella}}]{SalvadorSole2022}
{Salvador-Sol{\'e}}, E., {Manrique}, A., \& {Botella}, I. 2022,
  \href{https://doi.org/10.1093/mnras/stab2668}{\mnras}, 509, 5316,
  [\href{https://arxiv.org/abs/2109.06490}{2109.06490}].

\bibitem[{{San Roman} {et~al.}(2018){San Roman}, {Cenarro},
  {D{\'\i}az-Garc{\'\i}a}, {L{\'o}pez-Sanjuan}, {Varela}, {Gonz{\'a}lez
  Delgado}, {S{\'a}nchez-Bl{\'a}zquez}, {Alfaro}, {Ascaso}, {Bonoli},
  {Borlaff}, {Castander}, {Cervi{\~n}o}, {Fern{\'a}ndez-Soto}, {M{\'a}rquez},
  {Masegosa}, {Muniesa}, {Povi{\'c}}, {Viironen}, {Aguerri}, {Ben{\'\i}tez},
  {Broadhurst}, {Cabrera-Ca{\~n}o}, {Cepa}, {Crist{\'o}bal-Hornillos},
  {Infante}, {Mart{\'\i}nez}, {Moles}, {del Olmo}, {Perea}, {Prada}, \&
  {Quintana}}]{SanRoman2018}
{San Roman}, I., {Cenarro}, A.~J., {D{\'\i}az-Garc{\'\i}a}, L.~A., {et~al.}
  2018, \href{https://doi.org/10.1051/0004-6361/201630313}{\aap}, 609, A20,
  [\href{https://arxiv.org/abs/1707.07991}{1707.07991}].

\bibitem[{{San Roman} {et~al.}(2019){San Roman}, {S{\'a}nchez-Bl{\'a}zquez},
  {Cenarro}, {D{\'\i}az-Garc{\'\i}a}, {L{\'o}pez-Sanjuan}, {Varela},
  {Vilella-Rojo}, {Akras}, {Bonoli}, {Chies Santos}, {Coelho}, {Cortesi},
  {Ederoclite}, {Jim{\'e}nez-Teja}, {Logro{\~n}o-Garc{\'\i}a}, {Lopes de
  Oliveira}, {Nogueira-Cavalcante}, {Orsi}, {V{\'a}zquez Rami{\'o}},
  {Viironen}, {Crist{\'o}bal-Hornillos}, {Dupke}, {Mar{\'\i}n-Franch}, {Mendes
  de Oliveira}, {Moles}, \& {Sodr{\'e}}}]{SanRoman2019}
{San Roman}, I., {S{\'a}nchez-Bl{\'a}zquez}, P., {Cenarro}, A.~J., {et~al.}
  2019, \href{https://doi.org/10.1051/0004-6361/201832894}{\aap}, 622, A181,
  [\href{https://arxiv.org/abs/1804.03727}{1804.03727}].

\bibitem[{{S{\'a}nchez}(2020)}]{Sanchez2020}
{S{\'a}nchez}, S.~F. 2020,
  \href{https://doi.org/10.1146/annurev-astro-012120-013326}{\araa}, 58, 99,
  [\href{https://arxiv.org/abs/1911.06925}{1911.06925}].

\bibitem[{{S{\'a}nchez} {et~al.}(2018){S{\'a}nchez}, {Avila-Reese},
  {Hernandez-Toledo}, {Cortes-Su{\'a}rez}, {Rodr{\'\i}guez-Puebla},
  {Ibarra-Medel}, {Cano-D{\'\i}az}, {Barrera-Ballesteros}, {Negrete},
  {Calette}, {de Lorenzo-C{\'a}ceres}, {Ortega-Minakata}, {Aquino},
  {Valenzuela}, {Clemente}, {Storchi-Bergmann}, {Riffel}, {Schimoia}, {Riffel},
  {Rembold}, {Brownstein}, {Pan}, {Yates}, {Mallmann}, \&
  {Bitsakis}}]{sanchez2018sdss}
{S{\'a}nchez}, S.~F., {Avila-Reese}, V., {Hernandez-Toledo}, H., {et~al.} 2018,
  \href{https://doi.org/10.48550/arXiv.1709.05438}{\rmxaa}, 54, 217,
  [\href{https://arxiv.org/abs/1709.05438}{1709.05438}].

\bibitem[{{S{\'a}nchez} {et~al.}(2023){S{\'a}nchez}, {Galbany}, {Walcher},
  {Garc{\'\i}a-Benito}, \& {Barrera-Ballesteros}}]{CALIFA2023}
{S{\'a}nchez}, S.~F., {Galbany}, L., {Walcher}, C.~J., {Garc{\'\i}a-Benito},
  R., \& {Barrera-Ballesteros}, J.~K. 2023,
  \href{https://doi.org/10.1093/mnras/stad3119}{\mnras}, 526, 5555,
  [\href{https://arxiv.org/abs/2304.13022}{2304.13022}].

\bibitem[{{S{\'a}nchez} {et~al.}(2016){S{\'a}nchez}, {Garc{\'\i}a-Benito},
  {Zibetti}, {Walcher}, {Husemann}, {Mendoza}, {Galbany}, {Falc{\'o}n-Barroso},
  {Mast}, {Aceituno}, {Aguerri}, {Alves}, {Amorim}, {Ascasibar},
  {Barrado-Navascues}, {Barrera-Ballesteros}, {Bekerait{\`e}},
  {Bland-Hawthorn}, {Cano D{\'\i}az}, {Cid Fernandes}, {Cavichia}, {Cortijo},
  {Dannerbauer}, {Demleitner}, {D{\'\i}az}, {Dettmar}, {de
  Lorenzo-C{\'a}ceres}, {del Olmo}, {Galazzi}, {Garc{\'\i}a-Lorenzo}, {Gil de
  Paz}, {Gonz{\'a}lez Delgado}, {Holmes}, {Igl{\'e}sias-P{\'a}ramo}, {Kehrig},
  {Kelz}, {Kennicutt}, {Kleemann}, {Lacerda}, {L{\'o}pez Fern{\'a}ndez},
  {L{\'o}pez S{\'a}nchez}, {Lyubenova}, {Marino}, {M{\'a}rquez},
  {Mendez-Abreu}, {Moll{\'a}}, {Monreal-Ibero}, {Ortega Minakata},
  {Torres-Papaqui}, {P{\'e}rez}, {Rosales-Ortega}, {Roth},
  {S{\'a}nchez-Bl{\'a}zquez}, {Schilling}, {Spekkens}, {Vale Asari}, {van den
  Bosch}, {van de Ven}, {Vilchez}, {Wild}, {Wisotzki}, {Y{\i}ld{\i}r{\i}m}, \&
  {Ziegler}}]{CALIFA2016}
{S{\'a}nchez}, S.~F., {Garc{\'\i}a-Benito}, R., {Zibetti}, S., {et~al.} 2016,
  \href{https://doi.org/10.1051/0004-6361/201628661}{\aap}, 594, A36,
  [\href{https://arxiv.org/abs/1604.02289}{1604.02289}].

\bibitem[{{S{\'a}nchez} {et~al.}(2012){S{\'a}nchez}, {Kennicutt}, {Gil de Paz},
  {van de Ven}, {V{\'\i}lchez}, {Wisotzki}, {Walcher}, {Mast}, {Aguerri},
  {Albiol-P{\'e}rez}, {Alonso-Herrero}, {Alves}, {Bakos}, {Bart{\'a}kov{\'a}},
  {Bland-Hawthorn}, {Boselli}, {Bomans}, {Castillo-Morales}, {Cortijo-Ferrero},
  {de Lorenzo-C{\'a}ceres}, {Del Olmo}, {Dettmar}, {D{\'\i}az}, {Ellis},
  {Falc{\'o}n-Barroso}, {Flores}, {Gallazzi}, {Garc{\'\i}a-Lorenzo},
  {Gonz{\'a}lez Delgado}, {Gruel}, {Haines}, {Hao}, {Husemann},
  {Igl{\'e}sias-P{\'a}ramo}, {Jahnke}, {Johnson}, {Jungwiert}, {Kalinova},
  {Kehrig}, {Kupko}, {L{\'o}pez-S{\'a}nchez}, {Lyubenova}, {Marino},
  {M{\'a}rmol-Queralt{\'o}}, {M{\'a}rquez}, {Masegosa}, {Meidt},
  {Mendez-Abreu}, {Monreal-Ibero}, {Montijo}, {Mour{\~a}o}, {Palacios-Navarro},
  {Papaderos}, {Pasquali}, {Peletier}, {P{\'e}rez}, {P{\'e}rez}, {Quirrenbach},
  {Rela{\~n}o}, {Rosales-Ortega}, {Roth}, {Ruiz-Lara},
  {S{\'a}nchez-Bl{\'a}zquez}, {Sengupta}, {Singh}, {Stanishev}, {Trager},
  {Vazdekis}, {Viironen}, {Wild}, {Zibetti}, \& {Ziegler}}]{CALIFA2012}
{S{\'a}nchez}, S.~F., {Kennicutt}, R.~C., {Gil de Paz}, A., {et~al.} 2012,
  \href{https://doi.org/10.1051/0004-6361/201117353}{\aap}, 538, A8,
  [\href{https://arxiv.org/abs/1111.0962}{1111.0962}].

\bibitem[{{S{\'a}nchez} {et~al.}(2013){S{\'a}nchez}, {Rosales-Ortega},
  {Jungwiert}, {Iglesias-P{\'a}ramo}, {V{\'\i}lchez}, {Marino}, {Walcher},
  {Husemann}, {Mast}, {Monreal-Ibero}, {Cid Fernandes}, {P{\'e}rez},
  {Gonz{\'a}lez Delgado}, {Garc{\'\i}a-Benito}, {Galbany}, {van de Ven},
  {Jahnke}, {Flores}, {Bland-Hawthorn}, {L{\'o}pez-S{\'a}nchez}, {Stanishev},
  {Miralles-Caballero}, {D{\'\i}az}, {S{\'a}nchez-Blazquez}, {Moll{\'a}},
  {Gallazzi}, {Papaderos}, {Gomes}, {Gruel}, {P{\'e}rez}, {Ruiz-Lara},
  {Florido}, {de Lorenzo-C{\'a}ceres}, {Mendez-Abreu}, {Kehrig}, {Roth},
  {Ziegler}, {Alves}, {Wisotzki}, {Kupko}, {Quirrenbach}, {Bomans}, \& {CALIFA
  Collaboration}}]{Sanchez2013}
{S{\'a}nchez}, S.~F., {Rosales-Ortega}, F.~F., {Jungwiert}, B., {et~al.} 2013,
  \href{https://doi.org/10.1051/0004-6361/201220669}{\aap}, 554, A58,
  [\href{https://arxiv.org/abs/1304.2158}{1304.2158}].

\bibitem[{{S{\'a}nchez} {et~al.}(2021){S{\'a}nchez}, {Walcher},
  {Lopez-Cob{\'a}}, {Barrera-Ballesteros}, {Mej{\'\i}a-Narv{\'a}ez},
  {Espinosa-Ponce}, \& {Camps-Fari{\~n}a}}]{Sanchez2021}
{S{\'a}nchez}, S.~F., {Walcher}, C.~J., {Lopez-Cob{\'a}}, C., {et~al.} 2021,
  \href{https://doi.org/10.22201/ia.01851101p.2021.57.01.01}{\rmxaa}, 57, 3,
  [\href{https://arxiv.org/abs/2009.00424}{2009.00424}].

\bibitem[{{S{\'a}nchez-Bl{\'a}zquez} {et~al.}(2007){S{\'a}nchez-Bl{\'a}zquez},
  {Forbes}, {Strader}, {Brodie}, \& {Proctor}}]{SanchezBlazquez2007}
{S{\'a}nchez-Bl{\'a}zquez}, P., {Forbes}, D.~A., {Strader}, J., {Brodie}, J.,
  \& {Proctor}, R. 2007,
  \href{https://doi.org/10.1111/j.1365-2966.2007.11647.x}{\mnras}, 377, 759,
  [\href{https://arxiv.org/abs/astro-ph/0702572}{astro-ph/0702572}].

\bibitem[{{S{\'a}nchez-Bl{\'a}zquez}
  {et~al.}(2006{\natexlab{a}}){S{\'a}nchez-Bl{\'a}zquez}, {Gorgas}, \&
  {Cardiel}}]{SanchezBlazquez2006c}
{S{\'a}nchez-Bl{\'a}zquez}, P., {Gorgas}, J., \& {Cardiel}, N.
  2006{\natexlab{a}}, \href{https://doi.org/10.1051/0004-6361:20064846}{\aap},
  457, 823, [\href{https://arxiv.org/abs/astro-ph/0604571}{astro-ph/0604571}].

\bibitem[{{S{\'a}nchez-Bl{\'a}zquez}
  {et~al.}(2006{\natexlab{b}}){S{\'a}nchez-Bl{\'a}zquez}, {Gorgas}, {Cardiel},
  \& {Gonz{\'a}lez}}]{SanchezBlazquez2006b}
{S{\'a}nchez-Bl{\'a}zquez}, P., {Gorgas}, J., {Cardiel}, N., \& {Gonz{\'a}lez},
  J.~J. 2006{\natexlab{b}},
  \href{https://doi.org/10.1051/0004-6361:20064845}{\aap}, 457, 809,
  [\href{https://arxiv.org/abs/astro-ph/0604568}{astro-ph/0604568}].

\bibitem[{{S{\'a}nchez-Bl{\'a}zquez} {et~al.}(2014){S{\'a}nchez-Bl{\'a}zquez},
  {Rosales-Ortega}, {M{\'e}ndez-Abreu}, {P{\'e}rez}, {S{\'a}nchez}, {Zibetti},
  {Aguerri}, {Bland-Hawthorn}, {Catal{\'a}n-Torrecilla}, {Cid Fernandes}, {de
  Amorim}, {de Lorenzo-Caceres}, {Falc{\'o}n-Barroso}, {Galazzi}, {Garc{\'\i}a
  Benito}, {Gil de Paz}, {Gonz{\'a}lez Delgado}, {Husemann},
  {Iglesias-P{\'a}ramo}, {Jungwiert}, {Marino}, {M{\'a}rquez}, {Mast},
  {Mendoza}, {Moll{\'a}}, {Papaderos}, {Ruiz-Lara}, {van de Ven}, {Walcher}, \&
  {Wisotzki}}]{SanchezBlazquez2014}
{S{\'a}nchez-Bl{\'a}zquez}, P., {Rosales-Ortega}, F.~F., {M{\'e}ndez-Abreu},
  J., {et~al.} 2014, \href{https://doi.org/10.1051/0004-6361/201423635}{\aap},
  570, A6, [\href{https://arxiv.org/abs/1407.0002}{1407.0002}].

\bibitem[{{S{\'a}nchez-Portal} {et~al.}(2015){S{\'a}nchez-Portal},
  {Pintos-Castro}, {P{\'e}rez-Mart{\'\i}nez}, {Cepa}, {P{\'e}rez Garc{\'\i}a},
  {Dom{\'\i}nguez-S{\'a}nchez}, {Bongiovanni}, {Serra}, {Alfaro}, {Altieri},
  {Arag{\'o}n-Salamanca}, {Balkowski}, {Biviano}, {Bremer}, {Castander},
  {Casta{\~n}eda}, {Castro-Rodr{\'\i}guez}, {Chies-Santos}, {Coia}, {Diaferio},
  {Duc}, {Ederoclite}, {Geach}, {Gonz{\'a}lez-Serrano}, {Haines}, {McBreen},
  {Metcalfe}, {Oteo}, {P{\'e}rez-Fourn{\'o}n}, {Poggianti}, {Polednikova},
  {Ram{\'o}n-P{\'e}rez}, {Rodr{\'\i}guez-Espinosa}, {Santos}, {Smail}, {Smith},
  {Temporin}, \& {Valtchanov}}]{SanchezPortal2015}
{S{\'a}nchez-Portal}, M., {Pintos-Castro}, I., {P{\'e}rez-Mart{\'\i}nez}, R.,
  {et~al.} 2015, \href{https://doi.org/10.1051/0004-6361/201525620}{\aap}, 578,
  A30, [\href{https://arxiv.org/abs/1502.03020}{1502.03020}].

\bibitem[{{Santini} {et~al.}(2017){Santini}, {Fontana}, {Castellano}, {Di
  Criscienzo}, {Merlin}, {Amorin}, {Cullen}, {Daddi}, {Dickinson}, {Dunlop},
  {Grazian}, {Lamastra}, {McLure}, {Micha{\l}owski}, {Pentericci}, \&
  {Shu}}]{Santini2017}
{Santini}, P., {Fontana}, A., {Castellano}, M., {et~al.} 2017,
  \href{https://doi.org/10.3847/1538-4357/aa8874}{\apj}, 847, 76,
  [\href{https://arxiv.org/abs/1706.07059}{1706.07059}].

\bibitem[{{Schaefer} {et~al.}(2017){Schaefer}, {Croom}, {Allen}, {Brough},
  {Medling}, {Ho}, {Scott}, {Richards}, {Pracy}, {Gunawardhana}, {Norberg},
  {Alpaslan}, {Bauer}, {Bekki}, {Bland-Hawthorn}, {Bloom}, {Bryant}, {Couch},
  {Driver}, {Fogarty}, {Foster}, {Goldstein}, {Green}, {Hopkins},
  {Konstantopoulos}, {Lawrence}, {L{\'o}pez-S{\'a}nchez}, {Lorente}, {Owers},
  {Sharp}, {Sweet}, {Taylor}, {van de Sande}, {Walcher}, \&
  {Wong}}]{Schaefer2017}
{Schaefer}, A.~L., {Croom}, S.~M., {Allen}, J.~T., {et~al.} 2017,
  \href{https://doi.org/10.1093/mnras/stw2289}{\mnras}, 464, 121,
  [\href{https://arxiv.org/abs/1609.02635}{1609.02635}].

\bibitem[{{Schaefer} {et~al.}(2019){Schaefer}, {Tremonti}, {Pace}, {Belfiore},
  {Argudo-Fernandez}, {Bershady}, {Drory}, {Jones}, {Maiolino}, {Stark},
  {Wake}, \& {Yan}}]{Schaefer2019}
{Schaefer}, A.~L., {Tremonti}, C., {Pace}, Z., {et~al.} 2019,
  \href{https://doi.org/10.3847/1538-4357/ab43ca}{\apj}, 884, 156,
  [\href{https://arxiv.org/abs/1909.04738}{1909.04738}].

\bibitem[{{Schawinski} {et~al.}(2014){Schawinski}, {Urry}, {Simmons},
  {Fortson}, {Kaviraj}, {Keel}, {Lintott}, {Masters}, {Nichol}, {Sarzi},
  {Skibba}, {Treister}, {Willett}, {Wong}, \& {Yi}}]{Schawinski2014}
{Schawinski}, K., {Urry}, C.~M., {Simmons}, B.~D., {et~al.} 2014,
  \href{https://doi.org/10.1093/mnras/stu327}{\mnras}, 440, 889,
  [\href{https://arxiv.org/abs/1402.4814}{1402.4814}].

\bibitem[{{Scott} {et~al.}(2018){Scott}, {van de Sande}, {Croom}, {Groves},
  {Owers}, {Poetrodjojo}, {D'Eugenio}, {Medling}, {Barat}, {Barone},
  {Bland-Hawthorn}, {Brough}, {Bryant}, {Cortese}, {Foster}, {Green}, {Oh},
  {Colless}, {Drinkwater}, {Driver}, {Goodwin}, {Gunawardhana}, {Federrath},
  {Harischandra}, {Jin}, {Lawrence}, {Lorente}, {Mannering}, {O'Toole},
  {Richards}, {Sanchez}, {Schaefer}, {Sealey}, {Sharp}, {Sweet}, {Taranu}, \&
  {Varidel}}]{Scott2018}
{Scott}, N., {van de Sande}, J., {Croom}, S.~M., {et~al.} 2018,
  \href{https://doi.org/10.1093/mnras/sty2355}{\mnras}, 481, 2299,
  [\href{https://arxiv.org/abs/1808.03365}{1808.03365}].

\bibitem[{{Scoville} {et~al.}(2007){Scoville}, {Aussel}, {Brusa}, {Capak},
  {Carollo}, {Elvis}, {Giavalisco}, {Guzzo}, {Hasinger}, {Impey}, {Kneib},
  {LeFevre}, {Lilly}, {Mobasher}, {Renzini}, {Rich}, {Sanders}, {Schinnerer},
  {Schminovich}, {Shopbell}, {Taniguchi}, \& {Tyson}}]{Scoville2007}
{Scoville}, N., {Aussel}, H., {Brusa}, M., {et~al.} 2007,
  \href{https://doi.org/10.1086/516585}{\apjs}, 172, 1,
  [\href{https://arxiv.org/abs/astro-ph/0612305}{astro-ph/0612305}].

\bibitem[{{S{\'e}rsic}(1963)}]{Sersic1963}
{S{\'e}rsic}, J.~L. 1963, Boletin de la Asociacion Argentina de Astronomia La
  Plata Argentina, 6, 41.

\bibitem[{{Sersic}(1968)}]{Sersic1968}
{Sersic}, J.~L. 1968, {Atlas de Galaxias Australes}

\bibitem[{{Shen} {et~al.}(2003){Shen}, {Mo}, {White}, {Blanton}, {Kauffmann},
  {Voges}, {Brinkmann}, \& {Csabai}}]{Shen2003}
{Shen}, S., {Mo}, H.~J., {White}, S. D.~M., {et~al.} 2003,
  \href{https://doi.org/10.1046/j.1365-8711.2003.06740.x}{\mnras}, 343, 978,
  [\href{https://arxiv.org/abs/astro-ph/0301527}{astro-ph/0301527}].

\bibitem[{{Shin} {et~al.}(2021){Shin}, {Ly}, {Malkan}, {Malhotra}, {de los
  Reyes}, \& {Rhoads}}]{shin2021metal}
{Shin}, K., {Ly}, C., {Malkan}, M.~A., {et~al.} 2021,
  \href{https://doi.org/10.1093/mnras/staa3307}{\mnras}, 501, 2231,
  [\href{https://arxiv.org/abs/1910.10735}{1910.10735}].

\bibitem[{{Silva} \& {Bothun}(1998)}]{Silva1998}
{Silva}, D.~R. \& {Bothun}, G.~D. 1998,
  \href{https://doi.org/10.1086/300642}{\aj}, 116, 2793.

\bibitem[{{Singh} {et~al.}(2019){Singh}, {Gulati}, \& {Bagla}}]{Singh2019}
{Singh}, A., {Gulati}, M., \& {Bagla}, J.~S. 2019,
  \href{https://doi.org/10.1093/mnras/stz2523}{\mnras}, 489, 5582,
  [\href{https://arxiv.org/abs/1909.02744}{1909.02744}].

\bibitem[{{Smith} {et~al.}(2008){Smith}, {Marzke}, {Hornschemeier}, {Bridges},
  {Hudson}, {Miller}, {Lucey}, {V{\'a}zquez}, \& {Carter}}]{Smith2008}
{Smith}, R.~J., {Marzke}, R.~O., {Hornschemeier}, A.~E., {et~al.} 2008,
  \href{https://doi.org/10.1111/j.1745-3933.2008.00469.x}{\mnras}, 386, L96,
  [\href{https://arxiv.org/abs/0803.0327}{0803.0327}].

\bibitem[{{Sobral} {et~al.}(2014){Sobral}, {Best}, {Smail}, {Mobasher},
  {Stott}, \& {Nisbet}}]{Sobral2014}
{Sobral}, D., {Best}, P.~N., {Smail}, I., {et~al.} 2014,
  \href{https://doi.org/10.1093/mnras/stt2159}{\mnras}, 437, 3516,
  [\href{https://arxiv.org/abs/1311.1503}{1311.1503}].

\bibitem[{{Somerville} \& {Primack}(1999)}]{Sommerville1999}
{Somerville}, R.~S. \& {Primack}, J.~R. 1999,
  \href{https://doi.org/10.1046/j.1365-8711.1999.03032.x}{\mnras}, 310, 1087,
  [\href{https://arxiv.org/abs/astro-ph/9802268}{astro-ph/9802268}].

\bibitem[{{Sparre} {et~al.}(2015){Sparre}, {Hayward}, {Springel},
  {Vogelsberger}, {Genel}, {Torrey}, {Nelson}, {Sijacki}, \&
  {Hernquist}}]{Sparre2015}
{Sparre}, M., {Hayward}, C.~C., {Springel}, V., {et~al.} 2015,
  \href{https://doi.org/10.1093/mnras/stu2713}{\mnras}, 447, 3548,
  [\href{https://arxiv.org/abs/1409.0009}{1409.0009}].

\bibitem[{{Speagle} {et~al.}(2014){Speagle}, {Steinhardt}, {Capak}, \&
  {Silverman}}]{Speagle2014}
{Speagle}, J.~S., {Steinhardt}, C.~L., {Capak}, P.~L., \& {Silverman}, J.~D.
  2014, \href{https://doi.org/10.1088/0067-0049/214/2/15}{\apjs}, 214, 15,
  [\href{https://arxiv.org/abs/1405.2041}{1405.2041}].

\bibitem[{{Springel} {et~al.}(2005){Springel}, {White}, {Jenkins}, {Frenk},
  {Yoshida}, {Gao}, {Navarro}, {Thacker}, {Croton}, {Helly}, {Peacock}, {Cole},
  {Thomas}, {Couchman}, {Evrard}, {Colberg}, \& {Pearce}}]{Springel2005}
{Springel}, V., {White}, S. D.~M., {Jenkins}, A., {et~al.} 2005,
  \href{https://doi.org/10.1038/nature03597}{\nat}, 435, 629,
  [\href{https://arxiv.org/abs/astro-ph/0504097}{astro-ph/0504097}].

\bibitem[{{Sreejith} {et~al.}(2023){Sreejith}, {Slosar}, \&
  {Wang}}]{Sreejith2023}
{Sreejith}, S., {Slosar}, A., \& {Wang}, H. 2023,
  \href{https://doi.org/10.48550/arXiv.2310.19605}{arXiv e-prints},
  arXiv:2310.19605, [\href{https://arxiv.org/abs/2310.19605}{2310.19605}].

\bibitem[{{Starck} {et~al.}(2002){Starck}, {Pantin}, \& {Murtagh}}]{Starck2002}
{Starck}, J.~L., {Pantin}, E., \& {Murtagh}, F. 2002,
  \href{https://doi.org/10.1086/342606}{\pasp}, 114, 1051.

\bibitem[{{Steinhauser} {et~al.}(2016){Steinhauser}, {Schindler}, \&
  {Springel}}]{Steinhauser2016}
{Steinhauser}, D., {Schindler}, S., \& {Springel}, V. 2016,
  \href{https://doi.org/10.1051/0004-6361/201527705}{\aap}, 591, A51,
  [\href{https://arxiv.org/abs/1604.05193}{1604.05193}].

\bibitem[{{Stone}(1996)}]{Stone1996}
{Stone}, R. P.~S. 1996, \href{https://doi.org/10.1086/192369}{\apjs}, 107, 423.

\bibitem[{{Strateva} {et~al.}(2001){Strateva}, {Ivezi{\'c}}, {Knapp},
  {Narayanan}, {Strauss}, {Gunn}, {Lupton}, {Schlegel}, {Bahcall}, {Brinkmann},
  {Brunner}, {Budav{\'a}ri}, {Csabai}, {Castander}, {Doi}, {Fukugita},
  {Gy{\H{o}}ry}, {Hamabe}, {Hennessy}, {Ichikawa}, {Kunszt}, {Lamb}, {McKay},
  {Okamura}, {Racusin}, {Sekiguchi}, {Schneider}, {Shimasaku}, \&
  {York}}]{Strateva2001}
{Strateva}, I., {Ivezi{\'c}}, {\v{Z}}., {Knapp}, G.~R., {et~al.} 2001,
  \href{https://doi.org/10.1086/323301}{\aj}, 122, 1861,
  [\href{https://arxiv.org/abs/astro-ph/0107201}{astro-ph/0107201}].

\bibitem[{{Swanson} {et~al.}(2008){Swanson}, {Tegmark}, {Hamilton}, \&
  {Hill}}]{Mangle2008}
{Swanson}, M.~E.~C., {Tegmark}, M., {Hamilton}, A. J.~S., \& {Hill}, J.~C.
  2008, \href{https://doi.org/10.1111/j.1365-2966.2008.13296.x}{\mnras}, 387,
  1391, [\href{https://arxiv.org/abs/0711.4352}{0711.4352}].

\bibitem[{{Takahashi} {et~al.}(2007){Takahashi}, {Shioya}, {Taniguchi},
  {Murayama}, {Ajiki}, {Sasaki}, {Koizumi}, {Nagao}, {Scoville}, {Mobasher},
  {Aussel}, {Capak}, {Carilli}, {Ellis}, {Garilli}, {Giavalisco}, {Guzzo},
  {Hasinger}, {Impey}, {Kitzbichler}, {Koekemoer}, {Le F{\`e}vre}, {Lilly},
  {Maccagni}, {Renzini}, {Rich}, {Sanders}, {Schinnerer}, {Scodeggio},
  {Shopbell}, {Smol{\v{c}}i{\'c}}, {Tribiano}, {Ideue}, \&
  {Mihara}}]{Takahashi2007}
{Takahashi}, M.~I., {Shioya}, Y., {Taniguchi}, Y., {et~al.} 2007,
  \href{https://doi.org/10.1086/518037}{\apjs}, 172, 456,
  [\href{https://arxiv.org/abs/astro-ph/0703065}{astro-ph/0703065}].

\bibitem[{{Tamura} \& {Ohta}(2003)}]{Tamura2003}
{Tamura}, N. \& {Ohta}, K. 2003, \href{https://doi.org/10.1086/376469}{\aj},
  126, 596, [\href{https://arxiv.org/abs/astro-ph/0304404}{astro-ph/0304404}].

\bibitem[{{Tanaka} {et~al.}(2024){Tanaka}, {Shimasaku}, {Tacchella}, {Ando},
  {Ito}, {Yesuf}, \& {Matsui}}]{Tanaka2024}
{Tanaka}, T.~S., {Shimasaku}, K., {Tacchella}, S., {et~al.} 2024,
  \href{https://doi.org/10.1093/pasj/psad076}{\pasj}, 76, 1,
  [\href{https://arxiv.org/abs/2307.14235}{2307.14235}].

\bibitem[{{Tasca} {et~al.}(2017){Tasca}, {Le F{\`e}vre}, {Ribeiro}, {Thomas},
  {Moreau}, {Cassata}, {Garilli}, {Le Brun}, {Lemaux}, {Maccagni},
  {Pentericci}, {Schaerer}, {Vanzella}, {Zamorani}, {Zucca}, {Amorin},
  {Bardelli}, {Cassar{\`a}}, {Castellano}, {Cimatti}, {Cucciati}, {Durkalec},
  {Fontana}, {Giavalisco}, {Grazian}, {Hathi}, {Ilbert}, {Paltani}, {Pforr},
  {Scodeggio}, {Sommariva}, {Talia}, {Tresse}, {Vergani}, {Capak}, {Charlot},
  {Contini}, {de la Torre}, {Dunlop}, {Fotopoulou}, {Guaita}, {Koekemoer},
  {L{\'o}pez-Sanjuan}, {Mellier}, {Salvato}, {Scoville}, {Taniguchi}, \&
  {Wang}}]{Tasca2017}
{Tasca}, L.~A.~M., {Le F{\`e}vre}, O., {Ribeiro}, B., {et~al.} 2017,
  \href{https://doi.org/10.1051/0004-6361/201527963}{\aap}, 600, A110,
  [\href{https://arxiv.org/abs/1602.01842}{1602.01842}].

\bibitem[{{Taylor} {et~al.}(2014){Taylor}, {Mar{\'\i}n-Franch}, {Laporte},
  {Santoro}, {Marrara}, {Cepa}, {Cenarro}, {Chueca}, {Cristobal-Hornillos},
  {Ederoclite}, {Gruel}, {Moles}, {Rueda}, {Rueda}, {Varela}, {Yanes},
  {Benitez}, {Dupke}, {Fern{\'a}ndez-Soto}, {Jorden}, {Lousberg}, {Molino
  Benito}, {Palmer}, {Mendes de Oliveira}, \& {Sodr{\'e}}}]{Taylor2014}
{Taylor}, K., {Mar{\'\i}n-Franch}, A., {Laporte}, R., {et~al.} 2014,
  \href{https://doi.org/10.1142/S2251171713500104}{Journal of Astronomical
  Instrumentation}, 3, 1350010,
  [\href{https://arxiv.org/abs/1301.4175}{1301.4175}].

\bibitem[{{Thomas} {et~al.}(2005){Thomas}, {Maraston}, {Bender}, \& {Mendes de
  Oliveira}}]{Thomas2005}
{Thomas}, D., {Maraston}, C., {Bender}, R., \& {Mendes de Oliveira}, C. 2005,
  \href{https://doi.org/10.1086/426932}{\apj}, 621, 673,
  [\href{https://arxiv.org/abs/astro-ph/0410209}{astro-ph/0410209}].

\bibitem[{{Thomas} {et~al.}(2010){Thomas}, {Maraston}, {Schawinski}, {Sarzi},
  \& {Silk}}]{Thomas2010}
{Thomas}, D., {Maraston}, C., {Schawinski}, K., {Sarzi}, M., \& {Silk}, J.
  2010, \href{https://doi.org/10.1111/j.1365-2966.2010.16427.x}{\mnras}, 404,
  1775, [\href{https://arxiv.org/abs/0912.0259}{0912.0259}].

\bibitem[{{Tinsley}(1981)}]{Tinsley1981}
{Tinsley}, B.~M. 1981, \href{https://doi.org/10.1093/mnras/194.1.63}{\mnras},
  194, 63.

\bibitem[{{Tojeiro} {et~al.}(2017){Tojeiro}, {Eardley}, {Peacock}, {Norberg},
  {Alpaslan}, {Driver}, {Henriques}, {Hopkins}, {Kafle}, {Robotham}, {Thomas},
  {Tonini}, \& {Wild}}]{tojeiro2017}
{Tojeiro}, R., {Eardley}, E., {Peacock}, J.~A., {et~al.} 2017,
  \href{https://doi.org/10.1093/mnras/stx1466}{\mnras}, 470, 3720,
  [\href{https://arxiv.org/abs/1612.08595}{1612.08595}].

\bibitem[{{Tortora} {et~al.}(2010){Tortora}, {Napolitano}, {Cardone},
  {Capaccioli}, {Jetzer}, \& {Molinaro}}]{Tortora2010}
{Tortora}, C., {Napolitano}, N.~R., {Cardone}, V.~F., {et~al.} 2010,
  \href{https://doi.org/10.1111/j.1365-2966.2010.16938.x}{\mnras}, 407, 144,
  [\href{https://arxiv.org/abs/1004.4896}{1004.4896}].

\bibitem[{{Trager} {et~al.}(2000){Trager}, {Faber}, {Worthey}, \&
  {Gonz{\'a}lez}}]{Trager2000}
{Trager}, S.~C., {Faber}, S.~M., {Worthey}, G., \& {Gonz{\'a}lez}, J.~J. 2000,
  \href{https://doi.org/10.1086/301442}{\aj}, 120, 165,
  [\href{https://arxiv.org/abs/astro-ph/0004095}{astro-ph/0004095}].

\bibitem[{{Trager} {et~al.}(1998){Trager}, {Worthey}, {Faber}, {Burstein}, \&
  {Gonz{\'a}lez}}]{Trager1998}
{Trager}, S.~C., {Worthey}, G., {Faber}, S.~M., {Burstein}, D., \&
  {Gonz{\'a}lez}, J.~J. 1998, \href{https://doi.org/10.1086/313099}{\apjs},
  116, 1, [\href{https://arxiv.org/abs/astro-ph/9712258}{astro-ph/9712258}].

\bibitem[{{Trayford} {et~al.}(2016){Trayford}, {Theuns}, {Bower}, {Crain},
  {Lagos}, {Schaller}, \& {Schaye}}]{Trayford2016}
{Trayford}, J.~W., {Theuns}, T., {Bower}, R.~G., {et~al.} 2016,
  \href{https://doi.org/10.1093/mnras/stw1230}{\mnras}, 460, 3925,
  [\href{https://arxiv.org/abs/1601.07907}{1601.07907}].

\bibitem[{{Tremonti} {et~al.}(2004){Tremonti}, {Heckman}, {Kauffmann},
  {Brinchmann}, {Charlot}, {White}, {Seibert}, {Peng}, {Schlegel}, {Uomoto},
  {Fukugita}, \& {Brinkmann}}]{Tremonti2004}
{Tremonti}, C.~A., {Heckman}, T.~M., {Kauffmann}, G., {et~al.} 2004,
  \href{https://doi.org/10.1086/423264}{\apj}, 613, 898,
  [\href{https://arxiv.org/abs/astro-ph/0405537}{astro-ph/0405537}].

\bibitem[{{Trussler} {et~al.}(2020){Trussler}, {Maiolino}, {Maraston}, {Peng},
  {Thomas}, {Goddard}, \& {Lian}}]{Trussler2020}
{Trussler}, J., {Maiolino}, R., {Maraston}, C., {et~al.} 2020,
  \href{https://doi.org/10.1093/mnras/stz3286}{\mnras}, 491, 5406,
  [\href{https://arxiv.org/abs/1811.09283}{1811.09283}].

\bibitem[{{Urrutia} {et~al.}(2019){Urrutia}, {Wisotzki}, {Kerutt}, {Schmidt},
  {Herenz}, {Klar}, {Saust}, {Werhahn}, {Diener}, {Caruana}, {Krajnovi{\'c}},
  {Bacon}, {Boogaard}, {Brinchmann}, {Enke}, {Maseda}, {Nanayakkara},
  {Richard}, {Steinmetz}, \& {Weilbacher}}]{MuseWide2019}
{Urrutia}, T., {Wisotzki}, L., {Kerutt}, J., {et~al.} 2019,
  \href{https://doi.org/10.1051/0004-6361/201834656}{\aap}, 624, A141,
  [\href{https://arxiv.org/abs/1811.06549}{1811.06549}].

\bibitem[{{van de Voort} {et~al.}(2017){van de Voort}, {Bah{\'e}}, {Bower},
  {Correa}, {Crain}, {Schaye}, \& {Theuns}}]{VandeVoort2017}
{van de Voort}, F., {Bah{\'e}}, Y.~M., {Bower}, R.~G., {et~al.} 2017,
  \href{https://doi.org/10.1093/mnras/stw3356}{\mnras}, 466, 3460,
  [\href{https://arxiv.org/abs/1611.03870}{1611.03870}].

\bibitem[{{van den Bosch} {et~al.}(2008){van den Bosch}, {Aquino}, {Yang},
  {Mo}, {Pasquali}, {McIntosh}, {Weinmann}, \& {Kang}}]{Bosch2008}
{van den Bosch}, F.~C., {Aquino}, D., {Yang}, X., {et~al.} 2008,
  \href{https://doi.org/10.1111/j.1365-2966.2008.13230.x}{\mnras}, 387, 79,
  [\href{https://arxiv.org/abs/0710.3164}{0710.3164}].

\bibitem[{{van der Burg} {et~al.}(2018){van der Burg}, {McGee}, {Aussel},
  {Dahle}, {Arnaud}, {Pratt}, \& {Muzzin}}]{vanderburg2018}
{van der Burg}, R. F.~J., {McGee}, S., {Aussel}, H., {et~al.} 2018,
  \href{https://doi.org/10.1051/0004-6361/201833572}{\aap}, 618, A140,
  [\href{https://arxiv.org/abs/1807.00820}{1807.00820}].

\bibitem[{{Veilleux} \& {Osterbrock}(1987)}]{Veilleux1987}
{Veilleux}, S. \& {Osterbrock}, D.~E. 1987,
  \href{https://doi.org/10.1086/191166}{\apjs}, 63, 295.

\bibitem[{Vilella-Rojo {et~al.}(2021)Vilella-Rojo, Logro{\~n}o-Garc{\'\i}a,
  L{\'o}pez-Sanjuan, Viironen, Varela, Moles, Cenarro, Crist{\'o}bal-Hornillos,
  Ederoclite, Hern{\'a}ndez-Monteagudo, {et~al.}}]{vilella2021j}
Vilella-Rojo, G., Logro{\~n}o-Garc{\'\i}a, R., L{\'o}pez-Sanjuan, C., {et~al.}
  2021, arXiv preprint arXiv:2101.04062

\bibitem[{{Vilella-Rojo} {et~al.}(2015){Vilella-Rojo}, {Viironen},
  {L{\'o}pez-Sanjuan}, {Cenarro}, {Varela}, {D{\'\i}az-Garc{\'\i}a},
  {Crist{\'o}bal-Hornillos}, {Ederoclite}, {Mar{\'\i}n-Franch}, \&
  {Moles}}]{VilellaRojo2015}
{Vilella-Rojo}, G., {Viironen}, K., {L{\'o}pez-Sanjuan}, C., {et~al.} 2015,
  \href{https://doi.org/10.1051/0004-6361/201526374}{\aap}, 580, A47,
  [\href{https://arxiv.org/abs/1505.07115}{1505.07115}].

\bibitem[{{Villar} {et~al.}(2008){Villar}, {Gallego}, {P{\'e}rez-Gonz{\'a}lez},
  {Pascual}, {Noeske}, {Koo}, {Barro}, \& {Zamorano}}]{Villar2008}
{Villar}, V., {Gallego}, J., {P{\'e}rez-Gonz{\'a}lez}, P.~G., {et~al.} 2008,
  \href{https://doi.org/10.1086/528942}{\apj}, 677, 169,
  [\href{https://arxiv.org/abs/0712.4150}{0712.4150}].

\bibitem[{{Voigt} \& {Bridle}(2010)}]{Voigt2010}
{Voigt}, L.~M. \& {Bridle}, S.~L. 2010,
  \href{https://doi.org/10.1111/j.1365-2966.2010.16300.x}{\mnras}, 404, 458,
  [\href{https://arxiv.org/abs/0905.4801}{0905.4801}].

\bibitem[{{von der Linden} {et~al.}(2010){von der Linden}, {Wild}, {Kauffmann},
  {White}, \& {Weinmann}}]{vonderLinden2010}
{von der Linden}, A., {Wild}, V., {Kauffmann}, G., {White}, S. D.~M., \&
  {Weinmann}, S. 2010,
  \href{https://doi.org/10.1111/j.1365-2966.2010.16375.x}{\mnras}, 404, 1231,
  [\href{https://arxiv.org/abs/0909.3522}{0909.3522}].

\bibitem[{{Wake} {et~al.}(2017){Wake}, {Bundy}, {Diamond-Stanic}, {Yan},
  {Blanton}, {Bershady}, {S{\'a}nchez-Gallego}, {Drory}, {Jones}, {Kauffmann},
  {Law}, {Li}, {MacDonald}, {Masters}, {Thomas}, {Tinker}, {Weijmans}, \&
  {Brownstein}}]{Wake2017}
{Wake}, D.~A., {Bundy}, K., {Diamond-Stanic}, A.~M., {et~al.} 2017,
  \href{https://doi.org/10.3847/1538-3881/aa7ecc}{\aj}, 154, 86,
  [\href{https://arxiv.org/abs/1707.02989}{1707.02989}].

\bibitem[{{Walcher} {et~al.}(2014){Walcher}, {Wisotzki}, {Bekerait{\'e}},
  {Husemann}, {Iglesias-P{\'a}ramo}, {Backsmann}, {Barrera Ballesteros},
  {Catal{\'a}n-Torrecilla}, {Cortijo}, {del Olmo}, {Garcia Lorenzo},
  {Falc{\'o}n-Barroso}, {Jilkova}, {Kalinova}, {Mast}, {Marino},
  {M{\'e}ndez-Abreu}, {Pasquali}, {S{\'a}nchez}, {Trager}, {Zibetti},
  {Aguerri}, {Alves}, {Bland-Hawthorn}, {Boselli}, {Castillo Morales}, {Cid
  Fernandes}, {Flores}, {Galbany}, {Gallazzi}, {Garc{\'\i}a-Benito}, {Gil de
  Paz}, {Gonz{\'a}lez-Delgado}, {Jahnke}, {Jungwiert}, {Kehrig}, {Lyubenova},
  {M{\'a}rquez Perez}, {Masegosa}, {Monreal Ibero}, {P{\'e}rez}, {Quirrenbach},
  {Rosales-Ortega}, {Roth}, {Sanchez-Blazquez}, {Spekkens}, {Tundo}, {van de
  Ven}, {Verheijen}, {Vilchez}, \& {Ziegler}}]{Walcher2014}
{Walcher}, C.~J., {Wisotzki}, L., {Bekerait{\'e}}, S., {et~al.} 2014,
  \href{https://doi.org/10.1051/0004-6361/201424198}{\aap}, 569, A1,
  [\href{https://arxiv.org/abs/1407.2939}{1407.2939}].

\bibitem[{{Walcher} {et~al.}(2011){Walcher}, {Groves}, {Budav{\'a}ri}, \&
  {Dale}}]{Walcher2011}
{Walcher}, J., {Groves}, B., {Budav{\'a}ri}, T., \& {Dale}, D. 2011,
  \href{https://doi.org/10.1007/s10509-010-0458-z}{\apss}, 331, 1,
  [\href{https://arxiv.org/abs/1008.0395}{1008.0395}].

\bibitem[{{Wang} {et~al.}(2023){Wang}, {Sreejith}, {Lin}, {Ramachandra},
  {Solsar}, \& {Yoo}}]{Wang2023}
{Wang}, H., {Sreejith}, S., {Lin}, Y., {et~al.} 2023,
  \href{https://doi.org/10.21105/astro.2210.01666}{The Open Journal of
  Astrophysics}, 6, 30, [\href{https://arxiv.org/abs/2210.01666}{2210.01666}].

\bibitem[{{Wang} \& {Yang}(2022)}]{Wang2022}
{Wang}, S.-C. \& {Yang}, H. Y.~K. 2022,
  \href{https://doi.org/10.1093/mnras/stac788}{\mnras}, 512, 5100,
  [\href{https://arxiv.org/abs/2201.05298}{2201.05298}].

\bibitem[{{Weaver} {et~al.}(2024){Weaver}, {Cutler}, {Pan}, {Whitaker},
  {Labb{\'e}}, {Price}, {Bezanson}, {Brammer}, {Marchesini}, {Leja}, {Wang},
  {Furtak}, {Zitrin}, {Atek}, {Chemerynska}, {Coe}, {Dayal}, {van Dokkum},
  {Feldmann}, {F{\"o}rster Schreiber}, {Franx}, {Fujimoto}, {Fudamoto},
  {Glazebrook}, {de Graaff}, {Greene}, {Juneau}, {Kassin}, {Kriek}, {Khullar},
  {Maseda}, {Mowla}, {Muzzin}, {Nanayakkara}, {Nelson}, {Oesch}, {Pacifici},
  {Papovich}, {Setton}, {Shapley}, {Shipley}, {Smit}, {Stefanon}, {Taylor},
  {Weibel}, \& {Williams}}]{Weaver2024}
{Weaver}, J.~R., {Cutler}, S.~E., {Pan}, R., {et~al.} 2024,
  \href{https://doi.org/10.3847/1538-4365/ad07e0}{\apjs}, 270, 7,
  [\href{https://arxiv.org/abs/2301.02671}{2301.02671}].

\bibitem[{{Weinmann} {et~al.}(2006){Weinmann}, {van den Bosch}, {Yang}, \&
  {Mo}}]{Weinmann2006}
{Weinmann}, S.~M., {van den Bosch}, F.~C., {Yang}, X., \& {Mo}, H.~J. 2006,
  \href{https://doi.org/10.1111/j.1365-2966.2005.09865.x}{\mnras}, 366, 2,
  [\href{https://arxiv.org/abs/astro-ph/0509147}{astro-ph/0509147}].

\bibitem[{{Wen} {et~al.}(2012){Wen}, {Han}, \& {Liu}}]{Wen2012}
{Wen}, Z.~L., {Han}, J.~L., \& {Liu}, F.~S. 2012,
  \href{https://doi.org/10.1088/0067-0049/199/2/34}{\apjs}, 199, 34,
  [\href{https://arxiv.org/abs/1202.6424}{1202.6424}].

\bibitem[{{Werner} {et~al.}(2022){Werner}, {Hatch}, {Muzzin}, {van der Burg},
  {Balogh}, {Rudnick}, \& {Wilson}}]{Werner2022}
{Werner}, S.~V., {Hatch}, N.~A., {Muzzin}, A., {et~al.} 2022,
  \href{https://doi.org/10.1093/mnras/stab3484}{\mnras}, 510, 674,
  [\href{https://arxiv.org/abs/2111.14624}{2111.14624}].

\bibitem[{{Wester} \& {Dark Energy Survey Collaboration}(2005)}]{Wester2005}
{Wester}, W. \& {Dark Energy Survey Collaboration}. 2005, in Astronomical
  Society of the Pacific Conference Series, Vol. 339, Observing Dark Energy,
  ed. S.~C. {Wolff} \& T.~R. {Lauer}, 152

\bibitem[{{Wetzel} {et~al.}(2012){Wetzel}, {Tinker}, \& {Conroy}}]{Wetzel2012}
{Wetzel}, A.~R., {Tinker}, J.~L., \& {Conroy}, C. 2012,
  \href{https://doi.org/10.1111/j.1365-2966.2012.21188.x}{\mnras}, 424, 232,
  [\href{https://arxiv.org/abs/1107.5311}{1107.5311}].

\bibitem[{{Wetzel} {et~al.}(2015){Wetzel}, {Tollerud}, \& {Weisz}}]{Wetzel2015}
{Wetzel}, A.~R., {Tollerud}, E.~J., \& {Weisz}, D.~R. 2015,
  \href{https://doi.org/10.1088/2041-8205/808/1/L27}{\apjl}, 808, L27,
  [\href{https://arxiv.org/abs/1503.06799}{1503.06799}].

\bibitem[{{Whitaker} {et~al.}(2011){Whitaker}, {Labb{\'e}}, {van Dokkum},
  {Brammer}, {Kriek}, {Marchesini}, {Quadri}, {Franx}, {Muzzin}, {Williams},
  {Bezanson}, {Illingworth}, {Lee}, {Lundgren}, {Nelson}, {Rudnick}, {Tal}, \&
  {Wake}}]{Whitaker2011}
{Whitaker}, K.~E., {Labb{\'e}}, I., {van Dokkum}, P.~G., {et~al.} 2011,
  \href{https://doi.org/10.1088/0004-637X/735/2/86}{\apj}, 735, 86,
  [\href{https://arxiv.org/abs/1105.4609}{1105.4609}].

\bibitem[{{White} {et~al.}(1987){White}, {Davis}, {Efstathiou}, \&
  {Frenk}}]{White1987}
{White}, S. D.~M., {Davis}, M., {Efstathiou}, G., \& {Frenk}, C.~S. 1987,
  \href{https://doi.org/10.1038/330451a0}{\nat}, 330, 451.

\bibitem[{{Wilkinson} {et~al.}(2015){Wilkinson}, {Maraston}, {Thomas},
  {Coccato}, {Tojeiro}, {Cappellari}, {Belfiore}, {Bershady}, {Blanton},
  {Bundy}, {Cales}, {Cherinka}, {Drory}, {Emsellem}, {Fu}, {Law}, {Li},
  {Maiolino}, {Masters}, {Tremonti}, {Wake}, {Wang}, {Weijmans}, {Xiao}, {Yan},
  {Zhang}, {Bizyaev}, {Brinkmann}, {Kinemuchi}, {Malanushenko}, {Malanushenko},
  {Oravetz}, {Pan}, \& {Simmons}}]{Wilkinson2015}
{Wilkinson}, D.~M., {Maraston}, C., {Thomas}, D., {et~al.} 2015,
  \href{https://doi.org/10.1093/mnras/stv301}{\mnras}, 449, 328,
  [\href{https://arxiv.org/abs/1503.01124}{1503.01124}].

\bibitem[{{Williams} {et~al.}(2009){Williams}, {Quadri}, {Franx}, {van Dokkum},
  \& {Labb{\'e}}}]{Williams2009}
{Williams}, R.~J., {Quadri}, R.~F., {Franx}, M., {van Dokkum}, P., \&
  {Labb{\'e}}, I. 2009,
  \href{https://doi.org/10.1088/0004-637X/691/2/1879}{\apj}, 691, 1879,
  [\href{https://arxiv.org/abs/0806.0625}{0806.0625}].

\bibitem[{{Wolf} {et~al.}(2001){Wolf}, {Dye}, {Kleinheinrich}, {Meisenheimer},
  {Rix}, \& {Wisotzki}}]{COMBO2001}
{Wolf}, C., {Dye}, S., {Kleinheinrich}, M., {et~al.} 2001,
  \href{https://doi.org/10.1051/0004-6361:20011142}{\aap}, 377, 442,
  [\href{https://arxiv.org/abs/astro-ph/0012474}{astro-ph/0012474}].

\bibitem[{{Wolf} {et~al.}(2003{\natexlab{a}}){Wolf}, {Meisenheimer}, {Rix},
  {Borch}, {Dye}, \& {Kleinheinrich}}]{Wolf2003}
{Wolf}, C., {Meisenheimer}, K., {Rix}, H.~W., {et~al.} 2003{\natexlab{a}},
  \href{https://doi.org/10.1051/0004-6361:20021513}{\aap}, 401, 73,
  [\href{https://arxiv.org/abs/astro-ph/0208345}{astro-ph/0208345}].

\bibitem[{{Wolf} {et~al.}(2003{\natexlab{b}}){Wolf}, {Wisotzki}, {Borch},
  {Dye}, {Kleinheinrich}, \& {Meisenheimer}}]{COMBO2003}
{Wolf}, C., {Wisotzki}, L., {Borch}, A., {et~al.} 2003{\natexlab{b}},
  \href{https://doi.org/10.1051/0004-6361:20030990}{\aap}, 408, 499,
  [\href{https://arxiv.org/abs/astro-ph/0304072}{astro-ph/0304072}].

\bibitem[{{Woo} {et~al.}(2017){Woo}, {Carollo}, {Faber}, {Dekel}, \&
  {Tacchella}}]{Woo2017}
{Woo}, J., {Carollo}, C.~M., {Faber}, S.~M., {Dekel}, A., \& {Tacchella}, S.
  2017, \href{https://doi.org/10.1093/mnras/stw2403}{\mnras}, 464, 1077,
  [\href{https://arxiv.org/abs/1607.06091}{1607.06091}].

\bibitem[{{Woo} {et~al.}(2015){Woo}, {Dekel}, {Faber}, \& {Koo}}]{Woo2015}
{Woo}, J., {Dekel}, A., {Faber}, S.~M., \& {Koo}, D.~C. 2015,
  \href{https://doi.org/10.1093/mnras/stu2755}{\mnras}, 448, 237,
  [\href{https://arxiv.org/abs/1406.5372}{1406.5372}].

\bibitem[{{Woo} {et~al.}(2013){Woo}, {Dekel}, {Faber}, {Noeske}, {Koo},
  {Gerke}, {Cooper}, {Salim}, {Dutton}, {Newman}, {Weiner}, {Bundy}, {Willmer},
  {Davis}, \& {Yan}}]{woo2013}
{Woo}, J., {Dekel}, A., {Faber}, S.~M., {et~al.} 2013,
  \href{https://doi.org/10.1093/mnras/sts274}{\mnras}, 428, 3306,
  [\href{https://arxiv.org/abs/1203.1625}{1203.1625}].

\bibitem[{{Worthey}(1994)}]{Worthey1994}
{Worthey}, G. 1994, \href{https://doi.org/10.1086/192096}{\apjs}, 95, 107.

\bibitem[{{Worthey}(1999)}]{Worthey1999}
{Worthey}, G. 1999, in Astronomical Society of the Pacific Conference Series,
  Vol. 192, Spectrophotometric Dating of Stars and Galaxies, ed. I.~{Hubeny},
  S.~{Heap}, \& R.~{Cornett}, 283

\bibitem[{{Wright} {et~al.}(2022){Wright}, {Lagos}, {Power}, {Stevens},
  {Cortese}, \& {Poulton}}]{Wright2022}
{Wright}, R.~J., {Lagos}, C. d.~P., {Power}, C., {et~al.} 2022,
  \href{https://doi.org/10.1093/mnras/stac2042}{\mnras}, 516, 2891,
  [\href{https://arxiv.org/abs/2205.08414}{2205.08414}].

\bibitem[{{Wu} {et~al.}(2005){Wu}, {Shao}, {Mo}, {Xia}, \& {Deng}}]{Wu2005}
{Wu}, H., {Shao}, Z., {Mo}, H.~J., {Xia}, X., \& {Deng}, Z. 2005,
  \href{https://doi.org/10.1086/427821}{\apj}, 622, 244,
  [\href{https://arxiv.org/abs/astro-ph/0404226}{astro-ph/0404226}].

\bibitem[{{Wuyts} {et~al.}(2013){Wuyts}, {F{\"o}rster Schreiber}, {Nelson},
  {van Dokkum}, {Brammer}, {Chang}, {Faber}, {Ferguson}, {Franx}, {Fumagalli},
  {Genzel}, {Grogin}, {Kocevski}, {Koekemoer}, {Lundgren}, {Lutz}, {McGrath},
  {Momcheva}, {Rosario}, {Skelton}, {Tacconi}, {van der Wel}, \&
  {Whitaker}}]{wuyts2013}
{Wuyts}, S., {F{\"o}rster Schreiber}, N.~M., {Nelson}, E.~J., {et~al.} 2013,
  \href{https://doi.org/10.1088/0004-637X/779/2/135}{\apj}, 779, 135,
  [\href{https://arxiv.org/abs/1310.5702}{1310.5702}].

\bibitem[{{Yasuda} {et~al.}(2001){Yasuda}, {Fukugita}, {Narayanan}, {Lupton},
  {Strateva}, {Strauss}, {Ivezi{\'c}}, {Kim}, {Hogg}, {Weinberg}, {Shimasaku},
  {Loveday}, {Annis}, {Bahcall}, {Blanton}, {Brinkmann}, {Brunner}, {Connolly},
  {Csabai}, {Doi}, {Hamabe}, {Ichikawa}, {Ichikawa}, {Johnston}, {Knapp},
  {Kunszt}, {Lamb}, {McKay}, {Munn}, {Nichol}, {Okamura}, {Schneider},
  {Szokoly}, {Vogeley}, {Watanabe}, \& {York}}]{Yasuda2001}
{Yasuda}, N., {Fukugita}, M., {Narayanan}, V.~K., {et~al.} 2001,
  \href{https://doi.org/10.1086/322093}{\aj}, 122, 1104,
  [\href{https://arxiv.org/abs/astro-ph/0105545}{astro-ph/0105545}].

\bibitem[{{Yoon} {et~al.}(2023){Yoon}, {Kim}, \& {Ko}}]{Yoon2023}
{Yoon}, Y., {Kim}, J.-W., \& {Ko}, J. 2023,
  \href{https://doi.org/10.3847/1538-4357/acfed5}{\apj}, 957, 59,
  [\href{https://arxiv.org/abs/2310.07498}{2310.07498}].

\bibitem[{{York} {et~al.}(2000){York}, {Adelman}, {Anderson}, {Anderson},
  {Annis}, {Bahcall}, {Bakken}, {Barkhouser}, {Bastian}, {Berman}, {Boroski},
  {Bracker}, {Briegel}, {Briggs}, {Brinkmann}, {Brunner}, {Burles}, {Carey},
  {Carr}, {Castander}, {Chen}, {Colestock}, {Connolly}, {Crocker}, {Csabai},
  {Czarapata}, {Davis}, {Doi}, {Dombeck}, {Eisenstein}, {Ellman}, {Elms},
  {Evans}, {Fan}, {Federwitz}, {Fiscelli}, {Friedman}, {Frieman}, {Fukugita},
  {Gillespie}, {Gunn}, {Gurbani}, {de Haas}, {Haldeman}, {Harris}, {Hayes},
  {Heckman}, {Hennessy}, {Hindsley}, {Holm}, {Holmgren}, {Huang}, {Hull},
  {Husby}, {Ichikawa}, {Ichikawa}, {Ivezi{\'c}}, {Kent}, {Kim}, {Kinney},
  {Klaene}, {Kleinman}, {Kleinman}, {Knapp}, {Korienek}, {Kron}, {Kunszt},
  {Lamb}, {Lee}, {Leger}, {Limmongkol}, {Lindenmeyer}, {Long}, {Loomis},
  {Loveday}, {Lucinio}, {Lupton}, {MacKinnon}, {Mannery}, {Mantsch}, {Margon},
  {McGehee}, {McKay}, {Meiksin}, {Merelli}, {Monet}, {Munn}, {Narayanan},
  {Nash}, {Neilsen}, {Neswold}, {Newberg}, {Nichol}, {Nicinski}, {Nonino},
  {Okada}, {Okamura}, {Ostriker}, {Owen}, {Pauls}, {Peoples}, {Peterson},
  {Petravick}, {Pier}, {Pope}, {Pordes}, {Prosapio}, {Rechenmacher}, {Quinn},
  {Richards}, {Richmond}, {Rivetta}, {Rockosi}, {Ruthmansdorfer}, {Sandford},
  {Schlegel}, {Schneider}, {Sekiguchi}, {Sergey}, {Shimasaku}, {Siegmund},
  {Smee}, {Smith}, {Snedden}, {Stone}, {Stoughton}, {Strauss}, {Stubbs},
  {SubbaRao}, {Szalay}, {Szapudi}, {Szokoly}, {Thakar}, {Tremonti}, {Tucker},
  {Uomoto}, {Vanden Berk}, {Vogeley}, {Waddell}, {Wang}, {Watanabe},
  {Weinberg}, {Yanny}, {Yasuda}, \& {SDSS Collaboration}}]{York2000}
{York}, D.~G., {Adelman}, J., {Anderson}, John~E., J., {et~al.} 2000,
  \href{https://doi.org/10.1086/301513}{\aj}, 120, 1579,
  [\href{https://arxiv.org/abs/astro-ph/0006396}{astro-ph/0006396}].

\bibitem[{{Zahid} {et~al.}(2012){Zahid}, {Dima}, {Kewley}, {Erb}, \&
  {Dav{\'e}}}]{zahid2012census}
{Zahid}, H.~J., {Dima}, G.~I., {Kewley}, L.~J., {Erb}, D.~K., \& {Dav{\'e}}, R.
  2012, \href{https://doi.org/10.1088/0004-637X/757/1/54}{\apj}, 757, 54,
  [\href{https://arxiv.org/abs/1207.5509}{1207.5509}].

\bibitem[{{Zhang} {et~al.}(2023){Zhang}, {Li}, {Leja}, {Whitaker}, {Nersesian},
  {Bezanson}, \& {van der Wel}}]{Zhang2023}
{Zhang}, J., {Li}, Y., {Leja}, J., {et~al.} 2023,
  \href{https://doi.org/10.3847/1538-4357/acd84a}{\apj}, 952, 6,
  [\href{https://arxiv.org/abs/2211.10450}{2211.10450}].

\bibitem[{{Zheng} {et~al.}(2017){Zheng}, {Wang}, {Ge}, {Mao}, {Li}, {Li}, {Mo},
  {Goddard}, {Bundy}, {Li}, {Nair}, {Lin}, {Long}, {Riffel}, {Thomas},
  {Masters}, {Bizyaev}, {Brownstein}, {Zhang}, {Law}, {Drory}, {Roman Lopes},
  \& {Malanushenko}}]{Zheng2017}
{Zheng}, Z., {Wang}, H., {Ge}, J., {et~al.} 2017,
  \href{https://doi.org/10.1093/mnras/stw3030}{\mnras}, 465, 4572,
  [\href{https://arxiv.org/abs/1612.01523}{1612.01523}].

\bibitem[{{Zibetti} \& {Gallazzi}(2022)}]{Zibetti2022}
{Zibetti}, S. \& {Gallazzi}, A.~R. 2022,
  \href{https://doi.org/10.1093/mnras/stac370}{\mnras}, 512, 1415,
  [\href{https://arxiv.org/abs/2202.03975}{2202.03975}].

\bibitem[{{Zibetti} {et~al.}(2017){Zibetti}, {Gallazzi}, {Ascasibar},
  {Charlot}, {Galbany}, {Garc{\'\i}a Benito}, {Kehrig}, {de
  Lorenzo-C{\'a}ceres}, {Lyubenova}, {Marino}, {M{\'a}rquez}, {S{\'a}nchez},
  {van de Ven}, {Walcher}, \& {Wisotzki}}]{Zibetti2017}
{Zibetti}, S., {Gallazzi}, A.~R., {Ascasibar}, Y., {et~al.} 2017,
  \href{https://doi.org/10.1093/mnras/stx251}{\mnras}, 468, 1902,
  [\href{https://arxiv.org/abs/1701.06570}{1701.06570}].

\bibitem[{{Zinger} {et~al.}(2018){Zinger}, {Dekel}, {Kravtsov}, \&
  {Nagai}}]{Zinger2018}
{Zinger}, E., {Dekel}, A., {Kravtsov}, A.~V., \& {Nagai}, D. 2018,
  \href{https://doi.org/10.1093/mnras/stx3329}{\mnras}, 475, 3654.

\end{thebibliography}
\newpage 
\begin{spacing}{1.5}
\end{spacing}

	\begin{appendices}
		\appendixpage
		\noappendicestocpagenum
		\addappheadtotoc
        \chapter{AMICO versions}
\label{appendix:AMICOversions}

\begin{figure}
    \centering
    \includegraphics[width=\textwidth]{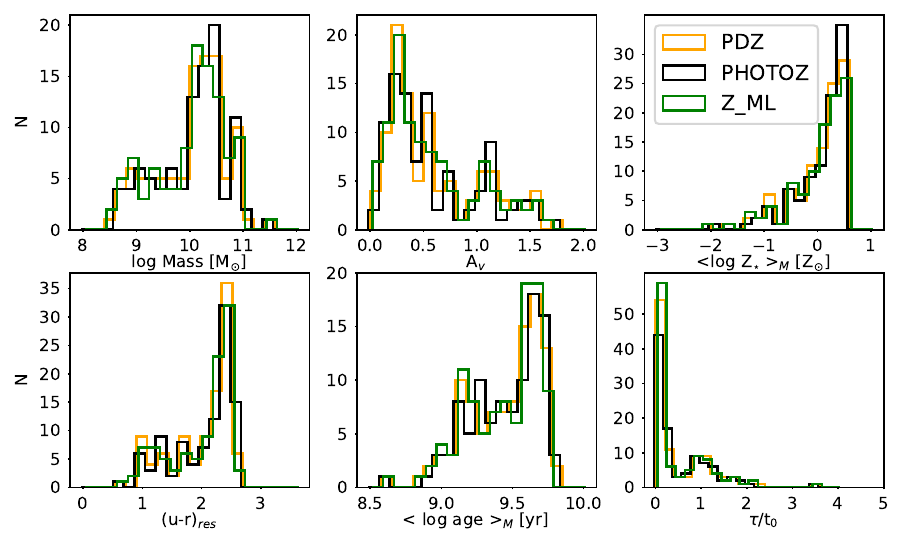}
    \caption{Stellar population properties of the galaxies belonging to the cluster, using different AMICO catalogues.}
    \label{fig:AMICO_comparison_05}
\end{figure}

As mentioned in Sect.\ref{sec:galaxymembers}, at the moment of writing of the paper \cite{Julio2022}, in which Chapter~\ref{chapter:cluster} is based, there were three versions of the code AMICO under study. The difference among them resides in the redshift used for the computation of the amplitude, the association probability, and the rest of the parameters. One of these versions uses the best \photozbest{} (the absolute maximum of the probability density function of the redshift or $z\mathrm{PDF}$) as input, however, other version uses $z_{\mathrm{ml}}$ (the median redshift of the $z\mathrm{PDF}$) and the last one uses the $z\mathrm{PDF}$ itself (see \citealt{HC2021} for more information about the redshifts).  The final version  (including a group catalogue for \mjp) was finally published in \citep{Maturi2023}. Since those results and the details about AMICO escape the scope of this work, we refer the reader to that work (althoud a summary is provided in Sect.~\ref{sec:mjp:AMICO}) as well as to the works \cite{2005A&A...436...37M, AMICO}, and restrict ourselves here to the study of the stellar population properties.

The parameter  we used to select our set of galaxies is the association probability. When changing the redshift,  the association probability is affected as well. As a consequence, some galaxies may be considered members or not of the cluster. In fact, the total number of galaxies changes (99 galaxies using the \photozbest{}, 95 using the $z_{\mathrm{ml}}$ and 114 using the $z\mathrm{PDF}$. Not only does the number of galaxies change, but also the set of the common galaxies between two different versions of the code does not equal the smallest set of the two, although they are very similar overall. There are 84 galaxies in common among the three sets, and the larger discrepancies are  12 galaxies that appear with \photozbest{} and $z\mathrm{PDF}$ but not with $z_\mathrm{ml}$, and other 12 galaxies that only appear with $z\mathrm{PDF}$ .

In Fig. \ref{fig:AMICO_comparison_05}, we show the distribution of the stellar properties obtained with the three different versions. These distributions are practically the same, and the differences that appear are mainly due to the different number of galaxies in each set, but the whole image of the cluster remains the same. This is an important result for us, since our results remain valid independently of the final version that may be chosen, as well as for AMICO, since it proves that the cluster catalogue is robust regardless of the redshift definition being used.

\chapter{Comparison of \jp \ data with GMOS spectroscopy.}
\label{appendix:GMOS}

\begin{figure*}%
    \centering
    \subfloat{{\includegraphics[width=0.45\textwidth]{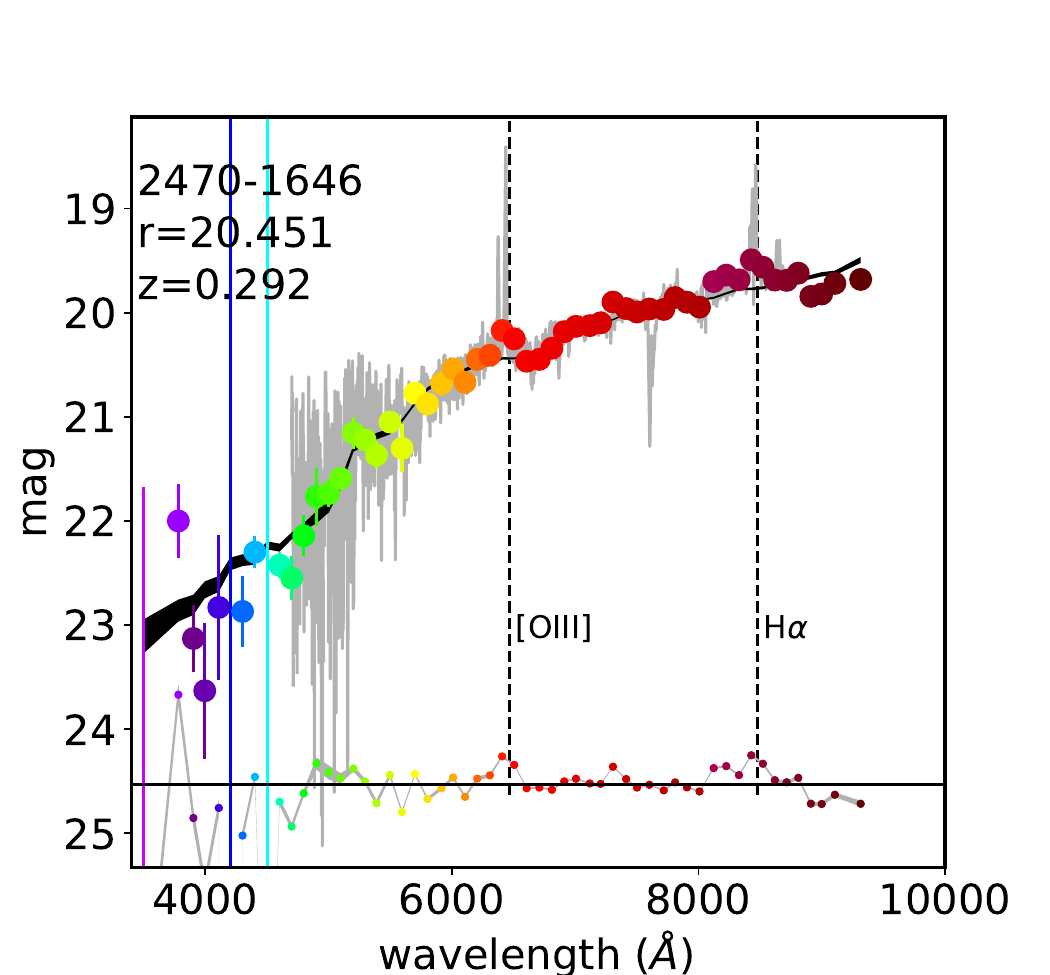} }}%
    \qquad
    \subfloat{{\includegraphics[width=0.45\textwidth]{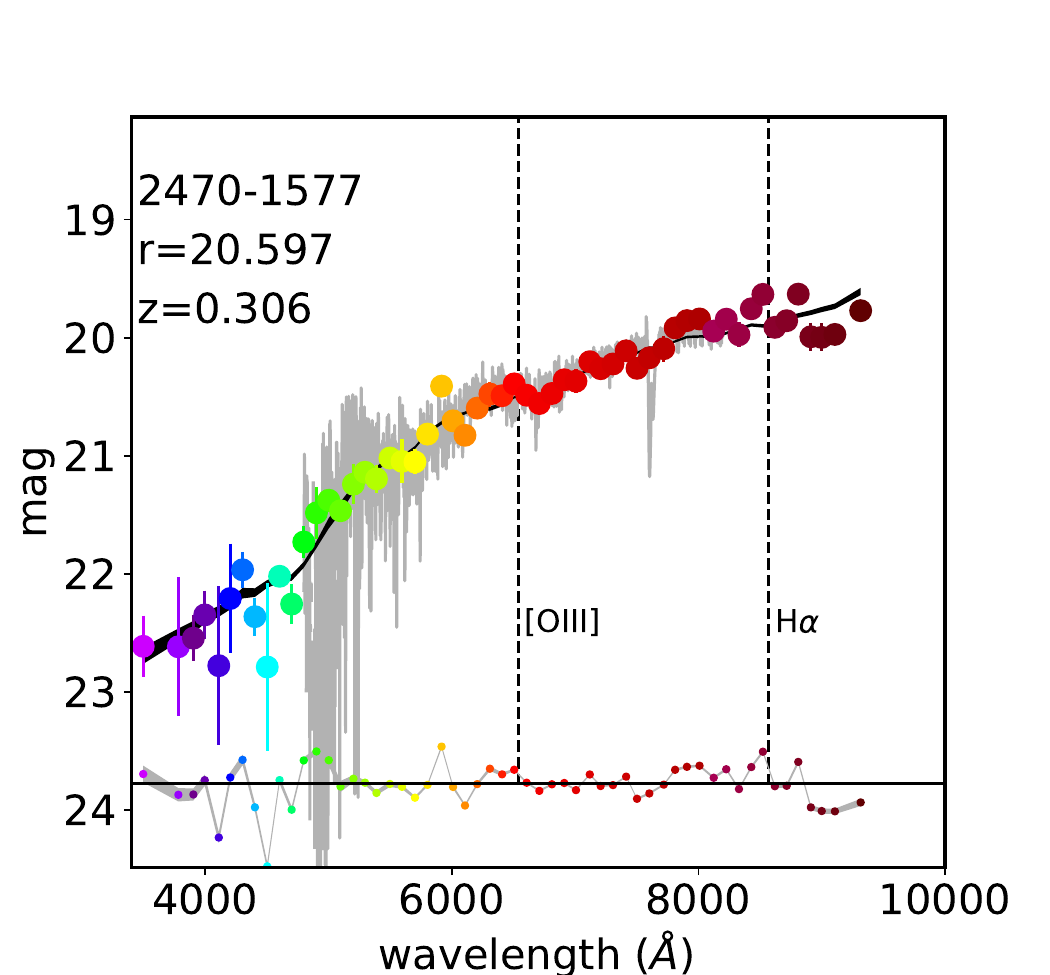} }}%
    \caption[Comparison among the \js \ and the GMOS spectra of two galaxies belonging to the cluster]{Comparison among the \js \ and the GMOS spectra of two galaxies belonging to the cluster. Colour points represent the \js{}. Grey lines represent the GMOS spectra. The black band shows the magnitudes of the mean model $\pm$ one $\sigma$ uncertainty level. The difference between the model and the mean model fitted magnitudes are plotted as a small coloured points around the black bottom line. Vertical, black dashed lines show the wavelengths corresponding to [OIII] and H$\alpha$ emission lines.}%
    \label{fig:GMOS_JPAS}%
\end{figure*}

\begin{table*}[]
    \centering
    \caption[Results of the emission line classification method]{Results of the emission line classification method. First column shows the \jp \ ID of the galaxies with H$\alpha$ emission in GMOS. Second and third columns show if the galaxy was classified as an emission line one with the median error method and the ANN method, respectively. Fourth and fifth columns show some comments on why each method failed to classify the galaxy properly. }
    
    \begin{tabular}{p{0.15\textwidth} p{0.1\textwidth} p{0.1\textwidth} p{0.25\textwidth} p{0.25\textwidth}}
         \hline
         \hline
         Galaxies with H$\alpha$ emission in GMOS & Median error classification & ANN classification & Median error notes&  ANN notes\\
         \hline
         2470-1168 & Yes & Yes& - & -\\ 
         2470-494 & Yes & Yes& - & -\\ 
         2470-536 & ? & ?& J-spectra shows no emission & Unable to calculate values\\ 
         2470-1646 & Yes & Yes& - & - \\
         2470-1744 & No & No& Classified as line-emission when using the spectroscopic redshift & EW$_{\mathrm{min}}$=35.31, EW=2.44$\pm$6.32\\ 
         2470-2129 & Yes & Yes& - & -\\ 
         2470-2328 & Yes & Yes& - & -\\ 
         2470-2524 & Yes & Yes& - & -\\ 
         2470-2401 & Yes & Yes& - & -\\
         2470-2667 & ? & ?& J-spectra shows no emission & Unable to calculate values\\ 
         2470-1920 & Yes & Yes& - & -\\ 
         2470-1625 & Yes & Yes& - & -\\ 
         2470-2403 & Yes & Yes& - & -\\ 
         \hline
    \end{tabular}
    \label{tab:GMOS_ha_emission_summary}
\end{table*}

We carried out a spectroscopic follow up of the cluster with the Gemini Multi Object Spectrometers \citep[GMOS][]{Hook2004} mounted on the Gemini North telescope (Gemini program ID: GN-2020A-DD-203, PI: Carrasco). In total, we measured the spectroscopic redshifts for 53 galaxies observed with GMOS, of which 38 galaxies are members of the cluster. Figure~\ref{fig:GMOS_JPAS} shows a comparison among the \js{} and the spectra obtained from GMOS for two galaxies belonging to the cluster. This comparison shows the power of \jp{} photometric system to provide information equivalent to the spectroscopy data.
We have already shown that \js{} can retrieve the stellar population properties with similar precision to spectroscopic datasets with S/N$\geq$10 \citep{Rosa2021}. With respect to GMOS data, \js{} have a better S/N ratio and it covers a larger wavelength range that is not affected by calibration issues at the wavelengths limits. Furthermore, many of the GMOS spectra do not cover H$\alpha$ wavelength range; and it is in only seven of these galaxies that  H$\alpha$ is observed. The right panel of  Fig.~\ref{fig:GMOS_JPAS} shows an example of a galaxy with measured H$\alpha$ emission through the ANN, and clearly detected in the \js, that is not covered in the GMOS spectrum. In addition, \mjp{} data allowed us to observe most of the galaxies of the cluster brighter than 22.5 magnitude in the \rb{} band. In contrast, MOS spectroscopy is limited by the minimum distance between fibers, which prevent the simultaneous observations of galaxies that are close in the sky, and some fibers are contaminated by several close objects. Thus, \jp \ data is more suitable than GMOS spectra for our analysis. 

In addition to the galaxies in the clusters, our GMOS observations include more galaxies of \mjp{}. To test our ELG detection methods, we included all the galaxies from \mjp{} that were observed in the $H\alpha$ wavelength range. This includes seven spectra of the galaxies belonging to the cluster plus another six galaxies outside the cluster that show H$\alpha$ emission in their GMOS spectra.  Table~\ref{tab:GMOS_ha_emission_summary} summarises the results obtained with both methods for these galaxies. We note that none of the methods classify as ELG any of the galaxies that show no H$\alpha$ emission in their spectra (covering the corresponding wavelength). The different criteria for red and blue galaxies is shown to be required in this table. If we choose the same $\sigma=1$ detection level for blue or red galaxies, we find that red galaxies that actually show no emission line are classified as ELG.  If we choose the same $\sigma=3$ detection level, we end up selecting too few blue ELG which do in fact show H$\alpha$ emission (seen in the GMOS spectra). This is a consequence of the different brightness of red and blue galaxies: the mean and standard deviation values of the \rb \ magnitudes is $20.37 \pm 0.82$~mag for red galaxies and $21.58 \pm 0.89$~mag for blue galaxies. This also produces the difference between the S/N ratios at the H$\alpha$ wavelength range for the red and blue galaxies. The median value of the S/N of the three filters closer to H$\alpha$ at the cluster redshift is 14.67 for red galaxies and 5.27 for blue galaxies, which is almost three times better for red galaxies than for blue galaxies.

\chapter{ELG classification}
\label{sec:ELG_class}

\begin{figure*}
    \centering
    \includegraphics[width=0.95\textwidth]{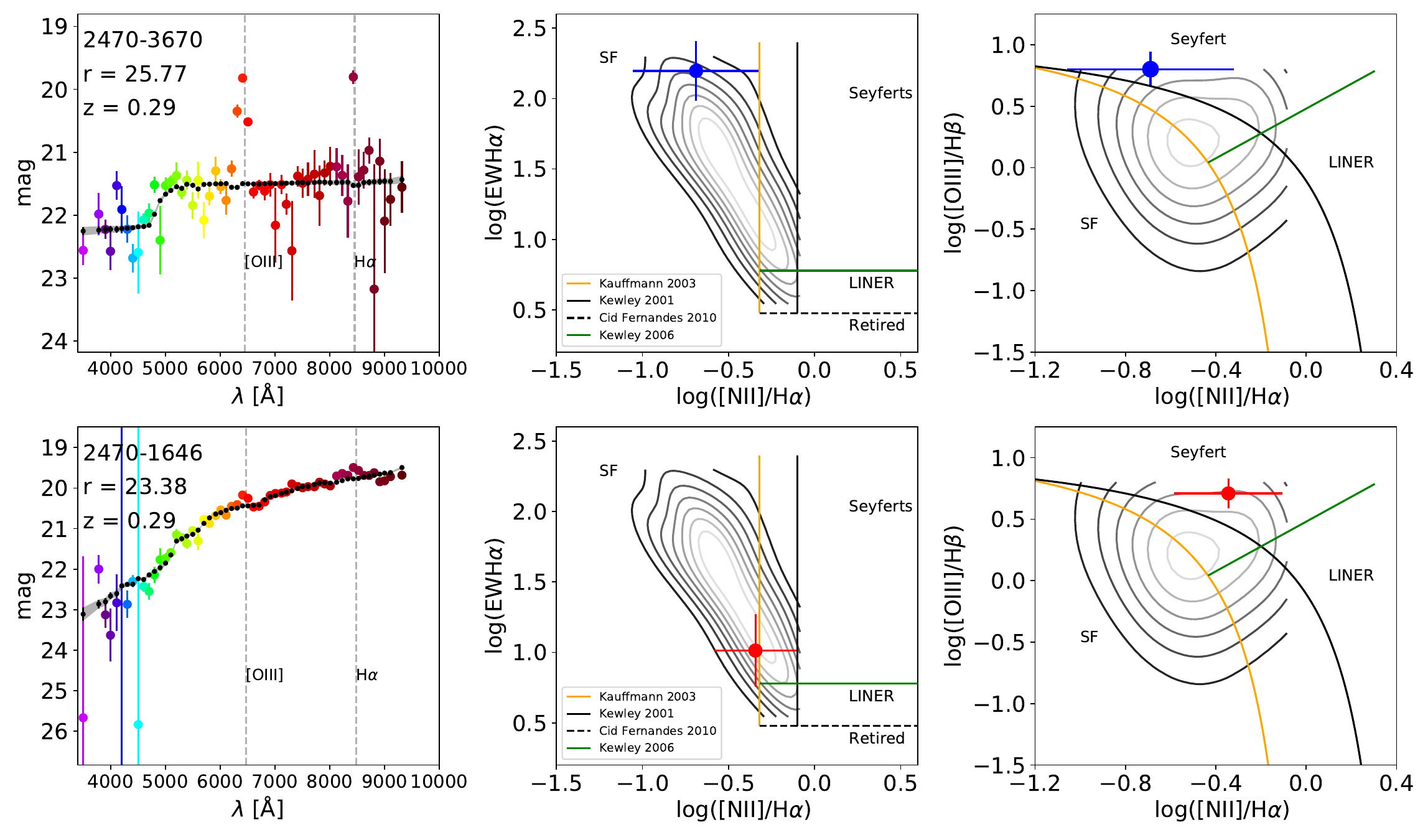}
    \caption[Example of a star-forming galaxy (top row) and a Seyfert galaxy (bottom row)]{Example of a star-forming galaxy (top row) and a Seyfert galaxy (bottom row). From left to right, the panels in each galaxy show the spectra and its position in the WHAN and BPT diagrams (where the contours represent this work's galaxy density in each diagram). The solid orange and black lines represent the \cite{WHAN_1} transposition of the \cite{Kauffmann2003} and \cite{Kewley2001} SF-AGN distiction criteria, and the green solid line represents the transposition of the \cite{Kewley2006} made by \cite{WHAN_1}. The dashed black line represents the distinction between retired galaxies and LINERs \citep{WHAN_2}.}
    \label{fig:Examples_AGN}
\end{figure*}

\begin{longtable}{c c c c }

    \hline
    \hline
        ID & WHAN & BPT & Final classification  \\
        \hline
2470-1030 & SF-Seyfert & Seyfert & AGN\\
2470-1117 & SF & SF-Seyfert & SF\\
2470-1168 & SF & SF & SF\\
2470-1174 & SF & SF-Seyfert & SF\\
2470-1205 & SF-Seyfert & Seyfert & AGN\\
2470-1287 & SF & SF-Seyfert & SF\\
2470-1344 & SF & SF & SF\\
2470-1376 & SF & SF-Seyfert & SF\\
2470-1401 & SF & SF & SF\\
2470-1457 & SF & SF & SF\\
2470-1478 & SF-Seyfert & SF-Seyfert & SF-AGN?\\
2470-1506 & SF & SF & SF\\
2470-1587 & SF-Seyfert & Seyfert & AGN\\
2470-1593 & SF-LINER & Seyfert & AGN\\
2470-1646 & SF & Seyfert & AGN\\
2470-1650 & SF & SF-Seyfert & SF\\
2470-1695 & LINER & SF-LINER & AGN\\
2470-1757 & SF & SF-Seyfert & SF\\
2470-1771 & LINER & LINER & AGN\\
2470-1789 & SF & SF & SF\\
2470-1827 & SF & SF & SF\\
2470-1941 & SF & SF & SF\\
2470-2129 & SF & SF-Seyfert & SF\\
2470-2328 & SF & SF-Seyfert & SF\\
2470-2350 & SF & SF & SF\\
2470-2446 & SF & SF & SF\\
2470-2493 & SF & SF & SF\\
2470-2524 & SF & SF & SF\\
2470-2693 & SF-Seyfert & SF-LINER & SF-AGN?\\
2470-2791 & SF & SF & SF\\
2470-2799 & SF & SF & SF\\
2470-2832 & SF-LINER & Seyfert & AGN\\
2470-2910 & SF & SF-Seyfert & SF\\
2470-2949 & SF & SF & SF\\
2470-2964 & SF-LINER & Seyfert & AGN\\
2470-3255 & SF-Seyfert & SF-LINER & SF-AGN?\\
2470-3345 & SF & SF & SF\\
2470-3670 & SF & Seyfert & SF\\
2470-3712 & SF-Seyfert & Seyfert & AGN\\
2470-3848 & SF-Seyfert & SF-Seyfert & SF-AGN?\\
2470-4414 & SF & SF & SF\\
2470-4691 & SF-Seyfert & Seyfert & AGN\\
2470-492 & SF & SF & SF\\
2470-494 & SF & SF & SF\\
2470-5523 & SF & SF & SF\\
2470-587 & SF & SF & SF\\
2470-701 & SF & SF & SF\\
2470-734 & SF-Seyfert & SF-LINER & SF-AGN?\\
2470-823 & SF-Seyfert & Seyfert & AGN\\
\hline

    \caption[Classification of the ELG population attending to the WHAN and BPT diagrams]{Classification of the ELG population attending to the WHAN and BPT diagrams. First column shows the ID of the galaxy. Second and third columns show the  classification on the WHAN and BPT diagrams, respectively. Fourth column shows our final consideration using both diagrams and taking into account the errors of the predictions. Question marks indicate the possibility of the galaxy belonging to such category, but express our insecurity due to errors being to big, the spectrum being too noisy or to a great discrepancy in the WHAN and BPT classes.}
    \label{tab:AGN}
\end{longtable}

In Sect.~\ref{sec:AGN}, we describe how we classified the ELG into SF galaxies and AGNs using the WHAN and BPT diagrams, and we explain that it is difficult to uniquely classify each galaxy. Figure~\ref{fig:Examples_AGN} illustrates this difficulty and the uncertainty  that we face in this classification. 

Figure~\ref{fig:Examples_AGN} shows two galaxies that are classified as SF in the WHAN diagram and as Seyfert galaxies in the BPT. The \js{} of the galaxy 2470--3670 shows that it is probably an extreme emission line galaxy \citep[EELG, see ][]{Iglesias2022}, due to its weak continuum and its strong emission lines. However, its Seyfert classification is poorly constrained by the [NII]/H$\alpha$ ratio and error. However, its probability of being a SF galaxy in the WHAN diagram is equal to 1, and so, this galaxy is classified as SF. On the contrary, 2470--1646 is classified as well-constrained in the BPT as a Seyfert. In addition, its GMOS spectrum shows [OIII]/H$\beta$ and [NII] /H$\alpha$ ratios of Seyfert galaxies. In the WHAN diagram, the \js{} data place it close to the SF region, and the probability of being classified as a SF is 0.55. A deep inspection of the \js{} images at the $H\alpha$ and close continuum filters shows that it is a spiral galaxy with extended H$\alpha$ emission that could be produced by the AGN and young stars. This galaxy is finally classified as an AGN, although it has SF properties. 

These cases motivate our simpler classification according with their probability in the WHAN diagram as SF or BPT diagram as AGN. Other cases that are between the 'composite' regions in the WHAN or BPT are classified as SF-AGN.  Table~ \ref{tab:AGN} shows the position on each diagram for every galaxy and the final classification given to it, attending to the criteria described in Sect.~\ref{sec:AGN}.

	\end{appendices}
\end{document}